\documentclass[vecphys]{svmono}

\usepackage{makeidx}         
\usepackage{graphicx}        

\usepackage{multicol}        
\usepackage[bottom]{footmisc}
\usepackage{eqalignno}

\usepackage{setspace}
\usepackage{fancyhdr}%
{%
}%
\usepackage{hyperref}
\usepackage[all]{hypcap}

\makeindex             
\usepackage{color}
\usepackage{lipsum}
\usepackage{setspace}
\font\titolo=cmbx12
 2%
\def\txt{\textstyle}%

\def\Ba   {{\mbox{\boldmath$ \alpha$}}}
\def\Bb   {{\mbox{\boldmath$ \beta$}}}

\def\Bd   {{\mbox{\boldmath$ \delta$}}}

\def\Bh   {{\mbox{\boldmath$ \eta$}}}

\def\Bx   {{\mbox{\boldmath$ \xi$}}}

\def\Bp   {{\mbox{\boldmath$ \pi$}}}

\def\Bff  {{\mbox{\boldmath$ \varphi$}}}

\def\BF   {{\mbox{\boldmath$ \Phi$}}}

\def\BDpr {{\mbox{\boldmath$ \partial$}}}

\newdimen\xshift \newdimen\xwidth \newdimen\yshift \newdimen\ywidth
\def\fline{\hbox to\hsize}
\def\ins#1#2#3{\vbox to0pt{\kern-#2pt\hbox{\kern#1pt #3}\vss}\nointerlineskip}

\def\eqfig#1#2#3#4#5{
\par\xwidth=#1pt \xshift=\hsize \advance\xshift
by-\xwidth \divide\xshift by 2
\yshift=#2pt \divide\yshift by 2
\fline{\hglue\xshift \vbox to #2pt{\vfil
#3 \includegraphics{figures/#4.eps}
}\hfill\raise\yshift\hbox{#5}}}

\def\8{\write12}

\let\a=\alpha \let\b=\beta  \let\g=\gamma  \let\d=\delta \let\e=\varepsilon
\let\z=\zeta  \let\h=\eta   \let\th=\theta \let\k=\kappa \let\l=\lambda
\let\m=\mu    \let\n=\nu    \let\x=\xi     \let\p=\pi    \let\r=\varrho
\let\s=\sigma \let\t=\tau   \let\f=\varphi \let\ch=\chi
\let\ps=\psi   \let\o=\omega \let\ch=\chi
\let\G=\Gamma \let\D=\Delta  \let\Th=\Theta\let\L=\Lambda \let\X=\Xi
\let\P=\Pi    \let\Si=\Sigma \let\F=\Phi    \let\Ps=\Psi
\let\O=\Omega 

\def\media#1{{\langle#1\rangle}}
\let\dpr=\partial
\def\tende#1{\,\vtop{\ialign{##\crcr\rightarrowfill\crcr
 \noalign{\kern-1pt\nointerlineskip} \hskip3.pt${\scriptstyle
 #1}$\hskip3.pt\crcr}}\,}
\def\otto{\,{\kern-1.truept\leftarrow\kern-5.truept\to\kern-1.truept}\,}
\def\fra#1#2{{#1\over#2}}

\def\AA{{\cal A}}\def\BB{{\cal B}}\def\CC{{\cal C}}\def\DD{{\cal D}}
\def\EE{{\cal E}}\def\FF{{\cal F}}\def\HH{{\cal H}}
\def\II{{\cal I}}\def\KK{{\cal K}}\def\LL{{\cal L}}
\def\NN{{\cal N}}\def\PP{{\cal P}}
\def\TT{{\cal T}}

\def\V#1{{\bf#1}}%
\def\defi{\,{\buildrel def\over=}\,}%
\def\lhs{{\it l.h.s.}\ }\def\rhs{{\it r.h.s.}\ }%
\def\ig{\int}\def\io{\infty}%
\def\*{{\vskip2mm}}\def\0{\noindent}%
\def\lis#1{\overline{#1}}%
\def\ap{{\it a priori}\ }

\def\pagina{\vfill\eject}

\renewcommand{\theequation}{\arabic{chapter}.\arabic{section}.\arabic{equation}}%
\def\be{\begin{equation}}%
\def\ee{\end{equation}}%
\def\wt{\widetilde}
\def\wh{\widehat}
\def\tto{\Rightarrow}
\def\iniz{\setcounter{equation}{0}\sectionmark{%
\ifodd\thepage\SEC\hfill\else\SEC\hfill\fi}}

\def\ie{{\it i.e.\ }}\def\etc{{\it etc.}}\def\eg{{\it  e.g.\ }}
\def\T#1{{#1_{\kern-3pt\lower7pt\hbox{$\widetilde{}$}}\kern3pt}}
\def\W#1{#1_{\kern-3pt\lower7.5pt\hbox{$\widetilde{}$}}\kern2pt\,}
\def\VV#1{{\underline #1}_{\kern-3pt%
\lower7pt\hbox{$\widetilde{}$}}\kern3pt\,}

\def\ifnextchar#1#2#3{\let\tempe #1\def\tempa{#2}\def\tempb{#3}\futurelet
\tempc\ifnch}
\def\ifnch{\ifx\tempc\tempe\let\tempd\tempa\else\let\tempd\tempb\fi\tempd}
\def\gobble#1{}
\font\tengr=grreg10
\font\eightgr=grreg8
\def\greekmode{%
\catcode`\<=13
\catcode`\>=13
\catcode`\'=11
\catcode`\`=11
\catcode`\~=11
\catcode`\"=11
\catcode`\|=11
\lccode`\<=`\<%
\lccode`\>=`\>%
\lccode`\'=`\'%
\lccode`\`=`\`%
\lccode`\~=`\~%
\lccode`\"=`\"%
\lccode`\|=`\|%
\tengr\def\bf{\tengrbf}
}
\newcount\vwl
\newcount\acct
\def\lt{<}

{
  \greekmode
  \gdef>{\ifnextchar `{\expandafter\smoothgrave\gobble}{\char\lq\>}}
  \gdef<{\ifnextchar `{\expandafter\roughgrave\gobble}{\char\lq\<}}
  \gdef\smoothgrave#1{\acct=\rq137 \vwl=\lq#1 \dobreathinggrave}
  \gdef\roughgrave#1{\acct=\rq103 \vwl=\lq#1 \dobreathinggrave}
  \gdef\dobreathinggrave{\ifnum\vwl\lt\rq140    
    \char\the\acct\char\the\vwl\else\expandafter\testiota\fi}
      \gdef\testiota{\ifnextchar |{\addiota\doaccent\gobble}{\doaccent}}
        \gdef\addiota{\ifnum\vwl=\lq a\vwl=\rq370
            \else\ifnum\vwl=\lq h\vwl=\rq371 \else\vwl=\rq372 \fi\fi}
              \gdef\doaccent{\accent\the\acct \char\the\vwl\relax}
              }

\newif\ifgreek\greekfalse

\def\begingreek{\bgroup\greektrue\greekmode}
\def\endgreek{\egroup}

\let\math=$
{\catcode`\$=13
}

\relax
\begingreek

\endgreek
\relax

\def\bgr{\begingreek}
\def\egr{\endgreek}

\newcounter{appendice}

\def\equ#1{(\ref{#1})}
\def\Eq#1{\label{#1}}
\def\bea{\begin{eqnarray}}
\def\eea{\end{eqnarray}}

\def\uu{{\bf u}}\def\kk{{\bf k}}\def\pp{{\bf p}}
\def\xx{{\bf x}}\def\hh{{\bf h}}\def\qq{{\bf q}}
\def\yy{{\bf y}}

\def\Cite#1{\cite{#1}\write13{{#1}@{\thepage}}}
\def\annotaa#1{{\footnote{NoA: #1}}}
\def\alert#1{{\color{red}#1}}

\begin{document}
\frontmatter
\pagenumbering{Roman}

\author{Giovanni Gallavotti}
\title{\bf\color{red}%
Nonequilibrium and Irreversibility}
\maketitle

\0Giovanni Gallavotti\\
INFN, Accademia dei Lincei\\
Universit\`a di Roma ``La Sapienza''\\
Pl. Moro 2\\
00185, Roma, Italy

\0e-mail: giovanni.gallavotti@roma1.infn.it\\
web: https://ipparco.roma1.infn.it/pagine/2024.html

\vglue1cm
\0{\small\it “This is a preprint of the following work: Giovanni
Gallavotti, Nonequilibrium and Irreversibility, 2024, submitted to
Springer. It is the version of the author’s manuscript prior to
acceptance for publication and has not undergone editorial and/or
peer review on behalf of the Publisher.  The final authenticated
version will be available online, at the Spriner-Nature web-site,
if, and when, accepted.}

\vfill\eject

\vglue1cm
.\kern6cm\vbox{\halign{\bf#&\bf#&\bf#\cr
A\kern.2cm& Daniela\kern.2cm&\&\kern.2cm Barbara\cr
\&\kern.2cm& Camomilla\kern.2cm&\&\kern.2cm Olim\cr}
}
 
\vfill\eject
\phantom{.}
\vfill\eject



%
%
%

         
\0{\titolo Preface to the second edition}
\vskip1cm

The book has been entirely revised and includes several
corrections. I kept the informal style, as a choice to avoid that
first readers about chaos give up considering the proposed
theories as too mathematical.

The first Chapter gives an historical perspective and a
motivation for the classic works partially translated in
Ch.\ref{Ch6}. The ``heat theorem'' section \ref{sec:IV-1} is new
and presents in modern form the proof of Boltzmann in Clausius'
version: it replaces Boltzmann's elementary example, now shifted
to appendix (\ref{appA}).  In the historical part the
interpretation of Boltzmann's example on the heat theorem in
Sec.\,(\ref{sec:III-6}) is updated and includes the new results.

Chapters 2,3,4 are an attempt to a consistent presentation of the
main ideas and tools that were developed since the 1970's toward
the intrepretation of the new experimental works and simulations
involving chaotic motions. The ideas are founded on the ground
breaking work by Sinai, Bowen and Ruelle and on the need to
interpret the results of the simulations performed on the new
computer hardware since it became available to ``every one''. I
have tried to stress, while illustrating several examples, that
stationary non equilibrium and equilibrium can be viewed from a
unifying viewpoint. Many changes have been apported: in the years
since the first edition I realized the need of a substantial
revision: still I refrained from adopting a really formal
style. Formal expositions exist and are always quoted: but my
intention was to introduce the problems while communicating the
surprise that I felt at the unification of basic theoretical
ideas of Statistical Mechanics, Dynamical Systems, Fluids, Probability
Theory and Computer simulations.

The main changes affect part of Chapter 4, Chapter 5 (on
conjectures and suggestions), and the Appendices.  I kept 
the heuristic and phenomenological character, with several
variations due to the evolution of the subjects since the first
edition: several sections have been rewritten and the order of
presentation has been changed. The long appendices on the BBGKY
hierarchy, whose purpose was to formulate, as an open problem,
some difficulties on its use to describe a nonequilibrium
stationary state, have been eliminated because the approach
resulted, in the meantime, too ambitious (\ie the problem did not
progress). Likewise the appendices
on the pendulum subject to noise, have also been suppressed: the
proposal was to present a possible path to the solution of a problem on
the convergence of a formal series representing the stationary
state. Since then, new results have been obtained following a
different path. The theory of Lyapunov exponents pairing,
appendix \ref{appI}, has been written following the original
derivations and only mentioning the simpler, successive, ones.
An appendix on the large deviations theorem for SRB distributions
is added for completeness, \ref{appN}.

The part dedicated to the fluids (viscous, incompressible and
with periodic boundary) in Sec. 4 is deeply changed and takes
into account recent ideas, on the regularized Navier-Stokes
equations: with entirely new conjectures (and related
simulations) on the properties that might survive the
regularization removal.

The applications to fluid mechanics, in Chapter 5 (Sec.,4,5,6)
and appendices J,K,L also replace and update the old
versions. Comparison between the Navier-Stokes equation and
reversible equations associated with it and with the Euler
equation, is illustrated via several conjectures, or heuristic
arguments: with special attention to the possible equivalence
between irreversible and reversible equations.

The continuing debate about reversibility and irreversible
behaviour is repeatedly commented, making also use of heuristic
digressions.

Irreversibility in systems governed by reversible
equations and reversibility in systems governed by irreversible
equations are interesting when the equations model the same
physical system: the resulting conflict is discussed when
possible, not only in fluids but also in mechanical 
or more general dynamical systems.

The above are changes which are remarkable at first sight: but
innumerable changes have been implemented where I felt that the
previous version was incomplete or misleading or, simply, I
thought it could be improved.

There are several sources of errors hidden, in the literature, in
fast applications of mathematical statements, for instance in
applications of the fluctuation theorem: I have tried to put them
in evidence, and to make them more clear by frequent reference to
other book sections where the same problems arise.

The main message that I would like to deliver is that this work is
intended to stress, via an informal or heuristic style, the
breakthrough proposal by Ruelle to unify the study of stationary
nonequilibrium and of equilibrium via the paradigm offered
by the simplest chaotic systems (namely the Anosov systems): seen
as playing for disordered motion the role that
the simplest ordered systems play for ordered motions (namely the
systems of harmonic oscillators, \ie systems animated by quasi
periodic motions, or more generally by integrable systems).

\vskip0.5cm

\hfill\halign{\hglue7.8cm#\hfill&#\hfill\cr
{\it Giovanni Gallavotti}&\cr
Roma,\ 5 December 2024&\cr}

\pagina

\kern-2cm
\0{\titolo Preface to the first edition}
\vskip1cm

.
\0{\it Every hypothesis must derive indubitable results from
mechanically well-defined assumptions by mathematically correct methods. If
the results agree with a large series of facts, we must be content, even if
the true nature of facts is not revealed in every respect. No one
hypothesis has hitherto attained this last end, the Theory of Gases not
excepted, \rm Boltzmann, \Cite{Bo909}[p.536,\#112].\vfil}

\kern10mm In recent years renewed interest grew about the problems of
nonequilibrium statistical mechanics. I think that this has been
stimulated by the new research made possible by the availability of
simple and efficient computers and of the simulations they make
possible.

The possibility and need of performing systematic studies has naturally led
to concentrate efforts in understanding the properties of states which are
in stationary nonequilibrium: thus establishing a clear separation between
properties of evolution towards stationarity (or equilibrium) and
properties of the stationary states themselves: a distinction which until
the 1970's was rather blurred.

A system is out of equilibrium if the microscopic evolution involves non
conservatives forces or interactions with external particles that can be
modeled by or identified with dissipative phenomena which forbid indefinite
growth of the system energy.  The result is that nonzero currents are
generated in the system with matter or energy flowing and dissipation being
generated. In essentially all problems the regulating action of the
external particles can be reliably modeled by non Hamiltonian forces.

Just as in equilibrium statistical mechanics the stationary states are
identified by the time averages of the observables. As familiar in measure
theory, the collections of averages of any kind (time average, phase space
average, counting average ...) are in general identified with probability
distributions on the space of the possible configurations of a system; thus
such probability distributions yield the natural formal setting for the
discussions with which we shall be concerned here. Stationary states will
be identified with probability distributions on the microscopic
configurations, {\it i.e.} on phase space which, of course, have to be
invariant under time evolution.

A first problem is that in general there will be a very large number
of invariant distributions: which ones correspond to stationary states
of a given assembly of atoms and molecules? {\it i.e.} which ones
lead to averages of observables which can be identified with time
averages under the time evolution of the system?

This has been a key question already in equilibrium: Clausius, Boltzmann,
Maxwell (and others) considered it reasonable to think that the microscopic
evolution had the property that, in the course of time, every configuration
was reached from motions starting from any other.

Analyzing this question has led to many developments since the early

stationary distributions that it had become clear would be
generically associated with any mildly chaotic dynamical system.

It seems that this fact is not (yet) universally recognized and the SRB
distribution is often shrugged away as a mathematical nicety. %
\footnote{\tiny It is possible to find in the literature heroic
efforts to avoid dealing with the SRB distributions by
essentially attempting to do what is actually done (and better)
in the original works.\label{heroic}\vfil}
 I dedicate a large part of Chapter \ref{Ch2} to trying to
 illustrate the physical meaning of the SRB distribution relating
 it to what has been called (by Cohen and me) ``chaotic
 hypothesis''. It is also an assumption which requires
 understanding and some open mindedness: personally I have been
 influenced by the ergodic hypothesis (of which it is an
 extension to non equilibrium phenomena) in the original form of
 Boltzmann, and for this reason I have proposed here rather large
 portions of the original papers by Boltzmann and Clausius, see
 Chapter \ref{Ch6}. The reader who is perplex about the chaotic
 hypothesis can find some relief in reading the mentioned
 classics and their even more radical treatment, of what today
 would be chaotic motions, via periodic motions. Finally the role
 of dissipation (in time reversible systems) is discussed and its
 remarkable physical meaning of entropy production rate is
 illustrated (another key discovery due to the numerical
 simulations with finite reversible thermostats mentioned above).

In Chapter \ref{Ch3} theoretical consequences of the chaotic
hypothesis are discussed: the leading ideas are drawn again from
the classic works of Boltzmann see
Sec.\,\ref{sec:II-6},\ref{sec:XII-6}\,: the SRB distribution
properties can conveniently be made visible if the Boltzmann
viewpoint of discreteness of phase space is adopted. It leads to
a combinatorial interpretation of the SRB distribution which
unifies equilibrium and non equilibrium relating them through the
coarse graining of phase space made possible by the chaothic
hypothesis (CH). The key question of whether it is possible to define
entropy of a stationary non equilibrium state is discussed in
some detail making use of the coarse grained phase space:
concluding that while it may be impossible to define a non
equilibrium entropy it is possible to define the entropy
production rate and a function that in equilibrium is the
classical entropy while out of equilibrium is ``just'' a Lyapunov
function (one of many) maximal at the SRB distribution.

In Chapter \ref{Ch4} several general theoretical consequences of the
chaotic hypothesis are enumerated and illustrated: particular attention is
dedicated to the role of the time reversal symmetry and its implications on
the universal (\ie widely model independent) theory of large fluctuations:
the fluctuation theorem by Cohen and myself, Onsager reciprocity and
Green-Kubo formula, the extension of the Onsager-Machlup theory of patterns
fluctuations, and an attempt to study the corresponding problems in a
quantum context. Universality is, of course, important because it partly
frees us from the non physical nature of the finite thermostats.

In Chapter \ref{Ch5} I try to discuss some special concrete applications,
just as a modest incentive for further research. Among them, however, there
is still a general question that I propose and to which I attempt a
solution: it is to give a quantitative criterion for measuring the degree of
irreversibility of a process, \ie to give a measure of the quasi static
nature of a process.  

In general I have avoided technical material preferring heuristic arguments
to mathematical proofs: however, when possible references have been given
for the readers who find some interest in the topics treated and want to
master the (important) details. The same applies to the appendices (A-K).

In Chapter \ref{Ch6} several classic papers are presented, all but one in
partial translation from the original German language. These papers
illustrate my personal route to studying the birth of ergodic theory and
its relevance for statistical mechanics,
\Cite{Ga989}\Cite{Ga995}\Cite{Ga995a}\Cite{Ga000}\Cite{Ga005a},
and, implicitly, provide motivation for the choices (admittedly
very personal) made in the first five chapters and in the
Appendices.

The Appendices A-P contain a few complements, and the remaining
appendices deal with technical problems which are still
unsolved. Appendix M gives an example of the work that may be
necessary in actual constructions of stationary states in the
apparently simple case of a forced pendulum in presence of noise.
Appendices Q-T discuss an attempt ({\it work in progress}) at
studying a stationary case of BBGKY hierarchy with no random
forces but out of equilibrium. I present this case because I
think that is it instructive although the results are deeply
unsatisfactory: it is part unpublished work in strict
collaboration with G. Gentile and A. Giuliani.  \*

The booklet represents a viewpoint, my personal, and does not
pretend to be exhaustive: many important topics have been left
out (like \Cite{BDGJL01}\Cite{DLS002}\Cite{GDL010}\Cite{BK013},
 just to mention a few works that have
led to further exciting developments). I have tried to present a consistent
theory including some of its unsatisfactory aspects.  \*

The Collected papers of Boltzmann, Clausius, Maxwell are freely
available: about Boltzmann I am grateful (and all of us are) to
Wolfgang Reiter, in Vienna, for actively working to obtain that
{\it \"Osterreichische Zentralbibliothek f\"ur Physik} undertook
and accomplished the task of digitizing the ``Wissenschaftliche
Abhandlungen'' and the ``Popul\"are Schriften'' now at

{\tt https://phaidra.univie.ac.at/detail/o:63668}

{\tt https://phaidra.univie.ac.at/o:63637}

\0respectively, making them freely available.
\*

\0{\it Acknowledgments:} I am indebted to D. Ruelle for his
teaching and examples. I am indebted to E.G.D. Cohen for his
constant encouragement and stimulation as well as, of course, for
his collaboration in the developments in our common works and for
supplying many ideas and problems.  To Guido Gentile and
Alessandro Giuliani for their close collaboration in an attempt
to study heat conduction in a gas of hard spheres. Finally a
special thank is to Professor Wolf Beigl\"ock for his constant
interest and encouragement.

\vspace{0.3cm}
\vspace{0.3cm}
\halign{#\hfill&\ #\hfill\cr
{\it Giovanni Gallavotti}&\ \cr
Roma,\ 28 October 2013&\ \cr}
\*

This is \ {\it Version {\bf 3.0}: 19 July 2024}
\hfill\eject

\tableofcontents
\setcounter{page}{1}
\mainmatter
\setcounter{page}{1}
\pagenumbering{arabic}

\chapter{Equilibrium}
\label{Ch1}

\chaptermark{\ifodd\thepage
Equilibrium\hfill\else\hfill 
Equilibrium\fi}
\kern2.3cm

\section{Many particles systems: kinematics, timing}
\def\SEC{Many particles systems: kinematics, timing}
\label{sec:I-1}\iniz
\lhead{\small\ref{sec:I-1}.\ \SEC}

Mechanical systems in interaction with thermostats will be often
modeled by evolution equations describing the time evolution of
the point $x=( X,\dot
X)=(x_1,\ldots,x_N,\dot{x}_1,\ldots,\dot{x}_N)\in R^{6N}$
representing positions and velocities of $N$ particles in the
ambient space $R^3$.

It will be useful to distinguish between positions and velocities
of the $N_0$ particles in the ``system proper'' (or ``test system'' as in
\Cite{FV963}), represented by $( X^{(0)},\dot
X^{(0)})=(x^{(0)}_1,\ldots,x^{(0)}_{N_0},$ $\dot{ x}^{(0)}_1,\ldots,\dot{
  x}^{(0)}_{N_0})$ and by the positions and velocities of the $N_j$
particles in the various thermostats $(X^{(j)},\dot
X^{(j)})=(x^{(j)}_1,\ldots,x^{(j)}_{N_j},\dot{
  x}^{(j)}_1,\ldots,\dot{x}^{(j)}_{N_j})$, $j=1,\ldots,m$: for a total of
$N=\sum_{j=0}^m N_j$ particles.

Time evolution\index{time evolution} is traditionally described by {\it
  differential equations}:
\be\dot x=F(x)\label{e1.1.1}\ee
whose solutions (given initial data $x(0)=(X(0),\dot X(0))$)
yield a trajectory $t\to x(t)= (X(t),\dot X(t))$ representing
motions developing in continuous time $t\in (-\infty,+\infty)$ in
``phase space'' (\ie the space where the coordinates of $x$
dwell). 

A description of an evolution in terms of {\it maps} gives the
state of the motion at discrete times $t_n$. The point
representing the state at time $t$ is denoted $S_tx$ in
the continuous time representation or, at the $n$-th observation, $S^n\xi$ in
the discrete time  representation. The maps $S_t,S^n$ verify
$S_tS_{t'}=S_{t+t'}$ and $S^nS^m=S^{n+m}$ for all $t,t'\in
(-\infty,\infty)$  and all integers $n,m$: evolutions defined
only for $t\ge0,n\ge0$ will also be considered but explicitly declared.
 
The connection between the two representations of motions is illustrated by
means of the following notion of {\it timing event\index{timing event}}.

Physical observations are always performed at discrete times:
{\it i.e.}  when some special, prefixed, {\it timing} event
occurs, typically when the state of the system is in a surface
$\Xi\subset R^{6N}$ and triggers the action of a ``measurement
apparatus'', {\it e.g.}  shooting a picture after noticing that a
chosen observable assumes a prefixed value. If $\Xi$ comprises
the collection of the timing events, {\it i.e.}  of the states
$\xi$ of the system which induce the act of measurement, the
system motion can also be represented as a map $\xi\to S\xi$
defined on $\Xi$.\,\footnote{\tiny Sometimes the observations can
be triggered by a clock arm indicating a chosen position on the
dial: in this case the phase space will coincide with
$R^{6N}$. But in what follows will be considered measurements
triggered by some observable taking a prefixed value, unless
otherwise stated.}

For this reason mathematical models are often maps which associate with a
timing event $\xi$, {\it i.e.} a point $\xi$ in the manifold $\X$ of the
measurement inducing events, the next timing event $S\xi$.

Here $x,\xi$ will not be necessarily points in $R^{6N}$ because
it is possible, and sometimes convenient, to use other
coordinates: therefore, more generally, $x,\xi$ will be points on
general manifolds $M$ or $\X$ called the {\it phase space}, or
space of the states. The dimension of the space $\Xi$ of the
timing events is one unit less than that of $M$: because, by
definition, timing events correspond to a prefixed value of some
observable $f(x)$; and it is convenient to allow also cases when
motion is defined just as a map $S$ on $\X$, without following timed
observations of a motion in continuous time.  Furthermore
sometimes the system admits conservation laws, or constraints,
allowing a description of the motions in terms of fewer
coordinates than the ones defining the system.

Of course the ``{\it section}'' $\Xi$ of the timing events has to be
  chosen so that every trajectory, or at least all trajectories but a
  set of $0$ probability with respect to the random choices that are
  supposed to generate the initial data, crosses infinitely many
  times the set $\Xi$, which in this case is also called a {\it Poincar\'e's
  section} and has to be thought of as a codimension $1$ surface drawn
  on phase space.

There is a simple relation between the evolution in continuous time
$x\to S_tx$ and its discrete representation $\xi\to S^n\xi$ in
discrete integer times $n$, between successive timing events: namely
$S\xi\equiv S_{\tau(\xi)}\xi$, if $\tau(\xi)$ is the time elapsing
between the timing event $\xi$ and the subsequent one $S\xi$.

Choice of timing observations at the realization of special or
``intrinsic'' events (\ie $x\in \Xi$), rather than at
``extrinsic'' events like at regularly spaced time intervals, is
for good reasons: namely to discard information that might be of
little relevance.

It is clear that, fixed $\t>0$, two events $x\in M$ and $S_\t x$
will evolve in a strongly correlated way. \index{intrinsic
  events} \index{extrinsic events} It will forever be that the
event $S_\t^nx$ will be followed $\t$ later by the next event;
which often is an information of little interest: unlike when
observations are timed upon the occurrence of dynamical events
$x\in\Xi$ which (usually) occur at ``random'' times, \ie such
that the time $\t(x)$ between an event $x\in\Xi$ and the
successive one $S_{\t(x)}x$ has a nontrivial distribution when
$x$ is randomly selected by the process that prepares the initial
data. This is quite generally so when $\Xi$ is a codimension $1$
surface in phase space $M$ which is crossed transversely by the
continuous time trajectories.

The discrete time representation, timed on the occurrence of
intrinsic dynamical events, can be particularly useful
(physically and mathematically) in cases in which the continuous
time evolution can lead $x$ close to singularities: the latter
can be avoided by choosing timing events which occur when the
point representing the system is neither singular nor too close
to a singularity, hence avoids situations in which the physical
measurements become difficult or impossible, see also
\Cite{BGGZ005}).

Very often, in fact, models idealize the interactions as due to potentials
which become infinite at some {\it exceptional} configurations. For instance the Lennard-Jones interparticle potential,
for the pair interactions between molecules of a gas, diverges as $r^{-12}$
as the pair distance $r$ tends to $0$; or the model of a gas representing
atoms as elastic hard spheres, with a potential becoming $+\infty$
at their contacts.

An important, paradigmatic, example of timed observations and of their
power to disentangle sets of data arising without any apparent order has
been given by Lorenz, \Cite{Lo963}: similar cases will be often
discussed here.

A first aim of the Physicist is to find relations, which are
general and model independent, between time averages of a few
(typically {\it very few}) observables. Time average of an
observable $F$ on the motion starting at $x\in M$, or in the
discrete time case starting at $\x\in\X$, is defined as
\be
\media{F}=
\lim_{T\to\io} \frac1T\ig_0^T F(S_tx)dt\qquad{\rm or}\ 
\qquad 
\media{F}=
\lim_{n\to\io} \frac1n\sum_{j=0}^{n-1} F(S^j\x)
\label{e1.1.2}\ee
and in principle the averages might depend on the starting point (and might
even not exist). There is a simple relation between timed averages and
continuous time averages, provided the observable $\t(\x)$ admits an
average $\lis\t$, namely if $\wt F(\x)\defi \ig_0^{\t(\x)} F(S_t\x)dt$
\be\media{F}=\lim_{n\to\io}\frac1{n\lis\t}\sum_{j=0}^{n-1} \wt
F(S^n\x)\equiv \frac1{\lis \t}\media{\wt F}\label{e1.1.3}\ee
{\it if} the limits involved exist.

\section{Birth of kinetic theory}
\def\SEC{Birth of kinetic theory}
\iniz\label{sec:II-1}
\lhead{\small\ref{sec:II-1}.\ \SEC}

The classical example of general, model independent, results is
offered by Thermodynamics: its laws establish general relations
which are completely independent of the detailed microscopic
interactions or structures (to the point that it may be not even
necessary to suppose that bodies consist of atoms).

It has been the first task of Statistical Mechanics to show that
Thermodynamics, under suitable assumptions, is or can be regarded
as a consequence of simple, but very general, mechanical models
for the motions of the elementary constituents of matter.

The beginnings go back to the classical age\index{classical age},
\Cite{Lu-050}: however in modern times {\it atomism} can be
traced to the discovery of Boyle's law\index{Boyle's law} (1660)
for gases, \Cite{Br003}[p.43], which could be explained
by imagining the gas as consisting of individual particles linked
by elastic springs.  This was a static theory in which particles
moved only when their container underwent compression or
dilatation. The same static view was to be found in Newton
\index{Newton} (postulating nearest neighbor interactions),
\Cite{Ne723}[p.265]\,, and later in Lavoisier and Laplace,\Cite{La821},
(postulating existence, and somewhat peculiar properties, of {\it
caloric} \index{caloric} to explain the nature of the nearest
neighbor molecular interactions).

A correct view, assigning to the atoms the possibility of free
motion was heralded by D. Bernoulli\index{Bernoulli D.},
\Cite{Br003}[p.57], (1738).  In his theory molecules
move and exercise pressure through their collisions with the
walls of the container: the pressure is not only proportional to
the density but also to the average of the square of the
velocities of the atoms, as long as their size can be
neglected. Very remarkably he introduces the definition of
temperature via the gas law for air: the following discovery by
Avogadro\index{Avogadro} (1811), ``law of equivalent volumes'',
on the equality of the number of molecules in equal volumes of
rarefied gases in the same conditions of temperature and
pressure, \Cite{Av811}, made the definition of temperature
independent of the particular gas employed, thus allowing a
macroscopic definition of absolute temperature.

The work of Bernoulli was not noticed until much later, and the same fate
befell on the work of Herapath,\index{Herapath} (1821), who was ``unhappy''
about Laplace's caloric hypotheses\index{caloric hypotheses} and proposed a
kinetic theory of the pressure deriving it as proportional to the average
velocity rather than to its square, but {\it that was not} the reason why
it was rejected by the {\it Philosophical Transactions of the Royal Society},
\Cite{Br976}, and sent to temporary oblivion.

A little later Waterston,\index{Waterston} (1843), proposed a kinetic
theory of the equation of state of a rarefied gas in which temperature was
proportional to the average squared velocity. His work on gases was first
published inside a book devoted to biology questions (on the physiology of
the central nervous system) and later written also in a paper submitted to
the {\it Philosophical Transactions of the Royal Society}, but rejected,
\Cite{Br976}, and therefore it went unnoticed.  Waterston went further by
adopting at least in principle a model of interaction between molecules
proposed by Mossotti\index{Mossotti}, (1836) \Cite{Mo836}, holding the view
that it should be possible to formulate a unified theory of the forces that
govern microscopic as well as macroscopic matter.

The understanding of heat as a form of energy transfer rather
than as a substance, by Mayer,\index{Mayer} (1841), and
Joule,\index{Joule} (1847), provided the first law of
Thermodynamics, on {\it internal energy}, and soon after
Clausius,\index{Clausius} (1850) \Cite{Cl850}, formulated the
second law, see Sec.\,(\ref{sec:VIII-1})\,, on the impossibility
of cyclic processes whose only outcome would be the transfer of
heat from a colder thermostat to a warmer one and showed the
earlier fundamental Carnot's efficiency\index{Carnot's
  efficiency} theorem, (1824) \Cite{Ca824}, to be a consequence
and the basis for the definition of the {\it
  entropy}.\index{entropy etymology} %
\footnote{\tiny The meaning of the word
       was explained by Clausius himself, \Cite{Cl865}[p.390]: 
``I propose
       to name the quantity $S$ the entropy of the system, after the Greek
       word {\bgr\eightgr <h trop'h\egr}, ``the transformation'', \Cite{LS968}, [{\sl
           in German {\it Verwandlung}}]. I have deliberately chosen the
       word entropy to be as similar as possible to the word energy: the
       two quantities to be named by these words are so closely related in
       physical significance that a certain similarity in their names
       appears to be appropriate.''  More precisely the German word really
       employed by Clausius, \Cite{Cl865}[p.390], is {\it
         Verwandlungsinhalt} or ``transformation content''.}

 And at this point the necessity of finding a connection
     between mechanics and thermodynamics had become clear and urgent.

The kinetic interpretation of absolute temperature as
proportional to the average kinetic energy by
Kr\"onig,\index{Kr\"onig} (1856) \Cite{Kr856}, was a real
breakthrough (because the earlier work of Bernoulli and
Waterstone\index{Bernoulli D.}\index{Waterston} had gone
unnoticed). The speed of the molecules in a gas had been linked
to the speed of sound but, from Kr\"onig, the mean square
velocity $u$ (in a rarefied gas) could be more reliably linked to
the temperature. It could be computed from the knowledge of the gas law
$pV=nRT$, with $n$ the number of moles, via Kr\"onig law of
proportionality between $nRT=pV$ and $n mu^2=Mu^2$,
\Cite{Kr856}, with $M$ being the mass of the gas
enclosed in the volume $V$ at pressure $p$: used by Clausius to
derive $u^2=\frac{3\,p V}M$.

The result was speeds $u$ of the order of $500$m/sec, as estimated by
Clausius,\index{Clausius} (1857) \Cite{Cl857}[p.124].
Therefore the value of $u$ was {\it too fast} to be compatible
with the known properties of diffusion of gases.

Clausius noted that compatibility could be restored by taking
into account the collisions between molecules: he introduced,
(1858), the {\it mean free path} $\l$ given in terms of the
atomic radius $a$ as $\l=\frac 1{n \p a^2}$ with $n$ the
numerical density of the gas. Since Avogadro's number and the
atomic sizes were not yet known this could only show that $\l$
could be expected to be much smaller than the containers size
$L$: however it opened the way to explaining why breaking an
ampulla of ammonia in a corner of a room does not give rise to
the sensation of its smell to an observer located in another
corner, not in a time as short as the sensation of the sound of
the broken glass.

The size $a$, estimated as early as 1816 by T. Young\index{Young T.} and
later by Waterston,\index{Waterston} (1859), to be of the order of
$10^{-8}$cm, can be obtained from the ratio of the volume $L^3$ of a gas
containing $N$ molecules to that of the liquid into which it can be
compressed (the ratio being $\r=\frac3{4\p}\frac{L^3}{a^3 N}$) and by the
mean free path ($\l=\frac{L^3}{N 4\p a^2}$) which can be found (Maxwell,
\index{Maxwell} 1859, \Cite{Ma860-a}[p.386]) from the liquid
dynamical viscosity ($\h=\frac {c}{Na^2}\sqrt{3n R T}$ with $n$ number of
moles, and $c$ a numerical constant of order $1$, 
\Cite{Ga000}[Eq.(8.1.4).(8.1.8)]); thereby
{\it also} expressing $N $ and $a$ in terms of macroscopically measurable
quantities $\r,\h$, measured carefully by Loschmidt,\index{Loschmidt}
(1865), \Cite{Br976}[p.75].

Knowledge of Avogadro's number\index{Avogadro's number} $N$ and
of the molecular radius $a$ allows to compute the diffusion
coefficient of a gas with mass density $\r$ and average speed
$V$: following Maxwell\index{Maxwell} in \Cite{Ma890t}[Vol.2, p.60 and
  p.345] it is $D=\frac{RT}{c N a^2 V \r}$,
with $c$ a constant of order $1$. This made quantitative
Clausius'\index{Clausius} explanation of diffusion in a
gas.\footnote{\tiny The value of $D$ depends sensitively on the
assumption that the atomic interaction potential is proportional
to $r^{-4}$ (hence at constant pressure $D$ varies as $T^2$). The
agreement with the few experimental data available (1866 and
1873) induced Maxwell to believe that the atomic interaction
would be proportional to $r^{-4}$ (hard core interaction would
lead to $D$ varying as $T^{3/2}$ as in his earlier work
\Cite{Ma860-a}).}

\section{Heat theorem\index{heat theorem} and Ergodic
  hypothesis\index{ergodic hypothesis}}
\def\SEC{Heat theorem\index{heat theorem} and Ergodic
  hypothesis\index{ergodic hypothesis}}
\iniz\label{sec:III-1}
\lhead{\small\ref{sec:III-1}.\ \SEC}

The new notion of entropy had obviously impressed every physicist
and the young Boltzmann\index{Boltzmann} attacked immediately the
problem of finding its mechanical interpretation: studying his
works is as difficult as it is rewarding. Central in his approach
is what I will call here {\it heat theorem} (abridging the
original diction ``main theorem of the theory of heat'').
\index{heat theorem}\index{Clausius theorem}\footnote{\tiny For a
precise formulation see below, p.\pageref{second principle}.}

The interpretation of absolute temperature as average kinetic
energy was already spread: inherited from the earlier works of
Kr\"onig and Clausius, it will play a key role in his subsequent
developments. But Boltzmann provides a kinetic theory argument
for this.\footnote{\tiny \Cite{Bo866}, see also
Sec. \ref{sec:I-6} below.}

With this key knowledge Boltzmann published his first attempt at
reducing the heat theorem to mechanics: he makes the point that it is a
form of the least action principle. Actually he considers an extension
of the principle: the latter compares close motions which, in a given time
interval, develop and connect fixed initial and final points. The
extension considered by Boltzmann compares close periodic
motions, see comments in Sec.\,(\ref{sec:IV-1})\,.

The reason for considering only periodic motions has to be found
in the basic philosophical conception that the motion of a system
of points makes it wander in the part of phase space compatible
with the constraints, visiting it entirely. It might take a long
time to do the travel but eventually it will be repeated. In his
first paper on the subject Boltzmann mentions that in fact motion
might, sometimes, be not periodic but even so it could possibly
be regarded as periodic with infinite period,
\Cite{Bo866}[\#2,p.30]: an interpretation of the
statement is discussed in Sec.\,\ref{sec:I-6}\,, see also Appendix
\ref{appB} below.

This is what is still sometimes called the {\it ergodic hypothesis}:
\index{ergodic hypothesis} Boltzmann will refine it more and more in his
later memoirs but it will remain essentially unchanged.

There are obvious mathematical objections to imagine a point
representing the system wandering through all phase space points
without violating the regularity or uniqueness theorems on
differential equations (as a space filling continuous curve
cannot be continuously differentiable and must self intersect on
a dense set of points). However it becomes soon clear that
Boltzmann does not really consider the world ({\it i.e.} space
and time) continuous: continuity is an approximation; derivatives
and integrals are approximations of ratios of increments and of
sums.
\footnote{\tiny From \Cite{Bo974}[p.227] {\it
    Differential equations require, just as atomism does, an initial idea
    of a large finite number of numerical values and points ...... Only
    afterwards it is maintained that the picture never represents phenomena
    exactly but merely approximates them more and more the greater the
    number of these points and the smaller the distance between them. Yet
    here again it seems to me that so far we cannot exclude the possibility
    that for a certain very large number of points the picture will best
    represent phenomena and that for greater numbers it will become again
    less accurate, so that atoms do exist in large but finite number}. For
  other relevant quotations see Sec.(1.1) and (5.2) in
  \Cite{Ga000}.\label{n1.1}}

Thus motion was considered periodic: a view, at the time shared
by Boltzmann, Maxwell, Clausius,\footnote{\tiny Today it seems
unwelcome because we have adjusted, under social pressure, to
think that chaotic motions\index{chaotic motions} are non
periodic and ubiquitous, and their images fill both scientific
journals and popular magazines.  It is interesting however that
the ideas and methods developed by the mentioned Authors have
been the basis of the chaotic conception of motion and of the
possibility of reaching some understating of it.}

The paper \Cite{Bo866}[\#2] has the ambitious title
``On the mechanical meaning of the fundamental theorem of heat
theory'':\index{heat theorem} under the assumption that,
\Cite{Bo866}[\#2,p.24], ``{\it an arbitrarily selected
  atom visits every site of the region occupied by the body in a
  given time (although very long) of which the times $t_1$ and
  $t_2$ are the beginning and end of the time
  interval\kern1mm\footnote{\tiny The recurrence
  time.\label{n1.2}} when motions velocities and directions
  return to themselves in the same sites, describing a closed
  path, thence repeating from then on their
  motion}'',\footnote{\tiny For Clausius' view see
  p.\,\pageref{Clausius ergodicity}\,. See also Maxwell's
  view,\,\Cite{Ma879}, {\it .. but we may with considerable
  confidence assert that except for particular forms of the
  surface of the fixed obstacle, the system will sooner or later,
  after a sufficient number of encounters, pass through every
  phase consistent with the equation of energy.}\label{Maxwell's
  ergodicity}} Boltzmann shows that the variation of the sum of
  average kinetic and potential energies in two close motions
  (interpreted as energy input on the system) divided by the
  average kinetic energy is an exact differential\index{exact
  differential}.

The two close motions are two periodic motions\index{periodic
  motion} corresponding to two equilibrium states of the system
that are imagined as end products of an infinitesimal step of a
quasi static thermodynamic transformation. The external potential
does not enter into the discussion: therefore the class of
transformations to which the result applies is very restricted
(in the case of a gas it would restrict to transformations
without work from external sources, as later stressed by
Clausius, see Sec.\,\ref{sec:VII-6}\,).\index{Clausius} The time
scale necessary for the recurrence\index{recurrence} is not taken
into account.

The ``revolutionary idea'' is that the states of the system are
identified with a periodic trajectory\index{periodic motion} and
from it the average values of physical quantities are computed:
this is the concept of state as a stationary
distribution\index{stationary distribution}. An equilibrium
state\index{equilibrium state} is thus identified with the
average values that the observables have in it: in the language
of modern measure theory this is a probability distribution, with
the property of being invariant under time evolution.

The states considered are equilibrium states: and thermodynamics
is\index{thermodynamics} viewed as the theory of the relations
between equilibrium averages of observables; {\it not as a theory
  of the transformations from one equilibrium state to another or
  leading to an equilibrium state} as will be done a few years
later with the Boltzmann's equation\index{Boltzmann's equation},
\Cite{Bo872}[\#22], in the case of gases, related to
Maxwell's\index{Maxwell} theory of the approach to equilibrium
(in rarefied gases) and improving it, \Cite{Ma867-b}, see also
Sec.\,(\ref{sec:XIV-6}) below.

The derivation of the heat theorem however might force the
reader to use hindsight to understand it: the Boltzmann versus
Clausius controversy\index{Boltzmann-Clausius controversy} due to
Clausius' ``Revisiting the second fundamental theorem of the
theory of heat and the general
principles of Mechanics'', \Cite{Cl871}, makes all this very clear, for
more details see the following Sec.\,(\ref{sec:V-6},\ref{sec:VII-6}\,).
\,\footnote{\tiny The second fundamental theorem is not the second law but a
  logical consequence of it, see Sec.(\,(\ref{sec:I-6}\,).} \index{heat theorem}

Receiving, \Cite{Bo871-0}[\#17], the comment that
his results were essentially the same as Boltzmann's own, in
\Cite{Bo866}[\#2], Clausius reacted politely but
firmly,\Cite{Cl872}. He objected to Boltzmann the obscurity of his derivation,
obliging the reader to suitably interpret formulae that in
reality are not explained (and ambiguous). Even accepting the
interpretation which makes the statements correct he stresses
that Boltzmann establishes relations between properties of
motions not subject to external forces: thus
limiting the analysis very strongly (and to a case in which, in
thermodynamics, the exactness of the differential $\frac{dQ}T$,
main result in Boltzmann's paper, would be obvious because the
heat exchanged would be function of the
temperature).\index{temperature}

Boltzmann acknowledged the point rather than profiting from the
fact that his formula remains the same, under further
interpretations, even if the external forces change, as
explained by Clausius himself through his exegesis of Boltzmann's
work: and Boltzmann promised, \Cite{Cl872}[p.271], to
take the critique into account in later works.  

In particular he discusses the subject in the third of the
impressive series of papers published in 1871,
\Cite{Bo871-a}[\#18],
\Cite{Bo871-b}[\#19],
\Cite{Bo871-c}[\#20]\,,
referred here as the ``trilogy'',\index{trilogy} see
Sec.(\,\ref{sec:VIII-6},\ref{sec:IX-6},\ref{sec:X-6}\,) below, just
before the formulation of the Boltzmann's
equation\index{Boltzmann's equation} which will turn him into
other directions. Boltzmann kept coming back to the more
fundamental heat theorem, ergodic hypothesis\index{ergodic
  hypothesis} and ensembles theory\index{ensembles theory} in
several occasions, and mainly in 1877 and 1884,
\Cite{Bo877-b}[\#42],\Cite{Bo884}[\#73].

The work of Clausius,\index{Clausius} \Cite{Cl871}, for details
see Sec.\,(\ref{sec:IV-1},\ref{sec:IV-6},\ref{sec:V-6}\,) below, is
very clear from a mathematical viewpoint: no effort of
interpretation is necessary and his analysis is clear and
convincing, with mathematical concepts (and using notations of
Lagrange's\index{Lagrange} calculus of variations,\,
\Cite{La867}[vol.I, p.336])\, carefully
defined and employed.  Remarkably he also goes back to the
principle of least action, an aspect of the heat theorem which is
now forgotten, and furthermore considers the ``complete problem''
taking into account the variation (if any) of the external
forces.

He makes a (weaker) ergodicity\index{weaker ergodicity} assumption: each atom
or small group of atoms undergoes a periodic motion and the statistical
uniformity follows from the large number of evolving units, \Cite{Cl871},
\*

\0''{\it ..temporarily, for the sake of simplicity we shall assume, as
  already before, that all points describe closed trajectories. For all
  considered points, and that move in a similar manner, we suppose, more
  specifically, that they go through equal paths with equal period,
  although other points may run through other paths with other periods.  If
  the initial stationary motion is changed into another, hence on a
  different path and with different period, nevertheless these will be
  still closed paths each run through by a large number of points.}''

And later, \Cite{Cl871},\label{Clausius ergodicity}

\0``{\it .. in this work we have supposed, until now, that all points move
  along closed paths. We now want to give up this special hypothesis and
  concentrate on the assumption that motion is stationary.

For the motions that do not follow closed paths, the notion of
recurrence, literally taken, is no longer useful so that it is
necessary to analyze it in another sense. Consider, therefore, first
the motions which have a given component in a given direction, for
instance the $x$ direction in our coordinate system. Then it is clear
that motions proceed back and forth, for the elongation, speed and return
time.  The time interval within which we find again every group of
points that behave in the same way, approximately, admits an average
value...}''
\*

He does not worry about the time scales needed to reach statistical
equilibrium: which, however, by the latter assumption are strongly reduced.
The first systematic treatment of the time scales will appear a short time
later, \Cite{Bo872}[\#22], in Boltzmann's theory of diluted
gases: via the homonym equation, time scales\index{time scale} are evaluated
in terms of the free flight time and become reasonably short and observable
compared to the super-astronomical recurrence times\index{recurrence time}.

Clausius'\index{Clausius} answer to Boltzmann, \Cite{Cl872} see
also Sec.(\,\ref{sec:VI-6}\,), is also a nice example on how a
scientific discussion about priority, accompanied by a strong
critique of various aspects of the work of a fellow scientist,
can be conducted without transcending out of reasonable bounds:
the paper provides an interesting and important clarification of
the original work of Boltzmann, which nevertheless remains a
breakthrough.

Eventually Boltzmann,\index{Boltzmann} after having discussed the
mechanical derivation of the heat theorem and obtained the theory
of\index{theory of ensembles} ensembles (the ones today called
microcanonical and\index{microcanonical ensemble} canonical)
and\index{canonical ensemble} Boltzmann's equation, finds it
necessary to rederive it via a combinatorial procedure
(pursuing its first use in \Cite{Bo868-a}[\#5]) in which
every physical quantity is regarded as discrete\index{discrete
  viewpoint}, \Cite{Bo877-b}[\#42], see also
Sec.\,(\ref{sec:XII-6}\,) below, and remarkably showing that the
details of the motion (\eg periodicity) are completely irrelevant
for finding that equilibrium statistics implies macroscopic
thermodynamics.

\section{Least action and heat theorem}
\def\SEC{Least action and heat theorem}
\iniz\label{sec:IV-1}
\lhead{\small\ref{sec:IV-1}.\ \SEC}
\*

It is useful to devote some time to a presentation of the
original idea for the heat theorem.

Boltzmann and Clausius theorems are based on a version of the action
principle for periodic motions. If $t\to x(t)$ is a
periodic motion developing under the action of internal forces with potential energy
$V(x)$ and of external forces with potential $V_{ext}(x)$, and with kinetic energy
$K(x)$, then the action of $x$ is defined, if its period is $i$
(adopting here the original notation for the period), by

\be{\cal A}(x)=\ig_0^i \big(\fra{m}2\dot x(t)^2
-V(x(t))\big)\,dt\label{e1.4.1}\ee
We are interested in periodic variations $\d x$ that can be represent
as:
\be\d x(t)=x'(\fra{i'}i t)-x(t)\defi x'(i'\f)-x(i\f)\hfill\label{e1.4.2}\ee
where $\f\in[0,1]$ is the {\it phase}, as introduced by Clausius,
\Cite{Cl871}, see also Sec.(\,ref{sec:V-6})\, below. The role of $\f$ is simply
to establish a correspondence between points on the initial trajectory $x$
and on the varied one $x'$: it is manifestly an arbitrary correspondence
(which could be defined differently without affecting the final result)
convenient to follow the algebraic steps. It should be noted that
Boltzmann does not care to introduce the phase and this makes his
computations difficult to follow (in the sense that once realized which
the final result is, it may be easier to reproduce it rather than to follow his
calculations).

Following Clausius version, \Cite{Cl871}, the heat theorem is
deduced from a few identities about the average values
$\lis{F}=i^{-1}\ig_0^i F(x(t))dt$ for generic observables $F$ and
$\lis K= i^{-1}\ig_0^i\fra12\dot x(t)^2dt$ for the average of the
kinetic energy.

Let {$x(t)=\x(\frac{t}i)$ with $\x(\f)$ 1-periodic}; it is $\lis
F\equiv \int_0^1 F(\x(\f))d\f$, and $\lis K=\frac12\int_0^1
\frac{\x'(\f)^2}{i^2} d\f$.  Changing $x$ by $\d x$ means to
change $\x$ by $\d\x$ and $i$ by $\d i$. So variation of an
average is the sum of the variation due to $\d\x$ and that due to
$\d i$: namely {$\d \lis F=\d_\x {\lis F}+\d_i \lis
  F$}. Therefore
\be{\d\lis V\equiv\d_\x{\lis
    V}},\qquad \d\lis{V_{ext}}=\d_\x{\lis{V_{ext}}},\qquad {\d
  \lis K=-2\frac{\d i}i \lis K +\d_\x \lis K}\label{e1.4.3}\ee
The variation $\d (\lis K +\lis V+\lis{V_{ext}} )\defi dU+dL$, with
$U\defi\lis K +\lis V$ and $dL\defi d \lis{V_{ext}}$, has a natural
interpretation of heat $\d Q$ in/out in the process $x\to
x'$.\, \footnote{\tiny Because $\d (\lis K +\lis V)\defi dU$ and
$\d \lis{V_{ext}}\defi \d L$ are the variation of the internal
energy $U$ amd $dL$ is the work that the system performs: \ie $\d
Q=\d U+\d L$.}\\

Then using the above Eq.\,\equ{e1.4.3}: $\d_\x \lis K=\d \lis K+2\frac{\d i}i \lis K $
\be\eqalign{
&\d Q\equiv { \d (\lis K +\lis V+\lis{V_{ext}} )
=-2\frac{\d i}i\lis K+\d_\x ({\lis K}+\lis V+\lis{V_{ext}} )}
\cr
&
={\bf-}2\frac{\d i}i\lis K+{2\d_\x{\lis K}}+
\Bd_\x {\bf ({-{\lis
K}} +\lis V+\lis{V_{ext}})}\equiv {\bf +}2\frac{\d i}i\lis K+2\d {\lis K} +{\bf 0}\cr}
\label{e1.4.4}\ee
where the last $0$ is due to {\it Maupertuis' principle},
as the motion follows the equations of motion (at fixed $i$).
Therefore, ({\it Heat theorem}):
\be{\frac{\d Q}{\lis K}}=+2\frac{\d i}{i}+2\frac{\d\lis K}{\lis K}=
{2\d\log (i\lis K)}\defi{\bf \d S}\label{e1.4.5}\ee
is an {exact} differential.
\*

In reality it is somewhat strange that both Boltzmann and
Clausius call Eq.\equ{e1.4.5} a ``generalization of the action
principle'': the latter principle uniquely determines a motion,
{\it i.e.} it determines its equations; Eq.\equ{e1.4.5} instead
does not determine a motion but it only establishes a relation
between the variation of average kinetic and potential energies
of close periodic motions under the assumption that they satisfy
the equations of motion; and it does not establish a variational
property (unless coupled with the second law of thermodynamics,
see footnote at p.\pageref{Kelvin-Planck}).

Still to derive it one proceeds as in the analysis of the action
principle and this seems to be the only connection between the
Eq.\equ{e1.4.5} and the mentioned principle.  Boltzmann
formulates explicitly, \Cite{Bo866}[\#2,sec.IV], what
he calls a generalization of the action principle \index{action
  principle generalization} and which is the Eq.\equ{e1.4.5}
(with $V_{ext}=0$ in his case): \*
 
\0``{\it If a system of points under the influence of forces, for
  which the ``vis viva'' principle\index{principle of vis viva}
  holds [{\sl \ie the kinetic energy variation equals the work of
      the acting forces}], performs some motion, and if all
  points undergo an infinitesimal change of the kinetic energy
  and if they are constrained to move on a trajectory close to
  the preceding one, then $\d\sum \fra{m}2\,\ig c\, ds$ equals
  the total variation of the kinetic energy times half the time
  during which the motion takes place, provided the sum of the
  products of the infinitesimal displacements of the points times
  their velocities and the cosine of the angle at each of the
  extremes are equal, for instance when the new limits are
  located on the lines orthogonal to the old trajectory limits}''
\*

It would be, perhaps, more appropriate to say that Eq.\equ{e1.4.5}
follows from $\V f= m \V a$ or, also, that it follows from the action
principle because the latter is equivalent to $\V f= m \V a$.
\footnote{\tiny This is an important point: the condition Eq.\equ{e1.4.5} does
  not give to the periodic orbits describing the state of the system any
  variational property (of minimum or maximum): the consequence is that it
  does not imply $\int \frac{\d Q}T\le 0$ in the general case of a cycle
  but only  $\int \frac{\d Q}T=0$ in the (considered) reversible cases of
  cycles. This comment also applies to Clausius' derivation. The inequality
  seems to be derivable only by  applying the second law in Clausius
  formulation.\label{Clausius inequality} It proves existence of entropy,
  however, see comment at p.\pageref{heat inequality}.}

A simple explicit example of Eq.\equ{e1.4.5} is in Appendix
\ref{appA}: from Boltzmann himself although it came somewhat
later, \Cite{Bo877a}[\#39], see also
Sec.(\,\ref{sec:XI-6}\,) below.

In Appendix \ref{appC} an interpretation of the proof of the
above theorem in a general monocyclic system is analyzed.  See
Appendix \ref{appD} for the extension to Keplerian
motion\index{keplerian motion}, \Cite{Bo884}.

Both Boltzmann and Clausius were not completely comfortable with the
periodicity.  As mentioned, Boltzmann imagines that each point follows the
same periodic trajectory which, if not periodic, ``can be regarded as
periodic with infinite period'', \Cite{Bo866}[p.30,\#2], see also
Appendix \ref{appB} below: a statement not always properly interpreted
which, however, will evolve, thanks to Boltzmann's own essential contributions,
into the ergodic hypothesis \index{ergodic hypothesis} of the XX-th century
(for the correct meaning see comment at p.\pageref{infinite period} in
Sec.(\,\ref{sec:I-6}\,), see also Appendix \ref{appB}).

Clausius worries about such a restriction more than Boltzmann does; and he
is led to think the system as consisting of many groups of points which
closely follow an essentially periodic motion.\index{periodic motion}
This is a conception close to the Ptolemaic conception
\index{Ptolemaic conception} of motion via cycles and epicycles, 
\Cite{Ga001b}.

\section{\index{ensembles}Ensembles and heat Theorem\index{heat theorem} }
\def\SEC{\index{ensembles}Ensembles and heat Theorem\index{heat theorem} }
\iniz\label{sec:V-1}
\lhead{\small\ref{sec:V-1}.\ \SEC}
\*

The identification of a thermodynamic
equilibrium\index{thermodynamic equilibrium} state with the
collection of time averages of observables was made, almost
without explicit comments {\it i.e as if it neither needed
  discussion nor justification}, in Boltzmann's\index{Boltzmann}
paper of 1866, \Cite{Bo866}[\#2], see also
Sec.(\,\ref{sec:I-6}\,) below.
As stressed above the analysis relied on the assumption that motions are
periodic.\index{periodic motion}

A first attempt to eliminate such hypothesis is already in the
work of 1868, \Cite{Bo868-a}[\#5], see also
Sec.(\,\ref{sec:II-6}\,) below, where the Maxwellian
distribution\index{Maxwellian distribution} is derived first for
a gas of hard disks and then for a gas of atoms interacting via
very short range potentials (so that interparticles collisions
duration in time and space can be neglected, although an external
potential can have a non zero range).

In this remarkable paper the canonical distribution
\index{canonical ensemble} for the velocity  of the
atoms in a single molecule is obtained from the microcanonical
distribution\index{microcanonical distribution} of the entire
gas. The ergodic hypothesis\index{ergodic hypothesis} appears
initially in the form: the molecule goes through all possible
[internal] states because of the collisions with the
others. However the previous hypothesis (\ie periodic motion
covering the energy surface) appears again to establish, as a
starting point, the microcanonical distribution for the entire
gas.

The argument is based on the fact that the collisions, assumed of
negligible space-time duration in a rarefied gas\index{rarefied gas}, see
p.\,\pageref{low density} below, change the coordinates via a
transformation with Jacobian determinant\index{Jacobian
  determinant} $1$ (because it is a canonical map\index{canonical
  map}) and furthermore, since the two colliding atoms are in
arbitrary configurations, then the distribution function, being
invariant under time evolution, must be a function of the only
relevant conserved quantity for the two atoms, \ie the sum of their
energies (while the sum of their momenta is not considered
because of translation and rotation invariance).

Also remarkable is that the derivation of the Maxwellian
distribution for a single particle from the uniform distribution
on the $N$-particles energy surface\index{energy
  surface} \footnote{\tiny Microcanonical in absence of external
forces, \ie just the uniform distribution of the kinetic energy
as the interactions are assumed instantaneous.} is performed \\
(1) by decomposing the possible values of the kinetic energy of
the system into a sum of values of individual particle kinetic
energies
\\
(2) each of which susceptible of taking finitely many values
(with degeneracies, dimension dependent, accounted in the $2$ and
$3$ dimensional space cases)
\\
(3) solving the combinatorial problem\index{combinatorial
  problem} of counting the number of ways to realize the given
values of the total kinetic energy and particles number
\\
(4) taking the limit in which the energy levels become dense,
integrating over all but one particles velocities, letting the
total number increase to $\infty$.

The combinatorial analysis has attracted a lot of attention
particularly, \Cite{Ba990}, if confronted with the later similar
(but {\it different}) analysis in \Cite{Bo877-b}[\#42],
see the comments in Sec.(\,(\ref{sec:XII-6}\,) below.

At this point Boltzmann felt the need to strengthen the implications
of his new general theory, foundation of the theory of equilibrium
ensembles, by producing at least an example of a
systems whose motions would visit the set of all configurations
with a given energy.

The idea that simple perturbations can lead to ergodicity (in the
sense of uniformly dense covering of the energy surface by a
single orbit) was illustrated in an example in a subsequent
paper, \Cite{Bo868-b}[\#6], see also
Sec.(\,\ref{sec:III-6}\,) below. However the example, chosen because
all calculations could be done explicitly and therefore should
show how ergodicity implies the microcanonical distribution, is a
mechanical problem with $2$ degrees of freedom which, however, is
{\it not ergodic} on {\it all} energy surfaces: see comments in
Sec.(\,\ref{sec:III-6}\,) below.\,\footnote{\tiny It is an example
similar to the example on the two body problem examined in 1877,
\Cite{Bo877a}[\#39], and deeply discussed in the
concluding paper \Cite{Bo884}[\#73].}

In 1871 Clausius\index{Clausius} also made an attempt to
eliminate the periodicity assumption, as discussed in
\Cite{Cl871}\Cite{Cl872}, see also Sec.(\,\ref{sec:V-6}\,) below. 

\def\SEC{Periodicity and ergodicity}
\section{Periodicity and ergodicity}
\iniz\label{sec:VI-1}
\lhead{\small\ref{sec:VI-1}.\ \SEC}

Research on the consequences of the 1868 paper continued as,
clearly, the results were so deep that further analysis was
nececessary. In 1871 Boltzmann\index{Boltzmann},
\Cite{Bo871-a}[\#18], considered a gas of polyatomic
molecules and related the detailed structure of the dynamics to
the determination of the invariant probability distributions on
phase space that would yield the time averages of observables in
a given stationary state without relying on the periodicity.

The assumption that in the collisions the range of the
interparticle forces is neglegible as well as the duration of the
collisions is eliminated in three 1871 papers.

Under the assumption that ``{\it the different molecules take all
  possible states of motion}'' Boltzmann undertakes again,
\Cite{Bo871-a}[\#18], see also Sec.(\,\ref{sec:VIII-6}\,)
below, the task of determining the number of atoms of the $N=\r
V$ ($\r=$ density, $V=$ volume of the container) molecules which
have given momenta $\V p$ and positions $\V q$ ($\V p$ are the
momenta of the $r$ atoms in a molecule and $\V q$ their
positions) determined within $d\V p,d\V q$, denoted $f(\V p,\V q)
d\V p d\V q$, greatly extending Maxwell's derivation of

\be f(p,q)=\r \frac{ e^{-h  p^2/2m}}{\sqrt{2m\,\p^3 h^{-3}}} d^3 p
d^3 q\label{e1.6.0}\ee
for monoatomic gases (and elastic rigid  bodies), \Cite{Ma860-a}.

The main assumption is no longer that the motion is periodic:
only the individual molecules assume in their motion all possible
states; and that is not supposed to happen periodically but
it is achieved thanks to the collisions with other molecules; no
periodicity any more.

Furthermore, \Cite{Bo871-a}[\#18,p.240]:
\*

\0``{\it Since the great regularity shown by the thermal phenomena
induces to suppose that $f$ is almost general and that it should be
independent from the properties of the special nature of every gas;
and, even, that the general properties depend only weakly from the form
of the equations of motion, with the exception of the cases in which
the complete integration does not present insuperable difficulties.}''
\*

\0and in fact Boltzmann develops an argument that shows that in presence of
binary collisions\index{binary collisions} in a rarefied gas the function
$f$ has to be $f=N e^{-h U}$ where $U$ is the total energy of the molecule
(kinetic plus potential). This is a consequence of Liouville's theorem and
of the conservation of energy in each binary collision.

The binary collisions assumption troubles Boltzmann, \Cite{Bo871-a}[p.255]:

\*\0``{\it An argument against is that so far the proof that such
distributions are the unique that do not change in presence of
collisions is not yet complete. It remains nevertheless
established that a gas in the same temperature and density state can
be found in many configurations, depending on the initial conditions,
{\it a priori} improbable and even that will never be experimentally observed.
}''.
\*

The (unresolved) doubt is the price paid by considering treating
the interparticle as pair collisions  (done in
the 1868 paper where the analysis is based on the realization
that in binary collisions, involving two molecules of $n$ atoms
each, with coordinates $(\V p_i,\V q_i), i=1,2$,\footnote{\tiny
Here $\V p_1=(p^{(1)}_1,\ldots p^{(n)}_1)$, $\V
q_1=(q^{(1)}_1,\ldots q^{(n)}_1), \V p_2=(p^{(1)}_2,\ldots
p^{(n)}_2)$, {\it etc.}, are the momenta and positions of the
atoms $1,\ldots,n$ in each of the two molecules.}\, only the
total energy and total linear and angular momenta of the pair are
constant (by the second and third Newtonian laws\index{Newtonian
  laws}) and, furthermore, the volume elements $d\V p_1 d\V
q_1d\V p_2d\V q_2$ do not change (by the Liouville's
theorem\index{Liouville's theorem}).

Visibly unhappy with the nonuniqueness of the results of
multibody collisions, Boltzmann resumes the analysis in a more
general context: {\it after all a molecule is a collection of
  interacting atoms}. Therefore one can consider a gas as a giant
molecule and apply to it the above ideas.

Under the assumption that there is only one constant of motion he
rederives in the subsequent paper, \Cite{Bo871-b}[\#19],
see also Sec.(\,\ref{sec:IX-6}\,) below, that the probability
distribution has to be what we call today a {\it microcanonical
  distribution} and that it implies a canonical distribution for
a (small) subset of molecules.

The derivation is the same that we present today to the
students. It has been popularized by Gibbs,\index{Gibbs}
\Cite{Gi902}, who acknowledges Boltzmann's work but curiously
quotes it as \Cite{Bo871-d}, {\it i.e.}  with the title of the
first section of Boltzmann's paper
\Cite{Bo871-b}[\#19], see also Sec.(\,\ref{sec:IX-6}\,)
below, which refers to a, by now, somewhat mysterious ``principle
of the last multiplier\index{last multiplier} of Jacobi''. The
latter is that in changes of variable the integration element is
changed by a ``last multiplier'' that we call now the {\it
  Jacobian determinant\index{Jacobian determinant}} of the
change. The true title (``{\it A general theorem on thermal
  equilibrium}'') is less mysterious although quite unassuming
given the remarkable achievement in it: this is the first work in
which the general theory of the ensembles is discovered
simultaneously with the equivalence between canonical and
microcanonical distributions for large systems.

Of course Boltzmann does not solve the problem of showing
uniqueness of the distribution (we know that this is essentially
never true in presence of chaotic dynamics,
\Cite{Ga000}\Cite{GBG004}). {\it Therefore to attribute a physical
  meaning to the distributions he has to show that they allow to
  define average values related by the laws of thermodynamics:
  \ie he has to go back to the derivation of a result like that
  of the heat theorem proving that $\frac{dQ}T$ is an exact
  differential}.

The periodicity assumption\index{periodicity assumption gone} is
long gone and the result might be not deducible within the new
context. He must have felt relieved when he realized, few days
later, \Cite{Bo871-c}[\#19], see also
Sec.(\,\ref{sec:XI-6}\,) below, that a heat theorem\index{heat theorem}
could also be deduced under the same assumption (uniform density
on the total energy surface) that the equilibrium distribution is
the microcanonical one\index{microcanonical distribution},
without reference to the dynamics.

Defining the heat $dQ$ received by a system as the variation of
the total average energy $dE$ plus the average of the work $dW$
performed by the system on the external
particles\,\footnote{\tiny {\it I.e.} the average variation, in
time, of the potential energy due to its variation generated by a
change in the potential of the external forces parameters.} it is
shown that $\fra {d Q}T$ is an exact differential if $T$ is
proportional to the average kinetic energy, see
Sec.(\,\ref{sec:XIII-6}\,) for the details.

This made statistical mechanics of equilibrium independent of the ergodic
hypothesis\index{ergodic hypothesis}; it will be further developed in
the 1884 paper, see Sec.(\,\ref{sec:XIII-6}\,) below, into a very general theory
of statistical ensembles extended and made popular by Gibbs.\index{Gibbs}

In the arguments used in the ``trilogy''\index{Boltzmann's trilogy},
\Cite{Bo871-a}[\#18],\Cite{Bo871-b}[\#19],%
\Cite{Bo871-c}[\#20], dynamics intervenes, as commented
above, only through binary collisions treated in detail: the analysis
will be employed a little later to imply, via the conservation
laws of Newtonian mechanics and Liouville's theorem, particularly
developed in \Cite{Bo871-a}[\#18], the new well known
Boltzmann's equation, which is presented explicitly immediately
after the trilogy, \Cite{Bo872}[\#22].

\section{Boltzmann's equation\index{Boltzmann's equation},
  entropy\index{entropy}, Loschmidt's paradox\index{Loschmidt's paradox}}
\def\SEC{Boltzmann's equation\index{Boltzmann's equation},
  entropy\index{entropy}, Loschmidt's paradox\index{Loschmidt's paradox}}
\iniz\label{sec:VII-1}
\lhead{\small(\,\ref{sec:VII-1}\,).\ \SEC}
\*

Certainly the result of Boltzmann most well known and used in technical
applications is the {\it Boltzmann's equation}, 
\Cite{Bo872}[\#22]: his
work is often identified with it (although the theory of
ensembles could well be regarded as his main achievement).  It is
a consequence of the analysis in \Cite{Bo871-a} (and of his
familiarity, since his 1868 paper, see Sec.(\,\ref{sec:II-6}\,), with
the work of Maxwell\index{Maxwell}
\Cite{Ma867-b}). It attacks a completely new problem: namely it does not deal
with determining the relation between properties of different equilibrium
states, as done in the analysis of the heat theorem\index{heat theorem}.

The subject is to determine how equilibrium is reached and shows that the
evolution of a very diluted $N$ atoms gas %
\footnote{\tiny Assume here for simplicity
  the gas to be monoatomic.} 
from an initial state, which is not time invariant, towards a
final equilibrium state admits a ``Lyapunov
function''\index{Lyapunov function}, if evolution occurs in
isolation from the external world: {\it i.e} a function $H(f)$ of
the {\it empirical distribution} $N f(p,q) d^3 p d^3 q$, giving
the number of atoms within the limits $d^3 p d^3 q$, which
evolves monotonically towards a limit $H_\io$ which is the value
achieved by $H(f)$ when $f$ is the canonical
distribution\index{canonical ensemble}.\index{theorem H}\index{H
  theorem}

There are several assumptions and approximations, some hidden. Loosely, the
evolution should keep the empirical distribution smooth: this is necessary
because in principle the state of the system is a precise configuration of
positions and velocities (a ``delta function'' in the $6N$ dimensional
phase space), and therefore the empirical distribution\index{empirical
  distribution} is a reduced description of the microscopic state of the
system, which is supposed to be sufficient for the macroscopic description
of the evolution; and only binary collisions take
place and do so losing memory of the past collisions (the {\it molecular
  chaos\index{molecular chaos hypothesis} hypothesis}). For a precise
formulation of the conditions under which Boltzmann's equations can be
derived for, say, a gas microscopically consisting of elastic hard balls
see \Cite{La974}\Cite{Ga000}\Cite{Sp006}.

The hypotheses are reasonable in the case of a rarefied gas: however the
consequence is deeply disturbing (at least to judge from the number of
people that have felt disturbed). It might even seem that chaotic motion is
against the earlier formulations, adopted or considered seriously not only
by Boltzmann, but also by Clausius and Maxwell, linking the heat theorem to
the periodicity of the motion and therefore to the recurrence of the
microscopic states.

Boltzmann had to clarify the apparent contradiction, first to himself as he
initially might have not realized its need it while under the enthusiasm of the
discovery. Once challenged he easily answered the critiques (``sophisms'',
see Sec.(\,\ref{sec:XI-6}\,) below), although his answers very frequently have
been missed, \Cite{Bo877a} and for details see Sec.(\,\ref{sec:XI-6}\,)
below. 

The answer relies on a more careful consideration of the time
scales: already Thomson\index{Thomson (Lord Kelvin)}, \Cite{Th874}, had
realized and stressed quantitatively the deep difference between the time
(actually, in his case, the number of observations) needed to see a
recurrence\index{recurrence time} in a isolated system and the time to
reach equilibrium. See also the ``{\it Christmas night, 1897}'' paper
of Boltzmann, \Cite{Sh024}\Cite{Bo898}[\#128]\,.

The latter time is a short time measurable in ``human units''
(usually from microseconds to hours) and in rarefied gases it is
of the order of the average free flight time, as implied by the
Boltzmann's equation which therefore also provides an explanation
of why approach to equilibrium is observable at all.

The first time scale, that will be called $T_\io$, is by far
longer than the age of the Universe already for a very small
sample of matter, like $1\,cm^3$ of hydrogen in normal conditions
which Thomson, and later Boltzmann, estimated to be of about
$10^{10^{19}}$ times the age of\index{age of Universe} the
Universe, \Cite{Bo896a}[Vol.2, Sec.88] (or
``equivalently''(!)  times the time of an atomic collision,
$10^{-12}sec$).

The above mentioned function $H(f)$ is simply 

\be H(f)=-k_B\,N\,\ig f(  p,  q) \log (f( 
p,  q)\d^3) {d^3  p d^3  q}\label{e1.6.1}\ee
where $\d$ is an arbitrary constant with the dimension of an
action and $k_B$ is an arbitrary constant. If $f$ depends on time
following the Boltzmann equation then $H(f)$ is monotonic non
decreasing, and becomes constant if and only if $f$ is the one
particle reduced distribution of a canonical
distribution\index{canonical ensemble}.

It is also important that if $f$ has an equilibrium value, given
by a canonical distribution, and if the system is a rarefied gas
so that the potential energy of the molecules mutual interaction
can be neglected, then $H(f)$ coincides, up to an arbitrary
additive constant, with the entropy per mole of the corresponding
equilibrium state provided the constant $k_B$ is suitably chosen
and given the value $k_B=R N_{A}^{-1}$, with $R$ the gas constant
and $N_A$ Avogadro's number.\index{Avogadro's number}

This induced Boltzmann to define $H(f)$ also as {\it the
entropy of the evolving state} of the system, thus extending the
definition of entropy far beyond its thermodynamic definition (where it
is a consequence of the second law of thermodynamics). Such extension
is, in a way, arbitrary: because it seems reasonable only if the system
is a rarefied gas.

In the cases of dense gases, liquids or solids, the analogue of Boltzmann's
equation, when possible, has to be modified as well as the formula in
Eq.\equ{e1.6.1} and the modification is not obvious as there is no
natural analogue of the equation. Nevertheless, after Boltzmann's analysis
and proposal, {\it in equilibrium}, {\it e.g.}  canonical or microcanonical
not restricted to rarefied gases, the entropy can be identified with $k_B
\log W$, $W$ being the volume (normalized via a dimensional constant) of
the region of phase space consisting of the microscopic configurations with
the same empirical distribution $f(p,q)\defi\frac{\r(p,q)}{\d^3}$ and it
is shown to be given by the ``Gibbs entropy''%
\index{Gibbs entropy}

\be S=-k_B\ig \r(\V p,\V q)\log \r(\V p,\V q)\,\frac{d^{3N}\V p
  d^{3N}\V q}{\d^{3N}}\label{e1.6.2}\ee
where $\r(\V p,\V q)\frac{d^{3N}\V p\, {d^{3N}\V q}}{\d^{3N}}$ is the
equilibrium probability for finding the microscopic configuration $(\V
p,\V q)$ of the $N$ particles in the volume element $d^{3N}\V p\,
{d^{3N}\V q}$ (made adimensional by the arbitrary constant $\d^{3N}$).

This suggests that in general (\ie not just for rarefied gases)
there could also exist a simple Lyapunov function\index{Lyapunov
function} controlling the approach to stationarity, with the
property of approaching a maximum value when the system
approaches a stationary state.

It has been recently shown in \Cite{GL003}\Cite{GGL005} that $H$,
defined as proportional to the logarithm of the volume in phase
space, divided by a constant with same dimension as the above
$\d^{3N}$, of the configurations that attribute the same
empirical distribution to the few observables relevant for
macroscopic Physics, is monotonically increasing, if regarded
over time scales short compared to the above defined recurrence
time scale $T_\io$ and provided the initial configuration is not
extremely special.\footnote{\tiny For there will always exist
configurations for which $H(f)$ or any other extension of it
decreases, although this can possibly happen only for a very
short time (of ``human size'') to start again increasing forever
approaching a constant (until a time $T_\io$ is elapsed and in
the unlikely event that the system is still enclosed in its
container where it has remained undisturbed and if there is still
anyone around to care).}\, The so defined function $H$ has been
called ``Boltzmann's entropy''\index{Boltzmann's entropy}.

However there may be several such functions besides the just
defined Boltzmann's entropy. Any of them would play a fundamental
role similar to that of entropy in equilibrium thermodynamics if
it could be shown to be independent of the arbitrary choices made
to define it: like the value of $\d$, the shape of the volume
elements $d^{3N}\V p \,d^{3N}\V q$ or the metric used to measure
volume in phase space: this however does not seem to be the case,
\Cite{Ga001}, except in equilibrium and this point deserves
further analysis, see Sec.(\,\ref{sec:XI-3}\,).

The analysis of the physical meaning of Boltzmann's equation has led to
substantial progress in understanding the phenomena of irreversibility of
the macroscopic evolution controlled by a reversible microscopic dynamics
and it has given rise to a host of mathematical problems, most of which are
still open, starting with controlled algorithms of solution of Boltzmann's
equation or, what amounts to the same, theorems of existence/uniqueness of the
solutions: for a discussion of some of these aspects see \Cite{Ga000}.

The key conceptual question is, however, how is it possible that
a microscopically reversible motion could rigorously lead to an evolution
described by an irreversible equation\index{irreversible
  equation} ({\it i.e.} an evolution in which there is a
monotonically evolving function of the state).

One of the first to point out the incompatibility of a monotonic approach
to equilibrium, as foreseen by the Boltzmann's equation, was Loschmidt. And
Boltzmann started to reply very convincingly in \Cite{Bo877a}, for details
see Sec.(\,\ref{sec:XI-6}\,) below, where Sec.II is dedicated to the so called
{\it Loschmidt's paradox\index{Loschmidt's paradox}}: which remarks that if
there are microscopic configurations in which the $H(f)$, no matter how it
is defined, is increasing at a certain instant there must also be others in
which at a certain instant $H(f)$ is decreasing and they are obtained by
reversing all velocities leaving positions unchanged, {\it i.e.} by
applying the {\it time reversal} operation.

This is inexorably so, no matter which definition of $H$ is
posed. In the paper \Cite{Bo877a}[p.121,\#39] a very
interesting analysis of irreversibility is\index{irreversibility}
developed. I point out here the following citations:

\* \0``{\it In reality one can compute the ratio of the numbers
  of different initial states which determines their probability,
  which perhaps leads to an interesting method to calculate
  thermal equilibria. Exactly analogous to the one which leads to
  the second main theorem. This has been checked in at least some
  special cases, when a system undergoes a transformation from a
  non uniform state to a uniform one. Since there are infinitely
  many more uniform distributions of the states than non uniform
  ones, the latter case will be exceedingly improbable and one
  could consider practically impossible that an initial mixture
  of nitrogen and oxygen will be found after one month with the
  chemically pure oxygen on the upper half and the nitrogen in
  the lower, which probability theory only states as not probable
  but which is not absolutely impossible}'' \*

To conclude later, at the end of Sec.II of
\Cite{Bo877a}[p.122,\#39]:

\*
\0''{\it But perhaps this interpretation relegates to the domain of
probability theory the second law, whose universal use appears very
questionable, yet precisely thanks to probability theory will be
verified in every experiment performed in a laboratory.}''  \*
\*

The work is partially translated and commented in the following
Sec.(\,\ref{sec:XI-6}\,).

\section{Conclusion}
\def\SEC{Conclusion}
\iniz\label{sec:VIII-1}
\lhead{\small\ref{sec:VIII-1}.\ \SEC}

Equilibrium statistical mechanics is born as an attempt to find the
mechanical interpretation of the second law of equilibrium
thermodynamics,
\footnote{\tiny \label{second principle}\index{heat theorem}
``The entropy of the universe is always increasing'' {\it is not
  a very good statement of the second law},\index{second law}
\Cite{Fe963}[Sec. 44.12].  The second law in
Kelvin-Planck's version ``A process whose {\it only} net result
is to take heat from a reservoir and convert it to work is
impossible''; and entropy is defined as a function $S$ such that
if heat $\D Q$ is added reversibly to a system at temperature
$T$, the increase in entropy of the system is $\D S=\frac{\D
  Q}T$, \Cite{Fe963}\Cite{Ze968}.\label{Kelvin-Planck} The Clausius'
formulation of the second law is ``It is impossible to construct
a device that, operating in a cycle will produce no effect other
than the transfer of heat from a cooler to a hotter body'',
\Cite{Ze968}[p.148]. In both cases the existence of
entropy follows as a theorem, Clausius' ``fundamental theorem of
the theory of heat'', here called ``heat theorem''} or at least
to find a mechanical interpretation of heat theorem (which is its
logical consequence).\index{heat theorem}

This leads, via the ergodic hypothesis, to establishing a connection
between the second law and the least action principle. The latter suitably
extended, first by Boltzmann and then by Clausius, is indeed related to the
second law:\index{second law} 
more precisely to the existence of the entropy function (\ie to
$\oint \frac{dQ}T=0$ in a reversible cycle, although not to $\oint
\frac{dQ}T\le 0$ in general cycles).

It is striking that all, Boltzmann, Maxwell, Clausius, Helmholtz, ...
tried to derive thermodynamics from mechanical relations valid for all
mechanical system, whether with few or with many degrees of
freedom. This was made possible by more or less strong 'ergodicity'
assumptions. And the heat theorem becomes in this way an identity
always valid. This is a very ambitious viewpoint and the outcome is
the Maxwell--Boltzmann  distribution on which, forgetting the details of the
atomic motions, the modern equilibrium statistical mechanics is
developing.

The mechanical analysis of the heat theorem for
equilibrium thermodynamics stands {\it independently} of the parallel
theory for the approach to equilibrium based on Boltzmann's equation:
therefore the many critiques towards the latter do not affect the
equilibrium statistical mechanics as a theory of thermodynamics.
Furthermore the approach to equilibrium is often studied under the much more
restrictive assumption that the system is a rarefied gas.

The apparently obvious contradiction between the theoren and the
basic equations assumed for the microscopic evolution, was
brilliantly resolved (but rarely understood at the time) by
Boltzmann, \Cite{Bo877a}[\#39], and Thomson,
\Cite{Th874}, ... who realized the probabilistic nature of the
second law as the dynamical law of entropy increase.

A rather detailed view of the reception that the work of Boltzmann received
and is still receiving can be found in \Cite{Uf008} where a unified view of
several aspects of Boltzmann's work are discussed, not always from the same
viewpoint followed here, and in the recent  exposition
of Boltzmann's work in \Cite{Da018}.
 
The possibility of extending the $H$ function {\it even when the
  system is not in equilibrium} and interpreting it as a state
function defined on stationary states or as a Lyapunov function,
is questionable and will be discussed in what follows. In fact
(out of equilibrium) the very existence of a well defined
function of the ``state'' of the system which (even if restricted
to stationary states) deserves to be called entropy is a problem:
for which no physical basis seems to exist indicating the
necessity of a solution one way or another.

The next natural question, and not as ambitious as understanding the
approach to stationary states (equilibria or not), is to develop a
thermodynamics for the stationary states of systems. These are states which
are time independent but in which currents generated by non conservative
forces, or other external actions,\footnote{\tiny Like temperature differences
  imposed on the boundaries.} occur.

Is it possible to develop a general theory of the relations
between time averages of various relevant quantities, thus
extending thermodynamics? To this question most of the following
will be devoted.

\chapter{Stationary Nonequilibrium}
\label{Ch2} 

\chaptermark{\ifodd\thepage
Stationary Nonequilibrium\hfill\else\hfill 
Stationary Nonequilibrium\fi}
\kern2.3cm
\section{Thermostats and infinite models}
\def\SEC{Thermostats and infinite models}
\label{sec:I-2}\iniz
\lhead{\small\ref{sec:I-2}.\ \SEC}

The essential difference between equilibrium and nonequilibrium is
that in the first case time evolution is conservative and Hamiltonian,
while in the second case time evolution takes place under the action
of external agents which could be, for instance, external
nonconservative forces or forces due to thermostats actions.

Non-conservative forces perform work which tends to increase the
kinetic energy: therefore a system subject only to this kind of
forces cannot reach a stationary state. For this reason in
nonequilibrium problems there must exist other forces which have
the effect of extracting energy from the system balancing, in
average, the work done or the energy injected in the system.

This is achieved, in experimental tests as well as in theory, by
adding thermostats to the system. Empirically a thermostat is a
device (consisting also of particles, like atoms or molecules)
which maintains its own temperature constant while interacting
with the system of interest.\index{thermostat}

In an experimental apparatus thermostats usually consist of large systems
whose particles interact with those of the system of interest: so large
that, for the duration of the experiment, the heat that they receive from
the system affects negligibly their temperature.

However it is clear that locally, near the boundary of separation between
system and thermostat, there will be variations of temperature which will
not increase indefinitely, because heat will flow away towards the far
boundaries of the thermostats containers. But eventually the thermostats temperature
\index{temperature} will start changing and the
experiment will have to be interrupted: so it is necessary that
the system reaches a 'satisfactorily' stationary
state\index{stationary state} before the halt of the experiment.
This is a situation that can be achieved by suitably large
thermostatting systems.

There are two main ways to model thermostats. At first, the simplest would
seem to imagine the system enclosed in a container $C_0$ in contact,
through separating walls, with other containers $\Th_1,\Th_2,\ldots,\Th_n$
as illustrated in Fig.(2.1.1).

\eqfig{100}{85}{\ins{90}{60}{$\V X_0,\V X_1,\ldots,\V X_n$}
\ins{60}{27}{$m_0\ddot{\V X}_{0i}=-\partial_i U_0(\V X_0)-\sum_{j}
\partial_i U_{0,j}(\V X_0,\V X_j)+\V E_i(\V X_0)$}
\ins{60}{10}{$m_j\ddot{\V X}_{ji}=-\partial_i U_j(\V X_j)-
\partial_i U_{0,j}(\V X_0,\V X_j)
$} }{fig2.1.1}{\kern-5pt(fig2.1.1)} 
\*
{\begin{spacing}{0.8}\tiny\0Fig.(2.1.1): $C_0$ represents the
    system container and $\Th_j$ the thermostats containers whose
    temperatures are denoted by $T_j$, $j=1,\ldots,n$. The
    thermostats are infinite systems of interacting (or free)
    particles which at all time are supposed to be distributed,
    far away from $C_0$, according to a Gibbs' distribution at
    temperatures $T_j$. All containers have elastic walls and
    $U_j(\V X_j)$ are the potential energies of the internal
    forces while $U_{0,j}(\V X_0,\V X_j)$ is the interaction
    potential between the particles in $C_0$ and those in the
    infinite thermostats.\end{spacing}} \*\*

The box $C_0$ contains the ``{\it system of interest}'', or
  ``{\it test system\index{test system}}'' to follow the
  terminology of the pioneering work \Cite{FV963}, by
  Feynman\index{Feynman} and \index{Vernon}Vernon, consisting of
  $N_0$ particles while the containers, labeled
  $\Th_1,\ldots,\Th_n$, are {\it infinite} and contain particles
  with average densities $\r_1,\r_2,\ldots,\r_n$ and temperatures
  at infinity $T_1,T_2,,\ldots,T_n$: they constitute the ``{\it
  thermostats}'', or ``{\it interaction systems}'', to
  follow \Cite{FV963}.  Positions and velocities are denoted $\V
  X_0,\V X_1,\ldots,\V X_n$, and $\dot{\V X}_0,\dot{\V
  X}_1,\ldots,\dot{\V X}_n$ respectively, particles masses are
  $m_0,m_1,\ldots,m_n$. The $\V E$ denote external, conservative
  or not, forces.

Thermostats temperatures are defined by requiring that initially
particles in each thermostat have a distribution
which, far from the test system, is
asymptotically a Gibbs distribution\index{Gibbs distribution}
with given densities $\r_1,\r_2,\ldots$, inverse
temperatures $(k_B T_1)^{-1},\ldots,$ $(k_B T_n)^{-1}$ and interaction
potentials $U_j(\V X_j)$ generated (for simplicity) by a short range pair
potential\index{pair potential} $\f$ (with at least the usual stability
properties)\,.\,\footnote{\tiny {\it I.e.} enjoying the lower boundedness
property $\sum_{i<j}^{1,m}\f(q_i-q_j)\ge -B m,\ {\rm
  with}\ B\ge0\ \forall m$,\,
\Cite{Ga000}[Sec.2.2]\,. Long range interactions are
excluded from the thermostat models comsidered here.
\label{stability}\index{stability}}

Likewise $U_0(\V X_0)$ denotes the potential energy of the pair
interactions of the particles in the test system and finally $U_{0,j}(\V
X_0,\V X_j)$ denotes the interaction energy between particles in $C_0$ and
particles in the thermostat $\Th_j$, also assumed to be generated by a pair
potential (\eg the same $\f$, for simplicity).

The interaction between thermostats and test system are supposed
to be {\it efficient} in the sense that the works done by the
external forces (if any) and by the thermostats forces (always
present) will balance, in the average, and keep the test system
within a bounded domain in phase space. Or at least keep its
distribution in spatially finite regions essentially concentrated
on bounded phase space domains; \ie assign, to the configurations
outside a ball (centered in $C_0$) of radius $P$ in momentum
and $L$ in position, a probability distribution which goes to zero
at a time independent rate as $R,L\to\infty$: thus being
compatible with the realization of a stationary state.

The above model, first proposed in \Cite{FV963} in a quantum mechanical
context, is a typical model that seems to be accepted widely as a
physically sound model for thermostats.

However it is quite unsatisfactory; not because infinite systems
are unphysical: after all we are used to consider $10^{19}$
particles in a container of $1\,cm^3$ as an essentially infinite
system; but because it is very difficult to develop a theory of
the motion of infinitely many particles distributed with positive
density. So far the cases in which the model has been pushed
beyond the definition assume that the systems in the thermostats
are free systems, examples already considered in \Cite{FV963}, (``free
thermostats''\index{free thermostats}).\label{free thermostat}

A further problem with this kind of thermostats, that will be
called ``{\it Newtonian}'' or ``{\it conservative}'', is that,
even in the cases of free thermostats, they are not suited for
simulations (although they might be studied theoretically). And
it is a fact that, in the last half century or so, new ideas and
progress in nonequilibrium have come from the results of
numerical simulations. However the simulations are performed on
systems interacting with {\it finite
thermostats}.\label{Newtonian thermostat}

Last, but not least, a realistic thermostat should be able to
maintain a temperature gradient because in a stationary state
only the temperature at infinity can be exactly constant: but in
infinite space and with the appropriate boundary conditions on
the container walls (\ie Neuman boundary conditions) this may be
impossible \footnote{\tiny Because heuristically it is tempting
to suppose that, in a stationary state, temperature should be
locally defined, and should tend to the value at infinity
following a kind of heat equation: but the heat equation with
reflecting boundary conditions might fail to have bounded solutions
in an infinite domain, like a line or an hyperboloid, with
 values at points tending to $\infty$ different from the values
 near $C_0$ {\it if the dimension of the container is $1$ or
 $2$}.\label{heat equation in 2D}}

\section{Finite thermostats}
\def\SEC{Finite thermostats}
\label{sec:II-2}\iniz
\lhead{\small\ref{sec:II-2}.\ \SEC}

The simplest finite thermostat\index{finite thermostat} models
can be illustrated in a similar way to that used in Fig.2.1.1:

\eqfig{300}{85}{\ins{90}{60}{$\V X_0,\V X_1,\ldots,\V X_n$}
\ins{60}{27}{$m_0\ddot{\V X}_{0i}=-\partial_i U_0(\V X_0)-\sum_{j}
\partial_i U_{0,j}(\V X_0,\V X_j)+\V E_i(\V X_0)$}
\ins{60}{10}{$m_j\ddot{\V X}_{ji}=-\partial_i U_j(\V X_j)-
\partial_i U_{0,j}(\V X_0,\V X_j)
-\a_j \V{{\dot X}}_{ji} $}}
{fig2.1.1}{\kern-5pt(fig2.2.1)}\label{Fig.2.2.1}
\*
{\begin{spacing}{0.8}\tiny\0Fig.2.2.1: Finite thermostats model
(Gaussian thermostats\index{Gaussian thermostat}): the containers
$\scriptstyle\Th_j$ are finite and contain $N_j$ particles. The
thermostatting effect is modeled by an extra force
$\scriptstyle-\a_j\dot{\V X}_j$ so defined that the {\it total}
kinetic energies $\scriptstyle K_j={m_j}\dot{\V X}_j^2/{2} $ are
{\it exact constants
of motion} with values $\scriptstyle K_j\defi 3 N_j k_B
T_j/2$.\end{spacing}}
\*\*

The difference with respect to the previous model is that the
containers $\Th_1,\ldots,\Th_n$ are now {\it finite} and contain
$N_1,\ldots,N_n$ particles respectively, (while $\CC_0$ still contains
$N_0$ particles): the $\Th_j$ can be
imagined as obtained by bounding the regions external to $C_0$ at
distance $\ell$ from the origin, by adding a spherical (for
definiteness) elastic boundary $\O_\ell$ of radius $\ell$,
centered in $C_0$, delimiting the regions $\Th_j$.

The condition that the thermostats temperatures $T_j$ should be fixed
is imposed by imagining that there is an extra force $-\a_j
\dot{\V X}_j$ acting on all particles of the $j$-th thermostat
and the multipliers $\a_j$ are so defined that the kinetic
energies $K_j=\frac12 {m_j} \dot{\V X}_j^2$ are {\it exact constants} of
motion (with values fixed by the initial thermostats data as $K_j\defi \frac32
N_j k_B T_j$, $k_B=$ Boltzmann's constant,
$j=1,\ldots,n$). Setting $m_j\equiv1$, the multipliers $\a_j$ are
then found to be:\footnote{\tiny Simply multiplying the both
sides of each equation in Fig.2.2.1 by $\dot{\V X}_j$ and
imposing, for each $j=1,\ldots,n$, that the \rhs vanishes.}
\be\a_j=
\frac{(Q_j+\dot U_j)}{2K_j}
\qquad\hbox{with}\quad
Q_j\defi  -\dot{\V
X}_j\cdot\partial_{{\V X}_j}U_{0,j}(\V X_{0},{\V X}_j)
\Eq{e2.2.1}\ee
where $Q_j$, which is the work per unit time performed by the particles in
the test system upon those in the container $\Th_j$, is naturally interpreted
as the {\it heat} ceded per unit time to the thermostat $\Th_j$.

The energies $U_0,U_j,U_{0,j},\,j>0,$ should be imagined as
generated by pair potentials $\f_0,\f_j,\f_{0,j}$ short ranged,
stable\,\footref{stability}\,, smooth or with a singularity
like a hard core or a high power of the inverse distance, and by
external potentials modeling the containers elastic
walls.

One could also imagine that thermostat forces may similarly act
within the system in $C_0$: \ie that there is an extra force
$-\a_0\dot{\V X}_0$ which also keeps the kinetic energy $K_0$
constant ($K_0=\frac12\dot{\V X_0}^2$ fixed by the initial
conditions as $\defi N_0\frac32 k_B T_0$): which could be called
an ``autothermostat'' \index{autothermostat} force on the test
system. This is relevant in several physically important
problems: for instance in electric conduction models the
thermostatting is due to the interaction of the electricity
carriers with the oscillations (phonons) of an underlying
lattice, and the latter can be modeled (if the masses of the
lattice atoms are much larger than those of the carriers),
\Cite{Ga996}, by a force keeping the total kinetic
energy (\ie temperature) of the carriers constant. In this case
defining $W_0=\V E(\V X_0)\cdot\dot{\V X}_0$ the multiplier
$\a_0$ would be defined by

\be
\a_0=
\frac{(W_0+Q_0+\dot U_0)}{2K_0}\qquad\hbox{with}\quad
Q_0\defi
-\sum_{j>0}\dot{\V
X}_0\cdot\partial_{{\V X}_0}U_{0,j}(\V X_{0},{\V
  X}_j)\label{e2.2.2}\ee
In the autothermostatted case, in which the total kinetic energy
is kept constant and no external thermostat is present, $\a_0$
would be $\a_0=({W_0+\dot U_0})/{2K_0}$.\label{auto}

Certainly there are other models of thermostats that can be
envisioned: all, including the above, were conceived in order to
make possible simulations. The first ones have been the
``Nos\'e-Hoover''\index{Nos\'e-Hoover thermostats} thermostats,
\Cite{No984}\Cite{Ho985}\Cite{EM990}. 
However they are not really different from the above in
Fig.2.2.1, or from the similar model in which the multipliers
$\a_j$ are fixed so that the {\it total energy} $K_j+U_j$ in each
thermostat is a constant; for instance, in the latter case $\a_j$
has to be defined as:
\be \a_j = \frac{Q_j}{2 K_j},\qquad 2 K_j=3N_j k_B \wt T_j
\label{e2.2.3}\ee
but here $\wt T_j$ are not constants and fluctuate with
time: their averages are determined by the initial
conditions, which fix instead the constant values of $K_j+U_j$.
\footnote{\tiny Likewise in a autothermostatted case in which the total
energy of the system is kept constant and no external thermostat
is present $\a_0$ would be $\a_0={W_0}/{3N_0 k_B \wt T_0}$, see also
Eq.\equ{e2.2.2}.}

The above examples of thermostats will be called {\it Gaussian
  isokinetic \index{Gaussian isokinetic thermostat}} if
$K_j=const, \,j\ge1,$ (hence $\a_j= \frac{Q_j+\dot U_j}{2K_j}$,
Eq.\equ{e2.2.1}) or {\it Gaussian isoenergetic \index{Gaussian
    isoenergetic thermostat}} if $K_j+U_j=const$ (hence $\a_j=
\frac{Q_j}{2 K_j}$, Eq.\equ{e2.2.3}).

Summarizing: Newtonian thermostats are infinite in size and
are acted upon only by Newtonian forces: \ie $\a_j\equiv 0$;
while Gaussian thermostats have finite size and include non
Newtonian forces. Other thermostats can be considered, \eg
imposing that the total kinetic energy is conserved in some
thermostats while in others is conserved the total energy; and
several other kinds of thermostats have been considered.
\*
\0{\it Remark:} Since $\a_j$'s are odd functions of $\dot{\V
  X}_j$, Gaussian thermostats generate a {\it reversible
  dynamics}\,\footnote{\tiny {\it I.e.} the phase space map
  $I(\dot{X},X)=(-\dot{X}, X)$ is such that the flow $S_t$ and
  $I$ ``anticommute'' $S_t I \equiv I S_{-t}$.}\,: this is {\it
  important} as it shows that Gaussian thermostats do not miss
  the {\it essential feature of Newtonian mechanics which is the
  time reversal symmetry}. Time reversal\index{time reversal}
  (TRS) is a symmetry of nature and any model close or equivalent
  to providing a faithful representation of nature should enjoy
  the same symmetry or should be supported by arguments to
  explain how reversibility remains an essential feature even in
  irreversible phenomena.\index{reversible dynamics} \*
It is interesting to keep in mind the reason for the attribute
``Gaussian'' to some of the above models. It is due to the
interpretation of the constancy of the kinetic energies $K_j$ or
of the total energies $K_j+U_j$, respectively, regarded as {\it
non holonomic constraints} imposed on the particles
motions. Gauss had proposed to call {\it ideal} the constraints
realized by forces satisfying his principle \index{Gauss'
principle} of {\it least constraint} and the forces $-\a_j\dot{\V
X}_j$, Eq.\equ{e2.2.1} or \equ{e2.2.3}, do satisfy the
prescription. The principle is reminded in Appendix \ref{appE}\,.
For definitness here attention will mainly be concentrated on the
latter Newtonian \index{Newtonian thermostat} and Gaussian
thermostats\index{Gaussian thermostat}.

\*
Of course it will be important to focus on results and
properties which: \*

\noindent
(1) do not depend on the thermostat model, at least if the numbers of
particles $N_0,N_1,\ldots,N_n$ are large,
\\
(2) have a physical interpretation
\*

The above view of the thermostats and the idea that purely Hamiltonian (but
infinite) thermostats could be represented equivalently by finite Gaussian
termostats, external to the system of interest, originated
in \Cite{No984}; it is discussed again by several Authors, see also
\Cite{WSE004}\Cite{Ch996}\Cite{Ga006c}\,. 

\section{Examples of nonequilibrium problems}
\def\SEC{Examples of nonequilibrium problems}
\label{sec:III-2}\iniz
\lhead{\small\ref{sec:III-2}.\ \SEC}

Some simple, concrete, examples of nonequilibrium systems are
illustrated in the following figures.

\eqfig{290}{70}{
\ins{24}{63}{${\V E}\ \to$}
\ins{70}{50}{periodic boundary (``{\it wire}'')}
\ins{70}{32}{$m\ddot{\V x}=\V E -\a \dot{\V x}$}
}{fig2.3.1}{(Fig.2.3.1)\label{fig2.3.1}}
\*

{\begin{spacing}{0.8}\0\tiny Fig.2.3.1: A model for electric
    conduction\index{electric conduction}. The container
    $\scriptstyle C_0$ is a box with opposite sides identified
    (periodic boundary). $\scriptstyle N$ particles, hard disks
    ($\scriptstyle N=2$ in the figure), collide elastically with
    each other and with other ($4$ in the figure) fixed hard
    disks: the mobile particles represent electricity carriers
    subject also to an electromotive force $\V{\scriptstyle E}$;
    the fixed disks model represents an underlying lattice whose
    phonons are phenomenologically represented by the force
    $\scriptstyle -\a \dot{\V x}$. This is an example of an
    autothermostatted system in the sense of
    Sec.\ref{sec:II-2}.\end{spacing}} \*\*

The multiplier $\a$ is $\a=\frac{\V E\cdot\dot{\V x}}{\dot{\V
    x}^2}$ and it forces constancy of the total kinetic energy
    $\frac12 m \dot{\V x}^2$: this is an electric conduction
    model of $N$ charged particles ($N=2$ in the figure, $\V x\in
    C_0$) in a constant electric field $\V E$ and interacting
    with a lattice of obstacles; it is ``autotermostatted''
    (because the particles in the container $C_0$ do not have
    contact with any ``external'' thermostat). This is a model
    that appeared since the early days (Drude,
    1899\index{Drude}, \Cite{Be964}[Vol.2, Sec.35]) in a slightly
    different form (\ie in dimension $3$, with point particles
    and with the thermostatting realized by replacing the
    $-\a\dot{\V x}$ force with the prescription that, after
    collision of a particle with an obstacle, its velocity is
    rescaled to a fixed $v=(\frac{3}m k_B T)^{\frac12}$).

The thermostat forces are a model of the effect of the
interactions between the particles (electrons) and a background
lattice (phonons).  The model is remarkable because it is the
first nonequilibrium problem that has been treated with real
mathematical attention and for which the analog of Ohm's
law\index{Ohm's law} for electric conduction has been (recently)
proved if $N=1$,
\Cite{CELS993}.

Another example is a model of thermal conduction, Fig.2.3.2:

\eqfig{360}{65}{
\ins{90}{61}{$T_1$}
\ins{170}{61}{$C_0$}
\ins{250}{61}{$T_2$}
}{fig2.3.2}{(Fig.2.3.2)\label{fig2.3.2}}
\* 

{\begin{spacing}{0.8}\0\tiny Fig.2.3.2: A model for thermal conduction%
\index{thermal conduction in gas} in a gas: particles in the 
central container $C_0$ are $N_0$ hard disks and the particles in the two
thermostats are also hard disks, some of which may be fixed and
represent an underlying lattice; collisions occur whenever the centers of
two disks are at distance equal to their diameters. Collisions with the
separating walls or bounding walls occur when the disks centers reach
them. All collisions are elastic.
  \end{spacing}}
\*\*

\0In the model $N_0$ hard disks interact by elastic collisions with each
other and with other hard disks ($N_1=N_2$ in number) in the containers
labeled by their temperatures $T_1,T_2$: the latter are subject to elastic
collisions between themselves and with the disks in the central container
$C_0$; the separations reflect elastically the particles when {\it their
  centers} touch them, thus allowing interactions between the thermostats
and the main container particles.  Interactions with the thermostats take
place only near the separating walls.

If one imagines that the upper and lower walls of the {\it central}
container are identified (realizing a periodic boundary
condition)\,%
\footnote{\tiny Reflecting boundary conditions on all walls of the side
  thermostat boxes are imposed to avoid that a current would be induced by
  the collisions of the ``flowing'' particles in the central container with
  the thermostats particles.}\,
  and that a constant field of intensity $E$ acts in the vertical direction
  then two forces conspire to keep it out of equilibrium, and the
  parameters $\V F=(T_2-T_1,E)$ characterize their strength: matter and
  heat currents flow.

The case $T_1=T_2$ has been studied in simulations to check that the
thermostats are ``efficient'', at least in the few cases examined: \ie that
the simple interaction, via collisions taking place across the boundary, is
sufficient to allow the systems to reach a stationary state, \Cite{GG007}. %
A mathematical proof of the above efficiency (at $E\ne0$), however, seems
difficult (and desirable).

To insure that system and thermostats can reach a stationary
state a further thermostat could be added $-\a_0\dot{\V X}_0$
that keeps the total kinetic energy $K_0$ constant and equal to
some $\frac32 N_0k_B T_0$: this would model a situation in which
the particles in the central container exchange heat with a
background at temperature $T_0$. This autothermotatted model is
subject to two external ``forces'', with parameters $T_2,T_1$ and
$E$, and could be considered to study Onsager reciprocity
properties, \Cite{Ga996}, between $T_2-T_1$ and $E$.

\section{Initial data}
\def\SEC{Initial data}
\label{sec:IV-2}\iniz
\lhead{\small\ref{sec:IV-2}.\ \SEC}

Any set of observations starts with a system in a state $X$
prepared. in phase space, by some well defined procedure: it has
become common to say ``{\it following a fixed protocol}''. In
nonequilibrium problems systems are always large, because the
thermostats and, often, the test systems are always supposed to
contain many particles: therefore any physically realizable
preparation procedure aimed at studying the statistical
properties of a test system will not produce, upon repetition,
the same initial state.\footnote{\tiny Even in numerical studies
the initial state is very often, on purpose, generated with some
random features to trust that the results have statistical
relevance.}

Here and in the following a {\it basic assumption} will be
adopted: whatever physically realizable preparation procedure is
employed, it will produce initial data whose {\it random}
microscopic coordinates have a probability distribution which has
a density, in the region of phase space allowed by the external
constraints.

In simulations the situation is more delicate: the initial data
are often generated by computer programs supposed to
implement the constraints on the system: which will be denoted
symbolically $\G$.

In the first two examples in the preceding section, the data
should be on a surface in the particles phase space, and in a
domain in the third (which can also be thought as surface with
boundary). The results have to be interpreted assuming that the
constraints are exactly implemented: typically this is done by
fitting them in theoretical models in which the constraints are
mathematically defined as surfaces (in $R^{6N}$ for systems of
$N$ particles, in the examples above). On such surfaces the
motions follow differential equations: here we shall deal only
with such theoretical systems.

In such idealized context it will
be supposed, to reflect the experimental reality, that the
initial data are generated by selecting them randomly with a
distribution that has a density over the surfaces in phase space
defined by the constraints.  \*

{\it The latter assumption about the initial data is very
  important and should not be considered lightly}.

\*
Mechanical systems as complex as systems of many
point particles interacting via short range pair potentials, {\it in
  general}, do admit {\it uncountably many} probability distributions $\m$
which are invariant, ``stationary'', under time evolution, \ie such that
for all measurable sets $E$ in phase space,
\be \m(S_{-t} E)=\m(E)\label{e2.4.1}\ee
where $S_t$ is the time evolution flow\index{evolution flow}, and
``measurable set'' means {\it ``any reasonable
  set''}.\,\footnote{\tiny For instance all open sets or even
sets that can be obtained from them by a countable combination of
operations of union and complemntation of sets.  In a discrete
representation of motion the full phase space $M$ is replaced by
the timing events surface $\Xi$, the flow $S_t$ by a map $S$ and
the invariance condition becomes $\m(S^{-1} W)=\m(W)$ for all
reasonable sets $W$.}\, All such distributions {\it could
  qualify} for representing an equilibrium or a stationary
nonequilibrium state.
\*

Therefore {\it a fundamental problem is: how to identify and
select, among the stationary probability distributions, the one
(or the few) which yield the observed statistical properties of
the stationary states} ?
\*

In the equilibrium cases the answer is provided by the ergodic
hypothesis: which can be formulated in various ways (and all of them
agree, if properly interpreted).  One distribution that gives the
statistical poperties of the equilibrium states is the
``canonical distribution'', and others are equivalent to it.  If
position and velocity coordinates of the $N_0$-particles, located
in a container $C_0$, are denoted $x=(\V X_0,\dot{\V X}_0)\in
R^{6 N_0}$, such distributions have the form
\be \m_0(dx)=  const\,\d(\G(x))\, d{\V X}_0d\dot{\V X}_0
\label{e2.4.2}\ee
where $dx=d{\V X}_0d\dot{\V X}_0$ is the ``Liouville volume
element'' and $\d(\G(x))$ {\it symbolically} imposes the
constraints $\G$ to which the motions are subject.\footnote{\tiny For
instance, for the canonical distribution the constraints on $x$ are that
$X_{0,j}\in C_0$ and the sum of kinetic and potential energy of
$x$ have a fixed sum $E$, $\G(x)=K(x)+U(x)-E=0$; $\d(K+U-E)$ is the delta
function.}

However, as discussed in the last three sections in Chapter
\ref{Ch1}\,, supplementary assumptions are needed to derive
Eq.\equ{e2.4.2} from the ergodic hypothesis, which requires
considerable mathematical care to be rigorously
formulated. Furthermore the hypothesis {\it cannot} be applied to
study non equilibrium stationary states: because the {\it naive
guess} that the stationary distribution can simply remain of the
form $\r(x)\d(\G(x))dx$, just replacing the constant in
Eq.\,\equ{e2.4.2}\, by a suitable function $\r(x)$, {\it has to
be discarded}, because in general there is {\it no probability
distribution} of this form which is
invariant, \Cite{ER985}\Cite{Ru999}.

The physical importance of the choice of the initial data, in
phase space $M$, in relation to the study of stationary states
has been proposed, stressed and formalized
by \index{Ruelle}Ruelle,
\Cite{Ru978b}\Cite{Ru980}\Cite{Ru999}. And it leads to a simple
hypothesis that unifies the equilibrium and nonequilibrium
theory. For the purpose of underlining the specificity of the
assumption on the initial data, denoted $({\V X},\dot{\V X})\in M$, it
will be formalized as\,\Cite{Ru999}\,:

\* \0{\bf Initial data hypothesis}\index{initial data hypothesis}
\label{initial data hypothesis}\label{initial data}
     {\it In a finite mechanical system the physically relevant
       stationary states correspond to invariant probability
       distributions $\m$, on phase space data $x=({\V X},\dot{\V
       X})$, which are time averages of distributions resulting
       by the evolution (following the equations of motion) of an
       initial distribution \be\m_0(dx)=\r(x)\d(\G(x)) dx, \qquad
       dx\defi d{\V X}d\dot{\V X}\label{e2.4.3}\ee with a
       continuous density $\r(x)\ge0$ and is compatible with the
       constraints}\,. \*

The $\r(x)$ is a {\it vastly arbitrary} function: assuming that
physically interesting initial data are generated on phase space
by {\it any} among the probability distributions $\m_{0}$, means
that the very few, often just unique, stationary distributions
$\m$ that we consider {\it physically relevant} to describe the
stationary states, are the $\m$ that can be obtained as limits of
time averages of iterates of distributions $\m_{0}$.

The average $\m$, hence the statistical properties of 
continuous observables $F(x)$ on the phase space $M$, will be:
\be \int F(y)\m(dy)=\lim_{T\to\infty} \frac1T\int_0^T dt \int_M
F(S_tx)\m_0(dx)
\label{e2.4.4}\ee
where the limit $\m$ is still possibly dependent on the initial
data, as it will be discussed later\,\footnote{\tiny Such
dependence will be related to phenomena like phase transitions
and intermittency, \index{phase transition}
\index{intermittency}}\,. In the discrete time cases $t$ is replaced
by integers $n$, $S_t$ by iterates $S^n$ of the evolution map
$S$ and $M$ by the timing events surface $\X$.

\section{Finite or infinite thermostats? Equivalence?}
\def\SEC{Finite or infinite thermostats? Equivalence?}
\label{sec:V-2}\iniz
\lhead{\small\ref{sec:V-2}.\ \SEC}

In the following we shall study mainly {\it finite} thermostats.%
\index{thermostats finite} It is clear that this can be of
interest only if the results can, in some convincing way, be
related to thermostats in which particles interact via Newtonian
forces, called in the preceding sections ``Newtonian
thermostats''.

As noted in Sec.(\,\ref{sec:I-2}\,) the only way to obtain thermostats
of this type is to make them infinite: because the work $Q$ that
the test system\index{test system} performs per unit time over
the thermostats (heat ceded to the thermostats) will change the
kinetic energy of the thermostats, when non conservative forces act on
the test system, and the only way to avoid indefinite heating is
that the heat flows away towards infinity, hence the necessity of
infinite thermostats. Newtonian forces and finite thermostats
will result eventually in an equilibrium state in which all
thermostats temperatures have become equal.

It is useful to deal with an example about the kind of problems met
when trying to analyze properties of systems like Gaussian thermostats
(isokinetic or isoenergetic) or of corresponding Newtonian
thermostats, just  introduced in Sec.(\,\ref{sec:I-2},\ref{sec:II-2}\,).

Consider a test system in $\CC_0$ and its Gaussian thermostats
$\{\Th_j\}_{j=1}^n$; if $x=(\V X_j,\dot{\V
X}_j)_{j=0}^n\defi(x_0,x_1,\ldots x_n)$: let the initial
data distribution be:
\be \wt\m_0(dx)=\r(x)\prod_{j=0}^n\m_j(dx_j)\label{e2.5.1}\ee
where the $\m_j,j\ge0,$ are defined as in Eq.\equ{e2.4.3} with the
appropriate constraint $\G_0$ on $x_0$ and with $\G_j$
constraining the kinetic energy of the $j$-th thermostat,
$j\ge1$, to be $K(x_j)=\frac12\dot{\V X_j}^2\defi\frac32 N_j k_B
T_j$, \ie its temperature to be a fixed value $T_j$.

The distribution $\wt\m_0$, in conformity with the initial data
hypothesis (see Eq.\equ{e2.4.3}), is rather arbitrary because of
the arbitrariness of the functions
$\r(x),\r_j(x_j)$.\,\footnote{\tiny Of course here the $\r_j(x)$ could
all be taken $=1$ leaving only $\r(x)$ arbitrary.}

In the case of infinite (``Newtonian''\, \ref{Newtonian
  thermostat}) thermostats the initial data distribution has to
  be properly defined, because the symbol $d{\V X}d\dot{\V X}$ in
  Eq.\equ{e2.4.2}, hence $dx$ above, has to be defined.  For
  this purpose imagine to ``regularize'' a Newtonian thermostat
  by enclosing it into an elastically reflecting sphere $\O_\ell$
  of radius $\ell$ and enclose $N_j^\ell$ particles in the
  regions $\Th_j\cap
\O_\ell\defi\Th_j^\ell$. Then define $\m^\ell_0(dx)$ as in
Eq.\equ{e2.5.1} adding to the constraints $\G_j$ that, for given
$(r_j)_{j=1}^n$, the densities in $\Th_j^\ell$ are fixed to
$r_j=\frac{N^\ell_j}{|\Th^\ell_j|}$. Finally define $\m_0$ as the
``thermodynamic limit'' $\lim_{\ell\to\infty}\m^\ell_0(dx)$,
interpreting it as a probability distribution on the space of the
configurations of the infinitely extended system.  \*

\0{\it Remarks:} (1) To establish existence of the latter
``thermodynamic limit'', conditions have to be imposed on the
interaction potentials and on the densities; here we suppose (for
simplicity) to consider only cases in which the limit existence
can be proved. There are a few instances of thermostats with only
conservative, short range and stable\footref{stability} forces,
in which this can be done: see \Cite{GP010b} for a
proof adapted to the distributions in Eq.\equ{e2.5.1} and to the
thermostats examples discussed.

\0(2) A warning is necessary: in special cases the preparation of
the initial data is, out of purpose or of necessity, such that
with probability $1$ it produces data which lie in a set of $0$
phase space volume, hence of $0$ probability with respect to any
distribution of the initial data like Eq.\equ{e2.5.1} (\ie
conform to the initial data hypothesis in
Sec.(\,\ref{sec:IV-2}\,)). In this case, {\it of course}, the
initial data hypothesis above does not apply: the averages will
still exist quite generally, but the corresponding stationary
state will be different from the one associated with data chosen
with a distribution among those of the form $\wt\m_0$,
Eq.\equ{e2.5.1}. Examples are easy to construct as it will be
discussed, for instance, in Sec.(\,\ref{sec:IX-3}\,) below.  \*

It is important to establish a relation between
infinite Newtonian thermostats and finite Gaussian
thermostats with additional {\it ad hoc} forces,
illustrated in Sec.(\,\ref{sec:II-2}\,).
To compare the evolutions in infinite Newtonian thermostats
\index{Newtonian-Gaussian comparison} and in large Gaussian
thermostats it is natural to choose the initial data in a
consistent way (\eg coincident) in the two cases.

Hence in both cases (Newtonian and Gaussian) it will be natural
to evolve a datum obtained with the distribution $\wt\m_{0}(dx)$,
Eq.\equ{e2.5.1}, and follow its evolution in the two systems:
imagining that in the Gaussian case the particles {\it outside}
the finite region $\O_\ell$, occupied by the thermostats and
bounded by a reflecting sphere of radius $\ell$, are ``frozen''
in the initial positions and velocities of $x$.

Probably the first objection is that a relation between the above
two dynamics seems doubtful because the equations of motion, and
therefore the motions, are different in the two cases. Hence a
first step would be to show that {\it instead} in the two cases
the motions of the particles are very close at least if the
particles are in, or close to, the test system and the finite
thermostats are large enough.

A heuristic argument is that the non Newtonian forces
$-\a_j\dot{{\V X}}_j$, Eq.\equ{e2.2.1}, are formally of the
order of the inverse number of particles $N_j$,\footnote{\tiny
Being inversely proportional to the {\it total} kinetic energy in the
thermostats $\Th_j$.}while the other factors ({\it i.e} $Q_j$ and
$\dot U_j$) are expected to be of order $O(1)$, \ie of the number
of particles present in a layer of size $\sim$twice the interparticle
interaction range: hence in large systems their effect should be
small (and zero in the limits, $N_j\to\infty$, of infinite
thermostats). This has been discussed, in the case of a single
self-thermostatted test system, in \Cite{ES993}, and more
generally (also accompanied by simulations) in \Cite{WSE004}\,.

It is possible to go quite beyond a theoretical heuristic
argument. However this requires first establishing existence and
some properties of the dynamics of systems of infinitely many
particles in the Newtonian\index{infinite systems dynamics}
infinite thermostats dynamics; and also
in the Gaussian dynamics with thermostats bounded by reflecting
spheres of radius $\ell\defi 2^kR$, multiple of the diameter $R$
of the test region. This can be partially done as described below.

The best that can be hoped about the Newtonian evolution is that
initial data $\dot{{\V X}},\V X$ chosen randomly with a
distribution $\wt\m_{0}$, as described after
Eq.\equ{e2.5.1},\,\footnote{\tiny Which is a Gibbs
distribution\index{Gibbs distribution} with given temperatures
and density for the infinitely many particles in each thermostat
and with any density for the finitely many particles in the test
system.} will generate a solution of the equations in
Fig.(2.1.1), {\it i.e.} a $\dot{{\V X}}(t),\V X(t)$ for which
both sides of the equations make sense and are equal for all
times $t\ge0$, {\it with the exception of motions starting from a
  set of initial data which has $0$ $\wt\m_{0}$-probability}.

Supposing the interaction potentials smooth, repulsive and short
range such a result is known in the literature and can be proved
in the present context, \Cite{GP010a}\Cite{GP010b}, in the
geometry of Fig.(2.1.1) in space dimension $2$ and, in at least
one special case of the same geometry, in space dimension $3$.
In such cases, if initial data $x=(\dot{{\V X}}(0),\V X(0))$ are
chosen randomly with the probability $\wt\m_{0}$, the equation in
Fig.(2.1.1) admits, with probability $1$, a solution $x(t)$, with
coordinates of each particle smooth functions of $t$.

Furthermore, in the same references quoted, the finite Gaussian
thermostats model, {\it either isokinetic or isoenergetic}, can
be realized in the geometry of Fig.(2.2.1) by terminating the
thermostats containers within a spherical surface $\O_\ell$,
centered in $C_0$, of radius $\ell=2^k R$, with $R$ being the
linear size of the test system and $k\ge1$ integer.

For purpose of comparison, the evolutions in the two models
will be started in the same configuration $x$, randomly chosen
with distribution $\wt\m_0$,\,\equ{e2.5.1}\,. In the Newtonian
thermostats models all particles will evolve, following the
infinitely many equations in Fig.(2.1.1), on a trajectory $t\to
x(t)=(\dot{{\V X}}(t),{\V X}(t))$, as guaranteed by the quoted
\Cite{GP010b}.  In the Gaussian model particles external to the
ball $\O_\ell$ will be imagined to keep positions and velocities
``frozen'' in time and not contributing to the Gaussian
constraint, while inside $\O_\ell$ the evolution of the finitely
many particles will be defined adding elastic reflections on the
spherical boundaries of $\O_\ell$. The followed trajectory will
be $\dot{{\V X}}^{[k]}(t),{\V X}^{[k]}(t)$, depending on the
radius $\ell\defi 2^kR$ delimiting the system and on the
isokinetic or isoenergetic nature of the thermostatting
forces.

The two evolutions will be called ``corresponding''.  Then it is
possible to prove, for the class of interaction potentials
studied in \Cite{GP010b} (Lennard-Jones potentials), the
property: \*
\0{\bf Theorem:} {\it Fixed arbitrarily a time $t_0>0$ there
  exist two constants $C,c>0$ ($t_0$--dependent) such that
  the Newtonian evolution $x_j(t)$ and the
  corresponding isokinetic (or isoenergetic) motions $x_j^{[k]}(t)$
  are related as:
\be |x_j(t)-x_j^{[k]}(t)|\le C e^{-c 2^k}, \qquad {\rm if}\ |x_j(0)|
<2^{k-1}R\label{e2.5.2}\ee
for all $(t\le t_0,j)$, with $\wt\m_0$-probability $1$ with respect to the
initial data choice.}  \*

In other words the Newtonian motion and the Gaussian
thermostatted motions are indistinguishable, {\it up to any
  prefixed time $t_0$}, if the thermostats are large (\ie $k$ is
large) and if we look at particles initially located within a
ball half the size of the confining sphere of radius $\ell=2^kR$,
where the spherical thermostats boundaries are located, \ie
within the ball of radius $2^{k-1}R$.

This theorem is only a beginning, although in the right direction
towards a physically satisfactory comparison of the corresponding
evolutions, as one would really like to prove that the evolution
of the initial distribution $\wt\m_0$ leads to a stationary
distribution in both cases and that the stationary distributions
for the Newtonian and the Gaussian thermostats {\it coincide in
  the ``thermodynamic limit''} $k\to\infty$.

\0{\bf Remarks:}\\ \0(1) At this point a key observation has to
be made: it is to be expected that in the thermodynamic limit,
once a stationary state is reached starting from $\wt\m_{0}$,
described after Eq.\equ{e2.5.1}, the thermostats temperature
locally at a point $q$ (to be suitably defined, \eg as the
average kinetic energy in a unit box around $q$) should vary
smoothly as $q\to\infty$ toward a value at infinity. A value
which, in each thermostat $\Th_j$, equals the initially
prescribed temperature (appearing in the random selection of the
initial data with the given distribution
$\wt\m_{0}$).\index{thermostats dimension dependence}
\\
Hence the temperature variation should be described, at least
approximately, by a solution of the heat equation $\D T(q)=0$ and
$T(q)$ not constant and tending to $T_j$ as $q\in\Th_j, \,
q\to\infty$. However, see \footref{heat equation in 2D}\,,
if the space dimension is $1$ or $2$ there is no such
harmonic function (if the boundary condition at the side walls of
the thermostats are reflecting, \ie demand that the normal
derivative of $T$ vanishes so that thermostats exchange energy
only with the system).

\0(2) In Sec.(\,\ref{sec:VIII-2}\,) the physical interpretation of the
average of the quantity $\sum_{j=1}^n3N_j\a_j$, Eq.\,\equ{e2.2.1}\,,
will be discussed and interpreted as rate of {\it entropy
  production} that the system generates into the thermostats.  It
is therefore interesting to note that $\sum_{j=1}^n3N_j\a_j$ can
be defined (in principle) as a mechanical observable in both
Gaussian or Newtonian thermostats.\footnote{\tiny Because the
\rhs in the quoted formulae are expressed in terms of mechanical
quantities $\scriptstyle Q_j,\dot U_j$ and the temperatures at
infinity $\scriptstyle T_j$ (furthermore the $\scriptstyle
{\dot U_j}/{k_B T_j}$ do not contribute to the time averages
being a total time derivative, see Sec.(\,\ref{sec:VIII-2}\,).}

If equivalence \index{Newtonian-Gaussian equivalence} between
corresponding (as defined in the above theorem) Newtonian
(irreversible) and Gaussian (reversible) thermostats could be
shown, in the sense that the average of the mechanical observable
$\sum_{j=1}^n3N_j\a_j$ would be the same in corresponding systems,
then physical consequences would follow. Indeed suggesting such
equivalence makes clear that it is suggested as {\it possible}
that a {\it Newtonian evolution produces entropy}: {\it i.e.}
entropy production\index{entropy production} in a stationary
state is compatible with the time reversibility of Newton's
equations, \Cite{WSE004}.

\section{Hyperbolicity, attracting surfaces, Anosov maps, chaotic data}
\def\SEC{SRB, attracting surfaces, Anosov's evolution and chaotic data}
\label{sec:VI-2}\iniz
\lhead{\small\ref{sec:VI-2}.\ \SEC}

The limit probability distributions in Eq.\,\equ{e2.4.4} will be
called {\it SRB distributions}, from Sinai,Ruelle,Bowen who
investigated, and solved in important cases,
\Cite{Si968a}\Cite{Bo970a}\Cite{BR975}, the more difficult question of
finding conditions under which, for motions $x\to S_t x=x(t)$ of
points $x$ on a phase space $M$, the following {\it pointwise}
limits:
\be \lim_{\t\to\infty}\frac1\t\int_0^t F(S_t x)\,dt=\int_M F(y)\m(dy)
\label{e2.6.1}\ee
exist for all continuous observables $F$, and {\it for all} $x\in
M$ chosen randomly according to the initial data hypothesis
(Sec.(\,\ref{sec:IV-2}\,)).\footnote{\tiny SRB will be used
as an adjective attached to the distributions that can be
obtained as limits in Eq.\equ{e2.4.4} or \equ{e2.6.1}: following
the usual informal usage of the adjective ``ergodic''.}

In timed observations the same question becomes to find
conditions under which, for all continuous observables $F$, the
following limits:
\be \lim_{\t\to\infty}\frac1k\sum_{q=0}^{k-1} F(S^kx)\,dt=\int_\Xi F(y)\m(dy)
\label{e2.6.2}\ee
exist {\it for all} $x$ chosen according to the initial data
hypothesis on the timing surface $\X$,, \ref{initial data}.

It is convenient to pose the question in a context more general
than that of the examples of the previous sections and consider,
in the following, a more general definition of phase space and of
dynamical systems on it: \*

\0{\bf Definition 0:} {\rm (System)} {\it (1)Phase space will be a
  general smooth Riemannian manifold, \ie a smooth bounded
  surface $M$ or $\X$ on which a smooth metric tensor defines
  distance $\d(x,y)$, between close points $x,y$, and an area element
  $dx$.\\ (2) Points $x\in M$ are imagined to move following a
  differential equation $\dot x= f(x)$ whose solutions $x\to S_t
  x=x(t)$ exist for all $t\in(-\infty,+\infty)$ and are
  smooth. \\ (3) Alternatively points $x\in\X$ move following
  iterations $S^n$, $n\in Z$, of a smooth map $S$ of $\X$.  \\
  (4) The initial data are always supposed to be generated
  with (any) probability distribution $\m_0(dx)=\r(x)dx$ with
  $\r(x)\ge0$ continuous, extending Sec.(\,\ref{sec:IV-2}\,).
  \\
  (5) In both cases $M$ and $\X$ will be called ``phase space'', and
  each pair $(M,S_t)$ or $(\X,S)$ will be called a ``dynamical
  system''.}\index{phase space}\index{dynamical
  system}\label{phase space}\label{dynamical system} \*

The Gaussian thermostats and the timed observations in the
previous sections are examples (unlike the Newtonian thermostats
in which $M$ is neither a surface nor it is bounded): provided
the collisions with the walls or between the particles are not
``hard'' but are imagined mediated by smooth potentials of very
short range.

{\it Warning:} In the following it will be often necessary to
distinguish between sets $E$ of positive surface area in the full
phase space $M$ or $\X$, as a Riemannian surface, and subsets $E$
of positive area on a surface $\AA$ of dimension smaller that of
$M$ or $\X$ .  Consistently the $\int_E dx$, if $dx$ denotes the
Riemannian surface on $\AA$, in the first case will be called
``volume'' of $E$ and in the second case ``area'' of $E$.
\label{volume-area}\index{volume \& area}

Particular attention will be dedicated to {\it attracting}
surfaces (or sets)\,\footnote{\tiny Here and in the following
distinguished from {\it attractors}.} for motions on a phase
space $M$ or $\X$,\index{attracting set}\label{attracting
  surface}\index{attractor}
\* \0{\bf Definition 1:} (Attracting surface) {\it Let $\AA$ be a
  smooth bounded surface in a phase space, $M$ for a flow $S_t$,
  or $\X$ for a map $S$. Suppose existence of constants
  $\e_0,\e,c>0$ such that if $x\in M$ and $\d(x,\AA)\le \e_0$ then
  $\d(S_tx,\AA)\le \e e^{-c t},\,t\ge0$, or if $x\in \X$ then
  $\d(S^nx,\AA)\le \e e^{-c n},\,n\ge0$, furthermore $\AA$
  contains a point with orbit dense in $\AA$.\\  $\AA$ will be
  called an {\it attracting surface} (or an {\it attracting set})
  and the set $U=\{x,\d(x,\AA)<\e_0\}$ will be an {\it
    attraction domain} for $\AA$.\index{attracting surface}} \*

In many interesting cases the full phase space $M$ or $\X$ may be
attracting surfaces (\eg in equilibrium, under the ergodic
hypothesis).

The limits \equ{e2.6.1},\equ{e2.6.2} express properties
stronger than those in the above Eq.\equ{e2.4.4}: no dependence,
and no average, on the initial data are demanded.  Existence of
the limits, outside a set of $0$ area in phase space, can be
established within a class of systems called {\it hyperbolic},
often met in studying stationary states of dynamical systems
$(\X,S)$ or $(M,S_t)$. A definition, in the case of a {\it
discrete time evolution}, is:
\*
\0{\bf Definition 2:} {\rm (Hyperbolic map)} {\it The 
  system $(\X,S)$, with $S$ a smooth invertible evolution map, is
  hyperbolic if the infinitesimal displacements $dx$, at any
  point $x\in \X$, have the properties:\index{hyperbolic
    system}\label{hyperbolicity} \\
(1) $dx$ can be decomposed as a sum $dx^s+dx^u$ along two
transverse tangent planes $V^s(x)$ and $V^u(x)$\,\footnote{\tiny
Transversality refers to the positivity of the angle $0<\f(x)<\p$
between the planes.\label{transversality}\index{transversality}}
which {\it depend continuously}\, on $x$,
\\
(2) $V^\a(x),\,\a=u,s$, are {\it covariant} under time evolution,
in the sense that $(\partial S)(x) V^\a(x)=V^\a(Sx)$, where $\dpr
S(x)$ is the linearization at $x$ of $S$ (``Jacobian matrix''),
\\
(3) under iteration of $S$ the vectors $S^k dx^s$ and $S^{-k} dx^u$ contract
exponentially fast in time $k$:
in the sense that
$|\frac{\partial S^k(x)}{\partial x}dx^s|\le C
e^{-\l k}|dx^s|$ and $|\frac{\partial S^{-k}(x)}{\partial x}
dx^u|\le C e^{-\l k}|dx^u|$, $k\ge0$, for some $x$-independent
$C,\l>0$.}\index{hyperbolicity} \*

For systems $(M,S_t)$ {\it developing in continuous time} on a
manifold $M$ a correspomding definition of hyperbolicity is
obtained by requiring on $S_t$, beyond the conditions 1-3 in the
above definition, that there is no fixed
point (\ie $|\dot x|>0$)  and  that the infinitesimal
displacements $dx$ pointing in the flow direction neither expand
nor contract with time (as they must remain bounded).\label{hyperbolic flow}

Hence, in flows, the generic $dx$ is required to be covariantly
decomposed as a sum $dx^s+dx^u+dx^0$ with $|\dpr S_t dx^0|\le C
|dx^0|$ as $t\to\pm\infty$ and $dx^s,dx^u$ exponentially
contracting respectively under $S_{\pm t}$ (\ie for some $C,\l>0$
it is $|\dpr S_t dx^s|\le C e^{-\l t} |dx^s|$ and $|\dpr S_{-t}
dx^u|\le C e^{-\l t}|dx^u|$ as $t\to+\infty$).

{\it Also systems in which time evolves only forward, \ie
$n\ge0,t\ge0,$ will be considered:} in such cases hyperbolicity
should be defined by demanding existence of the stable planes
$V^s(x)$ as above and of the unstable planes $V^u(x)$ on which
$\dpr S^n$ or $\dpr S_t$ expand exponentially vectors in $V^u(x)$
as $n,t\to +\infty$, without demanding existence of $S^n$ or
$S_t$ for $n,t<0$.
Examples arise often, for instance in
Navier-Stokes fluids, see Sec.(\,\ref{sec:III-5}\,).

It is easy to exhibit physically interesting dynamical systems
that are not hyperbolic. However here we shall only consider
hyperbolic systems: because it is for such systems that a general
theory can be built as a paradigm of nonequilibrium, \Cite{Ru999}.

Among hyperbolic systems and about the limits in
Eq.\equ{e2.6.1},\equ{e2.6.2}, a special role will be played
by the Anosov systems: \*

\0 {\bf Definition 3:} {\rm (Anosov system)} {\it A hyperbolic system
  $(M,S_t)$ or $(\X,S)$ with a point $x$ whose trajectory is
  dense (a property called ``transitivity''), and with a dense
  set of periodic points is called a Anosov's
  system.}\index{Anosov system}\label{Anosov system}\index{Anosov
  property}\label{Anosov
  flow}\label{transitivity}\index{transitivity} \*

Anosov systems have many properties whose physical meaning will
be discussed in the following and that make the evolutions
associated with such maps a paradigm of chaotic motions. Of
particular interest are systems with attracting sets
$\AA_1,\AA_2,\ldots$ on which $(\AA_j,S_t)$ are Anosov flows or
$(\AA_j,S)$ are Anosov maps.

For the moment we just mention a remarkable property,
namely:\index{SRB distribution} \*

\0{\bf Theorem:} (SRB) {\it Let $(M,S_t)$ or $(\X,S)$ be a
  hyperbolic dynamical system with an attracting surface
  $\AA\subset\X$, and let $(\AA,S_t)$ or $(\AA,S)$ be an Anosov
  system. Let $U$ be any attraction domain for $\AA$ and $\r(x)$
  be any continuous density over the volume elements $dx$ of $U$.
  There is a unique probability distribution $\m$ on $\AA$ such
  that, for all $U$ and $\r(x)$, the limits in Eq.\equ{e2.6.1}
  or, respectively, \equ{e2.6.2} exist and are $x$-independent
  for all continuous observables $F$ and for all $x$ outside a
  set of zero volume with respect to
  $\m_0(dx)=\r(x)dx$,\, \Cite{Si968a}\Cite{Bo970}\Cite{BR975}\Cite{Ru976}\,.}
\*

Under the assumption on the initial data, Sec.(\,\ref{sec:IV-2}\,)\,, it
follows that in Anosov systems the probability distributions that
give the statistical properties of the stationary states are {\it
  uniquely} determined as functions of the parameters on which
$S_t$ or $S$ depend.

Note that the arbitrariness of the density $\r(x)$, and the
$\r(x)$-independence of the averages, allows us to
introduce the notion of {\it chaotic data}: \*

\0{\bf Definition 4:} {\rm (Chaotic initial data)}
\label{chaotic data}{\it Let $(\X,S)$
  or $(M,S_t)$ be a hyperbolic dynamical system with an
  attracting surface $\AA$ such that $(\AA,S)$ or $(\AA,S_t)$ is
  an Anosov system. Initial data selected by any protocol with a
  probability distribution $\m_0(dx)=\r(x)dx$ with $\r(x)$
  continuous and concentrated on an attraction domain $U$ of
  $\AA$ will be called ``chaotic data''.}\index{chaotic data} \*

It is possible  that in a dynamical system there are
several attracting surfaces $\AA_1,\AA_2,\ldots$: initial data
will generate unique stationary states only if located in the
respective domains of attraction $U_1,U_2,\ldots$.

Furthermore it is still possible that the picture is incomplete
because, for instance, of the existence of attracting surfaces on
which the evolution is not hyperbolic, or which are periodic orbits.

To exclude the latter possibility an hypothesis will be
introduced in the next section, and systematically assumed in the
rest of the book, to adhere to the purpose of dealing with cases
that build a paradigm of chaotic motions, and to discuss
systematically consequences of the above definitions and of the
SRB theorem.

\section{Chaotic Hypothesis}
\def\SEC{Chaotic Hypothesis}
\label{sec:VII-2}\iniz
\lhead{\small\ref{sec:VII-2}.\ \SEC}

The latter mathematical results on Anosov maps and flows suggest a daring
assumption inspired by the, also certainly daring, assumption that all motions are
periodic, used by Boltzmann and Clausius to discover the relation between
the action principle and the second principle of thermodynamics, see
Sec.(\,\ref{sec:III-1}\,).

The assumption deals with properties of attracting surfaces in
hyperbolic systems with Anosov's property, see definitions in
Sec.(\,\ref{sec:VI-2}\,). It is an interpretation of a proposal
advanced by Ruelle,\index{Ruelle} \Cite{Ru978b}, in the context
of the theory of turbulence and in statistical mechanics,
\Cite{Ru999}. It is presented here as interpreted in \Cite{GC995}
and called ``{\it chaotic hypothesis}'' (CH). For empirically chaotic
evolutions, given by a map $S$ on $\Xi$, or by a flow $S_t$ on
$M$, as in Sec.(\,\ref{sec:VI-2}\,), it can be formulated
as:\label{chaotic hypothesis}\label{CH}\*

\0{\bf Chaotic hypothesis (CH):} {\it The evolution on phase space, a
  map $S$ on $\X$ or a flow $S_t$ on $M$, is hyperbolic. If
  restricted to a transitive attracting set $\AA$, it can be regarded
  as an Anosov system for the purpose of studying statistical
  properties of the stationary states. Furthermore the phase
  space contains at most a finite number of attracting sets
  $\AA_i, i=1,2,\ldots$ and all points in $M$, but a set of $0$
  volume, are attracted by $\cup_i \AA_i$.}\index{chaotic
  hypothesis} \*

This means that attracting surfaces $\AA$ can be considered ``for
practical purposes'' as smooth surfaces on which the evolution,
map $S$ or flow $S_t$, has the properties that characterize the
Anosov systems. In general there might be more than one
attracting set: $\AA_1,\AA_2,\ldots$: this will be related to the
{\it intermittency} phenomena,
Sec.(\,\ref{sec:V-5}\,).\index{intermittency}\label{intermittency}

To avoid misundertandings it is important to keep in mind that
the hypothesis is formulated for systems which exhibit
phenomenologically chaotic evolution: and not necessarily with large
number of degrees of freedom. However, as in the case of the
various forms of the ergodic hypothesis, simple counterexamples
exist. Once again the idea is to use the hypothesis to build a precise
intuition about properties of a rather general chaotic motion by
identifying it as generated by an Anosov's flow or map.
\*

\0{\it Remarks:} 
(1) The Anosov property might fail to hold for the system
$(\X,S)$ when the CH \index{chaotic hypothesis} is verified:
simply when in $\X$ there is no point with a dense orbit because
the system admits more than one invariant, closed, attracting
surfaces $\AA_i\subset \X$ which are not dense in $\X$. Still the
dynamical systems $(\AA_i,S)$ are Anosov systems: motions
starting on $\AA_i$ stay there and those starting close enough to
$\AA_i$ evolve approaching $\AA_i$ exponentially fast.
\\
(2) If a system satisfies the (CH) and
is transitive (\ie it has a dense orbit), then it is an Anosov
system.\label{tansitivity}
\\
(3) Beyond the CH more general assumptions could be
considered: however the basic proposal, {\it and the ratio behind
  the hypothesis}\index{chaotic hypothesis}, is to build
intuition about chaotic motions upon smooth and uniformly
hyperbolic dynamics. The viewpoint adopted here is that
systems with attracting sets on which motion verifies Anosov
property play a role in developing intuition about chaos
analogous to the role that in regular motions is played by the
harmonic oscillators.
\\
(4) 
Therefore CH implies that the attracting sets
$\AA_i$ should be visualized as smooth surfaces which attract
exponentially fast the nearby points: the interesting properties
of the dynamics will be related to the motions of points on such
surfaces.
\\
(5) Existence of attracting periodic orbits is excluded in the CH
by the hyperbolicity condition: however the restriction is set
only for simplicity, as adding attracting periodic orbits would
imply to impose hyperbolicity only in the attraction domains of
the other attracting surfaces and no sustantial changes.
\*

Under the (CH) the following corollary of the SRB
theorem holds (and completes the motivation of the definitions and
statements of the previous sections):
\*

\0{\bf Theorem:} {\it Consider initial data chosen with a
  probability distribution $\m_0(dx)$ with a density, on
  phase space, $\r(x)$  concentrated on an attraction domain $U$ of an
  attracting surface $\AA$. Their evolution, under the chaotic
  hypothesis, is such that the limit in Eqs.\equ{e2.6.1} or
  \equ{e2.6.2} exists for all densities $\r(x)$ and all initial
  data $x$, aside a set of $\m_0$-probability $0$ (\ie
  $0$ volume).  For all smooth $F$ the time averages are
  given by the integrals of $F$ with respect to the unique
  invariant SRB distribution $\m$ defined on
  $\AA$\,.\footnote{\tiny See the SRB theorem in
  Sec.(\,\ref{sec:VI-2}\,)}} \*

This would still hold under much weaker assumptions, see
\Cite{Ru999}, which, however, will not be discussed
given the heuristic role that is adopted here to be played by the
chaotic hypothesis.\,
\footnote{\tiny For instance if the attracting set satisfies the
property ``Axiom A'', \Cite{ER985}\Cite{Ru995}, the above theorem holds
as well as the key results, presented in the following: on
existence of Markov partitions, coarse graining and fluctuation
theorem, which are what is really wanted for our purposes, see
Sec.(\,\ref{sec:III-3},\ref{sec:VII-3},\ref{sec:VI-4}\,). The heroic
efforts mentioned in the footnote\footref{heroic} in the preface
reflect a misunderstanding of the physical meaning of the CH.}  
Concisely: the statistical properties of chaotic
initial data\index{chaotic initial data} (see the definition 4 in
Sec.(\,\ref{sec:VI-2}\,)) are independent of the protocol generating
them.

As the ergodic hypothesis is used to justify the use of the
microcanonical distributions to compute the statistical
properties of the equilibrium states, and to realize the
mechanical interpretation of the heat theorem (\ie existence of
the thermodynamic entropy function, see Ch.\ref{Ch1}\,), {\it
  likewise} the chaotic hypothesis will be used to infer features
of the probability distributions that describe statistical
properties of the more general stationary nonequilibrium states.

This is a nontrivial task, as it will be realized in the next
sections, because in nonequilibrium the SRB probability
distribution $\m$ generically will have to be concentrated on a
set of zero area {\it even when the attracting sets coincide with
the whole phase space}.

In the case in which the volume is conserved, \eg in the
Hamiltonian case, see Sec.(\,\ref{sec:X-2}\,), {\it the chaotic
hypothesis implies the consequences of the ergodic hypothesis}:
if the attracting set is the entire phase space (\ie the energy
surface, in absence of other constants of motion) the SRB
distribution is the microcanonical, which under the CH is the
only invariant distribution which has a density on phase
space.\index{chaotic \& ergodic hypotheses} This shows that
assuming the new hypothesis cannot immediately lead to a
contradiction between equilibrium and nonequilibrium statistical
mechanics. The hypothesis name has been chosen precisely because
of its assonance with the ergodic hypothesis of which it is
regarded here as an extension.

\section{Phase space contraction in continuous time}
\def\SEC{Phase space contraction in continuous time}
\label{sec:VIII-2}\iniz
\lhead{\small\ref{sec:VIII-2}.\ \SEC}

Stationary distributions for nonequilibrium systems are
concentrated on sets of zero surface area on the attracting
surfaces (even in the case in which the attracting surface is the
full phase space).

Understanding why is the same as realizing that the area
(generically) contracts under non Hamiltonian time evolution. It
is therefore convenient to discuss the phase space contraction
for a few ``simple'' evolutions and to add appropriate comments
to clarify its deep physical interpretation.

If we consider the surface element $d x$ on a phase space $M$ of
a flow $(M,S_t)$ then,
under a time evolution governed by a
differential equation $\dot x=f(x)$, the surface $d x$ is changed
into $S_t d x$ and its rate of change at $t=0$ is given by the
divergence of the equations of motion, which we denote
$\s(x)\defi-\sum_j\frac{\dpr \dot{x}_j}{\dpr x_j}$.

Given the equations of motion the surface contraction $\s_\G$, on
each surface determined by the values of the constraints $\G$,
will be computed in a few examples using its {\it identity} with
$\s$ when the constraints are constants of
motion.\,\footnote{\tiny Consider a surface $\scriptstyle P$ in
phase space $\scriptstyle M$ defined by the values of
$\scriptstyle d$ {\it constants of motion} $\G$ of an equation
$\scriptstyle \dot x=F(x)$ on $M$. The area element $\scriptstyle
d \x$ of the surface $P$ determined by given values of constants
of motion, changes per unit time by $\scriptstyle \s(x) d\x=-$
${\scriptscriptstyle \sum_j}$ $\scriptstyle \frac{\dpr\dot
x_j}{\dpr x_j}\,d\x$, \ie as much as $\scriptstyle d\V x$ itself:
there is no need, in this case, of computing parametric equations
of the surface $\scriptstyle P$ to compute the contraction on the
surface determined by $\G$.\label{surface contraction}} \*


\0{\bf(a)} {\it Isoenergetic} Gaussian thermostats model: the equation
of motion is in Fig.2.2.1 with $\a_j$ in Eq.\equ{e2.2.3}\, leads
to motions for which $K_j+U_j$ is a constant of motion. Let
$K_j\defi\frac12\dot{\V X}_j^ 2$ be the total kinetic energy in
the $j$-th thermostat, and define $\wt T_j$ by
$K_j\defi\frac32N_jk_B \wt T_j$. The constraint is
$K_j+U_j=const$ for $j\ge1$ and leads, after a brief calculation,
to a phase space contraction value:
\be \s(x)=\sum_{j>0} \frac{Q_j}{k_B \wt T_j}(1-\frac1{3N_j}),\quad
Q_j=-\partial_{\V X_j} W_j(\V X_0,\V X_j)\cdot\dot{\V X}_j,
\label{e2.8.1}\ee
The expression of $\s$, that will be called the {\it phase space
  contraction rate} of the 'Liouville volume', has the interesting
feature that $Q_j$ can be naturally interpreted as the heat that
the reservoirs receive per unit time, therefore the phase space
\index{phase space contraction} contraction contains a
contribution that can be identified as the entropy production
\index{entropy production} per unit time.
The constant factor $(1-\frac1{3N_j})$ is usually negligible if the
number of particles $N_j$ is very large as in most of the
following: if the numbers of particles are small it may play,
however, a role.
\\
Note that the name is justified {\it without any need to extend the notion
of entropy to nonequilibrium situations}: the thermostats keep the same
energy all along and may be regarded as systems receiving heat at
constant state.
\*
\0{\bf(b)} In the {\it isokinetic thermostat} case, denoting
$K_j\defi\frac32N_jk_B \wt T_j$,  $\s$ will contain a
further term equal to $\sum_{j>0} \frac{\dot U_j}{k_B
\wt  T_j}(1-\frac1{3N_j})$, which forbids us to give the naive
interpretation of entropy production rate to the phase space
contraction. In this case $K_j$ and $\wt T_j$ are constants.
\*
\0{\bf(c)} The difference betweem $\s$ in the above cases (a) and (b)
merits some comments.
\\
It has to be remarked that the above $\s$ {\it is not
  unambiguously defined}.  In fact the notion of phase space
contraction depends on what we call surface element, which above
was called $dx$. If if we look at a Gaussian thermostat with
phase space contraction $\s(x)$ with respect to the surface
element $dx$ then, given a smooth function $g(x)$, the phase
space contraction with respect to the surface element defined as
$e^{-g(x)}dx$ becomes $\s(x)+\dot g(x)$. Hence in the case of
isokinetic thermostats using as volume element $e^{-g(x)}dx$ with
$g(x)=\sum_j \frac{U_j(\V X_j)}{k_B \wt T_j}$ will give a
$\s(x)=\sum_j \frac{Q_j}{k_B\wt T_j}$ (up to the correction
factors $1-\frac1{3N_j}$).
\*
\0{\bf(d)} The remark (c) illustrates a special case of the
   {\it general property} that if the surface is measured using a
   different density, or a different Riemannian
   metric\index{Riemannian metric} on phase space, the new
   surface element contracts at a rate differing form the
   original one $\s(x)$ by a {\it time derivative} of some
   function $D$ on phase space and becomes $\s+\dot D$, see items
   (b),(c).\index{metric and contraction} \\ In other words
   $\sum_{j>0} \frac{Q_j}{k_B \wt T_j}$, in the above examples,
   does not depend on the system of coordinates, while $D$ does,
   {\it but the induced correction $\dot D$ has $0$ time
     average}, being a time derivative.

An immediate consequence is that $\s$ should be considered as
defined {\it up to a time derivative} and therefore only its time
averages over long times can possibly have a physical meaning;
the limit as $\t\to\infty$:
\be \s_+\defi \lim_{\t\to\infty}\frac1\t \int_0^\t \s(S_tx)
dt\label{e2.8.2}\ee
is independent of the metric and the density used to define the measure
of the surface elements.

In the nonequilibrium models considered in Sec.(\,\ref{sec:II-2}\,) the
value of $\s(x)$ differs from $\e(x)=\sum_{j>0} \frac{Q_j}{k_B
  \wt T_j}$ by a time derivative (neglecting, for simplicity, the
  factors $1-\frac1{N_j}$, when present) so that, at least under the
chaotic hypothesis, the {\it average phase space
  contraction\index{phase space contraction} equals the average
  entropy production\index{entropy production} rate of a
  stationary state}, \ie:
\be \s_+\equiv \e_+: \tto \hbox{\small\rm average\ 
  contraction\ $\equiv$\ average\ entropy\ production} \label{e2.8.3}\ee
An important general remark is that $\s_+\ge0$, \Cite{Ru996}, if the thermostats
are efficient in the sense that motions remain confined in phase space, see
Sec.(\,\ref{sec:I-2}\,)\,; the intuition is that it is so because $\s_+<0$ would
mean that any surface in phase space will grow larger and and larger with
time, thus revealing that the thermostats are not efficient (``it is not
possible to inflate a balloon inside a safe'').

Furthermore if $\s_+=0$ it can be shown, quite generally, that
the phase space contraction is the time derivative of an
observable,\, \Cite{Ru996}\Cite{Ru999}\,, and by choosing
conveniently the measures of the phase space surface elements, a
probability distribution will be obtained which admits a density
over phase space and which is invariant under time
evolution.\label{zero average contraction}

The remarks justify why the {\it time average} of the phase space
contraction will be often called, in the following, ``entropy creation
rate''.\index{entropy creation }

A special case is when $\s(x)\equiv0$: in this case the
normalized surface measure is an invariant distribution (as in
the case of the canonical ensemble).
\*

\0{\bf (e)} A more
interesting example is a single Gaussian thermostat with $\wt
T_1\defi (1-\frac1{3N_1})T$, see comment following
Eq.\equ{e2.8.1}, and $\V E=\V0, U_1=0$ (\ie no interaction inside
the thermostat, free thermostat,(\ref{free thermostat}) and no
external force on the system proper). Choose to measure phase
space surface elements with $\m_0(dx)=const\,\lis {dx}$ and:
\be \lis {dx}=e^{-\b(K_0+U_0+W_1(\V X_0,\V X_1)))}\prod_{j=0}^1
\d(K_j,\wt T_j)\prod_{j=0}^1 d{\V X}_j d\dot{\V
  X}_j\label{e2.8.4}\ee
with $\b=\frac1{k_B T},\wt T_0=T, \wt T_1=T (1-\frac1{3N_1})$.
The phase space contraction can be immediately computed as
$\frac{Q_1}{\wt T_1}+\b\dot H_0$, where $H_0\defi K_0+U_0+W_1(\V
X_0,\V X_1)$. However $\dot H_0=\dpr_{\V X_1} W_1(\V X_0,\V
X_1)\dot {\V X_1}=-Q_1$ and $\wt T_1=T (1-\frac1{3N_1})$ on the
surface on which the states of the system are
considered. Consequently, by (c) above, the total contraction is
$\s+\b\dot H_0=0$.

Furthermore remark that $\m_0(dx)$ is a ``isokinetic Gibbs
distribution'', \ie a probability distribution in which the total
kinetic energy of each thermostats is constant.  Therefore $\m_0$
provides an appropriate distribution for an equilibrium state,
\Cite{EM990} and \Cite{Ga000}\Cite{Ga006c}\,, and the above comments
provide a proof of stationarity and an exact solution to a
nonequilinrium problem, \Cite{EM990}: \*
\0{\bf Theorem: \it If only one free thermostat acts on the test
  system, with $N_1$ particles and temperature $k_B
  T_1=\b^{-1}(1-\frac1{3N_1})$, see comment following
  Eq.\equ{e2.8.1}, then the distribution in Eq.\equ{e2.8.4} is
  stationary.}\index{Theorem of Evans-Morriss} \* \*

The same argument can be applied to prove,\,\Cite{EM990}\,, the similar result
for the autothermostatted model in Eq.\equ{e2.2.2}: leading to
\be \lis {dx}=e^{-\b(K_0+U_0)}\d(K_0,T)\prod d{\V X}_0 
d\dot{\V X}_0\label{e2.8.5}\ee
as an exact stationary state.

The above remarkable exact results suggest to extend the notion
of equilibrium states to invariant probability distributions for
which $\s_+=0$, \Cite{Ru999}\,. In this way {\it a state is in
stationary nonequilibrium if the entropy production rate
$\e_+>0$}. The microcanonical distributions are examples and the
distributions Eq.\equ{e2.8.4} are very likely to be equivalent to
corresponding microcanonical ones.
\*

\0{\bf (f)} Several general properties on nonequilibrium stationary
states will be based, in the following Chapter \ref{Ch4}, on
properties of quantities that will be often called {\it
  dimensionless entropy\index{dimensionless entropy production}
  production rate} and {\it phase space contraction} defined as
$\frac{\e(x)}{\e_+}$ and $\frac{\s(x)}{\s_+}$, with
$\e,\s,\e_+,\s_+$ defined in (d) above. Since $\e$ and $\s$
differ by a time derivative of some function $D(x)$, it is
$\e_+=\s_+$ and the finite time averages:
\be p=\frac1\t\int_0^\t \frac{\e(S_tx)}{\e_+}\,dt \qquad\hbox{and}\quad
p'=\frac1\t\int_0^\t \frac{\s(S_tx)}{\s_+}\,dt\label{e2.8.6}\ee
will differ by $\frac{D(S_\t x)-D(x)}\t$: a quantity which will
tend to $0$ as $\t\to\infty$ (in Anosov systems or under the
chaotic hypothesis as $D(x)$, being smooth, is bounded).
Therefore for large $\t$ the statistics of $p$ and $p'$ in the
stationary state will be close, at least if the function $D$ is
bounded (as in Anosov systems).
\*

\0{\bf(g)} It has to be remarked that the value of the constant $\s_+$
is a well known quantity in the theory of dynamical systems:
associated with an attracting surface it is related to the Lyapunov
spectrum.\index{Lyapunov spectrum}
\\
Let $S_t$ be a flow, verifying the CH, on the $N$-dimensional manifold
$M$, with an attracting surface $\AA$; let $\dpr S_t$ be the $N\times
N$-dimensional Jacobian matrix of $S_t$:
\* \0{\bf Definition:} (Lyapunov's exponent) {\it The $N\times N$ matrices $\log (\dpr
  S_t(x)^*\dpr S_t(x))^{\frac1{2t}}$ have a spectrum
  $\l_0(x,t)\ge\l_1(x,t)\ge\ldots\ge\l_N(x,t)$ such that
  $\l_j=\lim_{t\to\infty}\l_j(x,t)$ exist for chaotic initial
  data in the attraction domain of $\AA$, (see
  Sec.(\,\ref{sec:VI-2}\,)).  They define the local Lyapunov
  spectrum of $S_t$ and, respectively, the average Lyapunov
  spectrum on $\AA$ with respect to the SRB
  distribution.}\index{Lyapunov spectrum}
\footnote{\tiny The Lyapunov exponents can more generally be
associated with any invariant probability distribution: therefore
it is in general necessary to specify the distribution to which
they are associated.\label{Lyapunov spectrum}}  \*

Then:
\be\s_+=-\sum_i \l_i\label{e2.8.7}\ee
$\l_i$ being the SRB Lyapunov exponents of $S_t$ on the attracting set for
$S_t$.

In the case of discrete time systems (not necessarily arising
from timed observations of a continuous time evolution) the
Lyapunov sectrum can be analogously defined in terms of the
Jacobian matrices $\dpr S^n$ for time evolutions defined by maps
$S$. Likewise there is a direct relation between $\s_+$ and the
average Lyapunov spectrum.

 \*

\section{Timed observations phase space contraction \index{timed observation}}
\def\SEC{\index{Timed observations phase space contraction}}
\label{sec:IX-2}\iniz
\lhead{\small\ref{sec:IX-2}.\ \SEC}

The average phase space contraction
(per timing interval) can be naturally defined for a system
$(\X,S)$ as:

\be \s_+=\lim_{n\to+\infty}-\frac1n \sum_{j=1}^n \log|\det\frac{\dpr S}{\dpr
  x}(S^jx)|
\label{e2.9.1}\ee 
in analogy with the corresponding definition for flows. The
definitions and results mentioned in the case of flows can be
immediately adapted to the case of timed observations.

It is worth to mention that it is desirable to try to use timed
observations when a flow $(M,S_t)$ might be not smooth on the
full phase space (before taking into account the constraints) so
that the evolution of a datum might visit sites with very large
surface contraction or expansion, difficult to follow. For instance
there are several interesting interaction models in which the pair
potential is unbounded above: like models in which the molecules
interact via a Lennard-Jones potential. As mentioned in
Sec.(\,\ref{sec:I-1}\,) these are cases in which observations timed to
suitable events may become particularly useful.

In the case of unbounded potentials (and finite thermostats) a
convenient timing could possibly be when the minimum distance
between pairs of particles reaches, while decreasing, a prefixed
small value $r_0$; the next event will be when all pairs of
particles, after separating from each other by more than $r_0$,
come back again with a minimum distance equal to $r_0$ and
decreasing: thus a timing events surface $\Xi$ is defined in the
phase space $M$ and the relative Poincar\'e's section avoids
systematically measuring quantities difficult to access.

An alternative Poincar\'e's\index{Poincar\'e's section} section
could be proposed as the set $\Xi\subset M$ of configurations in
which the total potential energy $W=\sum_{j=1}^n W_j(\V X_0,\V
X_j)$ equals a prefixed bound $\lis W$ with a derivative $\dot
W>0$.

Studying a system $(M,S_t)$, it may be possible to proceed as
follows.  Let $\t(x), \,x\in\Xi$ be the time interval from the
realization of the event $x$ to the realization of the next one
$x'=S_{\t(x)}x$. The phase space contraction is then $ \exp
-\int_0^{\t(x)} \s(S_t x)dt$, in the sense that the
(infinitesimal) surface element $\D_x$ in the point $x\in\Xi$,
becomes in the time $\t(x)$ a surface element $\D_{x'}$ around
$x'=S_{\t(x)}x$ with:
\be \frac{|\D_{x'}|}{|\D_x|}=\,e^{-\int_0^{\t(x)}
\s(S_t x)dt} \label{e2.9.2}\ee
Therefore if:
\\ (1)
$\s(x)=\e(x)+ \dot D(x)$ with $D(x)$ {\it bounded on $\Xi$} (but possibly
with large variations on the full phase space $M$), see remark
(d) in Sec.(\,\ref{sec:VIII-2}\,)\,.
\\ (2) $\t(x)$ is bounded and,
outside a set of zero surface on $\X$, has average $\lis\t>0$ 
\\ 
then, setting $\wt \e(x)= \int_0^{\t(x)} \e(S_t x)dt$, it follows
that the entropy production rate and the phase space contraction
have the same average $\e_+=\s_+$ and likewise
\be p=\frac1m\sum_{k=0}^{m-1} \frac{\wt\e(S^k
  x)}{\wt \e_+}\qquad\hbox{and}\quad
p'=\frac1m\sum_{k=0}^{m-1} \frac{\wt\s(S^k
  x)}{\wt \s_+}
\label{e2.9.3}\ee
will differ by 
$\fra1m \big(D(S^mx)-D(x)+\log v_{S^mx} -\log v_x\big)\tende{m\to\infty}0$.

Therefore in cases in which $D(x)$ might have large variations in
phase space, but on a timing section $\Xi$ it has 
smaller variation, while properties (1)-(2) above hold,
it might be convenient to suppose the CH hypothesis for the
evolution $S$ timed on $\Xi$ rather than trying to apply the CH
to evolutions $S_t$ in continuous time on the full phase space.
Because the actual contraction $e^{-\int_0^{\t(x)} \s(S_t x)dt}$
can be measured in simulations via the ratio Eq.\equ{e2.9.2},
without the need of computing $\s(S_tx)$ for $t\in[0,\t(x)]$, and
$D(x)$ does not affect the values of $p,p'$ in long time averages.

An example of application of the above considerations will be
found in Sec.(\,\ref{sec:IX-5}\,) and studies a case in which
they can be used in a concrete problem to test CH.

\section{Conclusions}
\def\SEC{Conclusions}
\label{sec:X-2}\iniz
\lhead{\small\ref{sec:X-2}.\ \SEC}

Nonequilibrium systems like the ones modeled in
Sec.(\,\ref{sec:II-2}\,) undergo, in general, motions which are
empirically chaotic at the microscopic level. The
CH, (\,\ref{chaotic hypothesis}\,), means that we may as well
assume that the chaos is maximal, \ie it arises because the
evolution has the Anosov property on the attracting
surfaces.\ref{Anosov system}

The evolution can be studied through timing events and is therefore
described by a map $S$ on a ``Poincar\'e's section'' $\X$ in the
phase space $M$.

It is well known that in systems with few degrees of freedom the attracting
sets are in general fractal sets: the CH implies that
instead one can neglect the fractality (at least if the number of degrees
of freedom is not very small) and consider the attracting sets as smooth
surfaces on which motion is strongly chaotic in the sense of Anosov.

The hypothesis implies (therefore) that the statistical
properties of the stationary states are those exhibited by
motions \\
(1) that follow initial data randomly chosen with any
distribution with density over phase space ({\it chaotic data})
\\
(2) that are strongly chaotic (\ie have the {\it Anosov property} on
the attracting surfaces).
\\
(3) and the stationary states of the system are described by the
SRB distributions $\m$ which are {\it uniquely} associated with {\it each}
attracting set.

Systems in equilibrium (which in our models means that neither
non conservative forces nor thermostats act) satisfying the CH
usually have no attracting set other than the whole phase space,
which is the energy surface,\footnote{\tiny Excluding, as usual,
specially symmetric cases, like spherical containers with elastic
boundary and cases of phase transitions.}  and have, as unique
SRB distribution, the Liouville distribution, \ie the CH implies
for such systems that the equilibrium states are described by
microcanonical distributions\index{chaotic \& ergodic
  hypotheses}. This means that non equilibrium statistical
mechanics based on the CH cannot enter into conflict with the
equilibrium statistical mechanics based on the ergodic
hypothesis.

The main difficulty of a theory of nonequilibrium is that
whatever model is considered, {\it e.g.}  any of the models in
Sec.(\,\ref{sec:II-2}\,)\,, there will be dissipation which manifests
itself through the non vanishing divergence of the equations of
motion: this means that phase space volume is not conserved, no
matter which metric we use for it, unless the time average $\s_+$
of the phase space contraction vanishes. Furthermore introduction
of non Newtonian forces can only be avoided by considering
infinite thermostats.

Since the average $\s_+$ cannot be negative in the nonequilibrium
stationary systems, \Cite{Ru996}, its positivity is identified with the
signature of a genuine nonequilibrium, while the cases in which
$\s_+=0$ are equilibrium systems, possibly ``in disguise'' \,
\ref{zero average contraction}\,. If $\s_+>0$ there cannot be any
stationary distribution which has a density on phase space: the
stationary states give probability $1$ to a set of configurations
which have $0$ volume in phase space (yet they may be dense in
phase space, and often are if $\s_+$ is small, see comments below).

Therefore any stationary distribution controlling the statistical
properties of a nonequlibrium state {\it cannot be described} by
a suitable density on phase space or on the attracting surfaces,
thus obliging us to develop methods to study such singular
distributions.

A point that will become important is that, in a system for which
the CH holds, {\it there might be several attracting surfaces}
$\AA_i$ contained in the phase space $M$. The case of a single
attracting surface $\AA$ which also coincides with $M$ is very
special: it is verified in equilibrium, if CH holds. And out of
equilibrium is verified when the non Hamiltonian and thermostatic
forces are sufficiently ``small'', because one of the key
properties is the ``{\it structural stability}'' of Anosov
evolutions on an attracting surface $\AA$: they retain, on the
same $\AA$, the Anosov property.\label{structural stability}
\footnote{\tiny At small perturbation an Anosov system can be
mapped via a change of coordinates back to the not forced system
(``structural stability of Anosov maps'').}

If CH is found too strong, one has to rethink
the foundations: the approach that Boltzmann used in his
discretized view of space and time, started in
\Cite{Bo868-a}[\#5] and developed in detail in
\Cite{Bo877-b}[\#42], could be a guide: it will be
attempted in the next Ch.\ref{Ch3}\,.


\chapter{Discrete phase space}
\label{Ch3} 

\chaptermark{\ifodd\thepage
Discrete phase space\hfill\else\hfill 
Discrete phase space\fi}
\kern2.3cm
\section{Recurrence}
\def\SEC{Recurrence}
\label{sec:I-3}\iniz
\lhead{\small\ref{sec:I-3}.\ \SEC}

Simulations have played a key role in recent studies on
nonequilibrium. And simulations operate on computers to
approximate solutions of differential equations: therefore phase
space points are given a digital representation which might be
very precise but rarely goes beyond $64$ bits per coordinate. If
the system contains a total of $N$ particles each of which needs
$d$ coordinates to be identified (in the simplest models $d=4$ if
$2$--dimensional, or $d=6$ if $3$-dimensional): this gives a phase
space (virtually) containing $\NN_{tot}=(2^{64})^{dN}$ points
which cover a phase space region of desired size, $V$ in velocity
and $L$ in position, with a lattice of mesh $2^{-64}V$ and
$2^{-64}L$ respectively.

Therefore the ``fiction'' of a discrete phase\index{discrete phase space}
space, used by Boltzmann in his foundational works,
\Cite{Bo877-b}[\#42,p.167], has been taken extremely seriously in modern
times, with the peculiarity that it is seldom even mentioned in the
numerical simulations.

A simulation is\index{simulation} a code that operates on discrete phase
space points transforming them into other points. In other words it is a
map $\lis S$ which associates with any point on phase space a new one with
a precise deterministic rule that can be called a {\it program}.

With very rare exceptions, \Cite{LS024}\Cite{LV993}\,, developed with
extreme care to avoid round-off errors and precisely for the
purpose of investigating irreversibilty and estimating recurrence
times, all programs for simulating solutions of ordinary
differential equations have some serious drawbacks. For instance
the map $\lis S$, defined by a program, might be not invertible:
different initial data might be mapped by $\lis S$ into the same
datum; unlike the true solution to a differential equation of
motion, which obeys a uniqueness theorem.

Since the number $\NN_{tot}$ is finite, all points will undergo a
motion, as prescribed by $\lis S$, which will become {\it
  recurrent}, \ie will become eventually a permutation of a
subset of the phase space points, hence {\it
  periodic}.\label{recurrent}\index{recurence}

The ergodic hypothesis\index{ergodic hypothesis}\index{ergodic
  permutation}, in the strict original formulation (interpreted
on discretized evolution, see Ch.\ref{Ch1} and, in particular,
Sec.(\,\ref{sec:III-1}\,))), was born out of the natural idea that the
permutation would be a {\it one cycle} permutation: every point,
representing a microscopic state, would recur and continue in a cycle,
\Cite{Ga995a}. 

In simulations, even if dealing with time
reversible systems, it would not be reasonable to assume that all
the phase space points are part of a permutation, because of the
mentioned non invertibility of essentially any program. It is
nevertheless possible that once the {\it transient} points (\ie
points that, being out of the permutation cycles, never recur)
are discarded, motion reduces to single cycle permutation
depending on the initial state.

So in simulations of motions of isolated systems the ergodic
hypothesis\index{ergodic hypothesis} really consists in two
parts: first, the non recurrent phase space points should be
``negligible'' and, second, the evolution of the recurrent points
consists in a single cycle permutation. Two comments:
\*
\0(a)
Periodicity is not in contrast with chaotic behavior: this is a
point that Boltzmann and others, \eg Thomson\index{Thomson (Lord
Kelvin)}, (\,\Cite{Bo877a}[\#39],\Cite{Th874}\,, clarified in
several papers, see also Sec.(\,\ref{sec:XI-6}\,) below) for the
benefit of the few that, at the time, listened.
\\ (b) The recurrence times\index{recurrence time} are beyond any
observable span of time (as soon as the particles number $N$ is
larger than a few units), \Cite{Bo896a}[Sec.88]:
whether in simulations or in real flows.
\*

Nevertheless in the above {\it discrete} form the ergodic
hypothesis, \ie existence of a periodic orbit which attracts
motions, can be formulated also for general nonconservative (\ie
dissipative) motions: assuming that {\it on attracting sets}
(discrete approximations of attracting surfaces) non
recurrent points are negligible and the recurrent points form a
one cycle permutation.  In other words, in this form, {\it the
ergodic hypothesis is the same for conservative and dissipative
systems provided phase space is identified with a transitive
attracting set}.  The statistics of the motions will be
uniquely determined by assigning a probability $\NN^{-1}$ to each
of the $\NN$ recurrent points on the attracting surface: and this
will explain why there is a {\it unique} stationary distribution
of physical interest, see Sec.(\,\ref{sec:VIII-3}\,).

In presence of dissipation the {\it non recurrent points will
  necessarily be ``most'' points}: reflecting the properties of
  dynamical systems $(M,S_t)$ or $(\X,S)$ which verify the
  chaotic hypothesis (CH), \index{chaotic
  hypothesis} Sec.(\,\ref{sec:VII-2}\,).  In the latter systems
  motions, in fact, develop approaching a set, which will have
  zero area (relative to the attracting surface, even when the
  attracting surface is the full phase space) because the area of
  any set contracts, \Cite{Ru996}, being asymptotically, hence
  forever, decreasing.

The suggested discretized picture is that contraction continues
  towards an attracting set: the latter is a periodic orbit that
  simulates a trajectory filling, in the non discretized phase
  space, densely a surface $\AA$ on which the map $(\AA,S)$ or
  the flow $(\AA,S_t)$ is an Anosov system (Sec.(\ref{sec:VI-2}),
  definition 3) and the periodic orbit may depend on the initial
  data (in the cases in which there may be more than one
  attracting surface $\AA$).

The statistical properties of the stationary states are carried by
uniform distributions on each of the recurrent, periodic, orbits: which
represent the discrete analogue of the SRB distributions 
supported (\ie attribute probability $1$) on an invariant subset
of $0$ area (on each of the attracting surfaces $\AA$, see
Sec.(\,\ref{sec:VII-2},\ref{sec:VIII-2}\,).  Even in the cases in
which the attracting set $\AA$ is the entire phase space, like in
the small perturbations of conservative Anosov systems, see
Sec.(\ref{sec:X-2}), the SRB distribution will, in general,
attribute probability $1$ to an invariant subset of $0$ area in
$\AA$ (to which would correspond a single periodic orbit)\,
\footnote{\tiny The concept of {\it attractor}, that will be
seldom used in the following, would be handy\label{attractor}
here as it distinguishes between what is called an attracting
surface and the invariant sets with minimal Hausdorff dimension
among those with SRB-probability $1$: and the latter could
naturally be identified with the recurrent points.}

Of course in simulations of chaotic motions periodicity is
usually not observable as the time scale for the recurrence will
remain out of reach (in equilibrium as well as out of
equilibrium). The chaotic nature of the motions hides a very
regular (and somewhat uninteresting) periodic
motion\index{periodic motion}, \Cite{Ga995a}.
\* \*

\0{\bf Remarks:} (1) It is by no means obvious that the motion
can be described by a permutation of the points of a regularly
discretized phase space, not even in equilibrium.  However in
simulations particles positions and velocities are often
digitally represented by points on a regular lattice.

\0(2) Occasionally an argument is found whereby, in equilibrium,
motion can be regarded as a permutation ``because of the volume
conservation due to Liouville's theorem''.\index{Liouville
  Theorem} But this {\it cannot} be a sensible argument due to the
chaoticity of motion: any volume element will be deformed under
evolution and stretched along certain directions while it will be
compressed along others. Therefore the points of the discretized
phase space should not be thought as small volume elements, with
positive volume, but precisely as individual ($0$ volume) points
which the evolution permutes.

\0(3) Boltzmann argued\index{Boltzmann}, in modern terms, that
after all we are interested in are very few observables: in their
averages and in fluctuations (over time) of their
values. Therefore we do not have to follow the details of the
microscopic motions and ``all'' we have to consider are the time
averages of a few physically important observables
$F_1,F_2,\ldots, F_n$, $n$ small. This means that we have to
understand what is now called a {\it coarse grained}
representation of motion, collecting together all points on
which the observables $F_1,F_2,\ldots, F_n$ assume the same
values. Such a collection of microscopic states has been called a
{\it macrostate}, \Cite{Le993}\Cite{GGL005}.  \\
The reason why motion appears to reach stationarity and to stay
in that situation is that for the overwhelming majority of the
microscopic states in the recurrent cycles, \ie recurrent points
of a discretized phase space, the interesting observables have
the same values.

Deviations of the observed $F_j$-values with respect to their
averages are observable over time scales that are most often of
human size and have nothing to do with the recurrence
times. Boltzmann gave a very clear and inspiring view of this
mechanism by developing the 
equation bearing his name,\index{Boltzmann's equation}
\Cite{Bo872}[\#22]: entering into its full implications
only a few years later when he had to face the conceptual
objections of Loschmidt and\index{Loschmidt} others,
\Cite{Bo877a}[\#39], and Sec.(\,\ref{sec:XI-6}\,).

\section{Stable \& unstable manifolds}
\def\SEC{Stable \& unstable manifolds}
\label{sec:II-3}\iniz
\lhead{\small\ref{sec:II-3}.\ \SEC}

The possibility of finding discrete representation of motions is
related to geometric features of ``hyperbolicity'', 
shared by the systems which verify the chaotic hypothesis,
see definition 2, Sec.(\,\ref{sec:VII-2}\,).

If the dynamics is represented by a map $S:\Xi\to\Xi$ and is
chaotic in the sense that the system is an Anosov system, see
defintion 3, Sec.(\,\ref{sec:VI-2}\,), then more detailed knowledge
is available about 
$(\X,S)$,\,\Cite{AS967}\Cite{KH997}[p.267],
and it will be used in the following.

Namely hyperbolicity implies that at each point $x\in\X$ there
are two tangent planes $V^s(x),V^u(x)$, continuosly dependent on
$x$, whose vectors are uniformly exponentially contracted,
respectively, by the action of the Jacobian matrices $\frac{\dpr
  S^n}{\dpr x}$ as $n\to\pm\infty$; nemely there are constants
$C,\l>0$ and:
\be \eqalign{
  |\partial S^n(x) \,dx|& \le C e^{-\l n}\,|dx|, \qquad
  n\ge0,\ dx\in V^s(x)\cr
    |\partial S^{-n}(x) \,dx|& \le C e^{-\l n}\,|dx|, \qquad
  n\ge0,\ dx\in V^u(x)\cr}  
\label{e3.2.1}\ee
Furthermore, if $(\X,S)$ is an Anosov system, the planes
$V^u(x),V^s(x)$ can be shown to be ``integrable'' families, \ie
$V^u(x)$ and $V^s(x)$ will be everywhere tangent to smooth
surfaces, $W^u(x)$ and $W^s(x)$, without boundary, continuously
dependent on $x$ and, for all $x\in\X$, dense on $\X$,
\Cite{AS967}.  \*

The surfaces $W^{\a}(x), \, \a=u,s$, are called stable
\index{stable manifold} and unstable manifolds\index{unstable
  manifold} through $x$ and their existence, as smooth, dense,
surfaces without boundaries,\, \Cite{Si968a}\,, will be taken
here as a property characterizing the kind of chaotic motions in
the system.

A simple but, at first, unintuitive property of the invariant
manifolds is that {\it although} they are locally smooth surfaces
(if $S$ is smooth) their tangent planes $V^{\a}(x)$ are not very
smoothly dependent on $x$ if $x$ is moved along a $W^\b$: if, in
dimension $2$, the $V^{\a}(x)$ are defined by assigning their
unit normal vectors $n^\a(x)$ (in a coordinate system) and
$d(x,y)$ is the distance between $x$ and $y$, then it is in
general $|n^\a(x)-n^{\a}(y)|\le L_c d(x,y)^c$ for $y\not\in
V^\a$, where $c<1$ can be prefixed arbitrarily close to $1$ at
the expenses of a suitably large choice of
$L_c$, \Cite{Ru989b}\,.
The $n^\a(x)$, in general, are not smooth, see
Appendix \ref{appF} for an argument explaining why, in spite of
smoothness of Anosov maps, only H\"older continuity of the planes
$V^\a(x)$ can, in general, be expected.\,\footnote{\tiny More generally call
$n^\a(x)$ a unit ball, centered in $x$, in the plane orthogonal
to $V^\a(x)$ and move $x$ to $y$ on $V^\b(x)$: then the angle
between $n^\a(x)$ and $n^\a(y)$ is $\le L_c d(x,y)^c$ with $c<1$.}

This implies that, although the Jacobian $\dpr S(x)$ of the map is
a smooth function of $x$, nevertheless the restriction $\dpr
S(x)|_\a$ of $\dpr
S(x)$ to vectors tangent to the manifolds $W^\a(x)$ or its {\it local
expansion rates}, \ie the logarithms of the determinants:
\be \l_\a(x)\defi \log|\det\dpr S(x)|_{V^\a(x)}|,
\,\a=s,u\label{e3.2.2}\ee
depend on $x$ {\it only H\"older continuously}. Namely constants
$L^\a_c$ exist such that\label{Holder continuity}\index{H\"older
continuity}:
\be \cases{
\|\dpr S(x)_{V^\a(x)}-\dpr S(y)_{V^\a(y)}\|\le L^\a_c |x-y|^\b\cr
|\l_\a(x)-\l_\a(y)|\le L^\a_c |x-y|^c\cr
  }
\label{e3.2.3}\ee
for $\a=u,s$ and $c<1$. 

This apparent anomaly%
\footnote{\tiny Naive expectation would have been, perhaps, that
if the manifold $\X$ and the map $S$ are smooth, say $C^\infty$,
also $V^\a(x)$ and $W^\a(x)$ depend smoothly on $x$.} manifests
itself formally because it is possible to give a formal expression for the
derivatives of $n^\a(x)$ which gives a finite value for the
derivatives of $n^\a(x)$ along $W^\a(x)$, but a value of the
derivatives along $W^{\a'}(x)$ undefined if
$\a'\ne\a$.\,\footnote{\tiny See Sec.(10.1) and Problem 10.1.5 in
\Cite{GBG004}, see also Appendix \ref{appF}.}

\section{Geometric aspects of hyperbolicity. Rectangles.}
\def\SEC{Geometric aspects of hyperbolicity. Rectangles.}
\label{sec:III-3}\iniz
\lhead{\small\ref{sec:III-3}.\ \SEC}

A hyperbolicity feature is that it leads to a
natural definition of coarse grained phase space partitions,
whose elements will be called {\it rectangles} and, after setting
up some geometrical definitions, their properties will be
discussed in the next sections.

Here it will be pointed out that in Anosov systems $(\X,S)$ the
manifolds $W^s,W^u$, with tangent planes $V^u(x),V^s(x)$ at every
$x$, can be used to build ``boxes'' (``rectangles'') with
boundaries on the manifolds $W^s,W^u$ which generate a digital
representation of the chaotic motions due to the action of $S$ on
$\X$.  \*

\0{\it Warning: \rm the analysis that follows will consider systems
  with $\X$ of arbitrary dimension: however, at
  first, consider the case of dimension $2$. The
  cases of dimension $\ge3$ present difficulties due to the
  mentioned non smoothness of the $x$ dependence of the tangent
  planes $V^u(x),V^s(x)$: for further details see
  \Cite{AS967}\Cite{Si968b}\Cite{Bo975}.}  \*

A geometric consequence of the Anosov's property of $(\X,S)$ is
that it is possible to give a natural definition of sets $E$
which have a boundary $\dpr E$ which consists of two parts one of
which, $\dpr^s E$, is compressed by the action of the evolution
$S$ and stretched under the action of $S^{-1}$ and the other,
$\dpr^uE$ to which the opposite fate is reserved. Such sets are
called {\it rectangles} and their construction is discussed in
this section.

Inside the ball $B_\g(x)$ of radius $\g$ centered at $x$ the surface
elements $W^s_\g(x)= W^s(x)\cap B_\g(x), W^u_\g(x)=W^u(x)\cap
B_\g(x)$ {\it containing $x$ and connected}\,
\footnote{\tiny Notice that, under the CH, motions
on the attracting sets $\AA$ are Anosov systems and the stable
and unstable manifolds of every point can be shown to be dense on
$\AA$, \Cite{Si968a}\,. Therefore $W^s(x)\cap B_\g(x)$ is a dense
family of ``layers'' in $B_\g(x)$, but only one is connected and
contains $x$, if $\g$ is small enough.\label{dense manifolds}}
have a simple geometric property.

Namely, if $\g$ is small enough (independently of $x$), near two
close points $\x,\h\in B_\g(x)$ on $\X$, by the hyperbolicity
implied by the CH, there will be a point 
$\z\defi[\x,\h]\in \X$,

\kern-6mm
\eqfig{130 }{110 }
{\ins{36}{69}{$\h$}
\ins{70}{60}{$\x$}
\ins{110}{52}{$W_\g^u(\x)$}
\ins{-5}{36}{$\z\defi[\x,\h]$}
\ins{12}{93}{$W^s_\g(\h)$}
\ins{63}{83}{$x$}}
{fig3.3.1}{fig3.3.1}

{\tiny\begin{spacing}{0.5}\0 Fig.(3.3.1): Representation of the
  operation that associates $\scriptstyle [\x,\h]$ with the two
  points $\scriptstyle \x$ and $\scriptstyle \h$: as the
  intersection of a short connected part $\scriptstyle
  W^u_\g(\x)$ of the unstable manifold of $\scriptstyle \x$ and
  of a short connected part $\scriptstyle W^s_\g(\h)$ of the
  stable manifold of $\scriptstyle \h$. The size $\g$ is short
  ``enough'', compared to the diameter of $\scriptstyle \Xi$, and
  it is represented by the segments to the right and left of
  $\scriptstyle \h$ and $\scriptstyle \x$. The ball $\scriptstyle
  B_\g(x)$ is not drawn.\end{spacing}} \*

\0whose $n$-th iterate in the past will be exponentially approaching
$S^{-n}\x$ while its $n$-th iterate in the future will be exponentially
approaching $S^n\h$, as $n\to+\infty$.  

This can be used to define special sets $E$ that will be called
{\it rectangles}, because\index{rectangle axes} they can be drawn
by giving two ``axes around a point $x$'', $C\subset W^u_\g(x)$,
$D\subset W^s_\g(x)$ ($\g$ small) as in Fig.(3.3.2):\index{axis}

\eqfig{200}{120}{
\ins{80}{85}{$C$}
\ins{74}{39}{$D$}
\ins{58}{69}{$x$}
\ins{120}{96}{$W^u_\g(x)$}
\ins{115}{27}{$W^s_\g(x)$}
\ins{22}{14}{$A$}}
{fig3.3.2}{fig3.3.2}\label{Fig.3.3.2}

\0{\tiny Fig.(3.3.2): rectangle $\scriptstyle A$ with axes
  $\scriptstyle C,D$ crossing at a center $\scriptstyle x$.\label{Fig.(3.3.2)}}  \*

\0The axes around $x$ will be simply connected surface elements,
intersecting transversally at $x$, \footref{transversality}\,, with a boundary
which has zero measure relative to the area on $W^s_\g(x)$ or
$W^u_\g(x)$ and which are the closures of their internal points (relative
to $W^s_\g(x)$ and $W^u_\g(x)$); an example are the connected parts,
containing $x$, of the intersections $C=W^u(x)\cap B_\g(x)$ and
$D=W^s(x)\cap B_\g(x)$. 

The boundaries of the surface elements will be either $2$ points, if the
dimension of $W^{s}(x),W^{u}(x)$ is $1$ as in the above figure or, more
generally, continuous connected surfaces each of dimension one unit lower
than that of $W^{s}(x),W^{u}(x)$.

Then define $E$ as the set
\be E=C\times D\equiv [C,D]\defi\cup_{{y\in C,\, z\in D}}\{
    [y,z]\},\label{e3.3.1}\ee
and call $x$ the {\it center} of $E$ with respect to the {\it pair of
  axes}, $C$ and $D$.\label{axes}  This is illustrated in Fig.(3.3.2). We shall say
that $C$ is an unstable axis and $D$ a stable one,\,\equ{e3.3.1}\,.

A given rectangle $E$ can be constructed as having any internal
point $y\in E$ as center, by choosing an appropriate pair of
axes.\footnote{\tiny If $C,D$ are chosen, as in the example, as
intersections between the ball $B_\g(x)$ with $W_u(x)$ and
$W_s(x)$ then their boundaries are smooth, whichever is the
dimension of $\X$. But the construction of $E$ involves, for
instance, moving a point $y$ on the boundary of $C\subset W_u(x)$
and drawing $W_s(y)$: therefore the latter boundary varies, in
general, only H\"older-continuously with $y$ so that, even in the
example, the boundary of $E$ is only H\"older continuous, unless
$\X$ has dimension $2$, making visualization of $E$ in higher
dimension more difficult.}

If $C,D$ and $C',D'$ are two pairs of axes for the same rectangle
we say that $C$ and $C'$, or $D$ and $D'$, are ``parallel'' (for
instance it is either $C\equiv C'$ or $C\cap
C'=\emptyset$).\,\footnote{\tiny The boundary of $C$,
intersection of $W_s(x)$ with $B_\g(x)$ is smooth, but the
boundary of $E$ traced by $W_s(y)$ while $y$ is over $D$ will
only be H\"older continuous. So only in the $2$-dimensional
systems the rectangle $E$ has a smooth boundary, as the figure
may suggest.}

The boundary of $E=C\times D$ is composed by sides $\partial^s
E$, $\partial^u E$, see Fig.(3.3.3), each not necessarily
connected as a set. For instance, in $2$-dimensions, the sides of
$E=C\times D$ break in disconnected parts: the
first parallel to the stable axis $C$ and the other to the
unstable axis $D$, see Fig.(3.3.3), and they can be defined in terms of the
boundaries $\partial C$ and $\partial D$ of $C$ and $D$
considered as subsets of unstable and stable lines that
contain them.
%
\eqfig{160}{135}{ \ins{85}{60}{$x$} \ins{55}{127}{$D$}
  \ins{112}{22}{$ D$} \ins{118}{100}{$C$} \ins{3}{43}{$C$}
  \ins{92}{115}{${\dpr}^sE$} \ins{21}{20}{$\dpr^s E$}
  \ins{80}{10}{$\dpr^u E$} \ins{18}{99}{$\dpr^u E$} }
      {fig3.3.3}{fig3.3.3}
\*

{\tiny\begin{spacing}{0.5}\0Fig.(3.3.3): The stable and unstable
  boundaries of a rectangle $\scriptstyle E=C\times D$ in the
  simple $\scriptstyle 2$-dimensional case in which the boundary
  really consists of two pairs of parallel axes.\end{spacing}}  \*

The stable and unstable parts of the boundary are defined as
\be\dpr^s E=[\dpr C,D] , \qquad
\dpr^u E=[C,\dpr D] .\label{e3.3.2}\ee
which in the $2$-dimensional case consist of two pairs of parallel
lines, as shown in Fig(3.3.3). 

If $C,C'$ are two parallel {\it stable} axes of a rectangle $E$,
a map $\th:C\to C'$ can be established by defining $\x'=\th(\x)$
if $\x,\x'$ are on the {\it same unstable} axis through $\x\in
C$: then it can be shown,\,\Cite{AS967}\,, that the map $\th$
maps sets of positive relative area on $C$ to sets of positive
relative area on $C'$. The corresponding property holds for the 
correspondence established between two parallel unstable axes
$D,D'$ by their intersections with the stable axes of $E$.

This property is called {\it absolute continuity}
\index{absolute continuity of foliations}\index{absolute continuity}
of the foliations $W^s(x)$ with respect to $W^u(x)$ and of
$W^u(x)$ with respect to $W^s(x)$.

\* \0{\bf Remark:} As mentioned any point $x'$ in $E$ is the
intersection of unstable and stable surfaces $C',D'$ so that $E$
can be written also as $C'\times D'$: hence any of its points can
be a center for $E$. It is also true that if $C\times D=C'\times
D'$ then $C\times D'=C'\times D$.  For this reason given a
rectangle any such $C'$ will be called an {\it unstable axis} of
the rectangle and any $D'$ will be called a {\it stable axis} and
the intersection $C'\cap D'$ will be a point $x'$ called the
center of the rectangle for the axes
$C',D'$.\label{stable-unstable-axes}\index{stable-unstable-axes}
The boundaries of $E$ are smooth surfaces in dimension $2$ and,
therefore, have $0$ area on $\X$: a property which holds also in
dimension larger than $2$ but requires a proof as the boundaries
are in general not smooth as mentioned above.

\section{Symbolic dynamics, chaos, mixing time}
\def\SEC{Symbolic dynamics, chaos, mixing time}
\label{sec:IV-3}\iniz
\lhead{\small\ref{sec:IV-3}.\ \SEC}

To visualize, and take advantage, of the chaoticity of motion
imagine that phase space $\X$, on which an Anosov map $S$ acts,
can be divided into {\it cells} $E_j,\, j=1,2,\ldots m$, with
pairwise disjoint interiors, determined by the dynamics. They
consist of rectangles $E_j=C_j\times D_j$ as in Fig.(3.3.2) with
the axes $C_j,D_j$ crossing at ``centers'' $x_j=C_j\cap D_j$, and
with the size of their diameters that can be supposed, by
construction, smaller than a prefixed $\d>0$.

The basic property of hyperbolicity and transitivity is that the
cells $E_1,E_2,\ldots,E_m$ can be so adapted to enjoy of the two
properties below: \*
\0(1) the ``stable part of the boundary'' of $E_j$, denoted
$\partial^s E_j$ has $0$ area and under the action of the
evolution map $S$ ends up as a subset of the union of all the
stable boundaries of the rectangles; likewise the unstable
boundary of $E_i$ has zero area and is mapped into the union of
all the unstable boundaries of the rectangles under
$S^{-1}$.\,\footnote{\tiny This follows from the remark that, by
construction, any intersection of two rectangles is again a
rectangle if it has internal points.  Hence if the attracting set
is covered by a finite number of rectangles a partition into
rectangle is obtained by intersecting all of them. Then by
successive approximations the rectangles can be modified to
obtain a partition with the extra property: the case of
$2$-dimensions is presented in Appendix\ref{appG}.}  In formulae:

\be S\dpr^s E_j\subset \cup_{k} \dpr^s E_k, \qquad
 S^{-1}\dpr^u E_j\subset \cup_{k} \dpr^u E_k,  \label{e3.4.1}\ee
\eqfig{300}{110}{
\ins{-8}{100}{$s$}
\ins{49}{90}{$E_i$}
\ins{110}{0}{$u$}
\ins{260}{0}{$u$}
\ins{142}{100}{$s$}
\ins{258}{19}{$S\, E_i$}}
{fig3.4.1}{fig.3.4.1}
\*
{\tiny\begin{spacing}{0.5}\0 Fig.3.4.1: The figures illustrate
  very symbolically, as $2$-dimensional squares, a few elements
  of a Markovian pavement (or Markov partition).  \index{Markov
    pavement}\index{Markov partition} An element $E_i$ of it is
  transformed by $S$ into $S E_i$ in such a way that the part of
  the boundary that contracts ends up exactly on a boundary of
  some elements among $E_1,E_2,\ldots,E_n$.\end{spacing}} \*
\0In other words no new stable boundaries are created if the
cells $E_j$ are evolved towards the future and no new unstable
boundaries are created if the cells $E_j$ are evolved towards the
past as visualized in the idealized figure Fig.3.4.1
(idealization due to the dimension $2$ and to the straight and
parallel boundaries of the rectangles).  \\
(2) Furthermore the intersections $E_i\cap S E_j$ with internal points have to
be connected sets (\ie $\d$ is small enough).
\*

A covering $\EE=\{E_i\}_{i=1}^m$ of $\X$ by rectangles with the
above properties will be called a ``Markovian partition'' or
``Markovian pavement''. It can be constructed via a sequence of
geometric approximations, \Cite{Si968a}\Cite{Si968b}\,: for 
a very elementary detailed description of the $2$-dimensional
case see Appendix \ref{appG}\,.\index{Markov partition}

Existence of partitions of this kind is important as it induces a
representation of the evolution which, in a sense, is the same
whatever is the (Anosov) dynamical system $(\X,S)$; the
representation will be called ``symbolic dynamics'':
\*

\0{\bf Definition: (Symbolic dynamics)}{\it If
  $\EE=\{E_1,\ldots,E_m\}$ is a Markovian partition for a Anosov
  system $(\X,S)$, the $m\times m$ matrix $M$ with $M_{ij}=1$ if
  the interior of $SE_i$ intersects the interior of $E_j$ and
  $M_{ij}=0$ otherwise, defines the ``compatibility matrix'' of
  $\EE$. If the diameter of the sets $E_j$ is sufficiently small
  and, with the exception of $x$'s in a set of $0$ area in $\X$,
  for each $x$ there is unique sequence of labels
  $\qq=\{q_j\}_{j=-\infty}^\infty$ such that $M_{q_k
    q_{k+1}}\equiv 1$, for all $k\in Z$, and with $S^k x\in
  E_{q_k}$: the sequence $\qq$ is called a {\it compatible
    history} of $x$ on $\EE$\label{symbolic history}.\index{symbolic
    history}\,\footnote{\tiny The exception is associated with
  points $x$ which are on the boundaries of the rectangles or on
  their iterates. In such cases it is possible to assign the
  symbol $q_0$ arbitrarily among the labels of the rectangles to
  which $x$ belongs: once made this choice a compatible
  history\index{compatible history} $\qq$ determining $x$ exists
  and is unique.\label{symbolic exception}} The sequence $\qq$
  uniquely determines $x$.}  \*

Given an Anosov system $(\X,S)$ and a Markovian partition $\EE$
with small enough elements, the correspondence
$x\ \otto\ $ 'compatible sequence $\qq$ on $\EE$' allows to
identify the Anosov dynamical system $(\X,S)$ with the dynamical
system in which $\X$ is the set $\wt\X$ of compatible
sequences $\qq$ on $\EE$ with $S$ identified with the one step
left translation $\wt S$ of $\qq$ (``symbolic
dynamics'').\, \footnote{\tiny The requirement of small enough
diameter is necessary to imply that the image of the interior of
any element $E_j$ intersects $E_i$ ({\it i.e.} if $M_{ji}=1$) in
a connected set: true only if the diameter is small enough
(compared to the minimum curvature of the stable and unstable
manifolds).}
\label{symbolic dynamics}\index{symbolic dynamics}

The dynamical system $(\wt \X,\wt S)$ is equivalent to the
original $(\X,S)$ for studying the SRB distribution as well as
any invariant probability distribution for $(\X,S)$ which
attributes $0$ probability to the points $x$ with compatible but
ambiguous symbolic representation:\,\footref{symbolic
  exception}\, sequences $\qq$ can be used to identify points of
$\X$ just as decimal digits are used to identify the coordinates
of points (where exceptions occur as well, and for the same
reasons, as ambiguities arise in deciding, for instance, whether
to use $.9999\cdots$ or $1.0000\cdots$), \footnote{\tiny Because
the exceptions in the $1\otto1$ correspondence $\X\otto\wt \X$
occurr only for points whose trajectory visits the boundaries of
the partition rectangles, which
form a set of $0$ area in $\X$}\,.

The dynamical system $(\wt \X,\wt S)$ will be called the
``symbolic dynamics of $(\X,S)$''.  The one-to-one correspondence
(aside for a set of zero area) between points of the surface $\X$
and compatible histories of $\wt\X$ is a key property of Anosov's
maps: it converts the evolution $x\to Sx$ into the trivial
translation $\wt S$ of the history $\qq=\{q_{k}\}_{k=-\infty}^\infty$ of
$x$, which becomes $\qq'\equiv \wt
S\qq\defi\{q_{k+1}\}_{k=-\infty}^\infty$.

The matrix $M$ will be called a ``compatibility matrix''%
\index{compatibility matrix}. Transitivity
(see Sec.\.(\ref{sec:VI-2}\,)\index{transitivity} implies that the
matrix $M$ admits an iterate $M^{n}$ which, for some $n>0$, has
{\it no vanishing entry}).  \index{symbolic dynamics}

Therefore the points $x\in\X$ can be thought as the {\it possible}
outputs of a Markovian process with transition matrix to $M$: for this
reason the partitions $\EE=\{E_j\}$ of $\X$ are called {\it Markovian}.
\label{def-markovian}\index{Markov partition}
\*

\0{\bf Remarks:} (1) The Markovian property\index{Markovian property} has a
geometrical meaning (seen from Fig.{3.4.1} above): imagine each $E_i$ as
the ``stack'' formed by all the connected unstable axes $\d(x)$,
intersections of $E_i$ with the unstable manifolds of its points $x$, which
can also be called unstable ``layers'' in $E_i$.
\\
Then if $M_{i,j}=1$, the expanding layers in each $E_i$ expand under the
action of $S$ and their images {\it fully cover} the layers of $E_j$ with
which they overlap.%
\footnote{\tiny Formally let $E_i\in\PP$, $x\in E_i$ and
$\d(x)=E_i\cap W_u(x)$: then if $M_{i,j}=1$, {\it i.e.}  if the interior of
$SE_i$ visits the interior of $E_j$, it is $\d(Sx)\subset S\d(x)$.}
A corresponding property holds for the stable layers.
\* 
\0(2) It is important to notice that once a Markovian pavement
$\EE=(E_1,\ldots,E_m)$ with elements with diameter $\le \d$ has been
constructed then it is possible to construct a new Markovian pavement $\EE_\t$
whose elements have diameter smaller than a prefixed quantity. It suffices
to consider the $\EE_\t$ whose elements are, for instance, the
sets which have interior points and have the form
\be E({\V q})\defi
E_{q_{-\t},\ldots,q_\t}\defi \cap_{i=-\t}^\t S^{i} E_{q_i}.\label{e3.4.2}\ee
Their diameters will be $\le 2C\d e^{-\t\l}$ if $C,\l$ are the
hyperbolicity constants, see Eq.\equ{e3.2.1}. The sets of the
above form with non empty interior are precisely the sets $E(\V
q)=\cap_{j=-\t}^{\t} S^{-j} E_{q_j}$ for which $\prod_{j=-\t}^{\t-1}
M_{q_j q_{j+1}}=1$.
\*
\0(3) If $x\in E(\V q)$, $\V q=(q_{-\t},\ldots,q_\t)$, then the symbolic
history of $x$ coincides at times $j\in [-\t,\t]$ with
$\V q$, \ie $S^{-j}x=q_j$ for $j\in [-\t,\t]$ (except for a set of $0$
area of $x$'s).
\*
\0(4) A rectangle $E(\V q)$ can be imagined as the stack of the
connected portions of unstable axes: $\d^u(\V q,x)=[W^u(x)]\cap E(\V q)$\,,
\footnote{\tiny The connectedness, symbolized by the brackets, is needed as
the $W_u(x)$ is dense on $\X$.\footref{dense manifolds}}\,; \ie
$E(\V q)=\cup_{x\in D} \d^u(\V q,x)$ (or as the stack $E(\V
q)=\cup_{x\in C} \d^s(\V q,x)$ with $\d^s$ defined similarly).
\*
\0(5) The symbolic representation of the connected portion of
unstable manifold $[W^u(x)] \cap E(\V q)$ simply consists of the
compatible sequences $\qq'$ with $q'_i=q_i$, $i\in[-\t,\t]$ and
which continue to $i<-\t$ into the sequence of symbols of $x$
with labels $i<-\t$ while for $i>\t$ are arbitrary.\label{rem4-3}
The connected portion of stable manifold has a corresponding
representation.
\*
\0(6) The smallest integer $n_0$ with the property that
$M^{n_0}_{ij}>0$ will be called the {\it symbolic mixing
  time};\label{def-symbolic}: it gives the minimum time needed to
  be sure that any symbol $i$ can be followed by any other $j$ in
  a compatible sequence with compatibility matrix given by the
  transitive matrix of the pavement $\EE$.\index{symbolic mixing
  time} \*
\0(7) Finally consider systems $(\X,S)$ for which CH holds and
there is one (for simplicity) attracting surface $\AA\ne \X$ with
$(\AA,S)$ an Anosov system. In this case a Markovian partition
$\EE$ can be built on $\AA$ and is formed by cells
$E_j\subset \AA$. The SRB distribution is the same for $(\X,S)$
and $(\AA,S)$, \Cite{BR975}.  The reason why the averages of
observables $F$ is the same is the following.\\
If $x$ is close
enough to $\AA$ it is on the stable manifold of a point $\lis
x\in\AA$; the continuity of $W_s(x)$ on $x$ implies continuity of
the map $x\to\lis x$ and an open set around $\lis x$ is image of
an open set around $x$; it is also $|S^nx-S^n\lis
x|\tende{n\to\infty}0$. Thus the average of a continuous $F$ will
be the same whether starting from $x$ or $\lis x$ and most $\lis
x$ will be chaotic, in the sense of
Sec.(\,\ref{sec:VI-2})\,.\footnote{\tiny Remark also that existence
of a point $\lis x\in \AA$ which has the same asymptotics of
$x\in\X$, close enough to $\AA$, is a consequence of the {\it
shadowing property}, \Cite{HK995}[18.1]: if $S^nx$ is a
trajectory starting at small enough distance $<\e_0$ from $\AA$,
then there is $\e>0$ and,\,(\ref{sec:VI-2})\,, for each $n$ there
is a point $\x_n\in\AA$ with $|S^n x -\x_n|<\e e^{-n \l}$;
therefore the sequence $\x_n$ is a '$\d$-pseudo orbit', because
$|\x_n-S\x_{n-1}|\le |\x_n-S^nx|+ |S(S^{n-1}x)-S\x_{n-1}|\le \e
e^{-n\l}+ C_0 \e e^{-(n-1)\l}\le C_1 e^{-\l n}\le \d$ for
suitable $C_0,C_1$. Hence shadowing property, which states that a
$\d$ pseudo orbit on $\AA$ is $C\d$-close to an orbit on $\AA$,
implies that there is $\lis x\in\AA$ such that $|S^nx-S^n\lis x|<
C e^{-n\l}$, for suitable $C$. However to obtain the SRB result
(when $\AA\ne\X$) the continuity of the map $x\to\lis x$ would
still be needed.}
\*

Simple examples will be discussed in Sec.(\,\ref{sec:V-3}\,).

\section{Examples of hyperbolic symbolic dynamics}
\def\SEC{Examples of hyperbolic symbolic dynamics}
\label{sec:V-3}\iniz\lhead{\small\ref{sec:V-3}.\ \SEC}

The paradigmatic example\index{paradigmatic example} is the
simple evolution on the $2$-dimensional torus $T^2=[0,2\p]^2$
defined by the transformation (called the ``Arnold cat
map''):\footnote{\tiny The map is not obtainable as a
Poincar\'e's section of the orbits of a $3$-dimensional manifold
simply because its Jacobian determinant is not $+1$.}
\be
S\Bff=S\pmatrix{\f_1\cr\f_2\cr}=
\pmatrix{1&1\cr1&0\cr}\pmatrix{\f_1\cr\f_2\cr}=
\pmatrix{\f_1+\f_2\cr\f_1\cr}\ \hbox{mod}\, 2\p\label{e3.5.1}\ee
It is possible to construct simple examples of Markovian partitions of $T^2$
because the stable and unstable directions
through a point $\Bff$ are everywhere the directions of the two
eigenvectors of the matrix $\pmatrix{1&1\cr1&0\cr}$.
 
Hence in the coordinates $\Bff$ they are straight lines (wrapping densely
over $T^2$, because the slope of the eigenvectors is irrational). For
instance Fig.(3.5.1) gives an example of a partition satisfying
the property (1) in Sec.(\,\ref{sec:IV-3}\,).

This is seen by remarking that in Fig.(3.5.1) the union of the stable
boundaries of the rectangles, \ie the lines with negative slope (irrational
and equal to $(-\sqrt5-1)/2$), are a connected part of the stable
manifold exiting on either side from the
origin; likewise the union of the unstable boundaries, \ie the lines with
positive slope (equal to $(\sqrt5-1)/2$) are a {\it connected
  part} of the unstable manifold exiting from the origin. 

\eqfig{80}{80}
{\ins{10}{40}{$1$}
\ins{65}{40}{$1$}
\ins{37}{73}{$2$}
\ins{40}{15}{$2$}
\ins{38}{43}{$3$} }
{fig3.5.1}{fig3.5.1}

{\begin{spacing}{0.7}\tiny\0Fig.(3.5.1): Three rectangles
    $\scriptstyle (E_1,E_2,E_3)$ pavement of torus $\scriptstyle
    T^2$: sides lie on two {\it connected} portions of stable and
    unstable manifold of origin (fixed point). Satisfies
    Eq.($\scriptstyle \equ{e3.4.1}$) {\it but is not
    Markovian}, because correspondence between histories-points
    is not $\scriptstyle 1-1$ even allowing for
    exceptions on a set of $0$ area: the three sets are too
    large. But the partition whose elements are $\scriptstyle
    E_j\cap S E_j$ has the desired properties, as it follows from
    the next figure.\end{spacing}\vfil}

Therefore under action of $S$ the union of the stable boundaries
will be still a part of the stable manifold through the origin
{\it shorter by a factor $(\sqrt5-1)/2<1$} hence it will be part
of itself, so that the first Eq.\equ{e3.4.1} holds. For the
unstable boundaries the same argument can be repeated using
$S^{-1}$ instead of $S$ to obtain the second Eq.\equ{e3.4.1}.

However it is checked that the partition in Fig.(3.5.1)
$\EE=(E_1,E_2,E_3)$ generates ambiguous histories in the sense
that $E_2$ is too large and in general the correspondence between
points and their symbolic history is $2-1$ (and more to $1$ on a
set of zero area).

By slightly refining the partition
(subdividing the set $E_2$ in Fig.(3.5.1)) a true Markovian
partition is obtained as shown in Fig.(3.5.2); from the figure
the transition matrix can be computed and the mixing time results
$n_0=7$.\,\ref{symbolic history}

\eqfig{216}{80}
{
}
{fig3.5.2}{fig3.5.2}

\*

{\begin{spacing}{0.8}\tiny\0Fig.(3.5.2): A Markovian pavement
    (left) for $S$ (``Arnold's cat \index{Arnold's cat} map'').
    The images under $S$ of the rectangles are shown in
    the right figure: corresponding rectangles are marked by the
    corresponding colors and numbers.\label{fig3.5.2}\end{spacing}}
\*

The examples above are particularly simple because of the
$2$-dimensionality of phase space and because the stable and
unstable manifolds of each point are straight lines.
\*

\0{\it Remark:} the Arnold cat is not reversible, but a
simple reversible Anosov system can be defined as $S_*=S\times
S^{-1}$ acting on $(\xx,\yy)\in T^2\times T^2\defi \AA$ as
$S_*(\xx,\yy)=(S\xx,S^{-1}\yy)$: if $I(\xx,\yy)=(\yy,\xx)$,
the system $(\AA,S_*)$ is reversible and Anosov.\label{reversible cat}
\*

If the dimension is higher and the manifolds are not flat, as
implied by different expansion (or contraction) in different
directions, the Markovian partition still exists but its elements
will, in general, have an irregular boundary (although H\"older continuous). For
this reason the reader is referred to the original papers
\Cite{Si968a}\Cite{Bo970a};. About the general $2$-di\-mensional
case, particularly simple, see Appendix \ref{appG}.

The dimensions of stable and unstable manifolds of an Anosov map
need not be equal; as when the total dimension is odd. As an
example consider the $3$-torus map described by the matrix:
\be S = \scriptstyle \pmatrix{\scriptstyle 0&\scriptstyle 0&\scriptstyle 1\cr
\scriptstyle 1&\scriptstyle 0&\scriptstyle -k\cr
\scriptstyle 0&\scriptstyle 1&\scriptstyle -1-k\cr},\scriptstyle\label{e3.5.2}\ee
with $k$ a suitably chosen integer, \eg $k=\pm1$ (its eigenvalues
equation is $\l^3+(1+k)\l^2+k\l-1=0$ and $k$ has to be so
chosen that $|\l|\ne1$).

\section{Coarse graining\index{coarse graining} and 
discrete phase\index{discrete phase space} space}
\def\SEC{Coarse graining\index{coarse graining} and 
discrete phase\index{discrete phase space} space}
\label{sec:VI-3}\iniz
\lhead{\small\ref{sec:VI-3}.\ \SEC}

Given smooth observables $F_1,\ldots.F_r$, let $\X$ be the phase
space of a timed evolution $S$, verifying CH,
with attracting set $\AA$, $\AA=\X$ for simplicity (so $(\X,S)$
is an Anosov map). Imagine $\X$ to be subdivided, for the purpose
of measurements of the observables $F_j$, in small regions in
which the $F_j$ can be supposed to have a constant value.

A convenient choice of the small regions will be to imagine them
constructed from a Markovian partition $\EE_0= (E_1,E_2,\ldots,E_m)$. Given
$\t\ge0$ consider the {\it finer Markovian partition} $\EE_\t$ whose
elements are the sets $E(\V q)\defi \cap_{j=-\t}^\t S^{-j}E_{q_{j}}$, $\V
q=(q_{-\t},\ldots, q_\t), \, q_j\in \{1,2,\ldots,m\}$ with non empty
interior: as discussed in Eq.\equ{e3.4.2} the elements $E(\V q)$ can be
made with diameter as small as pleased by choosing $\t$ large, because of
the contraction or stretching properties of their boundaries.

Choosing $\t$ large enough so that in each of the {\it cells}
$E(\V q)$ the ``interesting'' observables $F_1,\ldots,F_r$ have a
constant value (at least at a first approximation), we shall call
$\EE_\t$ a {\it coarse grained partition} of phase space into
``coarse cells'' adapted, or relative, to the observables
$F_1,\ldots,F_r$.\label{coarse cell}

Should we decide that higher precision is necessary we shall have
simply to increase the value of $\t$. But, given the precision
chosen, the time average of any observable $F$ of interest will
have the form:
\be \media{F}=\frac{\sum_{\V q} F(x_{\V q}) w(\V q)}{\sum_{\V q} w(\V q)}
\label{e3.6.1}\ee
where $\V q=(q_{-\t},\ldots,q_\t)$ are $2\t+1$ labels, among the
labels $1,2,\ldots,m$ of the Markovian pavement elements $E_j$
used to construct the coarse cells, and $x_{\V q}$ denotes a
point of the cell $E(\V q)$, and $w(\V q)$ are suitable {\it
  weights}.

The weights $w(\V q)$ will be expressed
via the $SRB$ distribution:
\be\frac{w(\V q)}{\sum_{\V q'} w(\V q')}=\m_{SRB}(E(\V
  q))\label{e3.6.2}\ee
and the problem is to determine, for systems of interest, the
weights (hence the SRB distribution).

To understand this point it is convenient to consider, to fix
ideas, a system of $N$ particles and a discretization of its
phase space $\X$ into {\it equally spaced} points $x$, centers of
tiny boxes \footnote{\tiny The name 'box' is chosen to mark the
distinction with respect to the rectangles of the coarse
partition.} of sides $\d p,\d q$ in the momentum and,
respectively, position coordinates and volume $(\d p\d
q)^{3N-1}\defi h^{3N-1}$, by far smaller than the diameter of the
largest coarse cell (as usual in simulations).\footnote{\tiny The
$2(3N-1)$ takes into account that the Poincar\'e's section $\X$
is here imagined to be performed by restricting one coordinate
with momentum dimension and one with dimension position and
drawing a section of dimension $6N-2$ on an energy surface of
dimension $6N-1$; $\d p,\d q$, hence $h$, should be imagined so
small that the $r$ observables $F_j$ are be approximately
constant in the Markovian cells (as in simulation codes).}

This will allow us to discuss time evolution in a way deeply
different from the usual: it has to be stressed that such
``points'' or ``microscopic'' cells, are not associated with any
particular observable; they can be thought of as tiny $6N-2$
dimensional boxes and represent the highest microscopic
resolution that can be achieved: they will be called {\it
  microcells} or {\it discrete points} of the discretized phase
space {\it and have nothing to do with the above coarse cells
  $E_j$ of the Markovian partition, which must be thought as much
  larger, each containing very large numbers of
  microcells}.\index{microscopic cell}\index{microcell}

Let $\NN_0$ be the total number of microcells {\it regularly
  spread} on $\X$. The dynamics $S$ will be thought as a map
$\lis S$ of microcells into themselves: it will then be
eventually periodic (with period possibly depending on the
initial microcell).  The recurrent points will be, in general,
$\NN\ll \NN_0$, \ie much less than the number $\NN_0$ of points
in the discretization of $\X$.

No matter how small coarse cells\index{coarse cells} $E(\V q)$
are chosen, as long as the number of discrete points inside them
is very large, it will be impossible to represent the motion as a
permutation: not even in the conservative case in which the
volume of the cells $S^n E(\V q)$ remains constant.  Simply
because the coarse cells are deformed by the evolution, being
stretched in some direction and compressed in others, if the
motion has nonzero Lyapunov exponents ({\it i.e.} is chaotic).

The next section will address the question: how can this be
reconciled with the numerical simulations, and with the naive
view of motion, as a permutation of microcells?  The phase space
volume will generally contract with time: yet we want to describe
the evolution in terms of an evolution permuting microscopic
states, hence the question: {\it how to determine the weights
$w(\V q)$} of the coarse cells?

\section{Microcells, phase space points and simulations}
\def\SEC{Microcells, phase space points and simulations}
\label{sec:VII-3}\iniz
\lhead{\small\ref{sec:VII-3}.\ \SEC}

The microcells (introduced in the previous section) should be
considered as realizations of objects alike to those arising in
computer simulations\index{simulation}: where phase space points
$x$ are ``digitally represented'' (``discretized'') on a regular
lattice with coordinates given by a string of integers and the
evolution $S$ becomes a {\it code}, or {\it program}, $\lis S$
simulating the solution of equations of motion for the model
under study. The code $\lis S$ operates {\it exactly} on the
coordinates (deterministic round-offs, enforced by the particular
computer hardware and software, should be considered part of the
program).

Assume that the evolution map $S$ on phase space $\X$ is an
Anosov map (hence smooth hyperbolic and with a dense orbit) so
that $\X$ is itself an attracting surface.  Then the general
properties analyzed in the previous sections will hold. In
particular there will be a partition $\EE_0$ of $\X$ into coarse
cells (``rectangles'') $\EE_0=\{E_1,E_2,\ldots,E_m\}$ with the Markovian
property, see (\,\ref{sec:IV-3}\,).  \index{rectangle}

The evolution $S$, considered in the approximation $\lis S$ in
which it acts on the discretized phase space\index{discrete phase
  space}, will produce (for approximations careful enough, see
below) a chaotic evolution ``for all practical purposes'': the
approximation is realized via the prescription $\lis S$ on the
evolution of the microcells (to be regarded as a computer
simulation, because the ``microcells'' are not real physical
objects). For the purposes of the {\it heuristic analysis} below,
attention is directed at:
\* 
\0(1) looking only at ``macroscopic observables''
\index{macroscopic observables} which are constant on the coarse
graining scale $\g= 2C e^{-\l \t}$, see Eq.\equ{e3.4.2}, of a
Markovian partition $\EE_\t=\{E({\V q}\})$ obtained from a
partition $\EE_0=\{E_1,E_2\ldots,E_m\}$ by intersecting its $2\t$
iterates $E({\V q})=\cap_{i=-\t}^{\t} S^i E_{q_i}$; and
\\
(2) looking only at phenomena accessible on time
scales\index{time scale} far shorter than the recurrence
times\index{recurrence time} (which are finite, in finite
representations of the evolution, but of size large enough to
make the recurrence phenomenon irrelevant).\,
\footnote{\begin{spacing}{.4}\0\tiny To get an idea of the orders
  of magnitude consider a rarefied gas of $\scriptstyle N$ mass
  $\scriptstyle m$ particles of density $\scriptstyle \r$ at
  temperature $\scriptstyle T$: the metric on phase space will be
  $\scriptscriptstyle ds^2=\Si(d\V p_i^2/{(m k_B T)^{3N}}+d\V
  q_i^2/\r^{2N})$;\label{phase space metric} each coarse cell
  will have size at least $\scriptscriptstyle \sim({mk_B
    T})^{1/2}$ in momentum and $\scriptstyle \sim\r^{-1/3}$ in
  position; this is the minimum precision required to give a
  meaning to the particles as separate entities. Each microcell
  could have coordinates represented with $\scriptstyle 32$ bits
  will have size of the order of $\scriptscriptstyle({mk_B
    T})^{1/2}2^{-32}$ in momentum and $\scriptstyle
  \r^{-1/3}2^{-32}$ in position and the number of {\it
    theoretically possible} phase space points representable in
  the computer will be $\scriptscriptstyle O((2^{32})^{6N})$
  which is obviously far too large to allow anything being close
  to a recurrence in essentially any simulation of a chaotic
  system involving more than $\scriptstyle N=1$ particle. For
  recent study on this problem see
  \Cite{LS024}.\end{spacing}\label{n3-3}} \*

It is supposed that:

\0(a) the division into microcells is fine enough to
allow us to describe evolution $S$ via a map $\lis S$ of the
microcells (hence via meaningful simulations); the map
$\lis S$ is ``careful enough'' in the sense that microcells that
are on an ustable axis of a coarse cell $E$ are mapped on the
union of unstable axes of the cells that intersect the image
$\lis S E$. This is not an innocent assumption: it will be clear
that other maps $\lis S$ can be conceived which lead to very
different conclusions.

\0(b) the map $\lis S$ approximating the evolution map
$S$ cannot be, in general, a permutation of microcells: in
simulations it will happen, {\it essentially always}, that it
will send distinct microcells into the same one. It does
certainly happen in nonequilibrium systems in which, in the
average, phase space contracts;\footnote{\tiny With extreme care
it is sometimes, and in equilibrium, possible to represent a
chaotic evolution $S$ with a code $\lis S$ which is a true
permutation: the only example that I know, dealing with a
physically relevant model, is
in \Cite{LV993}\Cite{LS024}\,.\label{LVLS}}

\0(c) The set of recurrent points for $\lis S$ is a {\it discrete
  representation} of $\X$, that we call $\lis \X$.  Even though
the map $\lis S$ may not be one-to-one, particularly in the
nonequilibrium cases, nevertheless it will be such on $\lis\X$,
\ie {\it eventually}: because any map on a finite space is a {\it
  permutation} of the points which are recurrent.

The discrete set $\lis\X$ will be imagined as a collection of
recurrent microcells\,\footnote{\tiny Recurrent are the
microcells with periodic evolution, as defined in
Sec.(\,\ref{sec:I-3}\,).} lying on, and approximating, unstable
manifolds\index{unstable manifold} of $\X$. More precisely in the
discretized evolution the phase space points will move towards an
attracting set, $\lis\X$, which will be a finite approximation of
$\X$, consisting of recurrent points; and $\lis\X$ will be
imagined as an array of points regularly spaced (as in the
digital realizations of the microcells) located, in each cell
(``rectangle'') $E({\V q})$, on finitely many among the unstable
axes of $E({\V q})$, see Sec.(\,\ref{sec:III-3}\,).

In each rectangle $E(\V q)$ of the Markovian partition (``coarse
grained cell'')\index{coarse graining}, such arrays approximate
some (finitely many) parallel unstable axes of $E(\V q)$,
see Fig(3.7.1), that will be called $\d(\V q)$ while their union
will be called $\D(\V q)$.

\eqfig{200}{60}{\ins{-20}{40}{$E(\V q)$}}{fig3.7.1}{Fig3.7.1}
\vskip3mm
    {\tiny\begin{spacing}{0.5}\0Fig(3.7.1):\label{Fig(3.7.1)} A
      very schematic and idealized drawing of the few
      intersections $\d(\V q)$ with $E(\V q)$, of the unstable
      surfaces which contain the microcells remaining, after a
      transient time, inside a coarse cell $E(\V q)$ (the
      infinitely many of the others being left out of the
      recurrent cycle). The second drawing (indicated by the
      arrow) represents schematically the collections of
      microcells which are on the unstable surfaces which, in
      $E(\V q)$, form the stack of unstable surface elements $\d(\V
      q)$, whose union will be called $\D(\V q)$. The
      recurrent microcells are imagined to regularly cover the
      attracting set unstable but recurrent surface
      elements, as supposed in (a) above.  \end{spacing} } \*
This is not an innocent assumption as discussed in the next section.

\0(d) The evolution $\lis S$ will permute the discrete arrays of
recurrent microcells on $\lis \X$ into themselves. Every
permutation can be decomposed into cycles: and a simple minded
assumption is that the recurrent microcells in the arrays
$\cup_{\V q}\D(\V q)$ take part in the same {\it one cycle
  permutation}, thus defining the discretized attracting set
$\lis \X$.

It is an analogue of the ergodic hypothesis for equilibrium: in
the stronger form that not only every recurrent microcell visits
all the other recurrent ones (\ie at stationarity there is a
single periodic orbit), but also the recurrent points are
regularly covering the discrete attracting set supposed
consisting of a few unstable axes, that in each cell $E(\V q)$
will be called $\d(\V q)$, collected into $\D(\V q)\subset E(\V
q)$. In the end, {\it however}, it implies the analytical
identification of the SRB distribution, see below.  \*

{\it Consistency} between expansion of the unstable surfaces
and existence of a cyclic permutation\index{permutation cyclic}
of the microcells regularly spaced on the recurrent set, imagined
to consists of unstable axes $\d(\V q)$, in $E(\V q)$, see
Fig.3.7.1, {\it puts severe restrictions} on the number of
recurrent microcells in each coarse grained cell $E(\V q)$.

The number $\NN(\V q)$ of recurrent microcells out of the total
number $\NN$ of microcells in $\lis\X$ contained in $E(\V q)$, by
Eq.\equ{e3.6.2}, is $\NN(\V q)=\NN \frac{w(\V q)}{\sum_{\V q}
w(\V q)}=\NN
\m_{SRB}(E({\V q}))$ and it will determine an approximation of
the weights $w(\V q)$
and of the SRB distribution, as
discussed in the next section.

The above viewpoint can be found in 
\Cite{Ga995a}\Cite{Ga999}\Cite{Ga001}\Cite{GBG004}.

\section{The SRB distribution\index{SRB distribution meaning}: 
its physical meaning}
\def\SEC{The SRB distribution\index{SRB distribution meaning}: 
its physical meaning}
\label{sec:VIII-3}\iniz
\lhead{\small\ref{sec:VIII-3}.\ \SEC}

The determination of the weights $w(\V q)$ determining the fraction
of the total number $N(\V q)$ of recurrent microcells in $E(\V q)$ can be
found through the following {\it heuristic argument},
\Cite{Ga995a}\Cite{Ga008a}.

Let $2\t$ be larger than the mixing time (defined in remark (6) in
Sec.(\,\ref{sec:IV-3}\,)) and call $n(\V q')$ the
number of unstable axes $\d(\V q')$ forming the approximate
attracting set $\lis\X\cap E(\V q')$ 
inside $E(\V q')$, denoted $\D(\V q')$ in the previous
section(see Sec.(\,\ref{sec:VII-3}\,) Fig.3.7.1)
.
\*

\0(i) The numerical density of recurrent microcells on the attracting set will be
$$\frac{N(\V q')}{\D(\V q')}=\frac{N(\V q')}{n(\V q')|\d_u(\V
q')|}$$
because the $n(\V q')$ unstable axes $\d_u(\V q')$ have
(approximately) the same surface.

\0(ii) The expanding action of $S^{2\t}$ (hence of its discretized
$\lis S^{2\t}$) will expand the unstable axes by a factor that will be
written:
$$e^{\L_{u,\t}(\V q')}$$
hence the density of microcells mapped from $E(\qq')$ into $E(\V
q)$, will decrease by the same factor.

\0(iii) Thus the number of
microcells that are mapped from $E(\V q')$ to $E(\V q)$ equals
$$
\frac{N(\V q')}{\D(\V q')}e^{-\L_{u,\t}(\V q')} \D(\V q)
$$
provided $S^\t E(\V q')\cap E(\V q)$ has an interior point.

\0(iv) Consistency implies:
\be N(\V q)=\sum_{\V q'}^*
\frac{N(\V q')}{|\D(\V q')|}\frac1{e^{\L_{u,\t}(\V q')}}|\D(\V q)|
\label{e3.8.1}\ee
where the $*$ signifies that $S^t E(\V q')\cap E(\V q)$ must have
an interior point in common; in other words for each compatible
$\qq=(q_{-\t},\ldots,q_\t)$ the sum is over the compatible
sequences $\qq'=(q'_{-\t},\ldots,q'_\t)$.  \*

Conclusion: the density $\r(q)\defi\frac{N(\V q)}{\D(\V q)}$
satisfies:
\be\r(\V q)=\sum_{\V q'}^*
e^{-\L_{u,\t}(\V q')}\r(\V q')\defi (\LL\r)(\V q)\Eq{e3.8.2}\ee
closely related to the similar equation for invariant densities of
Markovian surjective maps of the unit interval, \Cite{GBG004}.
It has to be kept in mind that it has been supposed that the
observables considered have values constant in each cell $E(\qq)$
\*

\0{\bf Remark:} For later reference it is important to
remark that the expansion per time step\index{expansion rate} at
a point $x\in E(\V q)$ for the map $S$ along the unstable
manifold is given by the determinant of the matrix $\dpr_uS(x)$,
giving the action of the Jacobian matrix $\dpr S(x)$ on the
vectors of the unstable manifold at $x$ (\ie it is the
restriction of the Jacobian matrix to the space of the unstable
vectors), see Eq.\equ{e3.2.2}. It has a logarithm:
\be \l^u(x)\defi \log|\det\dpr_u S(x)|\Eq{e3.8.3}\ee
And the logarithm $\L_{u,\t}(\V q)$ of the expansion at $x$ for
the map $ S^{2\t}$ as a map from $ S^{-\t}x$ to
$S^{\t+1}x$\, \footnote{\tiny Here we are identifying the expansion
of the map $S$ and that of the discretized version $\lis S$.}
depends on the Jacobian matrix $\partial_u S^{2\t}$ restricted to
the expanding axes. It is:
\be \L_{u,\t}(\V q)=\sum_{j=-\t}^{\t} \l_u(S^j x)\label{e3.8.4}\ee
by composition of differentiations.  
\*

For $2\t$ larger than the symbolic mixing time, see
(\,\ref{sec:IV-3}\,),\index{symbolic mixing time} the matrix
$(\LL_{u,\t})_{\V q,\V q'}$ in Eq.\equ{e3.8.2} has all elements
$>0$ (because $S^{m} E(\V q')$ intersects all $E(\V q)$ for
$m\ge2\t$).

The general theory of equations like Eq.\equ{e3.8.2}
immediately yields, from the ``Perron-Frobenius
theorem'',\index{Perron-Frobenius theorem}\,\Cite{GBG004}[Problem
4.1.17]\,:

\0(1) the equation has a simple eigenvalue $\l>0$ with maximum modulus
\\
(2) to which corresponds a positive eigenvector $v(\qq)>0$.
Hence $v$, normalized so that $\sum_\qq v(\qq)=1$,
uniquely determines $\r(\qq)$, 
\be \r(\V q)=v_{\V q}=\m_{SRB}(E(\V q))=h_l(\V q))
{e^{-\L_{u,\t}(\V q)}} h_r(\V q)
\label{e3.8.5}\ee
where $h_l,h_r$ are functions of $\t$ and of the
symbols in $\V q$.

Furthermore from the Eq.\equ{e3.8.2} and from \Cite{Ru968}
the further properties emerge, and will be discussed in Ch.\ref{Ch4}:
\\
(a) for $\t$ large, $h_l,h_r$
essentially depend only on the first few symbols in the string
$\V q$ with labels close to ${-\t}$ or close to $\t$,
respectively, and are uniformly bounded {\it above and below} in
$\t$.\\
(b) an exact expression for
$v_{\V q}$ and $\m_{SRB}$, see Eq.\equ{e3.8.6} below.
\*

If more observables need to be considered it is always possible to refine
the coarse graining\index{coarse graining} and even take the limit of
infinitely fine coarse graining, obtaining an exact expression
for the SRB distribution acting on any continuus observable $F$:
\be \media{F}_{SRB}= \lim_{\t\to\infty}
\frac{ {\sum_{\qq_\t}^*}e^{-\L_{u,\t}(\qq_\t)}\,F(x_{\qq_\t})}{\sum_{\qq_\t} 
e^{-\L_{u,\t}(x_{\qq_\t})}}
\label{e3.8.6}\ee
where the sum${}^*$, as above, is over the compatible
$\qq_\t=(q_{-\t},\ldots,q_\t)$, and $x_{\qq_\t}$ is an {\it arbitrary} point
in $\cap_{j=-\t}^\t S^{-j}E_{q_j}$.

The Eq.\equ{e3.8.6} has been proved to be {\it exact for
  $\m_{SRB}$}, for all smooth observables
  $F$, \Cite{BR975}\Cite{Ru999}\,: the limit can be shown to exist
  for all choices of $F$, independently of the particular
  Markovian partitions $\EE_{2\t}$ used for the coarse graining and of
  the choice of the center $x_{\V q}$ in $E(\V q)$ (rather
  arbitrarily picked up), \Cite{GBG004}.

It should be noticed that the uniform boundedness of $h_l,h_r$
imply (from Eq.\equ{e3.8.5}):
\be \frac1\t\log {\sum_{\V q}}^* e^{-\L_{u,\t}(\V q)}=O(\frac1\t)
\label{e3.8.7}\ee
reflecting a further result, ``Pesin's formula'', on the theory
of SRB distributions, \Cite{JP998}[p.697]\Cite{GBG004}[Prop.6.3.4].
This completes the heuristic theory of SRB
distributions.\index{Pesin's formula}
\label{Pesin's formula}

The above viewpoint can be found in
\Cite{Ga995a}\Cite{Ga999}\Cite{Ga001}\Cite{Ga008a}\Cite{Ga004b}[p.684].

\section{Other stationary distributions}
\def\SEC{Other stationary distributions}
\label{sec:IX-3}\iniz
\lhead{\small\ref{sec:IX-3}.\ \SEC}

So the SRB distribution arises naturally from assuming that
dynamics can be discretized on an array of points
(``microcells'') regularly covering selected unstable axes of
elements of a Markovian partition and becomes a one cycle
permutation of the recurrent microcells. This is so under the CH
and {\it holds whether the dynamics is conservative (Hamiltonian)
  or dissipative}.

It is, however, well known that Anosov systems admit
(uncountably) many invariant probability distributions, besides
the SRB. This can be seen by remarking that the 
configurations space is identified with a space of compatible
sequences, Sec.(\,\ref{sec:IV-3}\,), on a Markovian partition, and from
Statistical Mechanics (and probability theory of Markov
processes) there are uncountably many different stochastic
processes (known as Gibbs distributions) which can be assigned as
probability distributions on the sequences compatible with a
matrix with positive mixing time: examples
will be discussed in the following Ch.\ref{Ch4}, see also
\Cite{GBG004}[Sec.5].

Yet the analysis just presented apparently singles out SRB as the
unique invariant distribution. This is due to our assumption
that, in the discretization, non recurrent microcells (\ie that
are left behind by the dissipation) are attracted in each cell to
a few recurrent ones located on unstable axes, on which are also
supposed located on a regular discrete lattice, and to generate a
periodic orbit (single, by transitivity). This setting seems well
adapted to study via computer simulations.

But other invariant distributions can be obtained by custom made
discretizations of phase space, which will not cover the
attracting manifolds in a regular way.  For instance when
defining the choice of the initial data with other distributions
``not absolutely continuous with respect to the phase space
volume'', which, studied in suitably designed simulations, lead
to ``anomalous'' invariant distributions.  Besides the choice of
the array of microcells which is regular in each cell $E(\V q)$
covering a few unstable axes, other choices are possible,
tweaking the simulation code to respect the alternative
choices.  \*

A 'physical' example occurs when the attracting surface $\AA$ is
not the full phase space $M$: then there will be the SRB
distribution $\m$ generated by data in the neighborhood of $\AA$
with probability $1$ with respect to the volume on $M$. However
there will also be the distribution $\m^-$ generated with
probability $1$ by data chosen {\it exactly} on $\AA$ with
probability $1$ with respect to the area on $\AA$ and evolved
{\it backward} in time: such data are not chaotic in the sense of
definition 4 in Sec.\,(\ref{sec:VI-2}\,)\footnote{\tiny Because
$\AA$ has zero measure in $\X$.} and in general $\m\ne\m^-$
(remark that in general $S_{-t}$ will not even have $\AA$ as an
attracting set, even if CH holds) although both distributions are
invariant as distributions on $\AA$ and attribute probability $1$
to the surface $\AA$ itself, see also comments in
Sec.\,(\ref{sec:IV-4}\,).

The choice of a regular array of the microcells, heuristically
leading to the SRB distributions, is only justified by the
physical consequences and the success in the unification of
equilibrium and non equilibrium theories; it rests on the
ground-breaking work of Ruelle on axiom attractors and their
physical interpretation, \Cite{Ru976}\,.  Just as the success of
statistical mechanics is the ultimate justification
of Boltzmann's ergodic hypothesis. Given the infinitely many
distinct stationary distributions on phase space the
equilibrium ensembles and the SRB distributions are privileged
because of their marvelous consequences, rather than by \ap reasons.

The physical representations of the SRB distribution just
obtained, see \Cite{Ga995a}\Cite{Ga000}, shows that there is no
conceptual difference between stationary states in equilibrium
and out of equilibrium. If motions are chaotic, in both cases
they are permutations of microcells and the {\it SRB distribution
  is simply equidistribution over the recurrent microcells},
provided the microcells are\index{microcells recurrent} {\it
  uniformly spread} in phase space. In equilibrium this gives the
Gibbs microcanonical distribution and out of equilibrium it gives
the SRB distribution (of which the Gibbs'
distribution\index{Gibbs distribution} is a very special case).

The heuristic argument analyzed above, Sec.\,(\ref{sec:VIII-3}\,),
is an interpretation of the mathematical proofs behind the SRB
distribution which can be found in \Cite{Bo975}\Cite{GBG004}. Once
Eq.\equ{e3.8.6} is given, the expectation values of the
observables in the SRB distributions can be formally written as
sums over suitably small coarse cells and symmetry properties
inherited from symmetries of the dynamics become transparent and
can (and will) be used in the following to derive universal
properties\index{universal properties} of the stationary states.
\footnote{\tiny In the case the $(\X,S)$ is an Anosov system the
argument can be applied also to $(\X,S^{-1})$ assuming (a further
assumption) that the microcells are also regularly located on
(few) stable axes of the partition $\EE_\t$. If CH holds but
$(\X,S$) has one (say) attracting surface $\AA\ne \X$ the
partition $\EE$ is a partition of $\AA$, while $(X,S^{-1}$ has a
different attracting surface $S^{-1}\AA$.}

For instance systems, verifying CH, will be shown to obey a
``fluctuation theorem'' valid in reversible systems close, but
not infinitesimally close, to equilibrium: the theorem implies
universal identities which can be considered extensions of
Onsager's reciprocity relations\index{Onsager's reciprocity},
since they imply the latter whenever both applicable.

Reciprocity was ``only'' derived on irreversible systems
infinitesimally close to equilibrium (relying on the basic time
reversal symmetry of equilibrium systems). 
Therefore its extension to reversible dissipative systems, in the
form of the fluctuation theorem, not only extends validity of
reciprocity to systems with reversible dissipation (not a
surprise) but, holding also at small but not infinitesimal
reversible dissipation, it can be seen as an instance of a more
general property, that was ``only'' known infinitesimally close
to equilibrium.

It should not escape to the reader the connection between the
selection of the initial data, the discrete realization of the
dynamics on a regular finite array of microcells and the ensuing
privileged nature of the SRB distribution: this certainly has
philosophical implications, hinted above, in which we do not dare to be
involved here.\index{philosophical implications}

\section{Phase space cells count; entropy}
\def\SEC{Phase space cells and entropy}
\label{sec:X-3}\iniz
\lhead{\small\ref{sec:X-3}.\ \SEC}

The discrete representation, in terms of coarse
grain\index{coarse graining} cell s and microcells, leads to the
possibility of attempting a count of the number $\NN$ of the
microcells on the attracting set and therefore to define a kind
of entropy function\index{entropy}: see \Cite{Ga001}. The
following analysis is heuristic, as the analysis of the SRB
distribution in Sec.(\,\ref{sec:VIII-3}\,) above, and is reported also
because it leads to point out a cancellation that I find as
remarkable as unexpected.

Consider an Anosov system $S$ on a smooth surface $\X$ (phase
space). Let $\m_{SRB}$ be the SRB distribution describing the
asymptotic behavior of almost all (in
the sense of the area) initial data in $\X$. As discussed above the SRB
distribution admits a rather simple representation which can be
interpreted in terms of ``{\it coarse graining}'' of the phase
space.

Let $\EE$ be a ``Markov partition'' of phase space
$\EE=(E_1,\ldots,E_m)$ with sets $E_j$, see
p.\,\pageref{def-markovian}\,.  Let $\t$ be a time such that the
size of the $E(\V q)=\cap_{j=-\t}^\t S^{-j}E_{q_{j}}$ is so small
that the physically interesting observables can be viewed as
constant inside $E(\V q)$, so that $\EE_\t=\cap_{j=-\t}^\t
S^{-j}\EE$ can be considered a coarse grained\index{coarse
partition} partition of phase space, see
Sec.(\,\ref{sec:VI-3}\,).\label{E(q)}

Then the SRB probability $\m(E(\V q))$ of $E(\V q)$ is described
in terms of the functions $\l^u(x)=\log|\det (\dpr S)_u(x)|$,
Eq.\equ{e3.8.3}, and of the expansion rates $\L_{u,\t}(x)$ in
Eq.\equ{e3.8.6}. Here $(\dpr S)_u(x)$ (resp. $(\dpr S)_s(x)$) is
the Jacobian of the evolution map $S$ restricted to the unstable
(stable) manifold through $x$ and mapping it to the unstable (stable)
manifold through $Sx$. Selecting a point $x_{\V q}\in E(\V q)$ for
each $\V q$, the SRB distribution is given approximately by
Eq.\equ{e3.8.6} if $\t$ is large enough and, exactly, in the
limit $\t\to\infty$.

Adopting the discrete viewpoint\index{discrete viewpoint} on the
structure of phase space, Sec.\,(\ref{sec:VII-3}\,), regard motion as
a cyclic permutation\index{cyclic permutation} of recurrent
microcells and ask on general grounds the question, \Cite{Ga001}:
\*

\0{\it Can we count the number of recurrent microcells in which
the asymptotic SRB state of the system can be realized
microscopically?}  \*

This extends the question asked by Boltzmann for\index{Boltzmann} the
equilibrium case in \Cite{Bo877a}[\#39], as \*

\0``{\it In reality one can compute the ratio of the numbers
  of different initial states which determines their probability, which
  perhaps leads to an interesting method to calculate thermal
  equilibria}''
\*

\0and answered in \Cite{Bo877-b}[\#42p.166], see
Sec.(\,\ref{sec:VII-1}\,) above and Sec.(\,\ref{sec:XII-6}\,) below.

For $N$ particles in an equilibrium state the (often) accepted
answer is simple: the number is $\NN_0$, just equal to the number of
microcells (a strict form of ``ergodic hypothesis''\index{ergodic
hypothesis}). This means that we think that dynamics will
generate a one cycle permutation of $\NN_0$ microcells regularly
arranged on phase space $\X$ (which in this case is also the
attracting set), each of which is therefore, representative of
the equilibrium state. And the average values of macroscopic
observables are obtained simply as:

\be \media{F}=\NN_0^{-1}\sum_{\V q\in\EE_\t}
F(x_{\V q})\sim\ig_\X F(y)\m_{SRB}(dy)\label{e3.10.1}\ee
Let $W$ denote the area in phase space of the region $\X$ forming
a $6N-2$ dimensional Poincar\'e's section of the $6N-1$ dimensional
energy surface $M$, where sum of kinetic energy $K$ plus 
potential energy $U$ has a value $E$ while the positions of the
particles are confined within a container of volume $V$.

Then, imagining  phase space
discretized into microcells of  volume $h^{3N-1}$,
according to Boltzmann, see for instance p.372 in
\Cite{Bo896a}, the quantity:

\be S_B\defi k_B \log{\frac{W}{h^{3N-1}}}\label{e3.10.2}\ee
\ie $k_B$ times the logarithm of the total number of microcells
is, under the hypothesis that each microcell cyclically
visits all the others, proportional to the {\it physical
  entropy}\index{physical entropy} of the equilibrium state with
$N$ particles and total energy $E$, see
\Cite{Bo877-b}[\#42], (up to an additive constant
independent of the state of the system).%

A simple extension to systems out of equilibrium is to imagine,
as done in the previous sections, that a similar kind of
``ergodicity'' holds: namely that the microcells that represent
the stationary state form a subset of all the microcells, on
which evolution acts as a {\rm one cycle permutation} and that
entropy is defined by $k_B \log\NN$, with $\NN$ being the number
of phase space microcells {\it on the attracting set}, which in
general will be $\ll \NN_0$, if $\NN_0$ is the number of
regularly spaced microcells fitting in the phase space region
compatible with the constraints.

To proceed it is necessary to estimate the ratio between the
fraction of $W$ occupied by a coarse cell $E(\V q)$, namely
$\frac{|E(\V q)|}{W}$, and the SRB probability $\m_{SRB}(E(\V
q))$.

The ratio between the fraction of available phase space $\frac{|E(\V
q)|}{W}$ and the SRB probability $\m_{SRB}(E(\V q))$ can be estimated
heuristically by following the ideas of the previous section
Sec.(\,\ref{sec:VIII-3}\,)\,: it will provide an estimate of the number
$\NN$ of microcells on the discrete realization of the attracting set,
compared to their total number $\NN_0$, \ie total number of microstates.

For this purpose remark that the elements $E(\V
q)=\cap_{j=-\t}^\t S^{-j} E_{q_j}$ generating the Markovian
partition $\EE_\t$ can be symbolically represented as
Fig.3.10.1 below.

The surfaces of the expanding axes\index{expanding axis} of
$S^{-\t}E_{q_\t}$ and of the stable axes\index{stable axis} of $S^\t
E_{q_{-\t}}$, indicated with $\d_u,\d_s$ in Fig.(3.10.1) are
(approximately)
\be
\d_u=e^{-\sum_{j=1}^{\t} \l_u(S^{j}x)} \d_u(q_{\t}),\qquad
\d_s=
e^{\sum_{j=1}^{\t} \l_s(S^{-j}x)}
\d_s(q_{-\t})\label{e3.10.3}\ee
where $\d_s(q_{-\t}), \d_u(q_\t)$ are the surfaces of
the stable axis of $E_{q_{-\t}}$, and of the unstable axis of
$E_{q_{\t}}$, respectively.

From the figure it follows that:
\be\frac{|E(\V q)|}{W}=\frac{\d_s(q_{-\t})\d_u(q_{\t})\sin\f}{W} \,e^
{-\sum_{j=0}^{\t} (\l_u(S^{j}x)-\l_s(S^{-j}x))}\label{e3.10.4}\ee
where $\f$ is the angle at $x$ between $W^s(x)$ and $W^u(x)$ while
$\m_{SRB}(\V q)$ is given by Eq.\equ{e3.8.5}.

\eqfig{200}{140}{
\ins{129}{136}{$S_\t E_{q_{-\t}}$}
\ins{150}{115}{$E_{q_0}$}
\ins{180}{65}{$S_{-\t}E_{q_\t}$}
\ins{128}{17}{$\d_s$}
\ins{10}{25}{$\d_u$}}
{fig3.10.1}{fig3.10.1}

\begin{spacing}{.5}\0\tiny Fig.3.10.1:\label{fig3.10.1} the shadowed region
  represents the intersection $\scriptstyle \cap_{-\t}^\t S^{-j}
  E_{q_j}$; the angle $\scriptstyle \f$ between the stable axis
  of $\scriptstyle S_\t E_{q_{-\t}}$ and the unstable axis of
  $\scriptstyle S_{-\t} E_{q_\t}$ is marked in the dashed region
  (in general it is not $\scriptstyle 90^{\rm o}$) around a
  corner of the rectangle $\scriptstyle E(\V q)$.\end{spacing}
\*\*

A nontrivial property which emerges from the above formula is
that ${\d_s(q_{-\t})\d_u(q_{\t})\sin\f}$ is bounded, at
fixed $\t$, above and below as soon as it is $\ne0$,\,
\footnote{\tiny Simply because $q_j$ have finitely many values
and the angle $\f=\f(x)$ between stable and unstable manifolds at
$x$ is bounded away from $0,\p$ because of the transversality a
nd continuity of the planes $V^s(x),V^u(x)$ (in Anosov maps).}\,,
 \ie if $2\t$ is not smaller than the symbolic mixing
 time. The identity $\sum_{\V q}\frac{|E(\V
   q)|}{W}=1$ implies again (see Eq.\equ{e3.8.7}\,) a kind of
 Pesin's formula\index{Pesin's formula}\,:%
\footnote{\tiny Informally Pesin's formula is $\sum_{q_0,q_1,\ldots,q_N}
  e^{-\Si_0^\t \l_u(S^jx)}=O(1)$, see Eq.\equ{e3.8.6} and, formally,
  $s(\m_{srb})-\m_{srb}(\l_u)=0$, where $s(\m)$ is the Kolmogorov-Sinai
  entropy, see p.\pageref{Kolmogorov-Sinai's entropy}, and $\m(\l)\defi
  \int \l\,d\m$.  Furthermore $s(\m)-\m(\l_u)$ is maximal at $\m=\m_{srb}$:
  ``Ruelle's variational principle''
\index{Ruelle's variational principle}. 
See p.\pageref{Pesin's formula} and \Cite{GBG004}[Proposition
    6.3.4].}
\be
\log \sum_{\V q} e^{-\sum_{j=0}^{\t} (\l_u(S^{j}x)-\l_s(S^{-j}x))}=O(1)
,\qquad\forall\t>0\label{e3.10.5}\ee
Then $\frac{|E(\V q)|}{h^{3N-1}}$ is larger than the number of
attracting set microcells located inside $E(\V q)$: \ie
$\frac{|E(\V q)|}{h^{3N-1}}\ge \NN\m_{SRB}(\V q)$ where $h^{3N-1}$ is
the size of the microcells: thus, Eq.\equ{e3.10.2}, $\NN_0 h^{3N-1}\equiv W$ implies
$\NN\le \NN_0 \frac{|E(\V q)|}{W \m_{SRB}(\V q)}$.

Therefore $\NN\le \NN_0$ $\times$ times the ratio
of the \rhs of Eq.\equ{e3.10.4} to Eq.\equ{e3.8.5}, \ie of
the first to the second quantities:
\be\eqalign{
  \frac{|E(\V q)|}{W}&=
  e^{-\sum_{j=0}^{\t} (\l_u(S^{j}x)-\l_s(S^{-j}x))+O(1)}\cr
\m_{SRB}&=e^{\sum_{j=-\t}^{\t}  (-\l_u(S^{-j}x)-\l_u(S^{j}x)+O(1))}\cr}
\label{e3.10.6}\ee
(copied again omitting factors bounded uniformly above and
below)\,\footnote{\tiny Note that the correction $O(1)$ in the
first is due to factors $\d_s(q_{-\t})\d_u(q_\t)\cos\f$ common to
the quoted equations also bounded, independently of $\V q$, away
from $0$ and $\infty$ for $\t$ larger than the mixing symbolic
time, because of the remarked consequences of Pesin's formula
Eq.\,\equ{e3.10.3}\,. The $O(1)$ in the second line is due to the
uniform bounds of the factors $h_r,h_s$.}\,.

Hence, {\it using the key cancellation of $-\sum_{j<0}
\l_u(S^{j}x)$, in the exponents}:
\be\NN\le \NN_0\min_{\V q} e^{\sum_{j=0}^{\t} (\l_s(S^{-j}x(\V
q))+\l_u(S^{-j}x(\V q)))}\le \NN_0 e^{-\s_+ \t +O(1)}
\label{e3.10.7}\ee
as, for $x$ on the attracting set, the quantity $-\sum_{j=0}^\t
(\l_s(S^{j}x)+\l_u(S^jx))$ has average $-\t\s_+$, to leading
order in $\t$, with $\s_+=$ the average phase space contraction.

The picture must hold for all Markovian pavements $\EE_\t$ and
for all $\t$'s fulfilling the condition that the coarse grain
cells $E(\qq)$ contain a large number of microcells whose typical
sizes in momentum and positions will be called $\d p$ and $\d
q$: \ie $(\d p\,\d q)^{3N-1}=h^{3N-1}$ is the size of a
microcell. Hence if $\d_p,\d_q$ are the typical sizes in momentum
or, respectively, in position of an element of the partition
$\EE$, the sizes of the sides of $\EE_\t$ will be not smaller
than $e^{-\l \t}\d_p\gg \d p, e^{-\l \t}\d_q\gg \d q$, with $\l$
the maximal contraction of the stable and unstable manifolds
under $S$ or, respectively, $S^{-1}$.

The above condition will be fulfilled if $\t\le -\l^{-1}\log \g$
with $\g=\min (\frac{\d p}{\d_p},\frac{\d q}{\d_q})$ and small
$\g$ (that could be called a condition on the microcells
'precision' or 'size'): and it will imply, by Eq.\equ{e3.10.7}, the
inequality on the number $S_{mc}$ of recurrent microcells:
\be S_{mc}\defi k_B\log \NN\,\le \, k_B(\log\NN_0-
\fra{\s_+}{\l}\log\g)\label{e3.10.8}\ee
This inequality does not prove, without extra assumptions, that $\NN$
will depend nontrivially on $\g,\l,\s_+$ when $\s_+>0$. It gives, however,
an indication\footnote{\tiny I would say a strong one.} that $\NN$
might be {\it not independent} of the precision $\g$ used in defining 
microcells and course grained cells; and the dependence might not be
simply an additive constant because $\s_+/ \l$ is a dynamical quantity;
changing $\g$ to $\g'$ (\ie our representation of the microscopic
motion) changes the value of $S_{mc}$ proportionally to $\s_+/ \l$ which
depends nontrivially on the stationary state.

This is in {\it sharp contrast with the equilibrium result},
because in it $\s_+=0$ so that
a change in precision would change $\log\NN_0$ by a constant
independent of the equilibrium state (as in equilibrium the
number of microcells changes by $3N\log \frac{h'}h$ if the size
$h^{3N}$ is changed to ${h'}^{3N}$).  \*

Given a precision $\g$ of the microcells, the quantity $\NN$
estimates how many \index{microcell recurrent} ``recurent''
microcells must be used, in a discretization of phase space, to
obtain a {\it faithful} representation of the attracting set and
of its statistical properties on scales $\d_p,\d_q\gg \d p, \d
q$. Here by ``faithful'' on scale $\d_p,\d_q$ it is meant that
all observables which are, or can be regarded, constant on such
scale will show the correct statistical properties, \ie that
coarse cells of size much larger than $\g^{-1}$ will be visited with
the correct SRB frequency.  \index{SRB frequency}

\section{\texorpdfstring{$\bf k_B\log\NN:$}\, Entropy or Lyapunov function?}
\def\SEC{\texorpdfstring{$\bf k_B\log\NN:$}\, Entropy or Lyapunov function?}
\label{sec:XI-3}\iniz
\lhead{\small\ref{sec:XI-3}.\ $k_B\log\NN$: Entropy or 
Lyapunov function?}

From the previous sections some conclusions can be drawn, relying
on the CH.
\*

\0(1) Although $S_{mc}$ (see Eq.\equ{e3.10.8}) estimates the
recurrent cells count, it does not seem to deserve to be taken
also as a definition of entropy for statistical states of systems
out of equilibrium, not even for systems simple enough to admit a
transitive Anosov map as a model for their evolution.  The reason
is that it might not change by a trivial additive constant if the
size of the microcells is varied (except in the equilibrium case,
because $\s_+=0$): at least the question requires further
investigation. It also seems to be a notion distinct from the recently
introduced ``Boltzmann's entropy''\index{Boltzmann's
  entropy}, \Cite{Le993}\Cite{Ei922}.  \*

\0(2) $S_{mc}$ is also different from the Gibbs' entropy
\index{Gibbs entropy}, to which it is equivalent only in
equilibrium systems: in nonequilibrium (dissipative) systems the
latter can only be defined as $-\io$ and perpetually decreasing;
because {\it in such systems is defined the rate at which
(Gibbs') entropy is\index{entropy} ``generated'', or ``ceded to
the thermostats'' by the system,} to be $\s_+$, \ie to be minus
the average phase\index{average contraction} space contraction,
see \Cite{An982}\Cite{Ru999}\,.
\*

\0(3) We also see, from the above analysis, that the SRB distribution
appears to be the equal probability distribution among the $\NN$ microcells
which are not transient.%
\footnote{\tiny In equilibrium all microcells are non transient
under the strict ergodic hypothesis and the SRB distribution
coincides with the Liouville distribution.\label{n8-3}}
 Therefore $S_{mc}=k_B\log\NN$ maximizes a natural functional of
 the probability distributions $(\p_x)_{x\in\lis\X}$ defined on
 the discretized approximation of the attracting surface
 $\lis\X$; namely the functional defined by
 $S(\p)=-k_B\sum_x \p_x\log\p_x$.\\
 If a distribution
 $(\p(t)_x)_{x\in\lis\X}$ represents a non stationary state
 such $S(\p(t))$ does
 not seem interpretable (by the analysis in
 Sec.(\,\ref{sec:X-3})\,) as a function, of an evolving state, 
 defined up to an additive constant. It might nevertheless play
 the role of a {\it Lyapunov function}\index{Lyapunov function},
 which estimates how far a distribution is from the SRB
 distribution, and reaches its maximum $S_{mc}$ on the SRB
 distribution.  \*

\0(4) If we could take $\t\to\io$ the distribution $\m$ which
attributes a total weight to $E(\V q)$ equal to $N(\V
q)=\NN\m_{SRB}(E(\V q))$ would become the exact SRB
distribution. However it seems conceptually more satisfactory to
suppose that $\t$ will be large but not infinite.

\chapter{Fluctuations}
\label{Ch4} 

\chaptermark{\ifodd\thepage
Fluctuations\hfill\else\hfill 
Fluctuations\fi}
\kern2.3cm
\section{SRB potentials}
\def\SEC{SRB potentials}
\label{sec:I-4}\iniz
\lhead{\small\ref{sec:I-4}.\ \SEC}

Intuition about the SRB distributions, hence about the statistics
of chaotic evolutions\index{chaotic evolutions}, requires an
understanding of their nature. The physical meaning discussed in
Sec.(\,\ref{sec:VII-3}\,) is not sufficient because the key notion of
{\it SRB potentials\index{SRB potentials}} is still missing.

It is introduced in this section: it might appear, at first,
formal and of little physical interest. However it achieves a
very surprising result: namely it sets up the technical tools to
study the SRB distribution by representing it rigorously as a
short memory {\it $1$-dimensional} Gibbs distribution,
\Cite{Ga000}, \ie essentially a Markov chain. Such distributions
are correctly considered the simplest and best known probability
distributions.

Given a smooth hyperbolic transitive evolution $S$ on a smooth
surface $\X$ (\ie an Anosov map) it is possible to identify (up
to exceptions contained in a set of $0$ area) the points of $\X$
with the set $\wt\X$ of infinite strings of symbols
$\qq=\{q_j\}_{j=-\infty}^{\infty}$ representing the labels of the
elements of a Markovian partition $\EE=(E_1,\ldots,E_m)$
successively containing $S^jx$, hence compatible with the
partition transition matrix $M$, \ie $M_{q_j,q_{j+1}}=1$. And to
identify the action of $S$ on $x\in\X$ with a translation $\wt
S$, one step to the left, of the string representing $x$, as in
the definition of symbolic dynamics in
Sec.\,(\ref{sec:IV-3}\,).\label{compatibility matrix}

The identification permits to describe unambiguosly the
stationary distributions that can be reached from initial data
selected outside a $0$-area set in $\X$: hence in particular the
SRB distributions (as they are generated by ``chaotic data'',
see, (\ref{sec:VI-2})\,, hence from data that can be coded
unambiguously into compatible strings via Markovian
partitions). To simplify the notations the symbolic
dynamical systems $(\wt \X,\wt S)$ and $(\X,S)$ will be often
identified.

Let $\EE=(E_1,\ldots,E_m)$ be a Markovian partition: the time
average of any smooth observable can be computed, see
Eq.\equ{e3.8.6}, as:
\be \media{F}_{SRB}= \lim_{\t\to\infty}
\frac{ {\sum_{\V q}^*}e^{-\L_{u,\t}(x_{\V q})}\,F(x_{\V q})}{\sum_{\V q} 
e^{-\L_{u,\t}(\V q)}}
\label{e4.1.1}\ee
which is an exact formula for $\m_{SRB}$: here the $\sum^*$ is
the sum over compatible $\qq=(q_{-\t},\ldots, q_\t)$, $x_\qq$ is
a point {\it arbitrarily} selected in $\cap_{j=-\t}^\t S^{-j}E_{q_j}$ and:
\be \L_{u,\t}(x)=\sum_{j=-\t}^{\t} \l_u(S^j x)\label{e4.1.2}\ee
is defined in terms of the function $\l_u(x)\defi \log|\det
\partial S(x)_u|$ which gives the expansion rate of the unstable
manifold through $x$ in one time step.  The function
$\L_{u,\t}(x)$ gives the expansion rate of the unstable manifold
at $S^{-\t}x$ when it is transformed into the unstable manifold
at $S^\t x$ by  $S^{2\t}$.

The notion of potential of the expansion rates of the Markovian
partition $\EE=(E_0,\ldots,E_m)$ is introduced to express
conveniently the rates $\l_u(x)$.

Defining the SRB
potentials\index{SRB potentials} is very simple in the case in
which the compatibility matrix $M$ as {\it no zero entry}, \ie
all transitions are allowed. In this (unrealistic) case a point
$x=\cap_{j=-\infty}^\infty E_{q_j}$  can be identified with its
symbolic history $\qq$\, \ref{symbolic history}\, and each of its substrings
$\qq_k=(q_{-k},\ldots,q_k)$, $k=0,1,2\ldots$,
can be extended to an infinite compatible sequence as
$\ldots,1,1,q_{-k},\ldots,q_k,1,1,\ldots$
obtained by extending $\qq_k$ to an infinite sequence by writing
the symbol $1$ (say) to the right and left of it.

If the compatibility matrix $M_{i,j}$ has zero entries it is
always possible to complete compatible finite strings
$\qq_k=(q_{-k},\ldots,q_k)$ into infinite compatible strings
by virtue of the finite mixing time $n_0$ of the matrix $M$ (so that
$M^{n_0}_{i,j}>0$ for all $i,j$, see Sec.(\,\ref{sec:IV-3}\,)),
continuing $\qq_k$ to the right and left into $\lis\qq_k$
following a standard rule, as in the example detailed in the
footnote\,
\footnote{\tiny \label{standard continuation} If $n_0$ is the
symbolic mixing time, see Sec.(\,\ref{sec:IV-3}\,), for the
compatibility matrix, for each of the $m$ symbols $q$ fix a
string $\Ba(q)$ of length $n_0$ of symbols leading from $q$ to a
prefixed (once and for all) symbol $\lis q$ and a string $\Bb(q)$
of length $n_0$ of symbols leading from $\lis q$ to $q$; This
means fixing for each $q$ two compatible strings $\V a(\V
q)=\{a_0=q,a_1,\ldots, a_{n_0-1},a_{n_0}=\lis q\}$ and $\Bb(\V
q)=\{ b_0=\lis q,b_1,\ldots,b_{n_0-1},b_{n_0}=q\}$ such that
$\prod_{i=0}^{n_0-1} M_{a_i,a_{i+1}}=\prod_{i=0}^{n_0-1}
M_{b_i,b_{i+1}}=1$.  Then continue the string
$q_{-\t},\ldots,q_{\t}$ by attaching to it $\V a(q_\t)$ to the
right and $\V b(q_{-\t})$ to the left; and finally continue the
so obtained string of length $2n_0+2\t+1$ to a compatible
infinite string in a prefixed way, to the right starting and to
the left ending with $\lis q$ , choosing, arbitrarily but once
and for all, two compatible sequences starting and ending with
$\lis q$: the simplest is to repeat indefinitely a string of
length $n_0$ beginning and ending with $\lis q$ to the right and to
the left.}\,.

Therefore to avoid irrelevant notational difficulties we proceed
imagining that every finite string $\qq_k$ is, once and for all,
{\it uniquely} continued into a {\it standard} infinite
compatible string $\lis\qq_k$.
Then if $x$ has history $\qq= \{q_j\}_{j=-\infty}^{j=\infty}$
let:
\be \qq_k\defi (q_{-k},\ldots,q_k),\qquad \lis \qq_k
= (\ldots ,q_{-k},\ldots,q_k,\ldots)
\label{e4.1.3}\ee
and remark that it is (trivially) possible to write
$\l_u(x)$ as a sum of ``finite range
potentials'' on the history $\qq$ of $x$:\index{expansion potential}

\be \l_u(x)=\sum_{k=0}^\infty \F(q_{-k},\ldots,q_k)= 
\sum_{k=0}^\infty \F(\qq_k),\label{e4.1.4}\ee
where the potentials $\F(\qq_k)$  are defined ``telescopically'':
\be \eqalign{
\F(q_0)=& \l_u(x_{\lis\qq_0}), \qquad x_{\lis\qq_0}=x\cr
\F(q_{-k},\ldots,q_k)=&\l_u(x_{\lis\qq_{k}})-\l_u(x_{\lis\qq_{k-1}}),
\qquad k\ge1\cr}\label{e4.1.5}\ee
Convergence of the series in Eq.\equ{e4.1.5} is implied by the remark
that the two strings $\lis\qq_{k}, \lis\qq_{k-1}$ coincide at the positions
labeled $-(k-1),\ldots,(k-1)$: hyperbolicity then implies that the
points with symbolic representations given by $\lis\qq_{k}, \lis\qq_{k-1}$
are close within $2C e^{-\l (k-1)}$, see comment following
Eq.\equ{e3.4.2}.

Define also $\F(\h_c,\ldots,\h_{c+d})=\F(\Bh)$,
$d=2k,k=0,1,\ldots$, by setting $\F(\Bh)$ equal to
$\F(\Bh^{tr})$ with $\Bh^{tr}$ obtained from $\Bh$ by translating
it to a string with labels centered at the origin
$\Bh^{tr}=(\h^{tr}_{-k},\ldots,\h^{tr}_{k})$
with $\h^{tr}_j=\h_{c+j}$. For all other finite strings
$\F$ will also be defined, but set $\equiv0$.

In this way $\F$ will have been defined as a ``translation
invariant potential'' for the expansion rates: it can be $\ne0$
only for strings $\Bh=(\h_c,q_{c+1},\ldots,\h_{c+2k})$.  From
Eq.\equ{e4.1.2},\equ{e4.1.4},\equ{e4.1.5} the expansion
rates $\l_u(x),\L_{u,\t}(x)$ can then be written simply as:
\be \l_u(x)\equiv\sum_{\hbox{$\Bh:\qq,[0]$}} \F(\Bh)\quad \tto
\L_{u,\t}(x)=\sum_{\hbox{$\Bh:\qq,[-\t,\t]$}} \F(\Bh)
\label{e4.1.6}\ee
where $\qq$ ist the history of $x$ and $\sum_{\Bh:\qq,[-\t,\t]}$
denotes sum over the finite substrings
$\Bh=(q_c,\h_{c+1},\ldots,q_{c+2k})$ of $\qq $ with labels
centered in $c\in [-\t,\t]$ (\ie at the points $c+k\in[-\t,\t]$).

By the smoothness of the map $S$ and the
H\"older continuity of $\l_u(x),\l_s(x)$,
Eq.\equ{e3.2.3}, there exists constants $B_0,B,\l>0$ such that

\be |\F(\Bh)|\le B_0 \,e^{-\l \,k},\qquad
\|\F\|=\sum_{k\ge1} \max_{\Bh_k}|\F(\Bh_k)|\le B\label{e4.1.7}\ee
if $k$ is the length of the string $\Bh$ and the $\max$ is over
$\Bh_k=(\h_{-k},\ldots,\h_k)$.  \*

Given a finite compatible string $\qq_\t=(q_{-\t},\ldots,q_\t)$ a
point $x_{\qq_\t}\in\cap_{j=-\t}^\t S^{-j} E_{q_j}$ can be (arbitrarily)
chosen, simply using the standard infinite compatible sequence
$\lis\qq_\t$ extending $\qq_\t$ to an infinite sequence (via 
standard continuation defined above \footref{standard continuation}):
the latter determine uniquely a point $x_{\qq_\t}$ that can be
used to express SRB averages as, with the notation of
Eq.\equ{e4.1.6}\,\footnote{\tiny Remark that if $\qq_\t=(q_{-t},\ldots,q_\t)$ is a
finite compatible string then if $\lis\qq_\t$ is an infinite
compatible string  obtained by the mentioned ``standard
choices'' to continue $\Bh$ to an infinite compatible
string then $\lis\qq_\t$ is the history of a point $x_{\qq_\t}\in
\cap_{t\in [-\t,\t]}S^{-t} E_{q_t}$. Note that the arbitrariness
of $x_{\qq_\t}$ depends on
the mentioned ``standard choices'' made to continue $\qq_\t$ to an
infinite compatible string.}\,:
\be \media{F}_{SRB}= \lim_{\t\to\infty}
\frac{\sum_{\qq_\t} e^{-\sum_{\hbox{$\scriptstyle\Bh: \qq,[-\t,\t]$}}
        \F(\Bh)} \,F(x_{\lis\qq_\t}) }
{\sum_{\V q_\t} e^{-\sum_{\hbox{$\scriptstyle\Bh: \qq,[-\t,\t]$}} \F(\Bh)}}
\label{e4.1.8}\ee
(and the result is independent on the noted
arbitrariness of $x_{\lis\qq_\t}$).

\section{Chaos and Markov processes}
\def\SEC{Chaos and Markov processes}
\label{sec:II-4}\iniz
\lhead{\small\ref{sec:II-4}.\ \SEC}

In the literature many works can be found which deal with nonequilibrium
theory and are modeled by Markov processes\index{Markov process}\,,
introduced either as fundamental models or as approximations of
deterministic models.

This is sometimes criticized because of the {\it a priori}
stochasticity assumption, which might sound as introducing {\it ex
machina} the key property that should, instead, be derived.

The proposal of Ruelle on\index{Ruelle} the theory of
turbulence\index{turbulence}, see \Cite{Ru978b}\Cite{Ru980}\,,
but in fact already implicit in his earlier works, \Cite{Ru976},
has been the inspiration of the CH (Sec.(\,\ref{sec:VII-2}\,)).  The
coarse graining theory, that follows from the CH, essentially
explains why there is little difference between Markov chains
evolutions and chaotic evolutions, and it is useful to establish
a precise, general, connection between the two.

From the general expression for the SRB distribution in terms of
its symbolic dynamics representation, see Sec.(\,\ref{sec:IV-3}\,),
it follows that {\it if $\F(\Bh)\equiv 0$ for strings of length
  $k>k_0$}, \ie if the potential $\F$ has finite range rather
than an exponential decay to $0$ as in Eq.\equ{e4.1.7}, then
the limit in Eq.\equ{e4.1.8} would exist (independently of the
arbitrariness of the above choice of the string representing
$x_{\lis{\V q}_\t}$). And it would be a finite memory,
transitive\footnote{\tiny Because of transitivity of the
compatibility matrix.} Markov process, \ie a probability
distribution on the strings $\qq$, invariant under translations
$\wt S$, defining a dynamical system
$(\wt\X,\wt S)$ equivalent to an ordinary Markov process.

The long range of the potential does not really affect the
picture: technically infinite range processes with potentials
decaying exponentially fast are probability distributions in the
larger class of stochastic processes known as $1$--dimensional
{\it Gibbs distributions}: they have essentially the same
properties as Markov chains. In particular the limit in
Eq.\equ{e4.1.8} does not depend on the arbitrariness of the
(arbitrary) selection of the string $\lis\qq_\t$ (completing
$\qq_\t=(q_{-\t},\ldots,q_\t)$ and determining
$x_{\lis\qq_\t}$, \Cite{GBG004}).  Furthermore they are
translation invariant and exponentially fast mixing, in the sense
that if $S^n G(x)\defi G(S^n x)$ and $F,G$ are smooth with maxima
$\|F\|,\| G\|$:
\be |\media{F S^n G}_{SRB}-\media{F}_{SRB}\media{G}_{SRB}|\le \g\,||F||\,
||G||\,e^{-\k |n|}
\label{e4.2.1}\ee
for all $n\in Z$ and for suitable constants $\g,\k$ which depend on the
regularity of the functions $F,G$.\footnote{\tiny  The mixing
property is just the mixing property of the SRB distribution seen as
a process on the symbolic dynamics: if $F$ is a smooth osevable
then in the symbolic representation $\V q$ of points $x_{\V q}$ 
the function $F(x_{\V q})$ can be represented by potentials just
as done for the $\l_u(x_{\V q}),\l_s(x_{\V q})$ in the previous
section, and the smoothness of $F$ will imply exponential decays
of the potential representing it and the Eq.\equ{e4.2.1} is
given by the well known mixing of Markov chains or Gibbs processes.}

The dimension $1$ of the Gibbs process corresponding to the SRB
process in $(\X,S)$ is {\it remarkable} because it is only in
dimension $1$ that the theory of Gibbs distributions with
exponentially decaying potential is {\it elementary and
easy}. And, nevertheless, a general deterministic evolution,
under the CH, has statistical properties identical to those of a
Markov process if, as usual, the initial data are chosen close to
an attracting set and outside a $0$ area set in phase space (\ie
the initial data are chaotic in the sense defined in
Sec.(\,\ref{sec:VI-2}\,)).

Thus it is seen that the apparently different approaches to
nonequilibrium based on Markovian models or on deterministic
evolutions are in fact completely equivalent, at least in
principle.  For any deterministic model, under the CH, a Gibbs
process with short range potential $\F$ (\ie which decays
exponentially) can be constructed equivalent to it, via an
algorithm which, in principle, is constructive (because the
Markov partitions can be constructed, in principle: as
illustrated in $2$-dimensional systems in appendix \ref{appG},
see also the simple examples in Sec.(\,\ref{sec:V-3}\,)).

The result might be at first sight surprising: however it should
be stressed that the sequences of random numbers are
precisely generated as symbolic histories of chaotic
maps. \footnote{\tiny For a simple, handy and well known although
not the ``best'', example see \Cite{KR988}[p.46].}

Furthermore truncating the potential $\F$ to its values of range
$<k_0$, with $k_0$ large enough, will approximate the Gibbs process
with a finite memory Markov process.

The approximation can be pushed as far as
wished in many senses, for instance in the sense of ``distribution and
entropy'', see \Cite{Or974},\label{Kolmogorov-Sinai's entropy} %
\,\footnote{
\0\tiny This means that given $\scriptstyle \e, n>0$
there is $k$ large enough so that the Gibbs'
distribution\index{Gibbs distribution} $\scriptstyle \m_{SRB}$
with potential $\scriptstyle \F$ and the Markov process
$\scriptstyle \m_k$ with potential $\scriptstyle \F^{[\le k]}$,
truncation of $\scriptstyle \F$ at range $\scriptstyle k$, will
be such that $\scriptstyle \Si_{\V q=(q_0,\ldots,q_n)}$
$\scriptstyle |\m_{SRB}(E(\V q))-\m_k(E(\V q))|<\e$ and the
Kolmogorov-Sinai's entropy \index{Kolmogorov-Sinai's entropy}
(\ie $s( \m)\defi$ $\mathop{\lim}\limits_{m\to\infty}
-m^{-1}\Si_{\V q=(q_0,\ldots q_{m})} \m(E(\V q))\log\m(E(\V
q))$) of $\m_{SRB}$ and $\m_k$ are close within
$\e$.
}%
\, which implies that the process is a ``Bernoulli's process'':
   {\it i.e.}  the symbolic histories $\qq$ could even be coded,
   outside a set of $\m_{SRB}$ probability $0$, into new
   sequences of symbols without any compatibility restriction and
   with independent probabilities, \Cite{Ga973}\Cite{Le973}\Cite{MR975}.

It is also remarkable that the expression for the SRB distribution is
the same for systems in equilibrium or in stationary
nonequilibrium: in the sense that it
has the form of a Gibbs' distribution of a $1$ dimensional ``spin''
system (where the labels of the histories $\qq$ are the 'spin values').

\section{Symmetries, time reversal and axiom C}
\def\SEC{Symmetries, time reversal and axiom C}
\label{sec:III-4}\iniz
\lhead{\small\ref{sec:III-4}.\ \SEC}

In chaotic systems the symbolic dynamics\index{symbolic dynamics}
inherits naturally symmetry properties (if any) enjoyed by the
time evolution. An interesting property is, as an example in
particle systems, the standard time reversal symmetry (velocity
reversal with positions unchanged): it is important because it is
a symmetry of nature and present also in Gaussian thermostat
models, \eg in the models considered in Sec.(\,\ref{sec:III-2}\,),
which is among the reasons why they attracted so much
interest.\index{time reversal}

Time reversal (TRS) is, in general, defined as a smooth {\it
  isometric} map $I$ of phase space which ``anticommutes'' with
the evolution $S$, namely $IS=S^{-1}I$, and its iteration
yields identity: $I^2=\bf1$. 

If $\EE_0$ is a Markovian partition then also $I\EE_0$ has the
same property, because $I W^s(x)=W^u(Ix)$ and $I W^u(x)=W^s(Ix)$
so that the conditions in \equ{e3.4.1} hold. Since a time
reversal anticommutes with evolution it will be called a
``reverse'' symmetry.

It is then possible to consider the new Markovian partition
$\EE=\EE_0\cap I \EE_0$ whose elements have the form $E_{ij}\defi
E_i\cap IE_j$. Then $I E_{ij}=E_{ji}$.  In each set $E_{ij}$ with
$i\le j$ let $x_{ij}$ be a selected center for $E_{ij}$ and
choose as center for $E_{ji}$ the point $x_{ji}=I x_{ij}$.\,
\footnote{\tiny Hence $x_{ii}=Ix_{ii}$ if $E_i\cap I E_i$ is a non empty
rectangle. Such a fixed point exists because a rectangle is
homeomorphic to a ball.}
Define $I(i,j)=(j,i)$ and suppose also (not restrictive but
useful to simplify the exposition) that the metric on $\X$ is
time reversal symmetric.

The partition $\EE$ will be called ``time reversal
symmetric''. Then the compatibility matrix will enjoy the
property\index{time reversal symmetric partition}
$M_{\a,\b}=M_{I(\b,\a)}$, for all pairs $\a=(i,j)$ and
$\b=(i',j')$; and the map $I$ will act on the symbolic
representation $\{q_k\}_{-\infty}^\infty$ of $x$ transforming it into the
representation $\{Iq_{-k}\}_{-\infty}^\infty$ of $Ix$ and:
\be
\l_{u}(Ix) = -\l_{s}(x),\qquad \L_{u,\t}(Ix)=-\L_{s,\t}(x)
\label{e4.3.1}\ee
where $\L_{c,\t}(Ix)=\sum_{j=-\t}^\t \l_c(S^jIx)$ for $c=u,s$;
Eq.\equ{e4.3.1} rely on the assumed isometric property of $I$
which implies $\l_u(Ix)=-\l_s(x)$.  A symmetry of this type can
also be called a ``reverse symmetry''.

Likewise if $P$ is a symmetry, \ie is a smooth {\it isometric} map of
phase space which squares to the identity $P^2=\bf1$ and {\it commutes} with
the evolution $PS=SP$, then if $\EE_0$ is a Markovian pavement the 
$\EE_0\cap P\EE_ 0$ is $P$-symmetric\index{symmetric partition}, \ie $P
E_{i,j}=E_{j,i}$.  And, if $x_{i,j}$ is chosen so that
$Px_{i,j}=x_{j,i}$, for all $\t$ and $c=u,s$, as
in the previous case:
\be \l_{c}(x) =\l_{c}(Px),\qquad \L_{c,\t}(Px) =\L_{c,\t}(x)
\label{e4.3.2}\ee
A symmetry of this type can be called a ``direct symmetry''.

If a system admits two {\it commuting} symmetries%
 \index{symmetries: commuting} 
one direct, $P$, and one reversed, $I$, then $PI$ is a reversed
  symmetry\index{symmetry reversed}, \ie a {\it new} time reversal.

This is interesting when a time evolution has the symmetry $I$
on the full phase space but not on the attracting set $\AA$ and maps the
latter into a disjoint set $I\AA\ne \AA$ (a ``repelling
set''\index{repelling set}). If the evolution admits also a direct symmetry
$P$ mapping the repelling set back onto the attracting one ($PI\AA=\AA$),
then the map $PI$ maps the attracting set into itself and is a time
reversal symmetry for the motions {\it on the attracting set} $\AA$.

The natural question is: when can it be expected that $I$ maps the
attracting set into itself ? 

If the system is in equilibrium the attracting set is, according
to the CH, the full phase space (in absence of other symmmetries
or, see below, phase transitions).  Furthermore often the
velocity inversion is a time reversal symmetry, as in the models
of Sec.(\,\ref{sec:III-2}\,). \\
At small perturbation of such models, switching-in action of
nonconservative forces and thermostats, the system keeps the same
time reversal symmetry and the full phase space as attracting surface
(by the ``structural stability of Anosov
maps'')\index{structural stability}.

The situation becomes very different when the forcing increases: the
attracting surface $\AA$ can become strictly smaller than the full phase space
and the velocity reversal $I$ will map it into a disjoint repelling surface
$I\AA$, so that $I$ is no longer a time reversal for the interesting motions,
\ie the ones that take place on an attracting set. Or more
generally a perturbation, even if small, may break the reversal symmetry.
\footnote{\tiny If the time reversal symmetry is broken by a
small pertutnation, the structural stability will imply the
existence of a coordinate transformation $\F$ of the perturbed
Anosov system back into the unperturbed one: hence $\F$ will transform
the unperturbed time reversal $I$ into $\wt I$.
But it is illusory to think that $\wt I$ will just replace $I$.
The transformation $\F$ of a perturbed Anosov map
into the unperturbed one is, in general, not a smooth change of
coordinates but just H\"older continuous. So that the image $\wt I$ of
$I$ in the new coordinates will not be smooth, {\it unless} $\wt I$
remains the same map of phase space, as in the models considered in
Sec.(\,\ref{sec:II-3}\,).}

Nevertheless it is possible to formulate a property for the
evolution $S$, introduced in \Cite{BG997} and called {\it Axiom
  C}, which\label{axiom C} has the following features: \*

\0(1) there is a smooth TRS (time reversal symmetry) $I$ on the
full phase space but the attracting set $\AA$ is not invariant and
$I\AA\ne\AA$ is, therefore, a repelling set,

\0(2) the attracting surface $\AA$ is mapped onto the repelling
surface\index{repelling set} by a smooth isometric map $P$ which
commutes with $I$ and $S$, and iterated yields the identity:
$P^2=\bf1$,

\0(3) it is structurally stable\index{structurally stable}: \ie if the
evolution $S$ is perturbed into $S'$ then, for small enough perturbations, the
properties (1),(2) remain valid for $S'$.  \*

For such systems the map $PI$ will be a TRS for  motions on
the attracting surface $\AA$.  A precise definition of the Axiom
C property and related comments are in Appendix \ref{appH}.

The interest of the Axiom C is that, expressing a structurally
stable property, it might hold quite generally: thus ensuring
that the original global TRS, often associated with the velocity
reversal operation, is a symmetry that ``cannot be
destroyed''. Initially, when there is no forcing, it is a natural
symmetry of the motions; increasing the forcing eventually the
TRS is {\it spontaneously broken} because the attracting surface
$\AA$ becomes no longer dense on phase space (\ie
$I\AA\cap \AA=\emptyset$, although the time reversal $I$ remains a
symmetry for the motion in the sense that on the full phase space
$IS=S^{-1}I$).\index{symmetry breakdown}
\\
However if the system satisfies Axiom C a new symmetry $P$ is spawned (by
virtue of the constraints posed by the geometric Axiom C)
  which maps the attracting set onto the repelling set and commutes with
  $I$ and $S$. Therefore the map $I^*=PI$ maps the attracting set into
  itself and is a time reversal for the evolution $S$ restricted to the
  attracting set.

This scenario can repeat itself, as long as the system remains an
Axiom C system: the attracting set can split into a pair of
smaller attracting and repelling sets and so on until, increasing
further and further the forcing, an attracting set may be reached
on which either motion is no longer chaotic (\eg it is a periodic
motion), \Cite{Ga998}, or the Axiom C property no longer holds.

The above is to suggest that time reversal symmetry might be quite
generally a symmetry of the motions on the attracting sets, as suggested
by the axiom C structural stability. The
axiom C will be discussed in more detail Appendix \ref{appH}\,.
\*

\0{\it Remark:} The appearance of a non time reversal invariant
attracting manifold in a time reversible system can be regarded
as a {\it spontaneous symmetry breaking\index{symmetry
breaking}}: the existence of $I^*$, in the mentioned axiom C
systems, means that in some sense time reversal symmetry of the
system cannot be broken. If it does spontaneously break, then it
is replaced by a lower symmetry ($I^*$) which ``restores
it''. The analogy with the symmetries $T$ (broken) and $TCP$
(valid) of Fundamental Physics would be remarkable and the
implications on irreversibility arising in reversible systems are
manifest.

\section{Pairing symmetry}
\def\SEC{Pairing symmetry}
\label{sec:IV-4}\iniz
\lhead{\small\ref{sec:IV-4}.\ \SEC}

The analysis of Sec.(\,\ref{sec:III-4}\,) acquires greater interest
when general relations can be established between phase space
properties and attracting surface properties. For instance the
entropy production\index{entropy production} is related to the
full phase space area contraction, but in cases in which the
attracting surface $\AA$ is not the full phase space the question
arises: should the entropy production rate be the average volume
contraction in the full phase space or the average contraction of
the surface elements of $\AA$ ?

Therefore finding further symmetries promises to help in
attacking questions like the one exemplified above. An example of
possible further symmetry is the {\it pairing symmetry},
discovered in \Cite{Dr988} and greatly extended in \Cite{DM996}. 

Let $2D$ be the number of the Lyapunov exponents,\index{Lyapunov
  exponents} see  definition preceding Eq.
\equ{e2.8.7}\, excluding possibly an
even number of vanishing ones, and order the first $D$ exponents,
$\l^+_0\ge\l^+_1 \ldots\ge\l^+_{D-1}$, in decreasing order, while
the next $D$, $\l^-_0\le\l^-_1 \ldots\le \l^-_{D-1}$, are ordered
in increasing order. Then {\it pairing symmetry} occurs if:
\be\frac12(\l^+_j+\l^-_j)= const\qquad {\rm for\ all}\
j=0,\ldots,D-1;\label{e4.4.1}\ee
the constant will be called ``{\it pairing level}'' or ``{\it pairing
  constant}'': and it  must be, see Eq.\equ{e2.8.7}\,,
$-\frac1{2D}\s_+$, where $-\s_+$ is
the average phase space contraction.\index{pairing rule}%

Systems for which Eq.\equ{e4.4.1} has been proved are subject
 to strong restrictions, excluding in particular the reversible
 nonequilibrum models of Sec.(\,\ref{sec:II-2}\,); but a substantial
 step has been a proof applicable to the latter general models:
 the extension holds also in a far {\it stronger} sense than in
 the first proof.
 \\
 Namely the {\it local Lyapunov exponents}, \ie the $2D$ non
 trivial eigenvalues\footnote{\tiny The models in which the
 stronger pairing holds are defined in continuous time and the
 matrix $(\dpr S_t)^*\dpr S_t$ has two eigenvalues with time
 average $1$ (\ie two $0$ Lyapunov exponents).\label{n4-1}} of
 the matrix $\frac1{2t}\log (\dpr S_t(x)^T\dpr S_t(x))$ can be
 ordered as above and, averaged over time, yield the Lyapunov
 exponents in the limit $t\to\infty$: furthermore for each $t>0$
 are paired to a $j$-independent constant $\s_{t,pair}(x)$,
 dependent on the point $x$ in phase space and on $t$; and, as
 $t\to+\infty$, each pair has average
 $\lis\s_{t,pair}=-\frac1{2D}\s_+$.  This property will be called
 the {\it strong pairing rule}, see appendix \ref{appI}.

\*
\0{\it Remarks:} (1) It should be kept in mind that, while a
pairing of the Lyapunov exponents is, when holding, independent
of the metric used on phase space, the strong pairing rule can
only hold, if at all, for special metrics.
\\
(2) Among the examples are the evolutions in continuous time for
the equations of the reversible model (1) in Sec.(\,\ref{sec:III-2}\,),
Fig.2.3.1.  In the latter systems, and more generally in systems
in which there is an integral of motion, like in thermostatted
systems with a isokinetic or isoenergetic Gaussian constraint
(see Chapter \ref{Ch2}\,), there will be a vanishing Lyapunov
exponent besides the vanishing one associated with the flow
direction: it will be associated with the variations of the
integral of motion; as discussed in Appendix \ref{appI}, in the
general theory of the pairing, the two vanishing exponents have
to be excluded in checking Eq.\equ{e4.4.1}\,.
\*

The question raised at the beginning of the section is an example
in which the pairing symmetry may help in cases
with $\AA\ne M$. Keeping in mind the Kaplan-Yorke proposal on the
fractal dimensions of attracting sets, it is a natural
assumption, to suppose  that the surface $\AA$
has dimension $d_{ky}$ defined as the first {\it integer} $k$
such that $\sum_{j=1}^{k} \l_j<0$, still imagining the exponents
ordered as above. The so defined $d_{ky}$ is, therefore, the first integer
above the Kaplan-Yorke fractal dimension.\,\footnote{\tiny
This does not preclude the possibility that the
``attractor'', defined as a set on $\AA$ with SRB probability $1$
and minimal 'fractal dimension',\,\Cite{ER985}\,, be a fractal
set: smoothness of an attracting surface has nothing to do with
fractal dimensionality of invariant subsets,
see \Cite{ER985}\Cite{GC995}\Cite{Ga995c}. The motion on this
lower dimensional surface $\AA$ might still have an attractor
with dimension {\it lower} than the dimension of the surface
$\AA$ itself, as suggested by the Kaplan--Yorke
formula, \Cite{CF983}.}

Thus an empirical criterion, given the CH, emerges for deciding whether $\AA=M$
(case, often called ``ergodic'', in which the phase space
contraction is interpreted as entropy production rate) or $\AA\ne
M$ (case in which, as mentioned above, a question arises about the interpretation):
and a criterion could be: $\AA\ne M$ if $d_{ky}<{\rm
dim}M$.\,\label{full dimension attractive set} If pairing holds
and the pairing level is small enough then $d_{ky}={\rm dim}M$: this
seems to be a condition for $\AA=M$, however a smaller
$d_{ky}$ does not necessarily mean that $\AA\ne M$.

A condition for the identity $\AA=M$, applying
to time reversible systems verifying CH and independent
on a pairing rule, is the applicability of the Fluctuation
theorem (that will be discussed in the follwing sections).

A tempting alternative interpretation of the
pairing rule, \Cite{BGG997}, could be that, observing the local
exponents averaged over a time $\t$, the pairs with elements
which eventually (as $\t\to+\infty$) have {\it opposite signs}
express the average (over time $\t$) area deformation {\it on the
  attracting manifold}.  While the $d_p$ pairs consisting of {\it
  two negative} average exponents describe contraction of phase
space {\it transversely to the manifold} towards which motion is
attracted.

However {\it strong criticism} can be raised towards the latter
proposal: there is no a priori reason to think that, in presence
of pairing and CH, Lyapunov exponents should split, as dictated
by the size of the exponents, between exponents due to the
evolution restricted to the attracting $\AA$ and exponents
associated with the attraction by $\AA$. \label{pairing
alternative meaning}

\section{Large deviations}
\def\SEC{Large deviations}
\label{sec:V-4}\iniz
\lhead{\small\ref{sec:V-4}.\ \SEC}

An interesting property of the SRB distribution for Anosov maps
$(\X,S)$ and flows $(M,S_t)$ is that a large deviation law
governs the fluctuations of finite time averages of observables.

Continuing to consider discrete time evolutions, the immediate
consequence is that, if $G$ is a smooth observable and if $S$ is an
evolution satisfying the CH (see Sec.(\,\ref{sec:VII-2})\,), the
finite time averages\index{large deviation}:
\be \wt g=\media{G}_\t= \frac1\t\sum_{j=0}^{\t-1} G(S^jx)\label{e4.5.1}\ee
satisfy a {\it large deviations law}: \ie fluctuations of
$\t\media{G}_\t$ off the average $\t\media{G}_\infty$ as large as
$\t$ itself, are controlled by a function $\z(\wt g)$ {\it convex
and smooth} in an interval $(\wt g_1,\wt g_2)$,
containing $\media{G}_\infty$, where it is {\it maximal},
\Cite{Si972a}\Cite{Si977}\Cite{Si994}.  This means, see
Appendix \ref{appN} for a few details on a derivation based
on \Cite{MS967}[Sec.3]\,:
\*

\0{\bf Theorem:} (Large deviations) {\it The probability that
$\wt g\in [a,b]$ satisfies
\be P_\t(\tilde g\in [a,b])\simeq_{\t\to\infty} \,
e^{\t\,\max_{\tilde g\in [a,b]}\z(\tilde g)},\qquad \forall
a,b\in (\wt g_1,\wt g_2)\label{e4.5.2}\ee
where $\simeq$ means that $\t^{-1}$ times logarithm of
the \lhs converges to $\max_{[a,b]}\z(\wt g)$ as $\t\to\infty$,
and the interval $(\wt g_1,\wt g_2)$ is non trivial if
$\media{G^2}_\infty-\media{G}^2_\infty>0$.}
\Cite{Si968a}\Cite{Si977}.
\*

Since $\z(\wt g)$ is quadratic at its maximum (\ie at $\media{G}_\infty$), then
a central limit\index{central limit} theorem  holds for the
fluctuations of $\sqrt{\t}\,\media{G}_\t\equiv
\frac1{\sqrt\t}\sum_{j=0}^{\t-1} G(S^jx)$; however Eq.\equ{e4.5.2} is a
stronger property.

The Anosov systems $(M,S_t)$ in continuous time follow the
analogous large deviations laws simply replacing
Eq.\equ{e4.5.1} by $\wt g=\frac1\t\int_0^\t G(S_t x)dt$.
 \*

\0{\it Remarks:} (1) If the observable $G$ has nonzero
SRB-average $\media{G}_\infty\ne0$ it is convenient to consider,
instead, the dimensionless observable $g={G}/{\media{G}_\infty}$.
\\
(2) If $(\X,S)$ is an Anosov system and is {\it reversible}, \ie
there is a smooth, isometric, map $I$ of $\X$ such that
$IS=S^{-1}I$, then any {\it time reversal odd} observable $G$,
with non zero average and nonzero dispersion
$\media{G^2}_\infty-\media{G}^2_\infty>0$, is such that the
interval $(g_1,g_2)$ of large deviations for
${G}/{\media{G}_\infty}$ contains at least $[-1,1]$.
\\
(3) The Gaussian thermostats models\index{thermostat} of
Sec.(\,\ref{sec:II-2}\,) are all reversible with $I$ being the
ordinary time reversal (\ie $I$ = sign change of velocity
leaving positions unaltered), and the phase space contraction
$\s(x)$ (defined as the divergence of the equation of motion,
$\dot x=f(x)$, or $\s(x)=-\sum_0\frac{\dpr f_i(x)}{\dpr x_i}$)
is odd under time reversal, see
Eq.\equ{e4.3.1}. Therefore if $\s_+=\media{\s}_\infty>0$ it
follows that the observable:
\be p'=\frac1\t\int_{j=0}^{\t}
\frac{\s(S_t x)}{\s_+} dt\label{e4.5.3}\ee
has domain of large deviations of the form $(-\lis g,\lis g)$,
containing $[-1,1]$.  
\\
(4) In the Gaussian thermostats models of Sec.(\,\ref{sec:II-2})\,
$\s(x)$ differs from the {\it entropy production}\label{entropy
production} $\e(x)=\sum_{j>0}\frac{Q_j}{k_B T_j}$ by the time
derivative of an observable, see Eq.\equ{e2.9.3}: it follows
that $\s_+\equiv\media{\s}_{SRB}=\media{\e}_{SRB}\defi\e_+$ and
the finite or infinite time averages of $\s$ and of $\e$ have,
for large $\t$, the same distribution. Therefore the same large
deviations function $\z$ for the fluctuations of $p'$ in
Eq.\equ{e4.5.3} controls the fluctuations of $p$:
\be p=\frac1\t\int_{j=0}^{\t} \frac{\e(S_tx)}{\e_+} dt\label{e4.5.4}\ee
\ie of the entropy production rate, which is more interesting
from the Physics viewpoint.

In the following section an application will be derived providing
information about the fluctuations of $p'$ in reversible
evolutions, hence in various models about the entropy production
fluctuations.

\section{Time reversal and fluctuation theorem}
\def\SEC{Time reversal and fluctuation theorem}
\label{sec:VI-4}\iniz
\lhead{\small\ref{sec:VI-4}.\ \SEC}

It has been shown, \Cite{GC995}\Cite{GC995b} (and interpreted in a
mathematical form in \Cite{Ga995b}\Cite{Ru999}), that in a time
reversible Anosov system $(\X,S)$ the function $\z(p)$ giving the
large deviation law for the dimensionless
phase space contraction $p'$ in SRB states
and therefore for the dimensionless entropy production $p$,
Eq.\equ{e4.5.4}, has the {\it symmetry property}:

\*\0{\bf Theorem:} (Fluctuation Theorem) {\it Given a time
  reversible Anosov map with $\s_+>0$, there is $\lis p>1$
  such that the large deviations function $\z(p)$, controlling
  the fluctuations of dimensionless entropy production $p$,
  Eq.\equ{e4.5.4}, verifies:
\be \z(-p)=\z(p)-p\s_+, \qquad\hbox{\rm for all}\ p\in(-\lis p,\lis
p)\label{e4.6.1}\ee
with $\z(p)$ convex and smooth.}
\*

The Eq.\equ{e4.6.1} expresses the {\it fluctuation
  theorem\index{fluctuation theorem}} of \Cite{GC995}\Cite{Ru999}.\,
\footnote{\tiny As discussed below, it requires a proof and
therefore it should not be confused with several identities to
which, later, the same name has been given, see \Cite{GC004} and
Appendix \ref{appO}.}
The interest of the theorem is that, as long as CH and TRS
hold and the system is transitive (\ie $(\X,S)$ is
an Anosov system), it is {\it universal, model independent} and
yields a {\it parameter free} relation which deals with a
quantity which, as mentioned, in the models in
Sec.(\,\ref{sec:II-2}\,), has the physical meaning of entropy
production rate because:
\be p\,\e_+=\fra1\t \sum_{j=0}^{\t-1} \frac {Q_j}{k_B T_j},\label{e4.6.2}\ee
and therefore, in particle systems, has an independent
macroscopic definition, see Sec.(\,\ref{sec:VIII-2}\,), it might
accessible to experiments (as it will be discussed in the following).

The expression for $\mu_{SRB}$, defined via Eq.\equ{e4.1.8}, can be used
to study some statistical properties of $p'$ (Eq.\equ{e4.5.3}\,),
hence of $p$ (Eq.\equ{e4.5.4}\,).

Given a Markovian partition $\EE=(E_1,\ldots,E_m)$ for the system
$(\X,S)$, denote $\qq_\t=(q_{-\t},\ldots, q_\t)$ a compatible string of
labels $q=1,\ldots,m$.
The ratio of the probability of the interval $ [p,p+dp]$ to that
of $ [-p-dp,-p]$, can be studied using the notations and the
finite $\t$ approximation $h_{\t,l}(\qq))
e^{-\L_{u,\t}(x_{\qq)}} h_{\t,r}(\qq)$ of the SRB probability of
the configurations $\qq=(q_{-\t},\ldots q_\t)$ derived in
Eq.\equ{e3.8.5},  and formally exact in the limit
$\t\to\infty$.
\*

In evaluating the ratio, factors $h_{\t,l}(\qq),h_{\t,r}(\qq)$ can
be neglected because they are bounded above and below for all
$\qq,\t$ as discussed in Sec.(\,\ref{sec:VIII-3}\,)\,\footnote{\tiny
Alternatively as Eq.\equ{e3.10.2} and Fig.3.10.1 show $h_l,h_r$
can be bounded uniformly in terms of the sizes of the stable and
unstable boundaries of the elements of the Markovian partition
$\EE=\{E_i\}_{i=1}^m$ and in terms of the angles $\f$,
Eq.\equ{e3.10.4}\,, between stable and unstable manifolds.}\,.
Therefore setting $a_\t(x)\defi \frac1{2\t+1}\sum_{j=-\t}^{\t}
\frac{\s(S^jx)}{\s_+}$, the mentioned ratio is, up to a factor
bounded by a constant independent of $\t,\qq$:
\be\frac
{\sum_{\V q_\t,\,a_\t(x_{\V q_\t})=p} e^{-\Lambda_{u,\tau}({x_{\qq_\t}})}} 
{\sum_{\V q_\t,\,a_\t(x_{\V q_\t})=-p}e^{-\Lambda_{u,\tau}({x_{\qq_\t}})}}
\label{e4.6.3}\ee
\0Eq.\equ{e4.6.3} is studied by establishing a
one-to-one correspondence between addends in the numerator and in
the denominator, aiming at showing that corresponding addends
have a {\it constant ratio} which will, therefore, be the value
of the ratio in Eq.\equ{e4.6.3} to leading order as $\t\to\infty$.

This is made possible by TRS which is the (simple) extra
information with respect
to \Cite{Si977}\Cite{Bo970a}\Cite{Ru976}.

In fact the time reversal symmetry $I$ allows us to suppose,
without loss of generality, that the Markovian partition $\EE$,
hence $\EE_\t=\cap_{-\t}^\t S^j\EE$, can be supposed time
reversible, see Sec.(\,\ref{sec:III-4}\,): \ie for each $j$ there is a
$j'$ such that $I E_j=E_{j'}$.
The time reversal symmetry $IS=S^{-1}I$ implies Eq.\equ{e4.3.1}
which 
leads to the identity:
\be\L_{u,\t}(Ix)=-\L_{s,\t}(x) 
\label{e4.6.4}\ee
%
The ratio
Eq.\equ{e4.6.3} can therefore be rewritten as:
\be\frac
{\sum_{\V q,\,a_\t(x_{\V q})=p} e^{-\Lambda_{u,\tau}({x_{\V q}})}} 
{\sum_{\V q,\,a_\t(x_{\V q})=-p}e^{-\Lambda_{u,\tau}({x_{\V q}})}}
\equiv
\frac 
{\sum_{\V q,\,a_\t(x_{\V q})=p} e^{-\Lambda_{u,\tau}({x_{\V q}})}} 
{\sum_{\V q,\,a_\t(x_{\V q})=p} e^{\Lambda_{s,\tau}({x_{\V
        q}})} }
\label{e4.6.5}\ee
%

Then the ratios between corresponding terms in Eq.\equ{e4.6.5} are equal
to $e^{-\Lambda_{u,\tau}(x_{\V q})-\Lambda_{s,\tau}(x_{\V q})}$ which
is almost $y=e^{-\sum_{j=-\t}^\t \s(S^{-j}x_{\V
q})}=e^{-(2\t+1)a_\t(x_{\V q})\s_+}$.  In fact, $y$ is the
reciprocal of the determinant of the Jacobian matrix of $S^{2\t}$
as a map from $S^{-\t}x$ to $S^{\t+1} x$, \ie the reciprocal of the
total phase space volume variation, while
$y'=e^{-\Lambda_{u,\tau}(x_{\V q})-\Lambda_{s,\tau}(x_{\V q})}$
is only the reciprocal of the product of the variations of two
surface elements tangent to the stable and to the unstable
manifold in $x_{\V q}$, see Fig.3.10.1 and the related
comments.\footnote{\tiny Hence $y$ and $y'$ differ by factors
related to the sizes of the boundaries of the elements of the
Markovian partition considered (which are finitely many) and the
sines of the angles between the manifolds at $S^{-\tau/2}{x}$
and at $S^{\tau/2}{x}$, see Eq.\equ{e3.10.2}: the latter is
$>0$ and continuous, hence bounded away from $0$ and $\p$.}

But the Anosov property of the motion implies transversality of
intersections at $x\in\AA$ (\ie a continuous angle $\p>\f(x)>0$),
so that the ratio $y/y'$ is bounded away from $0$ and $+\infty$
by $(\V q,\tau)$--independent constants.

Therefore the ratio Eq.\equ{e4.6.1} is equal to $e^{(2\t+1)\,p\, \s_+ }$
up to a factor bounded above and below by a $(\tau,p)$--independent
constant, \ie to leading order as $\tau\to\infty$, and the fluctuation
theorem for stationary SRB states, Eq.\equ{e4.6.1}, follows.  \*

\0{\it Remarks:} (a) The peculiarity of the result is the linearity in $p$: we
expect that $\z(p)-\z(-p)= c\,\langle\sigma\rangle \, (p+ s_3 p^3+s_5
p^5+\ldots)$ with $c>0$ and $s_j\ne0$, since there is no reason,
{\it a priori},
to expect a ``simple'' (i.e.  with linear odd part) multifractal
distribution.
\footnote{\tiny Actual computation of $\z(p)$ is a task possible
in the $N=1$ case considered in\, \Cite{CELS993a}\,, but it is
essentially beyond our capabilities in slightly more general
systems.}
Thus $p$--linearity (i.e. $s_j\equiv 0$) is a {\it key test of the theory
}, \ie of the CH, and a quite unexpected result from the
latter viewpoint.  
\\ 
Recall, however, that the exponent $(2\t+1)\s_+\,\,p$
is correct up to terms of $O(1)$ in $\tau$ (i.e. deviations at small $p$, or
small $\t$, must be expected).  
\\
(b) Eq.\equ{e4.6.1} requires time reversibility and the CH and
both are strong assumptions: this explains why a few papers have
appeared in the literature trying to get rid of CH or TRS.
\\
(c) Experimental tests can possibly be designed to study 
that the entropy production\index{entropy production}
$\s\defi\sum_j\frac{Q_j}{k_B T_j}$, defined in experimental situations by
the actual measurements of the heat ceded to the thermostats at temperature
$T_j$ or, in simulations, by the phase space contraction
$\s$. Their purpose being to check that $\s$ satisfies
what will be called the ``{\it fluctuation relation\index{fluctuation
    relation}}'':
\be \frac{Prob( p\in \D)}{Prob(-p\in\D)}=e^{p\s_+ \t+O(1)}\label{e4.6.6}\ee
where $\s_+$ is the ininite time average of $\s$ and $\D$ is an interval
small compared to $p$.\label{FR} A positive result could be interpreted as
a confirmation of the CH, {\it provided} reversibility was carefully tested. 
\\
(d) It should be stressed that under the CH the attracting sets
$\AA$ are such that $(\AA,S)$ are Anosov systems, but the time
reversal symmetry of the motions on the attracting sets is very
subtle. As discussed in Sec.(\,\ref{sec:IV-4}\,) the fundamental
symmetry of time reversal does not hold on the attracting sets
$\AA$ when the $\AA$'s have dimensionality lower than that of the
phase space, \eg at strong forcing.  Therefore in applying the
fluctuation theorem, or the fluctuation relation, particular care
has to be reserved to understanding whether a mechanism of
respawning, on $\AA$, of a time reversal symmetry works: as
discussed in Sec.(\,\ref{sec:III-4}\,) this is essentially asking
whether the system enjoys the property called there Axiom C. In
absence of TRS the FT does not apply unless other propertiesare
used. Among the several test of the fluctuation relation now
available in the literature this is, {\it unfortunately often},
not considered, so that very few actually test the fluctuation
theorem.  \\
(f) The ``fluctuation'' and the ``conditional reversibility''
theorems of the next section can be formulated for Anosov maps
and Anosov flows. The discrete case is simpler to study than the
corresponding Anosov flows\index{Anosov flow} because Anosov maps
do not have a trivial Lyapunov exponent (the vanishing one,
associated with the phase space flow direction); the technique to
extend the analysis to Anosov flows, is developed
in \Cite{BR975}\Cite{Bo970a}\Cite{Ru976}.
The conditional reversibility theorem will be presented, in the
next section, only in the version for flows: the explicit and
natural formulation for maps will be skipped (to avoid
repetitions).

\section{Fluctuation
  patterns\index{fluctuation patterns}}
\def\SEC{Fluctuation patterns\index{fluctuation patterns}}
\label{sec:VII-4}\iniz
\lhead{\small\ref{sec:VII-4}.\ \SEC}

The fluctuation theorem, Eq.\equ{e4.6.1} has several extensions
including a remarkable, parameter free relation that concerns the
relative probability of {\it patterns} of evolution of an
observable and of their reversed
patterns, \Cite{Ga997}\Cite{Bo970a}\Cite{Ga000}\Cite{Ga002},
related to the Onsager-Machlup fluctuations theory, discussed in
various forms and variations in the literature.

It is natural to inquire whether there are other physical interpretations
of the theorem (and of the meaning of the CH) when the
external forcing is really different from the value $0$. %
\footnote{\tiny {\it I.e}
  not infinitesimally close to $0$ as in the classical theory of
  nonequilibrium thermodynamics, \Cite{DGM984}.}
A result in this direction is the {\it conditional reversibility theorem},
assuming the CH, TRS and $\s_+>0$, discussed below.

Consider observables $F$ which, for simplicity, have a
well-defined time reversal parity: $F(Ix)=\e_F F(x)$, with
$\e_F=\pm1$.  Also for simplicity, suppose that their time average
({\it i.e.} their SRB average) vanishes, $F_+=0$. Let $t\to
\f(t)$ be a smooth function, called ``{\it pattern}'', vanishing for
$t<0$ and for $t$ large; define also $I\f$ as the ``time reversed''
pattern $I_\t\f(t)\defi \e_F \f(\t-t)$.

Look at the probability, $P_{\t;p,\f;\h,\h'}$, relative to the SRB distribution
({\it i.e.}  in the ``natural stationary state''), that
\be\eqalign{&|F(S_tx)-\f(t)|<\h',\qquad t\in(0,\t)\cr
&|p-\frac1\t\int_0^\t\frac{\s(S_tx)}{\s_+}dt|<\h\cr}\label{e4.7.1}\ee
which will be called the probability that, within tolerance $\h'$, $F$
follows the fluctuation pattern $\f(t), t\in (0,\t),$ while there is an
average entropy production $p$ within $\h$.  Then the following 
statement, discussed in \Cite{Ga997}\Cite{Ga999}\Cite{Ru999}, can
be derived essentially in the same way as the above fluctuation
theorem and is informally stated as:
\*

\0(Patterns revesibility) {\it Assume the evolution
to be a time reversible Anosov flow on the manifold $M$ with
  average contraction rate $\sigma_+>0$. Let $F$ and $G$ be
  observables (time reversal odd for definiteness), let $f,g$ be
  patterns for $F,G$ respectively and let $If,Ig$ be the time
  reversed patterns; then:\index{patterns reversibility}
\be\frac{P_{\t,p,f,\h,\h'}}{P_{\t,-p,If,\h,\h'}}
=\frac{P_{\t,p,g,\h,\h'}} {P_{\t,-p,Ig,\h,\h'}}\label{e4.7.2}\ee
approximately for large $\t$ and exactly as $\t\to\infty$.}\,%
\footnote{\tiny Colorfully:
  {\it A waterfall will go up, as likely as we see it going down, in a
    world in which for some reason, or by the deed of a Daemon, the entropy
    production rate has changed sign during a long enough time}
  \Cite{Ga002}[p.476].}
\* 

No assumption on the fluctuation size (\ie on the size of $\f$, see however
remark (e) at the end of Sec.(\,\ref{sec:VI-4}\,), nor on the
size of the forces keeping the system out of equilibrium, is
made. It should however be kept in mind that the statement supposes that
the system has the Anosov property on the phase space $M$. It does
not apply when the attracting surface $\AA\ne M$. Or it applies if
$\AA$ is considered the phase space: but in this case the phase
space contraction has to be $\s_\AA$ measured on $\AA$.\,\footnote{\tiny
Therefore to study $\s_\AA$ it will be necessary, in general, to
have access to the parametric equations of $\AA$.}

A more mathematical form of the above
result,\Cite{Ga997}\,,\Cite{Ru999}[sec.3.10]\,:\index{fluctuation
  patterns theorem} \*
\0{\bf Theorem:} (Fluctuation Patterns) {\it Under the assumptions of the
  preceding statement, let $\zeta(p,\f)$ the be {\it large deviation
    function} for observing in the time interval $[0,\t]$ an average
  contraction of phase space $\fra1\tau \int_{0}^{\t}
  \sigma(S_tx)dt=p\sigma_+>0$ and at the same time $F(S_tx)$ to follow a
  fluctuation $\f(t)$. Then there is $\lis p>1 $
\be \z(-p,\e_F I\f)-\z(p,\f)=-p\s_+,\quad p\in(-\lis p,\lis p)
\label{e4.7.3}\ee
for all $\f$, with $\z$ the joint large deviation rate for $p$ and $\f$
(see below).}
\* Here the rate $\z$ is defined as the rate that controls the $\m_{SRB}$
probability that the {\it dimensionless average entropy creation rate} $p$
is in an interval $\Delta=(a,b)$ and, at the same time,
$|f(S_tx)-\f(t)|<\h$. Hence $\z$ is defined by:
\be\sup_{p\in\Delta,|\f-\ps|<\h}
e^{-\tau\zeta(p,\ps)}\label{e4.7.4}\ee
to leading order as $\tau\to\infty$ ({\it i.e.} the logarithm of the
mentioned probability divided by $\tau$ converges as $\tau\to\infty$ to
$\sup_{p\in\Delta,|\f-\ps|<\h}\zeta(p,\varphi)$).
\*

\0{\it Remarks:} (1) The result can also be formulated if $F$ is
replaced by $m$ observables $\V F=(F_1,\ldots,F_m)$, each of well
defined parity under time reversal, and the pattern $\f$ is
correspondingly replaced by $m$ patterns
$\Bff=(\f_1,\ldots,\f_m)$. The \rhs of the relation analogous to
Eq.\equ{e4.7.3} remains unchanged; hence this extension
provides in principle arbitrarily many, parameter free, fluctuation
relations. Only few of them can be observed because the
difficulty of observing $m$ patterns, each obviously so rare, becomes
more and more hard with increasing $m$.
\\
(2) In other words, in these systems, while it is very difficult
to see an ``anomalous'' average entropy creation rate during a
time $\tau$ ({\it e.g.}  $p=-1$), it is also true that ``{\it
  such anomaly is the hardest thing to see}'', see also \Cite{Ru999}.
Once we see it {\it all the observables will behave strangely}
and the relative probabilities of time reversed patterns will
become as likely as those of the corresponding direct patterns
under ``normal'' (\eg $p=1$) average entropy creation regime,
\Cite{Ru999}.  \\
(3) It can also be said that the motion in a time reversal
symmetric Anosov system is reversible, even in the presence of
dissipation, once the dissipation is fixed.  Interesting
variations of this property keep being discovered, see for
instance \Cite{GPB008}.  \\
(4) No assumption on the fluctuation size (\ie on the size of $\f$), nor on the
size of the forces keeping the system out of equilibrium, is made,
besides the Anosov property and $\sigma_+>0$; the results hold no matter
how small $\sigma_+$ is. They make sense even if
$\sigma_+=0$ but, while the fluctuation theorem becomes trivial
(provided the $\lis p>1$ is replaced by 'near $p=0$'), the
fluctuation patterns theorem remains valid in the form in
Eq.\equ{e4.7.2}; and can be written as (corresponding to $g=0$):
\be\frac{P_{\t,0,f,\h,\h'}}{P_{\t,0,If,\h,\h'}}=1 \label{e4.7.5}\ee
which says that if $\s_+=0$ any pattern $f$ of fluctuation has
the same probability of its time reversed pattern. This seems to
be a special case of the results in Onsager-Machlup
theory,\,\Cite{OM953a}\Cite{OM953b}\,, (at equilibrium, as $\s_+=0$).  \\
(5) The comment (e) in the previous section, about the general
case of attracting surfaces with dimension lower than that of
phase space, has to be kept in mind as it might set serious
limits to experimental checks (not, of course, checks of the
theorems but of the physical assumption in the CH and TRS which implies
the theorems).  \*

There are other remarkable extensions of the fluctuation relation in presence
of other symmetries: see \Cite{HPPG011}.

\section{Onsager's reciprocity, Green-Kubo formula}
\def\SEC{Onsager's reciprocity, Green-Kubo formula}
\label{sec:VIII-4}\iniz
\lhead{\small\ref{sec:VIII-4}.\ \SEC}

Under the CH the fluctuation theorem degenerates in the limit in which
$\sigma_+$ tends to zero, \ie when the external forces  and
dissipation disappear (and the stationary state becomes the
equilibrium state).  \index{Onsager's reciprocity}
\index{Green-Kubo formula} \index{fluctuation theorem}

Since the theorem deals with systems that are time reversible
{\it at and outside} equilibrium, Onsager's hypotheses are
certainly satisfied and the system should obey reciprocal
response relations at vanishing forcing. It led to the
idea\footnote{\tiny Suggested by P. Garrido from the data in the
simulation in \Cite{BGG997}.} that there might be a connection
between the fluctuation theorem and Onsager's reciprocity and also
to the related (stronger) Green-Kubo formula.

A check will be presented below, switching to continuous time, to
simplify the analysis and only referring to the finite models of
Sec.(\,\ref{sec:II-2},\ref{sec:III-2}\,), although extensions to
other systems would be possible. Time reversibility with $I$
velocity sign change and unaltered positions holds
for such models and CH is assumed, at equilibrium and at small
forcing (not necessarily infinitesimal). The structural stability
of Anosov systems then implies that the attracting surface
remains the full phase space (as in equilibrium: the
systems are transitive, \ie admit dense orbits).

Consider a transitive system $(M,S_t^{\V E})$ satisfying the
CH (\ie an Anosov flow) parameterized by the {\it
  thermodynamic forces} ${\bf E}=\{E_k\}_{k=1}^n$ that generate
it (\ie parameters that measure the strength of the forcing): as
in examples in Sec.(\,\ref{sec:III-2}\,).

Let $\s_{\V E}(x)$ be the divergence of the equations of motion
and define the {\it microscopic thermodynamic flux} ${\bf j}(x)$,
associated  with the {\it thermodynamic forces} ${\bf
  E}$ that generate it via the relation:
\be
j_h(x)=\frac{\partial\sigma_{\V E}(x)}{\partial E_h}
\label{e4.8.1}\ee
not necessarily at ${\bf E}=0$; hence $j_h(x)$ depends on $\bf
E$.

Suppose (as in the Sec.(\,\ref{sec:III-2}\,) examples) time reversibility and:
\be\s_{\V E}(x)=\sum_h j(x)_h E_h=\sum_h J^0_h(x) E_h+O(|{\bf
  E}|^2)\label{e4.8.2}\ee
with $J^0(x)$ independent on $\bf E$.

Then in \Cite{Ga996a} is  shown that the limit as
${\bf E}\to0$ of the fluctuation theorem function $\z(p)$ (in the
continuous time case) becomes simply a property of the average
{\it fluxes}, with respect to the SRB distribution $\m_{\bf
  E}(dx)$, $\langle{j_h}\rangle\defi\int
j_h(x)\m_{\bf E}(dx)$. Namely: \*

\0{\bf Theorem:} ({\rm Reciprocity and fluctuation theorem}) {\it
  Let $(M,S^{\V E}_t)$ be Anosov flows with $\V E$-independent
 TRS $I$, and let $\mu_{\bf E}$ be the SRB distribution; let
 $\s(x)$ be the divergence of the flow as in
 Eq.\equ{e4.8.2}. Then, if $\int \s_{\V E}(x)\mu_{\bf
 E})dx)\ne0$ for $\V E\ne0$, it is:
  \be L_{hk}\defi\frac{\dpr \langle{j_h}\rangle_{\mu_{\bf E}}}
      {\dpr E_k}\Big|_{\bf E=0}\kern-2mm=\frac12
\int_{-\infty}^{\infty} dt\Big(\int j_h(S_tx)j_k(x)\mu_{\bf E}(dx)\Big)\Big|_{{\bf E}=0}
=L_{kh}\label{e4.8.3}\ee
for $\V E$ small enough (not necessarily infinitesimal) as a
consequence of the fluctuation theorem.}
\*

This is the Green-Kubo's formula and for $h\ne k$ it
yields Onsager's reciprocal relations.  Therefore, in time
reversible system introduced in Sec.(\,\ref{sec:II-2}\,), the fluctuation
theorem can be regarded as {\it an extension to nonzero forcing}
of Onsager's reciprocity and of the Green-Kubo formula.

It is not difficult to see, heuristically, how the fluctuation
theorem, in the limit in which the driving forces tend to $0$,
yields Eq.\equ{e4.8.3}\,. For $\V E\ne0$ the large deviations function
$\z(p)$ is smooth in $p$ and near $p=1$ (where it reaches its
maximum) its Laplace transform will be, if the SRB average of $\s_{\bf E}(x)$,
denoted $\langle \s_{\bf E}\rangle $,
at $\bf E\ne 0$ is $\ne0$:
\be \l(\b)\defi\lim_{\t\to\infty}\frac1\t\log \int
e^{-\b\t(p(x)-1)\langle \s_{\bf E}\rangle }\m_{\bf E}(dx) \label{e4.8.4}\ee
The transform is smooth in $\b$ and can be expanded as $\l(\b)=
\frac1{2!} C_2 \langle \s_{\bf E}\rangle^2 \b^2+
\frac1{3!}C_3\s_{\bf E}^3\b^3+\ldots$ ($C_1=0$ because the
average of $(p-1)$ is zero).  The coefficients $C_j$ are the
cumulants of the variables $\s_{\bf E}(x)-\langle \s_{\bf
  E}\rangle $ which, by Eq.\equ{e4.8.2}\,, can be expressed via
$j_h(x)$ up to $O(|{\bf E}|^3)$:
\be
\eqalign{
  C_2 =&\int_{-\infty}^\infty
  dt\Big(
  \int (\s_{\bf E}(S_tx)\s_{\bf E}(x)
  -\langle \s_{\bf E}\rangle^2)\m_{\bf E}(dx)
  \Big)+O(|{\bf E}|^3)\cr}
\label{e4.8.5}\ee
The $\zeta(p)$ is recovered via a Legendre transform $\z(p)=
\max_\b\Big(-\b\langle \s_{\bf E}\rangle(p-1)+\lambda(\b)\Big) $.
Hence: $\z(p)=-\frac{\langle \s_{\bf E}\rangle^2}{2 C_2}(p-1)^2+
O(|{\bf E}|^3(p-1)^3)$ which, together with the fluctuation
theorem $\z(p)-\z(-p)=p\langle \s_{\bf E}\rangle$ and
Eq.\equ{e4.8.4}, yields at fixed $p$, the key relation:
\be\langle \s_{\bf E}\rangle={1\over 2}C_2+ O(|\bf E|^3)\label{e4.8.6}\ee
To derive Green--Kubo formulae, \ie Eq.\equ{e4.8.3} with
$h=k$, look first at the r.h.s.  of the expression for $\z(p)$;
discarding $O(|\bf E|^3)$, it becomes:
\be
C_2=\sum_{i,j}E_i E_j\int_{-\infty}^\infty 
dt\, \int \Big(J^0_i(S_tx) J^0_j(x)-
\langle J^0_i\rangle\langle J^0_j\rangle\Big)
\m_{\bf E}(dx)\label{e4.8.7}\ee
On the other hand the expansion of $\langle\s_{\bf E}\rangle$ in
the l.h.s. of Eq.\equ{e4.8.6} to second order in $\bf E$ gives:

\be\langle\s_{\bf E}\rangle={1\over2}
\sum_{ij}\Big(\dpr_{E_i}\dpr_{E_j}\langle\s_{\bf E}\rangle
\Big) E_i E_j +O(|{\bf E}|^3)\label{e4.8.8}\ee

\noindent{}because the first order term vanishes (by
Eq. \equ{e4.8.6}).

The r.h.s. of \equ{e4.8.8} is the sum of ${1\over2}E_iE_j$
times $\dpr_{E_i}\dpr_{E_j} \int \s(x) \mu_{\bf E}(dx)$ which
equals the sum of the following three terms which have to be
multiplied by $E_iE_j$ and summed over $i,j$: namely
\\
$\int(\dpr_{E_i} \dpr_{E_j}\s_{\bf E}(x) )\mu_{\bf E}(dx)$,\\
$\int(\dpr_{E_i}\s_{\bf E}(x))\dpr_{E_j}\mu_{\bf E}(dx)+
(i\leftrightarrow j)$,\\
$\int\s_{\bf E}(x)\dpr_{E_i}\dpr_{E_j}\mu_{\bf E}(dx)$,
\\
{\it all evaluated at ${\bf E}=0$}.

The first addend is $0$ (by time reversal), the third
addend is also $0$ (as $\s_{\bf E}=0$ at ${\bf E}=0$).  The second term
needs to be evaluated to lowest order (\ie $O(1)$)
and is the sum of
\be (\int j_i(x) \dpr_{E_j} d\m_{\V E}(dx))|_{|\bf E|=0} \equiv
\dpr_{E_j}\int J^0_i(x) d\m_{\V E}(dx)|_{|\bf E|=0} = L_{ij}\label{e4.8.9}\ee
plus $L_{ji}$.  Hence:
\be\dpr_{E_i}\dpr_{E_j} \langle\s\rangle|_{{\bf E}=0}=
\Big(\dpr_{E_j}\langle{J^0_i}\rangle_++
\dpr_{E_i}\langle{J^0_j}\rangle_+\Big)|_{{\bf E}=0}=(L_{ji}+L_{ij})
\label{e4.8.10}\ee

\0Thus equating r.h.s and l.h.s.  of Eq.
\equ{e4.8.6}, as expressed respectively via Eq.\equ{e4.8.7}
and \equ{e4.8.10}, the matrix ${L_{ij}+L_{ji}\over2}$ is
obtained: at least if $i=j$ this is Green-Kubo's formula, also
sometimes called a "fluctuation dissipation theorem".

This essentially suffices to obtain Onsager's reciprocity {\it i.e.}
$L_{ij}=L_{ji}$ and the identity of $L_{ij}$ with the double
integral in Eq.\equ{e4.8.7}: the main remark is that the
fluctuation theorem can be extended to give properties of the {\it
joint} distribution of the average over $t\in[-\t,\t]$ of
$\s_{\bf E}(S_tx)$ and of the corresponding average of
$E_j\dpr_{E_j}\s_{\bf E}(S_tx)$.  See appendix \ref{appM} for
details which show that the fluctuation patterns
theorem, Sec.(\,\ref{sec:VII-4}\,), provides the extra information needed
to obtain the full reciprocity. \*

The above analysis is unsatisfactory because it interchanges,
quite freely, limits and derivatives, and derivatives are even
taken of $\mu_{\bf E}$: which seems to require some imagination,
if $\V E\ne0$, as $\mu_{\bf E}$ is concentrated on a set of zero
area in phase space $M$. See \Cite{GR997} for a rigorous
analysis.

Certainly assuming reversibility in a system out of equilibrium can be
disturbing: one can, thus, inquire if there is a more general connection
between the CH, Onsager reciprocity and the Green-Kubo
formula.

This is indeed the case: the last quoted result is based on the
more general results, \Cite{Ru997}, about expressions for
derivatives of averages of classes of observables, with repect to
parameters of the equations, as integrals over time. It also
provides us with a further consequence of the CH: it can be
shown, \Cite{GR997}\,, that the relations Eq.\equ{e4.8.3} follow
from the sole assumption that at $\V E=0$ the system is time
reversible and that it satisfies the CH for $\V E$ near $0$: at
$\V E\ne0$ it needs not be reversible, see Sec.(\,\ref{sec:IX-3}\,).

\section{Local fluctuations: an example}
\def\SEC{Local fluctuations: an example}
\label{sec:IX-4}\iniz
\lhead{\small\ref{sec:IX-4}.\ \SEC}

There are cases in which the phase space contraction is an
``extensive quantity'', because thermostats do not act only
across the boundaries but they act also in the midst of the
system.

For instance this is the case in the electric conduction models,
see Sec.(\,\ref{sec:III-2}\,), in which dissipation occurs through
collisions with the phonons of the underlying lattice. Then heat
is generated in the bulk of the system and, if a large part of
the system is considered, the amount of heat generated in the
bulk might exceed the acceptable amount that exits from the
boundaries of the sample, thus making necessary a dissipation
mechanism that operates also in the system bulk (to avoid
excessive heating).

In this situation it can be expected that it should be possible
to define a local phase space contraction and prove for it a
fluctuation relation. This question has been studied in
\Cite{Ga999b}\Cite{GBG004} where a model of a chain of $N$ coupled maps
has been studied. The results are summarized below.

Consider a collection of $N^3$ independent identical systems that
are imagined located at the sites $\x$ of a $N\times N\times N$
cubic lattice $\LL_N$ centered at the origin: the state of the
system located at $\x$ is determined by a point $\f_\x$ in a
manifold $\TT$ which, for simplicity, will be a torus of
dimension $2d\ge2$. The evolution will be a perturbation of
the free evolution $\mathcal S_0$ of a state
$\Bff=(\f_\x)_{\x\in\LL_N}\in \TT^{N^3}\defi\X$ is $\Bff\to
({\mathcal S}_0\Bff)=(S_0\f_\x)_{\x\in \LL_N}$, \ie the system in
each location evolves independently; and the perturbed evolution
will be:
\be ({\mathcal S}_\e\Bff)_\x=
S_0\f_\x+\e\Ps_\x(\Bff), \qquad \x\in\LL_N \label{e4.9.1}\ee
with $\Ps_\x(\Bff)=\ps((\f_\h)_{|\h-\x|< r})$ a smooth
``perturbation'' of range $r$. The motion on $\X$,
generically, will not be volume preserving in the sense that:
\be \s_{\LL_N}(\Bff)\defi -\log|\det(\dpr_\Bff {\mathcal
S}(\Bff))|\not\equiv0\label{e4.9.2}\ee
even when, as it will be assumed here, $S_0$ is volume preserving.

It can be shown that the basic results in
\Cite{PS991}\Cite{BK996}\Cite{JP998} imply that
if the ``unperturbed dynamics'' $S_0$ is smooth, hyperbolic,
transitive (\ie if $S_0$ is an Anosov map) then for $\e$ small
enough, {\it but independently of the system size $N$}, the
system remains an Anosov map (\ie its structural stability holds
uniformly in $N$, see \ref{structural stability}\,).

Therefore the system $(\X,{\mathcal S}_\e)$ admits a SRB distribution
which can also be studied explicitly by perturbation theory,
\index{perturbation theory} \Cite{BK995}\Cite{BK997}\,.
For instance it can be shown that the SRB average of the phase
space contraction is also {\it extensive}, \ie proportional to
the volume $N^3$ of the system up to ``boundary corrections'' of
$O(N^2)$:
\be \media{\s_{\LL_N}}_{SRB}=N^3\lis\s_+(\e)+ 
O(N^2)\label{e4.9.3}\ee
with $\lis\s_+(\e)$ analytic in $\e$ near $\e=0$ and generically $>0$
there, for small $\e\ne0$. 

By the general theory, the fluctuation theorem for $p=\frac1\t
\sum_{j=0}^{\t-1} \frac{\s({\mathcal S}_\e^j)}{\s_+(\e,N)}$ will hold
provided the map ${\mathcal S}_\e$ is time reversible, see also
\Cite{Ga996b}\Cite{BGG007},
\Cite{GBG004}[Sec.10.4]

This suggests that given a subvolume $\L\subset \LL_N$ and setting
\be\s_\L(\Bff)=-\log |\det(\dpr_\Bff {\mathcal
  S}(\Bff))_\L|\label{e4.9.4}\ee
where $(\dpr_\Bff {\mathcal S}(\Bff))_\L$ denotes the submatrix
$(\dpr_\Bff {\mathcal S}(\Bff))_{\x\x'}$ with $\x,\x'\in\L$ and
\be p\defi \frac1\t\sum_{j=0}^{\t-1} 
    {\s_\L({\cal S}^j\Bff)},
\label{e4.9.5}\ee
then the random variable $p$ might obey a large deviation law with
respect to the SRB distribution.

It has been {\it heuristically} shown,
\Cite{Ga999b}[Eq.5.13],\,, that
\be \media{\s_\L}_{SRB}=\lis\s_+ \,|\L|+O(|\dpr\L|)\label{e4.9.6}\ee
{\it and} the large deviation law for $p$ is, {\it for all
  $\L\subseteq\LL_N$}, 
a function $\z(p)$  which has there the form:
\be \z(p)=|\L|\,\lis\z(p) +O(|\dpr\L|).\label{e4.9.7}\ee
defined smooth and convex in an interval $(p_1,p_2)$: $\lis\z(p)$
will be called ``intensive fluctuation rate''.  However a formal
proof of Eq.\equ{e4.9.7} (or its refutation) is still missing.

The analogy with the more familiar density fluctuations in a low
density gas is manifest: consider the probability that the number of
particles $n$ in a volume $\L$, subset of the container $V$,
$\L\subseteq V$, in a Lennard-Jones (say) gas in equilibrium at
temperature $T_0=(k_B\b_0)^{-1}$ and small enough density
$\r_0=\frac{N}V$.
 
The probability is such that the random variable
$p=\fra{n}{|\L|\r_0}$ obeys a large deviations law, controlled by
a convex function $f_0(p)$ of $p$ which is extensive and smooth
in an interval $(\lis a,\lis b)$ containing $p=1$: it is related
to the free energy function $f(\b,p\r_0)$, \Cite{Ol988}, in :
\be {\rm Prob}_{\r_0,T_0}(p \in [a,b])=e^{-|\L| \max_{p\in[a,b]}\b_0f_0(\b_0,p
  \r_0)},\quad [a,b]\subset (\lis a,\lis b)\label{e4.9.8}\ee
to leading order as $\L\to\infty$, where:
\be f_0(\b_0,p \r_0)\defi \,f(\b,p \r_0)-f(\b_0,\r_0)- \frac{\dpr
  f(\b_0,\r_0)}{\dpr\r}\Big|_{\r=\r_0}\,(p\r_0-\r_0)\label{e4.9.9}\ee
is the difference between the Helmholtz free energy at density
$\r$ and its linear extrapolation from $\r_0$, \Cite{GLM002}.
The function $f_0(\b_0,p \r_0)$ is {\it independent} of
$\L\subseteq V$. Hence in this example the unobservable density
fluctuations in large volumes $V$ can be measured via density
fluctuations in finite regions $\L$.\index{local fluctuations in
  equilibrium}

If the map ${\mathcal S}_\e$ is {\it also} time reversible and
$\lis\s_+>0$, Eq.\equ{e4.9.7}, validity of the fluctuation
theorem for the full system implies
$\lis\z(-p)=\lis \z(p)-p\lis\s$, because of the extensivity of
$\z$ in Eq.\equ{e4.9.7}; hence the global fluctuation theorem
implies a corresponding property for the
``local'' intensive fluctuation rate $\lis\z(p)$.

\section{Generalities on local fluctuations}
\def\SEC{Generalities on local fluctuations}
\label{sec:X-4}\iniz
\lhead{\small\ref{sec:X-4}.\ \SEC}

The example in the previous section indicates a direction to
follow in discussing large fluctuations in extended systems. A
key difference that can be expected in most problems is that in
{\it stationary states} of extended systems dissipation is not a
bulk property because of the conservative nature of the internal
forces. For instance in most models in Sec.(\,\ref{sec:II-2}\,) no
dissipation occurs in the system proper, $C_0$, but it occurs
``at the boundary'' of $C_0$ where interaction with the
thermostats takes place.

Of course we are familiar with the dissipation in gases and
fluids modeled by constant friction manifested throughout the
system: as, for instance, in the incompressible
Navier-Stokes\index{Navier-Stokes equation}
equation.\footnote{\tiny Incompressibility should be considered
a constraint that fixes temperature as a function $T=F(P,\r)$
of pressure $P$ and density $\r$, instead of leaving $T,\r$ to be
transported by the flow governed by two equations.}

However this is a phenomenologically accounted friction. If the
interest is on stationary states of the fluid motion (under
stirring forces) then the friction coefficient\index{friction
  coefficient} takes into account phenomenologically that
stationarity can be reached because the heat generated by the
stirring is transferred across the fluid and dissipated at the
boundary, or where contact with external thermostats takes place.

Therefore in models like the general ones in Sec.(\,\ref{sec:II-2}\,), or in
the second in Sec.(\,\ref{sec:III-2}\,), the average dissipation has to be
expected to be a boundary effect rather than a bulk effect (as it is
in the example in Sec.(\,\ref{sec:VIII-4}\,) or in the modification of the
model in Fig.2.3.2, considered in \Cite{Ga996a}).

Consider an extended system $C_0$ in contact with thermostats: \ie a
large system enclosed in a volume $V$ large enough so that it makes
sense to consider subvolumes $\L\subset V$ which still contain many
particles. Suppose the system to satisfy the CH and to
be in a stationary state: then the part of the system inside a subvolume
$\L$ will also be in a stationary state, \ie the probability of finding a
given microscopic configuration in $\L$ will be time-independent.

It is natural to try to consider the subvolume $\L\subset V$ as a
container in contact with thermostats in the complementary
subvolume $V/\L$. However the ``wall'' of separation between the
thermostats and the system, \ie the boundary $\dpr\L$, is only a
virtual wall and particles can cross it and do so for two reasons.

First they may cross the boundary because of a macroscopic
current established in the system by the action of the stirring
forces or by convection; secondly, even in absence of stirring,
when the nonequilibrium is only due to differences in temperature
at various sectors of the boundary of the global container $V$,
and even if convection is absent, particles cross back and forth
the boundary of $\dpr\L$ in their microscopic motion.

It is important to consider also the time scales over which the
phenomena occur. The global motion takes often place on a time scale
much longer than the microscopic motions: in such case it could be
neglected as long as the observations times are short enough. If not
so, it may be possible, if the region $\L$ is small enough, to
follow it, 
\footnote{\tiny For a time long enough for being able to consider it as a
moving container: for instance while its motion can be considered
described by a linear transformation and at the same time long enough
to be able to make meaningful observations.}  
because the local ``Brownian'' motion takes place on a short time
scale and it could be neglected provided the free path is much
smaller than the size of $\L$.

There are a few cases in which the two causes above can be
neglected: then, with reference for instance to the models in
Sec.(\,\ref{sec:II-2}\,), the region $\L$ can be considered in
contact with (virtual) reservoirs which, at the point
$\x\in\dpr\L$, have temperature $T(\x)$ so that the phase space
contraction of the system enclosed in $\L$ is, see
Eq.\equ{e2.8.1}, up to a total time derivative, expressed by
the integral over the boundary surface elements $ds_\x$:

\be \e=\ig_{\dpr\L}\fra{Q(\x)}{k_B T(\x)} ds_{\x},\label{e4.10.1}\ee
The fluctuations over time intervals {\it much longer than the time of
free flight but much shorter than the time it takes to diffuse over a
region of size of the order of the size of $\L$} (if such time scales
  difference is existent) can then be studied by the large deviation laws
  and a fluctuation relation, Eq.\equ{e4.6.6}, could even hold
  because in $\L$ there is no friction, provided the time averages are not
  taken over times too long compared to the above introduced ones.

The situations in which the above idea has chances to work and be
observable are dense systems, like fluids, where the free path is
short: and an attempt to an application to fluids will be discussed in
the next chapter.

The idea on the possibility of local fluctuation\index{local
fluctuation} theorems has been developed and tested
numerically, \Cite{GP999}, and
theoretically, \Cite{Ga997c}\Cite{Ga999b}\,, at least in a few
models, finding encouraging results which, hopefully, may lead to
a formal mathematical theory.

\section{Quantum systems, thermostats and non equilibrium}
\def\SEC{Quantum systems, thermostats and non equilibrium}
\label{sec:XI-4}\iniz
\lhead{\small\ref{sec:XI-4}.\ \SEC}

Recent experiments deal with properties on mesoscopic and atomic scale. In
such cases the quantum nature of the systems cannot be always neglected,
particularly at low temperature, \Cite{Ga000}[Ch.1], and the
question is whether a fluctuation analysis parallel to the one just seen in
the classical case can be performed in studying quantum phenomena.

Thermostats have a macroscopic phenomenological nature: in a
way they could be regarded as classical macroscopic objects in which
no quantum phenomena occur.  Therefore it seems natural to model them
as such and define  their temperature as the average kinetic
energy of their constituent particles so that the question of how to define
it does not arise.

The point of view has been clearly advocated in several papers,  for
instance in \Cite{MCT993} just before the fluctuation theorem and the
CH were developed. Here the analysis is presented with
minor variations, that are:\*
\0(a) Gaussian thermostats are used instead of the Nos\'e-Hoover thermostats
\\
(b) several different thermostats are allowed to interact with the system,
\*
\0following \Cite{Ga008a}, {\it aiming at the application of the chaotic
  hypothesis to obtain a fluctuation relation for systems with an important
  quantum component}. 
\*

A version of the CH for quantum systems is already\footnote{\tiny
Writing the paper \Cite{Ga008a} I was unaware of these works: I
thank Dr. M. Campisi\index{Campisi} for recently pointing this
reference out.}  implicit in \Cite{MCT993} and in the references
preceding it, where the often stated incompatibility of chaotic
motions with the discrete spectrum of a confined quantum system
is criticized.

Consider the system in Fig.2.2.1 when the quantum nature of the particles
in the finite container with smooth boundary $\CC_0$ cannot be
neglected. Suppose for simplicity (see \Cite{Ga008a}) that the
nonconservative force $\V E(\V X_0)$ acting on $\CC_0$ vanishes, {\it i.e.}
consider the problem of heat flow through $\CC_0$.  Let $H$ be the operator
on $L_2(\CC_0^{3N_0})$, space of symmetric or antisymmetric wave functions
$\Ps(\V X_0)$, defined as:
\be H=
-\frac{\hbar^2}{2m}\D_{\V X_0}+ U_0(\V X_0)+\sum_{j>0}\Big(U_{0j}(\V X_0,\V
X_j)+U_j(\V X_j)+K_j\Big)\label{e4.11.1}\ee
where $K_j=\fra{m}{2}\sum_{j>0} \dot{{\V X}}_j^2$ and $\D_{\V X_0}$ is
the Laplacian with $0$ boundary conditions (say); and notice that at fixed
external configuration $\V X_j$ its spectrum consists of eigenvalues $E_n=$
$E_n(\{\V X_j\}_{j>0})$ (because the system in $\CC_0$ has finite size).

A system--reservoirs model can be the {\it dynamical system} on the
space of the variables $\Big(\Ps,(\{\V X_j\},$ $\{\V{{\dot
X}}_j\})_{j>0}\Big)$ defined by the equations (where
$\media{\cdot}_\Ps\,=$ expectation in the wave function $\Ps$)
\be \eqalign{
-i&\hbar {\frac{d}{dt}\Ps(\V X_0)}= \,(H\Ps)(\V X_0),\kern20mm{\rm and\ for}\
j>0\cr
&\V{{\ddot X}}_j=-\Big(\dpr_j U_j(\V X_j)+
\media{\dpr_j U_{0,j}(\cdot,\V X_j)}_\Ps\Big)-\a_j \V{{\dot X}}_j\cr
&\a_j\defi\frac{\media{W_j}_\Ps-\dot U_j}{2 K_j}, \qquad
W_j\defi -\V{{\dot X}}_j\cdot \V\dpr_j U_{0j}(\V X_0,\V
X_j)\cr
&\media{\dpr_j U_{0,j}(\cdot,\V X_j)}_\Ps\defi
\int_{\CC_0} d^{N_0}\V X_0|\Ps(\V X_0)|^2 W(\V
X_0,\V X_j)\cr
}\label{e4.11.2} \ee
here the first equation is Schr\"odinger's\index{Schr\"odinger's
equation} equation, the second is an equation of motion for the
thermostats particles similar to the one in Fig.2.2.1, (whose
notation for the particles labels and the conservative
interaction energies $U$ is adopted here too). The evolution is
time reversible because the map $I(\Ps(\V X_0),\{\dot{{\V
X}}_j,\V X_j\}_{j=1}^n\})=(\lis{\Ps(\V X_0)},\{-\dot{{\V X}}_j,\V
X_j\}_{j=1}^n\})$ is a time reversal (isometric in
$L_2(\CC^{3N_0})\times R^{6 \Si_{j>0}N_j}$).

The model, that can be called {\it Erhenfest dynamics},
\index{Erhenfest dynamics} differs from the model in
\Cite{MCT993}\Cite{ACCCEF011} because of the use of a Gaussian rather than a
Nos\'e-Hoover thermostat; it provides a representation of the motions in the
thermostats or of the interaction system-thermostats which is
intended to be equivalent to the Feynman-Vernon proposal;
furthermore very interesting alternative models have been
proposed and are briefly illustrated in the next
section.

Evolution maintains the thermostats kinetic energies $K_j\equiv
\frac12\V{{\dot X}}_j^2$ exactly constant, so that they will be used
to define the thermostats temperatures $T_j$ via $K_j=\frac32 k_B T_j
N_j$, as in the classical case.

Let $\m_0(\{d\Ps\})$  be the {\it formal} measure on
$L_2(\CC_0^{3N_0})$ 

\be \Big(\prod_{\V X_0} d\Ps_r(\V X_0)\,d\Ps_i(\V X_0)
\Big)\,\d\Big(\int_{\CC_0} |\Ps(\V Y)|^2\, d\V Y-1\Big)
\label{e4.11.3}\ee
with $\Ps_r,\Ps_i$ real and imaginary parts of $\Ps$.  The
meaning of \equ{e4.11.3} can be understood by imagining to
introduce an orthonormal basis in the Hilbert's space and to
``cut it off'' by retaining a large but finite number $M$ of its
elements, thus turning the space into a high dimensional space
$C^M$ (with $2M$ real dimensions) in which $d\Ps=d\Ps_r(\V
X_0)\,d\Ps_i(\V X_0)$ is simply interpreted as the limit
$M\to\infty$ of the normalized euclidean volume in $C^M$.

The formal phase space volume element $\m_0(\{d\Ps\})\times\n(d\V
X\,d\V{{\dot X}})$ with 

\be \n(d\V X\,d\V{{\dot X}})\defi\prod_{j>0} \Big(\d(\V{{\dot
X}}^2_j-3N_jk_B T_j)\,d\V X_j\,d\V{{\dot X}}_j\Big)
\label{e4.11.4}\ee
is conserved, by the unitary property of the wave
functions evolution, just as in the classical case, {\it up
to the volume contraction in the thermostats}, \Cite{Ga006c}. 

If $Q_j\defi\media{W_j}_\Ps$, as in Eq.\equ{e4.11.2}, then the
contraction rate $\s$ of the volume element in Eq.\equ{e4.11.4} can be
computed and is (again):

\be\s(\Ps,\V{{\dot X}},\V X)=\,\e(\Ps,\V{{\dot X}},\V X)+\dot R(\V X),\qquad
\e(\Ps,\V{{\dot X}}, \V X)=\sum_{j>0} \frac{Q_j}{k_B T_j},
\label{e4.11.5}\ee
with $R$ a suitable observable and $\e$ that will be called {\it entropy
  production rate}:

In general solutions of Eq.\equ{e4.11.2} {\it will not be quasi periodic}
and the CH, \Cite{GC995b}\Cite{Ga000}\Cite{Ga008}, can therefore be
assumed (\ie there is no {\it a priori} conflict between the quasi periodic
motion of an isolated quantum system and the chaotic motion of a non isolated
system): if so the dynamics should select an invariant distribution
$\m$. The distribution $\m$ will give the statistical properties of the
stationary states reached starting the motion in a thermostat configuration
$(\V X_j,\V{{\dot X}}_j)_{j>0}$, randomly chosen with ``uniform
distribution'' $\n$ on the spheres $m\V{{\dot X}}_j^2=3N_jk_B T_j$ and in a
random eigenstate of $H$. The distribution $\m$, if existing and unique,
could be named the {\it SRB distribution} corresponding to the chaotic
motions of Eq.\equ{e4.11.2}.

In the case of a system {\it interacting with a single thermostat} at
temperature $T_1$ the latter distribution should attribute expectation
value to observables for the particles in $\CC_0$, \ie for the test system
hence operators on $L_2(\CC_0^{3N_0})$, equivalent to the
canonical distribution at temperature $T_1$, up to boundary terms.

Hence an important {\it consistency check} for proposing
Eq.\equ{e4.11.2} as a model of a thermostatted quantum
system\index{quantum system} is that, if the system is in contact
with a single thermostat containing configurations $\dot{{\V
    X}}_1,\V X_1$, then there should exist at least one
stationary distribution equivalent to the canonical distribution
at the appropriate temperature $T_1$ associated with the
(constant) kinetic energy of the thermostat: $K_1=\frac32 k_B
T_1\,N_1$.  In the corresponding classical case this is an
established result, see comments to Eq.\equ{e2.8.4},
\Cite{EM990}.

A natural candidate for a stationary distribution could be to
attribute a probability proportional to $d\Ps\,d\V X_1\,d \dot{\V
X}_1$ times
\be 
\sum_{n=1}^\infty e^{-\b_1 E_n(\V X_1)}\d(\Ps-\Ps_n(\V
X_1)\,e^{i\f_n})\,{d\f_n}\,\d(\dot{\V X}_1^2-2K_1)\label{e4.11.6}\ee
where $\b_1=1/k_B T_1$, $\Ps$ are wave functions for the system in
$\CC_0$, ${\dot {\V X}_1, \V X_1}$ are positions and velocities of the
thermostat particles and $\f_n\in [0,2\p]$ is a phase for the eigenfunction
$\Ps_n(\V X_1)$ of $H(\V X_1)$ and $E_n=E_n(\V X_1)$ is the corresponding
$n$-th level. The average value of an observable $O$ for the system in
$\CC_0$ in the distribution $\m$ in \equ{e4.11.6} would be

\be \media{O}_\m=Z^{-1}\int {\rm Tr}\, (e^{-\b_1 H(\V X_1)} O)\,\d(\dot{\V
X}_1^2-2K_1)d\V X_1\,d \dot{\V X}_1\label{e4.11.7}\ee
where $Z$ is the integral in \equ{e4.11.7} with $1$ replacing $O$,
(normalization factor).  Here one recognizes that $\m$ attributes to
observables the average values corresponding to a Gibbs state at
temperature $T_1$ with a random boundary condition $\V X_1$.

But Eq.\equ{e4.11.6} {\it is not invariant} under the evolution
Eq.\equ{e4.11.2}
and it seems difficult to exhibit explicitly an
invariant distribution along the above lines without having recourse to
approximations. A simple approximation is possible and is discussed in the
next section essentially in the form proposed and used in \Cite{MCT993}.

The above analysis may suggest that the SRB distribution%
\footnote{\tiny Defined, for instance, as the limit of the distributions obtained
  by evolving in time the Eq.\equ{e4.11.6}.}
for the evolution in Eq.\equ{e4.11.2} is equivalent to the Gibbs
distribution at temperature $T_1$ with suitable boundary conditions, at
least in the limit of infinite thermostats. So the Eq.\equ{e4.11.6}, in
spite of its non stationarity, remains interesting; but {\it only as a conjecture}.

Invariant distributions can, however, be constructed following the
alternative ideas in \Cite{St966}, as done recently in
\Cite{ACCCEF011}, see remark (5) in the next section.

\section{Quantum adiabatic 
approximation and alternatives\index{adiabatic approximation}}
\def\SEC{Quantum adiabatic 
approximation and alternatives\index{adiabatic approximation}}
\label{sec:XII-4}\iniz
\lhead{\small\ref{sec:XII-4}.\ \SEC}

Nevertheless it is interesting to remark that under the {\it adiabatic
  approximation} the eigenstates of the Hamiltonian at time $0$ evolve by
simply following the variations of the Hamiltonian $H(\V X(t))$ due to the
motion of the thermostats particles, without changing quantum numbers
(rather than evolving following the Schr\"odinger equation and becoming,
therefore, {\it different} from the eigenfunctions of $H(\V X(t))$).

In the adiabatic limit (in which the classical motion of the
thermostat particles takes place on a time scale much slower than
the quantum evolution of the system of interest) the distribution
\equ{e4.11.6} {\it is invariant}, \Cite{MCT993}.\index{quantum
adiabatic thermostats}
\index{quantum fluctuation relation}

This can be checked by first order perturbation analysis%
\index{perturbation theory} which shows that, to first order in $t$, the
variation of the energy levels (supposed non degenerate) is compensated by
the phase space contraction in the thermostat, \Cite{Ga008a}.

Under time evolution, $\V X_1$ changes, at time $t>0$, into $\V X_1+t
\V{{\dot X}}_1+O(t^2)$ and, assuming non degeneracy, the eigenvalue
$E_n(\V X_1)$ changes, by perturbation analysis, into $E_n+t \,
e_n+O(t^2)$ with

\be e_n\defi t\media{\V{{\dot X}}_1\cdot\V\dpr_{\V X_1}
U_{01}}_{\Ps_n}+t \V{{\dot X}}_1\cdot\V\dpr_{\V X_1}
U_{1}=-t\,(\media{W_1}_{\Ps_n}+\dot R_1)=-\frac1{\b_1}\a_1
\label{e4.12.1}\ee
with $\a_1$ defined in Eq.\equ{e4.11.2}.

Hence the Gibbs' factor changes by $e^{-\b t e_n}$ and at the same time
phase space contracts by $e^{t \frac{3 N_1 e_n}{2K_1}}$, as it follows from
the expression of the divergence in Eq.\equ{e4.11.5}. {\it Therefore if
  $\b$ is chosen such that $\b=(k_B T_1)^{-1}$ the state with distribution
  Eq.\equ{e4.11.6} is stationary} in the considered approximation,
(again, for simplicity, $O(1/N)$ is neglected, see comment following
Eq.\equ{e2.8.1}).

This shows that, {\it in the adiabatic approximation}, interaction with
only one thermostat at temperature $T_1$ admits at least one stationary
state. The latter is, by construction, a Gibbs state of thermodynamic
equilibrium with a special kind (random $\V X_1,\V{{\dot X}}_1$) of
boundary condition and temperature $T_1$.  \*

\0{\it Remarks:} (1) The interest of the example is to show that even in
quantum systems the CH makes sense and the interpretation
of the phase space contraction in terms of entropy production
\index{entropy production} remains unchanged.  
\*

\0(2) In general, under the CH, the SRB distribution of
\equ{e4.11.2} (which in presence of forcing, or of more than one
thermostat, is certainly quite non trivial, as in the classical mechanics
cases) will satisfy the fluctuation relation because, besides the CH, the fluctuation theorem only depends on reversibility: so the
model \equ{e4.11.2} might be suitable (given its chaoticity) to simulate
the steady states of a quantum system in contact with thermostats.
\*

\0(3) It is certainly unsatisfactory that the simple Eq.\equ{e4.11.6} is
not a stationary distribution in the single thermostat case (unless the
above adiabatic approximation is invoked). However, according to the
proposed extension of the CH, the model does have a
stationary distribution which should be equivalent (in the sense of
ensembles equivalence) to a Gibbs distribution at the same temperature: the
alternative distribution in remark (5) has the properties of being
stationary and at the same time equivalent to the canonical Gibbs
distribution for the test system in $\CC_0$.
\*

\0(4) The non quantum nature of the thermostat considered here and the
    specific choice of the interaction term between system and
    thermostats should not be important: the very notion of thermostat
    for a quantum system is not at all well defined and it is natural
    to think that in the end a thermostat is realized by interaction
    with a reservoir where quantum effects are not
    important. Therefore what the analysis really suggests is
    that, {\it in experiments in which really microscopic systems are
    studied, the heat exchanges of the system with the external world
    should fulfill a fluctuation relation}\index{fluctuation relation}.
 
\* \0(5) An alternative approach can be based on the quantum mechanics
formulation in \Cite{St966} developed and subsequently implemented in
simulations, where it is called\index{Erhenfest dynamics}
{\it Erhenfest dynamics}, and more recently in \Cite{ACCCEF011}. It can be
remarked that the equations of motion Eq.\equ{e4.11.2} can be derived from
the Hamiltonian $\HH$ on $L_2(\CC_0^{3N_0})\times \prod_{j=1}^n R^{6N_j}$
imagining a function $\Ps(\V X_0) \in L_2(\CC_0^{3N_0})$ as $\Ps(\V
X_0)\defi \k(\V X_0)+i\p(\V X_0)$, with $\p(\V X_0),\k(\V X_0)$ canonically
conjugate and defining $\HH$ as:
\be\eqalign{&
\sum_{j=0}^n \frac{\dot{{\V X}}_j^2}2 +U(\V X_j)
+\int_{R^{3N_0}}\Big(
\frac{\dpr_{{\V X_0}}\p({{\V X_0}})^2+\dpr_{{\V X_0}}\k({{\V X_0}})^2}2
\cr&+\frac{(\p({{\V X_0}})^2+\k({{\V X_0}})^2)(U({{\V X_0}})
+W({{\V X_0}}, \V X_j))}2\Big) d\V X_0\,\defi\, \HH\cr}\Eq{e4.12.2}
\ee
where $W(\V X_0,\V X_0)\equiv0$ and adding to it the constraints $\int
|\Ps(\V X_0)|^2d\V X_0=1$ (which is an integral of motion) and $\frac12
{\dot{{\V X}}}^2_j= 3 N_j k_B T_j,\, j=1,\ldots,n$ by adding to the
equations of the thermostats particles $-\a_j \dot{{\V X}}_j$
with $\a_j$ as in Eq.\equ{e4.11.2}.
In this case, by the same argument leading to the theorem following
Eq.\equ{e2.8.4}, the formal distribution $const\, e^{-\b \HH}d\V X_0
d\dot{{\V X}}_0\,d\p d\k$ is stationary and equivalent to the canonical
distribution for the test system if the thermostats have all the same
temperature. {\it This avoids using the adiabatic approximation}. This
alternative approach is well suitable for simulations as shown for instance
in \Cite{ACCCEF011}. The above comment is due to M. Campisi
\index{Campisi}(private communication, see also \Cite{Ca013} where a
transient fluctuation relation is studied).  \*
\0(6) It would be interesting to prove for the evolution of the Hamiltonian
in Eq.\equ{e4.12.2} theorems similar to the corresponding ones for the
classical systems in Sec.(\,\ref{sec:II-5}\,), under the same assumptions on the
interaction potentials and with Dirichlet boundary conditions for the
fields $\p,\k$.

\chapter{Conjectures and suggested applications}
\label{Ch5} 

\chaptermark{\ifodd\thepage
Applications\hfill\else\hfill 
Applications\fi}


\section{Equivalent thermostats}
\def\SEC{Equivalent thermostats}
\label{sec:I-5}\iniz
\lhead{\small\ref{sec:I-5}.\ \SEC}

In Sec.(\,\ref{sec:III-2}\,) a few models $(M,S_t)$ for the electric conduction 
have been considered:
\*

\0(1) the classical model of Drude,\index{Drude's theory},
\Cite{Be964}[Vol.2,Sec.35],\Cite{Se987}[p.139], in
which at {\it every collision} velocity of the electron, of
charge $e=1$, is rescaled to an assigned velocity $v$ defined by
the 'temperature' $T=m v^2/(3 k_B)$ assumed for the system.

\0(2) the Gaussian model in which the total kinetic energy of the
$N$ particles is kept constant by a thermostat force:
\be  m \ddot {\V x}_i=
\V E -\,\frac{m \V E \cdot\,\V J}{3 N k_B T}
\,\dot{\V x}_i+ ``{\rm collisional\ forces}''\label{e5.1.1}\ee 
where $3 N k_B T/2$ is the total kinetic energy (a constant of motion in
this model) and ${\bf J}\defi \sum_i\dot \xx_i$, \Cite{HHP987}\Cite{EM990}.
A third model could be

\0(3) a ``friction model'' in which particles independently
    experience a constant friction $\n$:

\be  m \ddot {\V x}_i=
\V E-\,\n \,\dot{\V x}_i+ ``{\rm collisional\ forces}''\label{e5.1.2}\ee 
where $\n$ is a constant tuned so that the {\it average kinetic
energy} is $3N k_B T/2$, \Cite{Ga995c}\Cite{Ga996b}.
\*

The models are deterministic: the first and third are
``irreversible'' while the second is reversible because the
isometry $I({\V x}_i,\V v_i)=({\V x}_i,-\V v_i)$ anticommutes
with the time evolution flow $S_t$ defined by the equation
Eq.\equ{e5.1.1}: $I S_t=S^{-t}I$.

Here the models will be considered in a thermodynamic context in
which the number of particles and obstacles are proportional to
the system size $L^d$, which is large. The CH
will be assumed with the attracting surface $\AA$ coincident with the
phase space $M$, hence the systems $(M,S_t)$ will admit a unique SRB
distribution, (\,\ref{sec:VI-2}\,).

Let $\m_{\d,T}$ be the SRB distribution for Eq.\equ{e5.1.1} for
the stationary state that is reached starting from initial data,
chosen randomly as in Sec.(\,\ref{sec:IV-2}\,), with energy $3N k_B
T/2$\,\footnote{\tiny In this model the kinetic energy is a
constant of motion.}  and density $\d=\frac{N}{L^d}$. The collection
of the distributions $\m_{\d,T}$  define, as the kinetic energy $T$ and
the density $\d$ vary, a ``statistical ensemble'' $\EE$ of
stationary distributions associated with the equation
Eq.\equ{e5.1.1}.

Likewise we call $\wt \m_{\d,\n}$ the class of SRB distributions
associated with Eq.\equ{e5.1.2} which forms an ``ensemble'' $\wt \EE$.

A correspondence between distributions of the ensembles $\EE$ and
$\wt \EE$ can be established by associating $\m_{\d,T}$ with $\wt
\m_{\d',\n}$, to be called {\it corresponding elements}, if:
\be \d=\d',\qquad \frac{3}2Nk_B T=\ig \frac12(\sum_j m \dot{\V x}^2_j)\,\wt
\m_{\d,\n}(d{\V x}\,d\dot{\V x})
\label{e5.1.3}\ee
Call an observable $F$ on $M$ a {\it local
  observable}\label{local oservable} if $F$ is a smooth function
  depending solely on the microscopic state of the particles
  whose position is inside some box $V_0$ strictly contained in
  the system container $V$.  Then the
  following conjecture was\index{equivalence conjecture} proposed
  in
\Cite{Ga996b}\,: \*

\0{\bf Conjecture:} {\it (equivalence conjecture) If $\LL$
  denotes the local smooth observables, for all $ F\in \LL$, it
  is\index{equivalence conjecture}
\be \lim_{N,V\to\io, N/L^d=\d} \wt \m_{\d,\n}(F)=
\lim_{N,V\to\io, N/L^d=\d} \m_{\d,T}(F) \label{e5.1.4}\ee 
if $T$ and $\n$ are related by Eq.\equ{e5.1.3}.\index{local oservable}
}
\*

This conjecture has been discussed in
\Cite{GC995b}[Sec.8]\Cite{Ga995c}[Sec.5]\Cite{Ga999}[Sec.2,5], \Cite{Ru000}
and \Cite{Ga000}[Sec.9.11]. In various studies on fluids, \eg
\Cite{Ga020b}[Sec.17]\Cite{Ga019c}\Cite{Ga021}, a similar
conjecture arises, see Sec.(\,\ref{sec:V-5}\,) and sections following it,
and Appendix \ref{appJ}\,.

The idea of this kind of ensemble equivalence was present since
the beginning as a motivation for the use of
Nos\'e--Hoover's\index{Nos\'e-Hoover} or Gaussian thermostats.
\index{Gaussian thermostat} It is clearly introduced and analyzed
in \Cite{ES993}, where earlier works are quoted.  When a system
is large the microscopic evolution time scale becomes much
shorter than the macrosopic one and the Gaussian multiplier $\a$
becomes a sum of quantities rapidly varying (in time as well as
in space) and therefore could be considered as essentially
constant at least if only macroscopic time--independent
quantities are observed.

The conjecture is very similar to the equivalence, in equilibrium
cases, between canonical and microcanonical ensembles: here the
friction $\n$ plays the role of the canonical inverse temperature and
the kinetic energy that of the microcanonical energy.

And in the following similar conjectures will be proposed.
As an example consider the collection $\EE'$ of
stationary distributions for the original model (1) of Drude,
whose elements $\m'_{\d,T}$ can be parameterized by the quantities $T$,
temperature (such that $\frac12\sum_j m \dot{\V x}_j^2=\frac32 N k_B T$),
and density ($N/V=\d$). This is an ensemble $\EE'$ whose elements can be
put into one-to-one correspondence with the elements of, say, the ensemble
$\EE$ associated with model (2), \ie with Eq.\equ{e5.1.1}:
elements carrying the same labels $\d,T$
$\m'_{\d,T}\in \EE', \m_{\d,T}\in \EE$ will be called
correspondent.

Then a similar equivalence conjecture is: {\it if $\m_{\d,T}\in
  \EE$, $\m'_{\d,T}\in \EE'$ are corresponding elements then
\be \lim_{N,V\to\io, N/V=\d} \m_{\d,T}(F)=
\lim_{N,V\to\io, N/V=\d} \m'_{\d,T}(F)\label{e5.1.5}\ee 
for all local observables $F\in \LL$.}

It is remarkable that the above equivalences implicitly suggest
an extended equivalence (rather than identity) between a
``reversible statistical ensemble'', \ie the collection $\EE$ of
the SRB distributions\index{SRB distribution equivalence}
associated with Eq.\equ{e5.1.1} and an ``irreversible
statistical ensemble'', \ie the collection $\wt \EE$ of SRB
distributions associated with Eq.\equ{e5.1.2}.

Hence we see that it is possible to conceive statistical
equivalence between various kinds of irreversible
dissipation\index{irreversible dissipation}  and of 
reversible dissipation models\,,\index{reversible dissipation}
\footnote{\tiny {\it E.g.}  a system subject to a Gaussian
thermostat.}  at least as far as the stationary averages of a large class of
special observables are concerned.

\section{Processes time scale and irreversibility}
\def\SEC{Processes time scale and irreversibility}
\label{sec:II-5}\iniz
\lhead{\small\ref{sec:II-5}.\ \SEC}

Carnot's theorem shows that the most efficient machines must
operate running in a {\it reversible cycle} in which the vapor
(of water, alcohol or other liquids) evolves through a sequence
$\PP$ of equilibrium states in a process so slow that, for
instance, differences in temperature {\it ``can be considered as
  vanishing''}; realizing the oxymoron Carnot continues: {\it
  ``\`A la v\'erit\'e, les choses ne peuvent pas se passer
  rigoureusement comme nons l'avons suppos\'e ...},
\Cite{Ca824}[p.13-14]. {\it Question}: can one make 
quantitative the needed slowness and measure the reversibility of
a process ? \Cite{Ga006a}.

A {\it process}, denoted $\G$, transforming an initial stationary
state $\m_{ini}\equiv \m_0$ via an evolution like the one in
Fig.2.2.1,\,\ref{Fig.2.2.1}\,, $\dot x=F(x)+\F(x,t)$ under initial forcing
$\F_{ini}\equiv \F(x,0)$ into a final stationary state
$\m_{fin}\equiv \m_\io$ under final forcing $\F_{fin}\equiv
\F(x,\infty)$, will be defined by a smooth function
$t\to \F(t),\, t\in[0,+\io)$, varying between $\F(x,0)=\F_0(x)$
to $\F(x,+\io)=\F_\io(x)$.

For intermediate times $0<t<\io$ the time evolution $x=(\dot{\V
  X},\V X) \to x(t)=S^{0,t}x$ is generated by the equations $\dot
x=F(x)+\F(x,t)$ with initial state in phase space $\FF$: it is a
non autonomous equation.

The time dependence of $\F(t)$ could, for instance, be due to a
motion of the container walls which changes the volume from the
initial $\CC_0=V_0$ to $V_t$ to $\CC'_0=V_\infty$: hence the
points $x=(\dot{\V X},\V X)$ evolve at time $t$ in a space
$\FF(t)$ which also may depend on $t$.

During the process the initial state evolves into a $t$ dependent
state: imagining the process repeated enough times so that it
makes sense to have an average distribution $\m_t(dx)$, for the
state of the system at time $t$, which attributes to an
observable $F_t(x)$, defined on $\FF(t)$, an average value given
by:
\be \media{F_t}=\ig_{\FF(t)} \m_t(dx) F_t(x)\defi
\ig_{\FF(0)} \m_0(dx) F_t(S^{0,t} x)\label{e5.2.1}\ee
We shall also consider the probability distribution $\m_{SRB,t}$
which is defined as the SRB distribution of the dynamical system
obtained by ``freezing'' $\F(t)$ at the value that is taken at
time $t$, and letting the time to evolve further until the
stationary state $\m_{SRB,t}$ is reached: {\it in general}
$\m_t\ne \m_{SRB,t}$.

Forces and potentials will be supposed smooth in their
variables except, possibly, impulsive elastic forces describing shocks,
allowed here to model shocks with the containers walls and possible
shocks between hard core particles.  

Chaotic hypothesis (CH) will be assumed: this means that in the physical
problems just posed on equations of motion written symbolically $\dot
x=F(x)+\F(x,t)$ with $\F$ time dependent, the motions are so chaotic that
the attracting sets, on which their long time motion {\it would} take place if
$\F(t)$ was fixed at the value taken at time $t$, are
smooth surfaces on which motion is chaotic.

It is one of the basic tenets in Thermodynamics that all
(nontrivial) processes between equilibrium states are
``irreversible'':\index{irreversible process} only idealized
(strictly speaking nonexistent) ``quasi static''
processes\index{quasi static process} through equilibrium states
can be reversible.  The question addressed here is whether
irreversibility can be made a quantitative notion at least in
models based on microscopic evolution, like the models in
Fig.2.2.1,\,\ref{Fig.2.2.1}\,, and in processes developing
between equilibrium states (whether equal or not).  \*

Some examples:

\0{\bf(1)} Gas of $N_0$ particles in contact with reservoirs
$T_1,T_2,\ldots$ with different temperatures and $N_1,N_2,\ldots$
particles, see Sec.(\,\ref{sec:II-2}\,):

\eqfig{300}{80}{
\ins{100}{75}{$U_i=\sum_{jk} v(q_k-q_j)$: 
 internal energy of $T_i$ }
\ins{100}{55}{$W_{0i}=\sum_{j\in C_0\atop k\in T_i} v(q_k-q_j)$:
   interaction $T_i-C_0$ }
}{fig5.2.1}{}    

\kern-19mm
$$\kern15mm\eqalign{
& {\ddot {\V X}_0=-\dpr_{\V X_0} (U_0(\V X_0)+{\sum}_{i>0}
W_{0i}(\V X_0,\V X_i))+ \V E(\V X_0)}\cr
& {\ddot {\V X}_i=-\dpr_{\V X_i} (U_i(\V X_i)+
W_{0i}(\V X_0,\V X_i))}
-{\a_i\dot{{\V X}}_i}\cr}$$

\*\kern-1mm
\begin{spacing}{0.5}\0\tiny Fig.5.2.1: $\a_i$
  s.t. {$\frac{m}2\sum_{i>0}\dot {{\V X}}_i^2=\frac12
    N_ik_BT_i(t)$}:\ \ {$\a_i=\frac{Q_i-\dot U_i}{N_i k_B
      T_i(t)}$}, see Eq.\equ{e2.2.1}.\end{spacing}
\*
\*
\0{\bf(2)} Gas in a container with moving wall 

\eqfig{330}{73}
{\ins{80}{6}{$\scriptstyle L(t)$}
\ins{200}{6}{$\scriptstyle L(t)$}
\ins{90}{60}{$\scriptstyle \infty$}
\ins{16}{70}{$\scriptstyle V(x,t)$}
\ins{215}{50}{$\to$}
}
{fig5.2.2}{Fig.5.2.2} 
\*
\begin{spacing}{0.5}\0\tiny Fig.5.2.2: The piston extension
  is $L(t)$ and $V(x,t)$ is a potential
  modeling its wall. A sudden doubling of $L(t)$ would correspond
  to a Joule-Thomson expansion\index{Joule-Thomson
    expansion}.\end{spacing}
\*\*

\0{\bf(3)} Paddle wheels stirring a liquid

\eqfig{300}{70}
{\ins{60}{65}{$\o t$}
}
{fig5.2.3}{Fig.5.2.3} 

\begin{spacing}{0.5}\0\tiny Fig.5.2.3: The wavy lines
  symbolizes the surface of the water.
  Slow rotation for a long time would correspond to the Joule
  paddle wheels measurement of the heat-work conversion
  factor.\end{spacing} \index{Joule's conversion factor}
\*\*
 
\0In the examples the $t$ dependence of $\F(x,t)$ is supposed to vanishes as
$t$ becomes large.  Example 1 is a process transforming an initial
stationary state into a new stationary state (varying some
parameters in the equations, \eg $\V E$); while examples 2,3
are processes transforming a equilibrium state into an 
equilibrium state (in general different).  \*

The work $Q_a\defi\sum_{j=1}^{N_a} -\dot{{\V x}}_{a,j}\cdot\BDpr_{x_{a,j}}
U_{0,a}$ in example 1 will be interpreted as {\it heat} $ Q_a$ ceded, per
unit time, by the particles in $\CC_0$ to the $a$-th thermostat (because
the ``temperature'' of $\CC_a,\, a>0$ remains constant).  The phase space
contraction rate due to heat exchanges between the system and the
thermostats can, therefore, be naturally defined as in
Eq.\equ{e2.8.1}\,(see also the comments following it):
\be \s^\G(\dot{\V X},\V X)\defi\sum_{a=1}^{N_a}
\frac{Q_a}{k_B T_a}+\dot R\label{e5.2.2}\ee
where $R=\sum_{a>0} \frac{U_a}{k_B T_a}$ (discarding the factors
$(1-\frac1{3N_a})$ for notational simplicity).

Phase space volume can also change because new regions become accessible
(or inaccessible)%
\footnote{\tiny For instance this may mean that the
external potential acting on the particles undergoes a change,
\eg a moving container wall means that the external potential due
to the wall changes from $0$ to $+\infty$ or from $+\infty$ to
$0$, as in example 2. Or, since the total energy varies, as in
example 3.}
so that the total phase space contraction rate, denoted
$\s_{tot,t}$, in general will be different from $\s^\G_t$.

It is reasonable to suppose, and often it can even be proved,
that at every time $t$ the configuration $S^{0,t}x$ is a
``typical'' configuration of the ``frozen'' system, if the initial
$x$ was typical for the initial distribution $\m_0$.
{\it I.e.},  denoting 
$\FF_a$ is the phase space of the $a$-th thermostat,  $S^{0,t}x$ will
be a point in $\FF(t)=V_t^N\times R^{3N}\times \prod \FF_a$, whose
statistics, under evolution via equations parameters
imagined frozen at their values at time
$t$, will be $\m_{SRB,t}$, see comments following
Eq.\equ{e5.2.1}.%
\footnote{\tiny If for all $t$ the ``frozen'' system is Anosov, then any
  initial distribution of data which admits a density on phase space will
  remain such, and therefore with full probability its configurations will
  be typical for the corresponding SRB distributions. Hence if $\BF(t)$ is
  constant the claim holds.}

Since we must consider as accessible the phase space occupied by
the attracting set of a typical phase space point, the volume
variation contributes an extra $\s^v_t(x)$ to the phase space
variation, via the rate at which the phase space volume $|\FF_t|$
contracts, namely:

\kern-3mm
\be \s^v_t(x)=-\frac1{|\FF_t|}\frac{d \,|\FF_t|}{dt}=
-N\frac{\dot V_t}{V_t}\label{e5.2.3}\ee
which does not depend on $x$, as it is a property of the phase space
available to (any typical) $x$.

Therefore the total phase space contraction per unit time
can be expressed as, see Eq.\equ{e5.2.3},\equ{e5.2.2},
\be \s_{tot}(\dot{\V X},\V X)= \sum_a \frac{Q_a}{k_B
  T_a}-N\frac{\dot V_t}{V_t}
+\dot R(\dot{\V X},\V X)\label{e5.2.4}\ee
and there is a simple and direct relation between 
phase space contraction\index{phase space contraction} and 
entropy production\index{entropy production} rate, 
\Cite{Ga005c}.  Eq.\equ{e5.2.4}
shows that their difference is a ``total time derivative'' $\dot R$.  

In studying stationary states with a fixed forcing $F(x)+\F(x,t)$
frozen at the value it has at time $t$ it is $N\dot V_t/V_t=0$.
The term $\dot R$ depends on the coordinates used, as discussed
in Sec.(\,\ref{sec:VIII-2}\,): hence define the {\it entropy
production rate} in a process to be Eq.\equ{e5.2.4} {\it
without} the $\dot R$ term:
\be \e(\dot{\V X},\V X)= \sum_a \frac{Q_a}{k_B T_a}-N\frac{\dot
  V_t}{V_t},\quad
\e(\dot{\V X},\V X)_{SRB}\defi \e^{srb}_t(\dot{\V X},\V X)-N\frac{\dot V_t}{V_t}
\label{e5.2.5}\ee
where $\e^{srb}_t$ is defined by the last equality and the name
is chosen to remind that if there was no volume change
($V_t=const$) and the external forces were constant (at the value
taken at time $t$) then $\e^{srb}_t$ would be the phase space
contraction natural in the theory for the SRB distributions when
the external parameters are frozen at the value that they have at
time $t$.

It is interesting, and necessary, to remark that in a stationary
state the time averages of $\e$, denoted $\e_+$, and of $\sum_a
\frac{ Q_a}{k_B T_a}$, denoted $\sum_a \frac{
  \media{Q_{a}}_+}{k_B T_a}$, coincide because $N\dot V_t/V_t=0$,
as $V_t=const$ (and $\dot R$ would have zero time average, being a total
derivative). On the other hand under very general assumptions,
much weaker than the CH, the time average of the
phase space contraction rate is $\ge0$, \Cite{Ru996}\Cite{Ru997}, so
that in a stationary state: $\sum_a \frac{\media{Q_{a}}_+}{k_B
  T_a}\ge0$ (a consistency property to be
required by any proposal of definition of entropy production).
  \*

\0{\it Remarks:} (1) In processes starting an ending in
the same stationary states (special ``cycles'') the above models
are a realization of {\it Carnot's machines\index{Carnot's
machines}}: the machine being the system in $\CC_0$ on which
external forces $\F$ work leaving the system in the same
stationary state achieving a transfer of heat between the various
thermostats (in agreement with the second law only if the
transfer is $\e_{+}\ge0$).  \\
(2) The fluctuations of the
entropy production Eq.\equ{e5.2.5} become observable, over a
time scale independent of the thermostats size, because
the heat exchanged is a boundary effect (due to the interaction
of the test system particles and those of the thermostats in
contact with it, see Sec.(\,\ref{sec:VIII-2}\,).  \*

Coming back to the question of defining an irreversibility
degree\index{irreversibility degree} of a process $\G$ we
distinguish between the (non stationary) state $\m_t$ into which
the initial state $\m_0$ evolves in time $t$, under varying
forces and volume, and the state $\m_{SRB,t}$ obtained by
``freezing'' forces and volume at time $t$ and letting the system
settle to become stationary, see comments following
Eq.\equ{e5.2.1}.  We call $\e_t$ the entropy production rate
Eq.\equ{e5.2.5} and $\e^{srb}_t$ the entropy
production\index{entropy production} rate in the ``frozen'' state
$\m_{SRB,t}$, as in Eq.\equ{e5.2.5}, tipically of size of order $N$.

The proposal is to define, in a process $\G$
leading from an equilibrium state to an
equilibrium state, the {\it irreversibility time
  scale}\index{irreversibility time scale} $\II(\G)^{-1}$ 
by setting:
\be \II(\G)=\frac1{N^2}\ig_0^\io
\Big(\media{\e_t}_{\m_t}-\media{\e^{srb}_{t}}_{SRB,t}\Big)^2
dt\label{e5.2.6}\ee
If CH is assumed then $\m_t$ will evolve exponentially fast under
the ``frozen evolution'' to $\m_{SRB,t}$.\footnote{\tiny {\it
I.e.} the correlations of local observables decay exponentially.}
Therefore the integral in Eq.\equ{e5.2.6} will converge for
reasonable $t$ dependences of $\F,V$: examples are examined in
the next section.

\section{Quasi static processes time scale: examples}
\def\SEC{Quasi static processes time scale: examples}
\label{sec:III-5}\iniz
\lhead{\small\ref{sec:III-5}.\ \SEC}

A physical definition of ``quasi static'' transformation is a
transformation that is very slow. If the equations of motion are
$\dot x=\F_t(x)$, the 'very slow' can be translated
mathematically, for instance, into an evolution in which $\F_t$
evolves like, if not exactly, as:
\be \F_t=\F_0+ (1-e^{-\g t})(\F_\io-\F_0)\defi
\F_0+ (1-e^{-\g t})\D.\label{e5.3.1}\ee
An evolution $\G$ close to quasi static, but simpler for
computing $\II(\G)$, would proceed changing $\F_0$ into
$\F_\io=\F_0+\h(t)\D$ with $\h$ growing, from $0$ to $1$, by
$1/\d$ steps of size $\d$, each of which has a time duration
$t_\d$ long enough so that, at the $k$-th step, the evolving
system is very close (see below) to its stationary state at
forced $\F=\F_0+k\d\D$.
\*\*
\eqfig{320}{45}
{\ins{-6}{55}{$\scriptstyle \e(t)$}
\ins{-1}{18}{$\scriptstyle \d$}
\ins{248}{50}{$\scriptstyle \e(\infty)=1$}
\ins{255}{6}{$\scriptstyle t$}
\ins{70}{6}{$\scriptstyle t_1$}
\ins{130}{6}{$\scriptstyle t_2$}
\ins{190}{6}{$\scriptstyle t_3$}
}
{fig5.2.4}{\small Fig.5.2.4}
\begin{spacing}{0.5}\0\tiny Fig.5.2.4: An example of a process
  proceeding at jumps of size $\d$ at times $t_0=0, t_1,\ldots$:
  the final value $\h=1$ can be reached in a finite time or in an
  infinite time.\end{spacing}
\*\*

If the time scale to approach equilibrium at {\it fixed} external
forces can be taken $=\k^{-1}$, independent of the value of the
forces, then $t_\d$ can be defined by $e^{-\k t_\d}\ll \d $: and
then $\II(\G)= const\, \d^{-1}\d^2\log\d^{-1}$, because the
variation of the averages of the phase space contraction,
$\e_{(k+1)\d,+}-\e_{k\d,+}$ is, in general, of order $\d$ (as a
consequence of the differentiability of the SRB states with
respect to the parameters, \Cite{Ru997b}); the dimensional
constant will be proportional to $\k^{-1}\lis\e^2$ with $\lis\e$
of the order of the $\e(t)$'s (supposed of the same order0.  \*

\0{\it Remarks:} {\bf(1)} A drawback of the above definition
is that although $\media{\e^{srb}_t}_{SRB,t}$ is {\it
  independent} on the metric that is used to measure volumes in
phase space the quantity $\media{\e_t}_{\m_t}$ {\it depends} on
it. Hence the irreversibility degree considered here reflects
also properties of our ability or method to measure (or to
imagine to measure) distances in phase space. One can keep in
mind that a metric independent definition can be simply obtained
by minimizing over all possible choices of the metric: but the
above simpler definition seems, nevertheless, preferable.  \*

\0{\bf(2)} Suppose that a process takes place because of the variation of
an acting conservative force, for instance because a gravitational force
changes as a large mass is brought close to the system, while no change in
volume occurs and the thermostats have all the same temperature. Then the
``frozen'' SRB distribution, for all $t$, is such that
$\media{\e^{srb}}_{SRB,t}=0$ (because the ``frozen equations'', being
Hamiltonian equations, admit a SRB distribution which has a density in
phase space).  The isothermal process thus defined has {\it therefore} (and
{\it nevertheless}) $\II(\G)>0$.  \*

\kern2mm \0{\bf(3)} Consider a typical irreversible process. Imagine a
gas in an adiabatic cylinder covered by an adiabatic piston and
imagine to move the piston. The simplest situation arises if
gravity is neglected and the piston is suddenly moved at 
speed $w$.

Unlike the cases considered so far, the absence of thermostats
(adiabaticity of the cylinder) imposes a more detailed analysis. The
simplest situation arises when the piston is moved at speed so large
that no energy is gained or lost by the particles because of
the collisions with the moving wall (this is, in fact, a case in which
there are no such collisions). This is an extreme idealization of the
classic Joule-Thomson experiment.

Let $S$ be the section of the cylinder and $H_t = H_0+w\,t$ be the distance
between the moving lid and the opposite base. Let $\Omega = S\,H_t$ be the
cylinder volume. In this case, if the speed $w\gg \sqrt{k_BT}$ the volume
of phase space changes because the boundary moves and it increases by $N
\,w\,S\,\Omega^{ N-1}$ per unit time, \ie its rate of increase is
(essentially, see remark 5 below) $N\frac{w}{H_t}$.

Hence $\media{\e_t}_t$ is $-N \frac{w}{H_t}$, while $\e^{srb}_t\equiv0$. If
$\th=\frac{L}w$ is the duration of the transformation ("Joule-Thomson'' process)
increasing the cylinder length by $L$ at speed $w$, then
\be \txt
\II(\Gamma ) =\frac1{N^2}\ig_0^\th N^2\Big(\frac{w}{H_t}\Big)^2\,dt 
\tende{T\to\io} 
w\frac{L}{H_0(H_0+L)} \label{e5.3.2}\ee
\0and the transformation is irreversible. The irreversibility time scale
approaches $0$ as $w\to\io$, as possibly expected, \ie
irreversibility is immediately realized.  If $H_0 = L$,
i.e. if the volume of the container is doubled, then $I(\G) =
\frac{w}{2L}$ and the irreversibility time scale of the process coincides
with its ``duration". 
\*
\0{\bf(4)} If in the context of (3) the
piston is replaced by a sliding lid which divides the cylinder in two
halves of height $L$ each: one empty at time zero and the other
containing the gas in equilibrium. At time $0$ the lid is lifted and a
process $ \Gamma'$ takes place. In this case $\frac{dV_t}{dt} = V \d(t)$
because the volume $V = S\,L$ becomes suddenly double (this amounts at a
lid receding at infinite speed).  Therefore the
evaluation of the irreversibility scale yields
\be \II(\Gamma') = \ig_0^\io N^2\d(t)^2\,dt =
+\io\label{e5.2.9}\ee
so that the irreversibility becomes immediately manifest:
$\II(\Gamma') = +\io$, $\II(\G')^{-1}=0$. This idealized
experiment is rather close to the actual Joule-Thomson
experiment.

In the latter example it is customary to estimate the degree of
irreversibility at the lift of the lid by the {\it thermodynamic
equilibrium entropy} variation between initial and final states. It
would of course be interesting to have a general definition of entropy
of a non stationary state (like the states $\m_t$ at times
$t\in(0,\io)$ in the example just discussed) that would allow
connecting the degree of irreversibility to the thermodynamic entropy
variation in processes leading from an initial equilibrium state to a
final equilibrium state, see \Cite{GL003}.
\*
\0{\bf (5)} The above case (3) but with $w$ comparable
with $\sqrt{k_B T}\defi\b^{-1/2}$ can
also be considered. The lid mass being supposed infinite, a
particle hits it if its perpendicular speed $v_1>w$ and rebounds
with a kinetic energy decreased by $2v_1w$: if $\r$ is the
density of the gas and $N$ the number of particles, the colliding
particles number per unit time is $v_1L^2\r \ch(v_1>w) \,g_\b(\V
v)d^3(\V v)$ ($\ch(\cdot)$ being the characteristic function and
$g_\b(\V v)$ is the Maxwell distribution)). Hence the the phase
space contraction per unit time receives a contribution $\simeq
\int_{v_1>w}g_\b(\V v) d^2\V v^\perp dv_1\, (L^2\r v_1)  (\b 2 v_1 w)$
(and its square contributes  $\sim  O(e^{-2\b w^2})$ to the time scale
$\II^{-1}$ in Eq.\equ{e5.2.6}\,).  \*
\0{\bf(6)} The Joule experiment, modeled by the paddle wheel in
Fig.5.2.3, for the measurement of the conversion factor of
calories into ergs can be treated in a similar way: but there is
no volume change and the phase space contraction, depending on
the speed $\o$ of rotation of the paddles, is similar to
the ``extra'' contribution in remark (5).
\*
\0{\bf(7)} In processes $\G$ starting from a stationary
nonequilibrium state but not ending with a stationary state
(because the evolution time is not long enough for the evolution
to reach a stationary state) the time scale $\II$,
Eq.\equ{e5.2.6}, is also defined. However it would not be
appropriate to call it the ``irreversibility time scale'': the
process time scale is, in all cases, a measurement of how far a
process is from a quasi static one between stationary initial and
final states.

\* It might be interesting, and possible, to study a geodesic
flow on a surface of constant negative curvature under the action
of a slowly varying electric field and subject to a isokinetic
thermostat: the case of constant field is studied in
\Cite{BGM998}, but more work will be necessary to obtain results
(even in weak coupling) on the process in which the electric
field $E(t)$ varies.

\section{Fluids}
\def\SEC{Fluids}
\label{sec:IV-5}\iniz
\lhead{\small\ref{sec:IV-5}.\
\SEC}

The ideas in Sec.(\,\ref{sec:I-5}\,) show that the negation of the
notion of reversibility should not be ``irreversibility'': {\it
  it should instead be the property that the natural time
  reversal map\,\footnote{\tiny $I$ = velocities reversal at
  unchanged positions. Should another symmetry be considered
  fundamental (\eg TCP symmetry) it could simply play the role
  of $I$ in the following.}
  $I$ does not verify} $IS=S^{-1}I$, \ie does not anticommute
with the evolution map $S$. This is likely to generate
misunderstandings as the word irreversibility usually refers to
lack of velocity reversal symmetry in systems whose microscopic
description is, or should be, velocity reversal symmetric.

The typical phenomenon of reversibility\index{reversibility} (\ie the
indefinite repetition, or ``{\it recurrence}'',\index{recurrence} of ``{\it
  impossible}'' states) in {\it isolated} systems should indeed manifest itself,
but on time scales much longer and/or on scales of space much smaller than
those interesting for the class of motions considered here: where motions
of the system could be considered as motions of a continuous fluid.

The transport coefficients\index{transport coefficient} (such as viscosity
or conductivity or other) {\it do not have a fundamental nature}: rather
they must be thought of as macroscopic parameters related to the disorder
at molecular level.  \*

{\it Therefore it should be possible to describe in different ways the same
  systems}, simply by replacing the macroscopic coefficients with
quantities that vary in time or in space but rapidly enough to make it
possible identifying them with some average values, at least on suitable
scales of time and space on which {\it the equations thus obtained would then be
  physically equivalent to the previous}.  \*

Obviously we can {\it neither expect nor hope} that, by modifying
the equations and replacing some constant quantities with
variable ones , simpler or easier equations will result (on the
contrary!).  However checking that two equations that should describe
the same phenomena do give, actually, the same results can be
expected to lead to {\it nontrivial relations} between properties
of the solutions (of both equations).

And providing different descriptions of the same system is not only
possible but it can even lead to laws and deductions that would be
impossible (or at least difficult) to derive if one did confine
himself to consider just a single description of the
system (think for instance to the description of equilibrium by the
microcanonical or the canonical ensembles).

What just said {\it has not been systematically applied to the
  mechanics of fluids}, although by now there are several
deductions of macroscopic irreversible equations starting from
microscopic reversible dynamics, for instance Lanford's
derivation\index{Lanford} of the Boltzmann equation,
\Cite{La974}.

{\it Therefore} keeping in mind the above considerations we shall
imagine other equations that could be ``equivalent'' to the
Navier--Stokes incompressible equation (in a container $\O$ with
some boundary conditions). 

Viscosity will be regarded as a {\it phenomenological} quantity whose
role is to forbid to a fluid, subject to non conservative external forces,
to increase indefinitely its energy. Hence we regard
the incompressible Navier Stokes equations as obtained from the
incompressible Euler equations by adding the viscosity force, which
achieves (or should achieve) the physical consequence of
producing dissipation which in {\it average} balances the work
performed by other forces acting on the fluid.

The idea is that the same effect can be obtained by adding other
forces which imply that the dissipation rate per unit time has a
finite average or is even {\it constant}; and one way to do that
is by imposing the constraint of constant dissipation rate via
Gauss' least effort principle (remark that the constant
dissipation constraint is non holonomic).

The equivalence viewpoint between irreversible and reversible
equations in fluid mechanics is first suggested by the
corresponding equivalence, Sec.(\,\ref{sec:I-5}\,), for the
thermostatted systems and by the early work \Cite{SJ993}\,.

In the latter work it is checked in a special case, where
Navier-Stokes equations, incompressible and at periodic b.c., in
$3$ dimensions are simulated (with $128^3$ harmonics $\uu_\kk$,
also called ``modes'', used to represent the velocity field as
$\uu(\xx)=\sum_\kk \uu_\kk e^{-i\kk\cdot\xx}$ imagined in a
periodic container $[0,2\p]^3$, hence $0\ne\kk\in Z^3$): the
results are compared with corresponding ones on similar equations
with $0$ viscosity but subject to a constraint forcing the energy
content, of the velocity field in a momentum shell $2^{n-1}< \|\V
k\|<2^n$, to follow the Kolmogorov-Obukov $\frac53$-law.  Showing
remarkable agreement.

It has to be stressed that, aside from the reversibility question, the idea
that the Navier Stokes equations can be profitably replaced by equations
that should be equivalent to it is widely, \Cite{Sa006}, and
successfully used in computational approaches (even in engineering
applications); 
\footnote{\tiny ``The action of the subgrid scales on the resolved scales
  is essentially an energetic action, so that the balance of the energy
  transfers alone between the two scale ranges is sufficient to describe
  the action of the subgrid scales'', \Cite{Sa006}[p.104]. And
  about the large eddy simulations: ``Explicit modeling of the desired
  effects, i.e. including them by adding additional terms to the equations:
  the actual subgrid models'', \Cite{Sa006}[p.105].}
 a prominent example are the ``large eddy simulations'' where
 effective viscosities may be introduced (usually generating non
 reversible models, \Cite{GPMC991}) to adjust, empirically, the
 influence of terms neglected in the process of cut-off to
 eliminate the short wavelength modes. But a fundamental approach
 does not seem to have been developed.\index{large eddie
 simulations}
\footnote{\tiny About one of the many important methods: ``There is no
  particular justification for this local use of relations that are on
  average true for the whole, since they only ensure that the energy
  transfers through the cutoff are expressed correctly on the average, and
  not locally, \Cite{Sa006}[p.124].}

The basic difference between the large eddies simulations
approach and the equivalence idea, discussed in this and the
following sections, is that it is not meant as a method to reduce
the number of equations and to correct the reduction by adding
extra terms in the simplified equations; it deals with the full
equation and proposes to establish an equivalence with corresponding
reversible equations. In particular the new equations are not
computationally nor theoretically easier.

For simplicity consider an incompressible fluid in a container
$\CC_0=[0,L]^d$, $d=2,3$, with periodic
boundaries, subject to a non conservative volume force $\V g$:
for $L=2\p$ the fluid can be described by a velocity field $\V
u(\xx)$, $\xx\in [0,2\p]^d$ with zero divergence $\BDpr\cdot\V u=0$,
``Eulerian description''. Hence $\uu$ can be represented via its
Fourier's harmonics, \ie complex $d$-dimensional vectors
$\uu_\kk\in C^d$, $0\ne\kk\in Z^d$, as:
\be\eqalign{
  \V u(\V x)=&\fra1{(2\p)^d}\sum_{0\ne\V k} \uu_\V k e^{-i\V
k\cdot\V x},\quad
\uu_\V k=\int\V u(\V
x)e^{i\V k\cdot\V x}d\V
x,\quad \kk\cdot\uu_\kk=0,\cr
}
\label{e5.4.1}\ee
its energy $\EE(\uu)$ and enstrophy $\DD(\uu)$ can be defined as
$\EE(\uu)=\sum_{0\ne\V k\in Z^d} |\uu_\kk|^2,\\
\DD(\uu)=\sum_{0\ne\V k\in Z^d}
\kk^2|\uu_\kk|^2.\label{enstrophy}$.
It is useful to keep in mind that in dimension $2$ the Fourier's
harmonics $\uu_\kk$ can be taken of the form
$\uu_\kk=i\frac{\kk^\perp}{\|\kk\|} u_\kk$, with $u_\kk=\lis
u_{-\kk}$ complex scalars, and $\kk^\perp=(-k_2,k_1)$ if
$\kk=(k_1,k_2),\, \|\kk\|^2=k_1^2+k_2^2$.

In the periodic boundary conditions case, imposing a time
independent forcing $\V g=\BDpr\wedge \V f$\,\footnote{\tiny \ie
$\V g$ has no gradient part: which could always be included in
the pressure $p$.}\, and a viscosity $\n$ or, alternatively, the Gaussian
constraint $\DD(\uu)=D=const$, leads formally\,\footnote{\tiny Formally
here refers to the fact that in general the equations are not
known to admit unique solutions: the issue is discussed in the
next section.}  to the Navier-Stokes (NS) or to the ``Gaussian
Navier-Stokes'' (RNS)\index{Gaussian Navier-Stokes} equation
which in dimension $d=3$ are:
\be \eqalign{
&\dot{\V u}+\T {\V u}\cdot \T\dpr {\V u}=\a(\V u) \D\V u-\BDpr p+ \V
  g,\qquad \BDpr\cdot\V u=0,\cr
(1)\ \ \  &\a(\V u)\defi \n, \qquad NS\quad {\rm eq.\ or\ respectively} \cr
(2)\ \  &\a(\V u)\defi \frac
{\int\wh\BDpr\V u\cdot\wh\BDpr((\T{\V u}\cdot\T \BDpr)\V u) \, d \xx-\int
  \D\V u\cdot\V g\,d\xx}
{\int (\D\V
  u(x))^2\,d x}\qquad RNS\quad {\rm eq.} \cr}\label{e5.4.2}\ee
where $\V g=\BDpr\wedge \V f$, for a fixed $\V f$, is a time
independent forcing.

In dimension $d=2$ the RNS equation is
simpler, as the cubic term in $\uu$ cancels exactly.

The above formal Gaussian Navier-Stokes or RNS equation, is {\it
  reversible} and time reversal $I$ is simply $I\V u(x)=-\V
u(x)$, which implies that, upon velocity reversal, ``fluid
elements'' retrace their paths with opposite velocity, unlike the
classical, viscous, NS equation\index{Navier-Stokes equation}.

A further different equation is obtained by requiring that
$\EE(\uu)=\int \V u^2 dx=E$ and, for any $d\ge2$, it is:
\be 
  \dot{\V u}+\T {\V u}\cdot \T\dpr {\V u}=\a(\V u) \D\V
  u-\BDpr p+ \V g,\qquad
  \a(\V u)\defi -\frac {\int \V u\cdot\V g\,dx}
    {\int (\T\dpr \V u(x))^2\,d x}
      \label{e5.4.3}\ee
which is interesting, although it {\it does not follow} from Gauss'
principle, unlike the previous RNS in Eq.\equ{e5.4.2}.
Imposing, instead, $\EE(\uu)=E$ via Gauss principle on Euler's
equation would yield the equation:
\be\dot{\V u}+\underline {\V u}\cdot
\underline\dpr {\V u}=\a(\V u) \V u-\BDpr p+ \V g \quad{\rm
with}\quad
\a(\V
u)\defi-\frac{\int \V u\cdot\V g\,dx}{\int {\V u(x)}^2\,d x}\label{e5.4.4}\ee
remarkable also because the $\a(\uu)$ is a ratio between
a$\int \V g(x)\cdot\uu(x)dx$ and the {\it constant of motion}
$E=\EE(\uu)$ imposed by the Gaussian constraint, (energy). Therefore, if
the forcing $\V g$ has only a finite number of non zero
harmonics, then $\a(\uu)$ also depends only on the same finite
number of harmonics.
\*

The Eq\equ{e5.4.2},\equ{e5.4.3},\equ{e5.4.4} fit quite naturally
in the frame of the theory of non equilibrium statistical mechanics even
though the model is not based on particles systems.

This will be made more precise, for  NS and the RNS in
Eq.\equ{e5.4.2}\,, in the following section, where also the
role of the well known open problem about the existence and uniqueness of
solutions to  the  equations,\,\Cite{Fe006}\,, for
$3$-dimensional fluids, {\it will not} be set aside.

\section{Developed turbulence}
\def\SEC{Developed turbulence}
\label{sec:V-5}\iniz
\lhead{\small\ref{sec:V-5}.\ \SEC}

Consider the equations Eq.\equ{e5.4.2} 
supposing that the forcing field $\V g$ has 'large scale', in the
sense that non zero Fourier's harmonics $g_\kk$ are {\it finitely
  many and fixed} throughout: so that the equations contain only
  one free parameter, namely $\n$ for NS, $D$ for RNS. 

The aim is to find conditions under which they might be {\it
  equivalent}.  Of course a first problem is that the equations
are infinite dimensional and it is even unknown, at least in
$\ge3$-dimensions, whether they admit unique and smooth solutions:
so that extending the theory to cover their stationary statistics
is simply out of question. Here this can be partly bypassed
adopting a phenomenological approach focusing on the
$3$-dimensional incompressible case with periodic boundary, on
the torus $T^3=[0,L]^3$, $L=2\p$; the $2$-dimensional case could
be analyzed in the same way with analogous results and, by far,
would be simpler.

Fix $\V g$ with a finite number of non zero Fourier's harmonics:
it will be a ``large scale'' forcing of the equations throughout
this section. The equations will be ``regularized'' by imposing a
UV-cut-off $N$, thus regarding them as equations for fields $\uu$
represented as in Eq.\equ{e5.5.1} with Fourier's components
$\uu_\kk$ {\it vanishing} if $|\kk|\defi\max_j|k_j|>N$ or if
$|\kk|=0$.  Explicitly, restricting all $|\kk|,|\kk_1|,|\kk_2|$
to be $\le N$ and $>0$, the Fourier's transform of the
Eq.\equ{e5.4.2} become, in the respective cases NS and RNS:
\be\eqalign{
  \dot\uu_\kk&=i\sum_{\kk_1+\kk_2=\kk} (\uu_{\kk_1}\cdot \kk_2)
  \uu_{\kk_2}-\a(\uu)\kk^2 \uu_\kk +\V g_\kk,\qquad \kk\cdot\uu_\kk=0\cr
  \a(\uu)&=
  \cases{(NS):\ \ \ \a(\uu)=\n \cr
     (RNS):\ \a(\uu)=\frac{i\sum_{\kk_1+\kk_2=\kk}
     (\uu_{\kk_1}\cdot \kk_2)(\uu_{\kk_2}\cdot
     \lis\uu_{\kk}\kk^2)  +\sum_\kk \V g_\kk\cdot\lis
     \uu_{\kk}\kk^2}{\sum_\kk \kk^4|\uu_\kk|^2}}\cr}
\label{e5.5.1}\ee
The first equation is the regularized NS equattion at constant
viscosity $\n$, and the second is the regularized RNS which keeps
enstrophy $\DD(\uu)=\sum_\kk \kk^2 |u_\kk|^2$ constant and
equal to the value $D$ of the initial state; $\kk^4$ abridges
$(\kk^2)^2$.

The dimension of the phase space $M_N$ is $N_0=(2 N+1)^2-1$ in
dimension $2$ and $N_0=2((2 N+1)^3-1)$ in dimension
$3$.\,\footnote{\tiny Take into account that $k_j$ are integers,
$\V k\cdot\V u_{\V k}=0$, $\V u_0=0$ and $\V u_{\V k}=
\lis{\V u}_{-\V k}$.}

In this way the equations Eq.\equ{e5.4.2} become ODE's with
only one free parameter ($\n$ or $D$ repectively) for which
solutions $\uu(t)$ exist, smooth and unique, and define
respectively dynamical systems $(M_N,S^{\n,N}_t)$ and
$(M_N,S^{D,N}_t)$,
\footnote{\tiny \label{OK41} An example, inspired by the ``OK41
theory'' \index{OK41 theory} of inertial
  turbulence,\Cite{Ga002}[Chapter 7]\,, will be considered below,
  see Sec.(\,\ref{sec:VI-5}\,), in which $\uu_\kk=0$ unless $|\V
  k|>0, |k_j|\le c R^{3/4}\defi N$, with $c$ a constant of $O(1)$
  and $R\defi{\sqrt{|\V g|L^3}}/\n$ (Reynolds number), with $|\V
  g|=\max_j |g_j(x)|$.\label{Reynolds number}\index{Reynolds
  number}}\,, which will be supposed to  verify CH hence to be Anosov flows for
  $t\ge0$ on the attracting surfaces $\AA_j$.
  And call $\m^{j,N}_\n$ and $\wt \m^{j,N}_D$ the SRB
  distributions on the attracting surfaces.

Define the {\it local observables}\index{local observable} $F(\V
u)$ for the velocity field $\V u$ of an incompressible fluid in a
periodic container: they will be functions $F$ depending only
upon finitely many Fourier components of the velocity field $\V
u$ whatever the regularization $N$ is, {\it i.e.}  depending on
the ``large scale'' properties of $\V u$. Therefore $\V g$, as
fixed above, is in particular a local observable,\label{local
observable}

To compare the solutions to the $N$-regularized equations, NS and
RNS, define ``corresponding'' pairs $\m^{N}_\n$ and $\wt\m^{N}_D$
of SRB distributions if:
\*
\0{\bf Definition 0:\label{corresponding fluids}} {\it The
parameters $\n$ (for NS) and $D$ (for RNS), and the 
SRB-statistics $\m^{N}_\n$ and $\wt\m^{N}_D$
will be called ``corresponding'' if:
$\m_\n^N (\DD(\V u))=D$.}
\*
Then a first conjecture is,
\Cite{Ga997b}:\index{equivalence conjecture} \*
\0{\bf Conjecture 0:} {\it Fixed a UV-cut-off $N$ large
enough, suppose that, given $\n,D$, the full phase space $M^N$ is
an attracting surface, for NS and RNS, and the SRB distributions
$\m^{N}_{\n},\wt\m^{N}_D$ are corresponding in the sense of
definition 0, then in the limit $\n\to0$ they have the same
statistical properties, \ie local observables have the same
statistics.} \*

Here by {\it ``same statistics''} as $\n\to0$,
it is meant that the difference
$\media{F}_{\m^{N}_\n}-\media{F}_{{\tilde\m^{,N}}_D}\tende{\n\to0}0$,
for all 'local observables' $F$ (as defined above).

Accepting the conjecture of equivalence in the limit $\n\to0$
implies the identity
$\media{\a}_{\tilde\m_D}=\n$, between the average friction
$\media\a$ in the Gaussian case and the viscosity $\n$ in the
Navier-Stokes case: because multiplying both sides of the
equations by $u_{-\kk}$ and summing over $\kk$, the non linear
term cancels; hence both  $\lis \a D$ and
$\n\lis\DD$ are averages of $\Si_\kk g_{\kk}u_{-\kk}$,
which is a local observable by the locality of $\bf
g$. Therefore, since in corresponding states $D=\lis \DD$ by
definition, the conjecture implies $\lis \a D=\n
\lis\DD$, \ie $\lis\a=\n$; and this shows that corresponding
distributions could equivalently be defined by $\m_D(\a)=\n$.

Denote $\sum^*$ summation over a selected half of the $\kk=(k_1,k_2)$ (\eg
$(k_1>0,k_2=0)$ and $(k_2>0)$, to take into account that
$\uu_\kk=\lis \uu_{-\kk}$); remark that the
divergence of the equation Eq.\equ{e5.5.1}, in dimension $d\ge2$ is, if
$\uu_\kk\equiv\uu_{\kk,r}+i\uu_{\kk,i}$ and 
$\KK_2=2^{d-1}\sum^*_\kk\kk^2=2^4N^4+o(N^4)$:
\be \eqalign{
  \s(\uu) &= \n \,\KK_2
  \sim\,\n\, O(N^{d+2}),\qquad {\rm NS\ in}\ \equ{e5.5.1} \cr
\s(\uu)&=\a(\uu)
\KK_2 +\sum_{\kk,\b=r,i}^*
(\dpr_{\uu_{\kk,\b}}\a(\uu))\kk^2 \uu_{\kk,\b},\quad {\rm RNS\ in}\ (5.4.1) \cr}
  \label{e5.5.2}\ee 
Hence the identity $\n=\lis\a$ in corresponding distributions
does not imply, not even in dimension $2$, identity of the
average divergences because of the contribution to $\s(\uu)$ of
the second sum in Eq.\equ{e5.5.2}\,, see Eq.\equ{e5.6.7}\,.

The above conjecture can be extended (keeping in mind the above
definition of correponding states) in two ways. A more
interesting one is, \Cite{Ga020b}:
\*
\0{\bf Conjecture 1:} {\it Let $\m^N_\n$ and $\wt \m^N_D$
  be corresponding stationary distributions
  for the dynamical systems in Eq.\equ{e5.5.1} regularized with
  UV-cut-off $N$. Then:
\be \lim_{N\to\infty} (\m^N_\n(F)-\wt
\m^N_D(F))=0\label{e5.5.3}\ee
for all local observables $F$.}
\*

The deep difference with respect to the conjecture 0 is that there is
no condition $\n\to0$ (or $R\to\infty$), while  $N\to\infty$
instead.
\*

\0{\bf Remark:} The above conjectures cannot be applied when the
stationary states of NS or RNS verify the CH but have attracting
surfaces $\AA_j$ or $\wt\AA_{j'}$ different form
$M^N$. Nevertheless the conjectures can be naturally extended to
such more general cases. Conjecture 0 (or 1) can be extended by saying
that the number of corresponding SRB distributions $\m^{j,N}_\n$
for NS and $\wt\m^{j',N}_D$ for RNS is the same if $\n$ is small
enough (or if $N$ is large enough) and for each $j$ there is a $j'$ such that the
distributions $\m^{j,N}_\n$ and $\wt\m^{j',N}_D$ have the same
statistics in the limit $\n\to0$ at $N$ fixed (or in the limit
$N\to\infty$ at $\n$ fixed). The extensions are particularly
relevant as the existence of several attracting surfaces arises
in the intermittency phenomena (discussed in Sec.(\,\ref{sec:VI-5}\,).
\*

The conjecture 1 is very strong, because it can be shown to imply
that {\it if} for $N$ large the value $\a(\uu(t))$ remains $>\e$,
for some $\e>0$, as it appears in several simulations, in $2$ and
$3$ dimensions, the velocity field $\uu$ of the RNS equation
would remain {\it uniformly smooth} as $N\to\infty$,
\Cite{Co007}\Cite{MBCGL022}.

Furthermore in the simulation in \Cite{MBCGL022} for $N$ large
enough the values of $\a(\uu)$, apparently, fluctuate only little
around $\n$: therefore, if conjecture 1 were verified and
$\a(\uu)$ stayed close to $\n$, the problem of finding a
constructive approach to NS would be essentially solved: the use
of the regularized RNS equation could simply replace that of the
standard NS with equal UV-cut-off, without worrying at finding a
solution in the limit $N\to\infty$. The issue of using the
viscous unregularized NS equation would not even arise (from a
Physics viewpoint).

The latter comment suggests to test in simulations whether, even
when $N$ is very large, $\a(\uu)$ assumes a $<0$ value (at least
rarely, so that the regularity argument could not be applied). In
\Cite{MBCGL022} negative values of $\a(\uu)$ do not appear:
{\it however} the simulation reports, at times {\it multiples} of
$h^{-1}$ of the integration step $h$, the average of $\a(\uu(t))$
in the same time interval.

The step $h$, in the quoted reference, is often $h\sim 10^{-5}$
and $\a(\uu(t))$ should be tested for positivity at much smaller
time steps than $h^{-1}$ (and possibly using smaller $\n$ or
higher precision,
\ie smaller $h$). Few tests provide, so far, estimates of how often a
negative value of $\a$ can be expected and the result is that as
$N$ grows the probability of $\a<0$ becomes unappreciable. At the
moment I still think that either large deviations with
$\a(\uu)<0$ are possible or the conjecture is not valid: by the
{\it principle of difficulties conservation} I consider unlikely
unconditional positivity, in RNS evolutions, of $\a$ at $N$
large, if $\n$ is small: more precisely, at given forcing, if
$\n$ is smaller than the smallest viscosity at which the NS
equation can still be proved to admit a smooth solution for all
$t>0$).\index{conservation of difficulties}

To intepret the results in \Cite{MBCGL022} a third conjecture,
much weaker than conjecture 1, was formulated for a fluid in
$[0,L]^3$ with periodic boundary in a model with cut-off at the
'Kolmogorov scale',\index{Kolmogorov scale} defined as the scale
$\k_R=\frac{2\p}L R_L^{\frac34}$ in terms of the {\it Reynolds number}
$R_L=(|\V g|L^3)^{\frac12}\n^{-1}$:
\index{Reynolds number}
\*
\0{\bf Conjecture 2:} {\it Let $F(\uu)$ be a local observable
  depending on the harnonics $\uu_\kk$ with $|\kk|< c \k_{R_L}$.  Then
  there is $c=O(1)$ such that if $\m^{N}_\n$ and
  $\wt\m^{N}_D$ are corresponding (as defined above), it is:
\be \lim_{N\to\infty} (\wt\m^N_D(F)-\m^N_\n(F))=0\Eq{e5.5.4}\ee
}
\*

The conjecture can be extended as done  for the conjectures 0,1
in the above remark.
The difference between the conjectures 1 and 2 is that, in the
latter, locality of the observable $F$ is not sufficient but it
has to be strengthened by the requirement that $F$ must depend
only on $\uu_\kk$ with $|\kk|<c \k_R$, otherwise Eq.\equ{e5.5.4}
may fail: a case which is apparently encountered in simulations
\Cite{MBCGL022}. However in my view more evidence should still be
gathered before dismissing conjecture 1.

The identity $\lis\a=\n$ holds also after conjectures 1 and 2 for
the same reason discussed after conjecture 0.

Finally the above conjectures can be adapted to the
$2$-dimensional cases with the natural modifications and
simplifications (\eg the Kolmogorov scale should be
$\k_{R_L}=\frac{2\p}L R^{\frac12}_L$ and the multipliers $\a$ are
simpler). Furthermore they make sense, even in $3$ dimensions,
also for the Eq.\equ{e5.4.3} where the Gaussian constraint
fixes the energy rather than the enstrophy, \Cite{SDNKT018}.

The equivalence ideas can be adapted also to other equations like
the Burgers equations with various forms of friction or of
reversible Gaussian constraints, \Cite{SEMGS021}.  And to many
other systems familiar in continous mechanics including Burgers
equation \Cite{Ku024}, shell
models \Cite{BCDGL018} or their versions with $\uu_\kk\ne0$ on
'sparse' values of $\kk$ (called NS on log-lattices
,\Cite{PBCCMD023}). Or the ideas can be applied to more general
dynamical systems (see Sec.(\,\ref{sec:VII-5}\,) or an example in
\Cite{GL014}\,).

In the next section we discuss the possibility of drawing some
observable conclusions from the Gaussian Navier-Stokes equations
hence, possibly by the above equivalence conjectures, on the ordinary
Navier-Stokes equations.

\section{Intermittency, phase transitions}\index{intermittency}
\def\SEC{Intermittency, phase transitions}\index{intermittency}
\label{sec:VI-5}\iniz
\lhead{\small\ref{sec:VI-5}.\ \SEC}

By {\it intermittency}\index{intermittency} here is meant,
\Cite{Ga002b}, an event that is realized rarely and randomly:
rarity can be in time, ``time intermittency'',\label{time
intermittency} in the sense that the interval of time in which
the event takes place has a small frequency among time intervals
$\d t$ of equal length into which we divide the time axis; it can
be in space, ``spatial intermittency'',\label{spatial
intermittency} if the event is rarely and randomly verified
inside cubes $\d x$ of a lattice into which we imagine to divide
the ambient space, \eg $R^d$, ($d=2,3$). Rarity can be also in
space–time, ``space–time intermittency'',\label{space–time
intermittency} if the event is rarely and randomly verified
inside regions $\d t\times \d x$ forming a lattice into which we
imagine partitioned the space time, for imstance $R^+ \times R^3$ or, in the
case of discrete evolutions, $Z^+ \times
R^d$.\index{intermittency in time}\index{intermittency in space}

\* Now address from a {\it heuristic viewpoint} the question: ``is this
intermittency observable''? is its rate, in time or space,
measurable? \Cite{Ga006d}. This will be analyzed on a fluid model.

Imagine that the fluid is in a large container $\CC_0$ periodic
(or with smooth boundaries) of linear size $L$: like for instance
a fluid evolving via Eq.\equ{e5.4.2}, irreversible or reversible, or
Eq.\equ{e5.4.3}. Suppose validity of the CH for the equations in
the model obtained by fixing the UV cut-off $N$ in both viscous
and reversible equations to a value suggested for the
irreversible NS equation by a version of the OK41
theory\,, \Cite{Ga002}[Sec.6.2]\,, see below.

Recall the basic Kolmogorov's assumption of the OK41-theory, on a
NS fluid flow in a periodic container $[0,L]^3$: the average
dissipation rate and average velocity variation $v_\ell$ on scale
$\ell$ are related by $\frac{v_\ell^3}{\ell}=constant$. This
allows us to define the Reynolds number $R_\ell$ on scale $\ell$
in terms of the Reynolds number $R_L=(|\V
g|L^3)^{\frac12}\n^{-1}$,\,\footnote{\tiny Taking into account
that the NS equation in the considered geometry depend only the
forcing, $\V g$ on large scale, the container size and the
viscosity $R_L$ is the only dimensionless parameter inversely
proportional to $\n$.}\,, and to express $R_\ell=R_L
(\frac{\ell}L)^{\frac43}$, so that the length scale where
$R_\ell=1$ is $\ell_K\defi\frac{L}{2\p}R^{-\frac34}_L$.

The assumption suggests considering in the following the
NS and RNS models in Eq.\equ{e5.5.1} regularized by setting a
cut-off at $k_\n=c\frac{2\p}{L} R^{\frac34}_L$, with
$c=O(1)$, \ie setting it to the scale $\ell$ where $R_\ell\sim
1$, which implies $N=c R^{\frac34}_L$.

The models will be called
$NS_{41}$ and $RNS_{41}$ respectively (the constant $c$ is fixed
on a case by case basis, as convenient).

The $\uu_\kk$ with $|\kk|>k_\n$, the ``subgrid components'' of
$\uu$, are set $0$ and, rather than imagining that they are
'independently' described by other equations, it is supposed that
either the viscosity or the Gaussian thermostat are responsible
for the dissipation and for both equations the attracting surface
is the full phase space $M^N$.\footnote{\tiny Including subgrid
components $u_\kk$ would also imply as unreasonable (due to the
different role of viscosity on the motion of the subgrid
components) to assume $M^N=\AA$: which remains possible if $N$ is
less than the maximal $c R^{3/4}_L$ (however in the next section
cases with the attracting surfaces $\AA\ne M^N$ will be
considered).}

The entropy creation rate $\s(\uu)$ is identified with the
divergence of the equations and the parameters $\n,D$ of the two
equations determining the SRB distributions, denoted
$\m^N_\n,\wt\m^N_D$, will be considered to be {\it
corresponding}, see definition 0 in
Sec.(\,\ref{sec:V-5}\,), \ie $\m^N_\n(\DD)=D$.

An essential difference arises between the 2D and the 3D
equations: in the NS equations, hence in $NS_{41}$, the phase
space contraction is exactly constant, being defined simply as
the divergence of the equations.\footnote{\tiny the quadratic
term in the NS equation contributes $0$ to the divergence, unlike
the RNS case.}  Therefore the fluctuation theorem (FT) cannot
hold for NS\,\footnote{\tiny Unless the attracting surface $\AA$
is smaller than the phase space, which is not such by the
assumption in the model definition.}\,: there are no
fluctuations of the NS divergence in $M^N$, while in FT the
divergence measures the fluctuations of phase space contraction on the attracting
surfaces $\AA$, as remarked in Sec.(\,\ref{sec:II-2}\,),
see \footref{surface contraction}\,.

In $RNS_{41}$ the above difficulty does not arise and the
divergence receives contributions from the gradient of $\a(\uu)$
and the assumption that the system $(M^N,S^N_t)$ is Anosov
system dispenses us to worry about the difference of divergence on
the attracting surface and on the full phase space.

Actually in general RNS equations even if an attracting surface
$\AA$ is smaller than the full phase space the area contaction
$\s_\AA(\uu)$ would be the same as the divergence on the full
phase space because the enstrophy is a constant of motion, see
remark p.\pageref{surface contraction} following Eq.\,\equ{e2.8.1}\,.

Since the sequel relies on FT, the analysis will be restricted to
the $RNS_{41}$ in dimension $d=2,3$, see comment at the end of
the section about $NS_{41}$ in $d=3$.

Then the average entropy production $\s_+$ and a large deviations
function $\z(p)$ will be defined for the reversible equation
$RNS_{41}$, and control the fluctuations of entropy
production\index{entropy production} via the FT
symmetry, Eq.\equ{e4.6.1}.

The $\s_+$ and $\z(p)$ will depend on the size of the system,
\ie with the number of degrees of freedom, so that there should be
serious doubts about the observability of so rare fluctuations.

However if we look at a small subsystem in a little volume $V_0$
of linear size $L_0$ we can try, heuristically, to regard it
again as a fluid enclosed in a box $V_0$ described by the same
reversible Gaussian Navier-Stokes equations. For some discussion
of the physical conditions see Sec.(\,\ref{sec:X-4}\,).

We imagine, therefore, that this small system also verifies a
fluctuation relation. Then, defining $\KK_{L_0}\defi
2^{d-1}\sum_{|\V k|\le c R_{L_0}^{3/4}} \V k^2$, the fluctuating
viscosity term contributes to the average phase space
contraction:
\be 
  \lis\s_{V_0}= \KK_{L_0}\lis\a_{L_0}(\V u)+\ldots,\qquad
\label{e5.6.1}\ee
where $R_{L_0}$ is the Reynolds number on scale $L_0$ (\ie
$R_{L_0}=(L_0/L)^{4/3} R_L$)\,, and the $\ldots$ refer to the
contributions of the second sum in
Eq.\equ{e5.5.2} (see Eq.\,\equ{e5.6.7}\,).

Then the averages of $\s_V$ and $\s_{V_0}$ in the reversible
equation, averaged over a time span $\t$, and the rates of their
fluctuations are expected to be controlled for $V$ large by
functions proportional to volume independent quantities:
\be\eqalign{ \z_V(p)=&\,\lis\z(p)\,\KK_L(R), \qquad \z_{V_0}(p)=
  \lis\z(p)\,\KK_{L_0}(R),\cr \media{\s_V}=&\lis\s_+\,\KK_{L}(R),
  \kern2.2mm\qquad \media{\s_{V_0}}_+=\,\lis\s_+
  \,\KK_{L_0}(R_{L_0})\cr}\label{e5.6.2}\ee
Hence, if we consider observables dependent on what happens
inside $V_0$, if $L_0$ is small so that $\KK_{L_0}(R)$ is not too
large, and if we observe them in time intervals of size $\t$,
then the time frequency during which we can observe a deviation
``of size'' $(1-p)\lis\s_+ \KK_{L_0}(R_{L_0})$ of $p$, \ie of
entropy production, will be small of the order of:
\be e^{(\lis\z(p)-\lis\z(1))\,\t\,\lis\s_+
\,\KK_{L_0}(R_{L_0})}\label{e5.6.3}\ee
for $\t$ large, where the local fluctuation rate $\lis\z(p)$
verifies, assuming the CH for the reversible equation:
\be \lis\z(-p)=\lis\z(p)-\lis\s_+\,p\label{e5.6.4}\ee
Therefore by observing the frequency of intermittency one can gain
some access to the function $\lis\z(p)$.

Note that one {\it will necessarily observe a given fluctuation
  somewhere} in the fluid if $L_0$ is taken small enough and the
size $L$ of the container large enough.

Furthermore the above ``entropy driven intermittency'' takes
place not only in time but also in space. Thus we shall observe,
inside a box of size $L_0$ ``somewhere'' in the total volume $V$
of the system and in a time interval $\t$ ``sometime'' in a large
observation time $T$, a fluctuation of entropy production of size
$1-p$ with high probability if
\be (T/\tau) (L/L_0)^3 e^{(\lis\z(p)-\lis\z(1))\,\t\,K_{L_0}(R)}\simeq
1\label{e5.6.5}\ee
and the special event $p=-1$ will occur 'with high probability' if
\be (L/L_0)^3 e^{-\lis\s_+\,\t\,K_{L_0}(R)}\simeq 1\label{e5.6.6}\ee
by Eq.\equ{e5.6.5}. Once this event is realized the fluctuation
patterns will have relative probabilities as described in the
fluctuations pattern theorem, Eq.\equ{e4.7.2}, \Cite{Ga002b}.
\*

Hence the intermittency described above is an example of {\it
  space-time intermittency}. Another source of intermittency
  arises when the system $(M,S^N_t)$ or $(\X,S^N)$ satisfies the
  CH but is not transitive. In this case, that has been set aside
  so far, there may be more than one attracting surface:
  $\AA_1,\AA_2,\ldots,\AA_m$. Then there will be several SRB
  distributions, one per each $\AA_i$, and the asymptotic
  behavior will depend on where (\ie close to which $\AA_i$) the
  initial data are located.

The phenomena that develop are closely similar to those that are
met in studying phase transitions. The domains $U_i$ of
attraction of the $\AA_i$ have boundaries in common, because of
the CH, but the$\AA_i$ have positive mutual distance. Still as
the UV cut-off $N$ grows the surfaces $\AA_i$ may become
separated by tiny regions that can be easily bridged, in
simulations, because of round-off errors with the consequent
apparently erratic behavior, particularly because new attractive
surfaces, as $N$ grows or as the control parameters vary, may
arise or old ones disappear. For other possibilities
leading to such erratic behaviors see \Cite{PM980}.\index{phase
transitions}\label{phase transitions}

It is therefore possible to think that the latter intermittence,
in NS and RNS models, is due to near tangency of the surfaces
$\AA_i$ (getting closer, while remaining separate, as
$N\to\infty$). The previous intermittency picture shows that very
different fluctuations (or different entropy generation) can take
place 'sometime or somewhere' in a very large fluid while the
second picture shows that, even in a small fluid, jumps between
different fluctuations can take place yielding the ``erratic''
flows called, usually, ``intermittent''.
\*

\0{\it Comment:}
Consider NS and RNS models on $M^N$ with $N$
harmonics and attracting surface $M^N$. Let $\n,D$ be the
equations parameters and suppose that they are corresponding,
in the sense of Definition 0 Sec.(\,\ref{sec:V-5}\,).
\\
The equivalence conjectures of Sec.(\,\ref{sec:V-5}\,) suggest to
consider the divergence in RNS, see Eq.\equ{e5.5.2} or explictly:
\be\s(\uu)=\a(\uu) \Big(\KK_N-2\frac{\sum_\kk \kk^6
|\uu_\kk|^2}{\sum_\kk \kk^4 |\uu_\kk|^2}\Big)-\frac{\sum_\kk
\kk^4 f_{-\kk} u_\kk}{\sum_\kk \kk^4 |\uu_\kk|^2}\label{e5.6.7}\ee
as a {\it non local} observable for $NS_{41}$.

A natural question arises: since the conjectures propose
equivalence between corresponding RNS and NS, then taking into
account that, by CH, in RNS the observable $\s(\uu)$ has large
deviations obeing FT, does $\s(\uu)$ have, in spite of not being
local, in NS large deviations obeying FT and with the same rate
$\z(p)$ ?

The same question can be posed about the Lyapunov spectra of $NS$
and $RNS$, also typical {\it non local} observables: are the
spectra of corresponding SRB distributions equal or at least
related ?

It is suprising that, at least in a very simple 2D case (with $N$
small so that $|k_i|_{max}|=3$, \ie $48$ degrees of freedom) a
simulation offers a positive answer, \Cite{Ga020}, to {\it both
questions} and shows that equivalence between the equations might
be extendible to special not local observables, see
Appendix \ref{appK}\,.

\section{Stochastic evolutions}
\def\SEC{Stochastic evolutions}
\label{sec:VII-5}\iniz
\lhead{\small\ref{sec:VII-5}.\ \SEC}

Time reversal symmetry (TRS) plays an essential role in the
fluctuations theory. Therefore the question whether a kind of
fluctuation theorem could hold even in cases in which the
symmetry is absent has been studied in several works with
particular attention devoted to problems in which stochastic
forces act.

The first ideas and results appear in \Cite{Ku998}, followed by
\Cite{LS999}\Cite{Ma999}. The natural approach is  to consider stochastic
models as special cases of deterministic ones: taking the
viewpoint that noise can (and {\it should}) be thought as
generated by a chaotic evolution (as it is done in several
simulations where it is generated by a random number generator
which is a program, \eg see
\Cite{KR988}, that simulates a chaotic evolution).

The latter approach is more recent and has also given results,
\Cite{BGG997}\Cite{BGG007}\Cite{BK013}, showing that extensions of the
fluctuation theorem\index{fluctuation theorem} can be derived in
special examples, which although stochastic nevertheless can be
mapped into a reversible deterministic dynamical system, by
including among the phase space variables the coordinates
describing the noise generator system.

However the path followed in the literature has mostly been along
different lines although, of course, it has provided important
insights even allowing the treatment of problems in which a
phenomenological constant, \eg friction, unavoidably destroys
time reversal symmetry.

To illustrate a simple case in which stochastic forces are
explicitly present, and nevertheless can be studied with the
methods of the previous sections, consider the system
$(M,S^\g_t)$ defined by the equation:
\be \ddot{x}_i+\g \dot x_i+\dpr_{x_i} U(\V x)-f_i=q_{i}(t)\label{e5.7.1}\ee
%
where $i=1,2,\ldots,N$ and  $x\in [0,2\p]^N$, $f_i(x)$ are driving forces, $\g$ is a
friction coefficient and $q_i(t)$ is a stochastic noise acting on
$x\in M$.

The noise $\qq$ is imagined to be generated by a dynamical
system, Hamiltonian for definiteness, in which the variables
$(\pp_i,\qq_{i})\in X$, with $\dot\qq_i=\pp_i$ follow independent
chaotic motions, Hamiltonian for definitness, on a $2d$
dimensional manifold $X$, ($d\ge2$), under the action of a force
$F(\qq_i)$, and finally $q_{i}$ is, for instance, one component
of $\qq_i$ (or a function of it). The phase space $M$ will be
$[0,2\p]^N\times X^N$ and the pairs $(\pp_i,\qq_i)\in X$ are
pairs of conjugate variables (each of dimension $n$, $n\ge2$)\,
\footnote{\tiny For instance the
$(\pp_i,\qq_i)$, could be, for each $i$, the canonical
coordinates of a geodesic flow on a $2$-dimensional compact manifold of
constant negative curvature, \Cite{CEG984}\,: a classical chaotic
system. Or more generally $q_i$ can be the output of a noise generator.}.

The systems describing $(\pp_i,\qq_i)$ can be considered a model for a family of
random number generators. In this section the variables $x_i$ are angles, so
that the above system is a family of coupled pendulums subject to friction,
with a {\it phenomenological friction coefficient} $\g$, and stirred by the
torques $f_i$ and by a random force.

Eq.\equ{e5.7.1} is not time reversible
because of the friction $\g$:
however the equivalence conjectures of Sec.(\,\ref{sec:I-5}\,) suggest to
consider the model $(M,S^\a_t)$ defined by the equation:
\be \eqalign{
&\ddot{x}_i+\a(\V x,\dot {\V x},\pp,\qq) \dot x_i+\dpr_{x_i}
U(\xx)-f_i=q_{i,1}
\cr  &\a(\V x,\dot{\V x},\pp,\qq)=\frac{\V
f\cdot\dot{\V x}+\sum_i q_{i,1}\dot{x_i}-\dot U}{N T}\cr
&\dot \qq_i= \pp_i,\qquad \dot\pp_i=F(\qq_i)\cr}
\label{e5.7.2}\ee
which has $\frac12 m\dot{\V x}^2=\frac12N T$ as an exact constant
of motion, if the initial data are on the surface $\frac12
m\dot{\V x}^2=\frac12NT$. It will be supposed that 
the system $(M,S^\a_t)$ is an Anosov system.

The equivalence conjectures considered in several cases in the previous
sections indicate that in this case for small $\g$ it could be
$\media{\a}_{SRB}=\g$, if the initial $T$ for Eq.(5.6.2) is the average
value of the kinetic energy for Eq.\equ{e5.7.1}, and the corresponding
stationary states should be equivalent.

The model in Eq.\equ{e5.7.2} is reversible and $I(\V x,\V v,\V
q,\V p) =(\V x,-\V v,\V q,-\V p)$ is a time reversal symmetry:
\be S_t I=I S_{-t}\label{e5.7.3}\ee
The SRB distribution can be defined as the statistics of the data
chosen with distribution:
\be \m(d\V x\,d \dot{\V x}\,d\pp,d\qq)= \r(\V x,
\dot{\V x},\pp,\qq) \, \d(N\dot{\V x}^2-T) d\V x\,d \dot{\V x}\,d\pp\,d\qq
\label{e5.7.4}\ee
where $\r$ is an arbitrary (regular) function.

The CH is naturally extended to such systems and the $\m_{SRB}$
describes the statistics of almost all initial data chosen with
the distribution in Eq.\equ{e5.7.4}. And since
reversibility\index{reversibility} holds, assuming that
$(M,S^\a_t)$ is an Anosov system, the phase space contraction\,
\footnote{\tiny Remark that in this case no contraction occurs in
the $\pp,\qq$ space because the evolution there is Hamiltonian.}:
\be \s(\V x, \dot{\V x},\pp,\qq)= (N-1)\a(\V x, \dot{\V
x},\pp,\qq)
\label{e5.7.5}\ee
has in general a positive time average $\s_+= N\g$ and if
$\s_+>0$ its finite time average,
\be p=\frac1\t \int_0^\t dt \frac{\s(S_t(\V x, \dot{\V
x},\pp,\qq))}{\s_+}
\label{e5.7.6}\ee
obeys a fluctuation relation with large deviations rate $\z(p)$:

\be \z(-p)=\z(p)-p\s_+\label{e5.7.7}\ee
for $p$ in a suitable interval $(-p^*,p^*)$ containing $[-1,1]$.

The $\a(\V x, \dot{\V x},\pp,\qq)$ is an observable which makes
sense also for the irreversible system in Eq.\equ{e5.7.1}\,:
hence, on the basis of conjectures analogues to the ones of the
previous sections and supposing that the SRB average $\lis\a$ of
$\a$ in Eq.\equ{e5.7.2} equals $\g$ the question can be raised:
are the SRB states in some sense``equivalent'' in the limit $\g\to0$ (in
analogy with conjecture 0 in Sec.(\,\ref{sec:V-5}\,) ? For instance
whether the statistics of observables $O(\V x)$ which depend only on a
$N$-independent finite number of coordinates $x_i$ have the same
statistics ?

It is natural, therefore, to look in the evolutions developing under
the irreversible Eq.\equ{e5.7.1}
for fluctuations of the phase
space contraction, and its large deviations 
law: is the law the same as Eq.\equ{e5.7.7} ? If so this would suggest
that there might be cases in which he fluctuation relation can be
seen even in irreversible systems.

An example in which the above arguments are developed in some
detail is in \Cite{BGG007} and the similar question can be raised
in other irreversible systems see \Cite{GL014}.  More tests would
therefore be important.

The cases of Eq.\equ{e5.7.1} under the assumption that the
  potential energy of the driving forces $\V f(\xx)$ is bounded
  and due to short range potentials, the
  $x_i$ are not angles but real variables, and $q_i(t)$ is, for
  each $i$, a
  white noise have been treated
  in \Cite{Ku998}\Cite{LS999}.  They provide remarkable examples
  in which the fluctuation relation is obeyed and can be proved
  without using the CH. And in
\Cite{LS999}\Cite{Ma999} a general theory of the fluctuation relation
is developed (far beyond the idea of the ``Ising model analogy'',
\Cite{BGG997}[Sec.3]).

The above discussion makes clear, once more, that there is little
difference between the stochastic cases and the deterministic
ones. It can be said that the theory of Markov partitions and
coarse graining turns deterministic systems into stochastic ones
and, viceversa, the equivalence conjectures of
Sec.(\,\ref{sec:I-5},\ref{sec:V-5}\,) do the converse. The
Eq.\equ{e5.7.1} is a paradigmatic case in which a stochastic
system might appear to be equivalent (as far as the entropy
production fluctuations are concerned) to a deterministic
reversible one.

An interesting case is Eq.\equ{e5.7.1} in which $i=1$, $x$
is an angle and the noise is a white noise: \ie a forced pendulum
subject to white noise and torque, see \Cite{Ia018}\Cite{IOS019} for
the analysis of the stationary state.

\section{Very large\index{very large fluctuations} fluctuations}
\def\SEC{Very large\index{very large fluctuations} fluctuations}
\label{sec:VIII-5}\iniz
\lhead{\small\ref{sec:VIII-5}.\ \SEC}

The importance of the boundedness assumption on the potential energy $U$ in
the theory of the fluctuation relation has been stressed in
\Cite{BGGZ005}. In this section an interesting example in which an
unbounded potential acts and a kind of fluctuation relation holds is
analyzed, \Cite{CV003a}\Cite{CV003}, to exhibit the problems that may arise and
gave rise to \Cite{BGGZ005}.

The system is a particle trapped in a ``harmonic potential'' and subject to
random forcing due to a Brownian interaction with a background modeled by a
white noise $\z(t)$ and ``overdamped'', \ie described by a Langevin
equation:
\be \dot x=-(x-vt) +\z(t),\qquad x\in R\label{e5.8.1}\ee
in dimensionless units, with $\media{\z(t)\z(t')}=\d(t-t')$. Here $v$
is a constant ``drag velocity'' and the model represents a particle in
a harmonic well dragged by an external force at constant speed $v$: a
situation that can be experimentally realized as described in
\Cite{JGC007}.

The driving force that is exercised by external forces, balancing the
reaction due to the climbing of the harmonic well, is $x-vt$ and the
energy $U$ is $\frac12(x-vt)^2$. The work done by the external force
and the harmonic energy variation during the time interval $(0,t)$ are
therefore:
\be W=\ig_0^t v\cdot(x(\t')-vt') dt',\qquad
D=\frac12((x(t)-vt)^2-\frac12x(0)^2,
\label{e5.8.2}\ee
and the quantity $Q=W-D$ is the work that the system performs on
the ``outside''.

The model is quite different from the ones considered so far. It
was introduced in \Cite{CV003a} where, although the system has no
time reversal symmetry, a form of fluctuation relation has been
proposed for the dimensionless entropy production rate
$p=\frac1\t\int_0^\t\frac{ Q(t)}{\media{Q}}dt$. This model is
extremely simple due to the linearity of the force and to the
Gaussian noise so that the fluctuations of $p$ have a rate that
can be quite explicitly evaluated.

The interest of the model is that $U$ is unbounded: and the
finite time average $\frac1\t\int_0^\t Q(t)dt$ differs from that
of $W$, given by $\frac1\t\int_0^\t W(t) dt$, by a ``boundary
term'', namely $\fra1\t D $. If $D$ were bounded this would have
no effect on the fluctuation rate $\z(p)$: but since $D$ is not
bounded care is needed. An accurate analysis, possible because
the model can be exactly solved, \Cite{CV003a}, shows that
$\z(p)$ only satisfies the fluctuation relation $\z(-p)=\z(p)-p$
for $|p|$ in a range smaller than expected.

The remark has consequences in a much more general context
including the thermostatted models of Sec.(\,\ref{sec:II-2}\,) as it
appeared also in \Cite{ESR003}. The problem was ascribed
correctly to the large values contributed to the phase space
contraction by the part $D$ depending on the metric used in phase
space : and provides, as discussed in Sec.(\,\ref{sec:VIII-2}\,) remark
(d), the difference between entropy production $\e$ and phase
space contraction $\s$; a difference that can become very large,
even in stationary distributions, when phase space is unbounded.

The difference is irrelevant in evaluating the averages of $\s$
or $\e$, Eq.\equ{e2.8.6}, only if, as in the models of
Ch.\ref{Ch2}, phase space is bounded.

A similar problem arises even in the more general context of
Sec.(\,\ref{sec:II-2}\,): there the interparticle potentials may be of
Lennard-Jones type but the surfaces $M$ on which the stationary
states are studied are bounded, by virtue of the constraints;
nevertheless the particles can come very close to the
singularities and the divergence can become very large. The
divergence has the form $\s=\sum\frac{Q+\dot U}{3k_B T}$, see
example (d) Sec.(\,\ref{sec:VIII-2}\,), where $U$ can become very
large and cause, in simulations, important errors.

A detailed analysis in the more general context of
Sec.(\,\ref{sec:II-2}\,) is presented in \Cite{BGGZ005}: there a simple
solution is given to the ``problem'' of apparent violation of the
fluctuation relation in important cases in which the extra terms
$\dot U$ forces can be very large.

Rather than in continuous time, just study the motion via observations
timed to events $x$ in which $U(x)$ is below a prefixed bound. This means
studying the motion via a Poincar\'e's section which is not too close to
configurations $x$ where $U(x)$ is too large. In this case the contribution
to the phase space contraction due to ``boundary terms'' like $\fra1\t
(U(x(\t)-U(x(0))$ vanishes as $\t\to\infty$ (and in a controlled way) and a
fluctuation relation can be expected if the other conditions are met (\ie
chaoticity and reversibility).

More detailed analysis of the problem of the fluctuation relation
in cases in which unbounded forces can act is
in \Cite{Za007}. There a general theory of the influence of the
singularities in the equations of motion is presented: the most
remarkable phenomena are that the fluctuation relation should be
expected to hold, but only for a limited range $p\in (-p^*,p^*)$,
less than the maximal observable with positive probability. And
beyond it observable deviations occur (unlike the case in which
the fluctuation relation holds as a theorem, \ie for Anosov
systems); and it becomes even possible that the function $\z(p)$
can become non convex: this is a property that appears in various
attempts at testing fluctuation relations.

Finally it can be remarked that Lennard-Jones\index{Lennard-Jones
forces} forces are an idealization and in nature the true
singularities (if at all present) can be very difficult to
see. This is also true in simulations: no matter which precision
is chosen in the program (usually not very high) there will be a
cut off to the values of any observable, in particular of the
maximum value of the potential energy. This means that, if really
long observation times could be accessed, eventually the boundary
terms would become negligible (in experiments, because Nature
forbids singularities or, in simulations, because computers have
not enough digits).

Hence eventually the problem is not a real one: {\it but} time
scales far beyond interest could be needed to realize
this. Therefore the theory based on timed observations not only
might be more satisfactory, but it might also, naturally, deal
with properties that are closer to possible
observations, \Cite{Za007}.

\section{Thermometry}
\def\SEC{Thermometry}
\label{sec:IX-5}\iniz
\lhead{\small\ref{sec:IX-5}.\ \SEC}

\0The proposal that the model in Secs.(\,\ref{sec:XI-4},\ref{sec:XII-4}\,) can
represent correctly a thermostatted quantum system is based on the image
of a thermostat as a classical object in which details like the
internal interaction are not relevant. Thus the proposal has
avoids the problems that arise because the definition of
temperature is not really obvious, particularly at low temperatures or on
nano-scale systems, when quantum phenomena become relevant (furthermore in
quantum systems the identity between average kinetic energy and absolute
temperature ceases to hold, \Cite{Ga000}[Chapter 2]). 
Also a basic question is the very notion of temperature in non equilibrium
systems.

A new idea on the subject is in an earlier proposal for
using fluctuation measurements to define temperature in spin
glasses, \Cite{CKP997}\Cite{CR003}[p.216].

If the models can be considered valid at least until it makes sense to
measure temperatures via gas thermometers, \ie optimistically down to the
$\sim 3 ^oK$ scale, but certainly at room temperature or somewhat higher,
then the CH can probably be tested with present day
technology with suitable thermometric devices.

If verified it could be used to develop a ``fluctuation thermometer'',
\index{fluctuation thermometer} to perform temperature measurements below
$~3\, ^o\kern-1mm K$, which is {\it device independent} in the
same sense in which the gas thermometers are device independent
({\it i.e.} do not require, in principle, ``calibration'' of a
scale by ``comparison'' procedures).

To fix ideas a recent device, ``active scanning thermal microscopy''
\Cite{NS002}, to measure temperature of a test system can be used for
illustration purposes. The device was developed to measure temperature in a
region of $100\,nm$, linear size, of the surface of the test system
supposed in a stationary state (on a time scale $>10 \,ms$).
\index{scanning thermal microscopy}

Consider a sample in a stationary equilibrium state, and put it in contact
with a bowing arm (``cantilever'', see figure below): monitor, via a
differential thermocouple, the temperature at the arm extremes and signal
the differences to a ``square root device'', \Cite{Sm972}, which drives
another device that can inject or take out heat from the arm and keep, by
feedback, the temperature differences in the arm $\D T=0$. The arm
temperature is then measured by conventional methods (again through a
thermocouple): the method is called ``active scanning thermal microscopy'',
\Cite{NS002}[p.729].
\footnote{\tiny The method consists in ``{\it detecting the
    heat flow along the cantilever and feeding power proportional to it to
    the cantilever. Feedback with sufficient gain that keeps the arm at the
    same temperature as the sample contact point, then cantilever
    temperature is measured by another thermocouple on the middle of the
    cantilever}'', \Cite{NS002}[p.729].}

A concrete example of a nanoscale device to measure temperature (above room
temperature on a scale of $100\,nm=10^3\, A^o$ can be found in \Cite{NS002}\,).

With other earlier methods it is possible to measure temperatures, on a
scale of $30\,nm$ on a time scale $>1ms$, closer to the quantum regime: but
the technology (``passive scanning thermal microscopy'') seems more
delicate, \Cite{NS002}.
\*

The device is a bow arm (``cantilever'') with a sensor microscopic tip
which probes the surface of a sample whose temperature has to be measured:
the sensor is connected to a control system, formed basically by a
thermocouple sending amplified signals to a square root circuit, which by
feedback imposes that the temperature of the arm stays the same as that of
the test system (by sending signals to a ``heater'' in contact with the
bowing arm). Then the arm temperature is measured by conventional methods
(\ie via another thermocouple).  The test system is supposed in a
stationary state.

\eqfig{330}{119}{
\ins{3}{80}{\small sensor}
\ins{250}{23}{$\sqrt{\phantom{-}}$} 
\ins{153}{20}{\small dtc}
\ins{153}{80}{$B$} 
\ins{53}{80}{$A$} 
\ins{110}{88}{$V_+$}
\ins{110}{40}{$V_-$}
} {fig5.9.1}{Fig.5.9.1}   
\begin{spacing}{0.5}\tiny\0 Fig.5.9.1: The
  (microscopic) sensor is attached to the arm $AB$: a differential
  termocouple (dtc) is at the extremes of $AB$, and through $AB$ is
  also maintained a current at small constant voltage $V_+-V_-$; the
  thermocouple sends signals to a ``square root circuit'' which controls a
  ``coil'' (a device that can heat or cool the arm). The circuit that fixes
  the voltage is not represented, and also not represented are the
  amplifiers needed (there has to be one at the exit of the differential
  thermocouple and one after the square root circuit); furthermore there has
  to be also a device that records the output of the square root circuit
  hence the power fluctuations.\vfil\end{spacing}
\*\*

This device suggests a similar one, Fig.5.9.1, for a different
use: a very schematic description follows. A small electric
current could be kept flowing through the arm $AB$, by an applied
constant voltage difference $V_+-V_-$, to keep the arm in a
nonequilibrium steady state (varying the current heat can be
obtained or given to a reservoir around the arm); contact between
the arm and the sample is maintained via the sensor (without
allowing heat exchanges between the two, after their equilibrium
is reached) and the heat flow $Q$ (from the ``heater'', which
should actually be a pair of devices, heater + cooler) to keep
the arm temperature constant, at the value of the sample
temperature, via a feedback mechanism driven by the square root
circuit. The heat fluctuations could be revealed through
measurements of the electric current flowing out of the square
root circuit or by monitoring the heater output.\footnote{\tiny
This is a device turning the arm into a test system and the
attached circuits into a thermostat.}

The average heat output $Q_+$ can be compared to the instantaneous heat
output $Q$ and the statistics $P_\t(p)$ of the ratio $p=\frac{Q}{Q_+}$ over
a time span $\t$ might be measured. The temperature (of the bowing arm, hence
of the system) can be read from the slope of the function $\frac1\t
\log\frac{P_\t(p)}{P_\t(-p)}=p\frac{Q_+}{k_B T}$ from the fluctuation
relation: alternatively this could be a test of the fluctuation relation.

The arm and the sensor should be as small as possible: hence $~100\,nm$
linear size and $10ms$ for response time \Cite{NS002} are too large for
observing important fluctuations: hopefully the delicate technology could
be improved and it might be possible to build a working device,
carefully taking into account all the warnings mentioned about
the FT tests.

The idea is inspired by a similar earlier proposal for using
fluctuation measurements to define temperature in spin glasses,
\Cite{CKP997}, \Cite{CR003}[p.216]. 

\section{Granular materials and friction}
\def\SEC{Granular materials and friction}
\label{sec:X-5}\iniz
\lhead{\small\ref{sec:X-5}.\ \SEC}

The current interest in granular materials properties and the
consequent availability of experimental tests, {\it
e.g.} \Cite{FM004}, suggests trying to apply 
nonequilibrium statistics ideas to obtain possible experimental tests
of the CH, in the form of a check of fluctuations probabilities
agreement with the fluctuation relation\index{fluctuation
relation}, Eq.\equ{e4.6.6}.

The main problem is that in granular materials collisions are
intrinsically {\it inelastic}. In each collision particles heat
up, and the heat is subsequently released through thermal
exchange with the walls of the container, sound emission (if the
experiment is performed in air), radiation, and so on. If one
still wants to deal with a {\it reversible} system, such as the
ones discussed in the previous sections, all these sources of
dissipation should be included in the theoretical description.
Clearly, it might be very difficult to pursue such a task.

A simplified description, \Cite{BGGZ006}, of the system consists in
neglecting the internal degrees of freedom of the particles. In this
case the inelastic collisions between particles will represent
the only source of dissipation in the system. Still the CH is expected to hold, but in this case the entropy
production is strictly positive and there is no hope of observing a
fluctuation relation, see {\it e.g.} \Cite{PVBTW005}, if one looks at
the whole system.

Nevertheless, in presence of inelasticity, temperature gradients
may be present in the system
\Cite{GZN996}\Cite{BMM000}\Cite{FM004}, and
heat is transported through different regions of the container.
The processes of heat exchange between different regions could be
described assuming that, under suitable conditions, the
inelasticity of the collisions can be neglected, and a
fluctuation relation for a (suitably defined) entropy production
rate might become observable.  This could lead to an interesting
example of ``ensemble equivalence''\index{ensemble equivalence}
in nonequilibrium, \Cite{Ga000}, and its possibility will be
pursued in detail in the following.

As a concrete model for a granular material experiment, let $\Si$ be a
thin container bounded by two flat parallel vertical walls, covered at
the top, and with a piston at the bottom that is kept oscillating by a
motor so that its height is:
\be z(t)= A \cos\o t\label{e5.10.1}\ee 
The model can be modified by introducing a sawtooth 
moving piston as in \Cite{BMM000}, however the results should not 
depend too much on the details of the time dependence of $z(t)$.

The container $\Si$ is partially filled with millimeter size balls (a
typical size of the faces of $\Si$ is $10\ cm$ and the particle number
is of a few hundreds): the vertical walls are so close that the balls
almost touch both faces so the problem is effectively two dimensional.
The equations of motion of the balls with coordinates $(x_i,z_i), \,
i=1,\ldots,N$, $z_i\ge z(t)$, are 
\be \eqalign{
m \ddot{x}_i=&
f_{x,i}\hfill\cr m \ddot{z}_i=& f_{z,i} -m g +
m\d(z_i-z(t)) \, 2\, (\dot z(t)-\dot z_i)\cr}\label{e5.10.2}\ee
where $m$=mass, $g$=gravity acceleration, and the collisions between the
balls and the container oscillating base are assumed to be elastic
\Cite{BMM000} (possibly inelasticity of the walls can be included into the
model with negligible changes \Cite{PVBTW005}); $\V f_i$ is the force
describing the particle collisions and the particle-walls or
particles-piston collisions.

The force $\V f_i=(f_{x,i},f_{z,i})$ has a part describing the particles
collisions: the latter are necessarily inelastic and it will be assumed
that their ineslasticity is manifested by a restitution coefficient
$\a<1$. A simple model for inelastic collisions with inelasticity $\a$
(convenient for numerical implementation) is a model in which
collisions take place with the usual elastic collision rule but,
immediately after, the velocities of the particles that have collided
are scaled so that the kinetic energy of the pair is
reduced by a factor $1-\a^2$ 
\Cite{GZN996}\Cite{BMM000}\Cite{PVBTW005}\,\footnote{\tiny A
better model could be obtained by rescaling by $\a$ only the 
velocity component parallel to the relative velocity.} 

Keeping in mind the discussion of Sec.(\,\ref{sec:IX-4}\,), about the formulation of
a local fluctuation relation\index{local fluctuation relation}, the
simplest situation, that seems accessible to experimental tests as well as to
simulations, is to draw ideal horizontal lines at heights
$h_1>h_2\gg A$
delimiting a strip $\Si_{0}$ in the container and to look at the particles
in $\Si_{0}$ as a thermostatted system: the thermostats being the regions
$\Si_1$ and $\Si_2$ at heights larger than $h_1$ and smaller then $h_2$,
respectively.

After a stationary state has been reached, the average kinetic energy
of the particles will depend on the height $z$, and in particular will
decrease on increasing $z$.

Given the motion of the particles and a time interval $t$ it will be
possible to measure the quantity $Q_2$ of (kinetic) energy that
particles entering or exiting the region $\Si_{0}$ or colliding with
particles inside it from below (the ``hotter side'') carry out of
$\Si_{0}$ and the analogous quantity $Q_1$ carried out by the
particles that enter, exit or collide from above (the ``colder
side'').

Let $T_i,\,i=1,2,$ be the average kinetic energies of the particles in
small horizontal corridors above and below $\Si_{0}$: a connection
between the model of granular material, Eq.\equ{e5.10.2}, and the
general thermostat model in Sec.(\,\ref{sec:II-2}\,) can be established. The
connection cannot be exact because of the internal dissipation induced
by the inelasticity $\a$ and of the fact that the number of particles,
and their identity, in $\Si_0$ depends on time, as particles come and
go in the region.

Under suitable assumptions, that can be expected to hold on a specific
time scale, the stationary state of Eq. \equ{e5.10.2} is effectively
described in terms of a stationary SRB state of models like the one
considered in Sec.(\,\ref{sec:II-2}\,), as discussed below.

Real experimental tests cannot have an arbitrary duration, \Cite{FM004}: the
particles movements might be recorded by a digital camera and the number of
photograms per second can be of the order of a thousand, so that the memory for
the data is easily exhausted as each photogram may have a size of about $1$Mb in
current tests ($<$2008). The same holds for numerical simulations where
the accessible time scale is limited by the available computational
resources.

Hence each test lasts up to a few seconds, starting after the
system has been moving for a while so that a stationary state can
be supposed to have been reached. And data of the trajectory, in phase
space, of each individual particle inside the observation
frame, \Cite{FM004}, is recorded and, subsequently, reconstructed.

To keep the number of particles $N_0$ in $\Si_0$ to be
approximately constant for the duration of the experiment, the
vertical size $(h_1-h_2)$ of $\Si_0$ should be chosen large
compared to $(Dt)^{1/2}$, where $t$ is the duration of the
experiment and $D$ is the diffusion coefficient of the grains.
Hence we are assuming, see Sec.(\,\ref{sec:X-4}\,), that the particles
motion is diffusive on the scale of $\Si_0$.  Note that at low
enough density the motion could be not diffusive on the scale of
$\Si_0$ (\ie free path larger than the width of $\Si_0$): then it
would not be possible to divide the degrees of freedom between
the subsystem and the rest of the system and moreover the
correlation length would be comparable with (or larger than) the
size of the subsystem $\Si_0$. This would completely change the
nature of the problem: and violations of the fluctuation
relation\index{fluctuation relation violation} would occur,
\Cite{BCL998}\Cite{BL001}.

Given the remarks above suppose that in observations of stationary states
lasting up to a maximum time $\th$:
\*

\0{}(1) the CH is accepted,

\0{}(2) it is supposed that the result of the observations would be 
the same if the particles above $\Si_{0}$ and below $\Si_{0}$ were kept at
constant total kinetic energy by reversible thermostats ({\it e.g.} 
Gaussian thermostats), \Cite{ES993}\Cite{Ga000}\Cite{Ru000},

\0{}(3) dissipation due to inelastic collisions between
particles in $\Si_{0}$ is neglected,

\0{}(4) fluctuations of the number of particles in 
$\Si_0$ is neglected,

\0{}(5) dissipation is present in the sense that
\be \s_+\defi 
\frac1\th\Big(\frac{Q_1(\th)}{T_1}+\frac{Q_2(\th)}{T_2}\Big)>0\;,
\label{e5.10.3}\ee 
with $Q_i(t)$ is the total heat ceded to
the particles in $\Si_i$, $i=1,2$, in time $\th$.

Chaoticity is expected, at least if dissipation is small, and evidence for it
is provided by the experiment in \Cite{FM004} indicating that the
system evolves to a chaotic stationary state. 

For the purpose of checking a fluctuation relation for
$\s_0=\frac1\t(\frac{Q_1(\t)}{T_1}+\frac{Q_2(\t)}{T_2})$, where $Q_i(\t)$
is the total heat ceded to the particles in $\Si_i$, $i=1,2$, in time $\t$,
the observation time $\t\le \th$ should be not long enough that dissipation
due to internal inelastic collisions becomes important. So measurements,
starting after the stationary state is reached, can have a duration $\t$
which cannot exceed {\it a specific time scale} in order that the
conditions for a local fluctuation relation can be expected to apply to
model Eq.\equ{e5.10.2}, as discussed below.

Accepting the assumptions above, a fluctuation relation might hold
for fluctuations of
\be p=\frac1{\t\,\s_+}{\Big(\frac{Q_1(\t)}{T_1}+\frac{Q_2(\t)}{T_2}\Big)}
\label{e5.10.4}\ee 
in the interval $(-p^*,p^*)$ with $p^*$ equal (at least) to $1$,
but more discussion of the assumptions is needed, see next
section.

The latter is therefore a property that might be accessible to simulations as
well as to experimental test. Note however that it is very likely that the
hypotheses (2)-(4) above will not be {\it strictly} verified in real
experiments, see the discussion in next section, so that the analysis and
interpretation of the experimental results might be non trivial.
Nevertheless, a careful test would be rather stringent.

\section{Neglecting granular friction\index{granular friction}: 
the relevant time scales}
\def\SEC{Neglecting granular friction\index{granular friction}: 
the relevant time scales}
\label{sec:XI-5}\iniz
\lhead{\small\ref{sec:XI-5}.\ \SEC}

The above analysis assumes, \Cite{BGGZ006}, the existence of (at least) two
time scales.  One is the ``equilibrium time scale'', $\th_e$, which is the
time scale over which the system evolving at constant energy, equal to the
average energy observed, would reach equilibrium in absence of friction and
forcing.

An experimental measure of $\th_e$ would be the decorrelation
time of self--correlations in the stationary state, and it can be
assumed that $\th_e$ is of the order of the mean time between
collisions of a selected particle. Note that $\th_e$ also
coincides with the time scale over which finite time corrections
to the fluctuation relation \index{fluctuation relation} become
irrelevant \Cite{ZRA004}: this means that in order to be able to
measure the large deviations functional for the normalized
entropy production rate $p$ in Eq.~\equ{e5.10.4} one has to
choose $t\gg\th_e$, see also \Cite{GZG005} for a detailed
discussion of the first orders finite time corrections to the
large deviation functional.

A second time scale is the `
`inelasticity time\index{inelasticity time scale} scale'' $\th_d$, which
is the scale over which the system reaches a stationary state if the
particles are prepared in a random configuration and the piston is
switched on at time $t=0$.

Possibly a third time scale is present: the ``diffusion
time\index{diffusion time scale} scale'' $\th_{D}$ which is the scale over
which a particle diffuses beyond the width of $\Si_0$.

The analysis above applies only if the time $t$ in
Eq.~\equ{e5.10.4} verifies $\th_e\ll t \ll\th_d, \th_D$ (note
however that the measurement should be started after a time
$\gg \th_d$ since the piston has been switched on, in order to
have achieved a stationary state); in practice this means that
the time for reaching the stationary state has to be quite long
compared to $\th_e$. In this case friction is negligible {\it for
the duration of the measurement} if the measurement is performed
starting after the system has reached a stationary state and
lasts a time $\t$ between $\th_e$ and $\min(\th_D,\th_d)$.

In the setting considered here, the role of friction is ``just'' that of
producing the nonequilibrium stationary state itself and the
corresponding gradient of temperature: this is reminiscent of the role
played by friction in classical mechanics problems, where periodic
orbits (the ``stationary states'') can be dynamically selected by
adding a small friction term to the Hamilton equations. Note that, as
discussed below, the temperature gradient produced by friction will be
rather small: however smallness of the gradient does not affect the
``FR time scale'' over which the fluctuation relation is observable \Cite{ZRA004}.

If internal friction were not negligible (that is if $t\ge
\th_d$) the problem would change nature: an explicit model (and
theory) should be developed to describe the transport mechanisms
(such as radiation, heat exchange between the particles and the
container, sound emission, ...)  associated with the dissipation
of kinetic energy and new thermostats should be correspondingly
introduced. The definition of entropy production should be
changed, by taking into account the presence of such new
thermostats. In this case, even changing the definition of
entropy production it is not expected that a fluctuation relation
should be satisfied: in fact internal dissipation would not break
the CH, but the necessary time--reversibility assumption would be
lost, \Cite{BGGZ006}.
 
The possibility of $\th_e\ll t \ll\th_d, \th_D$ is not obvious, neither in
theory nor in experimental tests.  A rough estimate of $\th_d$ can be given as
follows: the phase space contraction in a single collision is given by
$1-\a$. Thus the average phase space contraction per particle and per unit
time is $\s_{+,d} = (1-\a)/\th_e$, where $1/\th_e$ is the frequency of the
collisions for a given particle. It seems natural to assume that $\th_d$ is
the time scale at which $\s_{+,d} \th_d$ becomes of order $1$: on this time
scale inelasticity will become manifest. Thus, we obtain the following
estimate:

\be
\th_d \sim \frac1{1-\a} \th_e
\label{e5.11.1}\ee
In real materials $\a\le .95$, so that $\th_d$ can be at most of
the order of $20 \,\th_e$. Nevertheless it might be already
enough to observe, at intermediate times, a fluctuation relation.

The situation is completely different in numerical
simulations\index{numerical simulations} where there is freedom
to choose the restitution coefficient $\a$ (it can be chosen very
close to one \Cite{GZN996}\Cite{BMM000}\Cite{PVBTW005}, in order
to have $\th_d\gg\th_e$) and the size of the container $\Si_0$
(it can be chosen large, in order to have $\th_D\gg\th_e$).

\section{Simulations for granular materials\index{granular materials}}
\def\SEC{Simulations for granular materials\index{granular materials}}
\label{sec:XII-5}\iniz
\lhead{\small\ref{sec:XII-5}.\ \SEC}

To check the consistency of the hypotheses in Sec.(\,\ref{sec:X-5}\,), it has to be
shown that it is possible to make a choice of parameters so that 
$\th_e$ and $\th_d,\th_D$ are separated by a large time window.
Such choices may be possible, as discussed below, \Cite{BGGZ006}. Let

\be \eqalign{
\d\defi&h_1-h_2 \quad\hbox{is the width of}\  \Si_0,\cr
\e\defi& 1-\a,\cr 
\g\defi&\ \hbox{is the temperature gradient in}\  \Si_0,\cr
D\defi&\hbox{is the diffusion coefficient}\cr}
\label{e5.12.1}\ee
the following estimates hold:
\*

\0(a) $\th_e=O(1)$ as it can be taken of the order of the inverse
collision frequency, which is $O(1)$ if density is constant and 
the forcing on the system is tuned to keep the energy constant as $\e\to0$. 
\\
(b) $\th_d=\th_e O(\e^{-1})$ as implied by Eq.~\equ{e5.11.1}.***
\\
(c) $\th_D=O(\frac{\d^2}D)=O(\d^2)$ because $D$ is a constant
(if the temperature and the density are kept constant).
\\
(d) $\g=O(\sqrt\e)$, as long as $\d\ll\e^{-1/2}$.  
\*

In fact if the density is high enough to allow us to consider the
granular material as a fluid, as in Eq.~(5) of Ref.\Cite{BMM000}, the
temperature profile should be given by the heat equation $\nabla^2
T+c\e T=0$ with suitable constant $c$ and suitable boundary conditions
on the piston ($T=T_0$) and on the top of the container ($\nabla
T=0$).  This equation is solved by a linear combination of $const
\,e^{\pm\sqrt{c\e} z}$, which has gradients of order $O(\sqrt\e)$, as
long as $\d \ll 1/\sqrt\e$ and the boundaries of $\Si_0$ are further
than $O(1/\sqrt\e)$ from the top.

Choosing  $\d=\e^{-\b}$, with $\b<\frac12$,
and taking $\e$ small enough, it is $\th_e \ll \min\{\th_d,\th_D\}$ and
$\d\ll O(\e^{-\frac12})$, as required by item (d). 

\*
\0{\it Remark:} The entropy production rate due to heat
transport into $\Si_0$, in presence of a temperature gradient $\g$, is
given by $\s_+=O(\g^2 \d)=O(\e\d)$ because the temperature difference
is $O(\g\d)$ and the energy flow through the surface is of order
$O(\g)$ (with $\g=O(\sqrt\e)$, see item (d)). The order of magnitude
of $\s_+$ is not larger then the average amount $\s_d$ of energy
dissipated per unit time in $\Si_0$ divided by the average kinetic
energy $T$ (the latter quantity is of order $O(\th_e^{-1}\e\d)$
because, at constant density, the number of particles in $\Si_0$ is
$O(\d)$); however the entropy creation due to the dissipative
collisions in $\Si_0$ has fluctuations of order $O(\e\d^{\frac12})$
because the number of particles in $\Si_0$ fluctuates by
$O(\d^{\frac12})$. This is consistent with neglecting the entropy
creation inside the region $\Si_0$ due to the inelasticity in spite of
it being of the same order of the entropy creation due to the heat
entering $\Si_0$ from its upper and lower regions.

\*

The argument supports the proposal that in numerical simulations a
fluctuation relation test might be possible by a suitable
choice of the parameters. Other choices will be possible: for instance in
the high-density limit it is clear that $\th_D \gg \th_e$ because the
diffusion coefficient will become small. To what extent this can be
applied to experiment is a further question.  \*

\0{\it Remarks} \0(1) An explicit computation of the large deviation
function of the dissipated power, in the regime $t \gg \th_d$
({\it i.e.}  when the dissipation is mainly due to inelastic
collisions) recently appeared in \Cite{VPBTW005}. However in the
model only the dissipation due to the collisions was taken into
account, \Cite{BGGZ006}: so that it is not clear how the heat
produced in the collisions is removed from the system, see the
discussion above. It turned out that in this regime no negative
values of $p$ were observed so that the fluctuation
relation\index{fluctuation relation}, Eq.\equ{e4.6.6},
p.\pageref{FR}, does not hold. This is interesting and expected
on the basis of the considerations above. It is not clear if,
including the additional thermostats required to remove heat from
the particles and prevent them to warm up indefinitely, the
fluctuation relation, Eq.\equ{e4.6.6}, is recovered.
\\
(3) There has also been some debate on the interpretation of the
experimental results of \Cite{FM004}. In \Cite{PVBTW005} a
simplified model, very similar to the one discussed above, was
proposed and showed to reproduce the experimental data
of \Cite{FM004}. The prediction of the model is that the
fluctuation relation is not satisfied. Note however that the
geometry considered in \Cite{FM004}\Cite{PVBTW005} is different
from the one considered here: the whole box is vibrated, so that
the the temperature profile is symmetric, and a region $\Si_0$ in
the center of the box is considered. Heat exchange is due to
``hot'' (fast) particles entering $\Si_0$ (\ie $Q_+$) and
``cold'' (slow) particles exiting $\Si_0$ (\ie $Q_-$). One has
$Q=Q_+ + Q_- \neq 0$ because of the dissipation in $\Si_0$.

In this regime, again, the fluctuation relation is not expected
to hold if the thermostat dissipating the heat produced in the collisions
is not included in the model: it is an interesting remark of
\Cite{PVBTW005} that partially motivated the present discussion. Different
experiments can be designed in which the dissipation is mainly due to heat
exchanges and the inelasticity is negligible, as the one proposed above as
an example.
\\
(4) Even in situations in which the dissipation is entirely due to
irreversible inelastic collisions between particles, such as the ones
considered in \Cite{PVBTW005}\Cite{VPBTW005}, the CH is expected
to hold, and the stationary state to be described by a SRB distribution.
But in these cases failure of the fluctuation relation is not
contradictory, due to the irreversibility of the equations of motion.
\\
(5) In cases like the Gaussian isoenergetic models\index{isoenergetic
  thermostat}, or in other models in which the kinetic energy fluctuates
(\eg in the proposal above) care has to be paid to measure the fluctuations
of the ratios $\frac{Q}{K}$ rather than those of $Q$ and $K$ separately
because there might not be an ``effective temperature'' of the thermostats
(\ie fluctuations of $K$ may be important).
\\
(6) Finally it is important to keep track of the errors due to
the size of $\D$ in the fluctuation relation, Eq.\equ{e4.6.6}
p.\pageref{FR}: the condition that $\D\ll {p}$ make it very
difficult to test the fluctuation relation near $p=0$: this may
lead to interpretation problems and, in fact, in many
experimental, or simulation, works the fluctuation relation is
written for the non normalized entropy production $A$ (identified
with phase space contraction)
\be \frac{Prob( A\in \D)}{Prob(-A\in\D)}\simeq e^{A}\label{e5.12.2}\ee
where $A$ is the total phase space contraction in the observation
time (\Cite{JES004}, and see Appendix \ref{appO}\,), \ie $A=\t
p \s_+$.  This relation may lead to illusory agreement with data,
unless a detailed error analysis is done, as it can be confused
with the linearity at small $A$
due to the extrapolation at $p=0$ of the central limit theorem%
\footnote{\tiny Which instead can be applied only to
   $\scriptstyle|p-1|{\lower 1mm\hbox{$\buildrel < \over
      \sim$}}1/({\t\s_+})^{\scriptscriptstyle1/2}$.}
 or just to the linearity of the large deviation function near
 $p=0$.  Furthermore the $A=p\t\s_+$ depends on the observation
 time $\t$ and on the dissipation rate $\s_+$, if $p=O(1)$: all
 this is hidden in Eq.\equ{e5.12.2}.
\vfill\eject

\kern2.3cm
\vfill\eject

\chapter{Historical comments}
\label{Ch6} 

\chaptermark{\ifodd\thepage
Historical comments and translations\hfill\else\hfill 
Historical comments and translations\fi} 

\section{Proof of the second fundamental theorem.}
\def\SEC{Proof of the second fundamental theorem.}
\label{sec:I-6}\iniz
\lhead{\small\ref{sec:I-6}.\ \SEC}

\0{Partial translation and comments of L. Boltzmann, {\it{\"U}ber die
    mechanische {B}edeutung des zweiten {H}aupsatzes der
    {W\"ag}rme\-theorie}, Wien. Ber. {\bf 53}, 195-220,
  1866. Wissenshaftliche Abhanlunger, Vol.{\bf1}, p.9-33, \#2,
  \Cite{Bo866}.}
\*

\0[{\sl The distinction between the ``second theorem'' and
    ``second law'' is important: the first is $\oint
    \frac{dQ}{T}=0$ in a reversible cycle, while the second is
    the inequality $\oint \frac{dQ}{T}\le0$ in any cycle. For a
    formulation of the second law here intend the Clausius
    formulation, see footnote at p.\pageref{Kelvin-Planck}. The
    law implies the theorem but not viceversa.}]\index{Boltzmann}

\*
\0[{\sl In Sec.I Boltzmann's aim is to explain the mechanical meaning of
temperature. Two bodies in contact will be in thermal equilibrium if the
average kinetic energy (or any average property) of the atoms of each will
not change in time: the peculiarity of the kinetic energy is that it is
conserved in the collisions. Let two atoms of masses $m,M$ and velocities
$\bf v,V$ collide and let $\bf c,C$ be the velocities after collision.
The kinetic energy variation of the atom $m$ is $\ell=\frac12m v^2-\frac12
m c^2$. Choosing as $z$-axis (called $G$) so that the momentum $m\bf v$ and
$m\bf c$ have the same projection on the $xy$ plane it follows from the
collision conservation rules (of kinetic energy and momentum) that if
$\f,\F$ are inclinations of $\bf v,\bf V$ over the $z$-axis (and likewise
$\f',\F'$ are inclinations of $\bf c,\bf C$ over the $z$-axis) it is

$$\ell=\frac{2 m M}{(m+M)^2}(M C^2 \cos^2\F-m c^2\cos^2\f+(m-M) c C 
\cos\f\,\cos \F)$$
which averaged over time and over all collisions between atoms of the two
bodies yield an average variation of $\ell$ which is
$L=\frac{4mM}{3(m+M)^2}
\media{\frac{ M C^2}2-\frac{m c^2}2}$ so that the average kinetic energies
of the atoms of the two species have to be equal hence $T= A\media{\frac12
m c^2}+B$. The constant $B=0$ if $T$ is identified with the absolute
temperature of a perfect gas. The identification of the average kinetic
energy with the absolute temperature was already a well established fact,
based on the theory of the perfect gas, see Sec.(\,\ref{sec:II-1}\,) above. The
analysis is somewhat different from the one presented by Maxwell few years
earlier in \Cite{Ma860-a}[p.383]}]

\*
\0[{\sl In Sec.II it is shown, by similar arguments, that the average
kinetic energy of the center of mass of a molecule is the same as the
average kinetic energy of each of its atoms}]
\*

\0[{\sl In Sec.III the laws called of Amp\`ere-Avogadro, of Dulong-Petit and of
Neumann for a free gas of molecules are derived}]
\*

\0{\bf Sec. IV: Proof of the second theorem of the mechanical theory of heat}
\*

The just clarified meaning of temperature makes it possible to undertake
the proof of the second fundamental theorem of heat theory, and of course
it will be entirely coincident with the form first exposed by Clausius.

$$\ig \fra{d Q}T\le0\eqno(20)$$
To avoid introducing too many new quantities into account, we shall right
away treat the case in which actions and reactions during the entire
process are equal to each other, so that in the interior of the body either
thermal equilibrium or a stationary heat flow will always be found.  Thus
if the process is stopped at a certain instant, the body will remain in its
state.%
\footnote{\tiny Here B. means a system in thermal equilibrium evolving in a
``reversible'' way, \ie ``quasi static'' in the sense of thermodynamics,
and performing a cycle (the cyclicity condition is certainly implicit even
though it is not mentioned): in this process heat is exchanged, but there
is no heat flow in the sense of modern nonequilibrium thermodynamics
(because the process is quasi static); furthermore the process takes place
while every atom follows approximate cycles with period $t_2-t_1$, of
duration possibly strongly varying from atom to atom, and the variations
induced by the development of the process take place over many
cycles. Eventually it will be assumed that in solids the cycles period is a
constant to obtain, as an application, theoretical evidence for the
Dulong-Petit and Neumann laws.}  
In such case in Eq.(20) the equal sign will hold. Imagine first that the
body, during a given time interval, is found in a state at temperature,
volume and pressure constant, and thus atoms will describe curved paths
with varying velocities.

We shall now suppose that an arbitrarily selected atom runs, whatever the
state of the body, in a suitable time interval (no matter if very long), of
which the instants $t_1$ and $t_2$ are the initial and final times, at the
end of it the speeds and the directions come back to the original value in
the same location, describing a closed curve and repeating, from this
instant on, their motion,%
\footnote{\tiny This is perhaps the first time that what will become the {\it
    ergodic hypothesis} is formulated.  It is remarkable that an analogous
  hypothesis, more clearly formulated, can be found in the successive paper
  by Clausius, \Cite{Cl871}[l.8, p.438] (see the following
  Sec.(\,\ref{sec:IV-6}\,), which Boltzmann criticized as essentially identical
  to Section IV of the present paper: this means that already at the time
  the idea of recurrence and ergodicity must have been quite
  common. Clausius imagines that the atoms follow closed paths, \ie he
  conceives the system as ``integrable'', while Boltzmann appears to think,
  at least at first, that the entire system follows a single closed
  path. It should however be noticed that later, concluding Sec. IV,
  Boltzmann will suppose that every atom will move staying within a small
  volume element, introduced later and denoted $dk$, getting close to
  Clausius' viewpoint.}
 possibly not exactly equal\,
\footnote{\tiny Apparently this contradicts the preceding
statement: because now he thinks that motion does not come back
exactly on itself; here B. rather than taking into account the
continuity of space (which would make impossible returning
exactly at the initial state) refers to the fact that his
argument does not require necessarily that every atom comes back
exactly to initial position and velocity but it suffices that it
comes back ``infinitely close'' to them, and actually only in
this way it is possible that a quasi static process can develop.}\,
 nevertheless so similar that the average kinetic energy over the time
 $t_2-t_1$ can be seen as the atoms average kinetic energy during an
 arbitrarily long time; so that the temperature of every atom
 is%
\footnote{\tiny Here use is made of the result discussed in the Sec.I of the
   paper which led to state that every atom, and molecule alike, has equal
   average kinetic energy and, therefore, it makes sense to call it
   ``temperature'' of the atom.}
$$T=\fra{\ig_{t_1}^{t_2} \fra{m c^2}2 dt}{t_2-t_1}.$$
Every atom will vary [{\sl in the course of time}] its energy by an
infinitely small quantity $\e$, and certainly every time the work done
and the variation of kinetic energy will be in the average redistributed
among the various atoms, without contributing to a variation of the average
kinetic energy. If the exchange was not eventually even, it would be
possible to wait so long until the thermal equilibrium or stationarity will
be attained%
\footnote{\tiny A strange remark because the system is always in thermodynamic
  equilibrium: but it seems that it would suffice to say ``it would be
  possible to wait long enough and then exhibit ...''. A faithful literal
  translation is difficult. The comment should be about the possibility
  that diverse atoms may have an excess of kinetic energy, with respect to
  the average, in their motion: which however is compensated by the
  excesses or defects of the kinetic energies of the other atoms. Hence $\e$
  has zero average because it is the variation of kinetic energy when the
  system is at a particular position in the course of a quasi static
  process (hence it is in equilibrium). However in the following paragraph
  the same symbol $\e$ indicates the kinetic energy variation due to an
  infinitesimal step of a quasi static process, where there is no reason
  why the average value of $\e$ be zero because the temperature may
  vary. As said below the problem is that this quantity $\e$ seems to have
  two meanings and might be one of the reasons of Clausius complanets, see
  footnote at p.\pageref{unclear}.}%
 and then exhibit, as work performed in average, an average
 kinetic energy increase greater over by what was denoted $\e$,
 per atom.

At the same time suppose an infinitely small variation of the volume and
pressure of the body.\label{volume change}%
\footnote{\tiny Clausius' paper is, however, more clear: for instance the
  precise notion of ``variation'' used here, with due meditation can be
  derived, as proposed by Clausius and using his notations in which $\e$
  has a completely different meaning, as the function with two parameters
  $\d i,\e$ changing the periodic function $x(t)$ with period $i\equiv
  t_2-t_1$ into $x'(t)$ with $x'(t)= x(i t/(i+\d i)) +\e\x(i t/(i+\d i))$
  periodic with period $i+\d i$ which, to first order in $\d x=-\dot
  x(t)\frac{\d i}{i}t+\e \x(t)$.}  %
Manifestly the considered atom will follow one of the curves, infinitely
close to each other.\label{volume and pressure}
\footnote{\tiny Change of volume amounts at changing the external forces (volume
  change is a variation of the confining potential): but no mention here is
  made of this key point and no trace of the corresponding extra terms
  appears in the main formulae. Clausius {\it essential critique} to
  Boltzmann, in the priority dispute, is precisely that no account is given
  of variations of external forces.  Later Boltzmann recognizes that he has
  not included external forces in his treatment, see p.\pageref{B on ergale},
  without mentioning this point.}
Consider now the time when the atom is on a point of the new path from
which the position of the atom at time $t_1$ differs infinitely little, and
let it be $t'_1$ and denote $t'_2$ the instant in which the atom returns in
the same position with the same velocity; we shall express the variation of
the value of $T$ via the integrals
$$\ig_{t_1}^{t_{2}}\fra{mc^2}2\,dt=\, \fra{m}2\ig_{s_1}^{s_2} c ds$$ 
where $ds$ is the line element values of the mentioned arc with as extremes
$s_1$ and $s_2$ the positions occupied at the times
$t_1$ and $t_2$. The variation is
$$\fra{m}2\d\ig_{s_1}^{s_2} c ds=\fra{m}2\ig_{s'_1}^{s'_2} c'\,ds-\fra{m}2
\ig_{s_1}^{s_2} c\,ds$$
where the primed quantities refer to the varied curve and $s'_1,
s'_2$ are the mentioned arcs of the new curve at the times $t'_1,t'_2$.
To obtain an expression of the magnitude of the variation we shall consider
also $ds$ as variable getting

$$\fra{m}2\,\d\ig_{s_1}^{s_2} c\,ds=\fra{m}2\ig_{s_1}^{s_2}(\d c\,d
s\,+\,c\,\d ds);
\eqno{(21)}$$
It is $\fra{m}2\ig_{s_1}^{s_2} \d\,c\,ds=
\ig_{t_1}^{t_{2}}\fra{dt}2\,\d\fra{mc^2}2$, and furthermore, if $X,Y,Z$
are the components on the coordinate axes of the force acting on the atom,
it follows:
$$\eqalign{ 
d\,\fra{m\,c^2}2=&X\,dx+Y\,dy+Z\,dz\cr
d\d\fra{mc^2}2=& \d X\,dx+\d Y\,d y+\d Z\, d z+X\,\d d x+ Y \d\,d
y+Z\,\d d z=\cr
=&d(X\d x+Y\d Y+z\d Z)\cr&
+(\d X dx-d X\d x+\d Y dy-d Y\d y+\d Z dz-d Z\d z).\cr}
$$

Integrate then,%
\footnote{\tiny This point, as well as the entire argument, may
  appear somewhat obscure at first sight: but it arrives at the same
  conclusions that four years later will be reached by Clausius, whose
  derivation is instead very clear; see the sections of the Clausius paper
  translated here in Sec.(\,\ref{sec:IV-6}\,) and the comment on the action
  principle below.} %
considering that for determining the integration constant it must be $\d
\fra{m c^2}2=\e$ when the right hand side vanishes%
\footnote{\tiny The initial integration point is not arbitrary: it should rather
  coincide with the point where the kinetic energy variation equals the
  variation of the work performed, in average, on the atom during the
  motion.}%
, one gets the

$$\eqalign{
\d\fra{m\,c^2}2-\e\,=&\,(X\d x+Y\d Y+z\d Z)\cr
&+\ig (\d X dx-d X\d x+\d Y dy-d Y\d y+\d Z dz-d Z\d z).\cr}$$
Here the term on the left contains the difference of the kinetic
energies, the expression on the right contains the work made on
the atom, hence the integral on the \rhs expresses the kinetic
energy communicated to the other atoms.  The latter certainly is
not zero at each instant, but its average during the interval
$t_2-t_1$ is, in agreement with our assumptions, $=0$ because the
integral is extended to this time interval. Taking into account
these facts one finds

$$\eqalign{
\ig_{t_1}^{t_2}\fra{dt}2
\d\fra{mc^2}2=&\fra{t_2-t_1}2\e+\fra12\ig_{t_1}^{t_{2}}
(X\d x+Y\d Y+z\d Z)dt\cr =& \fra{t_2-t_1}2\e +\fra{m}2\ig_{t_1}^{t_{2}}
\Big(\fra{d^2 x}{dt^2}\d x+\fra{d^2 y}{dt^2}\d y+\fra{d^2 z}{dt^2}\d
z\Big)dt,\cr}\eqno{(22)}$$
a formula which, by the way,\footnote{\tiny Here too the meaning of $\e$ is not
  clear. The integral from $t_1$ to $t_2$ is a line integral of a
  differential $d(X\d x+\ldots)$ but does not vanish because the
  differential is not exact, as $\d x,\d y,\d z$ is not parallel to the
  integration path.} also follows because $\e$ is the sum of the increase
in average of the kinetic energy of the atom and of the work done in
average on the atom.%
\footnote{\tiny In other words this is the ``vis viva'' theorem
  because the variation of the kinetic energy is due to two causes: namely
  the variation of the motion, given by $\e$, and the work done by the
  acting (internal) forces because of the variation of the trajectory,
  given by the integral.

Clausius considered the statement
  in need of being checked.} %
If $ds=\sqrt{dx^2+dy^2+dz^2}$ is set and
$c=\fra{ds}{dt}$,

$$\fra{m}2\ig_{s_1}^{s_2} c\,\d ds=\fra{m}2\ig_{t_1}^{t_{2}}
\Big(\fra{d x}{dt}\,d\d x+\fra{d y}{dt}\,d\d y+\fra{d z}{dt}\,d\d
z\Big).\eqno{(23)}$$
Inserting Eq.(22) and (23) in Eq.(21) follows:\footnote{\tiny This is very close to
  the least action principle according to which the difference between
  average kinetic energy and average potential energy is stationary within
  motions with given extremes. Here the condition of fixed extremes does
  not apply and it is deduced that the action of the motion considered
  between $t_1$ and $t_2$ has a variation which is a boundary term;
  precisely $\{m\V v\cdot\V \d \V x\}_{t_1}^{t_2}$ (which is $0$) is the
  difference $\e$ between the average kinetic energy variation
  ${m}\d\ig_{s_1}^{s_2} c\,ds$ and that of the average potential
  energy. Such formulation is mentioned in the following p.\pageref{least}.}
$$\eqalign{
\fra{m}2\d\ig_{s_1}^{s_2} c\,ds&= \fra{t_2-t_1}2\e+ \ig_{t_1}^{t_{2}}
d\,\Big(\fra{d x}{dt}\,\d x+\fra{d y}{dt}\,\d y+\fra{d
z}{dt}\,\d z\Big)\cr
=& \fra{t_2-t_1}2\e
+ \Big\{\fra{m}2(\fra{d x}{dt}\,\d x+\fra{d y}{dt}\,\d y+\fra{d
z}{dt}\d z)\Big\}_{t_1}^{t_2}.\cr}$$
However since the atom at times $t_1$ and $t'_1$ is in the same
position with the same velocity as at the times $t_2$ and
$t'_2$, then also the variations at time $t_1$ have the same value taken at
time $t_2$, so in the last part both values cancel and remains

$$\e=\fra{ m\d\ig_{t_1}^{t_2} c\,ds}{t_2-t_1}= 
\fra{ 2\d\ig_{t_1}^{t_2}\fra{m c^2}2 dt}{t_2-t_1},\eqno{(23a)}$$
which, divided by the temperature, yield:\footnote{\tiny It
  should be remarked that physically the process considered is a
  reversible process in which no work is done: therefore the only
  parameter that determines the macroscopic state of the system,
  and that can change in the process, is the temperature: so
  strictly speaking Eq.(24) might be not surprising as also $Q$
  would be function of $T$. Clausius insists that this is a key
  point which is discussed in full detail in his work by allowing
  also volume changes and more generally action of external
  forces, see p.\pageref{B on ergale} below.}

$$\fra\e{T}= \fra{ 2\d\ig_{t_1}^{t_2}\fra{m c^2}2 dt}{
\ig_{t_1}^{t_2}\fra{m c^2}2 dt}=2\,\d\,\log \ig_{t_1}^{t_2}\fra{m c^2}2
dt.$$
Suppose right away that the temperature at which the heat is exchanged
during the process is the same everywhere in the body, realizing in this way
the assumed hypothesis. Hence the sum of all the $\e$ equals the amount of
heat transferred inside the body measured in units of work. Calling the latter
$\d Q$, it is:

$$\eqalign{
\d Q=&\sum \e=  2\sum \fra{\d\ig_{t_1}^{t_2}\fra{m c^2}2
dt}{t_2-t_1}\cr
\fra{\d Q}T=& \fra1 T\sum\e = 2\,\d\,\sum \log \ig_{t_1}^{t_2}\fra{m c^2}2
dt.
\cr}\eqno(24)$$

If now the body temperature varies from place to place, then we can
subdivide the body into volume elements $dk$ so small that in each the
temperature and the heat transfer can be regarded as constant; consider then
each of such elements as external and denote the heat transferred from the
other parts of the body as $\d Q\cdot dk$ and, as before,

$$\fra{\d Q}T\,dk=2\d \sum \log \ig_{t_1}^{t_2}\fra{m c^2}2
dt,$$
if the integral as well as the sum runs over all atoms of the element $dk$.
From this it is clear that the integral

$$\int\int \frac{\d Q}T dk $$
where one integration yields the variation $\d$ of what Clausius would call
entropy, with the value

$$2\sum \log \int_{t_1}^{t_2} \frac{m c^2}2 dt +C,$$ 
[{\sl and the integration}] between equal limit vanishes, if pressure and
counter-pressure remain always equal.%
\footnote{\tiny {\it I.e.} in a cycle in which no work is done.
This is criticized by Clausius.}

Secondly if this condition was not verified it would be possible to
introduce all along 
a new force to restore the equality. The heat amount, which in the last
case, through the force added to the ones considered before, must be
introduced to obtain equal variations of volumes and temperatures in all
parts of the body, must be such that the equation

$$\int\int \frac{\d Q}T dk=0$$
holds; however the latter [{\sl heat amount}] is necessarily larger than
that [{\sl the heat amount}] really introduced, as at an expansion of the
body the pressure had to overcome the considered necessary positive
force;%
\footnote{\tiny The pressure performs positive work in an expansion.} %
at a compression, in the case of equal pressures, a part of the compressing
force employed, and hence also the last heat generated, must always be
considered.%
\footnote{\tiny In a compression the compressing force must exceed
  (slightly) the pressure, which therefore performs a negative work. In
  other words in a cycle the entropy variation is $0$ but the Clausius
  integral is $<0$.} 
It yields also for the necessary heat supplied no
longer the equality, instead it will be:%
\footnote{\tiny The latter comments do not seem to prove the inequality, unless
  coupled with the usual formulation of the second law (\eg in the Clausius
  form, as an inequality). On the other hand this is a place where external
  forces are taken into account: but in a later letter to Clausius, who
  strongly criticized his lack of consideration of external forces,
  Boltzmann admits that he has not considered external forces, see
  p.\pageref{B on ergale}, and does not refer to his comments
  above.\label{heat inequality} See also comment at p.\pageref{volume and
    pressure}.}

$$\int \int \frac{\d Q}T dk<0$$

\0[{\sl The following page deals with the key question of the need of closed
    paths: the lengthy argument concludes that what is really needed is
    that in arbitrarily long time two close paths remain close. The change
    of subject is however rather abrupt.}]\label{open paths}
\*
I will first of all consider times $t_1,t_2, t'_1$ and $t'_2$, and the
corresponding arcs, also in the case in which the atom in a given longer
time does not describe a closed path. At first the times $t_1$ and $t_2$
must be thought as widely separated from each other, as well separated as
wished, so that the average ``vis viva'' during $t_2-t_1$ would be the true
average ``vis viva''.  Then let $t'_1$ and $t'_2$ be so chosen that the
quantity
$$\frac{dx}{dt}\d x+\frac{dy}{dt}\d y+\frac{dz}{dt}\d z\eqno(25)$$
assumes the same value at both times. One easily convinces himself that
this quantity equals the product of the atom speed times the displacement
$\sqrt{\d x^2+\d y^2+\d z^2}$ times the cosine of the angle between them.
A second remark will also be very simple, if $s'_1$ and $s'_2$ are the
corresponding points, which lie orthogonally to the varied trajectory
across the points $s_1$ and $s_2$ on the initial trajectory, then the
quantity (25) vanishes for both paths.  This condition on the variation of
the paths, even if not be satisfied, would not be necessary for the
vanishing of the integrals difference, as it will appear in the following.
Therefore from all these arguments, that have been used above on closed
paths, we get 
 
$$\int\int \frac{\d Q}{T} dk=2\sum\log \frac{\int_{t_1}^{t_2} \frac{m c^2}2 dt}
{\int_{\t_1}^{\t_2} \frac{m c^2}2 dt},$$
if $\t_1$ and $\t_2$ are limits of the considered [{\sl varied}] path,
chosen in correspondence of the integral on the left.  It is now possible
to see that the value of this integral taken equally on both paths does not
vanish since, if one proceeds in the above described way, the normal plane
at $s_1$ at its intersection with the next path is again a normal plane and
in the end it is not necessary a return on the same curve again at the same
point $s_1$; only the point reached after a time $t_2-t_1$ will be found at
a finite not indefinitely increasing distance from $s_1$, hence

$$\ig_{t_1}^{t_2}\fra{m c^2}2 dt\quad {\rm and}\quad 
\ig_{\t_1}^{\t_2}\fra{m c^2}2 dt$$
now differ by finite amounts and the more the ratio

$$\fra{\ig_{t_1}^{t_2}\fra{m c^2}2 dt} {\ig_{\t_1}^{\t_2}\fra{m c^2}2 dt}$$
is close to unity the more its logarithm is close to zero\footnote{\tiny The role
  of this particular remark is not really clear (to me).}; the more
$t_2-t_1$ increases the more both integrals increase, and also more exactly
the average kinetic energy takes up its value; subdivide then both domains
of the integrals $\ig\ig \fra{\d Q}T dk$, so that one of the integrals
differs from the other by a quantity in general finite, thus the ratio and
therefore the logarithm does not change although it is varied by infinitely
many increments.

{\it This argument, together with the mathematical precision of the
  statement, is not correct in the case in which the paths do not close in
  a finite time, unless they could be considered closed in an
  infinite time} [{\sl italics added in the translation, p.30}].%
\label{infinite period}
\footnote{\tiny {\it I.e.} the distance between the points corresponding to $s'_2$
  and $s_2$ remains small forever: in other words, we would say, if no
  Lyapunov exponent is positive, \ie the motion is not
  chaotic.}

\*
\0[{\sl Having completed the very important discussion, which Clausius may
    have overlooked, see Sec.(\,\ref{sec:VII-6}\,), on the necessity of
    periodicity of the motion, Boltzmann returns to the conceptual analysis
of the results.}]
\*

It is easily seen that our conclusion on the meaning of the quantities that
intervene here is totally independent from the theory of heat, and
therefore the second fundamental theorem is related to a theorem of pure
mechanics to which it corresponds just as the ``vis viva'' principle
corresponds to the first principle; and, as it immediately follows from our
considerations, it is related to the\index{least action} 
least action principle, in form
somewhat generalized about as follows:\label{least}

``{\it If a system of point masses under the influence of forces, for which
the principle of the ``vis viva'' holds, performs some motion, and if
then all points undergo an infinitesimal variation of the kinetic energy and
are constrained to move on a path infinitely close to the precedent, then
$\d\sum \fra{m}2\,\ig c\, ds$ equals the total variation of the kinetic
energy multiplied by half the time interval during which the motion
develops, when the sum of the product of the displacements of the points
times their speeds and the cosine of the angles on each of the elements are
equal, for instance the points of the new elements are on the normal of the
old paths}''.

This proposition gives, for the kinetic energy transferred and if the
variation of the limits of integration vanishes, the least action principle
in the usual form.

It is also possible to interpret the problem differently; if the second
theorem is already considered sufficiently founded because of experiment
credit or other, as done by Zeuner [{\sl Zeuner G.,\Cite{Ze860}}], in his
new monograph on the mechanical theory of heat, and temperature is defined
as the integrating divisor of the differential quantity $d Q$, then the
derivation exposed here implies that the reciprocal of the value of the
average kinetic energy is the integrating factor of $\d Q$, hence
temperature equals the product of this average kinetic energy time an
arbitrary function of entropy.  Such entirely arbitrary function must be
fixed in a way similar to what done in the quoted case: it is then clear
that it will never be possible to separate the meaning of temperature from
the second theorem.

Finally I will dedicate some attention to the applicability of Eq.(24) to
the determination of the heat capacity.

Differentiation of the equality $T=\fra{\ig_{t_1}^{t_2}\fra{m
\,c^2}{2}\,dt}{t_2-t_1}$ leads to

$$\d T=\fra{\d\,\ig_{t_1}^{t_2}\fra{m \,c^2}{2}\,dt}{t_2-t_1}
-\fra{\ig_{t_1}^{t_2}\fra{m
\,c^2}{2}\,dt}{t_2-t_1}\cdot\fra{\d\,(t_2-t_1)}{t_2-t_1};
$$
and we shall look for the heat $\d \,H$ spent to increase temperature by
$\d T$ of all atoms

$$\d H=\sum \fra{\d\,\ig_{t_1}^{t_2}\fra{m \,c^2}{2}\,dt}{t_2-t_1}
-\sum \fra{\ig_{t_1}^{t_2}\fra{m
\,c^2}{2}\,dt}{t_2-t_1}\cdot\fra{\d\,(t_2-t_1)}{t_2-t_1};
$$
and combining with Eq.(24) it is found

$$\d Q=2\,\d\,H+2\sum \fra{\ig_{t_1}^{t_2}\fra{m
\,c^2}{2}\,dt}{t_2-t_1}\cdot\fra{\d\,(t_2-t_1)}{t_2-t_1};$$
and the work performed, both internal and external\footnote{\tiny See the 
Clausius' paper where this point is clearer; see also the final comment.}

$$\eqalign{
\d\,L=&\d H+2\sum \fra{\ig_{t_1}^{t_2}\fra{m
\,c^2}{2}\,dt}{t_2-t_1}\cdot\fra{\d\,(t_2-t_1)}{t_2-t_1}\cr
=& \sum \fra{\d\,\ig_{t_1}^{t_2}\fra{m \,c^2}{2}\,dt}{t_2-t_1}
+\sum \fra{\ig_{t_1}^{t_2}\fra{m
\,c^2}{2}\,dt}{t_2-t_1}\cdot\fra{\d\,(t_2-t_1)}{t_2-t_1}\cr}\eqno{(25a)}
$$
and the quantity

$$\d\,Z=\ig\fra{\d L}T\,dh=\sum
\fra{\d\,\ig_{t_1}^{t_2}\fra{m \,c^2}{2}\,dt}{\ig_{t_1}^{t_2}
\fra{m \,c^2}{2}\,dt}+\sum \fra{\d\,(t_2-t_1)}{t_2-t_1};$$
called by Clausius ``disgregation'' integral\footnote{\tiny It is the free
energy} has therefore the value

$$Z=\sum \log \ig_{t_1}^{t_2}\fra{m \,c^2}{2}\,dt
+\sum\log (t_2-t_1)+C.\eqno{(25b)}$$
In the case when $t_2-t_1$, which we can call period of an atom,
does not change it is: $\d\,(t_2-t_1)=0,\,\d Q=2\d H, \,\d
\,L=\d\,H$; \ie the heat transferred is divided in two parts, one for the
heating and the other as work spent.

Suppose now that the body has everywhere absolutely the same temperature
and also that it is increased remaining identical everywhere, thus
$\fra{\ig_{t_1}^{t_2}\fra{m \,c^2}{2}\,dt}{t_2-t_1}$ and
$\d\fra{\ig_{t_1}^{t_2}\fra{m \,c^2}{2}\,dt}{t_2-t_1}$ are equal for all
atoms and the heat capacity $\g$ is expressed by $\fra{\d Q}{p\d\,T}$, if
heat and temperature are expressed in units of work and $p$ is the weight
of the body:

$$\g= \fra{\d Q}{p\d\,T}= \fra{
2\d\ig_{t_1}^{T_2} \fra{m c^2}2 dt
}{\fra{p}N\Big[\d \ig_{t_1}^{T_2} \fra{m c^2}2 dt- 
\fra{\ig_{t_1}^{t_2}\fra{m
\,c^2}{2}\,dt}{t_2-t_1}\cdot\fra{\d\,(t_2-t_1)}{t_2-t_1}\Big]}$$
where $N$ is the number of atoms of the body and, if $a$ is the atomic
number or, in composed bodies, the total molecular weight and $n$ the
number of molecules in the atom, it will be $\fra{p}{N}=\fra{a}n$. In the
case $\d\,(t_2-t_1)=0$ it will also be $\fra{a\g}n=2$.
\footnote{\tiny The
hypothesis $\d\,(t_2-t_1)$ looks ``more reasonable''in the case of solid
bodies in which atoms can be imagined bounded to periodic orbits around the
points of a regular lattice.}
Therefore the product of the specific heat and the atomic weight is twice
that of a gas at constant volume, which is $=1$. This law has been
experimentally established by Masson for solids (see the published paper
``{\it Sur la correlation ...}'', Ann. de. Chim., Sec. III, vol. {\bf 53}),
\Cite{Ma858}; it also implies the isochrony of the atoms vibrations in
solids; however it is possibly a more complex question and perhaps I shall
come back another time on the analysis of this formula for solids; in any
event we begin to see in all the principles considered here a basis for the
validity of the Dulong-Petit's and Neumann's laws.

\section{Collision analysis and equipartition}
\def\SEC{Collision analysis and equipartition}
\label{sec:II-6}\iniz
\lhead{\small\ref{sec:II-6}.\ \SEC}

\0{Translation and comments of:  L. Boltzmann\index{Boltzmann},
{\it Studien {\"u}ber das Gleichgewicht der lebendigen Kraft zwischen
bewegten materiellen Punkten}, Wien. Ber., {\bf 58}, 517--560, 1868,
{W}is\-sen\-schaft\-li\-che {A}bhandlungen, ed. {F}. {H}asen\-{\"o}hrl,
{\bf 1}, \#5, (1868)},\,\Cite{Bo868-a}\,.
\*

All principles of analytic mechanics developed so far are limited to the
transformation of a system of point masses from a state to another,
according to the evolution laws of position and velocity when they are left
unperturbed in motion for a long time and are concerned, with rare
exceptions, with theorems of the ideal, or almost ideal, gas.  This might
be the main reason why the theorems of the mechanical theory of heat which
deal with the motions so far considered are so uncorrelated and
defective. In the following I shall treat several similar examples and
finally I shall establish a general theorem on the probability that the
considered point masses occupy distinct locations and velocities.

\*
\0{\bf I. The case of an infinite number of point masses}
\*

Suppose we have an infinite number of elastic spheres of equal mass and
size and constrained to keep the center on a plane. A similar more general
problem has been solved by Maxwell
(Phil. Mag. march 1868); however partly because of the non complete
exposition partly also because the exposition of Maxwell in its broad lines
is difficult to understand, and because of a typo (in formulae (2) and (21)
on the quantities called $dV^2$ and $dV$) will make it even more difficult, I
shall treat here the problem again from the beginning.

It is by itself clear that in this case every point of the plane is a
possibly occupied location of the center of one of the elastic spheres and
every direction has equal probability, and only the speeds remain to
determine.  Let $\f(c)dc$ be the sum of the time intervals during which the
speed of one of the spheres happens to have a value between $c$ and $c+dc$
divided by such very large time: which is also the probability that $c$ is
between $c$ and $c+dc$ and let $N$ be the average number of the spheres
whose center is within the unit of surface where the velocities are between
$c$ and $c+dc$.

\eqfig{190}{109}
{\ins{86}{98}{$\g$}
\ins{41}{33}{$\b$}
\ins{56}{38}{$\b'$}
\ins{58}{89}{$X$}
\ins{14}{19}{$O$}
\ins{50}{61}{$\g'$}
\ins{65}{44}{$A'_k$}
\ins{92}{80}{$A_k$}
\ins{105}{19}{$A_1$}
\ins{125}{47}{$A'_1$}
}
{wa1}{Fig.1}

\0Consider now a sphere, that I call $I$, with speed $c_1$ towards $OA_1$,
Fig.1, represented in size and direction, and let $OX$ the line joining the
centers at the impact moment and let $\b$ be the angle between the
velocities $c_1$ and $c_k$, so that the velocities components of the two
spheres orthogonal to $OX$ stay unchanged, while the ones parallel to $OX$
will simply be interchanged, as the masses are equal; hence let us
determine the velocities before the collision and let $A_1A'_1$ be parallel
to $OX$ and let us build the rectangle $A_1 A'_1 A_k A'_k$; $OA'_1$ and
$OA'_k$ be the new velocities and $\b'$ their new angle.  Consider now the
two diagonals of the rectangle $A_1A_k$ and $A'_1A'_k$ which give the
relative velocities of the two spheres, $g$ before and $g'$ after the
collision,  and call $\g$ the angle between the lines $A_1A_k$ and $OX$,
and $\g'$ that between the lines $A'_1A'_k$ and $OX$; so it is easily
found:
$$\eqalignno{
g^2=& c_1^2+c_2^2-2 c_1 c_k \sin\g\cdot\g\sin\b\cr
c^{\prime2}_1=&c_1^2\sin^2\g+c_k^2\cos^2\g -2 c_1
c_k\sin\g\cos\g\sin\b&(1)\cr
c^{\prime2}_2=&c_1^2\cos^2\g+c_k^2\sin^2\g +2 c_1
c_k\sin\g\cos\g\sin\b\cr
{\rm tan}\,\b'=&\fra{(c_1^2-c_k^2)\sin\g\,\cos\g - c_1c_k(\cos^2\g-\sin^2
\g)\sin\b}{c_1c_k\cos\b}\cr=&\fra{\textstyle\sqrt{\textstyle
c^{'2}_1c^{'2}_k-c_1^2c_k^2
\cos^2\b}}{c_1c_k\cos\b}
\cr
&c_1c_k\cos\b=c'_1c'_k\cos\b';\qquad \g'=\p-\g.\cr
}$$
We immediately ask in which domains $c'_1$ and $c'_k$ are, given $c_1,c_k$.
For this purpose .......
\*

\0{\bf[}{\sl A long analysis follows about the relation between the area
elements $d^2\V c_1 d^2\V c_k$ and the corresponding $d^2\V c'_1 d^2\V
c'_k$.  Collisions occur with an angle, between the collision direction and
the line connecting the centers, between $\b$ and $\b+d\b$ with probability
proportional to $\s(\b)d\b$ where $\s(\b)$ is the cross section (equal to 
$\s(\b)=\fra12r \sin\b$ in the case, studied here, of disks of radius $r$).
Then the density $f(\V c) d^2\V c$ must have the property $\f(\V c_1) f(\V
c_k)\s(\b)d\b= \f(\V c'_1) f(\V c'_k)$ $\s(\p-\b)d\b\,\cdot J$ where $J$ is
the ratio $\fra{d^2\V c'_1 d^2\V c'_k}{d^2\V c_1 d^2\V c_k}$, if the
momentum and kinetic energy conservation relations hold: $\V c_1+\V c_k=\V
c'_1+\V c'_k$ and $\V c_1^2+\V c_k^2=\V c^{'2}_1+\V c^{'2}_k$ and if the
angle between $\V c_1-\V c_k$ and $\V c'_k-\V c'_1$ is $\b$.

The analysis leads to the conclusion, well known,
that $J=1$ and therefore it must be  $f(\V c_1) f(\V c_k)= f(\V c'_1) f(\V
c'_k)$ for all four velocities that satisfy the conservation laws of
momentum and energy: this implies that $f(\V c)=const \,e^{-h c^2}$.  
Boltzmann uses always the directional uniformity supposing $f(\V c)=\f(c)$
and therefore expresses the probability that the modulus of the velocity is
between  $c$ and $c+dc$ as $\f(c) c dc$ and therefore
the result is expressed by $\f(c)=b\,c\,e^{-h c^2}$, with
$b=2h$ a normalization constant (keeping in mind that the
$2$-dimensional case is considered).

In reality if systematic account was taken of the volume preserving
property of canonical transformations (\ie to have Jacobian $1$) the
strictly algebraic part of evaluating the Jacobian, would save several
pages of the paper. It is interesting that Boltzmann had to proceed to
rediscover this very special case of a general property of Hamiltonian
mechanics.

Having established this result for planar systems of elastic disks (the
analysis has been {\it very} verbose and complicated and B. admits that
``Maxwell argument was simpler but he has chosen on purpose a direct approach
based on very simple examples'', p.58), Boltzmann considers the
$3$--dimensional case in which the interaction is more general than elastic
collision between rigid spheres, and admits that it is described by a
potential $\ch(r)$, {\it with short range}. However he says that, since
Maxwell has treated completely the problems analogous to the ones just
treated in the planar case, he will study a new problem. Namely:
\*

\0``{\rm Along a line $OX$ an elastic ball
 of mass $M$ is moving attracted by $O$ with a force depending only on the
distance. Against it are moving other elastic balls of mass $m$ and their
diverse velocities during disordered time intervals dart along the same
line, so that if one considers that all flying balls have run towards $O$
long enough on the line $OX$ without interfering with each other, the
number of balls with velocity between $c$ and $c+dc$, which in average are
found in the unit length, is a given function of $c$, \ie $N \f(c) dc$.

The potential of the force, which attracts $M$ towards $O$ be $\chi(x)$,
hence as long as the motion does not encounter collisions it will be

$$\frac{M C^2}2=\chi(x)+A\eqno{(9)}$$
where $C$ is the speed of the ball $M$ and $x$ the distance between its
center and $O$. Through the three quantities $x,A$ and $c$ the kind of
collision is fixed. The fraction of time during which the constant $A$ of
Eq.(9) will be between $A$ and $A+dA$ be $\F(A)dA$.  The time during which
again $x$ is between the limits $x$ and $x+dx$ behaves as $\frac{dx}C$, and
we shall call $t(A)$, as it is a function of $A$, the fraction of time
during which the segment $dx$ is run and $x$ grows from its smallest value
to its largest. Consider a variation of the above values that
is brought about by the collisions, we want to compare 
the time interval between two collisions with $t(A)$; this time is 
$$\frac{\F(A) dA\, dx}{C t(A)}
... $$}
\*

\0The discussion continues (still very involved) to determine the balance
between collisions that change $A,x,C$ into $A',x',C'$: it very much
resembles Maxwell's treatment in \Cite{Ma867-b}[XXVIII, vol.2] and is a
precursor of the later development of the Boltzmann's equation,
\Cite{Bo872}[\#22], and follows the same path. The analysis will reveal
itself very useful to Boltzmann in the 1871 ``trilogy'' and then in the
1872 paper because it contains all technical details to be put together to
obtain the Boltzmann's equation.  The result is

$$\f(c)= b e^{-h \cdot\frac{m c^2}2},\quad \frac{\F(A) dA dx}{C t(A)}= 
2 Be^{h[\chi(x)-\frac{M C^2}2]}$$
with $2B$ a normalization, and it has to be kept in mind that only events
on the line $OX$ are considered so that the problem is essentially
$1$--dimensional.

The $3$-dimensional corresponding problem it treated in the rest of Sec.I,
subsection 3, and a new related problem is posed and solved in subsections
4 (p.70) and 5 (p.73).  There a point mass, named I with mass $M$, is
imagined on the on a line $OX$ attracted by $O$ and a second kind point
masses, named II, with mass $m$, interacting with I via a potential with
short range $\ell$. It is supposed that the fraction of time the point II
has speed between $c$ and $c+dc$ (the problem is again $1$-dimensional) is
$N\f(c)dc$ and that events in which two or more particles II\label{multiple
collisions} come within $\ell$ of I can be neglected.  The analysis leads
to the ``same'' results of subsections 2 and 3 respectively for the $1$ and
$3$ dimensional cases.}{\bf ]}
\*

\0{\bf II. On the equipartition  of the ``vis viva'' for a finite number of
point masses} (p.80)
\*

In a very large, bounded in every direction, planar region let there be $n$
point masses, of masses $m_1,m_2,\ldots,m_n$ and velocity
$c_1,c_2,\ldots,c_n$, and 
between them act arbitrary forces, which {\it just begin
to act at a distance which vanishes compared to their mean
distance} [{\sl italics added}].\label{low density}\label{interaction
  range}
\footnote{\tiny Often it is stated that Boltzmann does not consider cases in which
  particles interact: it is here, and in the following, clear that he
  assumes interaction but he also assumes that the average distance between
  particles is very large compared to the range of interaction. This is
  particularly important also in justifying the later combinatorial
  analysis. See also below.}

Naturally all directions in
the plane are equally probable for such velocities. But the probability
that the velocity of a point be within assigned limits and likewise that
the velocity of the others be within other limits, will certainly not be
the product of the single probabilities; the others will mainly
depend on the value chosen for the velocity of the first point.  The
velocity of the last point depends from that of the other $n-1$, because
the entire system must have a constant amount of ``vis viva''.

I shall identify the fraction of time during which the velocities are so
partitioned that $c_2$ is between $c_2$ and $c_2+dc_2$,
likewise $c_3$ is between $c_3$ and $c_3+dc_3$ and so on until $c_n$, 
with the probability $\f_1(c_2,c_3,\ldots,c_n) dc_2\,dc_3\ldots
dc_n$ for this velocity configuration.

The probability that $c_1$ is between $c_1$ and $c_1+dc_1$and the
corresponding velocities different from $c_2$ are between analogous limits
be $\f_2(c_1,c_3,\ldots,c_n)\cdot$ $dc_1\,dc_3\ldots dc_n$, {\it etc.}.

Furthermore let
$$\fra{m_1 c_1^2}2=k_1,\ \fra{m_2 c_2^2}2=k_2,\ \ldots\ \fra{m_n
c_n^2}2=k_n$$
be the kinetic energies and let the probability that $k_2$ is between $k_2$
and $k_2+dk_2$, $k_3$ is between $k_3$ and $k_3+dk_3\,\ldots$ until $k_n$
be $\ps_1(k_2,k_3,\ldots,k_n)\,dk_2$ $dk_3\ldots dk_n$. And analogously
define $\ps_2(k_1,k_3,\ldots,k_n)\,dk_1\,dk_3\ldots dk_n$
\etc., so that
$$\eqalign{
&m_2 c_2\cdot m_3 c_3\ldots m_nc_n \,
\psi_1(\fra{m_2 c_2^2}2,\fra{m_3 c_1^3}2,\ldots,
\fra{m_n c_n^2}2)=\f_1(c_2,c_3,\ldots,c_n)\quad{\rm or}\cr
&\f_1(c_2,c_3,\ldots,c_n)=2^{\fra{n-1}2}\,\sqrt{m_2m_3\ldots m_n}\,
\sqrt{k_2 k_3\ldots k_n}\,\ps_1(k_2,k_3,\ldots,k_n)\cr}$$
and similarly for the remaining $\f$ and $\ps$.

Consider a collision involving a pair of points, for instance $m_r$ and
$m_s$, which is such that $c_r$ is between $c_r$ and $c_r+d c_r$, and $c_s$
is between $c_s$ and $c_s+dc_s$. Let the limit values of these quantities
after the collision be between $c'_r$ and $c'_r+d c'_r$ and $c'_s$ be
between $c'_s$ and $c'_s+dc'_s$.

It is now clear that the equality of the ``vis viva'' will remain valid
always in the same way when many point, alternatively, come into collision
and are moved within different limits, as well as the other quantities
whose limits then can be remixed, among which there are the velocities of
the remaining points. [{\sl Here it seems that the constancy of the total
kinetic energy is claimed to be clear: which seems strange since at the
same time a short range interaction is now present. The reason behind this
assumption seems that, as B. says at the beginning of Sec.II, (p.80), the
range of the forces is small compared to the mean interparticle distance.}]

The number of points that are between assigned limits of the velocities,
which therefore have velocities between $c_2$ and $c_2+dc_2\ldots$, are
different from those of the preceding problems because instead of the
product $\f(c_r)dc_r\f(c_s)dc_s$ appears the function
$\f_1(c_2,c_3,\ldots,c_n) dc_2\,dc_3\ldots$ $dc_n$. This implies that
instead of the condition previously found

$$\fra{\f(c_r)\cdot\f(c_s)}{c_r\cdot
c_r}=\fra{\f(c'_r)\cdot\f(c'_s)}{c'_r\cdot c'_r}$$
the new condition is found:
$$\fra{\f_1(c_2,c_3,\ldots ,c_n)}{c_r\cdot
c_r}=\fra{\f_1(c_2,\ldots,c'_r,\ldots ,c'_s,\ldots, c_n)}{c_r
\cdot c_s}$$
The same holds, of course, for $\f_2,\f_3,\ldots$. If the function $\f$ is
replaced by $\ps$ it is found, equally,
$$\ps_1(k_2,k_3,\ldots,
k_n)=\ps_1(k_2,k_3,\ldots,k'_r,\ldots,k'_s,\ldots, k_n),\qquad {\rm
if}\ k_r+k_s=k'_r+k'_s.
$$
Subtract the differential of the first of the above relations 
$\fra{d\ps_1}{dk_r}dk_r  +\fra{d\ps_1}{dk_s} dk_s=
 \fra{d\ps_1}{dk'_r}dk'_r+\fra{d\ps_1}{dk'_s}dk'_s$
that of the second  [$dk_r+dk_s=dk'_r+dk'_s$] multiplied by $\l$
and set equal to zero the coefficient of each differential,
so that it is found:
$$\l=\fra{d \ps_1}{dk_r}=\fra{d \ps_1}{dk_s}=\fra{d \ps_1}{dk'_r}=
\fra{d \ps_1}{dk'_s}.$$
{\it I.e.}, in general,
$\fra{d\ps_1}{dk_2}=\fra{d\ps_1}{dk_3}=\fra{d\ps_1}{dk_4}=
\ldots \fra{d\ps_1}{dk_n}$, hence $\ps_1$ is function of
$k_2+\ldots+k_n$. Therefore we shall write $\ps_1(k_2,\ldots,k_n)$ 
in the form $\ps_1(k_2+k_3+\ldots+k_n)$.  We must now find the meaning of the
equilibrium about $m_1$ and the other points. And we shall determine the
full $\ps_1$.

It is obtained simply with the help of the preceding $\ps$ of which of
course the $\ps_1$ must be a sum. But these are all in reciprocal
relations. If in fact the total ``vis viva'' of the system is $n\k$, it is

$$k_1+k_2+\ldots+k_n=n\k$$
It follows that $\ps_1(k_2+k_3+\ldots+k_n) dk_2dk_3\ldots dk_n$
can be expressed in the new variables\footnote{\tiny In the formula $k_2$ and
$k_1$ are interchanged.}

$$k_3,k_4,\ldots, n\k-k_1-k_3-\ldots-k_n=k_2$$
and it must be for $\ps_2(k_1+k_3+\ldots+k_n) dk_1dk_3\ldots
dk_n$. Hence $\ps_1(k_2+k_3+\ldots+k_n)$ can be converted in
$\ps_1(n\k-k_1)$
and $dk_2dk_3\ldots dk_n$ in $dk_1dk_3\ldots dk_n$. Hence also

$$\ps_1(n\k-k_1)=\ps_2(k_1+k_3+\ldots+k_n)=\ps_2(n\k-k_2)=\ps_2(n\k-k_2)$$
for all $k_1$ and $k_2$, therefore all the $\ps$ are equal to the same
constant $h$. This is also the probability that in equal time intervals it
is $k_1$ between $k_1$ and $k_1+dk_1$,  $k_2$ is between $k_2$ and $k_2+dk_2$
\etc, thus for a suitable $h$, it is $h\, dk_1\,dk_2\,\ldots\,dk_n$
were again the selected differential element must be absent. Of course this
probability that at a given instant $k_1+k_2+k_3+\ldots$ differs from $n\k$
is immediately zero.

The probability that $c_2$ is between $c_2$ and $c_2+dc_2$, $c_3$ between
$c_3$ and $c_3+dc_3\ldots$ is given by

$$\f_1(c_2,c_3,\ldots,c_n)\, dc_2\,dc_3\,\ldots\,dc_n=
m_2m_3\ldots m_n \cdot h \cdot c_2c_3\ldots c_n dc_2\,dc_3\,\ldots\,dc_n.$$
Therefore the point $c_2$ is in an annulus of area
$2\p c_2 dc_2$, the point $c_3$ in one of area $2\p c_3 dc_3$ \etc, 
that of  $c_1$ on the boundary of length $2\p c_1$ of a disk
and all points have equal probability of being in such annuli.

Thus we can say: the probability that the point $c_2$ is inside the area 
$d\s_2$, the point $c_3$ in $d\s_3$ \etc, while  $c_1$ is on a line element
$d\o_1$, is proportional to the product

$$\fra1{c_1} \,d\o_1\,d\s_2\,d\s_3\, \ldots\, d\s_n,$$
if the mentioned locations and velocities, while obeying the principle of
conservation of the ``vis viva'', are not impossible.

We must now determine the fraction of time during which the ``vis viva'' of
a point is between given limits $k_1$ and
$k_1+dk_1$, without considering the ``vis viva'' of the other points. For
this purpose subdivide the entire ``vis viva'' in infinitely small equal
parts $(p)$, so that if now we have two point masses, for $n=2$ the
probability that $k_1$ is in one of the $p$ intervals
$[0,\fra{2\k}p]$, $[\fra{2\k}p,\fra{4\k}p]$, $[\fra{4\k}p,\fra{6\k}p]$
\etc is equal and the problem is solved.

For $n=3$ if $k_1$ is in $[(p-1)\fra{3\k}p,p\fra{3\k}p]$, then
$k_2$ and $k_3$ must be in the interior of the  $p$ intervals. If $k_1$
is in the next to the last interval, \ie if

$$(p-2)\fra{3\k}p\le k_1\le (p-1)\fra{3\k}p$$
two cases are possible ....
\*

\0{\bf[}{\sl Here follows the combinatorial calculation of the number of
ways to obtain the sum of $n$ multiples $p_1,\ldots,p_n$ of a unit $\e$ and
$p_1\e=k_1$ such that $\sum_{i=2}^{n-1} p_i\e=n\k-p_1\e$, and B. chooses
$\e=\fra{2\k}p$ with $p$ ``infinitely large'': \ie

$$\sum_{p_2=0}^{n\k/\e- p_1}\ \sum_{p_3=0}^{n\k/\e-p_1-p_2}\ldots\ldots
\sum_{p_{n-1}=0}^{n\k/\e-p_1-\ldots-p_{n-2}} 1$$
the result is obtained by explicitly treating the cases $n=2$ and $n=3$
and inferring the general result in the limit in which $\e\to0$.

The ratio between this number and the same sum performed also on $p_1$
is, fixing $p_1\in [k_1/\e,(k_1+dk_1)/\e]$, 

$$\eqalign{&\fra{dk_1\ig_0^{n\k-k_1}dk_2\ig_0^{n\k-k_1-k_2} dk_3\,\ldots\,
\ig_0^{n\k-k_1-k_2-\ldots-k_{n-2}} dk_{n-1}}
{\ig_0^{n\k}dk_1\ig_0^{n\k-1}dk_2\ig_0^{n\k-k_1-k_2} dk_3\,\ldots\,
\ig_0^{n\k-k_1-k_2-\ldots-k_{n-2}} dk_{n-1}}\cr
=&\fra{(n-1) (n\k-k_1)^{n-2} dk_1}{(n\k)^{n-1}},\cr}$$
This is, however, remarked {\it after} the explicit combinatorial analysis
of the cases $n=2$ and $n=3$ from which the last equality is inferred in
general (for $\e\to0$).

Hence the ``remark'' is in reality a proof simpler than the
combinatorial analysis of the number of ways to decompose the
total energy as sum of infinitesimal energies. The choice of
B. is certainly a sign of his preference for arguments based on a
discrete view of physical quantities.  And as remarked in
\Cite{Ba990} this remains, whatever interpretation is given to
it, an important analysis of B.

In the successive limit, $n\to\io$, the Maxwell's distribution is obtained.

$$\fra1\k e^{-k_1/\k}dk_1$$
concluding the argument.
\*

In the next subsection 7 B. repeats the analysis in the $3$-dimensional
case obtaining again the Maxwellian distribution for the speed of a single
particle in a system of $n$ point masses in a finite container with
perfectly elastic walls.

Finally in Sec. III.8 the case is considered in which also an external
force acts whose potential energy is $\ch$ (not to be confused with the
interparticle potential energy, also present and denoted with the same
symbol; and which is always considered here as acting instantaneously at
the collision instant, as assumed at the beginning of Sec. II). 

The Sec. III is concluded as follows {\bf]}
\* }

\0{\bf p.96.} As special case from the first theorem it follows, as already
remarked in my paper on the mechanical interpretation of the second
theorem, that the ``vis viva'' of an atom in a gas is equal to that of the
progressive motion of the molecule.\footnote{\tiny to which the atom
  belongs} The latter demonstration also contains 
the solution of others that were left incomplete: it tells us that for such
velocity distributions the balance of the ``vires vivae'' is realized in a
way that could not take place otherwise.

An exception to this arises when the variables
$x_1,y_1,z_1,x_2,\ldots,v_n$ are not independent of each other.
This can be the case of all systems of points in which the variables have
special values, which are constrained by assigned relations that remain
unaltered in the motions, but which under perturbations can be destroyed
(weak balance of the ``vis viva''), for instance when all points and the
fixed centers are located on a mathematically exact line or plane.

A stable balance is not possible in this case, when the points of the
system are so assigned that the variables for all initial data within a
certain period come back to the initial values, without having consequently
taken all values compatible with the `principle of the ``vis viva''.
Therefore such way of achieving balance is always so infinitely more
possible that immediately the system ends up in the set of domains
discussed above when, for instance, a perturbation acts on a system of
points which evolves within a container with elastic walls or, also, on an
atom in free motion which collides against elastic walls.\annotaa{This
last paragraph seems to refer to lack of equipartition in cases in which
the system admits constants of motion due to symmetries that are not
generic and therefore are destroyed by ``any'' perturbation.}
\*

\0[{\sl Boltzmann is assuming that the potential energy could be given its
    average value, essentially $0$, and be replaced by a constant as a {\it
      mean field}: because he, Sec.II, (p.80), assumes that the range of
    the forces is small compared to the mean interparticle distance and
    (tacitly here, but explicitly earlier in subsection 4,
    p.\pageref{multiple collisions}) that multiple collisions can be
    neglected. The fraction of time in which the potential energy is
    sizable is supposed ``by far'' too small to affect averages: this is
    not so in the case of gases consisting of polyatomic molecules as he
    will discuss in detail in the first paper of the trilogy,
    \Cite{Bo871-a}[\#17]. The analysis of the problem in modern terms would
    remain essentially the same even not neglecting that the total kinetic
    energy is not constant if the interaction between the particles is not
    pure hard core: in modern notations it would mean studying (in absence
    of external forces, for simplicity)

$$  \frac{\int \d(\sum_j \frac12m c_j^2+\sum_{i,j} \chi(x_i-x_j)- Nu)
d^{3N-1}\V c d^{3N}\V x}{\int \d(\sum_j \frac12m c_j^2+\sum_{i,j} 
\chi(x_i-x_j)- Nu)
d^{3N}\V c d^{3N}\V x}
$$
where $u$ is the specific total energy, if the pair interaction is short
range and stable (in the sense of existence of a constant $B$ such that for
all $N$ it is $\sum_{i,j}^N \chi(x_i-x_j)>-BN$) and the integral in the
numerator is over all velocity components but one: the analysis would then
be easily reduced to the case treated here by B.

In \Cite{Ba990} the question is raised on whether Boltzmann would have
discovered the Bose-Einstein distribution before Planck\index{Planck}
referring to the way he employs the discrete approach to compute the number
of ways to distribute kinetic energy among the various particles, after
fixing the value of that of one particle, in
\Cite{Bo868-a}[\#5,p.84,85]. This is an interesting and argumented view,
however Boltzmann considered here the discrete view a ``fiction'', see also
\Cite{Bo877-b}[\#42,p.167] and Sec.(\,\ref{sec:XII-6}\,) below,
and the way the computation is done would not distinguish whether
particles were considered distinguishable or not: the limiting
case of interest would be the same in both cases (while it would
be quite different if the continuum limit was not taken, leading
to a Bose-Einstein like distribution). This may have prevented
him to be led to the extreme consequence of considering the
difference physically significant and to compare the predictions
with those that follow in the continuum limit with the
distribution found with distinguishable particles, discussed
later in \Cite{Bo877-b}, see below Sec.(\,\ref{sec:XII-6}\,) and
also \Cite{Ga000}[Sec.(2.2),(2.6)].}]

\section{Dense orbits: an example}
\def\SEC{Dense orbits: an example}
\label{sec:III-6}\iniz
\lhead{\small\ref{sec:III-6}.\ \SEC}

\0{Comments on:  L. Boltzmann,\index{Boltzmann}
\it L{\"o}sung eines mechanischen Problems}, Wis\-sen\-schaft\-li\-che 
Abhandlungen, ed. F. Hasen\"ohrl, {\bf1}, \#6, 97--105,
(1868), \Cite{Bo868-b}\,.
\*

\0{\sl The aim of this example it to exhibit a simple case in which the
difficult problem of computing the probability of finding a point mass
occupying a given position with given velocity.  

Here B. presents an example where the ideas of the previous work,
Sec.(\,\ref{sec:II-6}\,), can be followed via exact calculations. A point mass
[{\sl mass $=1$}] subject to a central gravitational force but with a
centrifugal barrier augmented by a potential $+\fra\b{2R^2}$ and which is
reflected by an obstacle consisting in a straight line, \eg $x=\g>0$.

The discussion is an interesting example of a problem in ergodic theory for
a two variables map. Angular momentum $a$ is conserved between collisions
and there motion is explicitly reducible to an elementary quadrature which
yields a function (here $r\equiv R$ and $A$ is a constant of motion equal
to twice the total energy, constant because collisions with the line are
supposed elastic):

$$F(r,a,A)\defi \frac{a}{\sqrt{a^2+\b}}{\rm
arccos}(\frac{\frac{2(a+\b)}r-\a}
{\sqrt{\a^2+4 A(a^2+\b)}})$$ 
such that the polar angle at time $t$ is $\f(t)-\f(0)=
F(r(t),a_0,A)-F(r(0),a_0,A)$. Let $\e_0\defi \f(0)-F(r(0),a_0,A)$, then if
$\f_0,a_0$ are the initial polar angle and the angular momentum of a motion
that comes out of a collision at time $0$ then $r(0)\cos\f_0=\g$ and
$\f(t)-\e_0=F(r(t),a_0,A)$ until the next collision. Which will take place
when $\f_1-\e_0=F(\frac\g{\cos\f_1},a_0,A)$ and if $a_1$ is the outgoing
angular momentum from then on $\f(t)-\e_1=F(r(t),a_1,A)$ with
$\e_1\defi \f_1-F(\frac\g{\cos\f_1},a_1,A)$.

Everything being explicit B. computes the Jacobian of the just defined map
$S: (a_0,\e_0)\to (a_1,\e_1)$ and shows that it it $1$ (which is carefully
checked without reference to the canonicity of the map). The map is
supposed to exist, \ie that the Poincar\'e's section defined by the timing
event ``hit of the fixed line'' is transverse to the solution flow (which
may be not true, for instance if $A<0$ and $\g$ is too large). Hence the
observations timed at the collisions has an invariant measure $d\e d a$: if
the allowed values of $a,\e$ vary in a bounded set (which certainly happens
if $A<0$) the measure $\frac{d\e da}{\int d\e da}$ is an invariant
probability measure, \ie the microcanonical distribution, which can be used
to compute averages and frequency of visits to the points of the plane
$\e,a$. The case $\b=0$ would be easy but in that case it would also be
obvious that there are motions in which $\e,a$ does not roam on a dense
set, so it is excluded here.

The interest of B. in the example seems to have been to show
that, unless the interaction was very special (\eg $\b=0$) the
motion would invade the whole energy surface, in essential
agreement with the idea of ergodicity.  In reality {\it neither
density nor ergodicity is proved}.  It is likely that the
confined motions of this system are quasi periodic unless $A$ has
special (dense) values corresponding to ``resonant'' (\ie
periodic) motions. B. does not make here comments about the
possible exceptional (``resonant'') values of $E$; assuming that
he did not even think to such possibilities, it is clear that he
would not have been shocked by their appearance: at least for
many value of $E$ (\ie but for a zero measure set of $E'$'s) the
system would admit only one invariant measure and that would be
the microcanonical one, and this would still have been his point.
}. The relation between the example and KAM theorem has been made
precise in the conjectures in \Cite{Ga016}\Cite{GJ020}: it is now
proved as a theorem by Felder, \Cite{Fe021}. The, little known, Boltzmann's
work has also been noticed in \Cite{Da018}.
\*

\section{Clausius' version of recurrence and periodicity}
\def\SEC{Clausius' version of recurrence and periodicity}
\label{sec:IV-6}\iniz
\lhead{\small\ref{sec:IV-6}.\ \SEC}

\0{Translation and comments on Sec.10 of: R. \index{Clausius},
{\it Ueber die Zur{\"u}ck\-f{\"u}hrung des zweites Hauptsatzes der
                mechanischen W{\"a}rmetheorie und allgemeine
                mechanische Prinzipien}, Annalen der Physik, {\bf
                142}, 433--461, 1871.}
\*

\0{\sl Sec.10 deals with the necessity of closed atomic paths 
in the derivation of the second theorem of thermodynamics from
mechanics.}
\*
\0{\bf 10.} So far we considered the simple case of an isolated point 
moving on a closed path and we shall now consider more complicated cases.

We want to consider a very large number of point masses, interacting by
exercising a reciprocal force as well as subject to an external force.  It
will be supposed that the points behave in a stationary way under the
action of such force. Furthermore it will be supposed that the forces have
an {\it ergale},\footnote{\tiny {\it I.e.} a potential energy.}\index{ergale} \ie
that the work, performed by all forces upon an infinitesimal displacement
be the differential, with sign changed, of a function of all
coordinates. If the initial stationary motion is changed into a varied
stationary motion, still the forces will have an ``ergale'' [potential
  energy], which does not depend only on the changed position of the point,
but which also can depend from other factors. The latter can be thought, from a
mathematical viewpoint, by imagining that the ergale [the potential energy] is
a quantity that in every stationary motion is constant, but it has a value
that can change from a stationary motion to another.

Furthermore we want to set up an hypothesis which will clarify the
following analysis and correlates its contents which concern the motion
that we call heat, that the system consists in only one chemical species,
and therefore all its atoms are equal, or possibly that it is composed, but
every species contains a large number of atoms.  It is certainly not
necessary that all these atoms are of the same species.  If for instance
the body is an aggregate of different substances, atoms of a part move as
those of the others. Then, after all, we can suppose that every possible
motion will take place as one of those followed by the large number of
atoms subject to equal forces, proceeding in the same way, so that also the
different phases\footnote{\tiny Here it is imagined that each atom moves on a
 possible orbit but different atoms have different positions on the orbit,
 at any given time,  which is called its ``phase''.} of such motions
will be realized.  What said means that we want to suppose that in our
system of point masses certainly we can find among the large number of the
same species a large number which goes through the same motion under the
action of equal forces and with different phases.

Finally temporarily and for simplicity we shall assume, as already done
before, that all points describe closed trajectories, For such points about
which we are concerned and which move in the same way we suppose, more in
particular, that they go through equal trajectories with equal periods.  If
the stationary motion is transformed into another, hence on a different
trajectory with different period, nevertheless it will still follow a
closed trajectories each of which will be the same for a large number of
points.\footnote{\tiny The assumption differs from the ergodic hypothesis
  and it can be seen as an assumption that all motions are quasi periodic
  and that the system is integrable: it is a view that {\it mutatis
    mutandis} resisted until recent times both in celestial mechanics, in
  spite of Poincar\'e's work, and in turbulence theory as in the first few
  editions of Landau-Lifschitz' treatise on fluid mechanics, \Cite{LL971}.}

\section{Clausius' mechanical proof of the heat theorem}
\def\SEC{Clausius' mechanical proof of the heat theorem}
\label{sec:V-6}\iniz
\lhead{\small\ref{sec:V-6}.\ \SEC}
\*

\0{Translation of \S13,\S14,\S15, and comments: R. Clausius\index{Clausius},
{\it Ueber die Zur{\"u}ck\-f{\"u}hrung des zweites Hauptsatzes der
mechanischen W{\"a}rmetheorie und allgemeine
mechanische Prinzipien}, Annalen der Physik, {\bf 142}, 433--461, 1871.}
\*

\0[{\sl The translation is here because I consider several sentences in it
    interesting to appear  within their context: the reader interested to a
    rapid self contained summary of Clausius' proof is referred to
    Sec.(\,\ref{sec:IV-1}\,) and Sec.(\,\ref{sec:I-4}\,) above.}]
\*
\0{\bf 13.} In the present work we have supposed until now that all points
move on closed paths. We now want to set aside also this assumption and
concentrate on the hypothesis that the motion is stationary.

For motions that do not run over closed trajectories the notion of
recurrence is no longer usable in a literal sense, therefore it is
necessary to talk about them in another sense. Consider therefore right
away motions that have a given component in a given direction, for instance
the $x$ direction in our coordinates system.  It is then clear that motions
go back and forth alternatively, and also for the elongation, speed and
return time, as it is proper for stationary motion, the same form of motion
is realized.  The time interval within which each group of points which
behave, approximately, in the same way admits an average value.

Denote with $i$ this time interval, so that without doubt we can consider
valid, also for this motion, the Eq.(28), [{\sl \ie the equality, in
average, of the two sides of}]

$$-\sum m \fra{d^2 x}{dt^2}\,\d x= \sum \fra{m}2 \d (\fra {d
x}{dt})^2+\sum m (\fra {d
x}{dt})^2 \d \log i$$
[{\sl here $\d x$ denotes a variation of a quantity $x$ between two
    infinitely close stationary states}].\footnote{\tiny In Clausius $\d
  x=x'(i'\f)-x(i\f)$, $t'=i'\f$ and $t=i\f$ is defined much more clearly
  than in Boltzmann, through the notion of {\it phase} $\f\in [0,1]$
  assigned to a trajectory, and calculations are performed up to
  infinitesimals of order higher than $\d x$ and $\d i=(i'-i)$.} The above
equation can also be written for the components $y,z$, and of course we
shall suppose that the motions in the different directions behave in the
same way and that, for each group of points, the quantity $\d\log i$
assumes the same value for the three coordinates.

If then with the three equations so obtained we proceed as above for the
Eq. (28),(28b),(28c), we obtain the Eq.(31):\footnote{\tiny $\d L$ is the
  work in the process.  It seems that here the integration of both sides is
  missing, or better the sign of average over $v^2$, which instead is
  present in the successive Eq.(32).}

$$\d L=\sum \fra{m}2 \d v^2 +\sum m v^2 \d\log i$$

\0{\bf 14.} To proceed further to treat such equations a difficulty arises
because the velocity $v$, as well as the return time interval $i$, may be
different from group to group and both quantities under the sum sign cannot
be distinguished without a label. But imagining it the distinction will be
possible allowing us to keep the equation in a simpler form.

Hence the different points of the system, acting on each other, interact so
that the kinetic energy of a group cannot change unless at the concomitant
expenses of another, while always a balance of the kinetic energies of the
different points has to be reached, before the new state can be stationary.
We want to suppose, for the motion that we call heat, that a balance is
established between the kinetic energies of the different points and a
relation is established, by which each intervening variation reestablishes
the kinetic energy balance. Therefore the average kinetic energy of each
point can be written as $m c T$, where $m$ is the mass of the points and
$c$ another constant for each point, while $T$ denotes a variable quantity
equal for all points.

Inserting this in place of $\fra{m}2 v^2$ the preceding equation becomes:

$$\d L=\sum m \,c\, \d T+\sum
2m\, c\, T \,\d\log i \eqno{(32)}$$
Here the quantity $T$ can become a common factor of the second sum. We can,
instead, leave the factor  $\d T$ inside the first sum. We get

$$\eqalign{\d L=&\sum m\, c \,\d T+ T\sum
2\,m\,c\,\d\log i\cr
=&T( \sum m\, c\, \fra{\d T}T+ \sum
2\,m\,c\,\d\log i)\cr
=&T (\sum m\, c\, \d \log T+ \sum
2m\,c\,\d\log i)\cr}\eqno{(33)}$$
or, merging into one both sums and extracting the symbol of variation

$$T \d \sum (m\, c\, \log T+ 2\,\log i)$$
from which we finally can write

$$\d L=T\, \d \sum m\, c\, \log (Ti^2)\eqno{(34)}$$

\0{\bf 15.} The last equation entirely agrees, intending for $T$ the
absolute temperature, with Eq.(1) for the heat

$$dL=\fra{T}A\, dZ$$
making clear its foundation on mechanical principles. The quantity
denoted $Z$ represents the {\it disgregation} [{\sl free energy}] of the
body
which after this is represented as

$$A\sum m \,c\, \log T i^2$$
And it is easy also to check its agreement
with another equation of the mechanical theory of heat.

Imagine that our system of moving point masses has a kinetic energy which
changes because of a temporary action of a force and returns to the
initial value. In this way the kinetic energy so communicated in part
increments the kinetic energy content and in part it performs mechanical
work.

If $\d q$ is the communicated average kinetic energy and $h$ is the kinetic
energy available in the system, it will be possible to write:

$$\d q= \d h+\d L=\d \sum m\,c\, T+\d L=\sum m\,c\, \d T+\d L$$
and assigning to $\d L$ its value Eq.(33), it is found

$$\eqalign{
\d q=&\sum 2m\, c \,\d T+ T\sum
2m\,c\,\d\log i\cr
=& T\,(\sum 2m\, c\, \d \log T +\sum 2m\,c \,\d \log i)=T\,\sum2 m\, c\, \log
(T\,i)\cr}$$
\ie also

$$\d q=T \d\sum 2\, m\, c\, \log (T\, i)\eqno{(35)}$$

This equation appears as the E.(59) of my 1865
paper \Cite{Cl865}. Multiply, in fact, both sides of the preceding equation
by $A$ (the caloric equivalent of the work) and interpret the product $A\d
q$ as the variation of the kinetic energy spent to increment the quantity
of heat transferred and let it be $\d Q$, defining the quantity $S$ by

$$S=A\sum 2 m\, c \,\log (T\,i)\eqno{(36)}$$
so the equation becomes

$$\d Q=T\,\d S\eqno{(37)}$$
where the quantity $S$ introduced here is the one I called
{\it entropy}.

In the last equation the signs of variation can be replaced by
signs of differentiation because both are auxiliary to the
argument (the variation between a stationary motion transient to
another) and the distinction between such two symbols will not be
any longer necessary because the first will no longer
intervene. Dividing again the equation by $T$, we get

$$\fra{dQ}T=dS$$
Imagine to integrate this relation over a cyclic process, and remark that
at the end $S$ comes back to the initial value, so we can establish:

$$\ig \fra{d Q}T=0\eqno(38)$$
This is the equation that I discovered for the first time as an expression
of the second theorem of the mechanical theory of heat for reversible
cyclic processes.\annotaa{Pogg. Ann. {\bf 93}, 481, 1854, and
Abhandlungen \"uber die mechanische W\"armetheorie, {\bf I}, 127,
\Cite{Cl854}[p.460]}  At the
time I set as foundation {\it that heat alone cannot be trasferred from a
colder to a warmer body}. Later\annotaa{``Ueber die Anwendung
des Satzes von der Aequivalenz der Verwandlungen auf die innere Arbeit'',
Pogg. Ann. {\bf 116}, 73--112, 1862, and Abhandlungen \"uber die mechanische
W\"armetheorie, {\bf I}, 242-279, \Cite{Cl862}.} I derived the same equation in a very
different way, \ie based on the preceding law {\it that the work that the
heat of a body can perform in a transformation is proportional to the
absolute temperature and does not depend on its composition}. I treated, in
this way, the fact that in other way it can be proved the equation
as a key consequence of each law. The present argument tells us, as well, that
each of these laws and with them the second theorem of the mechanical
theory of heat can be reduced to general principles of mechanics.

\section{Priority discussion of Boltzmann (vs. Clausius )}
\def\SEC{Priority discussion of Boltzmann (vs. Clausius )}
\label{sec:VI-6}\iniz
\lhead{\small\ref{sec:VI-6}.\ \SEC}

\*
\0{Partial translation  and comments: L. Boltzmann,\index{Boltzmann}
{\it Zur priorit\"at der auffindung der beziehung zwischen dem zweiten
hauptsatze der mechanischen w\"armetheo\-rie und dem prinzip
der keinsten wirkung}, Pogg. Ann. {\bf 143},
211--230, 1871, \Cite{Bo871-0}, and Wis\-sen\-schaft\-li\-che
Abhandlungen, ed. F. Hasen\"ohrl, {\bf1}, \#17 p. 228--236}
\*

Hrn. Clausius presented, at the meeting of 7 Nov. 1870 of the
``Niederrheinischen Gesellschaft für Natur und Heilkunde vorgetragenen''
and in Pogg. Ann. 142, S. 433, \Cite{Cl872}, a work where it is proved that
the second fundamental theorem of the mechanical theory of heat follows
from the principle of least action and that the corresponding arguments are
identical to the ones implying the principle of least action. I have
already treated the same question in a publication of the Wien Academy of
Sciences of 8 Feb. 1866, printed in the volume 53 with the title {\it On
  the mechanical meaning of the second fundamental theorem of the theory of
  heat}, \Cite{Bo866}[\#2] and Sec.(\,\ref{sec:VI-6}\,) below]; and I
believe I can assert that the fourth Section of my paper published four
years earlier is, in large part, identical to the quoted publication of
Hr. Clausius. Apparently, therefore, my work is entirely ignored, as well
as the relevant part of a previous work by Loschmidt. It is possible to
translate the notations of Hr. Clausius into mine, and via some very simple
transformation make the formulae identical.  I claim, to make a short
statement, that given the identity of the subject nothing else is possible
but some agreement.  To prove the claim I shall follow here, conveniently,
the fourth section of my work of 8 Feb. 1866, of which only the four
formulae Eq.(23a),(24a),(25a) and (25b), must be kept in mind.%
\footnote{\tiny Clausius answer, see Sec.(\,\ref{sec:VII-6}\,) below, was to
  apologize for having been unaware of Boltzmann's work but rightly pointed
  out that Boltzmann's formulae became equal to his own after a suitable
  interpretation, absent from the work of Boltzmann; furthermore his
  version was more general than his: certainly, for instance, his analysis
  takes into account the action of external forces. As discussed, the
  latter is by no means a minor remark: it makes Clausius and Boltzmann
  results deeply different.  See also p.\pageref{volume and pressure} and
  p.\pageref{B on ergale}}

\section{Priority discussion: Clausius' reply}
\def\SEC{Priority discussion: Clausius' reply}
\label{sec:VII-6}\iniz
\lhead{\small\ref{sec:VII-6}.\ \SEC}

\0{Translation and comments: R. Clausius\index{Clausius}
{\it Bemerkungen zu der priorit\"atreclama\-tion des Hrn. Boltzmann},
Pogg. Ann. {\bf 144}, 265--274, 1871.}
\*

In the sixth issue of this Ann., p. 211, Hr. Boltzmann claims to have
already in his 1866 paper reduced the second main theorem of the mechanical
theory of heat to the general principles of mechanics, as I have discussed
in a short publication. This shows very correctly that I completely missed
to remark his paper, therefore I can now make clear that in 1866 I changed
twice home and way of life, therefore naturally my attention and my action,
totally involuntarily, have been slowed and made impossible for me to
follow regularly the literature.  I regret overlooking this all the more
because I have subsequently missed the central point of the relevant paper.

It is plane that, in all point in which his work overlaps mine, the
priority is implicit and it remains only to check the points that agree.

In this respect I immediately admit that his expressions about
disgregation [{\sl free energy}] and of entropy overlap with mine on two
points, about which we shall definitely account in the following;
but his mechanical equations, on which such expressions are derived are not
identical to mine, of which they rather are a special case.

We can preliminarily limit the discussion to the simplest form of the
equations, which govern the motion of a single point  moving periodically
on a closed path.

Let $m$ be the mass of the point and let $i$ its period, also let
its coordinates at time $t$ be$x,y,z$, and the acting force
components be $X,Y,Z$ and $v$ its velocity. The latter quantities
as well as other quantities derived from them, vary with the
motion, and we want to denote their average value by over lining
them.  Furthermore we think that near the initially considered
motion there is another one periodic and infinitely little
different, which follows a different path under a different
force.  Then the difference between a quantity relative to the
first motion and the one relative to the varied motion will be
called ``variation of the quantity'', and it will be denoted via
the symbol $\d$. And my equation is written as:

$$\lis{X\,\d x+Y \d\, y+Z\, \d Z}=\fra{m}2\d\lis{v^2}+m\lis
{v^2}\d\log i\eqno{(I)}$$
or, if the force acting on the point admits and ergale [{\sl potential}],
that we denote $U$, for the initial motion,%
\footnote{\tiny For Clausius' notation used here see
  Sec.(\,\ref{sec:IV-1}\,). Here an error seems present because the (I) implies
  that in the following (Ia) there should be $\lis{\d U}$: but it is easy
  to see, given the accurate definition of variation by Clausius, see
  Eq.\equ{e1.4.2} for details, that the following
  (Ia) is correct because $\lis{\d U}=\d {\lis U}$. In reality the averages
  of the variations are quantities not too interesting physically because
  they depend on the way followed to establish the correspondence between
  the points of the initial curve and the points of its variation, and an
  important point of Clausius's paper is that it established a notion of
  variation that implies that the averages of the variations, in general of
  little interest because quite arbitrary, coincide with the variations of
  the averages.}

$$\d\lis U=\fra{m}2\,\d\lis{v^2}+m\,\lis
{v^2}\,\d\log i\eqno{(Ia)}$$
Boltzmann now asserts that these equations are identical to the equation
that in his work is Eq.(22), if elaborated with the help of the equation
denoted (23a). Still thinking to a point mass moving on a closed path and
suppose that it is modified in another for which the point has a kinetic
energy infinitely little different from the quantity $\e$,
then Boltzmann's equation, after its translation into my notations, is

$$\fra{m}2 \lis{\d v^2}=\e+ \lis{X\,\d x+Y \d\, y+Z\, \d Z}\eqno(1)$$
and thanks to the mentioned equation becomes:

$$\e=\fra{\d i}{i} m \lis {v^2}+ m\d{\lis v^2}\eqno(2)$$
The first of these Boltzmann's equations will be identical to my Eq.(I), if
the value assigned to $\e$ can be that of my equation.%
\footnote{\tiny {\it I.e.} to obtain the identity, as Clausius remarks
  later, it is necessary that $\d\lis U=\e -\fra{m}2\d\lis {v^2}$ which is
  obtained if $\e$ is interpreted as conservation of the total average
  energy, as in fact Boltzmann uses $\e$ after his Eq.(23a): {\it but}
  instead in Boltzmann $\e$ is introduced, and used first, as variation of
  the average kinetic energy. The problem is, as remarked in
  Sec.(\,\ref{sec:I-6}\,) that in Boltzmann $\e$ does not seem clearly
  defined.\label{unclear}}

I cannot agree on this for two reasons.

The first is related to a fact, that already Boltzmann casually mentions,
as it seems to me, to leave it aside afterwards. In his equations both
quantities $\lis{\d v^2}$ and $\d \lis{v^2}$ (\ie the average value of the
variation $\d v^2$ and the variation of the average value of $v^2$) are
fundamentally different from each other, and therefore it happens that his
and my equations cannot be confronted.%
\footnote{\tiny Indeed if in (1) $\e$ is interpreted as what it should
  really be according to what follows in Boltzmann, \ie $\e=(\d (\lis
  U+\lis K)) $ Eq.(I) becomes a trivial identity while Eq.(Ia) is non
  trivial. However it has to be kept in mind that Eq.(I) is not correct!}%
 Hence I have dedicated, in my research, extreme care to avoid leaving
 variations vaguely defined. And I use a special treatment of the
 variations by means of the notion of {\it phase}.  This method has the
 consequence that for every varied quantity the average of the variation is
 the variation of the average, so that the equations are significantly
 simple and useful. Therefore I believe that the introduction of such
 special variations is essential for the subsequent researches, and do not
 concern a point of minor importance.

If now my variations are inserted in Boltzmann's Eq.(1)
the following is deduced:

$$\fra{m}2 \d \lis{v^2}=\e+ \lis{X\,\d x+Y \d\, y+Z\, \d Z}\eqno(1a)$$
and if next we suppose tat the force acting on the point has an ergale
[{\sl potential}], which we denote $U$, the equation becomes
$\fra{m}2\d\lis {v^2}=\e -\d\lis U$, alternatively written as

$$\e=\fra{m}2\d \lis{v^2}+\d\lis{ U}.\eqno(1b)$$
If the value of  $\e$ is inserted in Eq.(2) my Eq.(I),(Ia) follow. 
In spite of the changes in Eq.(1a) and (1b) 
Boltzmann's equations so obtained are not identical to mine for a second
and very relevant reason.

{\it I.e} it is easy to recognize that both Boltzmannian equations and
Eq.(1) and (2) hold under certain limiting conditions, which are not
necessary for the validity of mine. To make this really evident, we shall
instead present the Boltzmannian equations  as the most general equations,
not subject to any condition. Therefore we shall suppose
more conveniently that they take the form taken when the force acting on
the point has an ergale [{\sl potential}].

Select, in some way, on the initial trajectory a point as initial point of
the motion, which starts at time $t_1$ as in
Boltzmann, and denote the corresponding values of $v$ and $U$ with
$v_1$ and $U_1$. Then during the entire motion the equation

$$\fra{m}2 v^2+U=\fra{m}2 v_1^2+U_1\eqno(3)$$
will hold; thus, likewise, we can set for the average values:

$$\fra{m}2 \lis {v^2}+\lis U=\fra{m}2 v_1^2+\lis U_1\eqno(4)$$
About the varied motion suppose that it starts from another point, with
another initial velocity and takes place under the action of other
forces. Hence we shall suppose that the latter have an ergale 
$U+\m V$, where $V$ is some function of the coordinates and $\m$ an
infinitesimal constant factor. Consider now again the two specified on the
initial trajectory and on the varied one, so instead of $v^2$ we shall have
in the varied motion the value $v^2+\d v^2$
and instead of $U$ the value $U+\d U+\m(V+\d V)$; therefore,
since $\m \,\d V$ is a second order infinitesimal, this can be written
$U+\d U+\m V$. Hence for the varied motion Eq.(3) becomes:

$$\fra{m}2v^2+\fra{m}2\d v^2+U+\d U+\m V=
\fra{m}2v_1^2+\fra{m}2\d v_1^2+U_1+\d U_1+\m V_1\eqno(5)$$
so that my calculation of the variation leads to the equation:

$$\fra{m}2\lis {v^2}+\fra{m}2\d \lis {v^2}+\lis U+\d \lis U+\m \lis V=
\fra{m}2v_1^2+\fra{m}2\d v_1^2+U_1+\d U_1+\m V_1\eqno(5)$$
Combining the last equation with the Eq.(4) it finally follows

$$ \fra{m}2\d v_1^2+\d U_1+\m (V_1-\lis V)=\fra{m}2\d \lis {v^2}+\d \lis
U.\eqno(7)$$ 
This the equation that in a more general treatment should be in place of
the different Boltzmannian Eq.(1b). Thus instead of the 
Boltzmannian Eq.(2) the following is obtained

$$\fra{m}2 \d v_1^2+\d U_1+\m(V_1-\lis V)=\fra{\d i}i m \lis {v^2}+m
\d{\lis v^2}.\eqno(8)$$
As we see, since such new equations are different from the Boltzmannian
ones, we must treat more closely the incorrect quantity $\e$. 
As indicated by the found infinitesimal variation of the ``vis viva'', due
to the variation of the motion, it is clear that in the variation $\e$ of
the ``vis viva'' at the initial time one must understand, and hence set:

$$\e=\fra{m}2\d v_1^2.$$
Hence the Boltzmannian equations of the three terms, that are to the left
in Eq.(7) and (8), should only contain the first.

Hr. Boltzmann, whose equations incompleteness I have, in my view, briefly
illustrated, pretends a wider meaning for $\e$ in his reply, containing at
the same time the ``vis viva'' of the motion and the work, and consequently
one could set

$$\e= \fra{m}2\d v^2_1+\d U_1.$$
But I cannot find that this is said anywhere, because in the mentioned
places where the work can be read it seems to me that there is a gain that
exchanges the ``vis viva'' with another property of the motion that can
transform it into work, which is not in any way understandable, and from
this it does not follow that the varied original trajectory could be so
transformed that is has no point in common it and, also, in the
transformation the points moved from one trajectory to the other could be
moved without spending work.

Hence if one wishes to keep the pretension that the mentioned meaning of
$\e$, then always two of the three terms appearing in Eq.(7) and (8) are
obtained, {\it the third of them, \ie $\m(V_1-\lis V)$ no doubt is missing
in his equations}.

On this point he writes: ``The term $\m(\lis V -V_1)$ is really missing in my
equations, because I have not explicitly mentioned the possibility of the
variation of the ergale. Certainly all my equations are so written that
they remain correct also in this case. The advantage, about the possibility
of considering some small variation of the ergale and therefore to have at
hand the second independent variable in the infinitesimal $\d U$ exists
and from now on it will not be neglected...''.\label{B on ergale}
\Cite{Cl872}[p.271].

I must strongly disagree with the remark in the preceding reply, that all
his equations are written so that also in the case in which the ergale 
varies still remain valid. The above introduced Eq.(1) and (2), even if the
quantity  $\e$ that appears there receives the extended meaning $\fra{m}2\d
v_1^2+\d U_1$, are again false in the case in which by the variation of the
motion of a point the ergale so changes that the term $\m(\lis V_1-V)$  has
an intrinsic value. [{\sl see p.\pageref{volume change}.}]

It cannot be said that my Eq.(I) is {\it implicitly}
contained in the Boltzmannian work, but the relevant equations
of his work represent, also for what concerns my method of realizing the
variations, only a special case of my equations.

Because I must remark that the development of the treatment of the case in
which the ergale so changes is not almost unessential, but for researches
of this type it is even necessary.

It is in fact possible to consider a body as an aggregate of very many
point masses that are under the influence of external and internal
forces. The internal forces have an ergale, depending only on the points
positions, but in general it stays unchanged in all states of the body; on
the contrary this does not hold for the external forces. If for instance
the body is subject to a normal pressure $p$ and later its volume $v$
changes by $dv$, then the external work $p \,dv$ will be performed. This
term, when  $p$ is varied independently of $v$, is not an exact
differential
and the work of the external force cannot, consequently, be
representable as the differential of an ergale. The behavior of this force
can be so represented. For each given state of the body in which its
components are in a state of stationary type it is possible to assign an
ergale also to the external forces which, however, does not stay unchanged,
unlike that of the internal forces, but it can undergo variations while the
body evolved into another state, independent of the change of position of
the points.\index{ergale changes}

Keep now in mind the equations posed in the thermology of the changes of
state to build their mechanical treatment, which have to be reconsidered to
adapt them to the case in which the ergale changes.\label{ergale change}

I can say that I looked with particular care such generalizations. Hence it
would not be appropriate to treat fully the problem, but I obtained in my
mechanical equations the above mentioned term $\m(V_1-\lis V)$, for which
the corresponding term cannot be found in the mechanical equations. I must
now discuss the grounds for this difference and under which conditions such
term could vanish. I find that this will be obtained if the ergale
variation is not instantaneous 
happening at a given moment, but gradual and uniform while an entire cycle
takes place, and at the same time I claim that the same result is obtained
if it is supposed that we do not deal with a single moving point  but with
{\it very large numbers of equal points}, all moving in the same way but
with different phases, so that at every moment the phases are uniformly
distributed and this suffices for each quantity to be evaluated at a point
where it assumes a different value thus generating the average value. The
latter case arises in the theory of heat, in which the motions, that we call
heat, are of a type in which the quantities that are accessible to our
senses are generated by many equal points in the same way, Hence the
preceding difficulty is solved, but I want to stress that such solution 
appears well simpler when it is found than when it is searched.

The circumstance that for the motions that we call heat those terms
disappear from the averages had as a result that Boltzmann could obtain for
the disgregation[{\sl free energy}] and the entropy, from his more
restricted analysis, results similar to those that I obtained with a more
general analysis; but it will be admitted that the real and complete
foundation of this solution can only come from the more general treatment.

The validity condition of the result, which remains hidden in the more
restricted analyses, will also be evident.

In every case B. restricts attention to motions that take place along
closed trajectories. Here we shall consider motions on non closed curves,
hence it now becomes necessary a special argument.%
\footnote{\tiny The case of motions taking place on non closed trajectories is,
  {\it however}, treated by Boltzmann, as underlined in p.\pageref{open
    paths} of Sec.(\,\ref{sec:I-6}\,), quite convincingly.}

Here too I undertook another way with respect to Boltzmann, and this is the
first of the two points mentioned above, in which Boltzmann's result on
disgregation and entropy differ. In his method taking into account of
time is of the type that I called {\it characteristic time of the period of
a motion}, essentially different. The second point of difference is found
in the way in which we defined temperature. The special role of these
differences should be followed here in detail, but I stop here hoping to
come back to it elsewhere.%
\annotaa{While this article was in print I found in a parallel research
  that the doubtful expression, to be correct in general, requires a change
  that would make it even more different from the Boltzmannian one.}

Finally it will not be superfluous to remark that in another of my published
works the theorem whereby in every stationary motion  
{\it the average ``{\it vis viva}'' equals the virial} remains entirely
outside of the priority question treated here. This theorem, as far as I
know, has not been formulated by anyone before me.

\section{On the ergodic hypothesis (Trilogy: \#1)}
\def\SEC{On the ergodic hypothesis (Trilogy: \#1)}
\label{sec:VIII-6}\iniz
\lhead{\small\ref{sec:VIII-6}.\ \SEC}

\0{Partial translation and comments: L. Boltzmann\index{Boltzmann}, a
{\it {\"U}ber das {W\"a}rme\-gleichge\-wicht zwischen mehratomigen
{G}asmolek{\"u}len}, 1871, in {W}is\-sen\-schaft\-li\-che {A}bhandlungen,
ed. {F}. {H}asen{\"o}hrl, {\bf 1}, \#18, 237-258, \Cite{Bo871-a}.}
\index{trilogy}
\*
\0[{\sl This work is remarkable particularly for its Sec.II where the
    Maxwell's distribution is derived as a consequence of the of the
    assumption that, {\it because of the collisions with other molecules}, the
    atoms of the molecule visit all points of the molecule phase space.  It
    is concluded that the distribution of the atoms inside a molecule is a
    function of the molecule energy.  So the distribution of the
    coordinates of the body will depend on the total energy (just kinetic
    as the distance between the particles in very large compared with the
    interaction range). The question of
    uniqueness of the microcanonical distribution is explicitly
    raised as a problem. Strictly speaking the results do not depend on the ergodic
    hypothesis. The relation of the ``Trilogy'' papers with Einstein's
    statistical mechanics is discussed in \Cite{Re997}.}]  \*

According to the mechanical theory of heat every molecule of gas while in
motion does not experience, by far for most of the time, any collision; and
its baricenter proceeds with uniform rectilinear motion through space.
When two molecules get very close, they interact via certain forces, so
that the motion of each feels the influence of the other.

The different molecules of the gas take over all possible
configurations%
\footnote{\tiny In this paper B. imagines that a molecule of gas, in
due time, goes through all possible states, but this is not yet the ergodic
hypothesis because this is attributed to the occasional interaction of the
molecule with the others, see below p.\pageref{random collisions}. The
hypothesis is used to extend the hypothesis formulated by Maxwell for the
monoatomic systems to the case of polyatomic molecules. For these he finds
the role of the internal potential energy of the molecule, which must
appear together with the kinetic energy of its atoms in the stationary
distribution, thus starting what will become the theory of statistical
ensembles, and in particular of the canonical ensemble.} 
 and it is clear that it is of the utmost importance to know the
 probability of the different states of motion.

We want to compute the average kinetic energy, the average potential
energy, the mean free path of a molecule \&tc. and, furthermore, also the
probability of each of their values. Since the latter value is not known we
can then at most conjecture the most probable value of each quantity, as we
cannot even think of the exact value.

If every molecule is a point mass, Maxwell provides the value of the
probability of the different states (Phil. Mag., March\footnote{\tiny Maybe
February?}  1868), \Cite{Ma868}\Cite{Ma867-b}. In this case the state of a molecule
is entirely determined as soon as the size and direction of its velocity
are known.  And certainly every direction in space of the velocity is
equally probable, so that it only remains to determine the probability of
the different components of the velocity.

If we denote $N$ the number of molecules per unit volume, Maxwell finds
that the number of molecules per unit volume and speed between $c$ and
$c+dc$, equals, \Cite{Ma868}[Eq.(26), p.187]:

$$4\sqrt{\fra{h^3}\p} N e^{-h c^2}c^2\, dc,$$
where $h$ is a constant depending on the temperature.
We want to make use of this expression: through it the velocity
distribution is defined, \ie it is given how many molecules have a speed
between $0$ and $dc$, how many between $dc$ and $2dc$, $2dc$ and $3dc$,
$3dc$ and $4dc$, etc. up to infinity.

Natural molecules, however, are by no means point masses. We shall get
closer to reality if we shall think of them as systems of more point masses
(the so called atoms), kept together by some force. Hence the state of a
molecule at a given instant can no longer be described by a single variable
but it will require several variables. To define the state of a molecule at
a given instant, think of having fixed in space, once and for all, three
orthogonal axes. Trace then through the point occupied by the baricenter
three orthogonal axes parallel to the three fixed directions and denote the
coordinates of the point masses of our molecule, on every axis and at time
$t$, with
$\x_1,\h_1,\z_1,\x_2,\h_2,\z_2,\ldots,\x_{r-1},\h_{r-1},\z_{r-1}$.  The
number of point masses of the molecule, that we shall always call atoms, be
$r$. The coordinate of the $r$-th atom be determined besides those of the
first $r-1$ atoms from the coordinates of the baricenter.  Furthermore let
$c_1$ be the velocity of the atom 1, $u_1,v_1,w_1$ be its components along
the axes; the same quantities be defined for the atom 2, $c_2,u_2,v_2,w_2$;
for the atom 3 let them be $c_3,u_3,v_3,w_3$ \&tc. Then the state of our
molecule at time $t$ is given when the values of the $6r-3$ quantities
$\x_1,\h_1,\z_1,\x_2,\ldots,\z_{r-1},u_1,v_1,w_1,u_2,\ldots,w_r$ are known
at this time. The coordinates of the baricenter of our molecule with respect
to the fixed axes do not determine its state but only its position.

We shall say right away, briefly, that a molecule is at a given place when
its baricenter is there, and we suppose that in the whole gas there is an
average number $N$ of molecules per unit volume. Of such $N$ molecules at a
given instant $t$ a much smaller number $dN$ will be so distributed that,
at the same time, the coordinates of the atom 1 are between $\x_1$ and
$\x_1+d\x_1$, $\h_1$ and $\h_1+d\h_1$, $\z_1$ and $\z_1+d\z_1$, those of
the atom 2 are between $\x_2$ and $\x_2+d\x_2$, $\h_2$ and $\h_2+d\h_2$,
$\z_2$ and $\z_2+d\z_2$, and those of the $r-1$-{th} between $\x_{r-1}$ and
$\x_{r-1}+d\x_{r-1}$, $\h_{r-1}$ and $\h_{r-1}+d\h_{r-1}$, $\z_{r-1}$ and
$\z_{r-1}+d\z_{r-1}$, while the velocity components of the atom 1 are
between $u_1$ and $u_1+du_1$, $v_1$ and $v_1+dv_1$, $w_1$ and $w_1+dw_1$,
those of the atom 2 are between $u_2$ and $u_2+du_2$, $v_2$ and $v_2+dv_2$,
$w_2$ and $w_2+dw_2$, and those of the $r-1$-th are between $u_{r-1}$ and
$u_{r-1}+du_{r-1}$, $v_{r-1}$ and $v_{r-1}+dv_{r-1}$, $w_{r-1}$ and
$w_{r-1}+dw_{r-1}$.

I shall briefly say that the so specified molecules  are in the domain
(A). Then it immediately follows that
$$dN=f(\x_1,\h_1,\z_1,\ldots,\z_{r-1},u_1,v_1,\ldots,
w_r)d\x_1d\h_1d\z_1\ldots d\z_{r-1}du_1dv_1\ldots dw_r.$$
I shall say that the function $f$ determines a distribution of the states
of motion of the molecules at time $t$. The probability of the different
states of the molecules would be known if we knew which values has this
function for each considered gas when it is left unperturbed for a long
enough time, at constant density and temperature. For monoatomic molecules 
gases Maxwell finds that the function $f$ has the value
$$4\sqrt{\fra{h^3}\p} N e^{-h c^2}c^2\, dc.$$
The determination of this function for polyatomic molecules gases seems
very difficult, because already for a three atoms complex it is not
possible to integrate the equations of motion. Nevertheless we shall see
that just from the equations of motion, without their integration, a value
for the function $f$ is found which, in spite of the motion of the
molecule, will not change in the course of a long time and therefore,
represents, at least, a possible distribution of the states of the
molecules.%
\footnote{\tiny Remark the care with which the possibility is not excluded
of the existence invariant distributions different from the one that will
be determined here.}

That the value pertaining to the function $f$ could be determined without
solving the equations of motion is not so surprising as at first sight
seems. Because the great regularity shown by the thermal phenomena induces
to suppose that $f$ be almost general and that it should be independent
from the properties of the special nature of every gas; and also that the
general properties depend only weakly from the general form of the
equations of motion, except when their complete integration presents
difficulties not unsurmountable.%
\footnote{\tiny Here B. seems aware that special behavior could show up in
  integrable cases: he was very likely aware of the theory of the solution
  of the harmonic chain of Lagrange, \Cite{La867}[Vol.I].}

Suppose that at the initial instant the state of motion of the molecules is
entirely arbitrary, \ie think that the function $f$ has a given
value.\footnote{\tiny This is the function called ``empirical distribution'',
\Cite{GL003}\Cite{GGL005}.}  As time elapses the state of each molecule,
because of the motion of its atoms while it follows its rectilinear motion
and also because of its collisions with other
molecules,\label{random collisions}
becomes steady; hence the form of the function $f$ will in
general change, until it assumes a value that in spite of the motion of the
atoms and of the collisions between the molecules will no longer change.

When this will have happened we shall say that the states of the molecules
are distributed in {\it thermal equilibrium}. 
From this immediately the problem is posed to find for the function $f$ a
value that will not any more change no matter which collisions take
place. For this purpose we shall suppose, to treat the most general case,
that we deal with a mixture of gases. Let one of the kinds of gas (the kind
G) have $N$ molecules per unit volume. Suppose that at a given instant $t$
there are $dN$ molecules whose state is in the domain (a). Then as before
$$dN=f(\x_1,\h_1,\z_1,\ldots,\z_{r-1},u_1,v_1,\ldots,
w_r)d\x_1d\h_1d\z_1\ldots d\z_{r-1}du_1dv_1\ldots dw_r.\eqno(1)$$
The function $f$ gives us the complete distribution
of the states of the molecules of the gas of kind G at the instant $t$. 
Imagine that a certain time $\d t$ elapses. At time $t+\d t$ the
distribution of the states will in general have become another, hence the
function $f$  becomes different, which I denote $f_1$, so that at time
$t+\d t$  the number of molecules per unit volume whose state in the domain
(A) equals:

$$f_1(\x_1,\h_1,\ldots ,w_r)\, d\x_1\,d
h_1\,\ldots\, dw_r.\eqno(2)$$
\*
\centerline{{\bf \S{}I.} Motion of the atoms of a molecule}
\*

\0[{\sl Follows the analysis of the form of $f$ in 
absence of collisions: via Liouville's theorem it is shown that if $f$ is
invariant then it has to be a function of the coordinates of the molecules
through the integrals of motion. This is a wide extension of the argument
by Maxwell for monoatomic gases, \Cite{Ma868}.}]

\*
\centerline{{\bf \S{}II.} Collisions between molecules}
\*

\0[{\sl It is shown that to have a stationary distribution also in presence
of binary collisions it must be that the function $f$ has the form $A
e^{-h \f}$ where $\f$ is the total energy, sum of the kinetic energy and of
the potential energy of the atoms of the molecule. Furthermore if the gas
consists of two species then $h$ must be the same constant for the
distribution of either kinds of molecules and it is identified with the
inverse temperature. Since a gas, monoatomic or not, can be considered as a
giant molecule it is seen that this is the derivation of the canonical
distribution. The kinetic energies equipartition and the ratios of the
specific heats is deduced. It becomes necessary to check that this
distribution ``of thermal equilibrium'' generates average values for
observables compatible with the heat theorem: this will be done in the
successive papers. There it will also be checked that the ergodic
hypothesis in the form that each group of atoms that is part of a molecule
passes through all states compatible with the value of the energy (possibly
with the help of the collisions with other molecules) leads to the same
result if the number of molecules is infinite or very large. The question
of the uniqueness of the equilibrium distribution is however left open as
explicitly stated at p. 255, see below.}]
\*

\0{\bf p.255, \rm(line 21)} Against me is the fact that, until now, 
the proof that these distributions are the only ones that do not change in
presence of collisions is not complete. Nevertheless remains the fact that
[the distribution shows] that the same gas with equal temperature and
density can be in many states, depending on the given initial conditions,
{\it a priori} improbable and which will even never be observed in
experiments.
\*

\0[{\sl This paper\label{Bo-Ma-Beq} is also important as it shows that
    Boltzmann was well aware of Maxwell's paper, \Cite{Ma868}: in which a
    key argument towards the Boltzmann's equation is discussed in great
    detail. One can say that Maxwell's analysis yields a form of ``weak
    Boltzmann's equation'', namely several equations which can be seen as
    equivalent to the time evolution of averages of one particle observable
    with what we call now the one particle distribution of the
    particles. Boltzmann will realize, \Cite{Bo872}[\#22], that the one
    particle distribution itself obeys an equation (the Boltzmann equation)
    and obtain in this way a major conceptual simplification of Maxwell's
    approach and derive the $H$-theorem.}]

\section{Canonical ensemble and ergodic hypothesis (Trilogy: \#2)}
\def\SEC{Canonical ensemble and ergodic hypothesis (Trilogy: \#2)}
\label{sec:IX-6}\iniz
\lhead{\small\ref{sec:IX-6}.\ \SEC}

\0{Partial translation and comments of: L. Boltzmann,\index{Boltzmann}
{\it Einige allgemeine s{\"a}tze {\"u}ber {W\"a}rme\-gleichgewicht},
(1871), in {W}is\-sen\-schaft\-li\-che {A}bhandlungen,
ed. {F}. {H}asen{\"o}hrl, {\bf 1}, 259--287, \#19,
\Cite{Bo871-b}}\index{trilogy}
\*

\0{\bf \S{}I.} {\it Correspondence between the theorems on the polyatomic
molecules behavior and Jacobi's principle of the last multiplier.}
\footnote{\tiny This title is quoted by Gibbs in the introduction to his
{\it Elementary principles in statistical mechanics}, \Cite{Gi902}, thus
generating some confusion because this title is not found in the list of
publications by Boltzmann.}
\*

The first theorem that I found in my preceding paper {\it\"Uber das
  {W\"a}rme\-gleich\-gewicht zwischen mehratomigen {G}asmolek{\"u}len},
1871, \Cite{Bo871-a}[\#17], is strictly related to a theorem,
that at first sight has nothing to do with the theory of gases, \ie with
Jacobi's principle of the last multiplier.

To expose the relation, we shall leave aside the special form that the
mentioned equations of the theory of heat have, whose relevant developments
will be generalized here later.

Consider a large number of systems of point masses (as in a gas containing a
large number of molecules of which each is a system of point masses). The
state of a given system of such points at a given time is assigned by $n$
variables $s_1,s_2,\ldots,s_n$ for which we can pose the following
differential equations:

$$\fra{d s_1}{dt}=S_1,\fra{d s_2}{dt}=S_2,\ldots,\fra{d s_n}{dt}=S_n.$$
Let the $S_1,S_2,\ldots,S_n$ be functions of the $s_1,s_2,\ldots,s_n$ and
possibly of time. Through these equations and the initial value of the $n$
variables $s_1,s_2,\ldots,s_n$ are known the values of such quantities at
any given time. To arrive to the principle of the last multipliers, we can
use many of the conclusions reached in the already quoted paper; hence we
must suppose that between the point masses of the different systems of
points never any interaction occurs. As in the theory of gases the
collisions between molecules are neglected, also in the present research
the interactions will be excluded.
\*

\0[{\sl Follows a discussion on the representation of a probability
distribution giving the number of molecules in a volume element of the
phase space with $2n$ dimensions. The Liouville's theorem is proved for the
purpose of obtaining an invariant distribution in the case of equations of
motion with vanishing divergence. 

Subsequently it is discussed how to transform this distribution into a
distribution of the values of $n$ constants of motion, {\it supposing their
existence}; concluding that the distribution is deduced by dividing the
given distribution by the Jacobian determinant of the transformation
expressing the coordinates $s$ in terms of the constants of motion and of
time: the last multiplier of Jacobi is just the Jacobian determinant of the
change of coordinates. 

Taking as coordinates $n-1$ constants of motion and
as $n$-th the $s_1$ it is found that a stationary distribution is such that
a point in phase space spends in a volume element, in which the $n-1$
constants of motion $\f_2,\f_3,\ldots,\f_n$ have a value in the set $D$ of
the points where $\f_2,\f_3,\ldots$ are between $\f_2$ and $\f_2+d\f_2$,
$\f_3$ and $\f_3+d\f_3\ldots$, and $s_1$ is between $s_1$ and $s_1+ds_1$, a
fraction of time equal to the fraction of the considered volume element
with respect to the total volume in which the constants have value in $D$
and the $n$-th has an arbitrary value. 

The hypothesis of existence of $n$ constants of motion is not realistic in
the context in which it is assumed that the motion is regulated by
Hamiltonian differential equations. It will become plausible in the paper
of 1877, \Cite{Bo877-b}[\#42], where a discrete structure is admitted for
the phase space and time.}]

\*
\0{\bf \S{}II.} {\it Thermal equilibrium for a finite number of point masses.}
\*

\0[{\sl In this section the method is discussed to compute the 
average kinetic energy and the average potential energy in a system with
$n$ constants of motion.}]

\*
\0{\bf \S{}III. {\rm p.284}} {\it Solution for the thermal equilibrium for
the molecules of a gas with a finite number of point masses under an
hypothesis.}
\*

Finally from the equations derived we can, under an assumption which it
does not seem to me of unlikely application to a warm body, directly access
to the thermal equilibrium of a polyatomic molecule, and more generally of
a given molecule interacting with a mass of gas. The great chaoticity of
the thermal motion and the variability of the force that the body feels
from the outside makes it probable that the atoms get in the motion, that
we call heat, all possible positions and velocities compatible with the
equation of the ``vis viva'', and even that the atoms of a warm body
can take all positions and velocities compatible with the last equations
considered.%
\footnote{\tiny Here comes back the ergodic hypothesis in the form
saying that not only the atoms of a single molecule take all possible
positions and velocities but also that the atoms of a ``warm body'' with
which a molecule is in contact take all positions and velocities.
\\
This is essentially the ergodic hypothesis.  The paper shows how, through
the ergodic hypothesis assumed for the whole gas it is possible to derive
the canonical distribution for the velocity and position distribution both
of a single molecule and of an arbitrary number of them. It goes beyond the
preceding paper deducing the {\it microcanonical} distribution, on the
assumption of the ergodic hypothesis which is formulated here for the first
time as it is still intended today, and finding as a consequence the {\it
canonical} stationary distribution of the atoms of each molecule or of an
arbitrary number of them by integration on the positions and velocities of
the other molecules.
\\
This also {\it founds the theory of the statistical ensembles}, as
recognized by Gibbs in the introduction of his treatise on statistical
mechanics, \Cite{Gi902}. Curiously Gibbs quotes this paper of Boltzmann
attributing to it a title which, instead, is the title of its first Section.
The Jacobi's principle, that B. uses in this paper, is the theorem that
expresses the volume element in a system of coordinates in terms of that in
another through a ``final multiplier'', that today we call ``Jacobian
determinant'' of the change of coordinates. B. derives already in the
preceding paper what we call today ``Liouville's theorem'' for the
conservation of the volume element of phase space and here he gives a
version that takes into account the existence of constants of motion, such
as the energy. From the uniform distribution on the surface of constant 
total energy (suggested by the ergodic hypothesis) the canonical
distribution of subsystems (like molecules) follows by integration and use
of the formula $(1-\fra{c}{\l})^\l=e^{-c}$ if $\l$
(total number of molecules) is large.
\\
Hence imagining the gas large the canonical distribution follows for
every finite part of it, be it constituted by $1$ or by $10^{19}$ molecules:
a finite part of a gas is like a giant molecule.}

Let us accept this hypothesis, and thus let us make use of the formulae to
compute the equilibrium distribution between a gas in interaction with a
body supposing that only $r$ of the mentioned $\l$ atoms of the body
interact with the mass of gas.

Then $\ch$ [{\sl potential energy}] has the form $\ch_1+\ch_2$ where
$\ch_1$ is a function of the coordinates of the $r$ atoms, $\ch_2$ is a
function
of the coordinates of the remaining $\l-r$. Let us then integrate 
formula (24) [{\sl which is

$$dt_4=\fra{(a_n-\ch)^{\fra{3\l}2-1}
dx_1\,dy_1\,\ldots \,dz_\l }
{\ig\ig (a_n-\ch)^{\fra{3\l}2-1}
dx_1\,dy_1\,\ldots \,dz_\l} \eqno(24)$$
expressing the time during which, in average, the coordinates are between
$x_1$ and $x_1+dx_1$ $\ldots$ $z_\l$
and $z_\l+dz_\l$.}]
\\
for $dt_4$ over all values of $x_{r+1},y_{r+1},\ldots, z_\l$ obtaining for
the time during which certain $x_1,y_1,\ldots,z_r$ are, in average, between
$x_1$ and $x_1+dx_1$ \etc; hence for the average time that the atom
$m_1$ spends in the volume element $dx_1 dy_1 dz_1$, $m_2$ spends in $dx_2
dy_2 dz_2\ldots $, the value is found of

$$dt_5=\fra{dx_1\,dy_1\,\ldots \,dz_r\, \ig\ig \ldots
(a_n-\ch_1-\ch_2)^{\fra{3\l}2-1} dx_{r+1}dy_{r+1}\ldots dz_\l}
{ \ig\ig \ldots
(a_n-\ch_1-\ch_2)^{\fra{3\l}2-1} dx_{1}dy_{1}\ldots dz_\l}$$
If the elements $dx_1 dy_1 dz_1, dx_2 dy_2 dz_2\ldots$ were chosen so that
$\ch_1=0$ gave the true value of $dt_5$ it would be

$$dt_6= \fra{dx_1\,dy_1\,\ldots \,dz_r\, \ig\ig \ldots
(a_n-\ch_2)^{\fra{3\l}2-1} dx_{r+1}dy_{r+1}\ldots dz_\l}
{ \ig\ig \ldots
(a_n-\ch_1-\ch_2)^{\fra{3\l}2-1} dx_{1}dy_{1}\ldots dz_\l}$$
And then the ratio is

$$\fra{dt_5}{dt_6}=\fra{\ig\ig \ldots
(a_n-\ch_1-\ch_2)^{\fra{3\l}2-1} dx_{1}dy_{1}\ldots dz_\l}
{\ig\ig \ldots
(a_n-\ch_2)^{\fra{3\l}2-1} dx_{1}dy_{1}\ldots dz_\l}.$$
The domain of the integral in the denominator
is, because of the unchanged presence of the function $\ch_2$, dependent
from $a_n$. The domain of the integral in the numerator, in the same way,
which does not contain the variable on which the integral has to be made.
The last integral is function of $(a_n-\ch_1)$. Let $\fra{a_n}\l=\r$, where
$\l$ is the number, naturally constant, of atoms, thus also the integral
in the denominator is a function of $\r$, that we shall denote $F(\r)$;
the integral in the numerator is the same function of 
$\r-\fra{\ch_1}\l$, hence equal to $F(\r-\fra{\ch_1}\l)$ and therefore

$$\fra{dt_5}{dt_6}=\fra{F(\r-\fra{\ch_1}\l)}{F(\r)}.$$
Let now $\l$ be very large, Hence also $r$ can be very large;
we now must eliminate $\l$. If  $dt_5/dt_6$ is a finite and continuous function
of $\r$ and $\ch_1$ then $\r$ and $\fra{\ch_1}r$ have the order of
magnitude of the average ``vis viva''  of an atom. Let
$\fra{dt_5}{dt_6}=\ps(\r,\ch_1)$, then

$$\ps(\r,\ch_1)=\fra{F(\r-\fra{\ch_1}\l)}{F(\r)}\eqno(28)$$
Hence

$$\eqalign{
\fra{F(\r-\fra{2\ch_1}\l)}{F(\r-\fra{\ch_1}\l)}=
&\ps(\r-\fra{\ch_1}\l)=\ps_1\cr
\fra{F(\r-\fra{3\ch_1}\l)}{F(\r-\fra{2\ch_1}\l)}=
&\ps(\r-\fra{2\ch_1}\l)=\ps_2\cr
\ldots&\ldots\cr
\fra{F(\r-\fra{\m
\ch_1}\l)}{F(\r-\fra{{\m-1}\ch_1}\l)}=&\ps(\r-\fra{(\m-1)}\l,
\ch_1)=\ps_{\m-1}.\cr}$$
Multiplying all these equations yields

$$\log F(\r-\fra{\m\ch_1}\l)-\log F(\r)=\log\ps+\log\ps_1+\ldots+\log\ps_{\m-1}$$
Let now $\m \ch_1=\ch_3$ then

$$\log F(\r-\ch_3)-\log (\r)=\l \log \Ps(\r,\ch_3)$$
where $\Ps$ is again finite and continuous and if $\r$ and $\fra{\ch_3}r$ are
of the order of the average ``vis viva'' also
$\fra{\l}\m$ is finite. It is also:

$$F(\r-\ch_3)=F(\r)\cdot [\Ps(\r,\ch_3)]^\l.$$

Let us now treat $\r$ as constant and $\ch_3$ as variable and set

$$\r-\ch_3=\s,\ F(\r)=C, \ \Ps(\r,\r-\s)=f(\s)$$
thus the last formula becomes 
$F(\s)=C\cdot[f(\s)]^\l,$ and therefore formula (28) becomes

$$\ps(\r,\ch_1)=\Big[\fra{f(\r-\fra{\ch_1}\l)}{f(\r)}\Big]^\l=
\Big[1-\fra{f'(\r)}{f(\r)}\cdot\fra{\ch_1}\l\Big]^\l=
e^{\fra{f'(\r)}{f(\r)}\cdot\ch_1}$$ 
and if we denote $\fra{f'(\r)}{f(\r)}$ with $h$ 

$$dt_5=C' e^{-h\ch_1} dx_1dy_1\ldots dz_r.$$
Exactly in the same way the time can be found during which the coordinates
of the  $r$ atoms are between $x_1$ and $x_1+dx_1$
$\ldots$  and their velocities  are between $c_1$ and $c_1+d C_1$
$\ldots$. It is found to be equal to

$$C'' e^{-h(\ch_1+\sum \fra{m c^2}2)} c_1^2 c_2^2\ldots c_r^2
dx_1dy_2\ldots dc_r.$$
These equations must, under our hypothesis, hold for an arbitrary body in a
mass of gas, and therefore also for a molecule of gas. In the considered
case it is easy to see that this agrees with the formulae of my work {\it
  {\"U}ber das {W\"a}rme\-gleichgewicht zwischen mehratomigen
  {G}asmolek{\"u}len}, 1871, \Cite{Bo871-a}[\#17]. We also arrive here in a
much easier way to what found there. Since however the proof, that in the
present section makes use of the hypothesis about the warm body, which
certainly is acceptable but it had not yet been proposed, thus I avoided it
in the quoted paper obtaining the proof in a way independent from that
hypothesis.%
\footnote{\tiny He means that he proved in the quoted reference the
  invariance of the canonical distribution (which implies the
  equidistribution) without the present hypothesis.  However even that was
  not completely satisfactory as he had also stated in the quoted paper
  that he had not been able to prove the uniqueness of the solution found
  there (that we know today to be not true in general).}

\section{Heat theorem without dynamics (Trilogy: \#3)}
\def\SEC{Heat theorem without dynamics (Trilogy: \#3)}
\label{sec:X-6}\iniz
\lhead{\small\ref{sec:X-6}.\ \SEC}

\0{Comment: L. Boltzmann,
{\it {A}nalytischer {B}eweis des zweiten {H}auptsatzes der
mechanischen {W\"a}rmetheorie aus den {S\"a}tzen {\"u}ber das
{G}leichgewicht des leben\-digen {K}raft}, 1871,
{W}is\-sen\-schaft\-li\-che {A}bhandlungen, ed. {F}. {H}asen{\"o}hrl,
{\bf 1}, 288--308, \#20, \Cite{Bo871-c}.}\index{trilogy}\index{Boltzmann}
\*

{\sl Here it is shown how the hypothesis that, assuming that the
  equilibrium distribution of the entire system is the microcanonical one,
  then defining the heat $dQ$ received by the body as the variation of the
  total average energy $dE$ plus the work $dW$ done by the system on the
  outside (average variation in time of the potential energy due to a
  change of the values of the external parameters) it follows that $\fra {d
    Q}T$ is an exact differential if $T$ is proportional to the average
  kinetic energy. This frees (apparently) equilibrium statistical mechanics
  from the ergodic hypothesis and will be revisited in the paper of 1884,
  \Cite{Bo884}[\#73] see also Sec.(\,\ref{sec:XIII-6}\,), with the general theory
  of statistical ensembles and of the states of thermodynamic equilibrium.
  Here dynamics enters only through the conservation laws and the
  hypothesis (molecular chaos, see the first trilogy paper
  \Cite{Bo871-a}[\#17]) that never a second collision (with a given
  molecule) takes place when one is still taking place: properties before
  and after the collision are only used to infer again the canonical
  distribution, which is then studied as the source of the heat
  theorem. The hypothesis of {\it molecular chaos} preludes to the paper,
  following this a little later, that will mark the return to a detailed
  dynamical analysis with the determination, for rarefied gases, of the
  time scales needed to reach equilibrium, based on the Boltzmann's
  equation, \Cite{Bo872}[\#22].}

\section{Irreversibility: Loschmidt and ``Boltzmann's sea''}
\def\SEC{Irreversibility: Loschmidt and ``Boltzmann's sea''}
\label{sec:XI-6}\iniz
\lhead{\small\ref{sec:XI-6}.\ \SEC}\index{Loschmidt}\index{Boltzmann's sea}

\0{Partial translation and comments: L. Boltzmann,
{\it Bemerkungen {\"u}ber einige Probleme der mechanischen
{W}{\"a}rmetheo\-rie}, 1877, in {W}is\-sen\-schaft\-li\-che {A}bhandlungen,
ed. {F}. {H}asen{\"o}hrl, {\bf 2}, 112--148, \#39,\Cite{Bo877a}.}
\*

\0{\bf \S{}II.} {\bf On the relation between a general mechanics theorem
  and the second main theorem%
\footnote{\tiny In this paper the discussion is really about the second law rather
  than about the second main theorem, see the previous sections.} of the
theory of heat} (p.116) \*

In his work on the states of thermal equilibrium of a system of bodies,
with attention to the force of gravity, Loschmidt formulated an opinion,
according to which he doubts about the possibility of an entirely mechanical
proof of the second theorem of the theory of heat.  With the same extreme
sagacity suspects that for the correct understanding of the second
theorem an analysis of its significance is necessary deeper than what
appears indicated in my philosophical interpretation, in which perhaps
various physical properties are found which are still difficult to
understand, hence I shall immediately here undertake their explanation with
other words.

We want to explain in a purely mechanical way the law according to which
all natural processes proceed so that

$$\ig \fra{dQ}T\le 0$$
and so, therefore, behave bodies consistent of aggregates of point masses.
The forces acting between these point masses are imagined as functions of
the relative positions of the points. If they are known as functions of
these relative positions we say that the interaction forces are
known. Therefore the real motion of the point masses and also the
transformations of the state of the body  will be known once
given the initial positions and velocities of the generic point mass. We
say that the initial conditions must be given.

We want to prove the second theorem in mechanical terms, founding it on
the nature of the interaction laws and without imposing any restriction on
the initial conditions, knowledge of which is not supposed.
We look also for the proof that, 
provided initial conditions are similar, the transformations of the body
always take place so that

$$\ig \fra{dQ}T\le 0.$$
Suppose now that the body is constituted by a collection of point like
masses, or virtually such.  The initial condition be so given that the
successive transformation of the body  proceed so that

$$\ig \fra{dQ}T\le 0$$
We want to claim immediately that, provided the forces stay unchanged, it
is possible to exhibit another initial condition for which it is

$$\ig \fra{dQ}T\ge 0.$$
Because we can consider the values of the velocities of all point masses
reached at a given time $t_1$. We now want to consider, instead of the
preceding initial conditions, the following: at the beginning all point
masses have the same positions reached, starting form the preceding initial
conditions, in time $t_1$ but with the all velocities inverted. We
want in such case to remark that the evolution of the state towards the future
retraces exactly that left by the preceding evolution towards the time
$t_1$.

It is clear that the point masses retrace the same states followed by the
preceding initial conditions, but in the opposite direction.
The initial state that before we had at time $0$ we see it realized at time
$t_1$ [{\sl with opposite velocities}]. Hence if before it was

$$\ig \fra{dQ}T\le 0$$
we shall have now $\ge0$. 

On the sign of this integral the interaction cannot have influence, but it
only depends on the initial conditions. In all processes in the world in
which we live, experience teaches us this integral to be $\le 0$, and
this is not implicit in the interaction law, but rather depends on the
initial conditions. If at time $0$ the state [{\sl of the velocities}] of
all the points of the Universe was opposite to the one reached after a very
long time $t_1$ the evolution would proceed backwards and this would imply
$$\ig \fra{dQ}T\le 0$$
Every experimentation on the nature of the body and on the mutual
interaction law, without considering the initial conditions, to check that
$$\ig \fra{dQ}T\le 0$$
would be vain. We see that this difficulty is very attractive and we must
consider it as an interesting sophism.\index{sophism} To get close to the
fallacy that is in this sophism we shall immediately consider a system of a
finite number of point masses, which is isolated from the rest of the
Universe.

We think to a very large, although finite, number of elastic spheres, which
are moving inside a container closed on every side, whose walls are
absolutely still and perfectly elastic. No external forces be supposed
acting on our spheres. At time $0$ the distribution of the spheres in the
container be assigned as non uniform; for instance the spheres on the right
be denser than the ones on the left and be faster if higher than if lower
and of the same order of magnitude.  For the initial conditions that we have
mentioned  the spheres be at time $t_1$ almost uniformly mixed. 
We can then consider instead of the preceding initial conditions, the ones
that generate the inverse motion, determined by the initial conditions
reached at time $t_1$.  Then as time evolves the spheres come back; and at
time $t_1$ will have reached a non uniform distribution although the
initial condition was almost uniform. We then must argue as follows: a
proof that, after the time $t_1$ the mixing of the spheres must be with
absolute certainty uniform, whatever the initial distribution, cannot be
maintained.
This is taught by the probability itself; every non uniform distribution,
although highly improbable, is not absolutely impossible.
It is then clear that every particular uniform distribution, that follows
an initial datum and is reached in a given time is as improbable as any
other even if not uniform; just as in the lotto game every five numbers
are equally probable as the five $1,2,3,4,5$. And then the greater or
lesser uniformity of the distribution depends on the greater size of the
probability that the distribution becomes uniform, as time goes.

It is not possible, therefore, to prove that whatever are the initial
positions and velocities of the spheres, after a long enough time, a
uniform distribution is reached, nevertheless it will be possible to prove
that the initial states which after a long enough time evolve towards a
uniform state will be infinitely more than those evolving towards
a nonuniform state, and even in the latter case, after an even longer time,
they will evolve towards a uniform state.\footnote{\tiny Today this important
discussion is referred to as the argument of the {\it Boltzmann's
sea}, \Cite{Ul968}.}

Loschmidt's proposition teaches also to recognize the initial states that
really at the end of a time $t_1$ evolve towards a very non uniform
distribution; but it does not imply the proof that the initial data that
after a time $t_1$ evolve into uniform distributions are infinitely many
more. Contrary to this statement is even the proposition itself which
enumerates as infinitely more uniform distributions than non uniform, for
which the number of the states which, after a given time $t_1$ arrive to
uniform distribution must also be as numerous as those which arrive to
nonuniform distributions, and these are just the configurations that arise
in the initial states of Loschmidt, which become non uniform at time $t_1$.

It is in reality possible to calculate the ratio of the numbers of the
different initial states which determines the probabilities, which perhaps
leads to an interesting method to calculate the thermal
equilibria.%
\footnote{\tiny Boltzmann will implement the idea in
\Cite{Bo877-b}[\#42], see also Sec.(\,\ref{sec:XII-6}\,)
below.} Exactly analogous to the one that leads to the second
theorem.  It is at least in some special cases successfully
checked, when a system undergoes a transformation from a
nonuniform state to a uniform one, then $\ig \fra{dQ}T$ will be
intrinsically negative, while it will be positive in the inverse
case. Since there are infinitely more uniform then nonuniform
distributions of the states, therefore the last case will be
extremely improbable: and in practice it could be considered
impossible that at the beginning a mixture of oxygen and nitrogen
are given so that after one month the chemically pure oxygen is
found in the upper part and that the nitrogen in the lower, an
event that probability theory states as improbable but not as
absolutely impossible.

Nevertheless it seems to me that the Loschmidtian theorem has a great
importance, since it tells us how intimately related are the second
principle and the calculus of probabilities. For all cases in which $\ig
\fra{dQ}T$ can be negative it is also possible to find an initial condition
very improbable in which it is positive.  It is clear to me that for closed
atomic trajectories $\ig\fra{dQ}T$ must always vanish. For non closed
trajectories it can also be negative. Now a peculiar consequence of the
Loschmidtian theorem which I want to mention here, \ie that the state of
the Universe at an infinitely remote time, with fundamentally equal
confidence, can be considered with large probability both as a state in
which all temperature differences have disappeared, and as the state in
which the Universe will evolve in the remote future.\footnote{\tiny
  reference to the view of Clausius which claims that in the remote future
  the Universe will be in an absolutely uniform state. Here B. says that
  the same must have happened, with equal likelihood in the remote past.}

This is analogous to the following case: if we want that, in a given gas at
a given time, a non uniform distribution is realized and that the gas
remains for a very long time without external influences, then we must
think that as the distribution of the states was uniform before so it will
become entirely uniform.

In other words: as any nonuniform distribution evolves at the end of a time
$t_1$ towards a uniform one the latter if inverted as the same time $t_1$
elapses again comes back to the initial nonuniform distribution (precisely
for the said inversion). The [{\sl new}] but inverted initial
condition, chosen as initial condition, after a time $t_1$
similarly will evolve to a uniform distribution.\footnote{\tiny {\it I.e.} if once
having come back we continue the evolution for as much time again a
uniform distribution is reached.}

But perhaps such interpretation relegates in the domain of probability
theory the second principle, whose universal use appears very questionable,
and nevertheless just because of the theory of probability it will be
realized in every laboratory experimentation.
\*

\0[{\sl \S{}III, p. 127 and following: a check of the heat theorem is
presented in the case of a central motion, which will be revisited in the
papers of 1884 by v. Helmholtz\index{Helmholtz} and Boltzmann.}]
\*

Let $M$ be a point mass, whose mass will be denoted $m$, and let $OM=r$ its
distance from a fixed point $O$. The point $M$ is pushed towards $O$ by a
force $f(r)$. We suppose that the work

$$\f(r)=\int_r^\infty f(r) dr,$$
necessary to bring the point $M$ to an infinite distance from $O$, is
finite, so that it is a function $\f$ whose negative derivative in any
direction gives the force acting in the same direction, also called the 
force function [{\sl minus the potential}].   Let us denote by $v$ the
velocity of $M$ and with $\th$ the angle that the radius vector $OM$ form
with an arbitrarily fixed line in the plane of the trajectory, thus 
by the principle of the ``vis viva''\index{vis viva principle}:
$$\frac{m}2(\frac{dr}{dt})^2+\frac{m r^2}2(\frac{d\th}{dt})^2=\a-\f(r),
\eqno{(1)} $$
and by the area law
$$ r^2\frac{d\th}{dt}=\sqrt\b\eqno{(2)}$$
$\a$ and $\b$ remain constant during the whole motion. From these equations
follows
$$dt=\frac{dr}{\sqrt{\frac{2\a}m-\frac{2\f(r)}m-\frac\b{r^2}}},\eqno{(3)}$$
We define for the average value of the force function the term
$\lis \f=\frac{z}n$ with
$$\eqalign{z=& \int \frac{\f(r)\,dr}
{\sqrt{\frac{2\a}m-\frac{2\f(r)}m-\frac\b{r^2}}}\cr
n=&\int \frac{dr}
{\sqrt{\frac{2\a}m-\frac{2\f(r)}m-\frac\b{r^2}}}\cr}$$
The integration is from a minimum to a maximum of the radius vector $r$, or
also it is extended from a smaller to the next larger positive root of the
equation

$$\a-\f(r)-\frac{\b m}{2 r^2}$$
the average ``vis viva'' $T_1$ is $\a-\lis\f$. The polynomial equation must
obviously have two positive roots. The total sum of the ``vis viva'' and of
the work, which must be done on the point mass to lead it on a path to
infinity and its velocity to a standstill is $\a$. The force function be
unchanged, so also the work $\d Q$ necessary for somehow moving the point
mass from a path into another is equal to the increment $\d\a$ of the
quantity $\a$. Change now the force function, \eg because of the
intervening constant parameters $c_1,c_2,\ldots$, so under the condition
that the change of the nature of the force function demands no
work, $\d\a$ is the difference of the works, which are necessary for bringing
the point mass standing and again at infinite distance on the other path
deprived of the mentioned velocity.
The amount of ``vis viva'' and work necessary to move
from an initial site $M$ of the old
path to the new initial site $M'$ of the varied path as well as its
velocity in $M$ to the one in $M'$ differs from $\d\a$ for the extra work
resulting from the change of the force function in the varied state
over the one originally necessary to bring the point mass to an infinite
distance from the initial $M$. It is therefore

$$\d\a+\sum \frac{\dpr \f(r)}{\dpr c_k} \d c_k$$
where for $r$ one has to set the distance of the initial location $M$ from
$O$.  The extra work, which is due to the change of the force function
because of the required infinitely small displacement, is now infinitely
small of a higher order.  According to Clausius' idea the average
work in the change from one to the other paths is
$dQ=\d\a+\frac\z{n}$, with

$$\z=\int \frac{\ \Big[\,\sum\,\Big]\ \frac{\dpr\f}{\dpr c_k} \d c_k \,dr}
{\sqrt{\frac{2\a}m-\frac{2\f(r)}m-\frac\b{r^2}}}
$$
and $n$ is the above defined value.\footnote{\tiny Here it seems that there is a
  sign incorrect as $\z$ should have a minus sign.  {\it I have not
    modified the following equations}; but this has to be kept in mind; in
  Appendix \ref{appD} the calculation for the Keplerian case is reported in
  detail.\label{Keplerian sign} See also the footnote to p.129 of
  \Cite{Bo871-c}[\#20].} To let the integration easier to
perform, one sets
$$\f(r)=-\frac{a}r+\frac{b}{2r^2}$$
with an attraction of intensity $a/r^2-n/r^3$ expressed in the distance
$r$, $a$ and $b$ play here the role of the constants $c_k$. By the
variation of the motion let $\a,\b,a$ and $b$ respectively change by
$\d\a,\d\b,\d a$ and $\d b$, then it is:
$$\sum \frac{\dpr \f(r)}{\dpr c_k} \d c_k=-\frac{d\a}r+\frac{\d b}{2r^2}.$$
And also $\d Q$ is the average value of
$$\d\a=-\frac{\d a}r+\frac{\d b}{2 r^2}$$
These can also be compared with the results found above [{\sl in the
previous text}].  The quantity denoted before with $\d_2 V$ is in this case

$$\d_2\Big(\a-\frac{a}r+\frac{b}{2r^2}\Big)=
\d\a-\frac{\d a}r+\frac{\d b}{2 r^2}
$$
We now set, for brevity, formally

$$\r=r^2,\ s=\frac1r,\ {\rm and}\ \s=\frac1{r^2}$$
Hence let the trajectory be real so its endpoints must be the maximum and
the minimum of the radius vector $r$ or the pair of roots of the equation
$$\frac{2\a}m-\frac{2\f(r)}m-\frac\b{r^2}=
\frac{2\a}m-\frac{2a}m\frac1r-\frac{b+m\b}m\frac1{r^2}=0$$
must be positive, and the polynomial must be for the $r$, which are between
the pair of roots, likewise positive, and also for $r$ infinitely large
negative; \ie $\a$ must be negative and  positive at the Cartesian
coordinates $a$ and $b+b$. We want always to integrate from the smallest to
the largest value of $r$; then $dr$, $d\r$ and 

$$\sqrt{\frac{2\a}m-\frac{2a}m\frac1r-\frac{b+m\b}m\frac1{r^2}}$$
on the contrary $ds$ and $d\s$ are negative. Remark that

$$\eqalign{\int_{w_1}^{w_2}&\frac{dx}{\sqrt{A+Bx+Cx^2}}=\frac{\p}{\sqrt{-C}},\cr
\int_{w_1}^{w_2}&\frac{x\,dx}{\sqrt{A+Bx+Cx^2}}=-\frac{\p B}{2C\sqrt{-C}},\cr}$$
if $w_1$ is the smaller, $w_2$ the larger root of the equation
$A+Bx +C x^2=0$, so for the chosen form of the force function it is found 

$$\eqalign{
z=&-a\int \frac{dr}
{\sqrt{\frac{2\a}mr^2-\frac{2a}mr-\frac{b+m\b}m\frac1{r^2}}}
+\frac{b}2\int \frac{-ds}
{\sqrt{\frac{2\a}m+\frac{2a}m
s-\frac{b+m\b}m s^2}}
\cr
=&-a\cdot\p\sqrt{\frac{m}{-2 \a}}+\frac{b}2 \p\sqrt{\frac{m}{b+m\b}},\cr
\z=&-\d\a\,\p \sqrt{\frac{m}{-2 \a}} +\frac{\d
b}2\cdot\p\sqrt{\frac{m}{b+m\b}}\cr
n=&  \int \frac{dr}
{\sqrt{\frac{2\a}mr^2-\frac{2a}mr-\frac{b+m\b}m\frac1{r^2}}}
=-
\frac\p2\frac{a}\a\sqrt{\frac{m}{-2 \a}}\cr} $$
As $\a$ is negative, so to have all integrals essentially positive all
roots must be taken positive. It is

$$\eqalign{
\lis\f=&\frac{z}{n}=2\a-\frac{b\a}a\sqrt{\frac{-2\a}{b+m\b}},\cr
T_1=&\a-\lis\f=\frac{b\a}a\sqrt\frac{-2\a}{b+m\b}-\a\cr
dQ=&\d\a+\frac\z{n}=\d\a+2\a\frac{\d a}a-\frac{\a\d b}a
\sqrt{\frac{-2\a}{b+m\b}}\cr}$$
it is built here the term $\d Q/T_1$, so it is immediately seen that it is
not an exact differential, since $\d\b$ does not appear; and, furthermore,
if $a$ and $b$ and also the force function stay constant also it is not an
exact differential. On the contrary if $b=\d b=0$ the trajectory is closed
and $\d Q/T_1$ is an exact differential.\footnote{\tiny The sign error mentioned in
the footnote at p.\pageref{Keplerian sign} does not affect the conclusion
but only some intermediate steps.}

As a second case, in which the integration is much easier, is if

$$\f(r)=-a r^2+\frac{b}{2 r^2}$$

\0[{\sl The analysis follows the same lines and the result is again that
    $\frac{\d Q}{T_1}$ is an exact differential only if $\d b=b=0$.  It
    continues studying various properties of central motions of the kinds
    considered so far as the parameters vary, without reference to
    thermodynamics. The main purpose and conclusion of Sec. III seems
    however to be that when there are other constants of motion it cannot
    be expected that the average kinetic energy is an integrating factor
    for $dQ=dE-\media{\dpr_{\V c} \f\cdot \d\V c}$. The Newtonian potential
    is a remarkable exception, see Appendix \ref{appD}. Other exceptions
    are the $1$--dimensional systems, obtained as special cases of the
    central potentials cases with zero area velocity, $\b=0$. However, even
    for the one dimensional case, only special cases are considered here:
    the general $1$-dimensional case was discussed a few years later by
    v. Helmoltz, \Cite{He884a}\Cite{He884b}, and immediately afterwards by
    Boltzmann, \Cite{Bo884}[\#73], see Sec.(\,\ref{sec:XIII-6}\,)}].

\section{Discrete phase space, count of its points and entropy.}
\def\SEC{Discrete phase space, count of its points and entropy.}
\label{sec:XII-6}\iniz
\lhead{\small\ref{sec:XII-6}.\ \SEC}
\index{discrete phase space}\index{entropy}

\0{\it Partial translation and comments: L. Boltzmann\index{Boltzmann},
{\it \"Uber die Bezie\-hung zwischen dem zweiten Hauptsatze der
mechanischen W\"arme\-theo\-rie und der Wahr\-scheinlichkeitsrechnung,
respektive den S\"atzen \"uber das W\"armegleichgewicht}, in
Wis\-sen\-schaft\-li\-che Abhandlungen, ed. F. Hasen\"ohrl, Vol.{\bf2}, \#42
p. 164-233, \Cite{Bo877-b}.}
\*

\0{\bf p.166} ... We now wish to solve the problem which on my above quoted
paper {\it Bemerkungen \"uber einige Probleme der mechanischen
  {W}{\"a}rmetheorie}, \Cite{Bo877a}[\#39], I have already formulated
clearly, \ie the problem of determining the ``ratios of the number of
different states of which we want to compute the probabilities''.

We first want to consider a simple body, \ie a gas enclosed between
absolutely elastic walls and whose molecules are perfect spheres absolutely
elastic (or centers of force which, now, interact only when their
separation is smaller than a given quantity, with a given law and
otherwise not; this last hypothesis, which includes the first as a special
case does not at all change the result).  Nevertheless in this case the use
of probability theory is not easy. The number of molecules is not infinite
in a mathematical sense, although it is extremely large. The number of the
different velocities that every molecule can have, on the contrary, should
be thought as infinite. Since the last fact renders much more difficult the
calculations, thus in the first Section of this work I shall rely on easier
conceptions to attain the aim, as I often did in previous works (for
instance in the {\it Weiteren Studien}, \Cite{Bo872}[\#22]).
\*
.....
\*
\0{\bf\S{}I.} {\bf The number of values of the ``vis viva'' is
  discrete.}(p.167) \*

We want first to suppose that every molecule can assume a finite number of
velocities, for instance the velocities

$$0,\fra1q,\fra2q,\fra3q,\ldots,\fra{p}q,$$
where $p$ and $q$ are certain finite numbers. At a collision of two
molecules will  correspond a change of the two velocities, so that the state
of each will have one of the above mentioned velocities, \ie

$$0, \ {\rm or}\ \fra1q, \ {\rm or}\ \fra2q, \ {\it \&tc\ until\
}\,\fra{p}q,$$
It is plane that this fiction is certainly not realized in any
mechanical problem, but it is only a problem that is much easier to treat
mathematically, and which immediately becomes the problem to solve if $p$
and $q$ are allowed to become infinite.

Although this treatment of the problem appears too abstract, also it very
rapidly leads to the solution of the problem, and if we think that all
infinite quantities in nature mean nothing else than going beyond a bound,
so the infinite variety of the velocities, that each molecule is capable of
taking, can be the limiting case reached when every molecule can take an
always larger number of velocities.\index{continuum limit}

We want therefore, temporarily, to consider how the velocities are related
to their ``vis viva''. Every molecule will be able to take a finite number
of values of the ``vis viva''. For more simplicity suppose that the values
of the ``vis viva'' that every molecule can have form an arithmetic
sequence,

$$0,\e,\ 2\,\e,\ 3\,\e,\ldots, p\,\e$$
and we shall denote with $P$ the largest of the possible values of $p$.

At a  collision each of the two molecules involved
will have again a velocity
$$0, \ {\rm or}\ \e, \ {\rm or}\ 2\,\e, \ {\rm etc.}\ldots ,\ p\,\e,$$
and in any case the event will never cause that one of the molecules will
end up with having a value of the ``vis viva'' which is not in the
preceding sequence.

Let $n$ be the number of molecules in our container. If we know how many
molecules have a ``vis viva'' zero, how many $\e$, \&tc, then we say that
the distribution of the ``vis viva'' between the molecules is given.

If at the initial time a distribution of the molecules states is given, it
will in general change because of the collisions. The laws under which such
changes take place have been often object of research.  I immediately
remark that this is not my aim, so that I shall not by any means depend on
how and why a change in the distribution takes place, but rather to the
probability on which we are interested, or expressing myself more
precisely I will search all combinations that can be obtained by
distributing $p+1$ values of ``vis viva'' between $n$ molecules, and hence
examine how many of such combinations correspond to a distribution of
states.  This last number gives the probability of the relevant
distribution of states, precisely as I said in the quoted place of my {\it
Bemerkungen \"uber einige Probleme der mechanischen W\"armetheorie}
(p.121), \Cite{Bo877a}[\#39].

Preliminarily we want to give a purely schematic version of the problem to
be treated. Suppose that we have $n$ molecules each susceptible of assuming
a ``vis viva''

$$0,\e,\ 2\e,\ 3\e,\ldots, p\e.$$
and indeed these ``vis viva'' will be distributed in all possible ways
between the $n$ molecules, so that the sum of all the ``vis viva''
stays the same; for instance is equal to $\l \e=L$.

Every such way of distributing, according to which the first molecule has a
given ``vis viva'', for instance $2\e$, the second also a given one, for
instance $6\e$, \&tc until the last molecule, will be called a
``complexion'', and certainly it is easy to understand that each single
complexion is assigned by the sequence of numbers (obviously after division
by $\e$) to which contribute the ``vis viva'' of the single molecules.
We now ask which is the number ${\cal B}$ of the complexions in which $w_0$
molecules have ``vis viva'' $0$, $w_1$ ``vis viva'' $\e$, $w_2$ ``vis
viva'' $2\e$, {\it \&tc}, $\ldots\ w_p$ ``vis viva'' $p\e$.

We have said before that when is given how many molecules have zero ``vis
viva'', how many $\e$ {\it \&tc}, then the distribution of the states among
the molecules is given; we can also say: the number ${\cal B}$ gives us how
many complexions express a distribution of states in which $w_0$ molecules
have zero ``vis viva'', $w_1$ ``vis viva'' $\e$, {\it \&tc}, or it gives us
the probability of every distribution of states. Let us divide in fact the
number ${\cal B}$ by the total number of all possible complexions and we
get in this way the probability of such state.

It does not follow from here that the distribution of the states gives
which are the molecules which have a given ``vis viva'', but only how many
they are, so we can deduce how many ($w_0$) have zero ``vis viva'', and how
many ($w_1$) among them have one unit of ``vis viva'' $\e$, {\it \&tc}. All
of those values zero, one, {\it \&tc} we shall call elements of the
distribution of the states.

(p.170) ........... (p.175)

We shall first treat the determination of the number denoted above ${\cal
B}$ for each given distribution of states, \ie the permutability of such
distribution of states. Then denote by $J$ the sum of the permutabilities
of all possible distributions of states, and hence the ratio $\fra{{\cal
B}}{J}$ gives immediately the probability of the distribution that from now
on we shall always denote $W$.

We also want right away to calculate the permutability ${\cal B}$ of the
distributions of states characterized by $w_0$ molecules with zero ``vis
viva'', $w_1$ with ``vis viva'' $\e$ {\it \&tc}. Hence evidently

$$w_0+w_1+w_2+\ldots+w_p=n\eqno(1)$$
$$w_1+2w_2+3 w_3+\ldots+p w_p=\l,\eqno(2)$$
and then $n$ will be the total  number of molecules and $\l\e=L$ their
``vis viva''.

We shall write the above defined distribution of states with the method
described, so we consider a complexion with $w_0$ molecules with zero ``vis
viva", $w_1$ with "vis viva" unitary {\it \&tc}. We know that the number
of permutations of the elements of this complexion, with in total $n$
elements distributed so that among them $w_0$ are equal between them, and
so $w_1$ are equal between them ... The number of such complexions is known
to be%
\footnote{\tiny Here particles are considered distinguishable and the total
number of complexions is $P^n$.}
$${\cal B}=\fra{n!}{(w_0)!\,(w_1)!\ldots}\eqno(3)$$
The most probable distribution of states will be realized for those choices
of the values of $w_0,w_1,\ldots$ for which ${\cal B}$ is maximal and
quantities $w_0,w_1,\ldots$ are at the same time constrained by the
conditions (1) and (2). The denominator of ${\cal B}$ is a product and
therefore it will be better to search for the minimum of its logarithm, \ie
the minimum of

$$M=\ell[(w_0)!]+\ell[(w_1)!]+\ldots\eqno(4)$$
where $\ell$ denotes the natural logarithm.

..............

\0[{\sl Follows the discussion of this simple case in which the energy
levels are not degenerate (\ie this is essentially a $1$-dimensional case)
ending with the Maxwell distribution of the velocities. In Sec.II, p.186,
B. goes on to consider the case of $2$-dimensional and of $3$--dimensional
cells (of sides $da,db$ or $da,db,dc$) in the space of the velocities (to
be able to take degeneracy into account),
treating first the discrete case and then taking the continuum limit:
getting the canonical distribution. Sec.III (p.198) deals with the case of
polyatomic molecules and external forces.  

In Sec.{IV} (p.204), concludes that it is possible to define and count in
other ways the states of the system and discusses one of them.

An accurate analysis of this paper, together with \Cite{Bo868-a}, is in
 \Cite{Ba990} where two ways of computing the distribution when
the energy levels are discrete are discussed pointing out that 
{\it unless the continuum limit, as considered by Boltzmann in the two papers,
 was taken} would lead to a distribution of Bose-Einstein type or of
 Maxwell-Boltzmann type: see the final comment in Sec.(\,\ref{sec:II-6}\,)
 above and also \Cite{Ga000}[Sec.(2.2),(2.6)].  

In Sec.V, p.215, the link of the probability distributions found in the
previous sections with entropy is discussed, dealing with examples like the
expansion of a gas in a half empty container; the example of the barometric
formula for a gas is also discussed. On p.218 the following celebrated
statement is made (in italics) about ``permutability'' (\ie number of ways
in which a given (positions-velocities) distribution can be achieved) and is
illustrated with the example of the expansion of a gas in a half empty
container:}]
\*

{\it Let us think of an arbitrarily given system of bodies, which undergo
an arbitrary change of state, without the requirement that the initial or
final state be equilibrium states; then always the measure of the
permutability of all bodies involved in the transformations continually
increases and can at most remain constant, until all bodies during the
transformation are found with infinite approximation in thermal
equilibrium.}%
\footnote{\tiny After the last word appears in parenthesis and still in italics
  {\it (reversible transformations)}, which seems to mean ``{\it or
    performing reversible transformations}''.}

\section{Monocyclic and orthodic systems. Ensembles}
\def\SEC{Monocyclic and orthodic systems. Ensembles}
\label{sec:XIII-6}\iniz
\lhead{\small\ref{sec:XIII-6}.\ \SEC}
\index{monocyclic systems}\
\index{orthodic systems}\index{ensembles}
\index{orthodic ensemble}

\0{\sl Partial translation and comments: L. Boltzmann\index{Boltzmann},
{\it {\"U}ber die {E}igenshaften monozyklischer und anderer damit
verwandter {S}ysteme", (1884), {W}issen\-schaft\-liche {A}bhandlungen,
ed. {F}. {H}asen{\"o}hrl, {\bf 3}, 122--152, \#73, \Cite{Bo884}.}}%
\footnote{\tiny The first three paragraphs have been printed almost unchanged in
Wien, Ber, {\bf 90}, p.231, 1884; ...}
\*

{\it The most complete proof of the second main theorem is manifestly based
  on the remark that, for each given mechanical system, equations that are
  analogous to equations of the theory of heat hold.} [{\sl italics added}]\\
Since on the one hand it is evident that the proposition, in this
generality, cannot be valid and on the other hand, because of our scarce
knowledge of the so called atoms, we cannot establish the exact mechanical
properties with which the thermal motion manifests itself, the task arises
to search in which cases and up to which point the equations of mechanics
are analogous to the ones of the theory of heat.  We should not refrain to
list the mechanical systems with behavior congruent to the one of the solid
bodies, rather than to look for all systems for which it is possible to
establish stronger or weaker analogies with warm bodies. The question has
been posed in this form by Hrn. von Helmoltz\footnote{\tiny Berl.Ber, 6 and 27
  March 1884.} and I have aimed, in what follows, to treat a very special
case, before proceeding to general propositions, of the analogy that he
discovered between the thermodynamic behavior and that of systems, that he
calls monocyclic, and to follow the propositions, to which I will refer, of
the mechanical theory of heat intimately related to monocyclic
systems.\annotaa{A very general example of monocyclic system is offered by
  a current without resistance (see Maxwell, ``Treatise on electricity'',
  5.10.580, where $x$ and $y$ represent the v. Helmholtzian $p_a$ and
  $p_b$).}
\*

\centerline{\bf\S1}

Let a point mass move, according to Newton's law of gravitation, around a
fixed central fixed body $O$, on an elliptic trajectory.  Motion is not in
this case monocyclic; but it can be made such with a trick, that I already
introduced in the first Section of my work ``{\it Einige allgemeine
  s{\"a}tze {\"u}ber {W\"a}rme\-gleichgewicht}$\,$''%
\annotaa{Wiener. Berl. {\bf 63}, 1871,
  \Cite{Bo871-b}[\#18], [\sl see also Sec.(\,\ref{sec:IX-6}\,) above]}
and that also Maxwell\annotaa{Cambridge Phil. Trans.  {\bf 12}, III, 1879
  (see also Wiedemanns Beibl\"atter, {\bf 5}, 403, 1881).} has again
followed.

Imagine that the full elliptic trajectory is filled with mass, which at
every point has a density (of mass per unit length) such that, while time
elapses, density in each point of the trajectory remains unchanged. As it
would be a Saturn ring thought as a homogeneous flow or as a homogeneous
swarm of solid bodies so that, fixed one of the rings among the different
possible ones a stationary motion would be obtained. The external force can
accelerate the motion or change its eccentricity; this can be obtained
increasing or diminishing slowly the mass of the central body, so that
external work will be performed on the ring which, by the increase or
diminution of the central body mass, in general is not accelerated nor
decelerated in the same way.  This simple example is treated in my work
{\it Bemerkungen {\"u}ber einige Probleme der mechanischen
{W}{\"a}rmetheo\-rie}\annotaa{Wien, Ber. {\bf 75}, [{\sl see
Appendix \ref{appD} below and Sec.(\,\ref{sec:XI-6}\,) above}].  See also
Clausius, Pogg. Ann. {\bf 142}, 433; Math. Ann. von Clebsch, {\bf4}, 232,
{\bf 6}, 390, Nachricht. d. G\"ott. Gesellsch. Jarhrg. 1871 and 1871;
Pogg. nn. {\bf 150}, 106, and Erg\"angzungs, {\bf 7},215.} where in Section
3 are derived formulae in which now we must set $b=0$ and for $m$ we must
intend the total mass of the considered ring (there, by the way, the
appropriate value of the work is given with wrong signs).  I denote always
by $\F$ the total potential energy, with $L$ the total "vis viva" of the
system, by $dQ$ the work generated by the increase of the internal motion
which, as Hrn, v. Helmholtz, I assume that the external forces always undergo
an infinitesimal variation of their value and that are immediately
capable to bring back the motion into a stationary state.%
\footnote{\tiny The ``direct'' increase of the internal motion is the amount of
work done on the system by the internal and external forces (which in
modern language is the variation of the internal energy) summed to the work
$dW$ done by the system on the outside: $dQ=dU+dW$; which would be
$0$ if the system did not absorb heat. If the potential energy $W$ due to
the external forces depends on a parameter $a$ then the variation of
$W$ changed in sign,
$-\dpr_a W_a da$, or better its average value in time, is the work that the
system does on the outside due to the only variation of $W$ while the
energy of the system varies also because the motion changes because of the
variation of $a$. Therefore here it has to be interpreted as the average
value of the derivative of the potential energy $W=-a/r$ with respect to
$a$ times the variation $da$ of $a$. 
Notice that in the  Keplerian motion it id $2L=a/r$ and
therefore $\media{-\dpr_a W\,da}=\media{da/r}=\media{2Lda/a}$, furthermore
the total energy is $L+\F=-L$ up to a constant and hence $dU=-dL$.}
Let then $a/r^2$ be the total attraction force that the central body
exercises on the mass of the ring if it is at a distance $r$, let $C$
be an unspecified global constant then  it is, [{\sl for more details see
Appendix \ref{appD} below}],

$$ \F=C-2L,\quad dQ=-dL+2L\,\fra{da}a=L \,d \log \fra{a^2}{L};$$
hence $L$ is also the integrating factor of $dQ$, and the consequent value
of the entropy $S$ is $\log a^2/L$ and the consequent value of the
characteristic function [{\sl free energy}] is

$$K= \F+L-LS=C-L-LS=C-L-L\log\fra{a^2}L$$
Let $2L\frac{\sqrt{L}}a=q,\quad \fra{a}{\sqrt{L}}=s$, so that it is $dQ=q
ds$ and the characteristic function becomes $H=\F+L-q s=C=3L$ and it is
immediately seen that

$$\Big(\fra{\dpr K}{\dpr a}\Big)_L=\Big(\fra{\dpr H}{\dpr a}\Big)_q=-A$$
is the gain deriving from the increase of $a$, in the sense that $A da$ is
how much of the internal motion can be converted into work when $a$ changes
from $a$ to $a+da$. Thus it is

$$\Big(\fra{\dpr K}{\dpr L}\Big)_a=-S,\quad 
\Big(\fra{\dpr H}{\dpr q}\Big)_a=-s$$
The analogy with the monocyclic systems of Hrn.  v. Helmholtz' with a
single velocity $q$ is also transparent.  

....

\*
\centerline{\bf\S2}
\*

\0[{\sl Detailed comparison with the work of Helmoltz follows, in
    particular the calculation reported in Sec.(\,\ref{sec:IV-1}\,),
    p.\pageref{heat theorem}, is explained. The notion of monocyclic
    system, \ie a system whose orbits are all periodic, allows to regard
    each orbit as a stationary state of the system and, extending Helmotz'
    conception, as the collection of all its points which is called a
    ``monode'', \index{monode} each being a representative of the
    considered state.

Varying the parameters of the orbit the state changes (\ie the orbit
changes). Some examples of monocyclic systems are worked out: for all of
their periodic orbits is defined the amount of heat $dQ$ that the system
receives in a transformation (``work to increase the internal motion'' or
``infinitesimal direct increment\label{direct increment} of the internal
motion, \ie the heat acquired by a warm body''), and the amount of work
$dW$ that the system does against external forces as well as the average
kinetic energy $L$; in all of them it is shown that $\frac{dQ}L$ is an
exact differential. For a collection of stationary motions of a system this
generates the definitions (p.129-130):}]
\*

I would permit myself to call systems whose motion is stationary in this
sense with the name {\it monodes}.\annotaa{With the name ``stationary''
Hrn. Clausius would denote every motion whose coordinates remain always
within a bounded region.} They will therefore be characterized by the
property that in every point persists unaltered a motion, not function of
time as long as the external forces stay unchanged, and also in no point
and in any region or through any surface mass or "vis viva" enters from the
outside or goes out.  If the "vis viva" is the integrating denominator of
the differential $dQ$, which directly gives the work to increase the
internal motion,\footnote{\tiny in an infinitesimal transformation,
  {\it i.e.} the
variation of the internal energy summed to the work that the system
performs on the outside, which defines the heat received by the system. The
notion of {\it monode} and {\it orthode} will be made more clear in the
next subsection 3.}  then I will say that the such systems are {\it
orthodes}.[{\sl Etymologies: monode=\bgr m'onos\egr +\bgr
e>~idos\egr=unique+aspect; orthode=\bgr >orj'os\egr + \bgr
e>~idos\egr=right + aspect.}]

....
\*
\centerline{\S3.}
\*

After these introductory examples I shall pass to a very general
case. Consider an arbitrary system, whose state is characterized by
arbitrary coordinates $p_1,p_2,\ldots, p_g$; and let the corresponding
momenta be $r_1,r_2,\ldots, r_g$. For brevity we shall denote the
coordinates by $p_{\bf g}$ and the momenta by $r_{\bf g}$. Let the internal
and external forces be also assigned; the first be conservative. Let $\ps$
be the "vis viva" and $\ch$ the potential energy of the system, then also
$\ch$ is a function of the $p_{\bf g}$ and $\ps$ is a homogeneous function
of second degree of the $r_{\bf g}$ whose coefficients can depend on the
$p_{\bf g}$.  The arbitrary constant appearing in $\ch$ will be determined
so that $\ch$ vanishes at infinite distance of all masses of the system or
even at excessive separation of their positions.  We shall not adopt the
restrictive hypothesis that certain coordinates of the system be
constrained to assigned values, hence also the external forces will not be
characterized other than by their almost constancy on slowly varying
parameters. The more so the slow variability of the external forces will
have to be taken into account either because $\ch$ will become an entirely
new function of the coordinates $p_{\bf g}$, or because some constants that
appear in $\ch$, which we shall indicate by $p_{\bf a}$, vary slowly.
\*
{\bf 1.} We now imagine to have a large number  $N$ of such systems,
of exactly identical nature; each system absolutely independent from all
the others.%
\footnote{\tiny In modern language this is an ensemble: it is the generalization of
the Saturn ring of Sec.1: each representative system is like a stone in a
Saturn ring. It is a way to realize all states of motion of the same
system.  Their collection does not change in time and keeps the same
aspect, if the collection is stationary, \ie is a ``monode''.}
The number of such systems whose coordinates and momenta are
between the limits $p_1$ and $p_1+dp_1$, $p_2$ and
$p_2+dp_2\ldots$, $r_g$ and $r_g+dr_g$ be

$$dN= N e^{-h(\ch+\ps)} \fra{\sqrt\D\, d\s\,d\t}{\ig\ig
e^{-h(\ch+\ps)}\sqrt\D\, d\s\,d\t},$$
where $d\s=\D^{-\fra12} dp_1dp_2\ldots dp_g, \ d\t=dr_1\,dr_2\ldots dr_g$
(for the meaning of $\D$ see Maxwell {\it loc. cit} p.556).\footnote{\tiny In
general the kinetic energy is a quadratic form in the $r_{\bf g}$ and then
$\D$ is its determinant: it is the Jacobian of the linear transformation
$r_{\bf g}\otto r_{\bf g}'$ that brings the kinetic energy in the form
$\fra12|r'_{\bf g}|^2$.}

The integral must be extended to all possible values of the coordinates and
momenta. The totality of these systems constitutes a {\it monode} in the
sense of the definition given above (see, here, especially Maxwell {\it
  loc. cit}) and I will call this species of monodes with the name {\it
  Holodes}. [{\sl Etymology: \bgr <'olos\egr\ + \bgr e>~idos \egr or
    ``global'' + ``aspect'' .}]%
\footnote{\tiny Probably because the canonical distribution deals with all
  possible states of the system and does not select quantities like the
  energy or other constants of motion.}

Each system I will call an {\it element} of the 
holode\index{holode}.\footnote{\tiny Hence a
monode is a collection of identical systems called {\it elements} of the
monode, that can be identified with the points of the phase space. The
points are permuted by the time evolution but the number of them near a
phase space volume element remains the same in time, \ie the distribution
of such points is stationary and keeps the same ``unique aspect''.
The just given canonical distributions are particular
kinds of monodes called holodes. An holode is therefore an element of a
species (``gattung''), in the sense of collection, of monodes that are
identified with the canonical distributions [of a given mechanical system].
A holode will be identified with a state of thermodynamic equilibrium,
because it will be shown to have correct properties.  For its successive
use an holode will be intended as a statistical ensemble, \ie the family of
probability distributions, consisting in the canonical distributions of a
given mechanical system: in fact the object of study will be the properties
of the averages of the observables in the holodes as the parameters that
define them change, like $h$ (now $\b$, the inverse temperature) or the
volume of the container.}  The total "vis viva" of a holode is%
\footnote{\tiny In the case of a gas the number $g$ must be thought as the
  Avogadro's number times the number of moles, while the number $N$ is a
  number {\it much larger} and equal to the number of cells which can be
  thought to constitute the phase space. Its introduction is not necessary,
  and Boltzmann already in 1871 had treated canonical and microcanonical
  distributions with $N=1$: it seems that the introduction of the $N$
  copies, adopted later also by Gibbs, intervenes for ease of comparison of
  the work of v. Helmholtz with the preceding theory of 1871. Remark that
  B. accurately avoids to say too explicitly that the work of v. Helmholz
  is, as a matter of fact, a different and particular version of his
  preceding work. Perhaps this caution is explained by caution of Boltzmann
  who in 1884 was thinking to move to Berlin, solicited and supported by
  v. Helmholtz. We also have to say that the works of 1884 by v. Helmholtz
  became an occasion for B. to review and systematize his own works on the
  heat theorem which, after the present work, took up the form and the
  generality that we still use today as ``theory of the statistical
  ensembles''.}

$$L=\fra{Ng}{2h}.$$
Its potential energy $\F$  equals $N$ times the average value
$\lis\ch$ of $\ch$, \ie:

$$\F=N\fra{\ig \ch\, e^{-h\ch}\,d\s}{\ig \, e^{-h\ch}\,d\s}.$$
The coordinates $p_{\bf g}$ correspond therefore to the v. Helmholtzian
$p_{\bf b}$, which appear in the "vis viva" $\ps$ and potential energy
$\ch$ of an element. The intensity of the motion of the entire
ergode\footnote{\tiny This is a typo as it should be holode: the notion of ergode
is introduced later in this work.} and hence also $L$ and $\F$ depend
now on $h$ and on $p_{\bf a}$, as for Hrn. v. Helmholtz on $q_{\bf
b}$ and $p_{\bf a}$.

The work provided for a direct increase, see 
p.\pageref{direct increment}, of internal motion is:

$$\d Q=\d \F+\d L-N\fra{\ig \d\ch\,e^{-h\ch}\,d\s}{\ig\,e^{-h\ch}\,d\s}$$
(see here my work\annotaa{Wien. Ber., {\bf63}, 1871, formula (17).}
{\it Analytischer Beweis des zweiten {H}auptsatzes der mechanischen
{W\"a}rme\-theorie aus den {S\"a}tzen {\"u}ber das {G}leichgewicht des
lebendigen {K}raft}), \Cite{Bo871-c}[\#19], [{\sl see also Sec.(\,\ref{sec:X-6}\,)
above}].  The amount of internal motion generated by the external work,
when the parameter $p_a$ varies\footnote{\tiny Here we see that Boltzmann considers
among the parameters $p_a$ coordinates such as the dimensions of the
molecules container: this is not explicitly said but it is often used in
the following.} by $\d p_a$, is therefore $-P\d p_a$, with

$$-P=\fra{N\ig\fra{\dpr\ch}{\dpr p_a} e^{-h \ch}\,d\s}{\ig e^{-h
\ch}\,d\s}$$
The "vis viva" $L$ is the integrating denominator of $\d Q$: all holodes
are therefore orthodic, and must therefore also provide us with
thermodynamic analogies. Indeed let\footnote{\tiny Here the argument in the
  original relies to some extent on the earlier paragraphs: a self contained 
check is therefore reported in this footnote for ease of the reader:
  $$F\defi-h^{-1}\log \int e^{-h(\ch+\f)}\defi -h^{-1}\log Z(\b,p_a),\quad
  T=h^{-1}$$ and remark that 
$$dF=(h^{-2} \log Z+h^{-1}
  (\F+L))dh-h^{-1}\dpr_{p_a}\log Z\, dp_a$$
Define $S$ via $F\defi U-TS$
  and $U=\F+L$ then 
$$dF=dU-T dS-S dT =-\frac{d T}T (-(U-TS)+U)+Pdp_a$$ 
hence $dU
  -TdS-SdT=-\frac{dT}T TS - P dp_a$, \ie $Td S=dU+Pdp_a$ and the factor
  $T^{-1}=h$ is the integrating factor for $dQ\defi dU+Pdp_a$, see 
\Cite{Ga000}[Eq.(2.2.7)].  } 
$$\eqalign{s=&\fra{1}{\sqrt{h}}\Big(\ig e^{-h\ch}d\s\Big)^{\fra1g}
e^{\fra{h\ch}{g}}=\sqrt{\fra{2L}{N g}}\Big(\ig
e^{-h\ch}d\s\Big)^{\fra1g} e^{\fra\F{2L}},\cr
q=&\fra{2L}s,\quad K=\F+L-2L\log s,\quad H=\F-L, \cr}$$
[{\sl the intermediate expression for $s$ is not right and instead of $\ch$
    in the exponential should have the average $\frac{\F}N$ of $\ch$\,}]

{\bf2.} Let again be given many ($N$) systems of the kind considered at the
beginning of the above sections; let all be constrained by the constraints

$$\f_1=a_1,\ \f_2=a_2,\ \ldots\ ,\f_k=a_k.$$
These relations must also, in any case, be integrals of the equations of
motion. And suppose that there are no other integrals. Let $dN$ be the
number of systems whose coordinates and momenta are between $p_1$ e
$p_1+dp_1$, $p_2$ and $p_2+dp_2$, $\ldots\ \ r_g$ and $r_g+dr_g$. Naturally
here the differentials of the coordinates or momenta that we imagine
determined by the equations $\f_1=a_1,\ldots$ will be missing.  These
coordinates or momenta missing be $p_c,p_d,\ldots,r_f$; their number
be $k$. Then if

$$dN=\fra{\fra{N dp_1 dp_2\ldots dr_g}{\sum\pm \fra{\dpr\f_1}{\dpr
p_c}\ldots\fra{\dpr\f_k}{\dpr
r_f}}}
{\ig\ig\ldots 
\fra{ dp_1 dp_2\ldots dr_g}{\sum\pm \fra{\dpr\f_1}{\dpr
p_c}\cdot
\fra{\dpr\f_2}{\dpr p_d}\ldots \fra{\dpr\f_k}{\dpr r_f}}}$$
the totality of the $N$ systems will constitute a monode, which is defined
by the relations $\f_1=a_1,\ldots$.  The quantities $a$ can be either
constant or subject to slow variations. The functions $\f$ in general may
change form through the variation of the $p_a$, always slowly. Each single
system is again called element. 

Monodes that are constrained through the only value of the equation of the
``vis viva''\footnote{\tiny The equation of the ``vis viva'' is the energy
conservation $\f=a$ with $\f=\psi+\chi$, if the forces are conservative
with potential $\ch$.} will be called {\it ergodes}, while if also other
quantities are fixed will be called {\it subergodes}.  The ergodes are
therefore defined by

$$dN=\fra{ 
\fra{N\,dp_1dp_2\ldots dp_g dr_1\ldots dr_{g-1}}
{\fra{\dpr \ps}{\dpr r_g}}
}
{ \ig \ig \fra{\,dp_1dp_2\ldots dp_g dr_1\ldots dr_{g-1}}{\fra{\dpr
\ps}{\dpr r_g}}}$$
Hence for the ergodes there is a $\f$, equal for all the identical systems
and which stays, during the motion, equal to the constant energy of each
system $\ch+\ps= \fra1N (\F+L)$. Let us set again $\D^{-\fra12}dp_1
dp_2\ldots dp_g=d\s$, and then (see the works cited above by me and by
Maxwell):

$$\eqalign{
\F=& N\fra{\ig \ch \ps^{\fra{g}2-1} d\s}{\ig
\ps^{\fra{g}2-1}d\s},\qquad
L= N\fra{\ig  \ps^{\fra{g}2} d\s}{\ig
\ps^{\fra{g}2-1}d\s},\cr
\d Q=& N\fra{\ig \d\ps \ps^{\fra{g}2-1} d\s}{\ig
\ps^{\fra{g}2-1}d\s}=\d(\F+L)-N 
\fra{\ig  \d\ch\,\ps^{\fra{g}2-1} d\s}{\ig
\ps^{\fra{g}2-1}d\s},\cr}$$
$L$ is again the integrating factor of $\d Q$,\footnote{\tiny The (elementary)
integrations on the variables $r_{\bf g}$ with the constraint $\ps+\ch=a$
have been explicitly performed: and the factor $\ps^{\fra{g}2-1}$ is
obtained, in modern terms, performing the integration $\ig\d(\ch-(a-\ps))
dr_{\bf g}$ and in the formulae $\ps$ has to be interpreted as
$\sqrt{a-\ch}$, as already in the work of 1871.}  and the entropy thus
generated is $\log (\ig \ps^{\fra{g}2} d\s)^{\fra2g}$, while it will also
be $\d Q=q\,\d s$ if it will be set:

$$s= (\ig \ps^{\fra{g}2} d\s)^{\fra1g},\qquad q=\fra{2L}s.$$
Together with the last entropy definition also the characteristic function
$\F-L$ is generated. The external force in the direction of the parameter
$p_a$ is in each system

$$-P= \fra{\ig \fra{\dpr\ch}{\dpr p_a} \ps^{\fra{g}2-1} d\s}{\ig
\ps^{\fra{g}2-1}d\s}.$$
Among the infinite variety of the subergodes I consider those in which for
all systems not only is fixed the value of the equation of "vis viva" [{\sl
value of the energy}] but also the three components of the angular
momentum. I will call such systems {\it planodes}. Some property of such
systems has been studied by Maxwell, {\it loc. cit.}. Here I mention only
that in general they are not orthodic.  

The nature of an element of the ergode is determined by the parameters
$p_{\bf a}$\footnote{\tiny In the text, however, there is $p_{\bf b}$: typo?}.
Since every element of the ergode is an aggregate of point masses and the
number of such parameters $p_{\bf a}$ is smaller than the number of all
Cartesian coordinates of all point masses of an element, so such $p_{\bf
a}$ will always be fixed as functions of these Cartesian coordinates, which
during the global motion and the preceding developments remain valid
provided these functions stay constant as the "vis viva" increases or
decreases.\footnote{\tiny Among the $p_{\bf a}$ we must include the container
dimensions $a,b,c$, for instance: they are functions of the Cartesian
coordinates which, however, are {\it trivial constant functions}. The
mention of the variability of the "vis viva" means that the quadratic form
of the ``vis viva'' must not depend on the $p_{\bf a}$.}.  If there was
variability of the potential energy for reasons other than because of the
mentioned parameters\footnote{\tiny I interpret: the parameters controlling the
external forces; and the ``others'' can be the coupling constants between
the particles.}, there would also be a slow variability of these functions,
which play the role of the v. Helmholtzian $p_{\bf a}$, and which here we
leave as denoted $p_{\bf a}$ to include the equations that I obtained
previously and the v. Helmholtzians ones.\footnote{\tiny It seems that B. wants to
say that between the $p_{\bf a}$ can be included also possible coupling
constants that are allowed to change: this permits a wider generality.} And
here is the place of a few considerations.

The formulae, that follow from formulae (18) of my work ``{\it Analytischer
Beweis des zweiten {H}auptsatzes der mechanischen {W\"a}rme\-theorie aus
den {S\"a}tzen {\"u}ber das {G}leichgewicht des lebendigen {K}raft}'',
1871, \Cite{Bo871-c}[\#19], see also Sec.(\,\ref{sec:X-6}\,) above, have not been
developed in their full generality, in fact there I first speak of {\it a}
system, which goes through all possible configurations compatible with the
principle of the ``vis viva'' and secondly I only use Cartesian
coordinates; and certainly this is seen in the very often quoted work of
Maxwell ``{\it On the theorem of Boltzmann ... {\it \&tc}}'',
[\Cite{Ma879}].  This being said these formulae must also hold for ergodes
in any and no matter how generalized coordinates. Let these be, for an
element of an ergode, $p_1,p_2,\ldots, p_g$, and thus it is\footnote{\tiny The dots
the follow the double integral signs cannot be understood; perhaps this is
an error repeated more times.}

$$\fra{d{\cal N}}N=\fra{\D^{-\fra12} \ps^{\fra{g}2-1} dp_1\ldots dp_g}{
\ig\ig \ldots \D^{-\fra12} \ps^{\fra{g}2-1} dp_1\ldots
dp_g},$$
where $N$  is the total number of systems of the ergode, $d{\cal N}$ the
number of such systems whose coordinates are between $p_1$ and
$p_1+dp_1$, $p_2$ and $p_2+dp_2\ldots$ $p_g$ and $p_g+dp_g$. Let here $\ps$ 
be the form of the "vis viva" of a  system. The relation
at the nine-th place of the quoted formula  (18) yields

....

\0[{\sl (p.136): Follows the argument that shows that the 
results do not depend on the
system of coordinates. Then a few examples are worked out, starting with the
case considered by Helmholtz, essentially one dimensional, ergodes ``with
only one fast variable'' (p.137):}]
\*

The monocyclic systems of Hrn. v. Helmholtz with a single velocity are not
different from the ergodes with a single rapidly varying coordinate, that
will be called $p_g$ which, at difference with respect to the
v. Helmhotzian $p_a$, is not subject to the condition, present in his
treatment, of varying very slowly. 

Hence the preceding formula is valid equally for monocyclic systems with a
unique velocity and for warm bodies, and therefore it has been clarified
the mentioned analogy of Hrn.  v. Helmholtz between rotatory motions and
ideal gases (see Crelles Journal, {\bf 07}, p.123,; Berl. Ber. p.170).

Consider a single system, whose fast variables are all related to
the equation of the ``vis viva'' ({\it isomonode}), therefore it is
$N=1$, $\ps=L$. For a rotating solid body it is $g=1$. Let $p$ be the
position angle $\th$ and $\o=\fra{d\th}{dt}$, then

$$\ps=L=\fra{T\o^2}2=\fra{r^2}{2T},\quad r=T\o;$$
where $\D=\fra1T$, and always $\D=\m_1\m_2\ldots$, while $L$ has the form

$$\fra{\m_1 r_1^2+\m_2 r_2^2+\ldots}2.$$
$T$  is  the inertia moment; $\ig\ig\ldots dp_1dp_2\ldots$ is reduced to
$\ig dp=2\p$ and can be treated likewise, so that the preceding general
formula becomes $\d Q=L \,\d \log (TL)$. If a single mass $m$ 
rotates at distance  $\r$ from the axis, we can set  $p$ equal
to the arc $s$ of the point where the mass is located; then it will be:

$$\eqalign{
\ps=&L=\fra{m v^2}2=\fra{r^2}{2m},\quad r=mv,\quad \D=\fra1m,\cr
=&\ig\ig \ldots dp_1 dp_2\ldots=\ig dp=2\p\r,\cr}$$
where $v=\fra{ds}{dt}$; therefore the preceding formula follows $\d
Q=L\,\d\log(m L \r^2)$. For an ideal gas of monoatomic molecules it is
$N=1$, $\ps=L$; $p_1,p_2,\ldots,p_g$ are the Cartesian coordinates
$x_1,y_1,\ldots,z_n$ of the molecules, hence $g=3n$, where $n$ is the total
number of molecules, $v$ is the volume of the gas and $\ig\ig \ldots
dp_1dp_2\ldots$ is $v^n$, $\D$ is constant, as long as the number of
molecules stays constant; hence the preceding general formula $\d
Q=L\,\d\log(L v^{\fra23})$ follows, which again is the correct value
because in this case the ratio of the specific heats is $\fra53$.

....

\0[{\sl p.138: Follow more examples. The concluding remark (p.140) in
Sec. 3 is of particular interest as it stresses that the generality of the
analysis of holodes and ergodes is dependent on the ergodic
hypothesis. However the final claim, below, that it applies to polycyclic
systems may seem contradictory: it probably refers to the conception of
Boltzmann and Clausius that in a system with many degrees of freedom all
coordinates had synchronous (B.) or asynchronous (C.) periodic motions, see
Sec.(\,\ref{sec:I-6},\ref{sec:IV-6},\ref{sec:V-6}\,).}]

...

The general formulae so far used apply naturally both to the monocyclic
systems and to the polycyclic ones, as long as they are ergodic, and
therefore I omit to increase further the number of examples.

\*
[{\sl Sec. 4,5,6 are not translated}]


\section{Maxwell 1866}
\def\SEC{Maxwell 1866}
\label{sec:XIV-6}\iniz
\lhead{\small\ref{sec:XIV-6}.\ \SEC}\index{Maxwell}

\0{Commented summary of: {\it On the dynamical theory of gases}, di
  J.C. Maxwell, {\it Philosophical Transactions}, {\bf157}, 49--88, 1867,
  \Cite{Ma867-b}[XXVIII, vol.2].}  \*

The statement {\it Indeed the properties of a body supposed to be a uniform
  {\it plenum} may be af\/firmed dogmatically, but cannot be explained
  mathematically}, \Cite{Ma867-b}[p.49], is in the overture of the
second main work of Maxwell on kinetic theory.\footnote{\tiny Page numbers
  refer to the original: the page numbers of the collected papers,
  \Cite{Ma867-b}, are obtained by subtracting 23.}

The work begins with a discussion on Friction phenomenology.
The first new statement is about an experiment that he performed on
viscosity of almost ideal gases: yielding the result that at pressure $p$
viscosity is independent of density $\r$ and proportional to temperature,
\ie to $\frac{p}\r$. This is shown to be possible if the collisions
frequency is also temperature independent or, equivalently, if the
collision cross section is independent of the relative speed. 

{\it Hence an
  interaction potential at distance $|x|$ proportional to $|x|^{-4}$} is
interesting and it might be a key case. The argument on which the
conclusion is based is interesting and, as far as I can see, quite an
unusual introduction of {\it viscosity}.%

Imagine a displacement $S$ of a body (think of a parallelepiped of sides
$a,b=c$ (here $b=c$ for simplicity, {\sl in the original $b,c$ are not set
  equal}) containing a gas, or a cube of metal and let $S=\d a$ in the case
of stretching or, in the case of deformation {\it at constant volume},
$S=\d b^2$). The displacement generates a ``stress'' $F$, \ie a force
opposing the displacement that is imagined proportional to $S$ via a
constant $E$.
%

If $S$ varies in time then the force varies in time as $\dot F=E \dot S$:
here arises a difference between the iron parallelepiped and the gas one;
the iron keeps being stressed as long as $\dot S$ stays fixed or varies
slowly. On the other hand the gas in the parallelepiped undergoes a stress,
\ie a  difference in pressure in the different directions, which goes away,
    {\it after some material-dependent time}, even if $\dot S$ is fixed,
    because of the equalizing effect of the collisions.

In the gas case (or in general viscous cases) the rate of disappearance of
the stress can be imagined proportional to $F$ and $F$ will follow the
equation $\dot F=E \dot S -\frac{F}\t$ where $\t$ is a constant with the
dimension of a time as shewn by the solution of the equation $F(t)=E\t \dot
S+const e^{-\frac{t}\t}$ and a constant displacement rate results for
$t\gg\t$ in a constant force $E\t \dot S$ {\it therefore $E\t$ has the
  meaning of a viscosity}.  
\*

The continuation of the argument is difficult to understand exactly; {\it
  my} interpretation is that the variation of the dimension $a$ accompanied
by a compensating variation of $b=c$ so that $ab^2=const$ decreases the
frequency of collision with the wall orthogonal to the $a$-direction by by
$-\frac{\d a}a$, relative to the initial frequency of collision which is
proportional to the pressure; hence the pressure in the direction $a$
undergoes a relative diminution $\frac{\d p}p=-\frac{\d a}a=2\frac{db}b$
[{\sl Maxwell gives instead $\frac{\d p}p=-2\frac{\d a}a$}], and this
implies that the force that is generated is $dF=d (b^2p)$: a force
(``stress'' in the above context) called in \Cite{Ma867-b} ``linear
elasticity'' or ``rigidity'' for changes of form. It disappears upon
re-establishment of equal pressure in all directions due to collisions
\Cite{Ma867-b}. So that $dF=2b p db+b^2 dp= 4p b db=2 p db^2$ and the
rigidity coefficient is $E=2p$ [{\sl Maxwell obtains $p$}].
%

Hence the elasticity constant $E$ is $p$ and by the above general
argument the viscosity is $p\t$: the experiment quoted by Maxwell yields a
viscosity proportional to the temperature and independent of the density
$\r$, \ie $p\t$ proportional to $\frac{p}\r$ so that $\t$ is temperature
independent and inversely proportional to the density.  

In an earlier work he had considered the case of a hard balls
gas, \Cite{Ma860-a}[XX] concluding density independence,
proportionality of viscosity to $\sqrt{T}$ and to the inverse $r^{-2}$ of
the balls diameter.

Collision kinematics technical analysis starts with the
derivation of the collision kinematics. Calling
$v_i=(\x_i,\h_i,\z_i)$ the velocities of two particles of masses
$M_i$ he writes the outcome $v'_i=(\x'_i,\h'_i,\z'_i)$ of a
collision in which particle $1$ is deflected by an angle $2\th$
as: \* \0$
\x'_1=
\x_1+\frac{M_2}{M_1+M_2}
\{(\x_2-\x_1)2(\sin\th)^2+\sqrt{(\h_2-\h_1)^2+
(\z_2-\z_1)^2}\,\sin2\th\,\cos\f\}
$

\eqfig{300}{90}{
\ins{40}{40}{$\a$}
\ins{14}{36}{$\g$}
\ins{31}{68}{$\b$}
\ins{11}{59}{$A$}
\ins{28}{19}{$B$}
\ins{55}{67}{$C$}
\ins{-8}{28}{$v'_1$}
\ins{63}{28}{$v_1$}
\ins{23}{87}{$x$}
\ins{90}{80}{$\g=C,\ \f=\a,\ 2\th=B,\ \th=A$ }
\ins{90}{60}{$\cos A\,=\,\cos B\,\cos C\,+\,\sin B\,\sin C\,\cos\a$}
\ins{90}{45}{$\cos \th'\,=\,\cos 2\th\,\cos \g\,+
\,\sin 2\th\,\sin \g\,\cos\f$}
\ins{90}{30}{$v'_1=\frac{M_1 v_1+M_2 v_2}{M_1+M_2}+ 
                    \frac{M_2}{M_1+M_2}(v'_1-v'_2)$}
}{fig6.1.2}{}

\kern-3mm
\0{\small Fig.6.1.2: spherical triangle for momentum conservation.}
\*

\0which follows by considering the spherical triangle with vertices on the
$x$-axis, $v_1$, $v'_1$, and 
$\cos\g=\frac{(\x_1-\x_2)}{|v_1-v_2|},\sin\g= \frac{\sqrt
{(\h_1-\h_2)^2 +(\z_1-\z_2)^2}}{|v_1-v_2|}$, 
with $|v'_1-v'_2|=|v_1-v_2|$, \Cite{Ma867-b}[p.59].

At this point the formalism is ready
to analyse the observables variation upon collision and the
next step is to evaluate the amount of a quantity $Q=Q(v_1)$ contained
per unit (space) volume in a velocity volume element $d\x_1\,d\h_1\,d\z_1$
around $v_1=(\x_1,\h_1,\z_1)$: this is $Q(v_1) dN_1$, with $dN_1= f(v_1)
d^3 v_1$. Collisions at impact parameter $b$ and relative speed
$V=|v_1-v_2|$ occur at rate $dN_1 \,V b db d\f dN_2$ per unit volume. They
change the velocity of particle $1$ into $v'_1$ hence change the total
amount of $Q$ per unit volume by 

$$(Q'-Q)\,V b db d\f dN_1 dN_2.\eqno{(2.1)}$$

Notice that here independence is assumed for the particles distributions as
in the later Boltzmann's {\it stosszahlansatz}.\index{stosszahlansatz}

Then it is possible to express the variation of the amount of $Q$ per unit
volume. Maxwell considers ``only''
$Q=\x_1,\x_1^2,\x_1(\x_1^2+\h_1^2+\z_1^2)$ and integrates $(Q'-Q)\,V b db
d\f dN_1 dN_2$ over $\f\in[0,2\p]$, on $b\in[0,\infty)$ and then over
$dN_1,dN_2$. The $\f$ integral yields, setting
$s_\a\defi\sin\a,c_\a\defi\cos\a$,

$$\kern-3mm\eqalign{
(\a):\ \int_0^{2\p}& d\f (\x_1'-\x_1) d\f= \frac{M_2}{M_1+M_2}
(\x_2-\x_1) 4\p s_\th^2
\cr
(\b):\ \int_0^{2\p}&
(\x_1^{'2}-\x_1^2)d\f= 
\frac{M_2}{(M_1+M_2)^2}\Big\{(\x_2-\x_1)(M_1\x_1+M_2\x_2)8\p s_\th^2\cr
&
+M_2\Big((\h_2-\h_1)^2+(\z_2-\z_1)^2 -2(\x_2-\x_1)^2\Big)\p
s_{2\th}^2\Big\}\cr
(\b'):\ \int_0^{2\p}&
(\x_1'\h'_1-\x_1\h_1)d\f= 
\frac{M_2}{(M_1+M_2)^2}
\Big\{\Big(M_2\x_2\h_2-M_1\x_1\h_1\cr
&\kern-5mm+\frac12(M_1-M_2)(\x_1\h_2+\x_2\h_1)
8\p s_\th^2\Big)
-3M_2\Big((\x_2-\x_1)(\h_2-\h_1)\p s_{2\th}^2\Big)%
\Big\}\cr
(\g):\ \int_0^{2\p}&\kern-3mm 
(\x'_1 V^{'2}_1-\x_1 V_1^2)d\f=\frac{M_2}{M_1+M_2} 4\p
s_\th^2 \Big\{(\x_2-\x_1) V_1^2+ 2\x_1(U-V_1^2)\Big\}
\cr
&\kern-5mm+(\frac{M_2}{M_1+M_2})^2
\Big((8\p s_\th^2-3\p s_{2\th}^2)2(\x_2-\x_1)(U-V_1^2)\cr
&\kern-5mm+(8\p s_\th^2+2\p
s_{2\th}^2\x_1) V^2\Big)
+(\frac{M_2}{M_1+M_2})^3(8\p s_\th^2-2\p s_{2\th}^2)2(\x_2-\x_1) V^2\cr
}
$$
where $V_1^2\defi(\x_1^2+\h_1^2+\z_1^2),\,
U\defi(\x_1\x_2+\h_1\h_2+\z_1\z_2)$, $V_2^2\defi(\x_2^2+\h_2^2+\z_2^2)$
$V^2\defi ((\x_2-\x_1)^2+(\h_2-\h_1)^2+(\z_2-\z_1)^2)$.

If the interaction potential is $ \frac{K}{|x|^{n-1}}$ the deflection $\th$
is a function of $b$.  Multiplying both sides by $V b db$ and integrating
over $b$ a linear combination with coefficients

$$\eqalign{
&B_1=\int_0^\infty 4\p b s_\th^2 db = \Big(\frac{K
(M_1+M_2)}{M_1M_2}\Big)^{\frac2{n-1}} V^{\frac{n-5}{n-1}} A_1\cr
&B_2=\int_0^\infty \p b s_{2\th}^2
db=\Big(\frac{K
(M_1+M_2)}{M_1M_2}\Big)^{\frac2{n-1}} V^{\frac{n-5}{n-1}} A_2\cr}$$
with $A_1,A_2$ dimensionless and expressed by a quadrature.

\index{precarious assumption}
To integrate over the velocities it is necessary to know the
distributions $dN_i$ and a ``precarious
assumption''\index{precarious assumption} is recalled. The only
case in which the distribution has been determined in
\Cite{Ma867-b}[p.62] is when the momenta distribution
is stationary: this was obtained in \Cite{Ma860-a}
under the assumption which ``{\it may appear precarious}'' that
``the probability of a molecule having a velocity resolved
parallel to $x$ lying between given limits is not in any way
affected by the knowledge that the molecule has a given velocity
resolved parallel to $y$''.  Therefore in \Cite{Ma867-b} a
different analysis is performed: based on the energy conservation
at collisions which replaces the independence of the distribution
from the coordinates directions. The result is that (in modern
notations), if the mean velocity is $0$,

$$dN_1=f( v_1) d^3v_1
=\frac{N_1}{(2\p(M_1 \b)^{-1})^{\frac32}} e^{-\b\frac{M_1}2 v_1^2}d^3v_1$$
where $N_1$ is the density.

\* \0{\it Remark: However the velocity distribution is not supposed, in the
  following, to be Maxwel\-lian but just close to a slightly off center
  Maxwellian. It will be assumed that the distribution factorizes over the
  different particles coordinates, and over positions and momenta.}  \*

The analysis is greatly simplified if $n=5$, \Cite{Ma867-b}[vol.2,
  p.65-67] and a balance of the variations of
key observablesis preformed. Consider the system as containing two kinds of
particles.  Let the symbols $\d_1$ and $\d_2$ indicate the effect produced
by molecules of the first kind and second kind respectively, and $\d_3$ to
indicate the effect of the external forces.  Let $\k\defi\Big( \frac{
  K}{M_1M_2(M_1+M_2)} \Big)^{\frac12}$ and let $\media{\cdot}$ denote the
average with respect to the velocity distribution.

$$\eqalignno{
(\a):\ & \frac{\d_2\media{\x_1}}{\d t}= 
\k
\,N_2M_2\, A_1\,\media{\x_2-\x_1}
\cr
(\b):\ & \frac{\d_2\media{{\x^2_1}}}{\d t} = 
\k\,
 \frac{N_2M_2}{(M_1+M_2)}
\Big\{2 A_1\media{(\x_2-\x_1)(M_1\x_1+M_2\x_2)}\cr
&
+M_2 A_2\Big(\media{(\h_2-\h_1)^2+(z_2-z_1)^2 -2(\x_2-\x_1)^2}\Big)\Big\}
\cr
(\b'):\ & \frac{\d_2\media{\x_1\h_1}}{\d t} = 
\k\,
 \frac{N_2M_2}{(M_1+M_2)}
\Big\{A_1\Big(\media{2M_2\x_2\h_2-2M_1\x_1\h_1}\cr
&+(M_1-M_2)\media{(\x_1\h_2+\x_2\h_1)}
\Big)
-3 A_2M_2\Big(\media{(\x_2-\x_1)(\h_2-\h_1)}\Big)
\Big\}\cr
(\g):\ & \frac{\d_2\media{\x_1V_1}}{\d t}= 
\k 
N_2M_2\Big\{A_1\media{(\x_2-\x_1) V_1^2+ 2\x_1(U-V_1^2)}\Big\}
\cr
&+\frac{M_2}{M_1+M_2}
\Big((2A_1-3 A_2)2\media{(\x_2-\x_1)(U-V_1^2)}\cr
&+(2A_1+2A_2)\media{\x_1 V^2}\Big)
+(\frac{M_2}{M_1+M_2})^2(2A_1-2A_2)2\media{(\x_2-\x_1) V^2}\Big\}\cr
}
$$
More general relations can be found if external forces are imagined to act
on the particles. If only one species of particles is present the relations
simplify, setting $M=M_1,N=N_1$ and
$\k\defi (\frac{K}{2M^3})^{\frac12}$, into

$$\eqalign{
(\a):\ & \frac{\d_1\media{\x}}{\d t}= 0
\cr
(\b):\ & \frac{\d_1\media{{\x^2}}}{\d t} = 
\k\,
 M N A_2 \Big\{(\media{\h^2}-\media{\h}^2)
+(\media{\z^2}-\media{\z}^2)-2(\media{\x^2}-\media{\x}^2)\Big\}
\cr
(\b'):\ & \frac{\d_1\media{\x\h}}{\d t} = 
\k\,M\,N\, 3\, A_2\,\Big\{\media{\x}\media{\h}-\media{\x\h}\Big\}\cr
(\g):\ & \frac{\d_1\media{\x_1V_1}}{\d t}= 
\k \,N\,M \,3 \,A_2 \Big\{\media{\x}\media{V_1^2}-\media{\x V_1^2}
\Big\}
\cr
}
$$

Adding an external force with components $X,Y,Z$
$$\eqalign{
(\a):\ & \frac{\d_3\media{\x}}{\d t}= X
\cr
(\b):\ & \frac{\d_3\media{{\x^2}}}{\d t} = 
2\media{\x X}
\cr
(\b'):\ & \frac{\d_3\media{\x\h}}{\d t} = 
(\media{\h X+\x Y})\cr
(\g):\ & \frac{\d_2\media{\x_1V_1}}{\d t}= 
2\media{\x(\x X+\h Y+\z Z)}+ X \media{V^2} \hbox{\hglue3.7cm}
\cr
}
$$

The problem of the continuum limit \index{continuum limit} will
be examined as restricted to the case of only one species the
change of the averages due to collisions and to the external
force is the sum of $\d_1+\d_3$.  Changing notation to denote the
velocity $u+\x,v+\h,w+\z$, so that $\x,\h,\z$ have $0$ average
and ``almost Maxwellian distribution'' while $u,v,w$ is the
average velocity, and if $\r\defi NM$, see comment following
Eq.(56) in \Cite{Ma867-b}[XXVIII, vol.2], it is
$$\eqalignno{
(\a):\ & \frac{\d u}{\d t}= X 
\cr
(\b):\ & \frac{\d\media{{\x^2}}}{\d t} = 
\k\,
 \r \,A_2 \,(\media{\h^2}
+\media{\z^2}-2\media{\x^2})
\cr
(\b'):\ & \frac{\d\media{\x\h}}{\d t} = 
-3\, \k\,\r\, A_2\, \media{\x\h}\cr
(\g):\ & \frac{\d\media{\x_1V_1}}{\d t}= 
-3\k\, \r\,A_2\,\media{\x V_1^2}+ X \media{3\x^2+\h^2+\z^2}
+2 Y\media{\x\h}+2Z\media{\x\z}
\cr
}
$$

Consider a plane moving in the $x$ direction with velocity equal to the
average velocity $u$: then the amount of $Q$ crossing the plane per unit
time is $\media{\x Q}\defi\int \x Q(v_1) \r f(v_1) d^3 v_1$.

For $Q= \x$ the momentum in the direction $x$ transferred through a plane
orthogonal to the $x$-direction is $\media{\x^2}$ while the momentum in the
direction $y$ transferred through a plane orthogonal to the $x$-direction
is $\media{\x\h}$

The quantity $\media{\x^2}$ is interpreted as pressure in the $x$
direction and the tensor $T_{ab}=\media{v_a v_b}$ is interpreted as stress
tensor.

Finally the ``Weak'' Boltzmann equation\index{Boltzmann Eq.:
  weak} is derived. Supposing the particles without structure and
point like the $Q=\x v_1^2\equiv \x(\x^2+\x\h^2+\x\z^2)$ is
interpreted as the heat per unit time and area crossing a plane
orthogonal to the $x$-direction.

The averages of observables change also when particles move without
colliding. Therefore $\frac{\d}{\d t}$ does not give the complete
contribution to the variations of the averages of observables. The complete
variation is given by

$$\dpr_t( \r\media{Q})= \r\frac{\d\media {Q}}{\d t}-
\dpr_x({\r\media{(u+\x)Q}})-
\dpr_y( {\r\media{(v+\h)Q}})-
\dpr_z( {\r\media{(w+\z)Q}})
$$
if the average velocity is zero at the point where the averages are
evaluated (otherwise if it is $(u',v'w')$ the $(u,v,w)$ should be replaced
by $(u-u',v-v',w-w')$). The special choice $Q=\r$ gives the ``continuity
equation''.

$$ \dpr_t\r+\r(\dpr_x u+\dpr_yv+\dpr_z w)=0$$
which allows us to write the equation for $Q$ as

$$\r \dpr_t\media{Q}+\dpr_x(\r\media{\x Q})+
\dpr_y (\r\media{\h Q})+
\dpr_z (\r\media{\z Q})=\r \frac{\d Q}{\d t}
$$

\0{\it Remark:} {\sl Notice that the latter equation is exactly Boltzmann's
  equation (after multiplication by $Q$ and integration, \ie in ``weak
  form'') for Maxwellian potential ($K |x|^{-4}$) if the expression for
  $\frac{\d\media {Q}}{\d t}$ derived above is used. The assumption on the
  potential is actually not used in deriving the latter equation as only
  the stosszahlansatz matters in the expression of $\frac{\d\media {Q}}{\d
    t}$ in terms of the collision cross section (see Eq.(2.1) above).}  \*

The work is concluded with examples among which
is the heat conduction.\index{heat conduction}
Tghe choice $Q=(u+\x)$ yields

$$\r\dpr_t u+\dpr_x (\r\media{\x^2})+
\dpr_y (\r\media{\h^2})+\dpr_z (\r\media{\z^2})= \r X$$
The analysis can be continued to study several other problems. As a last
example the heat conductivity in an external field $X$ pointing in the
$x$-direction is derived by choosing
$Q=M (u+\x)(u^2+v^2+w^2+2u\x+2v\h+2w\z+\x^2+\h^2+\z^2)$ which yields

$$\eqalign{
\r&\dpr_t\media{\x^3+\x\h^2+\x\z^2}+\dpr_x \r \media{\x^4+\x^2\h^2+\x^2\z^2}
\cr
&=-3 \k \r^2 A_2 \media{\x^3+\x\h^2+\x\z^2}+ 5X \media{\x^2}\cr}$$
having set $(u,v,w)=0$ and having neglected all odd powers of the velocity
fluctuations except the terms multiplied by $\k$ (which is ``large'': for
hydrogen at normal conditions it is $\k\sim1.65\cdot
10^{13}$, in cgs units, and $\k\r=2.82\cdot10^9\, s^{-1}$).

In a stationary state the first term is $0$ and $X$ is related to the
pressure: $X=\dpr_x p$ so that

$$\dpr_x \r \media{(\x^4+\x^2\h^2+\x^2\z^2)}-5  \media{\x^2}\dpr_x p
= -3 \k \r^2 A_2 \media{(\x^3+\x\h^2+\x\z^2)}$$
This formula allows us to compute
$\frac12 \r \media{(\x^3+\x\h^2+\x\z^2)}\defi -\ch
\dpr_x T$ as

$$\eqalign{
\ch=&\frac1{6\k \,\r \,A_2} \Big(
\r \dpr_x\media{(\x^4+\x^2\h^2+\x^2\z^2)}-5 \media{x^2} 
\dpr_x p\Big)\frac1{\dpr_x T}
\cr
=&\frac1{6\k A_2} \frac{5 \dpr_x( \b^{-2})-5 \b^{-1}\dpr_x
(\b^{-1})}{M^2\dpr_x T}=-\frac{5}{6\k
A_2} \b^{-3}\frac{\dpr_x\b}{M^2\dpr_x T}
=\frac{5 k_B}{6\k M^2 A_2} k_BT
\cr}$$
{\it without having to determine the momenta distribution but only assuming that
the distribution is close to the Maxwellian}. The dimension of $\ch$ is
$[k_B] cm^{-1} s^{-1}= g\,cm\,s^{-3}\,{}^o\kern-1mm K^{-1}$. The numerical
value is $\ch=7.2126\cdot 10^4\, c.g.s.$ for hydrogen at normal conditions
(cgs units).
\*

\0{\it Remarks:} (1) Therefore the conductivity for Maxwellian potential
turns out to be proportional to $T$ rather than to $\sqrt{T}$: in general
it will depend on the potential and in the hard balls case it is, according
to \Cite{Ma860-a}[Eq.(59)], proportional to $\sqrt{T}$. In all cases it is
independent on the density. The method to find the conductivity in this
paper is completely different from the (in a way elementary) one
in \Cite{Ma860-a}.
\\
(2) The neglection of various odd momenta indicates that the analysis is a
first order analysis in the temperature gradient.
\\
(3) In the derivation density has not been assumed constant: if its
variations are taken into account the derivatives of the density cancel if
the equation of state is that of a perfect gas (as assumed
implicitly). However the presence of the pressure in the equation for the
stationarity of $Q$ is due to the external field $X$: in absence of the
external field the pressure would not appear {\it but} in such case it
would be constant while the density could not be constant; the calculation
can be done in the same way and it leads to the {\it same} conductivity
constant.

\chapter{Appendices}
\label{ChA} 
\chaptermark{\ifodd\thepage
Appendix\hfill\else\hfill 
Appendix\fi}
\renewcommand{\thesection}{\Alph{section}}
\renewcommand{\theequation}{\Alph{section}.\arabic{equation}}

\section{Appendix: Heat theorem\index{heat theorem} (Boltzmann's example)}
\def\SEC{Appendix: Heat theorem\index{heat theorem} (Boltzmann's example)}
\iniz\label{appA}
\lhead{\small\ref{appA}: \SEC}\index{Boltzmann}

The example is built on a case in which all motions are really periodic,
namely a one-dimensional system with potential $\f(x)$ such that
$|\f'(x)|>0$ for $|x|>0$, $\f''(0)>0$ and $\f(x)\tende{x\to\io}+\io$. All
motions are periodic (systems with this property are called {\it
  monocyclic},\index{monocyclic} see Sec.{sec:XIII-6} below). We suppose
that the potential $\f(x)$ depends on a parameter $V$.\label{heat theorem}

Define {\it state} a motion with given energy $U$ and given
$V$. And:
\*

\halign{#\ $=$\ & #\hfill\cr
$U$ & total energy of the system $\equiv K+\f$\cr
$T$ & time average of the kinetic energy $K$\cr
$V$ & the parameter on which $\f$ is supposed to depend\cr
$p$ & $-$ average of $\dpr_V \f$.\cr}
\*
\0A state is parameterized by $U,V$ and if such parameters change by $dU,
dV$, respectively, let 
\be
dW=-p dV, \qquad dQ=dU+p dV,\qquad \lis K= T.\label{eA.1}\ee
Then the heat theorem\index{heat theorem} is in this case: 

\* \0{\sl Theorem} (\Cite{Bo868-b}[\#6],
\Cite{Bo877a}[\#39],\Cite{He884b}):
   {\it The differential $({dU+pdV})/{T}$ is exact.}  \*

In fact let $x_\pm(U,V)$ be the extremes of the oscillations of the motion
with given $U,V$ and define $S$ as:

\be S=2\log 2\ig_{x_-(U,V)}^{x_+(U,V)}\kern-3mm
\sqrt{K(x;U,V)}dx=2\log \ig_{x_-(U,V)}^{x_+(U,V)}\kern-3mm2
\sqrt{U-\f(x)}dx\label{eA2}\ee
so that $dS=\fra{\ig \Big(dU-\dpr_V\f(x) dV\Big)\, \fra{dx}{\sqrt{K}}}{ \ig
  K\fra{dx}{\sqrt{K}}}\equiv \frac{d Q}T$, and $S=2\log i\lis K$ if
$\fra{dx}{\sqrt K} =\sqrt{\fra2m} dt$ is used to express the period $i$ and
the time averages via integrations with respect to $\fra{dx}{\sqrt K}$.

Therefore Eq.\equ{e1.4.5} is checked in this case. 
This completes the discussion of the case in which motions are periodic.
The reference to Eq.\equ{e1.4.5} as {\it heat theorem}
abridges the original diction {\it second main theorem of the mechanical
theory of heat}, \Cite{Bo866}[\#2],\Cite{Cl871}). 

The above analysis admits an extension to Keplerian motions, discussed
in \Cite{Bo877a}[\#39], provided one considers only motions with a fixed
eccentricity.

\section{Appendix: Aperiodic Motions as Periodic with Infinite Period!}
\def\SEC{Appendix: Aperiodic Motions as Periodic with Infinite Period!}
\iniz\label{appB}
\lhead{\small\ref{appB}: \SEC}\index{periodic with infinite period}

The famous and criticized statement on the {\it infinite period of
aperiodic motions}, \Cite{Bo866}[\#2], is the heart of the application of
the heat theorem to a gas in a box and can be reduced to the above
ideas.  Imagine, \Cite{Ga000}, the box containing the gas to be covered
by a piston of section $A$ and located to the right of the origin at
distance $L$, so that the box volume is $V=AL$.

The microscopic model for the piston will be a potential
$\lis\f(L-\x)$ if $x=(\x,\h,\z)$ are the coordinates of a
particle. The function $\lis\f(r)$ will vanish for $r>r_0$, for some
$r_0<L$, and diverge to $+\io$ at $r=0$. Thus $r_0$ is the width of the
layer near the piston where the force of the wall is felt by the
particles that happen to roam there.

Noticing that the potential energy due to the walls is $\f=\sum_j
\lis\f(L-\x_j)$ and that $\dpr_V \f=A^{-1}\dpr_L\f$ we must evaluate
the time average of
\be\dpr_L \f(x)=-\sum_j \lis\f'(L-\x_j)\,.\label{eB.1}\ee
As time evolves the particles with $\x_j$ in the layer within $r_0$ of
the wall will feel the force exercised by the wall and bounce
back. Fixing the attention on one particle in the layer we see that it
will contribute to the average of $\dpr_L \f(x)$ the amount
\be\fra1{\rm total\ time} 2\ig_{t_0}^{t_1}- \lis\f'(L-\x_j)
dt\label{eB.2}\ee
if $t_0$ is the first instant when the point $j$ enters the layer and
$t_1$ is the instant when the $\x$-component of the velocity vanishes
``against the wall''. Since $-\lis\f'(L-\x_j)$ is the $\x$-component
of the force, the integral is $2m|\dot\x_j|$ (by Newton's law),
provided $\dot\x_j>0$ of course. One assumes that the density is low
enough so that no collisions between particles occur while the
particles travel within the range of the potential of the wall: {\it i.e.} 
the mean free path is much greater than the range of the potential
$\lis\f$ defining the wall.

The number of such contributions to the average per unit time is
therefore given by $\r_{wall}\, A\, \ig_{v>0} 2mv\, f(v)\, v\, dv$ if
$\r_{wall}$ is the (average) density of the gas near the wall and
$f(v)$ is the fraction of particles with velocity between $v$ and
$v+dv$. Using the ergodic hypothesis ({\it i.e.}  the microcanonical
ensemble) and the equivalence of the ensembles to evaluate $f(v)$ (as
$\frac{e^{-\frac\b2 mv^2}}{\sqrt{2\p\b^{-1}}}$) it
follows that:
\be p\defi -\media{\dpr_V\f}= \r_{wall} \b^{-1}\label{eB.3}\ee
where $\b^{-1}=k_B T$ with $T$ the absolute temperature and $k_B$
Boltzmann's constant. Hence we see that Eq.\equ{eB.3} yields the correct
value of the pressure, \Cite{MP972},\Cite{Ga000}[Eq.(9.A3.3)]; in fact it
is often even taken as the microscopic definition of the pressure,
\Cite{MP972}).

{\it On the other hand} we have seen in Eq.\equ{eA.1} that if all
motions are periodic the quantity $p$ in Eq.\equ{eB.3} is the right
quantity that would make the heat theorem work. Hence regarding all
trajectories as periodic leads to the heat theorem with $p,U,V,T$ having
the {\it right physical interpretation}. And Boltzmann thought, since the
beginning of his work, that trajectories confined into a finite region of
phase space could be regarded as periodic {\it possibly with infinite
  period}, \Cite{Bo866}[p.30], see p.\pageref{infinite period}.

\section{Appendix: The heat theorem\index{heat theorem} without dynamics}
\def\SEC{Appendix: The heat theorem\index{heat theorem} without dynamics}
\iniz\label{appC}
\lhead{\small\ref{appC}: \SEC}

The assumption of periodicity can be defended mathematically only in
a discrete conception of space and time: furthermore the Loschmidt
paradox had to be discussed in terms of ``{\it numbers of
different initial states which determines their probability, which
perhaps leads to an interesting method to calculate thermal
equilibria}'', \Cite{Bo877a}[\#39] and Sec.(\,\ref{sec:XI-6}\,).

The statement, admittedly somewhat vague, had to be made precise: the
subsequent paper, \Cite{Bo877-b}[\#42], deals with this problem and adds
major new insights into the matter.

It is shown that it is possible to forget completely the details of
the underlying dynamics, except for the tacit ergodic hypothesis in
the form that all microscopic states compatible with the constraints
(of volume and energy) are visited by the motions. The discreetness
assumption about the microscopic states is for the first time not
only made very explicit but it is used to obtain in a completely new way
once more that the equilibrium distribution is equivalently a
canonical or a microcanonical one. The method is simply a
combinatorial analysis on the number of particles that can occupy a
given energy level. The analysis is performed first in the one
dimensional case ({\i.e.} assuming that the energy levels are ordered
into an arithmetic progression) and then in the three dimensional case
({\it i.e.} allowing for degeneracy by labeling the energy levels with
three integers). The combinatorial analysis is the one that has
become familiar in the theory of the Bose-Einstein gases: if the
levels are labeled by a label $i$ a microscopic configuration is
identified with the occupation numbers $(n_i)$ of the levels $\e_i$
under the restrictions
\be \sum_i n_i=N,\qquad \sum_i n_i\e_i=U\label{eC.1}\ee
fixing the total number of particles $N$ and the total energy $U$.  The
calculations in \Cite{Bo877-b}[\#42] amount at selecting among the $N$
particles the $n_i$ in the $i$-level with the energy restriction in
Eq.\equ{eC.1}: forgetting the latter the number of microscopic states
would be $\frac{N!}{\prod_i n_i!}$ and the imposition of the energy value
would lead, as by now usual by the Lagrange's multipliers method, to an
occupation number per level

\be n_i=N\frac{e^{\m-\b \e_i}}{Z(\m,\b)}\label{eC.2}\ee
if $Z$ is a normalization constant.

The Eq.\equ{eC.2} is the canonical distribution which implies that
$\fra{dU+p\,dV}{T}$ is an exact differential if $U$ is the average energy,
$p$ the average mechanical force due to the collisions with the walls, $T$
is the average kinetic energy per particle, \Cite{Ga000}[Ch.1,2]: it is
apparently not necessary to invoke dynamical properties of the motions.

Clearly this is not a proof that the equilibria are described by the
microcanonical ensemble. However it shows that for most systems,
independently of the number of degrees of freedom, one can define a
{\it mechanical model of thermodynamics}. The reason we observe
approach to equilibrium over time scales far shorter than the
recurrence times is due to the property that on most of the energy
surface the actual values of the observables whose averages yield the
pressure and temperature assume the same value. This implies that this
value coincides with the average and therefore satisfies the {\it heat
theorem}, \ie the statement that $(d U+p\,dV)/T$ is an
exact differential if $p$ is the pressure (defined as the average
momentum transfer to the walls per unit time and unit surface) and $T$
is proportional to the average kinetic energy.
\vfill\eject

\section{Appendix: Keplerian motion\index{keplerian motion} and heat theorem}
\def\SEC{Appendix: Keplerian motion\index{keplerian motion} and heat theorem}
\iniz\label{appD}
\lhead{\small\ref{appD}: \SEC}

It is convenient to use polar coordinates $(\r, \th)$, so
that if $A=\r^2\dot\th, E=\fra12 m\dot{\V x}^2-\fra{g\,m}{\r}$, $m$
being the mass and $g$ the gravity attraction strength 
($g=k M$ if $k$ is the gravitational constant and $M$ is the central
mass) then
\be E= \fra12 m\dot{\r}^2+\fra{m A^2}{2\r^2}-\fra{mg}\r\, ,
\qquad \f(\r)=-{gm\over \r} \label{eD.1}\ee
and, from the elementary theory of the two body problem, if $e$ is the
eccentricity and $a$ is the major semiaxis of the ellipse

\be\eqalign{
&\dot\r^2=\fra2m(E-\fra{m A^2}{2\r^2}+\fra{mg}\r)\defi
A^2(\fra1\r-\fra1{\r_+})(\fra1{\r_-}-\fra1{\r})\cr
&\fra1{\r_+\r_-}=\fra{-2E}{mA^2}, \quad
\fra1{\r_+}+\fra1{\r_-}=\fra{2g}{A^2},
\quad\fra{\r_++\r_-}2\defi a=\fra{mg}{-2E}\cr
&\sqrt{\r_+\r_-}\defi a
\sqrt{1-e^2}, \qquad\sqrt{1-e^2}=\fra{A}{\sqrt{ag}}\,.
\cr}\label{eD.2}\ee
Furthermore if a motion with parameters $(E,A,g)$ is periodic (hence
$E<0$) and if $\media{\cdot}$ denotes a time average over a period
then
\be \eqalign{ E&=-\fra{mg}{2a}, \qquad \media{\f}=-\fra{mg}a, \qquad
\media{\fra1{\r^2}}=\fra{1}{a^2\sqrt{1-e^2}}\cr
\media{K}&=\fra{mg}{2a}=-E, \qquad T=\fra{mg}{2a}\equiv \media{K}
\cr}\label{eD.3}\ee

Hence if $(E,A,g)$ are regarded as state parameters then

\be\frac{dE -\media{\frac{\dpr_g \f(r)}{\dpr g} }dg}{T}=\frac{dE-2E\frac{d g}g}
{-E}=d\log \fra{g^2}{-E}\defi dS\label{eD.4}\ee
Note that the equations $pg=2 T$ and $E=-T$ can be interpreted as,
respectively, analogues of the ``equation of state'' and the ``ideal
specific heat'' laws (with the ``volume'' being $g$, the ``gas constant''
being $R=2$ and the ``specific heat'' $C_V=1$).

If the potential is more general, for instance if it is increased by
$\frac{b}{2 r^2}$, the analogy fails, as shown by Boltzmann,
\Cite{Bo871-c}[\#19], see Sec.(\,\ref{sec:XI-6}\,) above. Hence there may be cases
in which the integrating factor of the differential form which should
represent the heat production might not necessarily be the average kinetic
energy: essentially all cases in which the energy is not the only constant
of motion. Much more physically interesting examples arise in quantum
mechanics: there in the simplest equilibrium statistics (free Bose-Einstein or
free Fermi-Dirac gases) the average kinetic energy and the temperature do
not coincide, \Cite{Ga000}.

\vfill\eject

\section{Appendix: Gauss' least constraint principle}
\def\SEC{Appendix: Gauss' least constraint principle}
\iniz\label{appE}
\lhead{\small\ref{appE}.\ \SEC}\index{least constraint principle}

Let $\f(\dot {\V x},{\V x})=0$, $(\dot{\V x},{\V x})= \{\dot{\V
x}_j,{\V x}_j\}$ be a constraint and let $\V R(\dot{\V x},{\V x})$ be
the constraint reaction and $\V F(\dot{\V x},{\V x})$ the active
force.

Consider all the possible accelerations $\V a$ compatible with the
constraints and a given initial state $\dot{\V x},{\V x}$. Then $\V R$ is
{\it ideal} or {\it satisfies the principle of minimal constraint} if the
actual accelerations $\V a_i=\fra1{m_i} (\V F_i+\V R_i)$ minimize the
{\it effort} defined by:
\be\sum_{i=1}^N\fra1{m_i} (\V F_i-m_i\V a_i)^2\ \otto\
\sum_{i=1}^N (\V F_i-m_i\V a_i)\cdot\d \V a_i=0\label{eE.1}\ee
\0for all possible variations $\d \V a_i$ compatible with the
constraint $\f$. Since all possible accelerations following
$\dot{\V x},{\V x}$ are such that $\sum_{i=1}^N
\dpr_{\dot{\V x}_i}\f(\dot{\V x},{\V x})\cdot\d \V a_i=0$ we can write
\be\V F_i-m_i\V a_i-\a\,\dpr_{\dot{\V x}_i} \f(\dot{\V x},{\V
x})=\V0\label{eE.2}\ee
\0with $\a$ such that
\be\fra{d}{dt}\f(\dot{\V x},{\V x})=0,\label{eE.3}\ee
\ie
\be \a=\fra{\sum_i\,(\dot{\V x}_i\cdot\dpr_{{\V x}_i} \f+\fra1{m_i} \V
F_i\cdot\dpr_{\dot{\V x}_i}\f)}{\sum_i m_i^{-1}(\dpr_{\dot{\V x}_i}\f)^2}
\label{eE.4}\ee
\0which expresses analytically Gauss' principle,
see \Cite{Wh917} and \Cite{Ga000}[Appendix 9.A4].

Note that if the constraint is even in the $\dot{\V x}_i$ then $\a$ is
odd in the velocities: therefore if the constraint is imposed on a
system with Hamiltonian $H= K+V$, with $K$ quadratic in the velocities
and $V$ depending only on the positions, and if other purely
positional forces (conservative or not) act on the system then the
resulting equations of motion are reversible if time reversal is
simply defined as velocity reversal.
\*

The Gauss' principle has been somewhat overlooked in the Physics
literature in statistical mechanics: its importance has again only recently
been brought to the attention of researchers, see the review
\Cite{HHP987}.  A notable, though by now ancient, exception is a paper
of Gibbs, \Cite{Gi902}, which develops variational formulas which he
relates to Gauss' principle of least constraint.

Conceptually this principle should be regarded as a {\it definition}
of {\it ideal non holonomic constraint}, much as D'Alembert's
principle or the least action principle are regarded as the definition of
{\it ideal holonomic constraint}.

\section{Appendix: Non smoothness of stable/unstable manifolds}
\def\SEC{Appendix: Non smoothness of stable/unstable manifolds}
\iniz
\label{appF}
\lhead{\small\ref{appF}.\ \SEC}


A simple argument to understand why even in analytic Anosov
systems the stable and unstable manifolds might just be H\"older continuous
can be given.

Let $T^2$ be the two dimensional torus $[0,2\p]^2$, and let
$S_0$ be the Arnold's cat Anosov map on $T^2$; $(x',y')=S_0(x,y)$
and $S(x,y)=S_0(x,y)+\e f(x)$ be an
smooth perturbation of $S_0$:
\be\eqalign{&S_0\to\cases{x'=2x+y\ {\rm mod}\,2\p\cr y'=x+y\ {\rm
      mod}\,2\p\cr},\qquad
S(x,y)=S_0(x,y)+\e f(x,y)\cr}\label{eF.1}\ee
with $f(x,y)$ periodic.
Let $v^0_+,v^0_-$ be the eigenvectors of the matrix $\dpr
S_0=\pmatrix{2&1\cr1&1\cr}$: which are the stable and unstable tangent
vectors at all points of $T^2$; and let $\l_+,\l_-$ be the corresponding
eigenvalues. 

Abridging the pair $(x,y)$ into $\x$, try to define a change of coordinates
map $h(\x)$, analytic in $\e$ near $\e=0$, which transforms $S$ back into
$S_0$; namely $h$ such that
\be h(S_0\x)=S(h(\x))+\e f(h(\x))\label{eF.2}\ee
and which is {\it at least} mildly continuous, \eg H\"older continuous 
with $|h(\x)-h(\x')|\le B
|\x-\x'|^\b$ for some $B,\b>0$ (with $|\x-\x'|$ being the distance on $T^2$
of $\x$ from $\x'$).  If such a map exists it will transform stable or
unstable manifolds for $S_0$ into corresponding stable and unstable
manifolds of $S$.

It will now be shown that the map $h$ cannot be expected to be
differentiable but only H\"older continuous with an exponent $\b$ that can
be prefixed as close as wished to $1$ (at the expense of the coefficient
$B$) but {\it not} $=1$.
Write $h(x)=x+\h(x)$, with $\h(x)=\sum_{k=0}^\infty \e^k\h^{[k]}(x)$; then
\be\eqalign{&\h(S_0x)=S_0 \h(x)+\e f(x+\h(x))\cr
&\h^{[k]}(S_0x)-S_0 \h^{[k]}(x)=\Big(\e f(x+\h(x))\Big)^{[k]}\cr}
\label{eF.3}\ee
where the last term is the $k$-th order coefficient of the expansion.
If $h_\pm(\x)$ are scalars so defined that $h(\x)=h_+(\x) v_+ +
h_-(\x) v_-$ it will be
\be\eqalign{
h^{[k]}_+(\x)=&\l_+^{-1}\h^{[k]}_+(S_0\x)-\Big(\e f_+(\x+h(\x))\Big)^{[k]}
\cr
h^{[k]}_-(\x)=&\l_-\h^{[k]}_+(S_0^{-1}\x)+\Big(\e f_+(S_0^{-1}\x+
h(S_0^{-1}\x))\Big)^{[k]}\cr}\label{eF.4}\ee
Hence for $k=1$ it is
\be h^{[1]}_+(\x)=-\sum_{n=0}^\infty \l_+^{-n}
f_+(S_0^n\x),\qquad 
h^{[1]}_-(\x)=\sum_{n=0}^\infty \l_-^{n}
f_-(S_0^{-n-1}\x)\label{eF.5}\ee
the \rhs is a convergent series of differentiable functions as
$\l_+^{-1} =|\l_-|<1$. 

However if differentiated term by term the $n$-th term will be given by
$(\dpr S_0)^n (\dpr^n f_+)(S_0^n\x)$ or, respectively, $(\dpr S_0)^{-n}
(\dpr^n f_-)(S_0^{-1-n}\x)$ and these functions will grow as $\l_+^n$ or as
$|\l_-|^n$ for $n\to\infty$, respectively, unless $f_+=0$ in the first case
or $f_-=0$ in the second, so one of the two series cannot be expected to
converge, \ie the $h^{[1]}$ cannot be proved to be
differentiable. Nevertheless if $\b<1$ the differentiability of $f$ implies
$|f(S_0^{\pm n}\x)-f(S_0^{\pm n}\x')|\le B |S_0^{\pm n}(\x-\x')|^\b$ for
some $B>0$, hence

\be |h^{[1]}_+(\x)-h^{[1]}_+(\h)| \le
\sum_{n=0}^\infty \l_+^{-n}
B \l_+^{n\b}|\x-\h|^\b\le \frac{B}{1-\l_+^{1-\b}}|\x-\h|^\b
\label{eF.6}\ee
because $|S_0^{\pm n}(\x-\h)|\le \l_+^n |\x-\h|$  for all $n\ge0$, and
therefore $h^{[1]}$ is H\"older continuous.

The above argument can be extended to all order in $\e$ to prove rigorously
that $h(\x)$ is smooth in $\e$ near $\e=0$ and H\"older continuous in
$\x$, for details see \Cite{GBG004}[Sec. 10.1]. It can also lead to show
that locally the stable or unstable manifolds (which are the $h$-images of
those of $S_0$) are infinitely differentiable surfaces (actually lines, in
this $2$-dimensional case), \Cite{GBG004}[Sec. 10.1].

\section{Appendix: Markovian partitions construction}
\def\SEC{Appendix: Markovian partitions construction}
\label{appG}\iniz
\lhead{\small\ref{appG}.\ \SEC}\index{Markov partition}

This section is devoted to a mathematical proof of existence of
Markovian partitions of phase space for Anosov maps on a
$2$--dimensional manifold which at the same time provides an
algorithm for their construction.\footnote{\tiny It follows an
idea of M. Campanino.}. Although the idea can be extended to
Anosov maps in dimension $>2$ the (different) original
construction in arbitrary dimension is easily found in the
literature, \Cite{Si968a}\Cite{Bo970a}\Cite{GBG004}.

Let $S$ be a smooth (or analytic) map defined on a smooth compact manifold
$\X$: suppose that $S$ is hyperbolic, transitive. Let $\d>0$ be fixed, with
the aim of constructing a Markovian partition with rectangles of diameter
$\le \d$.

Suppose for simplicity that the map has a fixed point $x_0$. Let $\t$ be so
large that $\lis C= S^{-\t} W^s_\d(x_0)$ and $\lis D= S^{\t} W^u_\d(x_0)$
(the notation used in Sec.(\,ref{sec:III-3}\,) is $W_\d(x)\defi$ connected part
containing $x$ of the set $W(x)\cap B_\g(x)$) fill $\X$ so that no point
$y\in \X$ is at a distance $>\frac12\d$ from $\lis C$ and from $\lis D$.

The surfaces $\lis C,\lis D$ will repeatedly intersect forming, quite
generally, 'rectangles' $Q$: but there will be cases of rectangles $Q$
inside which part of the boundary of $\lis C$ or of $\lis D$ will end up
without {\it fully} overlapping with an axis of $Q$ ({\it i.e.} the stable
axis $w^s$ in $Q$ containing $\lis C\cap Q$ or, respectively, the unstable
axis $w^u$ in $Q$ containing $\lis D\cap Q$) thus leaving an incomplete
rectangle as in Fig.(3.A.1):
\vfill\eject
\eqfig{200}{60}{
\ins{68}{35}{$\to$}
\ins{28}{53}{$\lis C$}
\ins{120}{53}{$\lis C$}
}
{fig3-A.1}{fig3.A.1}
\begin{spacing}{0.5}\tiny\0 Fig.3.A.1: An incomplete rectangle:
$\lis C$ enters it but does not cover fully the stable axis
(completed by the dashed portion in the second
figure).\end{spacing}
\*\*

In such cases $\lis C$ will be extended to contain $w^s$, or
$\lis D$ will be extended to contain $w^u$: this means continuing
$W^s_\d(x_0)$ to $W^s_\d(x_0)'$ and $W^u_\d(x_0)$ to
$W^u_\d(x_0)'$ elongating them by at most $
Le^{-\l\t}$, where $L$ is the largest of the lengths of
$W^i_\d(x_0), i=u,s$. So doing the surfaces $\lis C'=S^{-\t}
W^s_\d(x_0)'$ and $\lis D'=S^{\t} W^u_\d(x_0)'$ will partition
$\X$ into complete rectangles and there will be no more
incomplete ones.

By construction the rectangles $\EE_0=(E_1,\ldots,E_n)$ delimited
by $\lis C',\lis D'$ will all have diameter $<\d$ and no pair of
rectangles will have interior points in common. Furthermore the
boundaries of the rectangles will be smooth (a peculiarity of
dimension $2$). Since by construction $S_\t\lis C'\subset \lis
C'$ and $S^{-\t}\lis D'\subset \lis D'$ if $Le^{-\l\t}<\d$,
as it will be supposed, the main property Eq.\equ{e3.4.1} will
be satisfied with $S_\t$ instead of $S$ and $\EE_0$ will be a
Markovian pavement for $S_\t$, if $\d$ is small enough.

But if $\EE=(E_1,\ldots,E_n),\EE' =(E'_1,\ldots,E'_m) $ are Markovian
pavements also $\EE''=\EE\vee \EE'$, the pavement whose elements are
$E_i\cap E'_j$, is Markovian and $\EE\defi \vee_{i=0}^{\t-1} S^i \EE_0$ is
checked to be Markovian for $S$ ({\it i.e.} the property in
Eq.\equ{e3.4.1} holds not only for $S_\t$ but also for $S$). 

Anosov maps may fail to admit a fixed point: however they certainly have
periodic points (actually a dense set of periodic points)\footnote{\tiny Let $E$ be
  a rectangle: then if $m$ is large enough $S^mE$ intersects $E$ in a
  rectangle $\d_1$, and the image $S^m\d_1$ intersects $\d_1$ in a
  rectangle $\d_2$ and so on: hence there is an unstable axis $\d_\infty$ of
  $E$ with the property $S^m\d_\infty\supset\d_\infty$. Therefore
  $S^{-km}\d_\infty\subset\d_\infty$ for all $k$ hence $\cap_k
  S^{-km}\d_\infty$ is a point $x$ of period $m$ inside $E$.}. If $x_0$ is
a periodic point of period $n$ it is a fixed point for the map $S^n$. The
iterates of an Anosov map are again Anosov maps: hence there is an Markov
pavement $\EE_0$ for $S^n$: therefore $\EE\defi \vee_{i=0}^{n-1} S^i \EE_0$
is a Markovian pavement for $S$.
\*

\section{Appendix: Axiom C}\index{Axiom C}
\def\SEC{Appendix: Axiom C}\index{Axiom C}
\iniz\label{appH}
\lhead{\small\ref{appH}.\ \SEC}

To begin with a simple example consider a Anosov map $S_*\times
S_*$ acting on $\AA$ and on a copy $\AA'$ of $\AA$; suppose that
$S_*$ admits a time reversal symmetry $I^*$: for instance $\AA$
could be the torus $T^2\times T^2$, and $S_*$ the map, in the
example in the end remarks in Sec.(\,\ref{sec:V-3}\,). If $x$ is a
point in $\AA$ the generic point of the phase space $\X$ will be
determined by $(x,z,y)$ where $x\in \AA, y\in\AA'$ and $z$ is a set of
transverse coordinates that tell us how far we are from the
$\AA$ and $\AA'$.

In the FigH.1 given $x\in\AA$ the {\it stable} manifold ${\cal
  V}_s(x)$ on $\AA$ through $x$ generates, together with the
vectors pointing towards $\AA$, a surface represented by the
``vertical'' surface through $x$ (intersecting $\AA$ on ${\cal
  V}_s(x)$).

Likewise in the second drawing in
FigH.1 the {\it unstable} manifold ${\cal V}_u(x)$ on $\AA$ through $x$
generates, together with the vectors pointing towards
$\AA$, a surface represented by the ``vertical'' surface through
$x$. The two vertical surfaces intersect near $\AA$ on a line $L(x)$.

The third drawing illustrates the possible geometric property
that the two surfaces and the line $L(x)$ above extend in phase
space to intersect the $\AA'$: the line $L(x)$ intersects on a
point $Px\in \AA'$ and the two vertical manifolds in the stable
and unstable manifolds on $L(x)$ on $\AA'$.

\eqfig{300}{100}
{\ins{28}{16}{$x$}
 \ins{120}{16}{$x$}
 \ins{181}{88}{$I'x$}
 \ins{188}{16}{$x$}
}
{figH1}{FigH.1}
\*

\begin{spacing}{0.5}\0{\tiny FigH.1:
    Each of the $3$ lower surfaces represent the attracting set
    $\AA$ and the upper the repellers $\AA'$. Motion $S$ on $\AA$
    and $\AA'$ is chaotic (Anosov). The system is reversible
    under the symmetry $I$ but motion on $\AA$ and $\AA'$ is not
    reversible because $I\AA=\AA'\ne \AA$. Given a point
    $x\in\AA$ its stable plane at $x$ together with the
    directions consisting of the contracting directions towards
    $\AA$ generates surfaces transversal to $\AA$ represented
    vertical rectangle in the first drawing. Likewise the
    unstable plane at $x$ together with the contracting
    directions towards $\AA$ generates surfaces transversal to
    $\AA$ as in the second drawing. The two surfaces intersect
    near $x$ in a line sticking out of $\AA$. The axiom-C
    property requires that the two vertical surfaces extend to
    cross transversally both $\AA,\AA'$ as in the third figure:
    thus establishing a one-to-one correspondence $x\otto I' x$
    between attracting and repelling sets.  The composed map
    $I\circ\,I'$ leaves $\AA$ (and $\AA'$) {\it invariant} and is a
    time-reversal for the restriction of the evolution $S$ to
    $\AA$ and $\AA'$.}%
  \end{spacing}
  \*

It is possible to think that in generic reversible systems
satisfying the CH the situation is the above: namely there is an
"irrelevant" set of coordinates $z$ that describes the departure
from the future and past attracting sets $\AA,\AA'$: the latter
are are copies of each other (via the global time reversal $I$:
$\I\AA=\AA'$) and on each of them is defined a map $I^*$ which
inverts the time arrow: such map will be naturally called the
{\it local time reversal}.

In the above case the map $I^*$ and the coordinates $(x,z,y)$ are
"obvious".  The problem is to see whether they are defined quite
generally under the only assumption that the system is reversible
and has a unique future and a unique past attracting set on which
motion is an Anosov system.  This is a problem that is naturally
solved in general when the system verifies the ``Axiom C''.\index{axiom C}

A formal definition of the axiom C property is:
\*
\0{\bf Definition: \rm(Axiom C)}\index{Axiom C} {\it A smooth system
  $(\CC,S)$ verifies Axiom C if it admits a unique attracting set
  $\AA$ on which $S$ is Anosov and a distinct repelling set
  $\AA'$ and:
\0(1) for every $x\in\CC$ the tangent space $T_x$ admits a
H\"older--continuous\footnote{\tiny One might prefer to require real smoothness,
  \eg $C^p$ with $1\le p\le \io$: but this would be too much for rather
  trivial reasons, like the ones examined in App.\ref{appG}.  On the other
  hand H\"older continuity might be equivalent to simple $C^0$--continuity
  as in the case of Anosov systems, see \Cite{AA966}\Cite{Sm967}.} decomposition
as a direct sum of three subspaces $T^u_x,T^s_x,T^m_x$ such that if
$\d(x)=\min(d(x,\AA),d(x,\AA'))$, there exist constants $C,c$ and:
\halign{\qquad#\quad&#           &#     \hfill      &\qquad#\hfill\cr
(a)& $dS\, T^\alpha_x$&$=T^\alpha_{Sx}$          &$\a=u,s,m$ \cr
(b)& $|dS^n w|$     &$\le C e^{-\l n}|w|$,       &$w\in T^s_x,\ n\ge0$\cr
(c)& $|dS^{-n} w|$  &$\le C e^{-\l n}|w|$,       &$w\in T^u_x,\ n\ge0$\cr
(d)& $|dS^n w|$     &$\le c \d(x)^{-1} e^{-\l |n|}|w|$,&$w\in T^m_x,\
\forall n$\cr}\label{anosov-def}
\0where the dimensions of $T^u_x,T^s_x, T^m_x$ are $>0$.
\\
\0(2) if $x$ is on the attracting set $\AA$ then $T^s_x\oplus
T^m_x$ is tangent to the stable manifold in $x$; if $x$ is on
the repelling set $\AA'$ then $T^u_x\oplus T^m_x$ is tangent to the
unstable manifold in $x$.} \*

Although $T^u_x$ and $T^s_x$ might not be uniquely determined, the
planes $T^s_x\oplus T^m_x$ and $T^u_x\oplus T^m_x$ are uniquely
determined for $x\in\AA$ and, respectively, $x\in\AA'$.

The geometrical and dynamical meaning can be commented as follows.
An Axiom C system is a system satisfying the CH in a form
which gives also properties of the motions away from the attracting set:
\ie it has a stronger, and more global, hyperbolicity property.

Namely, if $\AA$ and $\AA'$ are the attracting and repelling sets the
stable manifold of a periodic point $p\in \AA$ and the unstable manifold of
a periodic point $q\in\AA'$ not only have a point of transverse
intersection, but they intersect transversely {\it all the way} on a
manifold connecting $\AA$ to $\AA'$; the unstable manifold of a point in
$\AA'$ will accumulate on $\AA$ {\it without winding around it}.

It will be helpful to continue referring to FigH.1 to help intuition.
The definition implies that given $p\in\AA$ the stable and unstable
manifolds of $x$ intersect in a line which intersects $\AA'$ and establish
a correspondence $I'$ between $\AA$ and $\AA'$.

If there the map $S$ satisfies the definition above and it has  also a
 time reversal symmetry $I$ then the map $I^*=I\,I'$ on $\AA$ (and on
$\AA'$) has the property of a time reversal for the restriction of
$S$ to $\AA$ (or $\AA'$). The axiom C property will be useful in
 particular when it can also be proved that the map $I\otto I'$
 is smooth.

For more details see \Cite{BG997}. Axiom C systems can also be shown to
have the property of structural stability: they are $\O$--stable in the
sense of \Cite{Sm967}[p.749], see \Cite{BG997}.

\section{Appendix: Pairing theory}
\def\SEC{Appendix: Pairing theory}
\label{appI}\iniz
\lhead{\small\ref{appI}.\ Pairing theory}

Consider\index{pairing theory} the equations of motion, for an
evolution in continuous time of $N$ particles in $2N\,d\,$
dimensions, for $X=(\Bx,\Bp)\in R^{2Nd}$:
\be \dot{\Bx}=\Bp,\quad \dot\Bp=-\BDpr \f(\Bx) -\a\Bp,
\qquad \a=\ch \qquad {\rm or}\qquad \a=-\frac{\BDpr
\f(\Bx)\cdot\Bp}{\Bp^2}\Bp\label{eI.1}\ee
which correspond to motion with kinetic energy controlled by
constant viscosity or by a thermostat fixing it at a given value
$\frac12\Bp^2$, in a system of $N$ particles in dimension $d$
interacting via a potential $\f$ containing internal and external
conservative forces. Call $M$ the phase space where the motions
take place.  \*

{\it Throughout it will be assumed that the system has a single
attracting set $\AA$ verifying the CH and $\AA=M$.}
\*

The positions $\Bx=(\x_1,\ldots,\x_N)$ vary in $R^d$ (in
which case the external forces contain ``wall potential barriers'' which
confine the system in space, possibly to non simply connected regions) or
in a square box with periodic boundaries, $T^d$. The potential $\f$ could
possibly be not single valued, although its gradient is required to be
single valued. In the model in Eq.\equ{eI.1} with variable $\a$
the momentum variables $\Bp=(\p_1,\ldots,\p_N)$ are
constrained to keep $\Bp^2$ constant by the thermostat action
$-\a(\Bx,\Bp)\Bp$.

If $\a=\n>0$, let $J(X)$ be the $2N\,d\,\times 2N\,d\,$ matrix of
the derivatives of Eq.\equ{eI.1} so that the Jacobian matrix
$W(X,t)=\dpr_X S_t(X)$ of Eq.\equ{eI.1} is the solution of
$\dot W(X,t)=J(S_t(X)) W(X,t)$. For $\n=0$ Jacobian matrix
$J_0(X)$ can be computed yielding: $J(X)=\pmatrix{0&\d_{ij}\cr
  M_{ik}& 0\cr}=J_0$.

More generally the Jacobian matrix of a motion with equations
$\dot\Bx=-\dpr_\Bp H(\Bp,\Bx); \dot\Bp=-\dpr_\Bx
H(\Bp,\Bx)-\ch\Bp$, \ie the matrix of the derivatives $\frac{\dpr
  \dot X}{\dpr X}$, will have the form
\be
J(X)=J_0(X)-\pmatrix{0&0\cr0&-\ch\cr}=J_0(X)+
\pmatrix{+\frac12\ch&0\cr0&-\frac12\ch\cr}
  -\frac\ch2 I \label{eI.2}\ee
with $J_0$ an infinitesimal symplectic matrix and $I$ the identity matrix.
Remark that the sum of the first two matrices on the r.h.s is the
Jacobian matrix of the Hamiltonian $H'(\Bp,\Bx)=
H(\Bp,\Bx)+\frac\ch2\Bp\cdot\Bx$: hence it is symplectic
(infinitesimal). So $J(X)$ is the sum of a symplectic matrix and
$-\frac\ch2 I$.

Hence $\dot W(X(t))= (J(X)-\frac\ch2 I)W(X(t))$ and
$W(X(t))=e^{-t\frac\ch2} \lis W(X(t))$ with $\lis W(t)$
symplectic and the spectrum of $W(X(t))$ will be shifted by
$-\frac\ch2 t$ from that of $\lis W(t)$ symmetric with respect to
$0$: hence pairing follows at the level $-\frac\ch2$ and the same
holds for the spectrum of $(\lis W(t)^T\lis W(t))^\frac12$,
\Cite{Dr988}, because if $L$ is a symplectic matric such is
$L^TL$ (recall that $L,L^T$ are symplectic if $LJL^T=J$: hence
$L^TLJL^TLJ=-1$, \Cite{HZ994}).

The {\it above} pairing proof, \Cite{Dr988}, holds for systems obtained by
adding a constant friction $-\ch \Bp$ to {\it general Hamiltonian}
systems.

A {\it second proof}, \Cite{DM996}\Cite{WL998}\Cite{Ru999}, of
the pairing can be given and considerably extends, as suggested
in \Cite{ECM990}, the pairing proof to the {\it special case} of
Gaussian thermostatted systems in Eq.\equ{eI.1}, with
$\a \Bp=-\frac{\BDpr {\Bff}(\Bx)\cdot\Bp}{\Bp^2}\Bp$. In the
following the version in \Cite{DM996} is analyzed, although the
versions in \Cite{WL998}\Cite{Ru999} are considerably simpler.

The Jacobian is:
\be J(X)=\pmatrix{0&\d_{ij}\cr
F_{ik}(\d_{k,j}-\frac{\p_k\p_j}{\Bp^2})& -\a
  \d_{ij}+\frac{\p_i\dpr_j\f}{\Bp^2}-2\frac{\dpr_k\f\,\p_k}{\Bp^4}
  \p_i\p_j\cr}
\label{eI.3}\ee
with $F_{ik}\defi-\dpr_{ik}\f$; which shows explicitly the properties that
an infinitesimal vector $\pmatrix{\e\Bp\cr0}$ joining two initial data
$(\Bx,\Bp)$ and $(\Bx+\e \Bp,\Bp)$ is eigenvector of $J(X)$ with eigenvalue
$0$. 
This means that $W(X,t)$ maps $\pmatrix{\e\Bp\cr0\cr}$ into
$\pmatrix{\e\Bp_t\cr0\cr}$ if
$S_t(\Bx,\Bp)=(\Bx_t,\Bp_t)$.

And the time derivative of $\pmatrix{0\cr\Bp}$, \ie
$J\pmatrix{0\cr\Bp}=\pmatrix{\Bp\cr0}$,  is
orthogonal to $\pmatrix{0\cr\Bp}$: so that $\Bp^2$ is a constant
of motion, as directly consequence of the value $\a$ in
Eq.\equ{eI.1}.

In other words the ``tangent'' solution of $\dot W(t)=J(\uu(t))
W(t)$ will have two vectors on which it acts expanding or
contracting them less than exponentially: {\it hence the map
  $S_t$ will have two zero Lyapunov exponents}.

Given $\uu$ follow its trajectory $\uu(t)=(\Bx(t),\Bp(t))$, and study the 
linearized evolution equation $\dot W(t)=J(\uu(t)) W(t)$.
The preceding remarks show that it will be convenient two use a system of
coordinates which, in an infinitesimal neighborhood of $X=(\Bx,\Bp)$, is
orthogonal and describes infinitesimal $2N\,d\,$-dimensional 
vectors $(d\Bx,d\Bp)$: via the $N\,d\,$-dimensional components of
$d\Bx$ and $d\Bp$ in a reference frame with origin in $\Bx$ and with
axes $\V e_0,\ldots,\V e_{N\,d\,-1}$ with $\V e_0\equiv
\frac{\Bp}{|\Bp|}$.

Calling $J_\Bp$ the matrix $J$ in the new basis, it soon appears
to have the form:
\be
J_{\Bp}=\pmatrix{
0      &\cdot  &\cdot   &      1&\cdot          &\cdot       &\cdot&0 \cr
\cdot  &0      &\cdot   &      0&1              &\cdot       &\cdot&0 \cr
\cdot  &\cdot  &\cdot   &      0&\cdot          &\cdot       &\cdot&0 \cr
\cdot  &\cdot  &\cdot   &      0&\cdot          &\cdot       &1&0 \cr
\cdot  &\cdot  &0   &      0&\cdot          &\cdot       &\cdot&1 \cr
0 &M_{01} &\ \cdot &        0    &\frac{\dpr_1\f}{|\Bp|}&\frac{\dpr_2\f}{|\Bp|}&\cdot&\cdot\cr
0 &M_{11}          &\cdot   &0&-\a          &\cdot       &\cdot&0 \cr
0 &M_{21}          &\cdot   &0&0            &-\a         &\cdot&0 \cr
\cdot  &\cdot           &\cdot   &0&\cdot        &\cdot       &\cdot&0 \cr
\cdot  &\cdot           &M_{N\,d-1\,,N\,d-1\,} &     0&\cdot        &\cdot      &\cdot&-\a \cr
}\label{eI.4}\ee
where $M_{ij}\defi -\dpr_{ij}\f(\Bx)$. The cancellations that
appear in Eq.\equ{eI.4} allow to identify in the matrix $J_\Bp$
the following structure.

If $(x,\V u,y,\V v)$ is a $2N\,d\,$ column vector ($x,y\in R,\,
\V u,\V v\in R^{N\,d\,-1}$) and $P$ is the projection which sets
$x=y=0$, the matrix $J_\Bp$ becomes $J_\Bp=PJ_\Bp P+(1-P)J_\Bp
P+(1-P)J_\Bp(1-P)$ (since $PJ_\Bp(1-P)=0$) and, setting
$Q\defi(1-P)$,

\be J_\Bp^{n}=(PJ_\Bp P)^{n}+(QJ_\Bp Q)^{n}+(QJ_\Bp
Q)^{n-1}QJ_\Bp P\label{eI.5}\ee
The key remark, \Cite{Dr988}\Cite{DM996}, is that the $2\times2$ matrix
$QJ_\Bp Q$ is $\pmatrix{0&1\cr0&0\cr}$, and
  the $2(Nd-1)\defi 2D$ dimensional matrix $PJ_\Bp P$ has the
  form:
  \be\pmatrix{0&1\cr M'&-\a\cr}=
  \pmatrix{\frac\a2&1\cr M'&-\frac\a2\cr}-\frac\a2
\label{eI.6}\ee
with $M'$ a symmetric matrix and both matrices $QJ_\Bp Q$ and
$PJ_\Bp P +\frac\a2$
{\it are  infinitesimal symplectic matrices}.
\footnote{\tiny The first being the Jacobian matrix for the
$1$-dimensional Hamiltonian $H=\frac12 p^2$; the second for the
$D$-dimensional Hamiltonian $H(\pp,\qq)=\frac12
\pp^2-\frac12(\qq,M\qq)+\frac12 \a\, \pp\cdot\qq$.}
therefore the solution of the equation:

\be \dot W_\Bp(X,t)=J_\Bp(S_t X) W_\Bp(X,t), \qquad
W_\Bp(X,0)=1,\label{eI.7}\ee
will have, in the new coordinates, form
$W_\Bp(X,t)=W_{\Bp,0}(X,t) e^{-\frac12\int_0^t \a(S_t X)dt}$ with
$W_{\Bp,0}$ a {\it symplectic matrix}. Hence the spectral
properties of $W_\Bp(X,t)$ will be those of $W_{\Bp,0}(X,t)$
shifted by $-\frac12\int_0^t \a(S_t X)dt$.

Ordering in decreasing order its $2D=2(N\,d\,-1)$ eigenvalues $\l_j(X,t)$
of the product ${\fra1{2t}}\log (W_\Bp(X,t)^TW_\Bp(X,t))$ it is:
\be \l_j(X,t)+\l_{2D-j}(X,t)= -\int_0^\t\a(S_t
X)dt,\quad j=0,\ldots D\label{eI.8}\ee
that can be called ``local pairing rule''.

Since $(QJQ)^n=0$ for $n\ge2$ the $2\times2$ matrix $QJQ$ will give $2$
extra exponents which are $0$. 

Going back to the original basis we see that the eigenvalues of
the matrices $\dpr S_t(X)^T\dpr S_t(X)$ and
$W_\Bp(X,t)^TW_\Bp(X,t)$ coincide after discarding the two $0$
exponents, and the pairing rule holds, \Cite{DM996}.  This will
mean $\l_j+\l_{2D-j}=const$ or, if $\s_+$ is the average phase
space contraction:
\be \l_j+\l_{2(Nd-1)-j}=-\frac{\s_+}{D},\qquad D=Nd-1\Eq{eI.9}\ee
The first pairing proof given above holds for equations obtained
by adding a {\it constant} viscosity $-\ch {\Bp}$ to a {\it
  general} Hamiltonian system (not necessarily of the form
$\frac12\Bp^2+\f(\Bx)$ as in Eq.\equ{eI.1}).  But the second
proof {\it does not} apply immediately to more general Gaussian
constrained equations as {\it it assumes the form}
Eq.\equ{eI.1} for the Hamiltonian. Does the pairing also apply
to such cases? \Cite{WL998}\Cite{Ru999}: see Appendix \ref{appJ}.

\section{Appendix: Fluid equations, pairing and ergodicity}
\def\SEC{Appendix: Fluid equations, pairing and ergodicity}
\label{appJ}\iniz
\lhead{\small\ref{appJ}.\ \SEC}\index{Gaussian fluid equations}

In several applications the statement that the stationary state
is ``ergodic'' is mentioned: often intending that trajectories
are dense in phase space aside from a set of $0$ area in phase
space. The CH insures the statement validity in
the sense that motions, with chaotic initial data as in
Sec.(\,\ref{sec:VI-2}\,), are asymptotic to attracting sets and its
restriction to them is ergodic in the mathematical sense with
respect to the SRB distribution.

However it is often necessary to distinguish when an attracting
set is the entire phase space and when it is not. For instance
this is important when checking whether the system can be
considered an Anosov system: a question which arises in
simulations in which the fluctuation theorem is applied to
reversible fluid motions.

In the Navier-Stokes equations it seems that if the UV cut-off
$N$ is increased the negative Lyapunov exponents eventually
outnumber the positive ones. Hence the attracting set is, at $N$
large, likely to be a smooth but low-dimensional surface: and as
discussed, even in reversible systems, the time reversal symmetry
breaks; warning that further work is necessary before trying to
apply the fluctuation theorem or other consequences of
reversibiity.

From Sec.(\,\ref{sec:IV-4}\,) a test of the Anosov property can be
based on the equality of the Kaplan-Yorke dimension $d_{ky}$ with
the phase space dimension. In simulations on $2$-dimensional
fluids, incompressible and in periodic containers, it is not
difficult to find examples in which even at moderate
regularization the sum $-\sum_j\l_j$ of negative exponents is
sizably larger than the sum of positive ones. An example is in
the following figure, Fig.J1:

\eqfig{144}{92}{}{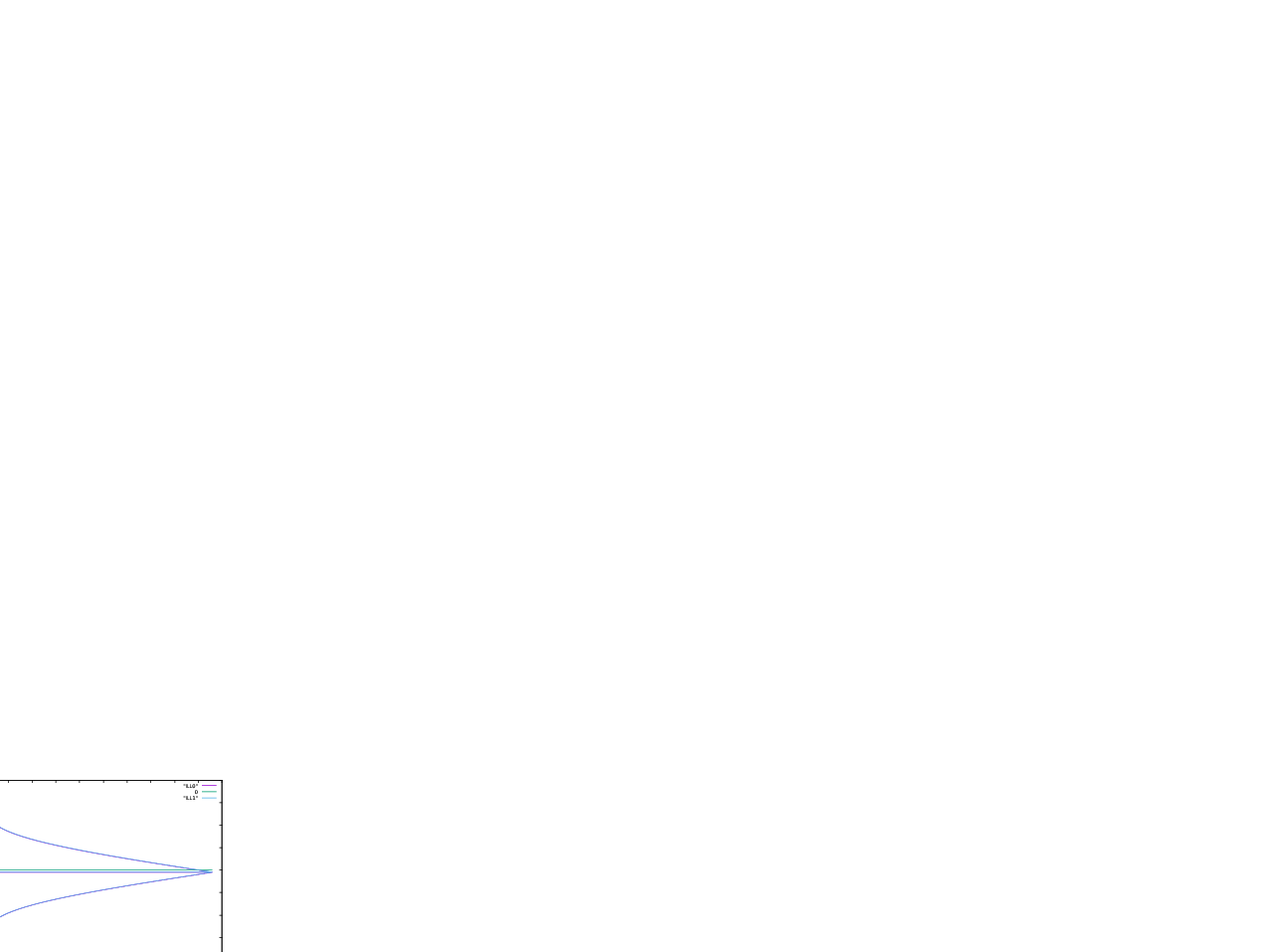}{Fig.J1}
\begin{spacing}{0.5}\0{\tiny Fig.J1: $R=2048$, {$N=960$}, local exponents
  $\l_k,\, 0\le k < d/2$ in decreasing order and $\l_{d-k}, 0\le
    k <d/2$ in increasing order e le linee
    $(\l_{k}+\l_{d-1-k})/2$ and the line $\equiv0$. Simulation
    step $2^{-19}$.  Reversible and irreversible case overlapping
    and apparent pairing.}\end{spacing}
\*

Apparent pairing and equivalence of exponents in corresponding
reversible and irreversible flows although unappreciable in the
scale of the figure becomes appreciable at much larger scale: it
is below $2\%$ except in a small interval around $\l_j\sim0$.
Nevertheless the result is surprising particularly because it is
expected to depend on the size of the UV regularization $N$ and
the pairing curve to bend downwards, at fixed forcing, as $N$
grows: it would be important to find theoretical support to the
apparent pairing at low $N$ and to understand whether it could be
related also to the precision in the calculations (simulation
step size $2^{-19}$ in Fig.J1).

In the next section a simple case in which the dimension
$d_{ky}$, defined in Sec.(\,\ref{sec:IV-4}\,), equals the phase space
dimension and consequences can be derived, in the light of
equivalence conjectures even about irreversible flows.

\section{Appendix: Fluctuation theorem and viscosity}
\def\SEC{Appendix: Fluctuation theorem and viscosity}
\label{appK}\iniz
\lhead{\small\ref{appK}.\ \SEC}

In reversible Navier-Stokes equations for $\V u$,
Eq.\equ{e5.4.2}\,(ii), the non constant multiplier $\a$ leads
to a ``phase space contraction'' (formally $-\sum_\kk
\frac{\dpr \dot u_\kk}{\dpr u_\kk}$):
\be\s(u)=\a(u)\Big(\KK_2-2\frac{E_6(u)}{E_4(u)}\Big)+\frac{F(u)}{E_4(u)}
\Eq{eK.1}\ee
with $\KK_2,E_4(u),E_6(u),F(u)$ are, see Eq.\equ{e5.6.7}:
\be \eqalign{
  \KK_2=&2^{d-2}\sum_\kk \kk^2, \ E_4(u)=\sum_\kk(\kk^2)^2
  |u_\kk|^2,\ \cr E_6(u)=&\sum_\kk(\kk^2)^3 |u_\kk|^2,\
  F(u)=\frac{\sum_\kk (\kk^2)^2\lis \f_\kk u_\kk}{E_4(u)}
\cr}\Eq{eK.2}\ee
where the sums run over the $\kk$ with $|k_i|\le N,\, i=1,2$ or
$i=1,2,3$. The expression is valid for $d=2,3$ dimensions {\it
provided} the appropriate $\a(\uu)$ is selected (if $d=3$
$\a(\uu)$ contains a contribution from the transport term): in
the following the $2$-dimensional case will be considered.

In {\it time reversible Anosov flows} the fluctuations of the
divergence satisfy the general symmetry relation, the
``fluctuation theorem'' discussed in Sec.(\,\ref{sec:VI-4}\,). Namely
if $S_t$ denotes the evolution and $\s_+$ the infinite time
average of $\s(S_t u)$ and
\be p(u)=\frac1\tau\int_0^\tau
{\s(S_\th u)}d\th\label{eK.3}\ee
then $p$ has a probability distribution in the stationary state
such that $p\in dp$ has density $P(p)=e^{\z(p)\tau +O(1)}$,
asymptotically as $\tau\to\infty$, with the {\it universal}
symmetry,\, \equ{e4.6.1}\,, $\z(-p)=\z(p)-p \s_+$, with $\s_+$
infinite time average os $\s(S_t u)$ called
fluctuation relation.

Before posing the main question: {\it is it meaningful to ask
  whether the fluctuation relation holds in irreversible
  evolutions ?} it is necessary to study, first:

1) the probability distribution $P$ of $p(\uu)$ is defined, by
Eq.(\,\equ{eK.3},\equ{eK.2}\,) both in the reversible and
irreversible flows: although $p(\uu)$ is not a local observable,
nevertheless it might be among the non local observables with
equal or close corresponding distributions,
Sec.(\,\ref{sec:I-5},\ref{sec:VII-5}\,).

2) the {\it local Lyapunov spectrum}: defined by considering the
Jacobian matrix of the evolution, formally the matrix
$J_{\kk,\hh}=\frac{\dpr \bf \dot {\rm u}_\kk} {\dpr u_\hh}$, and
then computing the eigenvalues of its symmetric part $J_s$, in
decreasing order, and averaging each one over the flow.

To apply the theorem, evolution should also be time reversible:
hence it cannot be applied, at least not without further
work.\footnote{\tiny If the attracting surface $\AA$ is
not the full phase space $M^N$ then the time reversal image
$\BB=I\AA$ is likely to be disjoint from $\AA$ and the motion
restricted to $\AA$ is not symmetric under the natural time
reversal $I$.\label{AA}}

Whether the spectra of the reversible and irreversible evolutions
are close is related to the key question: can the systems
$(M^N,S^N_t)$ be reasonably considered Anosov systems ?  as
mentioned above a first test is the equality of the dimension
$d_{ky}$ with the full phase space dimension of $M^N$.

The test turns out to be possible, in a reasonable computer time,
in simple cases with a small $N$ cut off: here the $N=48$ modes
(a modest $7\times7$ grid) and $\n=1/2048$: the local Lyapunov
spectrum can be equivalently defined as the Lyapunov exponents of
the trivial linear flow $S_tv = e^{J^s(\uu)t} v$ and can be
computed using the algorithm in \Cite{BGGS980b}.\footnote{\tiny
Fast in this case, if the time series $\uu(t)$ is available
because $\uu$, hence, $J(\uu)$ remain fixed: they are here
computed by iterating a large number $k$ of times (of the order
of $h^{-1}$) the matrix $(1+h J^s(\uu))$ and applying the quoted
method. To obtain $k$-independent results the time series should
be taken at time intervals large enough.}
\*
\eqfig{180}{115}{}{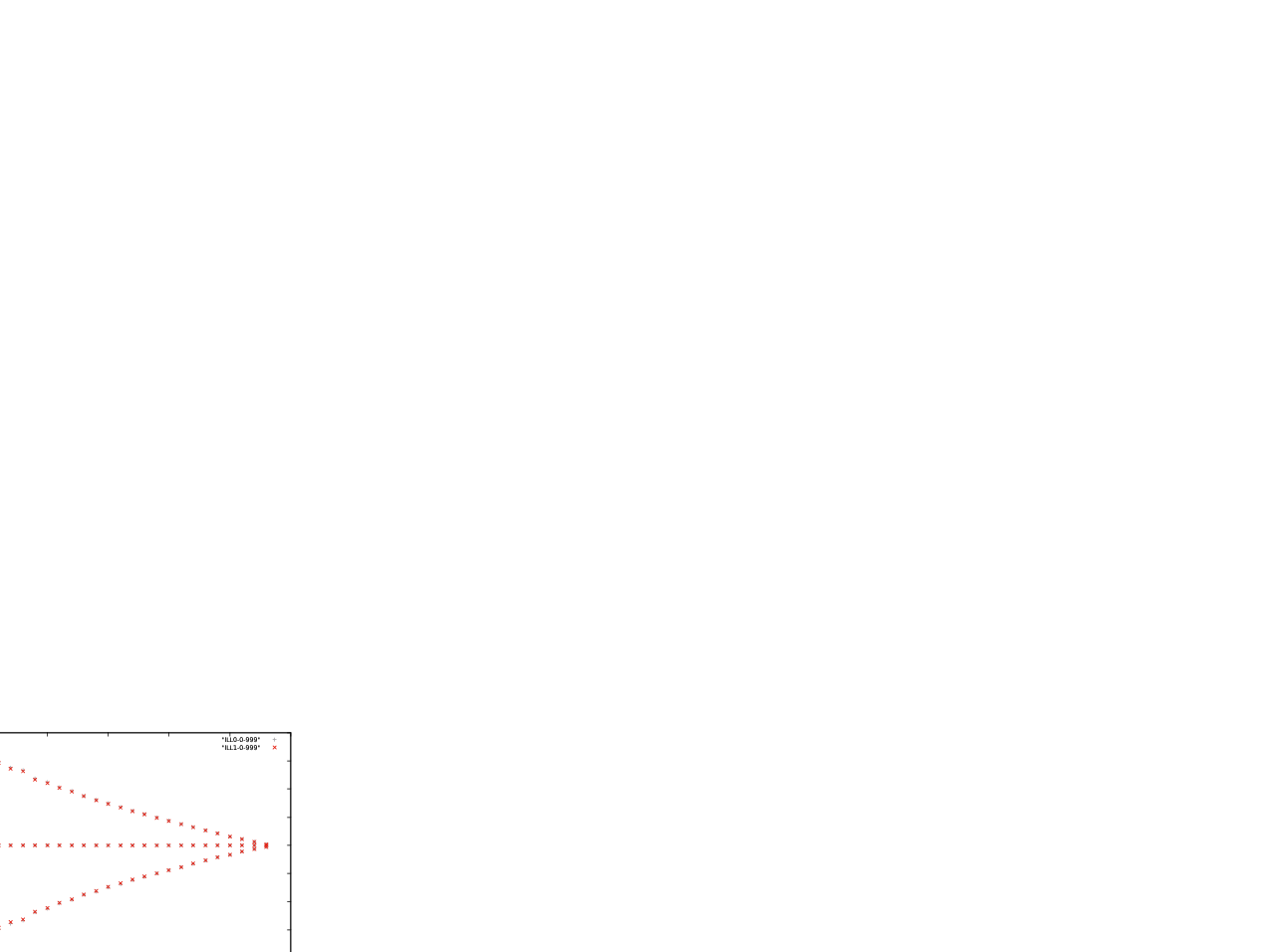}{}
\vskip-2mm
\begin{spacing}{0.6}\0{\tiny Fig.K1: Local Lyapunov spectra for {\it both}
  $NS_{irr}$ {\it and} $NS_{rev}$ flows (overlapping on the
  drawing scale) with $d=48$ modes, $R=2048$. Rapid computation
  with only $1000$ samples taken every $4/h$ time steps of time
  $h=2^{-13}$ and averaged: the upper and lower values give the
  $d/2$ exponents $\l_k$ and respectively $\l_{d-1-k}$, while the
  middle values are $\frac12(\l_k+\l_{d-1-k})$ not constant but
  close to $\simeq -.01$. This figure shows positive exponents to
  be equally numerous as the negative ones and the features {\bf
  a),b)} listed below and $d_kj=48$. }
\end{spacing}
\*\*

The quick check in Fig.K1 (see also \Cite{Ga019a}\Cite{Ga019c}) reports
$\l_k,k=0\ldots d/2-1$: the first half of the $d=(2N+1)^2-1$ exponents in
decreasing order and the second half $\l_{d-1-k},k=0\ldots d/2-1$ as
function of $k$ (upper and lower curves), as well as
$\frac12(\l_k+\l_{d-1-k})$ (intermediate line). 

It yields other somewhat surprising results besides showing the
equality of the numbers of positive and negative exponents which,
as mentioned above, provides evidence that the attracting set
fills densely phase space so that the time reversal symmetry
remains a symmetry on the attracting set.  Figure draws {\it in
  the same panel}, spectra from both $NS_{rev}$ and $NS_{irr}$
flows under equivalence conditions; they apparently ovelap and
show: \\
{\bf a)} ``coincidence'' of the spectra of the $NS_{rev}$ and $NS_{irr}$
evolutions: quite surprising and justifying an attempt to formulate and
check the fluctuation relation in the {\it irreversible} flows.  \\
{\bf b)} apparent ``pairing'': see appendix \ref{appI}\,, the
exponents appear ``paired'', \ie $\frac12(\l_k+\l_{d-1-k})$ is
$k$-independent.  Therefore the flow, being reversible and having
equal number of pairs of opposite sign, can be consistently
assumed to be a Anosov flow and \\
{\bf c)} The local Lyapunov spectrum is related to the actual
Lyapunov spectrum via interesting inequalities,
\Cite{Ru982}\Cite{Li984}: which could be also handy to test accuracy of
simulations.\*

The graph for $(\z(p)-\z(-p))/\s_+$ in Eq.\equ{e4.6.1} is studied for
both reversible and irreversible flows. It exhibits the main result: 

\eqfig{180}{100}{}{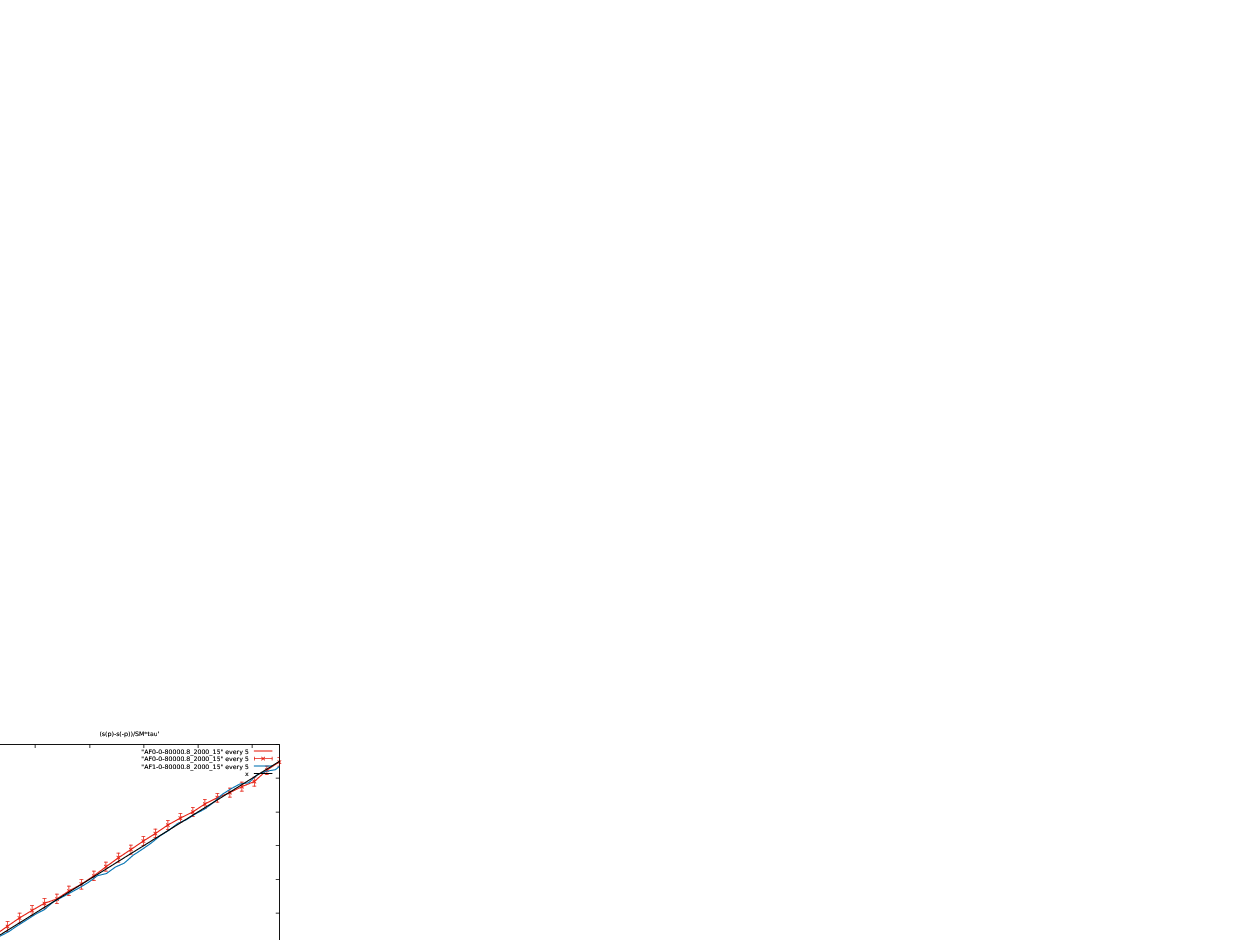}{}
\begin{spacing}{0.5}\0{\tiny Fig.K2: Test the fluctuation relation in the flow
  $NS_{irr}$ and $NS_{rev}$ flows with $48$ modes, $R=2048$. The
  $\tau$ is chosen $8$, the slope of the graph increases with
  $\tau$ reaching $1$ at $\tau=8$. The graph is built with
  $8\cdot 10^4$ data, divided into $2\cdot10^3$ bins, obtained
  sampling the flow every $4/h$ time steps of size
  $h=2^{-13}$. The keys AF0 and AF1 deal with $NS_{irr}$ (red
  o.l.) and, respectively, $NS_{rev}$ (blue o.l.)  and the error
  bars (red o.l.) deal with $NS_{irr}$; the line $f(x)=x$ is a
  visual aid.}
\end{spacing} 
\*

Fig.K1 and Fig.K2 also show that the proposed equivalence {\it
extends}, in this case, also to the phase space contraction
(``entropy production rate'', \Cite{Ga013b}\Cite{Ga019a}) as a
quantity defined for the reversible evolution but regarded as an
observable for the irreversible $NS$. The interest of the result
in the figures is to provide a fluctuation
relation in a {\it irreversible} evolution.\footnote{\tiny In
summary the prediction is based on CH, on the equality of numbers
of negative and non negative exponents and on the extension of
the equivalence hypothesis to the entropy production rate.}

The histogram of the PDF corresponding to Fig.K2 is very close to a Gaussian
centered at $1$ and width yielding the slope: 
\*
\eqfig{180}{115}{}{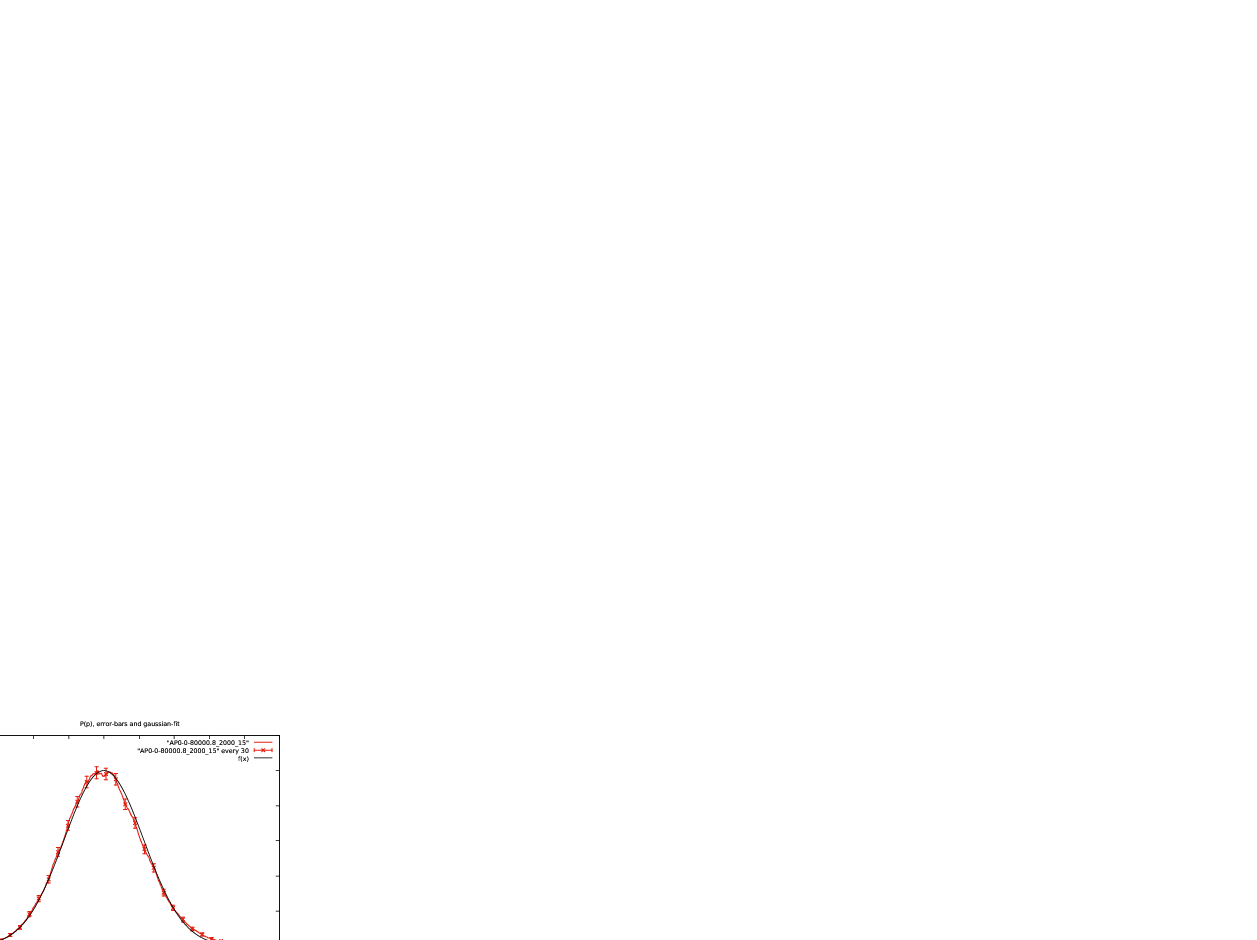}
.\hglue.5cm\raise 2cm \hbox{\kern3.5cm\tiny A}
%
\begin{spacing}{0.5}\0{\tiny Fig.K3: A histogram (with max
normalized to $1$), of the PDF {\it for the irreversible flow} of
  the variable $p$ (red o.l.), with $\tau=8$ generating the
  Fig.K3 out of the $8\cdot 10^4$ measurements of $\s(\uu)$ in
  the $NS_{rev},NS_{irr}$ equations. The $p$-axis is divided in
  $2000$ bins and for each $p$ the average of the number (and
  corresponding error bars) of points in $[p-\d,p+\d]$ is plotted
  (red o.l.)  with $\d=15/2000$ (corresponding to a small
  interval of $p$ compared to the width $2\sqrt{\s_+\tau} $) and
  the interpolating Gaussian (blue o.l.). The error bars for the
  reversible flow (not drawn) have the same
  sizes.}  \end{spacing} \*

Again showing an instance of a reversible-irreversible
equivalence.

A final comment could be made about the apparent pairing
of the exponents in Fig.K1 in the irreversible model: this has
been observed in \Cite{GRS004}: it is not surprising as the equal
number of positive and negative pairs is consistent with an
attractive surface identical with the full phase space $M^N$ and
therefore with an Anosov motion on $M^N$; but the cut-off $N$ is
very low ($N=7$) and it is expected that increasing $N$ at fixed
$R$ will soon lead to more negative Lyapunov exponents compared
to positive ones and the fluctuation relation will fail (because
the flow will no longer be transitive (\ie the attracting surface
$\AA$ will no longer be $M^N$).

\section{Appendix: Reversibilty and friction}
\def\SEC{Appendix: Reversibilty and friction}
\label{appL}\iniz
\lhead{\small\ref{appL}.\ \SEC}

Consider the OK41 regularization, which, in the simplest
formulation, studies the regularized equations with $N=
R^{\frac34}$, \footref{OK41}, and study the reversible equation.

Then the fluctuation theorem can be applied to it: so that the
SRB-probability to see, in motions following the reversible
equation in Eq.\equ{e5.5.1}, a ``{\it wrong}'' average friction
$\a=-\n$ for a time $\t$ (instead of the ``right'' $+\n$).
Assuming that the average divergence for the reversible equation
has at least the same order of magnitude as that of the
irreversible equation, and recalling Eq.\equ{e5.5.2}, it is
expected:
\be  {\rm Prob}_{srb}\sim \exp{\Big(-\t {2^6N^5}\n R^{\frac{15}{4}}
\Big)}\,\defi\, e^{-\g\t}\label{eL.1}\ee
The probability can be estimated in the situation considered
below for a flow in air with the parameters in the first line:
\be \left\{\eqalign{
  \n=&  1.5\,10^{-2}\,\frac{cm^2}{sec},\quad v=10.\,\frac{cm}{sec}\,
\quad L=100.\,cm\cr
   R=&   6.6\,10^{4},\quad  N=4.1\,10^3, \quad  \g\sim1.^{18}\, sec^{-1}\cr
   P\defi& {\rm Prob}_{srb}=
    e^{-\g\t}\sim e^{-1.18\,10^9},\qquad {\rm if}\quad 
\t=10^{-9} sec\cr}\right.\label{eL.2}\ee
where the first line are data of an example of fluid motion and
the other two lines follow from Eq.\equ{e5.7.5} and the
fluctuation relation.  They show that, by the fluctuation
relation, viscosity can be $-\n$ during $10^{-9}s$ ({\it say})
with probability $P$ as in Eq.\equ{eL.2}: the unlikelihood is
similar, in spirit, to the estimates about
Poincar\'e's\index{Poincar\'e's recurrence} recurrences,
\Cite{Ga002}.

\section{Appendix: Reciprocity and fluctation theorem}
\def\SEC{Appendix: Reciprocity and fluctation theorem}
\label{appM}\iniz
\lhead{\small\ref{appM}.\ \SEC}\index{Green-Kubo and fluctation theorem}

In the following is shown that reciprocity (\ie $L_{ij}=L_{ji}$)
can be immediately obtained, along the lines and following the notations of
Sec.(\,\ref{sec:VIII-4}\,), by applying the fluctuation patterns
theorem in the Sec.(\,\ref{sec:VII-4}).

For this purpose define, given $j$, the observable
$q(x)=E_j\dpr_{E_j} \s_{\V E}(x)$ where the factor $E_j$ is
there only to keep $\s_{\V E}$ and $E_j\dpr_{E_j}\s_{\V E}$ with the
same dimensions. The observable $q(x)$ is odd under time
reversal, as is also $p(x)=\s_{\V E}(x)$, and their averages
$p,q$ over time $[-\tau/2,\tau/2]$:
\be\kern-3mm{1\over\tau} \int_{-\tau/2}^{\tau/2} \s_{\V E}(S_tx)dt{\buildrel
  def\over =}p\langle\s_{\bf E}\rangle,\quad {1\over\tau} 
\int_{-\tau/2}^{\tau/2} E_j\dpr_{E_j} \s_{\V E}(S_tx)dt{\buildrel
  def\over =} q \, E_j\langle \dpr_{E_j}\s_{\V
  E}\rangle \label{eM.1}\ee
where $\langle\cdot\rangle$ denotes the SRB averagewill verify
the fluctuation patterns theorem, see Eq.\equ{e4.7.3}.

Then if $\pi_\tau(p,q)$ is the joint probability of $p,q$ the
fluctuation patterns theorem implies, See Eq.\equ{e4.7.3}\,:
\be
\lim_{\tau\to\infty} 
{1\over\tau}\log\pi_\tau(p,q)=-\zeta(p,q)\label{eM.2}\ee
The $\zeta(p,q)$ can be computed in the same way as $\zeta(p)$:
considering just two parameters for $\V E=(E_1,E_2)$, first
define the transform $\lambda(\beta_1,\beta_2)$:
\be
\lim_{\tau\to\infty} \textstyle
{1\over\tau}\log\kern-1pt\int e^{-\tau(\beta_1\, (p-1)
\langle\s_{\V E}\rangle+
\beta_2\,(q-1) E_2\dpr_{E_2}\langle \s_{\V E}\rangle)}\pi_\tau(p,q)dpdq
\label{eM.3}\ee
The Legendre trasform of $\l$ will be:
\be\eqalign{
  \zeta(p,q)=&\max_{\b_1,\b_2}\Big(
\b_1\, (p-1)\langle \s_{\V E} \rangle+
\b_2\,(q-1)
E_2\dpr_{E_2}\langle \s_{\V E} \rangle-\l(\b_1,\b_2)\Big)
\cr}\label{eM.4}
\ee
The  $\l(\vec\b)$, $\vec\b=(\b_1,\b_2)$, is 
evaluated via the cumulant expansion, as in Sec.(\,\ref{sec:VIII-4}\,), yields:
\be
\lambda(\vec\b)={1\over2}\,\Big(\vec\b, 
C\,\vec\b)+O(|{\bf E}^3|)\label{eM.5}\ee
\0{}where $C$ is the $2\times2$ symmetric matrix of the second order 
cumulants. The coefficient $C_{11}$ is given by $C_2$ appearing in 
\equ{e4.8.5}; $C_{22}$ is given by the same expression with $\s$ 
replaced by $E_2\dpr_{E_2}\s$, while $C_{12}=C_{21}$ is the mixed cumulant:
\be
C_{12}=\int_{-\infty}^\infty dt \int 
\Big(\s(S_tx)\,E_2\dpr_{E_2}\s(x)
-\langle \s_{\bf E}\rangle
\langle E_2\dpr_{E_2}\s_{\bf E}\rangle\Big) \m_{\bf E}(dx)\label{eM.6}\ee
where $\langle \cdot \rangle$ denotes the $\m_{\bf E}$ average.

Hence if $\vec w=\pmatrix{(p-1)\langle\s_{\bf E}\rangle\cr
  (q-1)\langle E_2\dpr_{E_2}\s\rangle_{\bf E}\cr}$ we get
$\z(p,q)={1\over2}\,\Big(C^{-1}\vec w,\vec w\Big)+O(|{\bf E}|^3)$
and the fluctuations patterns theorem, Sec.(\,\ref{sec:VII-4}\,),
implies that $\z(p,q)-\z(-p,-q)=p\langle \s_{\bf E}\rangle$ is {\it $q$
  independent}.

By simple algebra, this is $2C^{-1}{\langle\s_{\bf E}\rangle^2
  \choose \langle\s_{\bf E}\rangle E_2\langle\dpr_{E_2}\s_{\bf E}\rangle_{\bf
    E}}={\langle\s_{\bf E}\rangle\choose0}$, hence:
\be{\langle\s_{\bf E}\rangle\choose E_2\dpr_{E_2}\langle\s_{\bf
      E}\rangle}=\frac12C{1\choose0},\qquad
\langle\s_{\bf E}\rangle=\frac12C_{11},\
E_2\dpr_{E_2}\langle\s_{\bf E}\rangle=\frac12C_{21}
\label{eM.7}\ee
\0completely analogous to \equ{e4.8.6}. The first relation is
the same already examined to obtain $L_{11}$; hence proceeding 
in the same way expand both sides of the second of
\equ{eM.7} {\it to lowest order} in the $E_i$'s: the result is that
$L_{12}$ is given by $\frac12$ times the mixed
cumulant in \equ{eM.6}  and
the Green-Kubo and Onsager relations follow.
\*

\section{Appendix: Large deviations in SRB states}
\def\SEC{Appendix: Large deviations in SRB states}
\label{appN}\iniz
\lhead{\small\ref{appN}.\ \SEC}

The following is based on the approach
in \Cite{MS967}\Cite{MS968}\, and illustrates how a proof of the
theorem in Sec.\ref{sec:V-4} is reduced to a few bounds on
quantities familiar in the theory of statistical mechanics and
Markov processes.

In the original references the main difficulty was to prove the
needed bounds. However there is one case in which the bounds
can be obtained without very strong conditions on the parameters
of the systems. And the case occurs precisely among the probability
distributions on sequences $\qq=\{q_k\}_{-\infty}^\infty$,
$q_k=1,2,\ldots,n$: \ie to a class of distributions containing
the SRB ones and more generally of stochastic processes
on strings of symbols.

Consider the symbolic dynamical system describing a SRB
distribution: phase space $\X$ consists in sequences 
$\{q_x\}_{x\in(-\infty,\infty)}$, of labels and $q_x=1,
\ldots,n$, with a transitive compatibility matrix $M_{q,q'}$, see
(\,\ref{sec:IV-3}\,). 

The SRB distribution is constructed as
$\t\to\infty$, \equ{e4.1.1}, by considering compatible strings
$q_{-\t},\ldots, q_\t$ {\it continued to infinite strings $\qq$}
following a (arbitrarily) fixed standard rule\footnote{\tiny To respect
compatibility with the mixing matrix, see
\footref{standard continuation} in Sec.(\,\ref{sec:I-4}\,).}

Call $\X_V$ the collection of compatible strings
$\qq=\{q_x\}_{x\in V}\equiv\qq_V$ (their unique
continuation to infinite strings will {\it not} be mentioned
explicitly).

The SRB distribution for a system verifying CH has a form which
does not depend on the special system (its 'universality' is the
main feature of the theory) and assigns to infinite compatible
strings a probability which is the $V\to\infty$ limit of a
probability $P_V(\qq)$ on strings $\qq\in\X_V$ defined in terms of
``potentials'' $\f_I(\qq_I)$, (\,\ref{sec:I-4}\,).

The potentials $\f_I(\qq_I)$ are defined on intervals $I$
 centered\footnote{\tiny Center is defined unambiguously because
 they are chosen of odd size, see (\,\ref{sec:I-4}\,).} in
 $V=[-\t,\t]$ and are such that, the expansion rate
 $\L_V(x_\qq)=\sum_I\f_I(\qq_I)$, in terms of which the SRB is
 expressed in the limit $V\to\infty$, see \equ{e4.1.1}\,, is the
 distribution on $\X_V$ defined by:
\be P_V(\qq)=Z_V^{-1}
e^{-\sum_{I\cap V\ne\emptyset} \f_I(\qq_I)},\qquad\qq\in \X_V\label{eN.1}\ee
where $Z_V$ is a normalization factor (``{\it partition
function}'').\label{patition function}

Let $\L=[-L,L]\subset V$ and $M_\L(\qq)$ be an observable
localized in $\L\subset V$ of the (special) form
$M_\L(\qq)=\sum_{x\in \L} m(q_x)$: where $m(q)$ is
assigned\footnote{\tiny For instance $m(q)=q$ or $m(q)=q^2$ or
other.}\,.  This is restrictive, but the analysis could apply
with minor changes, to a very general class of observables, to
cover the observables quoted in the theorem in Sec.(\,\ref{sec:V-4}\,).

It is convenient, in view of the special role played below by the
``single site'' potential $\f(q)$, to define $\f'_I=\f_I$ if
$|I|>1$, to call $\f(q)=\m_0 m(q)$ and to adopt the notation
$\f=(\m_0,\f')$. The exponential in Eq.\equ{eN.1} is consequently
split as $\m_0 M_V(\qq)+\sum_{|I|>1, I\cap
V\ne\emptyset} \f'_I(\qq_I)\defi \m_0 M_V(\qq)+ U_V(\qq)$; so $P_V$ becomes:
\be P_{\m_0,V}(\qq) =Z_V(\m_0,\f')^{-1}{e^{-\m_0 M_V(\qq)-U_V(\qq)}}\label{eN.2}\ee

The large deviations problem is to estimate, {\it uniformly in
$V$}, the probability that $M_\L(\qq)\in [aL,bL]$, with
$[-L,L]\equiv \L\subset V$ in the distribution in which the ``chain''
$\qq\in \X_V$, has probability $P_V(\qq)\equiv P_{\m_0,V}(\qq)$.
\*

\0{\bf (1)} {\it Thermodynamic functions controlling the large
deviations of $M_L(\qq)$.}\,\Cite{GLM002}
\*

First prove that $\lim_{V\to\infty}\frac1V \log Z_V$ exists:
begin with the remark that, replacing the $\f_I$ by $0$ if
$I\not\subset V$ the exponential bound
$|\f_I|=\max_\qq|\f_I(\qq)|<B_0 e^{-\l |I|}$ on the size of
$I$ \equ{e4.1.7}\,, leads to bound the quantity
$\sum_{I\not\subset V} \f_I(\qq)$, out of the exponential in
\equ{eN.1},  by $B\|\f\||$ with $B$ independent on the
size of $V$ and $\|\f\|=\sum_{0\in 
I}|\f_I|<\infty$\,\footnote{\tiny By the exponential bound on
$\f$ only $I$'s close to boundary points of $V$
'really' contribute to the bound.}\,.
Therefore for a suitable constant $B_1$ and a suitable $\th$:
\be Z_V(\m_0,\f')\le e^{\|\f\| V},\quad
Z_V(\m_0,\f')=Z^0_{V}(\m_0,\f') e^\th,\qquad |\th|<B_1\label{eN.3}\ee
if $Z^0_{V}(\m_0,\f')$ is defined by replacing, in the definition
of $Z_V(\m_0,\f')$, the potentials $\f_I(\qq)$ by $\f_I(\qq)=0$
if $I\not\subset V$ .
\*

\0{\bf Remark:} Instead of setting $0$ the $\f_I$ with $I$ containing
points out of $V$, the same inequality holds for the
analogous $Z$  obtained by setting $0$ the
$\f_I$'s with points out of a corridor of width $D$ around the border
of $V$: the constant has to be increased by an amount independent
of the size of $V$ ($D$-dependent: $2B_1+2D\|\f\|$).
\*

If $V_k=2^k V_0$ denotes a sequence of intervals with sizes
$2^kL_0,L_0>0$ it is $Z^0_{V_k}=Z^0_{V_{k-1}}\kern-3mm\cdot
Z^0_{V_{k-1}}
\kern-3mm\cdot e^{\th' V}$ for a suitable $\th'$ and a constant $B'$, that can be
expressed in terms of $\|\f\|$ alone, and $|\th'|<B'$:
\be\frac1{V_k} \log Z^0_{V_k}=\frac1{V_{k-1}}\log
Z(V^0_{k-1}) +\frac{\th'}{V_k}\label{eN.4}\ee
so that $\lim_{k\to\infty} \frac1{V_k}\log Z^0_{V_k}$ exists and
so does, by \equ{eN.3}, $\lim_{k\to\infty} \frac1{V_k}\log
Z_{V_k}$ equal to it:  call both
$P(\m_0,\f')$.

Furthermore, given $\ell>\ell'$, let $2^{k-1}\ell=2^{k'+h}\ell'
+p\ell' 2^h,p<1$
and define the intervals $V_k,V'_{k'+h},\wt V$ of
respective length $2^k\ell,2^{k'+h}\ell',p 2^h\ell'$ then
$V_k=V_{k-1}\cup V'_{k'+h}\cup \wt V_p$ and, for suitable $B_2$, the above remark 
gives:
\be \frac1{V_k}\log
Z_{0,k}=\frac12\frac1{V_{k-1}}\log Z_{0,k-1}
+\frac{V'_{k'+h}}{V_{k}} \frac1{V'_{k'+h}}\log
Z_{0,k'+h}+B_2 \frac{2^h}{V_{k}}\label{eN.5}\ee
and, taking the limit as $k,k'\to\infty$, keeping $k'\gg h\to\infty$,
Eq.\equ{eN.4} becomes $P(\f)=\fra12 P(\f)+\fra12 P'(\f)$. Hence
$P(\f)=P'(\f)$ and 
$P(\m_0,\f')$ does not depend on the sequence $V$ used to compute
it; and $P_V(\m_0,\f')\defi \frac1V\log Z_V(\m_0,\f')$ and
$P^0_{V}(\f)=P^0_{V}(\m_0,\f')\defi \frac1{V}\log Z^0_V(\m_0,\f')$ both
converge to $P(\m_0,\f')$.\,\footnote{\tiny In statistical
mechanics $P$ could be called the ``free energy'' of the
chain confined in $V$.}.

The functions $P_V(\f),P^0_{V}(\f)$ are uniformly continuous in
$\f$: because, for instance,
if $\d\defi \ps-\f$:
\be\eqalign{|P^0_V(\f)-&P^0_V(\ps)|=
\frac1V |\int_0^1 \dpr_t \log Z^0_{V}(\f+t\d)\,dt|\cr
\le&
\frac1V \int_0^1  \Big|\frac{-\sum_{I} (\d_I(\qq))
\exp(-\sum_I (\f_I(\qq)+t\d_I(\qq)))}{Z^0_{V}(\f+t\d)}\Big|
dt\cr
\le&
\frac1V \int_0^1 V \|\d\| dt
= \|\f-\ps\|\cr}
\label{eN.6}\ee
and is convex in $\m_0$ (from the convexity of $x\to e^x)$). In
the same way $\lis m_\L=-\L^{-1}\dpr_\m \log Z_\L(\m,\f')$ is bounded by
$|\m|\max |m(q)|+\|\f'\|$.

A deeper property is that $P_\L(\f),P(\f)$ (as well as
$P^0_\L(\f))$) are $C^\infty$ as functions of the parameters
$z(q)=e^{-\m_0 m(q)}$ for $\m_0\in R$: and $\dpr^j
P_\L\tende{L\to\infty}\dpr^j P$ {\it uniformly} on bounded
intervals $\m_0\in R$ and $j\in Z^+$. This can be seen as a
property of 1D Gibbs distributions with exponentially decaying
potential, see item (3) below.
\*

\0{\bf(2)} {\it Analysis of the large deviations of $M_\L(\qq)$ around its
average $\lis m_\L \L$,} \Cite{GLM002}\,.
\*
The variable $M_\L(\qq)$ is studied via its Laplace transform:
\be Z_V(\m_0,\f';\l)=\frac{\sum_\qq e^{-\l M_\L(\qq)} e^{-\m_0
M_V(\qq)-U(\qq)}}{Z_V(\m_0,\f')}\label{eN.7}\ee
Adapting the above inequalities, for a constant $B$ independent
of $V$, it is: $Z_V(\m_0,\f';\l)=Z_\L(\m_0+\l) Z_{V/\L}(\m_0,\f')
e^{\lis \th}$ with $|\lis\th|<B$ and:
\be \frac{Z_V(\m_0,\f';\l)}{Z_V(\m_0,\f')}=
\frac{Z_\L(\m_0+\l,\f')}{Z_\L(\m_0,\f')}e^{\wt\th}\label{eN.8}\ee
Remark that, if $U_V(\qq) \defi -\sum_{I\cap\L\ne0}\f'_I(\qq)$,
and $\P_\L(E)$ is the probability of the event $E$ with respect
to $\P_\L(\qq)=e^{-\m_0 M_\L(\qq)-U_V(\qq)}/Z_\L(\m_0,\f';\l)$,
the average of $M_V(\qq)$ is $\lis m_\L=-\dpr P_\L(\m_0)$ and:
\be
\eqalign{
\log&\P_\L\Big(M_\L>(\lis m_\L+a)\L\Big)\cr
\le&
\log\frac1{Z_\L(\m_0,\f')}\Big(\sum_\qq e^{-\m_0 M_\L(\qq)- U(\qq) +\l
M_\L(\qq)-\l (\lis m_\L+a)\L}\Big)\cr
=&{\Big(P_\L(\m_0-\l)-P_\L(\m_0)+\l (\dpr P_\L(\m_0)-a)\Big)\L}\cr}\label{eN.9}\ee
for $a>0,\l>0$. {\it Likewise} a similar bound is obtained for
$M_\L<(\lis m_\L-a)\L, a>0$.

Hence, if $\ch_\L^{-1}\defi \dpr^2 P_\L(\m_0)< c_2\le 0$
(convexity in $\m$ of $P_\L(\m)$), $\frac1{3!}|\dpr^3P_\L|<C$
hold for all $\L$, and $\lis m_\L=-\dpr_{\m_0} P_\L(\f)$
converges to $\dpr_{\m_0} P(\f)$, choosing
$a=\e=\l\le \frac1{\sqrt{C}}$:
\be \kern-3mm\eqalign{
&\P_\L(|M_\L-\lis m_\L|>\e\L)\le
e^{(P_\L(\m_0+\e)-P_\L(\m_0)-\dpr_{\m_0}P_\L(\m_0)\, \e)\L}\cr
&\log \P_\L(|M_\L-\lis m_\L|>\e\L)\le
(\frac12 \ch^{-1}_\L \e^2)\L-a^2\L,\qquad \ch_\L<0\cr}
\label{eN.10}\ee
It is important
the keep in mind that $\ch_L,C_\L,\lis m_\L$ have to be
eventually checked to converge to limits $\ch\le0,C>0,\lis m$ as
$L\to\infty$, see below: so that in Eq.\equ{eN.10} $\e=\l=a$ can
be taken provided $\e< \fra1{\sqrt C}$, for $\L$ large.

Therefore the probability of one event $|M_\L-\lis m_\L \L|\le \e\L$
has to be $\ge \e\L^{-1}$: because the total probability of the
opposite event is $\le 2 e^{-{\e^2}{|\ch_\L|}\L}<1$ for 
$\e<\frac{1}{\sqrt{C}}$ if $\L$ is large. And even the choice 
$\l=a=\e<\frac{1}{\sqrt{C}}\L^{-\d}$
with $\d>0$ fixed and small (compared to $\frac12$) leads to the
same conclusion.

The limit as $\L\to\infty$ gives a result about the large
deviations around the average $\lis m=\lim_{\L\to\infty} \lis
m_\L$. Essentially the same argument leads to the large
deviations around the value $\wt m_\L$ corresponding to
$\wt \m=\m_0+\wt \l$.

Suppose existence, and uniformity in any bounded interval containing $\m_0$,  of
the limits as $\L\to\infty$ of $\wt m_\L,\wt\ch_\L\le0,\wt\l$ and of a
bound $C\ge |\dpr^3 P_\L(\m)|$. 

Remark that if $|M_\L-\wt m_\L \L|<\e \L$ then
\be e^{-\m_0M_\L(\qq)-U(\qq)}\le
e^{-\m_0M_\L(\qq)-U(\qq) -\wt\l(M_\L-\wt
m_\L \L)+2|\wt\l| \e\L}\label{eN.11}
\ee
then $\e>0$ small, by
Eq.\equ{eN.10}:
\be\eqalign{\P_\L&(|M_\L-\wt m_\L \L|<\e \L)
\cr&
\le\frac{Z_\L(\wt \m,\f')}{Z_\L(\m_0,\f')}
\Big(\frac{\sum_\qq e^{-\wt \m M(\qq)-U(\qq)+\wt\l \wt
m_\L \L}}{Z_\L(\wt \m,\f')}\Big)\,
e^{O(\e\L |\wt\l|)}\cr
&=\frac{Z_\L(\wt \m,\f')}{Z_\L(\m_0,\f')}
e^{O((\e|\wt\l|+\e^2/|\ch_+|)\L)}
\cr
&=e^{(P_\L(\m_0+\wt\l)-\wt\l\dpr_\m P_\L(\m_0+\wt\l)
-P_\L(\m_0)) + O(\e\L)}=e^{\z(\wt m)\L+o(\L)}
\cr}\label{eN.12}
\ee
where $\z(m)=P(\m_0+\l)-\l \dpr_\m P(\m_0+\l)-P(\m_0)$ if
$m=-\dpr_\m P(\m_0+\l)$ and, prefixed $\d>0$, $\e$ is chosen
$O(\L^{-\d}),\d>0$.

Then it follows that the limit as $\L\to\infty$ of the
probability of the events $M_\L-\wt m_\L \L\in [a\L,b\L]$ is
$e^{\max_{m\in (a+\e,b-\e)\L} \z(m)\L+o(\L)}$ for $\e$ small
enough. The ``large deviations rate'' at $m$ is expressed in terms
of a $\l$ determined by $m=\dpr_\m P(\m_0+\l)$, and is:
\be \z(m)=(P(\m_0+\l)-\l\dpr_\m P(\m_0+\l)
-P(\m_0), \qquad m=\dpr_\m P(\m_0+\l)\label{eN.13}\ee
\*

\0{\bf(3)} {\it Transfer matrix.}\index{transfer matrix}
\label{Perron-Frobenius theorem}\index{Perron-Frobenius' theorem}
\*

Limits of $\wt m_\L,\wt\ch_\L$ and of a bound $C_\L$ on $|\dpr^3
P_\L(\m)|$ have been assumed in item (2) as $\m\in R$ varies in
any bounded interval. Uniform bounds, and limits in
$\L\to\infty$, on all derivatives $\dpr^k_\m P_\L(\m,\f')$ can be
derived for $\m\in R$ in a bounded interval. At the same time the
convergence, also supposed in item (2), as $\L\to \infty$ to
$\lis m,\dpr^k_\m(P(\m,\f')$, to
$\ch=\dpr^2_\m(P(\m,\f')\le0$ and $\dpr^3_\m(P(\m,\f')$,
uniformly for $\m_0$ varying in a bounded interval, has to
be checked.

Smoothness can be obtained as a consequence of the general formalism for one
dimensional Gibbs distributions in  \Cite{Ru968}.
It can be deduced starting from the classical theory of the {\it
transfer matrix} operator $\LL$: it is the ``Perron-Frobenius''
operator acting on the space $C(Q_+)$ of continuous functions on
{\it semi-infinite compatible sequences}
$\qq=\{q_0,q_1,\ldots\}$,\,\footnote{\tiny Here continuity is
with respect to the product topology: \ie
$\qq_n\tende{n\to\infty} \qq$ if $q_{n,k}\tende{n\to\infty}
q_k$ for each $k$.}  defined, in the present context, by:
\be (\LL u)(q_1,q_2,\ldots)=\sum_{q_0} e^{-m(q_0)\m_0-\sum_{0\in I
}\f'_I(\qq)} u(q_0,q_1,\ldots)\label{eN.14}\ee
The operator $\LL$ iterated $\ge n_0$ times has a strictly
 positive kernel (by the transitivity of the compatibility
 matrix): it “immediately” follows (i.e. it follows from well
 known results on the theory of operators, or better of matrices,
 like the Perron-Frobenius theorem, see \Cite{Ru968}), that the
 largest eigenvalue $\g(\m,\f')$ is isolated, positive, simple
 and smooth as a function of $\m$ for $\m\in (-\infty,\infty)$.

A key, from which the properties needed in the above
 item (2) follow, can be seen by looking at the iterates $\LL^L
 u$ acting on the function $u(\qq)\equiv1$: and it results that
 $\frac{\LL^L 1}{Z_\L(\f)}= e^{O(1)}$ and means that the
 logarithm of a positive eigenvalue $\g(\m,\f')$ of $\LL$ should be
 $\log \g(\m,\f')=P(\m,\f')$, defined in \equ{eN.6}.

Uniform (in $\L$) smoothness of $P_\L(\m,\f')$, convergence to
 $P(\m_0,\f')$, strict convexity as function of $\m\in R$ require
 a more refined analysis.  The analysis is ``classical'' since the
 work \Cite{Ru968}, and will not be repeated: here  only
 will be reported the list of the main steps. Details on the
 proof developed carefully in similar problems on transfer
 operators can be found, for instance, also in
 \Cite{Ru968},\Cite{GM970},\Cite{CO981},\Cite{JP998}\,. A few
 hints follow.
\*

\0i) The  adjoint $\LL^*$  operates on the space of the measures  on $Q_+$,
\be(\LL^*\r)(f)=\int\r(d\qq)\sum_{q_0}e^{-\m_0(q_0)-U(q_0\qq)} 
f(q_0\qq)=
\r(\LL f)\label{eN.15}\ee
The map
$\r\to \frac{\LL^*\r}{|\LL^*\r(1)}$ maps
continuously\,\footnote{\tiny In the ``weak'' convergence of
measures on $Q_+$, in which $\r_n\to\r$ if $\r_n(f)\to\r(f)$ for
$f\in C(Q_+)$.}\, into itself the compact convex set $E$ of
positive measures $\r$ with $\r(1)=1$ (hence 
probability distributions $\r$).

Hence the map has eigenvector $\n$ and with positive eigenvalue
(by the fixed point theorem): the first step in the theory is to
prove $\n$ is the unique positive eigenvector. It is then proved
that also $\LL$ has an eigenvector $h$ positive and both
$\LL,\LL^*$ have the same eigenvalue $\g(\m_0,\f')\equiv
P(\m_0,\f')$ continuously depending on
$\m_0$. At an intermediate step the eigenvalue $\g$ of $\LL,\LL^*$ is
proved to be isolated, allowing to define $(\LL-\g)^{-1}$ on the
functions $f$ with $\n(f)=0$.

\0ii) Existence of $\n,\g,h$ and smoothness of $\g$ implies
smoothness of derivatives of $\g,h$ of any order: this is
suggested by the possibility of deriving formal expressions of
the higher $\m$-derivatives by simple algebra. For istance to
obtain smoothness of first order derivatives of $\g$ and $h$
proceed as follows:
\be\LL h=\g h  \tto \dpr \LL\, h-\dpr\g\,h+ (\LL-\g)\dpr h=0\label{eN.16}\ee
integrate with respect to $\n$ and use that $\g$ is eigenvalue of
$\LL,\LL^*$ and $\n(h)=1$, to obtain $\n(\dpr\LL\,h)=\dpr\g$ so
the smoothness of $\LL$ and $h$ implies that of $\dpr\g$.

It is then shown that $(\LL-\g)^{-1}$ is a bounded operator on
the space of the functions with $\n(f)=0$. The derivative of $h$
follows from Eq.{eN.16} as $\dpr h=-(\LL-\g)^{-1}(\dpr \LL\,
h-\dpr\g\,h)$: here $(\LL-\g)^{-1}$ is defined just because the
definition of $\dpr \g$ implies $\n(\dpr \LL\, h-\dpr\g\,h)=0$.
\*

\0iii) In the same way formal expressions can be derived recursively
for higher derivatives by differentiating more times $\LL h-\g
h=0$: $0=(\dpr^j(\LL-\g)h)=\sum_{k=0}^{j-1} {j\choose k}
(\dpr^{k-j}\LL-\dpr^{k-j}\g)\dpr^k h+(\LL-\g)\dpr^j h$;
integrating the results with respect to $\n$ an expression for
$\dpr^j \g$ is obtained, while applying $(\LL-\g)$ the same
relation gives $\dpr^j h$.

\*

The conclusion is that, for $\L$ large, the $\z(m)$ is smooth for
$m\in (m_-,m_+)$ where $m_\pm=\lim_{\m\to\pm\infty}\dpr_\m
P(\m,\f')$ and $m_-=\min m(q),m_+=\max m(q)$.
\*

\IfFileExists{appNN.exist}{\include appNN}{}

\section{Appendix: An exact formula}
\def\SEC{Appendix: An exact formula}
\label{appO}\iniz
\lhead{\small\ref{appO}.\ \SEC}

An immediate consequence of the fluctuation
theorem\index{fluctuation theorem}, recalling that in the
thermostatted models of Ch.\ref{Ch2} the dissipation per unit
time $\e(x)$ and the phase space contraction $\s(x)$ integrated
over a time $t$ give the same result up to a correction bounded
time independently, is:

\be\media{ e^{-\ig_0^\t \e(S_tx)\,dt}}_{SRB}=e^{O(1)}
\label{eO.1}\ee
{\it i.e.\ } $\media{ e^{-\ig_0^\t \e(S_tx)\,dt}}_{SRB}$ stays
bounded as $\t\to\io$. It follows simply from the fluctuation
theorem, Eq.\equ{e4.6.6}, relation between the probability
densities $\p_t(-p)=\p_t(p) e^{-t p\e_+ +O(1)}$: integrating over
$p$ both sides the Eq.\equ{eO.1} follows (private communication
from F.Bonetto, \Cite{Ga998b}[Eq.(16)])\,,
see \Cite{Ga000}[Eq.(9.10.4)]\,.

This relation bears resemblance to {\it Jarzynski's
  formula},\Cite{Ja997}, \index{Jarzinsky's formula} which deals
  with a canonical Gibbs distribution (in a finite volume)
  corresponding to a Hamiltonian $H_0(p,q)$ and temperature
  $T=(k_B
\b)^{-1}$, and with a time dependent family of Hamiltonians
$H(p,q,t)$ which interpolates between $H_0$ and a second
Hamiltonian $H_1$ as $t$ grows from $0$ to $1$ (in suitable
units) which is called {\it a protocol}.\index{protocol}

Imagine to extract samples $(p,q)$ with a canonical probability
distribution $\m_0(dpdq)= Z_0^{-1}e^{-\b H_0(p,q)}dpdq$, with
$Z_0$ being the canonical partition function, and let
$S^{0,t}(p,q)$ be the solution of the Hamiltonian {\it time
dependent} equations $\dot p=-\dpr_q H(p,q,t),\dot q=\dpr_p
H(p,q,t)$ for $0\le t\le1$. Then
\Cite{Ja997}\Cite{Ja999}, establish an identity as follows.
\*

Let $(p',q')\defi S^{0,1}(p,q)$ and let $W(p',q')\defi
H_1(p',q')-H_0(p,q)$, then the distribution $Z_1^{-1} e^{-\b
H_1(p',q')}dp'dq'$ is exactly equal to $\frac{Z_0}{Z_1} e^{-\b
W(p',q')}\m_0(dp dq)$. Hence
\be\eqalign{
&\media{e^{-\b W}}_{\m_0}=\frac{Z_1}{Z_0}=e^{-\b \D F(\b)}
\quad{\rm or\ equivalently}\cr
&\media{e^{\b(\D F-W)}}=1\cr}
\label{eO.2}\ee
where the average is with respect to the Gibbs distribution
$\m_0$ and $\D F$ is the free energy variation between the
equilibrium states with Hamiltonians $H_1$ and $H_0$
respectively.
\*

\0{\it Remarks:} (i) The reader will recognize in this {\it exact identity}
an instance of the Monte Carlo method\index{Monte Carlo method}
(analogically implemented rather than in a simulation). Its
interest lies in the fact that it can be implemented {\it without
actually knowing} neither $H_0$ nor $H_1$ nor the {\it protocol}
$H(p,q,t)$. It has to be stressed that the protocol, \ie the
process of varying the Hamiltonian, has an arbitrarily prefixed
duration which has {\it nothing to do} with the time that the
system will need to reach the equilibrium state with Hamiltonian
$H_1$ of which we want to evaluate the free energy variation.
\\
(ii) If one wants to evaluate the difference in free energy
between two equilibrium states at the same temperature in a
system that one can construct in a laboratory then ``all one has
to do'' is \*

(a) Fix a protocol, {\it i.e.\ } a procedure to transform the
forces acting on the system along a well defined {\it fixed once
and for all} path from the initial values to the final values in
a fixed time interval ($t=1$ in some units), and

(b) Measure the energy variation $W$ generated by the
``machines'' implementing the protocol. This is a really
measurable quantity at least in the cases in which $W$ can be
interpreted as work done on the system, or related to it.

(c) Then average of the exponential of $-\b W$ with respect to a
large number of repetition of the protocol. This can be useful
even, and perhaps mainly, in biological experiments.
\*

\0(iii) If the ``protocol'' conserves energy (like a Joule expansion of a
gas) or if the difference $W=H_1(p',q')-H_0(p,q)$ has zero
average in the equilibrium state $\m_0$ we get, by Jensen's
inequality ({\it i.e.\ } by the convexity of the exponential
function: $\media{e^A}\ge e^{\media A}$), that $\D F\le0$ as
expected from Thermodynamics.
\\
(iv) The measurability of $W$ is a difficult question, to be
discussed on a case by case basis. It is often possible to
identify it with the ``work done by the machines implementing the
protocol''.
\*

The two formulae Eq.\equ{eO.1} and Eq.\equ{eO.2} are however very
different:
\*

\0(1) the $\ig_0^\t \s(S_tx)\, dt$ is an entropy production in a non
equilibrium stationary state rather than $\D F-W$ in a {\it not
  stationary process} lasting a prefixed time {\ie two completely
  different situations}.  In Sec.(\,\ref{sec:VIII-4}\,) the
  relation between Eq.\equ{eO.1} and the Green-Kubo formula is
  discussed.

\0(2) the average is over the SRB distribution of a stationary state, in
    general out of equilibrium, rather than on a canonical
    equilibrium state.

\0(3) the Eq.\equ{eO.1}, says that $\media{ e^{-\ig_0^\t
    \e(S_tx)\,dt}}_{SRB}$ is bounded as $\t\to\io$ rather than
being $1$ exactly, \Cite{Ja999}.  \*

The Eq.\equ{eO.2} has proved useful in various equilibrium
problems (to evaluate the free energy variation when an
equilibrium state with Hamiltonian $H_0$ is compared to one with
Hamiltonian $H_1$); hence it has some interest to investigate
whether Eq.\equ{eO.2} can have some consequences.

If a system is in a steady state and produces entropy at rate
  $\e_+$ ({\it e.g.\ } a living organism feeding on a background)
  the fluctuation theorem Eq.\equ{e4.6.1} and its consequence,
  Eq.\equ{eO.2}, gives us informations on the fluctuations of
  entropy production, {\it i.e.\ } of heat produced, and {\it
  could be useful}, for instance, to check that all relevant heat
  transfers have been properly taken into account.

\section{Appendix: Transient FT}
\def\SEC{Appendix: Transient FT}
\label{appQ}\iniz
\lhead{\small\ref{appQ}.\ \SEC}\index{Evans-Searles' formula}

It has been remarked that time reversal $I$ puts some constraints
on fluctuations in systems that evolve {\it towards non
equilibrium} starting {\it from an equilibrium state} $\m_0$ or,
more generally from a state $\m_0$ which is proportional to the
volume measure on phase space and $\m_0(IE)\equiv\m_0(E)$ (but
not necessarily stationary).

For instance if the equations of motion are $\dot x=f(x)$ and
$-\s(x)=$ divergence of $f$, \ie $\s(x)=-\dpr\cdot f(x)$, where
$\dot x=f_0(x)+E g(x)$ with $\dot x=f_0(x)$ a volume preserving
evolution and $E$ a parameter. It is supposed that
$\s(Ix)=-\s(x)$ and $\m_0(I E)\equiv
\m_0(E)$.  Then one could pose the question, \Cite{ES994}, \*

``{\it Which is the probability that in time $t$ the volume
  contracts by the amount $e^A$ with $A=\int_0^t \s(S_t x)dt$,
  compared to that of the opposite event $-A$\em?}\kern1mm'' \*
If $\EE_A$ = set of points whose neighborhoods contract with
contraction $A$ in time $t$, then the set $\EE_A$ at time $t$
becomes (by definition) the set $S_t\EE_A$ with
$\m_0(S_t\EE_A)=e^{-A}\m_0(\EE_A)$, $A=\int_0^t \s(S_\t x)d\t$.

However $\EE^-_A\defi I S_t\EE_A$ is the set of points $\EE_{-A}$
which contract by $-A$ as:

\be\eqalign{
z&e^{-\int_0^\t \s(S_\t I S_tx)d\t}\equiv e^{-\int_0^t \s(S_\t
  S^{-t} I x)d\t}\equiv e^{-\int_0^t \s(I S^{-\t} S^{t} x)d\t}\cr
  &\equiv e^{+\int_0^\t \s(S^{t-\t} x)d\t}\equiv
  e^{+\int_0^\t \s(S^{\t} x)d\t}\equiv e^{A}\cr}\label{eQ.1}\ee
In other words the set $\EE_A$ of points which contract by $A$ in
time $t$ becomes the set of points whose time reversed images is
the set $\EE^-_A\defi I S_t\EE_A$ which expand by $A$. The
measures of such sets are $\m_0(\EE_A)$ and $\m_0(I
S_t\EE_A)\equiv\m_0(\EE_A)e^{-A}\equiv
\m_0(\EE^-_A)$ (recall that $I$ is measure preserving), hence
\be\frac{\m_0(\EE_A)}{\m_0(\EE^-_A)}\equiv e^A\label{eQ.2}\ee
for any $A$ (as long as it is
  ``possible'', \Cite{ES994}).\footnote{\tiny For instance in the
  Hamiltonian case $A\ne0$ would be impossible.}
\*

This\index{transient fluctuation theorem} has been called ``{\it
transient fluctuation theorem}''. It is extremely general and
does not depend on any chaoticity assumption. Just reversibility
and time reversal symmetry and the evolution of an initial
distribution $\m_0$ which is invariant under time reversal. It
says nothing about the SRB distribution (which is singular with
respect to the Liouville distribution).

For instance if a reversible system has an attractive surface
$\AA$ which has dimension {\it less} than that of phase space,
hence $I\AA\cap \AA=\emptyset$ the transient theorem hold but the
fluctiation theorem does not as it describes a property of the
surface of $\AA$ which, without extra assumptions, is of
difficult access and certainly different from the total phase
space volume.

Some claims that occasionally can be found in the literature that
the above relation is equivalent to the fluctuation theorem rely
on further assumptions.

The similarity with the conceptually completely different
expression of the fluctuation theorem Eq.\equ{e4.6.6} explain,
perhaps, why this is very often confused with the fluctuation
theorem, \Cite{GC004}.

{\it It is easy to exhibit examples of time reversible maps or
  flows, with many Lyapunov exponents, positive and negative, for
  which the transient fluctuation theorem holds but the
  fluctuation relation fails because the CH fails (\ie the
  fluctuation theorem cannot be applied).}

Relations of the kind of the transient fluctuation theorem have
appeared in the literature quite early in the development of non
equilibrium theories, perhaps the first have
been \Cite{BK981a}\Cite{BK981b}.



\def\SEC{References}
\lhead{\small\SEC}
\label{references}
\iniz

{\small Refences to Boltzmann quote the original paper and its
number in his Wisseshaften Abhandlungen}

\bibliographystyle{unsrt}
\bibliography{0Bib}

\begin{thebibliography}{100}

\bibitem{Bo909}
L.~Boltzmann.
\newblock {\em {W}is\-sen\-schaft\-li\-che {A}bhandlungen, ed. {F}.
  {H}asen\-{\"o}hrl}, volume 1,2,3
  (http://www.esi.ac.at/further/Boltzmann\_online.html).
\newblock Barth, Leipzig, 1909.

\bibitem{Ga989}
G.~Gallavotti.
\newblock {L' hypoth\`ese ergodique et Boltzmann}.
\newblock {\em {In ``Dictionnaire Philosophique'' (ed. K. Chemla), Presses
  Universitaires, Paris}}, pages 1081--1086, 1989.

\bibitem{Ga995}
G.~Gallavotti.
\newblock {\em Trattatello di Meccanica Statistica}, volume~50.
\newblock Quaderni del CNR-GNFM, Firenze, 1995.

\bibitem{Ga995a}
G.~Gallavotti.
\newblock Ergodicity, ensembles, irreversibility in {B}oltzmann and beyond.
\newblock {\em Journal of Statistical Physics}, 78:1571--1589, 1995.

\bibitem{Ga000}
G.~Gallavotti.
\newblock {\em Statistical Mechanics. A short treatise}.
\newblock Springer Verlag, Berlin, 2000.

\bibitem{Ga005a}
G.~Gallavotti.
\newblock Equilibrium statistical mechanics.
\newblock {\em Encyclopedia of Mathematical Physics, ed. J.P. Fran{\c{c}}oise,
  G.L. Naber, Tsou Sheung Tsun}, 1:51--87, 2006.

\bibitem{BDGJL01}
L.~Bertini, A.~De Sole, D.~Gabrielli, G.~Jona-Lasinio, and C.~Landim.
\newblock Fluctuations in stationary nonequilibrium states of irreversible
  processes.
\newblock {\em Physical Review Letters}, 87:040601, 2001.

\bibitem{DLS002}
B.~Derrida, J.~L. Lebowitz, and E.~R. Speer.
\newblock Exact free energy functional for a driven diffusive open stationary
  nonequilibrium system.
\newblock {\em Physical Review Letters}, 89:030601, 2002.

\bibitem{GDL010}
A.~Gerschenfeld, B.~Derrida, and J.~L. Lebowitz.
\newblock {Anomalous Fourier’s law and long range correlations in a 1D
  non-momentum conserving mechanical model}.
\newblock {\em Journal of Statistical Physics}, 141:757--766, 2010.

\bibitem{BK013}
J.~Bricmont and A.~Kupiainen.
\newblock Diffusion in energy conserving coupled maps.
\newblock {\em Communications in Mathematical Physics}, 321:311--369, 2013.

\bibitem{FV963}
R.P. Feynman and F.L. Vernon.
\newblock The theory of a general quantum system interacting with a linear
  dissipative system.
\newblock {\em Annals of Physics}, 24:118--173, 1963.

\bibitem{BGGZ005}
F.~Bonetto, G.~Gallavotti, A.~Giuliani, and F.~Zamponi.
\newblock Chaotic {H}ypothesis, {F}luctuation {T}heorem and {S}ingularities.
\newblock {\em Journal of Statistical Physics}, 123:39--54, 2006.

\bibitem{Lo963}
E.~Lorenz.
\newblock Deterministic non periodic flow.
\newblock {\em Journal of the Atmospheric Science}, 20:130--141, 1963.

\bibitem{Lu-050}
T.~Lucretius.
\newblock {\em {De Rerum Natura}}.
\newblock Rizzoli, Milano, 1976.

\bibitem{Br003}
S.G. Brush.
\newblock {\em History of modern physical sciences, Vol.I: The kinetic theory
  of gases}.
\newblock Imperial College Press, London, 2003.

\bibitem{Ne723}
I.~Newton.
\newblock {\em {Philosophiae Naturalis Principia Methematica}}.
\newblock Royal Society, Amsterdam, 1723.

\bibitem{La821}
P.S. Laplace.
\newblock Sur l'attraction des corps sphériques et sur la répulsion des
  fluides élastiques.
\newblock {\em Annales de Chimie}, 18:181--189, 1821.

\bibitem{Av811}
A.~Avogadro.
\newblock Essai d'une mani{\`e}re de determiner les masses relatives des
  molecules {\'e}l{\'e}mentaires des corps, et les proportions selon lesquelles
  elles entrent dans ces combinaisons.
\newblock {\em Journal de Physique, de Chimie et d'Histoire naturelle, {\rm
  translated} in lem.ch.unito.it/chemistry/essai.html}, 73:58--76, 1811.

\bibitem{Br976}
S.G. Brush.
\newblock {\em The kind of motion that we call heat, (I, II)}.
\newblock North Holland, Amsterdam, 1976.

\bibitem{Mo836}
O.F. Mossotti.
\newblock {\em Sur les forces qui r\'egissent la constitution int\'erieure des
  corps}.
\newblock Stamperia Reale, Torino, 1836.

\bibitem{Cl850}
R.~Clausius.
\newblock On the motive power of heat, and on the laws which can be deduced
  from it for the theory of heat.
\newblock {\em Philosophical Magazine}, 2:1--102, 1851.

\bibitem{Ca824}
S.~Carnot.
\newblock {\em {R\'eflections sur la puissance motrice du feu et sur les
  machines propres \`a d\'evelopper cette puissance}}.
\newblock {Online in https://gallica.bnf.fr; original Bachelier, 1824;
  reprinted Gabay, 1990.}, Paris, 1824.

\bibitem{Cl865}
R.~Clausius.
\newblock {{\"U}ber einige f{\"u}r Anwendung bequeme formen der
  Hauptgleichungen der mechanischen W{\"a}rme\-theorie}.
\newblock {\em {Annalen der Physik und Chemie}}, 125:353--401, 1865.

\bibitem{LS968}
H.G. Lidddell and R.~Scott.
\newblock {\em {A Greek-English Lexicon}}.
\newblock Oxford, Oxford, 1968.

\bibitem{Kr856}
A.~Kr{\"o}nig.
\newblock {Grundz{\"u}ge einer Theorie der Gase}.
\newblock {\em Annalen der Physik und Chemie}, XCIX:315--322, 1856.

\bibitem{Cl857}
R.~Clausius.
\newblock The nature of the motion which we call heat.
\newblock {\em Philosophical Magazine}, 14:108--127, 1865.

\bibitem{Ma860-a}
J.C. Maxwell.
\newblock Illustrations of the dynamical theory of gases.
\newblock {\em In: The Scientific Papers of {J.C. M}axwell, {Cambridge
  University Press}, Ed. {W.D. Niven}, Vol.1}, pages 377--409, 1860.

\bibitem{Ma890t}
J.C. Maxwell.
\newblock {\em The Scientific Papers of {J.C. M}axwell}.
\newblock {Cambridge University Press}, Ed. {W.D. Niven}, Vol.1,2, Cambridge,
  1964.

\bibitem{Bo866}
L.~Boltzmann.
\newblock {\"U}ber die mechanische {B}edeutung des zweiten {H}auptsatzes der
  {W\"a}rme\-theorie.
\newblock {\em Wiener Berichte}, 53, (W.A.,\#2):195--220, (9--33), 1866.

\bibitem{Bo974}
L.~Boltzmann.
\newblock {\em Theoretical Physics and philosophical writings, ed. B. Mc
  Guinness}.
\newblock Reidel, Dordrecht, 1974.

\bibitem{Ma879}
J.~C. Maxwell.
\newblock On {B}oltzmann's theorem on the average distribution of energy in a
  system of material points.
\newblock {\em Transactions of the Cambridge Philosophical Society},
  12:547--575, 1879.

\bibitem{Bo872}
L.~Boltzmann.
\newblock Weitere {S}tudien {\"u}ber das {W\"a}rmegleich\-gewicht unter
  {G}asmolek{\"u}len.
\newblock {\em Wiener Berichte}, 63, (W.A.,\#22):275--370, (316--402), 1872.

\bibitem{Ma867-b}
J.C. Maxwell.
\newblock On the dynamical theory of gases.
\newblock {\em In: The Scientific Papers of {J.C. M}axwell, {Cambridge
  University Press}, Ed. {W.D. Niven}, Vol.2}, pages 26--78, 1866.

\bibitem{Cl871}
R.~Clausius.
\newblock {Ueber die Zur{\"u}ckf{\"u}hrung des zweites Hauptsatzes der
  mechanischen W{\"a}rmetheorie und allgemeine mechanische Prinzipien}.
\newblock {\em Annalen der Physik}, 142:433--461, 1871.

\bibitem{Bo871-0}
L.~Boltzmann.
\newblock Zur priorit{\"a}t der auffindung der beziehung zwischen dem zweiten
  hauptsatze der mechanischen w{\"a}rmetheorie und dem prinzip der kleinsten
  wirkung.
\newblock {\em Poggendorf Annalen}, 143, (W.A.,\#17):211--230, (228--236),
  1871.

\bibitem{Cl872}
R.~Clausius.
\newblock Bemerkungen zu der priorit{\"a}t reclamation des hrn. boltzmann.
\newblock {\em Annalen der Physik}, 144:265--280, 1872.

\bibitem{Bo871-a}
L.~Boltzmann.
\newblock {\"U}ber das {W\"a}rmegleichgewicht zwischen mehratomigen
  {G}asmolek{\"u}len.
\newblock {\em Wiener Berichte}, 68, (W.A.,\#18:397--418, (237--258), 1871.

\bibitem{Bo871-b}
L.~Boltzmann.
\newblock Einige allgemeine s{\"a}tze {\"u}ber {W\"a}rme\-gleichgewicht.
\newblock {\em Wiener Berichte}, 63, (W.A.,\#19):679--711, (259--287), 1871.

\bibitem{Bo871-c}
L.~Boltzmann.
\newblock {A}nalytischer {B}eweis des zweiten {H}auptsatzes der mechanischen
  {W\"a}rme\-theorie aus den {S\"a}tzen {\"u}ber das {G}leichgewicht des
  lebendigen {K}raft.
\newblock {\em Wiener Berichte}, 63, (W.A.,\#20):712--732,(288--308), 1871.

\bibitem{Bo877-b}
L.~Boltzmann.
\newblock {\"U}ber die {B}eziehung zwischen dem zwei\-ten {H}aupt\-satze der
  mechanischen {W}{\"a}rmetheo\-rie und der {W}ahrscheinlichkeitsrechnung,
  respektive den {S}{\"a}tz\-en {\"u}ber das {W}{\"a}rme\-gleichgewicht.
\newblock {\em Wiener Berichte}, 76, (W.A.,\#42):373--435, (164--223), 1877.

\bibitem{Bo884}
L.~Boltzmann.
\newblock {\"U}ber die {E}igenshaften monozyklischer und anderer damit
  verwandter {S}ysteme.
\newblock {\em Crelles Journal}, 98, (W.A.,\#73):68--94, (122--152), 1884.

\bibitem{La867}
J.L. Lagrange.
\newblock {\em Oeuvres}.
\newblock Gauthiers-Villars, Paris, 1867-1892.

\bibitem{Bo868-a}
L.~Boltzmann.
\newblock Studien {\"u}ber das gleichgewicht der le\-bendigen kraft zwischen
  bewegten materiellen punkten.
\newblock {\em Wiener Berichte}, 58, (W.A.,\#5):517--560, (49--96), 1868.

\bibitem{Bo877a}
L.~Boltzmann.
\newblock Bemerkungen {\"u}ber einige probleme der mechanischen
  {W}{\"a}rme\-theo\-rie.
\newblock {\em Wiener Berichte}, 75, (W.A.,\#39):62--100, (112--148), 1877.

\bibitem{Ga001b}
G.~Gallavotti.
\newblock Quasi periodic motions from {H}ypparchus to {K}olmogorov.
\newblock {\em Rendiconti Accademia dei Lincei, Matematica e applicazioni},
  12:125--152, 2001.

\bibitem{Ba990}
A.~Bach.
\newblock {Boltzmann's} probability distribution of 1877.
\newblock {\em Archive for the History of exact Sciences}, 41:1--40, 1990.

\bibitem{Bo868-b}
L.~Boltzmann.
\newblock L{\"o}sung eines mechanischen problems.
\newblock {\em Wiener Berichte}, 58, (W.A.,\#6):1035--1044, (97--105), 1868.

\bibitem{Gi902}
J.~Gibbs.
\newblock {\em Elementary principles in statistical mechanics}.
\newblock Schribner, Cambridge, 1902.

\bibitem{Bo871-d}
L.~Boltzmann.
\newblock {Z}usammenhang zwischen den {S\"a}tzen {\"u}ber das {V}erhalten
  mehratomiger {G}asmolek{\"u}le mit {J}acobi's {P}rinzip des letzten
  {M}ultiplicators.
\newblock {\em Wiener Berichte}, 63, (W.A.,\#19):679--711, (259--287), 1871.

\bibitem{GBG004}
G.~Gallavotti, F.~Bonetto, and G.~Gentile.
\newblock {\em Aspects of the ergodic, qualitative and statistical theory of
  motion}.
\newblock Springer Verlag, Berlin, 2004.

\bibitem{La974}
O.~Lanford.
\newblock Time evolution of large classical systems.
\newblock {\em Dynamical systems, theory and applications, Lecture Notes in
  Physics, ed. J. Moser}, 38:1--111, 1974.

\bibitem{Sp006}
H.~Spohn.
\newblock On the integrated form of the {BBGKY} hierarchy for hard spheres.
\newblock {\em arxiv: math-ph/0605068}, pages 1--19, 2006.

\bibitem{Th874}
W.~Thomson.
\newblock The kinetic theory of dissipation of energy.
\newblock {\em Proceedings of the Royal Society of Edinburgh}, 8:325--328,
  1874.

\bibitem{Sh024}
J.W. Shim.
\newblock A commented translation of {B}oltzmann’s work, {\it ueber die
  sogenannte h-curve}.
\newblock {\em European Physics Journal H}, 49:21, 2024.

\bibitem{Bo898}
L.~Boltzmann.
\newblock Ueber die sogenannte h-curve.
\newblock {\em Mathematische Annalen, and W.A. \#128}, 50:325--332, 1898.

\bibitem{Bo896a}
L.~Boltzmann.
\newblock {\em {L}ectures on gas theory, English edition annotated by S.
  Brush}.
\newblock University of California Press, Berkeley, 1964.

\bibitem{GL003}
S.~Goldstein and J.~L. Lebowitz.
\newblock On the ({B}oltzmann) entropy of nonequilibrium systems.
\newblock {\em Physica D}, 193:53--66, 2004.

\bibitem{GGL005}
P.~L. Garrido, S.~Goldstein, and J.~L. Lebowitz.
\newblock Boltzmann entropy for dense fluids not in local equilibrium.
\newblock {\em Physical Review Letters}, 92:050602 (+4), 2004.

\bibitem{Ga001}
G.~Gallavotti.
\newblock Counting phase space cells in statistical mechanics.
\newblock {\em Communication in Mathematical Physics}, 224:107--112, 2001.

\bibitem{Fe963}
R.P. Feynman, R.B. Leighton, and M.~Sands.
\newblock {\em The Feynman lectures in Physics, Vol. I, II, III}.
\newblock Addison-Wesley, New York, 1963.

\bibitem{Ze968}
M.~W. Zemansky.
\newblock {\em Heat and thermodynamics}.
\newblock McGraw-Hill, New-York, 1957.

\bibitem{Uf008}
J.~Uffink.
\newblock Boltzmann's work in statistical physics.
\newblock In Edward~N. Zalta, editor, {\em The Stanford Encyclopedia of
  Philosophy}. The Stanford Encyclopedia of Philosophy, winter 2008 edition,
  2008.

\bibitem{Da018}
O.~Darrigol.
\newblock {\em {Atoms, Mechanics, and Probability}}.
\newblock Oxford University Press, New York, 2018.

\bibitem{Ga996}
G.~Gallavotti.
\newblock Chaotic hypothesis: {O}nsager reciprocity and
  fluctuation--dissi\-pation theorem.
\newblock {\em Journal of Statistical Physics}, 84:899--926, 1996.

\bibitem{No984}
S.~Nos\'e.
\newblock A unified formulation of the constant temperature molecular dynamics
  methods.
\newblock {\em Journal of Chemical Physics}, 81:511--519, 1984.

\bibitem{Ho985}
W.~Hoover.
\newblock Canonical equilibrium phase-space distributions.
\newblock {\em Physical Review A}, 31:1695--1697, 1985.

\bibitem{EM990}
D.~J. Evans and G.~P. Morriss.
\newblock {\em Statistical Mechanics of Non{\-}equilibrium Fluids}.
\newblock Academic Press, New-York, 1990.

\bibitem{WSE004}
S.R. Williams, D.J. Searles, and D.J. Evans.
\newblock Independence of the transient fluctuation theorem to thermostatting
  details.
\newblock {\em Physical Review E}, 70:066113 (+6), 2004.

\bibitem{Ch996}
Ph. Choquard.
\newblock Lagrangian formulation of nos{\'e}-hoover and of isokinetic dynamics.
\newblock {\em www.esi.ac.at}, 1996.

\bibitem{Ga006c}
G.~Gallavotti.
\newblock {Entropy, Thermostats and Chaotic Hypothesis}.
\newblock {\em Chaos}, 16:043114 (+6), 2006.

\bibitem{Be964}
R.~Becker.
\newblock {\em Electromagnetic fields and interactions}.
\newblock Blaisdell, New-York, 1964.

\bibitem{CELS993}
N.~I. Chernov, G.~L. Eyink, J.~L. Lebowitz, and {Ya.}~G. Sinai.
\newblock Steady state electric conductivity in the periodic {L}orentz gas.
\newblock {\em Communications in Mathematical Physics}, 154:569--601, 1993.

\bibitem{GG007}
P.~Garrido and G.~Gallavotti.
\newblock Boundary dissipation in a driven hard disk system.
\newblock {\em Journal of Statistical Physics}, 126:1201--1207, 2007.

\bibitem{ER985}
J.~P. Eckmann and D.~Ruelle.
\newblock Ergodic theory of chaos and strange attractors.
\newblock {\em Reviews of Modern Physics}, 57:617--656, 1985.

\bibitem{Ru999}
D.~Ruelle.
\newblock Smooth dynamics and new theoretical ideas in non-equilibrium
  statistical mechanics.
\newblock {\em Journal of Statistical Physics}, 95:393--468, 1999.

\bibitem{Ru978b}
D.~Ruelle.
\newblock What are the measures describing turbulence.
\newblock {\em Progress in Theoretical Physics Supplement}, 64:339--345, 1978.

\bibitem{Ru980}
D.~Ruelle.
\newblock Measures describing a turbulent flow.
\newblock {\em Annals of the New York Academy of Sciences}, 357:1--9, 1980.

\bibitem{GP010b}
G.~Gallavotti and E.~Presutti.
\newblock Thermodynamic limit for isokinetic thermostats.
\newblock {\em Journal of Mathematical Physics}, 51:0353303 (+9), 2010.

\bibitem{ES993}
D.~J. Evans and S.~Sarman.
\newblock Equivalence of thermostatted nonlinear responses.
\newblock {\em Physical Review E}, 48:65--70, 1993.

\bibitem{GP010a}
G.~Gallavotti and E.~Presutti.
\newblock Fritionless thermostats and intensive constants of motion.
\newblock {\em Journal of Statistical Physics}, 139:618--629, 2010.

\bibitem{Si968a}
{Ya.}~G. Sinai.
\newblock Markov partitions and {$C$}-diffeomorphisms.
\newblock {\em Functional Analysis and its Applications}, 2(1):64--89, 1968.

\bibitem{Bo970a}
R.~Bowen.
\newblock Markov partitions for axiom {A} diffeomorphisms.
\newblock {\em American Journal of Mathematics}, 92:725--747, 1970.

\bibitem{BR975}
R.~Bowen and D.~Ruelle.
\newblock The ergodic theory of axiom {A} flows.
\newblock {\em Inventiones Mathematicae}, 29:181--205, 1975.

\bibitem{Bo970}
R.~Bowen.
\newblock Markov partitions and minimal sets for axiom {A} diffeomorphisms.
\newblock {\em American Journal of Mathematics}, 92:907--918, 1970.

\bibitem{Ru976}
D.~Ruelle.
\newblock A measure associated with axiom {A} attractors.
\newblock {\em American Journal of Mathematics}, 98:619--654, 1976.

\bibitem{GC995}
G.~Gallavotti and D.~Cohen.
\newblock {Dynamical ensembles in nonequilibrium statistical mechanics}.
\newblock {\em Physical Review Letters}, 74:2694--2697, 1995.

\bibitem{Ru995}
D.~Ruelle.
\newblock {\em Turbulence, strange attractors and chaos}.
\newblock World Scientific, New-York, 1995.

\bibitem{Ru996}
D.~Ruelle.
\newblock Positivity of entropy production in nonequilibrium statistical
  mechanics.
\newblock {\em Journal of Statistical Physics}, 85:1--25, 1996.

\bibitem{LS024}
D.~Levesque and N.~Sourlas.
\newblock {Time irreversibility in Statistical Mechanics}.
\newblock {\em arXiv:2402.12910v3}, pages 1--3, 2024.

\bibitem{LV993}
D.~Levesque and L.~Verlet.
\newblock Molecular dynamics and time reversibility.
\newblock {\em Journal of Statistical Physics}, 72:519--537, 1993.

\bibitem{Le993}
J.~L. Lebowitz.
\newblock Boltzmann's entropy and time's arrow.
\newblock {\em Physics Today}, {\rm September}:32--38, 1993.

\bibitem{AS967}
D.V. Anosov and Y.G. Sinai.
\newblock Some smooth ergodic systems.
\newblock {\em Russian Mathematical Surveys}, 22:103--167, 1967.

\bibitem{KH997}
A.~Katok and B.~Hasselblatt.
\newblock {\em Introduction to the modern theory of dynamical systems},
  volume~54 of {\em Encyclopedia of Mathematics and its applications}.
\newblock Cambriidge University Press, Cambridge, 1997.

\bibitem{Ru989b}
D.~Ruelle.
\newblock {\em Elements of differentiable dynamics and bifurcation theory}.
\newblock Academic Press, New-York, 1989.

\bibitem{Si968b}
{Ya.}~G. Sinai.
\newblock Construction of {M}arkov partitions.
\newblock {\em Functional Analysis and its Applications}, 2(3):70--80, 1968.

\bibitem{Bo975}
R.~Bowen.
\newblock {\em Equilibrium states and the ergodic theory of Anosov
  diffeormorphisms}, volume 470 of {\em Lecture Notes in Mathematics}.
\newblock Springer-Verlag, Berlin-Heidelberg, 1975.

\bibitem{HK995}
A.~Katok and B.~Hasselblatt.
\newblock {\em Introductionto the modern theory of dynamical systems},
  volume~54 of {\em {Encyclopedia of Mathematics and its Applications}}.
\newblock Cambridge University Press, Cambridge, UK, 1995.

\bibitem{Ga999}
G.~Gallavotti.
\newblock New methods in nonequilibrium gases and fluids.
\newblock {\em Open Systems and Information Dynamics}, 6:101--136, 1999.

\bibitem{Ga008a}
G.~Gallavotti.
\newblock Heat and fluctuations from order to chaos.
\newblock {\em European Physics Journal B, EPJB}, 61:1--24, 2008.

\bibitem{Ru968}
D.~Ruelle.
\newblock Statistical mechanics of one--dimensional lattice gas.
\newblock {\em Communications in Mathematical Physics}, 9:267--278, 1968.

\bibitem{JP998}
M.~Jiang and {Ya.}~B. Pesin.
\newblock Equilibrium measures for coupled map lattices: Existence, uniqueness
  and finite-dimensional approximations.
\newblock {\em Communications in Mathematical Physics}, 193:675--711, 1998.

\bibitem{Ga004b}
G.~Gallavotti.
\newblock Entropy production in nonequilibrium stationary states: a point of
  view.
\newblock {\em Chaos}, 14:680--690, 2004.

\bibitem{Ei922}
E.~Einstein.
\newblock Zur {T}heorie des {R}adiometers.
\newblock {\em Annalen der Physik}, 69:241--254, 1922.

\bibitem{An982}
L.~Andrej.
\newblock The rate of entropy change in non--{H}amiltonian systems.
\newblock {\em Physics Letters}, 111A:45--46, 1982.

\bibitem{KR988}
B.W. Kernigham and D.M. Ritchie.
\newblock {\em {The {\bf C} Programming Language}}.
\newblock {Prentice Hall Software Series}. {Prentice Hall}, {Engelwood Cliffs,
  N.J.}, 1988.

\bibitem{Or974}
D.~Ornstein.
\newblock {\em Ergodic Theory, randomness and dynamical systems}, volume~5 of
  {\em Yale Mathematical Monographs}.
\newblock Yale University Press, New Haven, 1974.

\bibitem{Ga973}
G.~Gallavotti.
\newblock Ising model and {B}ernoulli shifts.
\newblock {\em Communications in Mathematical Physics}, 32:183--190, 1973.

\bibitem{Le973}
F.~Ledrappier.
\newblock Mesure d'equilibre sur un reseau.
\newblock {\em Communications in Mathematical Physics}, 33:119--128, 1973.

\bibitem{MR975}
G.~Monroy and L.~Russo.
\newblock A family of codes between some markov and bernoulli schemes.
\newblock {\em Communications in Mathematical Physics}, 43:155--159, 1975.

\bibitem{BG997}
F.~Bonetto and G.~Gallavotti.
\newblock Reversibility, coarse graining and the chaoticity principle.
\newblock {\em Communications in Mathematical Physics}, 189:263--276, 1997.

\bibitem{Ga998}
G.~Gallavotti.
\newblock Breakdown and regeneration of time reversal symmetry in
  nonequilibrium statistical mechanics.
\newblock {\em Physica D}, 112:250--257, 1998.

\bibitem{Dr988}
U.~Dressler.
\newblock {Symmetry property of the Lyapunov exponents of a class of
  dissipative dynamical systems with viscous damping}.
\newblock {\em Physical Review A}, 38:2103--2109, 1988.

\bibitem{DM996}
C.~Dettman and G.~Morriss.
\newblock Proof of conjugate pairing for an isokinetic thermostat.
\newblock {\em Physical Review E}, 53:5545--5549, 1996.

\bibitem{Ga995c}
G.~Gallavotti.
\newblock Topics in chaotic dynamics.
\newblock {\em Lecture Notes in Physics, ed. Garrido--Marro}, 448:271--311,
  1995.

\bibitem{CF983}
P.~Constantin and C.~Foias.
\newblock {Global Lyapunov Exponents, Kaplan-Yorke Formulas and the Dimension
  of the Attractors for 2D Navier-Stokes Equations}.
\newblock {\em Retrieved from the University of Minnesota Digital Conservancy},
  http://hdl.handle.net/11299/4254:1--47, 1983.

\bibitem{BGG997}
F.~Bonetto, G.~Gallavotti, and P.~Garrido.
\newblock Chaotic principle: an experimental test.
\newblock {\em Physica D}, 105:226--252, 1997.

\bibitem{Si972a}
{Ya.}~G. Sinai.
\newblock Gibbs measures in ergodic theory.
\newblock {\em Russian Mathematical Surveys}, 27:21--69, 1972.

\bibitem{Si977}
{Ya.G.} Sinai.
\newblock {\em Lectures in ergodic theory}.
\newblock Lecture notes in Mathematics. Princeton University Press, Princeton,
  1977.

\bibitem{Si994}
{Ya.G.} Sinai.
\newblock {\em Topics in ergodic theory}, volume~44 of {\em Princeton
  Mathematical Series}.
\newblock Princeton University Press, 1994.

\bibitem{MS967}
R.A. Minlos and J.G. Sinai.
\newblock The phenomenon of phase separation at low temperatures in some
  lattice models of a gas, i.
\newblock {\em Math. USSR Sbornik}, 2:335--395, 1967.

\bibitem{GC995b}
G.~Gallavotti and D.~Cohen.
\newblock {Dynamical ensembles in stationary states}.
\newblock {\em Journal of Statistical Physics}, 80:931--970, 1995.

\bibitem{Ga995b}
G.~Gallavotti.
\newblock Reversible {A}nosov diffeomorphisms and large deviations.
\newblock {\em Mathematical Physics Electronic Journal (MPEJ)}, 1:1--12, 1995.

\bibitem{GC004}
G.~Gallavotti and D.~Cohen.
\newblock {Note on nonequilibrium stationary states and entropy}.
\newblock {\em Physical Review E}, 69:035104 (+4), 2004.

\bibitem{CELS993a}
N.~I. Chernov, G.~L. Eyink, J.~L. Lebowitz, and {Ya.}~G. Sinai.
\newblock Derivation of {O}hm's law in a deterministic mechanical model.
\newblock {\em Physical Review Letters}, 70:2209--2212, 1993.

\bibitem{Ga997}
G.~Gallavotti.
\newblock Fluctuation patterns and conditional reversibility in nonequilibrium
  systems.
\newblock {\em Annales de l' Institut H. Poincar\'e}, 70:429--443, 1999 and
  chao-dyn/9703007.

\bibitem{Ga002}
G.~Gallavotti.
\newblock {\em Foundations of Fluid Dynamics}.
\newblock (second printing) Sprin\-ger Verlag, Berlin, 2005.

\bibitem{DGM984}
S.~de~Groot and P.~Mazur.
\newblock {\em Non equilibrium thermodynamics}.
\newblock Dover, Mineola, NY, 1984.

\bibitem{GPB008}
A.~Gomez-Marin, J.M.R. Parondo, and C.~Van den Broeck.
\newblock The footprints of irreversibility.
\newblock {\em European Physics Letters}, 82:5002+4, 2008.

\bibitem{OM953a}
L.~Onsager and S.~Machlup.
\newblock Fluctuations and irreversible processes.
\newblock {\em Physical Review}, 91:1505--1512, 1953.

\bibitem{OM953b}
S.~Machlup and L.~Onsager.
\newblock Fluctuations and irreversible processes. ii.
\newblock {\em Physical Review}, 91:1512--1515, 1953.

\bibitem{HPPG011}
P.~Hurtado, C.~P\'eres-Espigares, J.~Pozo, and P.~Garrido.
\newblock Symmetries in fluctuations far from equilibrium.
\newblock {\em Proceedings of tha National Academy of Science (PNAS)},
  108:7704–7709, 2011.

\bibitem{Ga996a}
G.~Gallavotti.
\newblock Extension of {O}nsager's reciprocity to large fields and the chaotic
  hypothesis.
\newblock {\em Physical Review Letters}, 77:4334--4337, 1996.

\bibitem{GR997}
G.~Gallavotti and D.~Ruelle.
\newblock {SRB} states and non\-equi\-li\-brium statistical mechanics close to
  equi\-li\-brium.
\newblock {\em Com\-mu\-ni\-ca\-tions in Mathematical Physics}, 190:279--285,
  1997.

\bibitem{Ru997}
D.~Ruelle.
\newblock Entropy production in nonequilibrium statistical mechanics.
\newblock {\em Communications in Mathematical Physics}, 189:365--371, 1997.

\bibitem{Ga999b}
G.~Gallavotti.
\newblock A local fluctuation theorem.
\newblock {\em Physica A}, 263:39--50, 1999.

\bibitem{PS991}
Y.B. Pesin and Y.G. Sinai.
\newblock Space--time chaos in chains of weakly inteacting hyperbolic mappimgs.
\newblock {\em Advances in Soviet Mathematics}, 3:165--198, 1991.

\bibitem{BK996}
J.~Bricmont and A.~Kupiainen.
\newblock High temperature expansions and dynamical systems.
\newblock {\em Communications in Mathematical Physics}, 178:703--732, 1996.

\bibitem{BK995}
J.~Bricmont and A.~Kupiainen.
\newblock Coupled analytic maps.
\newblock {\em Nonlinearity}, 8:379–396, 1995.

\bibitem{BK997}
J.~Bricmont and A.~Kupiainen.
\newblock Infinite dimensional srb measures.
\newblock {\em Physica D}, 103:18--33, 1997.

\bibitem{Ga996b}
G.~Gallavotti.
\newblock {Equivalence of dynamical ensembles and Navier Stokes equations}.
\newblock {\em Physics Letters A}, 223:91--95, 1996.

\bibitem{BGG007}
F.~Bonetto, G.~Gallavotti, and G.~Gentile.
\newblock A fluctuation theorem in a random environment.
\newblock {\em Ergodic Theory and Dynamical Systems}, 28:21--47, 2008.

\bibitem{Ol988}
S.~Olla.
\newblock {Large Deviations for Gibbs Random Fields}.
\newblock {\em Probability Theory and Related Fields}, 77:343--357, 1988.

\bibitem{GLM002}
G.~Gallavotti, J.~L. Lebowitz, and V.~Mastropietro.
\newblock Large deviations in rarefied quantum gases.
\newblock {\em Journal of Statistical Physics}, 108:831--861, 2002.

\bibitem{GP999}
G~Gallavotti and F.~Perroni.
\newblock An experimental test of the local fluctuation theorem in chains of
  weakly interacting anosov systems.
\newblock {\em https://arxiv.org/abs/chao-dyn/9909007}, pages 1--15, 1999.

\bibitem{Ga997c}
G.~Gallavotti.
\newblock Chaotic hypothesis and universal large deviations properties.
\newblock {\em Documenta Mathematica}, extra volume ICM98, vol. I:205--233,
  1998.

\bibitem{MCT993}
F.~Mauri, R.~Car, and E.~Tosatti.
\newblock {Canonical Statistical Averages of Coupled Quantum-Classical
  Systems}.
\newblock {\em Europhysics Letters}, 24:431--436, 1993.

\bibitem{ACCCEF011}
J.L. Alonso, A.~Castro, J.~Clemente-Gallardo, J.C. Cuchi, P.~Echenique, and
  F.~Falceto.
\newblock {Statistics and Nos\'e formalism for Ehrenfest dynamics}.
\newblock {\em {Journal of Phy\-sics A}}, 44:395004, 2011.

\bibitem{Ga008}
G.~Gallavotti.
\newblock {\em The Elements of Mechanics (II edition)}.
\newblock Gallavotti website, Roma, 2008 [I edition was Springer 1984].

\bibitem{St966}
F.~Strocchi.
\newblock {Complex Coordinates and Quantum Mechanics}.
\newblock {\em {Reviews of Modern Physics}}, 38:36--40, 1966.

\bibitem{Ca013}
M.~Campisi.
\newblock {Quantum Fluctuation Relations for Ensembles of Wave Functions}.
\newblock {\em New Journal of Physics}, 15:1--12, 2013.

\bibitem{Se987}
F.~Seitz.
\newblock {\em The modern theory of solids}.
\newblock Dover, Mineola, 1987 (reprint).

\bibitem{HHP987}
B.~D. Holian, W.~G. Hoover, and H.~A.~W. Posch.
\newblock Resolution of loschmidt paradox: The origin of irreversible behavior
  in reversible atomistic dynamics.
\newblock {\em Physical Review Letters}, 59:10--13, 1987.

\bibitem{Ru000}
D.~Ruelle.
\newblock A remark on the equivalence of isokinetic and isoenergetic
  thermostats in the thermodynamic limit.
\newblock {\em Journal of Statistical Physics}, 100:757--763, 2000.

\bibitem{Ga020b}
G.~Gallavotti.
\newblock {Nonequilibrium and Fluctuation Relation}.
\newblock {\em Journal of Statistical Physics}, 180:172--226, 2019.

\bibitem{Ga019c}
G.~Gallavotti.
\newblock {Reversible viscosity and Navier--Stokes fluids}.
\newblock {\em Springer Proceedings in Mathematics \& Statistics},
  282:569--580, 2019.

\bibitem{Ga021}
G.~Gallavotti.
\newblock {Viscosity, Reversibility, Chaotic Hypothesis, Fluctuation Theorem
  and Lyapunov Pairing}.
\newblock {\em {Journal of Statistical Physics}}, {185:21}:1--19, 2021.

\bibitem{Ga006a}
G.~Gallavotti.
\newblock Irreversibility time scale.
\newblock {\em Chaos}, 16:023130 (+7), 2006.

\bibitem{Ga005c}
G.~Gallavotti.
\newblock Nonequilibrium statistical mechanics (stationary): Overview.
\newblock {\em Encyclopedia of Mathematical Physics, ed. J.P. Fran{\c{c}}oise,
  G.L. Naber, Tsou Sheung Tsun}, 3:530--539, 2006.

\bibitem{Ru997b}
D.~Ruelle.
\newblock {Differentiation of SRB states}.
\newblock {\em Communications in Mathematical Physics}, 187:227--241, 1997.

\bibitem{BGM998}
F.~Bonetto, G.~Gentile, and V.~Mastropietro.
\newblock Electric fields on a surface of constant negative curvature.
\newblock {\em Ergodic Theory and Dynamical Systems}, 20:681--686, 2000.

\bibitem{SJ993}
Z.S. She and E.~Jackson.
\newblock Constrained {E}uler system for {N}avier-{S}tokes turbulence.
\newblock {\em Physical Review Letters}, 70:1255--1258, 1993.

\bibitem{Sa006}
P.~Sagaut.
\newblock {\em {Large Eddy Simulation for Incompressible Flows}}.
\newblock {Scientific computation}. Springer, Berlin, 2006.

\bibitem{GPMC991}
M.~Germano, U.~Piomell, P.~Moin, and W.H. Cabot.
\newblock A dynamic subgridscale eddy viscosity model.
\newblock {\em Physics of Fluids A}, 3:1760--1766, 1991.

\bibitem{Fe006}
C.~Fefferman.
\newblock {\em {Existence \& smoothness of the Navier–Stokes equation}}.
\newblock in {J.Carlson,A.Jaffe,A.Wiles}: {The millennium prize problems}.
  American Mathematical Society, Providence, RI, 2006.

\bibitem{Ga997b}
G.~Gallavotti.
\newblock Dynamical ensembles equivalence in fluid mechanics.
\newblock {\em Physica D}, 105:163--184, 1997.

\bibitem{Co007}
P.~Constantin.
\newblock On the euler equations of incompressible fluids.
\newblock {\em Bulletin of the American Mathematical Society}, 44:603--621,
  207.

\bibitem{MBCGL022}
G.~Margazoglu, L.~Biferale, M.~Cencini, G.~Gallavotti, and V.~Lucarini.
\newblock {Non-equilibrium Ensembles for the 3D Navier-Stokes Equations}.
\newblock {\em Physical Review E}, 105:065110, 2022.

\bibitem{SDNKT018}
V.~Shukla, B.~Dubrulle, S.~Nazarenko, G.~Krstulovic, and S.~Thalabard.
\newblock {Phase transition in time-reversible Navier-Stokes equations}.
\newblock {\em Physical Review E}, 100:043104, 2019.

\bibitem{SEMGS021}
K.~Seshasayanan, S.~Eswaran, M.~Maji, S.~Ghosh, and V.~Shukla.
\newblock Equivalence of nonequilibrium ensembles:turbulence with a dual
  cascade.
\newblock {\em Physical Review E}, 108:015102, 2021.

\bibitem{Ku024}
S.~Kuksin.
\newblock The k41 theory and turbulence in 1d burgers equation.
\newblock {\em Chaos}, 34:1--29, 2024.

\bibitem{BCDGL018}
L.~Biferale, M.~Cencini, M.~DePietro, G.~Gallavotti, and V.~Lucarini.
\newblock Equivalence of non-equilibrium ensembles in turbulence models.
\newblock {\em Physical Review E}, 98:012201, 2018.

\bibitem{PBCCMD023}
Q.~Pikeroen, A.~Barral, G.~Costa, C.~Campolina, A.~Mailybaev, and B.~Dubrulle.
\newblock Tracking complex singularities of fluids on log-lattices.
\newblock {\em arXiv.2312.01702}, 2023.

\bibitem{GL014}
G.~Gallavotti and V.~Lucarini.
\newblock {Equivalence of Non-Equilibrium Ensembles and Representation of
  Friction in Turbulent Flows: The Lorenz 96 Model}.
\newblock {\em Journal of Statistical Physics}, 156:1027--1053, 2014.

\bibitem{Ga002b}
G.~Gallavotti.
\newblock Intermittency and time arrow in statistical mechanics and turbulence.
\newblock {\em Fields Institute Communications}, 31:165--172, 2002.

\bibitem{Ga006d}
G.~Gallavotti.
\newblock Microscopic chaos and macroscopic entropy in fluids.
\newblock {\em Journal of Statistical Mechanics (JSTAT)}, 2006:P10011 (+9),
  2006.

\bibitem{PM980}
Y.~Pomeau and P.~Manneville.
\newblock {Different ways to turbulence in dissipative dynamical systems}.
\newblock {\em Physica D}, 1:219--226, 1980.

\bibitem{Ga020}
G.~Gallavotti.
\newblock {Ensembles, Turbulence and Fluctuation Theorem}.
\newblock {\em {European Physics Journal E}}, 43:37, 2020.

\bibitem{Ku998}
J.~Kurchan.
\newblock Fluctuation theorem for stochastic dynamics.
\newblock {\em Journal of Physics A}, 31:3719--3729, 1998.

\bibitem{LS999}
J.~Lebowitz and H.~Spohn.
\newblock A {G}allavotti--{C}ohen type symmetry in large deviation functional
  for stochastic dynamics.
\newblock {\em Journal of Statistical Physics}, 95:333--365, 1999.

\bibitem{Ma999}
C.~Maes.
\newblock The fluctuation theorem as a {G}ibbs property.
\newblock {\em Journal of Statistical Physics}, 95:367--392, 1999.

\bibitem{CEG984}
P.~Collet, H.~Epstein, and G.~Gallavotti.
\newblock {Perturbations of geodesic flows on surfaces of constant negative
  curvature and their mixing properties}.
\newblock {\em Communications in Mathematical Physics}, 95:61--112, 1984.

\bibitem{Ia018}
A.~Iacobucci.
\newblock {\em {Nonequilibrium stationary states of rotor and oscillator
  chains}}.
\newblock https://theses.hal.science/tel-01792503, Paris, 2018.

\bibitem{IOS019}
A.~Iacobucci, S.~Olla, and G.~Stoltz.
\newblock Convergence rates for nonequilibrium langevin dynamics.
\newblock {\em Annales Math\'ematiques du Qu\'ebec}, 43:73--98, 2018.

\bibitem{CV003a}
R.~{Van Zon} and E.~G.~D. Cohen.
\newblock Extension of the fluctuation theorem.
\newblock {\em Physical Review Letters}, 91:110601 (+4), 2003.

\bibitem{CV003}
R.~{Van Zon} and E.~G.~D. Cohen.
\newblock Extended heat-fluctuation theorems for a system with deterministic
  and stochastic forces.
\newblock {\em Physical Review E}, 69:056121 (+14), 2004.

\bibitem{JGC007}
S.~Joubaud, N.B. Garnier, and S.~Ciliberto.
\newblock Fluctuation theorems for harmonic oscillators.
\newblock {\em Journal of Statistical Mechanics}, 9:P09018, 2007.

\bibitem{ESR003}
D.~J. Evans, D.~J. Searles, and L.~Rondoni.
\newblock {Application of the Gallavotti--Cohen fluctuation relation to
  thermostated steady states near equilibrium}.
\newblock {\em Physical Review E}, 71:056120 (+12), 2005.

\bibitem{Za007}
F.~Zamponi.
\newblock Is it possible to experimentally verify the fluctuation relation? a
  review of theoretical motivations and numerical evidence.
\newblock {\em Journal of Statistical Mechanics}, page P02008, 2007.

\bibitem{CKP997}
L.F. Cugliandolo, J.~Kurchan, and L.~Peliti.
\newblock Energy flow, partial equilibration, and effective temperatures in
  systems with slow dynamics.
\newblock {\em Physical Review E}, pages 2898--3914, 1997.

\bibitem{CR003}
A.~Crisanti and F.~Ritort.
\newblock Violation of the fluctuation-dissipation theorem in glassy systems:
  basic notions and the numerical evidence.
\newblock {\em Journal of Physics A}, pages R181--R290, 2003.

\bibitem{NS002}
O.~Nakabeppu and T.~Suzuki.
\newblock Microscale temperature measurement by scanning thermal microscopy.
\newblock {\em Journal of Thermal Analysis and Calorimetry}, 69:727--737, 2002.

\bibitem{Sm972}
D.T. Smith.
\newblock A square root circuit to linearize feedback in temperature
  controllers.
\newblock {\em Journal of Physics E: Scientific Instruments}, 5:528, 1972.

\bibitem{FM004}
K.~Feitosa and N.~Menon.
\newblock A fluidized granular medium as an instance of the fluctuation
  theorem.
\newblock {\em Physical Review Letters}, 92:164301+4, 2004.

\bibitem{BGGZ006}
F.~Bonetto, G.~Gallavotti, A.~Giuliani, and F.~Zamponi.
\newblock Fluctuations relation and external thermostats: an application to
  granular materials.
\newblock {\em Journal of Statistical Mechanics}, 2006(05):P05009, 2006.

\bibitem{PVBTW005}
A.~Puglisi, P.~Visco, A.~Barrat, E.~Trizac, and F.~van Wijland.
\newblock Fluctuations of internal energy flow in a vibrated granular gas.
\newblock {\em Physical Review Letters (cond-mat/0509105)}, 95:110202 (+4),
  2005.

\bibitem{GZN996}
E.~L. Grossman, E.L., Tong Zhou, and E.~Ben-Naim.
\newblock Towards granular hydrodynamics in two-dimensions.
\newblock {\em cond-mat/9607165}, 1996.

\bibitem{BMM000}
J.~Brey, M.~J. Ruiz-Montero, and F.~Moreno.
\newblock Boundary conditions and normal state for a vibrated granular fluid.
\newblock {\em Physical Review E}, 62:5339--5346, 2000.

\bibitem{BCL998}
F.~Bonetto, N.~Chernov, and J.~L. Lebowitz.
\newblock (global and local) fluctuations in phase-space contraction in
  deteministic stationary nonequilibrium.
\newblock {\em Chaos}, 8:823--833, 1998.

\bibitem{BL001}
F.~Bonetto and J.~L. Lebowitz.
\newblock Thermodynamic entropy production fluctuation in a two-dimensional
  shear flow model.
\newblock {\em Physical Review E}, 64:056129, 2001.

\bibitem{ZRA004}
F.~Zamponi, G.~Ruocco, and L.~An\-gelani.
\newblock Fluctuations of entropy production in the isokinetic ensemble.
\newblock {\em Journal of Statistical Physics}, 115:1655--1668, 2004.

\bibitem{GZG005}
A.~Giuliani, F.~Zamponi, and G.~Gallavotti.
\newblock Fluctuation relation beyond linear response theory.
\newblock {\em Journal of Statistical Physics}, 119:909--944, 2005.

\bibitem{VPBTW005}
P.~Visco, A.~Puglisi, A.~Barrat, E.~Trizac, and {F. van} Wijland.
\newblock Fluctuations of power injection in randomly driven granular gases.
\newblock {\em Journal of Statistical Physics}, 125:529--564, 2005.

\bibitem{JES004}
O.~Jepps, D.~Evans, and D.~Searles.
\newblock {The fluctuation theorem and Lyapunov weights}.
\newblock {\em {Physica D}}, 187:326--337, 2004.

\bibitem{Ze860}
G.~Zeuner.
\newblock {\em {Grundgz\"uge der Mechanischen W\"armetheorie}}.
\newblock Buchandlung J.G. Engelhardt, Freiberg, 1860.

\bibitem{Ma858}
M.A. Masson.
\newblock Sur la corr\'elation des propri\'et\'es physiques des corps.
\newblock {\em Annales de Chimie}, 53:257--293, 1858.

\bibitem{Ga016}
G.~Gallavotti.
\newblock Ergodicity: a historical perspective. equilibrium and nonequilibrium.
\newblock {\em European Physics Journal H}, 41,:181--259, 2016.

\bibitem{GJ020}
G.~Gallavotti and I.~Jauslin.
\newblock {A Theorem on Ellipses, an Integrable System and a Theorem of
  Boltzmann.}
\newblock {\em arXiv}, 2008.01955:1--9, 2020.

\bibitem{Fe021}
G.~Felder.
\newblock {Poncelet Property and Quasi-periodicity of the Integrable Boltzmann
  System}.
\newblock {\em Letters in Mathematical Physics}, 111:1--19, 2021.

\bibitem{LL971}
L.D. Landau and E.M. Lifschitz.
\newblock {\em M\'ecanique des fluides}.
\newblock MIR, Moscow, 1971.

\bibitem{Cl854}
R.~Clausius.
\newblock Ueber eine ver{\" a}nderte form des zweiten hauptsatzes der
  mechanischen w{\"a}rmetheorie.
\newblock {\em Annalen der Physik und Chemie}, 93:481--506, 1854.

\bibitem{Cl862}
R.~Clausius.
\newblock On the application of the theorem of the equivalence of
  transformations to interior work.
\newblock {\em Philosophical Magazine}, 4-XXIV:81--201, 1862.

\bibitem{Re997}
J.~Renn.
\newblock {Einstein's controversy with Drude and the origin of statistical
  mechanics: a new glimpse from the ``Love Letters''}.
\newblock {\em Archive for the history of exact sciences}, 51:315--354, 1997.

\bibitem{Ma868}
J.C. Maxwell.
\newblock On the dynamical theory of gases.
\newblock {\em Philosophical Magazine}, XXXV:129--145, 185--217, 1868.

\bibitem{Ul968}
G.~E. Uhlenbeck.
\newblock {\em {An outline of Statistical Mechanics}, in {Fundamental problems
  in Statistical Mechanics, II}}.
\newblock ed. E. G. D. Cohen, North Holland, Amsterdam, 1968.

\bibitem{He884a}
H.~Helmholtz.
\newblock {\em Prinzipien der Statik monocyklischer Systeme}, volume III of
  {\em {W}is\-sen\-schaft\-li\-che {A}bhandlungen}.
\newblock Barth, Leipzig, 1895.

\bibitem{He884b}
H.~Helmholtz.
\newblock {\em Studien zur Statik monocyklischer Systeme}, volume III of {\em
  {W}is\-sen\-schaft\-li\-che {A}bhandlungen}.
\newblock Barth, Leipzig, 1895.

\bibitem{MP972}
C.~Marchioro and E.~Presutti.
\newblock Thermodynamics of particle systems in presence of external
  macroscopic fields.
\newblock {\em Communications in Mathematical Physics}, 27:146--154, 1972.

\bibitem{Wh917}
E.T. Whittaker.
\newblock {\em A treatise on the analytic dynamics of particles \& rigid
  bodies}.
\newblock Cambridge University Press, Cambridge, 1917 (reprinted 1989).

\bibitem{AA966}
V.I. Arnold and A.~Avez.
\newblock {\em {Ergodic probems of Classical Mechanics}}.
\newblock {Mathematical Physics Monographs}. Benjamin, 1968.

\bibitem{Sm967}
S.~Smale.
\newblock Differentiable dynamical systems.
\newblock {\em Bullettin of the American Mathematical Society}, 73:747--818,
  1967.

\bibitem{HZ994}
H.~Hofer and E.Zehnder.
\newblock {\em Symplectic invariants and {H}amiltonian dynamics}.
\newblock {Birkhauser Advanced Texts}. {Birkhauser Verlag}, Basel, 1994.

\bibitem{WL998}
M.P. Wojtkowski and C.~Liverani.
\newblock Conformally symplectic dynamics and symmetry of the lyapunov
  spectrum.
\newblock {\em Communications in Mathematical Physics}, 194:47--60, 1998.

\bibitem{ECM990}
D.~Evans, EGD Cohen, and G.~Morriss.
\newblock Viscosity of a simple fluid from its maximal lyapunov exponents.
\newblock {\em Physical Review A}, 42:5990--5997, 1990.

\bibitem{BGGS980b}
G.~Benettin, L.~Galgani, A.~Giorgilli, and J.~Strelcyn.
\newblock {Lyapunov charateristic exponents for smooth dynamical systems and
  for Hamiltonian systems; A method for computing all of them. Part 2,
  Numerical application}.
\newblock {\em Meccanica}, 15:21--30, 1980.

\bibitem{Ga019a}
G.~Gallavotti.
\newblock Navier-stokes equation: irreversibility turbulence and ensembles
  equivalence.
\newblock {\em arXiv:1902.09610}, page 09160, 2019.

\bibitem{Ru982}
D.~Ruelle.
\newblock Large volume limit of the distribution of characteristic exponents in
  turbulence.
\newblock {\em Communications in Mathematical Physics}, 87:287--302, 1982.

\bibitem{Li984}
E.~Lieb.
\newblock On characteristic exponents in turbulence.
\newblock {\em Communications in Mathematical Physics}, 92:473--480, 1984.

\bibitem{Ga013b}
G.~Gallavotti.
\newblock {\em Nonequilibrium and irreversibility}.
\newblock Theoretical and Mathematical Physics. Springer-Verlag, 2014.

\bibitem{GRS004}
G.~Gallavotti, L.~Rondoni, and E.~Segre.
\newblock Lyapunov spectra and nonequilibrium ensembles equivalence in 2d
  fluid.
\newblock {\em Physica D}, 187:358--369, 2004.

\bibitem{MS968}
R.A. Minlos and J.G. Sinai.
\newblock The phenomenon of phase separation at low temperatures in some
  lattice models of a gas, ii.
\newblock {\em Transactions of the Moscow Mathematical Society}, 19:121--196,
  1968.

\bibitem{GM970}
G.~Gallavotti and S.~Miracle-Sol\'e.
\newblock Absence of phase transitions in hard-core one dimensional systems
  with long range interactions.
\newblock {\em Journal of Mathematical Physics}, 11:147--154, 1970.

\bibitem{CO981}
M.~Cassandro and E.~Olivieri.
\newblock Renormalization group and analyticity in one dimension: A proof of
  dobrushin's theorem.
\newblock {\em Communications in Mathematical Physics}, 80:255--269, 1981.

\bibitem{Ga998b}
G.~Gallavotti.
\newblock Chaotic dynamics, fluctuations, non-equilibrium ensembles.
\newblock {\em Chaos}, 8:384--392, 1998.

\bibitem{Ja997}
C.~Jarzynski.
\newblock Nonequilibrium equality for free energy difference.
\newblock {\em Physical Review Letters}, 78:2690--2693, 1997.

\bibitem{Ja999}
C.~Jarzynski.
\newblock Hamiltonian derivation of a detailed fluctuation theorem.
\newblock {\em Journal of Statistical Physics}, 98:77--102, 1999.

\bibitem{ES994}
D.J. Evans and D.~Searles.
\newblock Equilibrium microstates which generate second law violating steady
  state.
\newblock {\em Physical Review E}, 50:1645--1648, 1994.

\bibitem{BK981a}
G.~N. Bochkov and Yu.~E. Kuzovlev.
\newblock Nonlinear fluctuation-dissipation relations and stochastic models in
  nonequilibrium thermodynamics: I. generalized fluctuation-dissipation
  theorem.
\newblock {\em Physica A}, 106:443--479, 1981.

\bibitem{BK981b}
G.~N. Bochkov and Yu.~E. Kuzovlev.
\newblock {Nonlinear fluctuation-dissipation relations and stochastic models in
  nonequilibrium thermodynamics: II. Kinetic potential and variational
  principles for nonlinear irreversible processes}.
\newblock {\em Physica A}, 106:480--520, 1981.

\end{thebibliography}

\def\SEC{Subject index}
\lhead{\small\SEC}
\label{Subject index}
\iniz

\printindex

\0List of frequent abbreviations

\vskip9mm
\halign{#\ \hfill& #\hfill\cr
CH& Chaotic Hypothesis\cr EH & Ergodic Hypothesis\cr NS&
Navier-Stokes equation\cr INS& Irreversible Navier-Stokes
equation\cr RNS& Reversible Navier-Stokes equation\cr SRB&
Sinai-Bowen-Ruelle distribution\cr smooth& $C^\infty$\cr TRS&
Time Reversal Symmetry\cr OK41& Obukov-Kolmogorov\cr UV&
Ultraviolet cut-off\cr {\bf 1}& Identity Map\cr}

\vfill\eject

\def\hfb{\hfill\break}
\def\hyp{\hyperpage}

\section{Citations index}
\def\SEC{Citations index} 
\hrule
\*\*
\0{Reference {\bf number} followed by \alert{\bf citation pages}}
\*

\vskip1cm

\begin{multicols}{2}
\parindent=0mm
\input NN.aus
\end{multicols}

\vfill\eject
\section{{Chapters abstracts}}
\def\SEC{{Chapters abstracts}}
\*

{\bf Chapter 1}

A brief survey of the development of Equilibrium theory, from
Carnot's theorem to the least action principle and the Ergodic
Hypothesis (EH), as a background for the study of stationary Non
Equilibrium.

Particular attention is devoted to the application of the EH and
of the Least Action principle to Boltzmann's derivation of the
second law 1866, (presented in Clausius' version in the new
Sec. (1.4)).

Identification of a thermodynamic equilibrium state with the
collection of time averages of observables was made, almost
without explicit comments and the analysis relied EH as the
assumption that motions are periodic.  A first attempt, already
in 1868, to eliminate such hypothesis appears initially in the
form: the molecule goes through all possible states because of
the collisions with the others. However the previous hypothesis
(i.e. periodic motion covers the energy surface) appears again to
use EH, as a starting point, to obtain the microcanonical
distribution for the entire gas.

The ``solution of a mechanical problem'' arises in this context
to present an example of ergodicity, and it is interesting to
examine its conflict (conjectured in the in the first edition,
and here developed in Sec. (6.3)) with KAM theory.

Litle later (1971) the ergodic hypothesis relation to periodic
orbits plays a minor role, and a general theory of the
'ensembles' is discovered simultaneously with the equivalence
between canonical and microcanonical distributions for large
systems, as explicitly recognized by Gibbs.

And soon the discovery of H-theorem and Boltzmann's equation
systematized the early work of Maxwell on the theory of gases.
However in this book the theme of the Boltzmann equation is not
really touched: attention is concentrated on stationary states in
and out of equilibrium, with the single exception of the attempt
to introduce the ``time scale'' of a process of transformation of
an equilibrium state into another aimed at quantifying the notion
of ``infinitely slow'' (thermodynamically reversible) process in
Sec. (5.2).

\vfill\eject
{\bf Chapter 2}

Thermostats are introduced, and illustrated via several
mechanical examples, as models and tools to study stationary
states in presence of driving forces balanced by dissipation.

This gives the opportunity of stressing the role of the initial
data randomness: Ruelle's viewpoint on the privileged
status of the initial data, as data generated by a protocol
producing them as probability-one samples of {\it any
distribution} with density over phase space.

In presence of chaotic motions the stationary states reached by
the evolution are determined, independently of the protocol for
the initial data, if chaoticity is intended in the precise sense
of hyperbolic evolution on the surfaces attracting the
motions. The latter property is formalized as ``Chaotic
hypothesis'' (CH).

And it is stressed that its adoption solves the problem of
identifying, in equilibrium as well as out of equilibrium, which
is the distribution, among the {\it infinitely many} stationary
ones, physically relevant to provide the time averages of the
observables: such remarkable property holds because, in systems
verifying the CH, on each attracting surface (often unique) there
is a unique stationary state, called SRB distribution,
controlling the average of physical observables. Multiplicity of
attracting surfaces is similar to phase coexistence of equilibria
and in Sec.(5.6) is proposed among the causes of
``intermittency''.

The idea emerges that the CH is a paradigm for chaotic motions
playing the role that the quasi periodic motions play for ordered
evolutions: the picture implied by the CH is that in phase space
one or more surfaces which attract nearby points exist and are
smooth and all their points (up to a set of zero area) have
trajectories that cover them densely. 

Furthermore in conservative systems undergoing chaotic
evolutions the CH implies the EH so that the two
hypotheses cannot contradict each other.

Still the CH has to be regarded as a law at the same level as the
EH: it also covers systems well beyond the equilibrium, but {\it
tertium datur!} as shown by easy counterexamples.

Accent is also on examples of non equilibrium
problems and of thermostats, finite or infinite, that control
dissipation, with wide discussion of 
the phase space contraction interpretation as heat ceded to thermostats.

Difficulties about the interpretation due to the dependence of
phase space volume on the metric used to measure it are
avoided by showing that the amount of heat ceded does not depend
on the metric.  At the same time it is stressed that it is different to
consider phase space contraction on the whole phase space or
on the attracting surfaces (often much smaller and of difficult
access).

\vfill\eject
{\bf Chapter 3}

Recurrence is discussed in detail under the perspective to
present it in the light of a discrete representation of motion,
in principle necessary in simulations and in their
interpretation.

A discrete viewpoint is important also in qualitative
understanding of the statistical properties of motions on the
attracting surfaces. It is stressed that studying recurrence
leads to a formulation which unifies entirely different dynamical
systems: under the Chaotic hypothesis (CH) the most diverse
dynamical systems can be digitally represented as shifts of a one
dimensional chain of symbols on which the invariant distributions
appear as stationary stochastic processes, natural
generalizations of Markov processes.

The ``universal'' representation of stationary states of
Ch-verifying systems as, essentially, Markov processes is
remarkable because stochasticity is not introduced {\it a priori}.

Representation of phase space as an array of points, regularly
filling it and moving according to a specific rule (``code''), is
deeply different from the phase space cells used, at times, in
equilibrium Statistical Mechanics: here the points are called
``microcells'', to be conceived as regularly disposed on a
lattice in phase space just as they are ideally conceived in
simulations codes.

CH and the hyperbolic nature of the time evolution allows to
collect the microcells in much larger groups, forming a partition
of phase space into ``coarse cells'', labeled by a finite number
of labels, small enough so that the (few) observables of interest
can be considered constant on each cell.

The motion will be dissipative, at least in non equilibrium
cases, and most microcells will not be recurrent: so the
statistical properties will be carried by recurrent ones, \ie by
the periodic points on each attracting set.  In discretized
representations all motions will be {\it eventually} periodic and
it is tempting to go back to the original proposal by Boltzmann
and think that there is only one attracting periodic orbit: and
the whole discussion (hinted in chaper 1) about this point and
the EH can be imagined and repeated.

Hyperbolicity makes it possible to design the coarse cells so
that each microcell is identified with the string formed by the
labels of the coarse cells successively visited in its motion: and the
regularity of the array of microcells can be seen as the source
of the physical relevance of SRB distribution, as the {\it heuristic}
interpretation of the SRB distribution shows in sec.(3.8).

Discrete interpretation of the SRB distribution leads to an
estimate of the number of microcells composing the attracting
periodic orbit: it is presented because it gives the opportunity
of showing the existence of a mathematical cancellation,
interesting in itself, and to argue that the logarithm of the
recurrent microcells number has the interpretation of entropy
only in the case of equilibrium states while in non equilibrium
cannot be seen as a function of the state.

\vfill\eject
{\bf Chapter 4}

The CH allows to represent as a standard stochastic process (one
dimensional ``Gibbs states') the stationary states (in
equilibrium and out of it). Therefore it is natural to ask
whether the universality of the representation of the SRB
distributions, for CH-complying stationary non-equilibrium
states, can be translated into universal laws.

The first candidates for such search are properties of the
fluctuations statistical properties, which can be expected to
inherit properties of the system is studied. Certainly the first
question is the compatibility between irreversibility and
microscopic reversibility.

Help comes from the simulations: since the eighties reversible
models for irreversible phenomena have been designed and
studied. And questions have been debated like how is
reversibility reflected on the properties of, say, transport
coefficients, which are constants in the corresponding
irreversible models.

The question admits a rather general answer in the ``fluctuation
theorem'', which gives some light on the statistics of the
phase space contraction (\ie on the statistics of the rate of
heat delivered to the thermostats, sec.(2.2)).

Or, still in reversible models of stationary non equilibrium
states, it allows a general answer to the relation, at stationarity,
between the probability in a stationary state of an event in
which, in a given time, a given observable evolves following a
given path and the probability that it evolves following the
reversed path.

The answer is that the two probabilities are the same if
conditioned to an opposite value of the average rate of entropy
production (identified with the rate of phase space contraction).

The fluctuation theorem can also be related to
simple extensions of Onsager's reciprocity and Green-Kubo
formulae. Finally an attempt is made towards an application of
the above ideas to quantum systems.

\vfill\eject
{\bf Chapter 5}

A possible application of previous sections ideas,
particularly on phase space discretization, leads to propose
``the degree of irreversibility'' of a process controlled by one
or more parameters assigned to vary in time, while the system
starts in a stationary state corresponding to  initial
parameters value and in time evolves, for instance, to a final
stationary state. In particular initial, intermediary, final
parameters values could be such that, if kept constant, would
lead to an equilibrium state (a model of ``thermodynamic
process'').

A definition of the rapidity of an evolution, or of the duration
of a process, can be taken as a measure of its
``irreversibility'', expected because, with Carnot, a reversible
process is ``infinitely slow'': several examples on well known
processes are provided. The analysis in sec.(5.2) relies on the
remark of existence of a cancellation in the formal expression of
the SRB distribution.

Applications of CH to fluctuations in reversible evolutions,
chapter 4, induces immediately to ask whether similar questions can
be studied in the case of {\it irreversible} evolutions.

Since reversible models are often studied in simulations while 
theory very often deals with irreversible models, it is
natural to try to establish more formally a correspondence rule
between reversible and irreversible models of the same evolution
problem.

A few conjectures are formulated that can be summarized in a
general rule concerning a system described by two models, both
fulfilling CH: one in which a parameter is a constant while in
the other the parameter is replaced by a quantity designed to
keep, say, dissipation rate constant at the average value
occurring in the first system.

A basic question, ``equivalence problem'', is whether, or
on which conditions, the statistics of the fluctuations of
classes of observables of the two models are the same.
Examples of conjectured equivalence are provided.

An example is the incompressible Navier-Stokes fluid with UV cutoff parameter
$N$ on the harmonics of  the velocity $\uu$; consider the classic
Navier-Stokes (NS) equation with viscosity $\n$ and the
reversible Navier-Stokes (RNS) equation with viscosity replaced
by a multiplier $\a(\uu)$ such that the enstrophy $En=\int (\dpr \uu(x))^2 dx$
is exactly constant.

If at fixed viscosity the average enstrophy is $En_N$, then the
statistics of the observables depending only on finitely many
Fourier's harmonics (``large scale harmonics''), evaluated
following the two equations, is the same in the limit
$N\to\infty$, \ie at {\it removal of the cut-off}.
The conjecture has been tested in dimension 2 and even in
dimension 3 and results are commented.

Other equivalence conjectures arise in mechanical systems
exemplified also in problems on granular materials or for
stochastic evolutions. And a proposal of employing
the fluctuation theorem, to design a thermometer
to measure  absolute temperature of very small regions of a
surface, is also discussed.

\vfill\eject
{\bf Chapter 6}
\parindent=0mm

Selected classic papers are partially translated with few comments.

1) {\it Heat theorem, action principle, EH}. The first
Boltzmann's work on the second principle mechanical basis (the
``Heat theorem''), \Cite{Bo866}.

2) {\it Collision analysis and equipartition}. By  a
combinatorial argument, in presence of hard core (or
instantaneous) interactions, Maxwell's distribution is derived and
the microcanonical ``ensemble'' is introduced, \Cite{Bo868-a}.

3) {\it Solution of a mechanical problem}. An attempt to give an
example of an ergodic system. Comments on the relation, and
conflict, with KAM theory,\,\Cite{Bo868-b}\,.

4,5) {\it Recurrence and periodicity}: Clausius 
version of the heat theorem,\,\Cite{Cl871}\,.

6,7) {\it Priority discussions Boltzmann-Clausius on heat
theorem},\,\Cite{Bo871-0}\Cite{Cl872}\,.

8) {\it On the ergodic hypothesis}. Collisions between molecules imply that
 the molecule atoms visit all points of the single molecule phase
 space; EH is not used. And microcanonical and canonical
 ensembles are introduced: for the {\it first time} the
 the microcanonical distribution uniqueness is
 questioned,\,\Cite{Bo871-a}\,.

9) {\it Canonical ensemble and ergodic hypothesis}.  Solution for
the thermal equilibrium for the molecules of a gas with a finite
number of point masses ``under an hypothesis''. The hypothesis
concerns the uniqueness of the stationary distribution, raised
already in 8). And ergodicity appears again in the form 
that not only the atoms of a single molecule take all possible
positions and velocities but also that the atoms of a ``warm
body'' with which a molecule is in contact take all positions and
velocities,\,\Cite{Bo871-b}\,.

10) {\it Heat theorem without dynamics}. It is shown how the hypothesis
that, assuming that the equilibrium distribution of the entire
system is the microcanonical one, then defining the heat $dQ$
received by the body as the variation of the total average energy
$dE$ plus the work $dW$ done by the system on the outside it
follows that $\frac{dQ}T$ is an exact differential if $T$ is
proportional to the average kinetic energy. This frees
(apparently) equilibrium statistical mechanics from EH
and will be revisited in 1884\, \Cite{Bo871-c}\,.

11) {\it Irreversibility: Loschmidt and ``Boltzmann's sea''}.
``{\sl We want to prove the second theorem in mechanical terms,
founding it on the nature of the interaction laws and without
imposing any restriction on the initial conditions, knowledge of
which is not supposed}'',\,\Cite{Bo877a}\,.

12) {\it Discrete phase space, count of its points and entropy},
The relation between the theorems on the mechanical theory of
heat and on the probability theory of equipartition and
problem of determining the “{\sl ratios of the number of
different states of which we want to compute the
probabilities}”,\,\Cite{Bo877-b}\,.

13) {\it Monocyclic and orthodic systems. Ensembles}: ``{\sl The most complete
proof of the second main theorem is manifestly based on the
remark that, for each given mechanical system, equations that are
analogous to equations of the theory of heat hold}'',\,\Cite{Bo884}\,.

14) Maxwell 1866: ``{\sl Indeed the properties of a body supposed to be
a uniform plenum may be affirmed dogmatically, but cannot be
explained mathematically}'',\,\Cite{Ma867-b}\,.

\vfill\eject
{\bf Chapter 7 (Appendices)}

\parindent=0mm

A) A simple example of the mechanical interpretation of heat theorem

B) Comments on Boltzmann's statement: ``aperiodic motions as
periodic with infinite period''.

C) The heat theorem without details on the dynamics: Comments on
the first paper in which the theory of the canonical ensemble
appears. The discreetness assumption about the microscopic states
is for the first time not only made very explicit but it is used
to obtain in a completely new way that the equilibrium
distribution is equivalently a canonical or a microcanonical one.

D) A further mechanical example realizing the heat theorem in the
frame of Keplerian motion.

E) Gauss' least constraint principle (as applied in thermostat
models in recent literature).

F) Non smoothness of stable/unstable manifolds: A qualitative
argument to understand why, even in analytic Anosov. systems they
may be non smooth

G) Example of construction of Markov partitions (in general 2
dimensional Anosov systems).

H) Axiom C: definition.

I) Lyapunov's pairing theory: Pairing theory in mechanical systems.

J) Fluid equations (pairing and ergodicity). Apparent pairing and
equivalence of exponents in corresponding reversible and
irreversible flows. 

K) Fluctuation theorem and viscosity. Main question: is it
meaningful to ask whether the fluctuation relation holds in
irreversible evolutions ? Then discussion about pairing of
Lyapunov exponents, FT, and equivalence RNS and NS.

L) Reversibilty and friction: Considers a regularization, which
studies the regularized NS equations with cut-off $R^{\frac34}$,
toghether with the corresponding reversible equation. Model is
inspired by the OK41 turbukence theory.

M) Reciprocity and fluctation theorem: it is shown that Onsager's
reciprocity can be also obtained, and extended, in reversible
systems verifying CH by applying the fluctuation patterns theorem.

N) Large deviations in SRB states: illustrates how a proof of the
theorem is reduced (assuming Perron-Frobenius' theorem extension
of Ruelle) to a few bounds on quantities familiar in the theory
of statistical mechanics and Markov processes.

O) An exact formula: An immediate consequence of the fluctuation
theorem.

P) Transient FT: “transient fluctuation theorem”. It is extremely
general and does not depend on any chaoticity assumption. Just
reversibility and time reversal symmetry and the evolution of an
initial distribution  which is invariant under time reversal. It says
nothing about the SRB distribution (which is singular with
respect to the Liouville distribution).

\end{document}